\documentclass[12pt]{article}
\usepackage{latexsym}
\usepackage{amssymb}
\usepackage{amsmath}
\usepackage{amsfonts}
\usepackage{tikz-cd}
\usepackage{tensor}
\usepackage{mathrsfs}
\usepackage[nottoc]{tocbibind}
\usepackage{tensor}
\usepackage{slashed}
\usepackage{indentfirst}
\usepackage{mathtools}
\usepackage{physics}
\usepackage{tikz}
\usepackage{etoolbox}
\usepackage{bbm}
\usepackage{tabularx}
\usepackage{multirow}
\usepackage{cancel}
\usepackage{bm}
\usepackage{graphicx}
\graphicspath{ {./images/} }
\usepackage[scr=boondox]{mathalpha}
\usepackage[scr]{rsfso}
\usepackage{appendix}
\usepackage{circledsteps}

\numberwithin{equation}{section}

\allowdisplaybreaks
\tikzset{
  curarrow/.style={
  rounded corners=8pt,
  execute at begin to={every node/.style={fill=red}},
    to path={-- ([xshift=-50pt]\tikztostart.center)
    |- (#1) node[fill=white] {$\scriptstyle d_*$}
    -| ([xshift=50pt]\tikztotarget.center)
    -- (\tikztotarget)}
    }}

\makeatletter
\newcommand{\foo@newbla}
    {$\hspace{1mm}\begin{tikzpicture}[scale=3]
        \filldraw[fill=black!, draw=black!50!black]  (0mm,0mm) -- (0.4mm,0.8mm) --(-0.4mm,0.8mm) -- cycle;
    \end{tikzpicture}$}
\newcommand{\newbla}[1]
    {\csname foo@#1\endcsname}
\makeatother

\makeatletter
\newcommand{\foo@newblatilde}
    {\begin{tikzpicture}[scale=3]
        $\tilde{\filldraw[fill=black!, draw=black!50!black]
        (0mm,-0.25mm) -- (0.4mm,0.5mm) --(-0.4mm,0.5mm) -- cycle;}$
    \end{tikzpicture}}
\newcommand{\newblatilde}[1]
    {\csname foo@#1\endcsname}
\makeatother

\newcommand{\longvertspacing}{~\\~\\~\\~\\~\\~\\}
\newcommand{\mylinespacing}{\vspace{1em}}

\newcommand{\mytitle}{On The General Projective Theory\\ of Matter and Gravitation}
\newcommand{\myname}{Michael J. Connolly}
\newcommand{\mydegree}{\capitalize{Doctor of Philosophy}}
\newcommand{\myprogram}{\capitalize{Physics}}
\newcommand{\mymonth}{August}
\newcommand{\myyear}{2025}
\newcommand{\myadvisor}{Thesis Supervisor}
\newcommand{\committeememberA}{Vincent G.J. Rodgers}
\newcommand{\committeememberB}{Wayne N. Polyzou}
\newcommand{\committeememberC}{Palle Jorgensen}
\newcommand{\committeememberD}{Yannick Meurice}

\usepackage{geometry}
\usepackage[utf8]{inputenc}

\usepackage{varioref}
\usepackage[linktocpage=true]{hyperref}
\usepackage[capitalise,noabbrev]{cleveref}

\usepackage{tocloft}

\usepackage[nottoc]{tocbibind}

\usepackage{graphicx}
\graphicspath{{figures/}}
\usepackage{boxhandler}

\fontsize{12}{1} \selectfont
\usepackage{setspace}
\usepackage[indentafter]{titlesec}

\titleformat{\section}[block]{\normalfont\large\bfseries\centering}{\thesection}{1em}{\MakeUppercase}

\titleformat{\subsection}[block]{\normalfont\normalsize\bfseries\centering}{\thesubsection}{1em}{}
\titleformat{\subsubsection}[block]{\normalfont\small\bfseries\centering}{\thesubsubsection}{1em}{}

\titlespacing*{\section}{0pt}{36pt}{18pt}
\titlespacing*{\subsection}{0pt}{24pt}{12pt}
\titlespacing*{\subsubsection}{0pt}{12pt}{6pt}

\newenvironment{tightcenter}{%
  \setlength\topsep{0pt}
  \setlength\parskip{0pt}
  \begin{center}
}{%
  \end{center}
}
\renewenvironment{quote}{%
   \list{}{%
     \leftmargin0cm   
     \rightmargin\leftmargin
   }
   \item\relax
}
{\endlist}

\usepackage{ifxetex}

\usepackage{xparse}

\ExplSyntaxOn
\NewDocumentCommand{\capitalize}{>{\SplitList{~}}m}
 {
  \seq_clear:N \l_capitalize_words_seq
  \ProcessList{#1}{\CapitalizeFirst}
  \seq_use:Nn \l_capitalize_words_seq { ~ }
 }
\NewDocumentCommand{\CapitalizeFirst}{m}
 {
  \capitalize_word:n { #1 }
 }

\sys_if_engine_pdftex:TF
 {
  \cs_set_eq:Nc \capitalize_tl_set:Nn { protected@edef }
 }
 {
  \cs_set_eq:NN \capitalize_tl_set:Nn \tl_set:Nn
 }

\cs_new_protected:Nn \capitalize_word:n
 {
  \capitalize_tl_set:Nn \l_capitalize_word_tl { #1 }
  \seq_if_in:NfTF \g_capitalize_exceptions_seq { \tl_to_str:n { #1 } }
   { \seq_put_right:Nn \l_capitalize_words_seq { #1 } } 
   { \seq_put_right:Nx \l_capitalize_words_seq { \tl_mixed_case:V \l_capitalize_word_tl } }
 }
\cs_generate_variant:Nn \tl_mixed_case:n { V }
\NewDocumentCommand{\AppendToList}{m}
 {
  \clist_map_inline:nn { #1 }
   {
    \seq_gput_right:Nx \g_capitalize_exceptions_seq { \tl_to_str:n { ##1 } }
   }
 }
\cs_generate_variant:Nn \seq_if_in:NnTF { Nf }
\seq_new:N \l_capitalize_words_seq
\seq_new:N \g_capitalize_exceptions_seq
\ExplSyntaxOff

\AppendToList{a,is,of,óf}

\title{\MakeUppercase{\normalsize\mytitle}\vspace{-2em}}
\author{\vspace{-5ex}}
\date{\vspace{-5ex}}

\begin{document}
\newcommand{\change}{\mbox{\textcent}}
\newcommand{\blcirc}{\tikz\draw[black,fill=black](-0.08,0) circle (.3ex);}

\maketitle

\pagenumbering{roman}
\thispagestyle{empty}
\longvertspacing
\begin{tightcenter}
    by
\end{tightcenter}
\mylinespacing
\begin{tightcenter}
\myname
\end{tightcenter}
\longvertspacing
\begin{tightcenter}
A thesis submitted in partial fulfillment\\
of the requirements for the\\
\mydegree~\\
degree in \myprogram~in the\\
Graduate College of \\
The University of Iowa
\end{tightcenter}
\mylinespacing\mylinespacing
\begin{tightcenter}
\mymonth~\myyear
\end{tightcenter}
\mylinespacing\mylinespacing\mylinespacing\mylinespacing
\renewcommand{\arraystretch}{1.25}
\hspace{2cm}
\begin{tabular}{rl}
Thesis Committee:&\committeememberA, \myadvisor\\
&\committeememberB\\
&\committeememberC\\
&\committeememberD
\end{tabular}


\pagenumbering{gobble} 
\vspace*{\fill} 
\begin{quote} 
\centering 
Copyright by\\ 
\mylinespacing
\myname \\
\mylinespacing
\myyear \\
\mylinespacing
All Rights Reserved \\
\mylinespacing
\end{quote}
\vspace*{\fill}

\pagenumbering{roman} 
\setcounter{page}{2}

\vspace*{\fill} 
\begin{quote}
To those I've lost along The Way.
\centering
\mylinespacing
\end{quote}
\vspace*{\fill}

\vspace*{\fill} 
\begin{quote}
\centering
\textit{\textit{“You hear the sound of water...\\ and that's quite as important as anything I've got to say.”}}
\mylinespacing\\
-Alan Watts

\mylinespacing
\mylinespacing
\end{quote}
\vspace*{\fill}

\begin{tightcenter}
\textbf{ABSTRACT}
\mylinespacing
\mylinespacing
\end{tightcenter}
\begin{doublespace}

We develop a generalized projective gauge theory of gravity and spinorial matter, incorporating both non-metricity and torsion. The work is divided into three parts. 

Part I provides a thorough review of affine geometry, decomposing arbitrary affine connections into Levi-Civita, disformation, and contorsion. We discuss curvature tensors, their contractions, and the role of projective and symmetric projective transformations. A review of the Einstein-Hilbert action and its field equations sets the stage for exploring projective deformations. Common modifications—matter coupling and a cosmological constant—are introduced, then recast in differential-form language (Palatini formalism). We also review the Metric-Affine gauge theory of gravity and its M\"{o}bius representation, along with topological terms (Euler, Pontrjagin, Nieh-Yan) and Bianchi identities. Motivations for a projective approach include the string theoretic derivation of the Diffeomorphism gauge potential, its role in analogy to the electromagnetic gauge potential, and the higher-dimensional volume bundle that unifies projective transformations and reparameterizations. The Thomas-Whitehead (TW) model is reviewed, showing how its projective Schouten tensor (Diffeomorphism field adjacent) can yield a Higgs-like potential that induces a cosmological constant and mass.

Part II constructs projective space from the affine tangent bundle, identifies its group of transformations (Projective General Linear Group), and derives the projective algebra under Lorentz decomposition, showing a group contraction results in a pseudo-affine group. We extend the volume bundle to incorporate translations of the contact point between spacetime and projective tangent spaces, and introduce a factoring-map whose minimal condition for a projective structure generates a new scalar field. The newly found projective symmetric teleparallel (PST) connections are then defined, yielding torsion-free, flat, projectively invariant connections. Generalizing to non-inertial frames, and using nonlinear gauge symmetry realizations to implement local Lorentz symmetry, we construct a general geometric framework that unifies TW and Metric-Affine gravity with projectively invariant spacetime torsion. We introduce projective generalized Higgs fields and show how certain gauge choices reduce these to fundamental projective fields, and how they may be used to define fundamental geometric objects such as the projective $2$-frame and spacetime connection. A Lovelock-inspired action is shown to support only curvature (metric) dynamics, implying a topologically constrained Schouten field via the Pontrjagin density. The projective Pontrjagin density is shown to contain a new topological invariant, not present in the literature. We analyze the solution space, revealing possible nontrivial projective torsion vector modes and degenerate co-frames, while recovering an (A)-dS description of spacetime, with bare cosmological constant modified by a rigid Schouten field. 

Part III formulates projective spinors by defining gamma matrices via the projective linear group metric, then employing nonlinear gauge symmetry realizations to ensure local Lorentz covariance. A general spinor metric introduces a complex phase of redundancy. Requiring a real, Hermitian action leads to a self-adjoint operator that eliminates any coupling to non-metricity, leaving torsion as the sole gravitational interaction with spinors—reducing effectively to an Einstein-Cartan-type theory. An induced chiral mass emerges without the CP violation plaguing TW theory, potentially aiding neutrino mass models. Projective spinor currents and chiral currents are briefly investigated, and the foundations of future work are built to explore the full coupling of projective gravity and matter, and a projective description of the chiral anomaly.

\mylinespacing
\mylinespacing
\end{doublespace}

\begin{doublespace}
\begin{tightcenter}
\textbf{PUBLIC ABSTRACT}
\mylinespacing
\mylinespacing
\end{tightcenter}

We propose a novel approach to gravity and spinning particles (spinors) that allows additional ``twisting” and ``stretching” effects in spacetime, beyond the curving that is captured by Einstein’s theory alone. These extra geometric features—beyond simple curvature—might help us solve some long-standing puzzles in physics, such as the cause of dark energy or how certain particles gain their mass. In particular, we include two types of geometric deformations—often called torsion (twisting spacetime) and non-metricity (changes in how spacetime distances are measured)—and treat them as part of a unified framework tied to projective transformations. These transformations are simply different descriptions of material motions that are all equivalent. The theory is presented in three parts.

Part I lays out essential background. We explain how spacetime geometry can be split into familiar features from Einstein’s work and pieces that capture the extra twisting and stretching. We show how standard results, such as Einstein’s field equations, emerge and then extend them to include dark energy and matter. We also introduce standard topological, or shape-preserving terms that may have interesting physical consequences yet do not affect everyday observations. We then reconstruct all of this in a format more suitable for the ``gauge," or locally redundant treatment of gravity. The purpose of this is to uncover how gravity fits in to our ``Standard Model" of particle physics.

Part II develops the general ``projective gauge” viewpoint in detail. Here, we treat the geometry of spacetime as living in a higher-dimensional setting, allowing us to interpret certain fields—linked to material path reparameterizations—as fundamental components. This leads to a richer picture of gravity, in which this new field acquires mass much like a Higgs field, produces a cosmological constant (dark energy), and leads to additional possibilities for how gravitational effects propagate throughout spacetime.

Part III constructs and incorporates material fields, or spinning matter fields (spinors). We discover that the twisting deformation of the geometry can directly influence material spin, while the stretching part—despite other attempts—has no material influence whatsoever. Interestingly, a chiral mass arises that can be used to provide neutrinos mass, linking small neutrino masses to higher-scale physics within this unified projective geometric framework.

Altogether, this thesis shows how an expanded picture of spacetime—with new geometric fields and symmetries—can both reproduce the successes of Einstein’s gravity and offer novel insights into dark energy, mass generation, and the behavior of spinning particles. This lays the groundwork for future studies of how these new geometric features might shape cosmology, high-energy physics, and possibly the nature of existence.

\end{doublespace}

\setcounter{tocdepth}{3} 

\begin{spacing}{0}
\tableofcontents
\addtocontents{toc}{\vspace{1.2em}}
\end{spacing}

\newpage

\phantomsection
\addcontentsline{toc}{section}{\hspace{-1.5em}List of Tables}

\begin{spacing}{1}
\begingroup
\renewcommand{\addcontentsline}[3]{}
\listoftables
\endgroup
\addtocontents{lot}{\vspace{1.2em}}
\end{spacing}




\vfill
\pagebreak

\pagenumbering{arabic}
\begin{onehalfspacing}


\pagebreak
\thispagestyle{empty}

\phantomsection
\addcontentsline{toc}{section}{\hspace{-1.5em}\textbf{PART I: MOTIVATIONS}}
\begingroup
\renewcommand{\addcontentsline}[3]{}
\vspace*{5cm}
\noindent
\makebox[\textwidth]{\Huge \textbf{PART I:}}\vspace{+3em}
\makebox[\textwidth]{\Huge \textbf{MOTIVATIONS}}
\vfill
\label{part-I}
\endgroup

\pagebreak

\section{Spacetime \& Geometry}

We begin with a review of the mathematical structures necessary to describe gravitational phenomena, following closely \cite{nakahara}. Beginning with a material path, or a curve through spacetime, we develop a notion of vectors, parallel transport, and the associated connection. From there, differential forms are introduced, along with a notion of the inner product via the pseudo-Riemannian metric. We then briefly develop arbitrary rank and weight tensor densities. 

With the basic underlying mathematical objects constructed, we transition to a decomposition of arbitrary affine connections into their Levi-Civita and non-Levi-Civita parts and discuss some properties. A curvature tensor is introduced, and its various traces outlined. The connection decomposition is reflected in its curvature tensor, and its various traces reviewed. This section closes with a short overview of projective transformations of the connection and the resulting transformation of the curvature.

\subsection{Fundamentals}
\label{sec:intro-fundamentals}

Consider an $m$-dimensional spacetime manifold $\mathcal{M}$ with coordinates $\{x^m\}$. A map $c(\tau):(a,b)\rightarrow\mathcal{M}$, parameterized by $\tau\in(a,b)$ on an open interval, defines a parameterized curve in $\mathcal{M}$. Let $f:\mathcal{M}\rightarrow\mathbb{R}$ be a smooth, invertible function. The directional derivative of $f(c(\tau))$ at $\tau=0$,
\begin{equation}
    \left.\frac{d(f(\tau))}{d\tau}\right|_{\tau=0},
\end{equation}
defines a vector at the point $c(0)\in\mathcal{M}$. If two curves $c_1(\tau)$ and $c_2(\tau)$ exist such that $c_1(0)=c_2(0)$, giving the same point $p_{\mathcal{M}}$ of $\mathcal{M}$, and their directional derivatives produce the same vector, then $c_1(\tau)$ is equivalent to $c_2(\tau)$, i.e., $c_1(\tau)\sim c_2(\tau)$. A tangent vector is then defined by this equivalence class of curves, $[c(\tau)]$. The collection of equivalence classes of curves at $p_{\mathcal{M}}$, i.e., the set of all tangent vectors at that point, forms an $m$-dimensional vector space $T_p\mathcal{M}$: the tangent space of $\mathcal{M}$ at the point $p_{\mathcal{M}}$. A basis for $T_p\mathcal{M}$ is provided by the set of holonomic vectors $\{\partial_m\}$, with
\begin{equation}
    \partial_m:=\frac{\partial}{\partial x^m}.
\end{equation}

Let the vector $X=X^m\partial_m:=\frac{dx^m}{d\tau}\partial_m$ denote an element of $T_p\mathcal{M}$. Then, the directional derivative of $f(c(\tau))$ at $\tau=0$ may be expressed as
\begin{equation}
    \left.\frac{d(f(\tau))}{d\tau}\right|_{\tau=0}=X^m\partial_mf=:X[f].
\end{equation}
Applying the differential operator $X=X^m\partial_m$ directly to the coordinate functions $\{x^m\}$ provides a \textit{velocity vector},
\begin{equation}
    V^m:=X[x^m]=\left.\frac{dx^m}{d\tau}\right|_{\tau=0}.
\end{equation}

The operator $X$ is formally independent of the coordinate system. In other words, $X=X'$. For this to be so, a change of coordinates $\{x^m\}\rightarrow \{x^{n'}\}$ must act on the vector components in a manner opposite to that of the basis vector transformation $\{\partial_m\}\rightarrow\{\partial_{n'}\}$. Using the technology described above, we therefore have
\begin{equation}
    X^{m'}=\frac{\partial x^{m'}}{\partial x^n}X^n,\quad\quad \partial_{m'}=\frac{\partial x^n}{\partial x^{m'}}\partial_n.
\end{equation}

A non-holonomic basis may also be used. The non-holonomic basis $\{e_m\}$ is related to $\{\partial_n\}$ via
\begin{equation}
    e_{m}=e^n{}_{m}\partial_n,
\end{equation}
where $e^n{}_m\in GL(m,\mathbb{R})$ is an element of the $m$-dimensional general linear group, i.e., the group of $m\times m$ invertible matrices. In this basis, the differential operator $X$ is expressed as
\begin{equation}
    X=X^m e_m.
\end{equation}
Acting on $X$ in the non-holonomic basis with the velocity operator gives rise to the directional \textit{covariant derivative}:
\begin{equation}\label{parallel transport-intro}
\begin{split}
    V[X]&=V^m\partial_m(X^ne_n)\\
    &=V^m(\partial_mX^n+\Gamma^n{}_{lm}X^l)e_n,
    \end{split}
\end{equation}
where
\begin{equation}\label{connection-gen-def}
    \Gamma^n{}_{lm}:=(e^{-1})^n{}_k\partial_me^k{}_l
\end{equation}
goes by many names: the \textit{Christoffel symbols}, \textit{connection coefficients}, \textit{affine connection}, or simply, the \textit{connection}. Eq. \eqref{parallel transport-intro} is known as \textit{parallel transport} since it is a measure of the change in the components of $X$ along $V$. We then define the covariant derivative with respect to the affine connection as
\begin{equation}\label{cov-der-def-general}
    \nabla_mX^n:=\partial_mX^n+\Gamma^n{}_{lm}X^l.
\end{equation}
The covariant derivative satisfies a few key properties. For example, taking vectors $X,Y$ and the scalar function $f$: 
\begin{equation}
    \nabla_l(X^m+Y^m)=\nabla_lX^m+\nabla_lY^m,
\end{equation}
\begin{equation}\label{product-rule-gen}
    \nabla_l(fX^mY^n)=(\nabla_lf)X^mY^n+f(\nabla_lX^m)Y^n+fX^m(\nabla_lY^n).
\end{equation}

Taking $X=V$ in the equation of parallel transport, Eq. \eqref{parallel transport-intro}, the \textit{geodesic equation} arises when the directional covariant derivative of the tangent vector $X^m=\frac{dx^m}{d\tau}$ along the curve $c(\tau)$ is proportional to itself:
\begin{equation}
    X^m\nabla_mX^n=f(\tau)X^n,
\end{equation}
for some function $f(\tau)$. Furthermore, the \textit{geodetic equation} is used to refer to the set of vanishing geodesic equations:
\begin{equation}\label{geodesic-intro}
    X^m\nabla_mX^n=0.
\end{equation}

Covariance of these expressions refers to their behavior under transformations of coordinates $\{x^m\}\rightarrow \{x^{n'}\}$. For example, covariance of Eq. \eqref{geodesic-intro} requires
\begin{equation}
     X^m\nabla_mX^n\rightarrow  \frac{\partial x^n}{\partial x^{n'}}\left(X^{m'}\nabla_{m'}X^{n'}\right).
\end{equation}
For this expression to hold, the connection $\Gamma$ must behave in a particular way under transformations of coordinates. The transformation behavior of the connection coefficients may be easily found by considering $\Gamma$ in the new system of coordinates $\{x^{m'}\}$, and relating its form to the initial set:
\begin{equation}\label{gamma-conn-trans}
\begin{split}
    \Gamma^{l'}{}_{m'n'}&=(e^{-1})^{l'}{}_a\partial_{n'}e^a{}_{m'}\\
    &=\frac{\partial x^{l'}}{\partial x^p}(e^{-1})^p{}_a\frac{\partial }{\partial x^{n'}}\left(\frac{\partial x^r}{\partial x^{m'}}e^a{}_r\right)\\
    &=\frac{\partial x^{l'}}{\partial x^p}(e^{-1})^p{}_a\left(\frac{\partial x^r}{\partial x^{m'}}\frac{\partial e^a{}_r}{\partial x^{n'}}+\frac{\partial^2 x^r}{\partial x^{n'}\partial x^{m'}}e^a{}_r\right)\\
    &=\frac{\partial x^{l'}}{\partial x^p}\frac{\partial x^r}{\partial x^{m'}}\frac{\partial x^q}{\partial x^{n'}}(e^{-1})^p{}_a\frac{\partial e^a{}_r}{\partial x^q}+\frac{\partial x^{l'}}{\partial x^r}\frac{\partial^2 x^r}{\partial x^{n'}\partial x^{m'}}\\
    &=\frac{\partial x^{l'}}{\partial x^p}\frac{\partial x^r}{\partial x^{m'}}\frac{\partial x^q}{\partial x^{n'}}\Gamma^p{}_{rq}+\frac{\partial x^{l'}}{\partial x^p}\frac{\partial^2 x^p}{\partial x^{n'}\partial x^{m'}}.
\end{split}
\end{equation}
An $m$-dimensional manifold $\mathcal{M}$, equipped with an affine connection $\Gamma^l{}_{nm}$, will be called an $L_m$.

As a vector space, $T_p\mathcal{M}$ admits a dual vector space, the \textit{co-tangent space} $T^*_p\mathcal{M}$, whose elements are linear maps from $T_p\mathcal{M}$ to $\mathbb{R}$. These linear maps $\omega:T_p\mathcal{M}\rightarrow\mathbb{R}$, which belong to $T^*_p\mathcal{M}$, are called \textit{dual vectors}, \textit{co-tangent vectors}, or \textit{$1$-forms}. For example, the differential $df$ of a function is a $1$-form. Recall the action of a vector $V$ on $f$:
\begin{equation}
    V[f]=V^m\frac{\partial f}{\partial x^m}\in\mathbb{R}.
\end{equation}
Therefore, the action of $df\in T^*_p\mathcal{M}$ on $V\in T_p\mathcal{M}$ is defined as
\begin{equation}\label{pre-inner-product-definition}
    \langle df, V\rangle:=V[f]=V^m\frac{\partial f}{\partial x^m}.
\end{equation}
Expressing $df$ in terms of the coordinates,
\begin{equation}\label{chain-rule}
    df=\frac{\partial f}{\partial x^m}dx^m,
\end{equation}
one finds the holonomic basis $\{dx^m\}$ for $T^*_p\mathcal{M}$. Using the definition above, one may see this basis as dual to the holonomic basis of $T_p\mathcal{M}$, since
\begin{equation}\label{basis-inner-product}
    \langle dx^m,\partial_n\rangle=\partial_n[x^m]=\frac{\partial x^m}{\partial x^n}=\delta^m{}_n.
\end{equation}
The non-holonomic basis of $T^*_p\mathcal{M}$ is easily found by requiring that its action on the non-holonomic basis of $T_p\mathcal{M}$ returns the identity. Letting $\{(e^{-1})^m\}$ denote such a non-holonomic basis of $T^*_p\mathcal{M}$, 
\begin{equation}
    (e^{-1})^m=(e^{-1})^m{}_ndx^n,
\end{equation}
and using Eq. \eqref{basis-inner-product} above, we obtain
\begin{equation}
    \langle(e^{-1})^k,e_l\rangle =(e^{-1})^k{}_ne^m{}_l\frac{\partial x^n}{\partial x^m}=\delta^k{}_l.
\end{equation}

An arbitrary $1$-form $\omega\in T^*_p\mathcal{M}$ is expressed with respect to the holonomic basis as
\begin{equation}
    \omega=\omega_mdx^m,
\end{equation}
where $\omega_m$ are the components of $\omega$ in the basis defined by $\{dx^m\}$. Using again, the definition in Eq. \ref{pre-inner-product-definition}, the \textit{inner product} $\langle\cdot,\cdot\rangle:T^*_p\mathcal{M}\times T_p\mathcal{M}\rightarrow\mathbb{R}$ between an arbitrary vector and $1$-form may be found:
\begin{equation}
    \langle\omega,V\rangle= \omega_mV^n\langle dx^m,\partial_n\rangle=\omega_mV^n\delta^m{}_n=\omega_mV^m.
\end{equation}
This inner product does not require any additional structure to be imposed on the system.

An inner product $g_p$ between two vectors $X,Y$ may be introduced as a map $T_p\mathcal{M}\otimes T_p\mathcal{M}\rightarrow \mathbb{R}$, denoted $g_p(X,Y)$, which is required to be symmetric, $g_p(X,Y)=g_p(Y,X)$, and negative (positive) semi-definite. The latter statement is signature-dependent (determined by the sign of its eigenvalues), and in this document, we choose to work with the mostly-minus convention. Mathematically,
\begin{equation}
    g_p(X,Y)=0\quad\forall \; X\in T_p\mathcal{M}\quad\Rightarrow \quad Y=0.
\end{equation}
When these requirements are satisfied, $g_p$ is a \textit{pseudo-Riemannian metric}, or simply, a \textit{metric}.

Conveniently, we may reorganize this metric inner product by considering $g_p(X,\;\cdot\;):T_p\mathcal{M}\rightarrow\mathbb{R}$ as a linear map, defined to take vectors $Y\in T_p\mathcal{M}$ to $Y\mapsto g_p(X,Y)$. Then, $\omega_X:=g_p(X,\;\cdot\;)\in T^*_p\mathcal{M}$ may be identified as a $1$-form. Similarly, one may consider the reverse process, where the $1$-form $\omega\in T^*_p\mathcal{M}$ gives rise to a vector $X_{\omega}\in T_p\mathcal{M}$ through
\begin{equation}
    \langle\omega,Y\rangle=g_p(X_{\omega},Y).
\end{equation}
These statements manifest an isomorphism between $T_p\mathcal{M}$ and $T^*_p\mathcal{M}$. Expressed in the holonomic basis of $T^*_p\mathcal{M}\otimes T^*_p\mathcal{M}$, we have
\begin{equation}
    g_p=g_{mn}(p_{\mathcal{M}})dx^m\otimes dx^n.
\end{equation}
The components of the metric may be used to map $1$-forms to vectors and vectors to $1$-forms:
\begin{equation}
    \omega_m=g_{mn}X^n,\quad\quad X^n=g^{nm}\omega_m,
\end{equation}
where $g_{mn}g^{nk}=\delta^{k}{}_m$. The above statement is exemplary of the isomorphism between $T_p\mathcal{M}$ and $T^*_p\mathcal{M}$. 

When the $L_m$ is given a metric structure, i.e., when it is endowed with a metric tensor $g_{mn}$, the $L_m$ is called a $\textit{Metric-Affine}$ geometry, denoted $(L_m,g)$. Much of this document assumes the existence of such a metric tensor, and therefore, unless otherwise stated, we will simply use $L_m$ to refer to the Metric-Affine geometry. As will be seen later, when particular relationships are imposed between the metric and connection, distinct phases emerge, to which the $L_m$ reduces. The most restrictive case is Minkowski space, $M_m$.

One may naturally extend the previous construction to arbitrary rank vectors and forms. We define a tensor of type-$(i,j)$ as a multi-linear object mapping $i$ elements of $T^*_p\mathcal{M}$, and $j$ elements of $T_p\mathcal{M}$ to $\mathbb{R}$. Denoting this set of $(i,j)$-objects as $\mathcal{T}^i{}_{j,p}(\mathcal{M})$, an element may be expressed in the coordinate basis as
\begin{equation}
    t=t^{m_1\dots m_i}{}_{n_1\dots n_j}\frac{\partial}{\partial x^{m_1}}\dots\frac{\partial}{\partial x^{m_i}}dx^{n_1}\dots dx^{n_j}.
\end{equation}
In this form, it is clear $t$ is a multi-linear function to $\mathbb{R}$, from 
\begin{equation}
    \otimes^i T^*_p\mathcal{M}\otimes^j T_p\mathcal{M}.
\end{equation}
From here, it is obvious that $1$-forms are tensors of type-$(0,1)$ and vectors are tensors of type-$(1,0)$. The action of $t$ on a collection of $i$ $1$-forms $\{\omega_{(i)}\}$ and $j$ vectors $\{X_{(j)}\}$ clearly produces a real number, since the natural inner product provides
\begin{equation}
t(\omega_{(1)},\cdots,\omega_{(i)};X_{(1)},\cdots,X_{(j)})=t^{m_1\dots m_i}{}_{n_1\dots n_j}\omega_{(1)m_1}\cdots\omega_{(i)m_i}X^{n_1}_{(1)}\cdots X^{n_j}_{(j)}.
\end{equation}

Of particular importance in this document will be the completely antisymmetric tensors of type-$(0,j)$. These are differential forms of order $j$, or $j$-forms. Define the completely antisymmetric (\textit{wedge}) product $\wedge$ of $j$ $1$-forms as
\begin{equation}
    dx^{m_1}\wedge\dots\wedge dx^{m_j}:=\sum_{P\in S_j}\text{sgn}(P) dx^{m_{P(1)}}\otimes\dots\otimes dx^{m_{P(j)}},
\end{equation}
where $S_j$ denotes the symmetric group of order $j$, $P$ a permutation, and $\text{sgn}(P)$ the sign of the permutation. For example, a $2$-form may be expressed as 
\begin{equation}
    dx^m\wedge dx^n=dx^m\otimes dx^n-dx^n\otimes dx^m.
\end{equation}
If the dimension of the space is $m$, and $j>m$, the result vanishes. An arbitrary $j$-form $\omega$ may thus be expressed as
\begin{equation}
    \omega=\frac{1}{j!} \omega_{m_1\dots m_j}dx^{m_1}\wedge\dots\wedge dx^{m_j},
\end{equation}
where the combinatorial factor accounts for the over-counting in permutations. Due to the antisymmetry of the wedge product, we have $\omega\wedge\omega=0$. Moreover, for any $k$-form $\nu$, we may construct the $k+j$-form
\begin{equation}
    \omega\wedge\nu=(-1)^{jk}\nu\wedge\omega.
\end{equation}

When $\omega$ is taken as the $m$-form
\begin{equation}
    \omega=\omega(p_{\mathcal{M}})dx^1\wedge\dots\wedge dx^m,
\end{equation}
we find that under a transformation of coordinates $\{x^n\}\rightarrow \{x^{n'}\}$, 
\begin{equation}
    \omega\rightarrow \omega'=\omega(p)\frac{\partial x^1}{\partial x^{n_1'}}dx^{n_1'}\wedge\dots\wedge\frac{\partial x^m}{\partial x^{n_m'}}dx^{n_m'}=\omega(p)\left|\frac{\partial x^n}{\partial x^{m'}}\right|dx^{1'}\wedge\dots\wedge dx^{m'},
\end{equation}
where $|A|:=\text{det}(A)$. This object may be understood as a \textit{tensor density} of weight $w=-1$. Since this document is primarily concerned with an extension of tensors to the more general \textit{tensor densities}, we now develop these further. Following \cite{tensor-density}, for $w\in\mathbb{R}$, a $w$-density on, for example, any $m$-dimensional vector space $E$ is a map
\begin{equation}
s:\otimes^mE\rightarrow \mathbb{R},
\end{equation}
with the property that for any linear map $f:E\rightarrow E$, and any set of $m$ vectors $\{X_{(m)}\}\in E$, 
\begin{equation}
    s(fX_{(1)},\dots,fX_{(m)})=|\det(f)|^ws(X_{(1)},\cdots,X_{(m)}).
\end{equation}
This holds just as well if $E$ is replaced by its dual vector space $E^{*}$, and the set of vectors is replaced with a set of $1$-forms. Thus, a tensor density $T$ of rank-$(i,j)^w$ may generally be expressed as a multi-linear map to $\mathbb{R}$ from
\begin{equation}
    \otimes^i E^*\otimes^j E\otimes^mE^{(*)}.
\end{equation}
Taking $E$ and $E^*$ to be $T_p\mathcal{M}$ and $T^*_p\mathcal{M}$, respectively, we may express a rank-$(i,j)^w$ tensor density $T$ in the coordinate basis as
\begin{equation}
    T=t^{m_1\dots m_i}{}_{n_1\dots n_j}\frac{\partial}{\partial x^{m_1}}\dots\frac{\partial}{\partial x^{m_i}}dx^{n_1}\dots dx^{n_j}\left|dx^1\wedge\dots\wedge dx^m\right|^w.
\end{equation}

We now move on to discussing the various geometric deformations associated with a general $(L_m,g)$.\\

\subsection{Connection Decomposition}

The covariant differentiation of arbitrary rank-$(i,j)^w$ tensor densities of weight $w$ may be easily found from Eq. \eqref{cov-der-def-general} by requiring that the resulting expression transform covariantly. The nonzero weight $w$ introduces a factor involving the ``internal" natural trace of the connection:
\begin{equation}
\begin{split}
    \nabla_mV^{p_1\dots p_i}_{\quad q_1\dots q_j}&=\partial_mV^{p_1\dots p_i}_{\quad q_1\dots q_j}-w\Gamma^k{}_{km}V^{p_1\dots p_i}_{\quad q_1\dots q_j}\\
    &\quad+\sum_{n=1}^{i}\Gamma^{p_n}{}_{km}V^{p_1\dots p_{n-1}kp_{n+1}\dots p_i}_{\quad q_1\dots q_j}-\sum_{n=1}^{i}\Gamma^{k}{}_{q_nm}V^{p_1\dots p_i}_{\quad q_1\dots q_{n-1}kq_{n+1}\dots q_j}.
    \end{split}
\end{equation}
For example, a rank-$(0,0)^w$ scalar density satisfies
\begin{equation}
    \nabla_mV=\partial_mV-w\Gamma^k{}_{km}V,
\end{equation}
which reduces to the ordinary partial derivative when $w=0$. The term \textit{natural trace} here refers to any trace that can be taken without the use of a metric. The ``internal" qualifier becomes evident when, for example, the connection is expressed as a matrix-valued $1$-form. This distinction will become clearer once the exterior formalism is developed. We therefore refer to $\Gamma^k{}_{mk}$ as the \textit{external natural} trace and $\Gamma^k{}_{mn}g^{mn}$ as the \textit{unnatural} trace.

In a general $L_m$, the connection coefficients may be uniquely decomposed into three distinct geometric objects \cite{Hehl}. To achieve this, we first define the \textit{non-metricity} tensor as
\begin{equation}\label{non-metricity-def-gen}
    Q_{lmn}:=-\nabla_lg_{mn}.
\end{equation}
Unfortunately, this is one of the few definitions used here that does not follow the convention of placing the form index farthest from the base character ($Q$). The non-metricity measures changes in lengths and angles under parallel transport \cite{Hehl}. This object depends on both the metric and the connection, since
\begin{equation}
    Q_{lmn}=-\partial_lg_{mn}+\Gamma^p{}_{ml}g_{pn}+\Gamma^p{}_{nl}g_{mp}.
\end{equation}
A \textit{compatibility} exists between the metric and the connection when
\begin{equation}
    Q_{lmn}=0.
\end{equation}
An ``index-up" counterpart of $ Q_{lmn}$ is easily found by using the product rule from Eq. \eqref{product-rule-gen}, satisfied by the covariant derivative:
\begin{equation}
\begin{split}
    Q_l{}^{pq}:&=g^{pm}g^{qn}Q_{lmn}\\
    &=-g^{pm}g^{qn}\nabla_lg_{mn}\\
    &=-\nabla_l(g^{pm}g^{qn}g_{mn})+g_{mn}\nabla_l(g^{pm}g^{qn})\\
    &=-\nabla_lg^{pq}+2\nabla_lg^{pq}\\
    &=\nabla_lg^{pq}.
    \end{split}
\end{equation}

In $m>2$ dimensions, the non-metricity has a unique decomposition provided by the Young's Tableaux into four distinct irreducible subspaces, invariant under both the linear and orthogonal groups \cite{Hehl}. However, for the purposes of this document, we do not require the full decomposition. Instead, we decompose it with respect to the internal natural trace as
\begin{equation}\label{non-metricity-decomp}
    Q_{lmn}=\frac{2}{m}Q_lg_{mn}+q_{lmn},
\end{equation}
where 
\begin{equation}
    Q_l:=\frac{1}{2}g^{mn}Q_{lmn},\quad\quad\quad q_{l[mn]}=0,\quad\quad\quad  g^{mn}q_{lmn}=0.
\end{equation}
The non-metricity co-vector $Q_l$ is known as the \textit{Weyl co-vector}. Here, \textit{internal} trace refers to a trace over the non-form indices, while \textit{unnatural} refers to the necessity of a metric to perform the trace. 

A nonzero Weyl co-vector contributes to a rescaling of vectors under parallel transport—this corresponds to a \textit{dilation} of vector lengths. In contrast, a nonzero, traceless $q_{lmn}$ contributes to \textit{shearing} effects experienced by a vector under parallel transport. A shear, in the context of a parallelogram, for example, is a volume-preserving deformation that changes the shape. Essentially, a shear defect alters both \textit{length} and \textit{angle} in a direction-dependent manner, which may lead to potential violations of causality \cite{mielke-geometro}. 

The combination of non-metricities, known as the \textit{disformation tensor}, is defined as
\begin{equation}\label{disformation-general}
    L^l{}_{mn}:=\frac{1}{2}g^{lr}\left(Q_{mnr}+Q_{nrm}-Q_{rmn}\right).
\end{equation}
Expanding the definition of $L^l{}_{mn}$ by substituting the covariant derivative contained in $Q_{lmn}$, one may solve the resulting expression for an arbitrary affine connection $\Gamma^l{}_{nm}$. Doing so, one finds
\begin{equation}\label{connection-partial-decomposition}
    \Gamma^l{}_{mn}=\hat{\Gamma}^l{}_{mn}+N^l{}_{mn},
\end{equation}
where $\hat{\Gamma}^l{}_{mn}$ is the \textit{Levi-Civita} connection associated with the metric $g_{mn}$,
\begin{equation}
    \hat{\Gamma}{}^l{}_{mn}:=\frac{1}{2}g^{lr}\left(\partial_mg_{nr}+\partial_ng_{rm}-\partial_rg_{mn}\right).
\end{equation}
The Levi-Civita connection is the unique \textit{symmetric} connection satisfying
\begin{equation}
    -\hat{Q}_{lmn}:=\hat{\nabla}_lg_{mn}=\partial_lg_{mn}-\hat{\Gamma}^k{}_{ml}g_{kn}-\hat{\Gamma}^k{}_{nl}g_{mk}=0.
\end{equation}
From the definition in Eq. \eqref{connection-gen-def}, it is straightforward to show that the Levi-Civita connection describes the parallel transport of the holonomic basis with respect to the affine connection $\Gamma$, since
\begin{equation}
    \nabla_n\partial_m=\hat{\Gamma}{}^l{}_{mn}\partial_l.
\end{equation}

The $N^l{}_{mn}$ term appearing in Eq. \eqref{connection-partial-decomposition} is known as the \textit{distortion tensor}, which represents all geometric distortions or deformations of $\mathcal{M}$. The distortion tensor consists of two distinct geometric objects:
\begin{equation}\label{distortion-def}
    N^l{}_{mn}:=K^l{}_{mn}+L^l{}_{mn}.
\end{equation}
The first,
\begin{equation}
    K^l{}_{mn}:= \frac{1}{2}g^{lp}\left(T_{mpn}+T_{npm}-T_{pmn}\right),
\end{equation}
is the \textit{contorsion tensor}, while the second, $L^l{}_{mn}$, is the disformation tensor given in Eq. \eqref{disformation-general}. The contorsion tensor is defined in terms of the \textit{torsion tensor}, which is the antisymmetric part of the connection:
\begin{equation}\label{torsion-def-gen}
    T^q{}_{mn}:=\Gamma^q{}_{[mn]}=\Gamma^q{}_{mn}-\Gamma^q{}_{nm}.
\end{equation}
Torsion measures the failure of parallelograms to close when formed by vectors under parallel transport \cite{Hehl}. This represents a \textit{twisting} deformation of the geometry comprising $\mathcal{M}$. In the definition of the contorsion tensor, we use the shorthand notation $T_{mpn}=g_{qm}T^q{}_{pn}$ to denote the torsion tensor with all indices lowered. While this document does not focus on the irreducible decomposition of torsion, a thorough discussion may be found in \cite{Hehl}.

With the generic decomposition of an arbitrary affine spacetime connection $\Gamma^l{}_{mn}$ in hand, decomposing its natural traces is straightforward. First, we determine the natural traces of distortion:
\begin{equation}
    N^{k}{}_{kn}=\frac{1}{2}Q_{n}{}^k{}_k=Q_n,\quad\quad\quad\quad N^k{}_{nk}=Q_m-T_m,
\end{equation}
where $T_m:=T^k{}_{mk}=-T^k{}_{km}$ is the torsion vector. Since the unnatural trace of torsion vanishes due to its antisymmetry, we define the torsion vector $T_n$ as the natural trace over the first and third index. Unfortunately, the more natural option differs from this by a minus sign. 

From here, the affine connection traces are simply
\begin{equation}\label{affine-conn-traces}
    \Gamma^k{}_{kn}=\hat{\Gamma}{}^{k}{}_{kn}+Q_n,\quad\quad\quad\quad\Gamma^{k}{}_{mk}=\hat{\Gamma}{}^{k}{}_{nk}+Q_m-T_m,
\end{equation}
where
\begin{equation}
   \hat{\Gamma}^k{}_{mk}=\hat{\Gamma}^k{}_{km}=\frac{1}{2}g^{pq}\partial_mg_{pq}=\partial_m\log\sqrt{(-1)^qg}
\end{equation}
is the natural trace of the Levi-Civita connection, and $g:=|g|=|g_{mn}|$ is the metric determinant. The factor of $(-1)^q$ follows from our choice of pseudo-Riemannian metric with a necessarily Lorentzian- or split-signature $(p,q)$, where $p$ counts the positive (timelike) entries and $q$ the negative (spacelike) entries, with $p+q=m$. Here, we are primarily interested in even $m=2k$-dimensional spacetimes of split-signature $(1,3)$, and will thus often write $-g$. 

We introduce the definition
\begin{equation}\label{lil-g-intro-def}
    g_n:=\frac{-1}{m+1}\partial_n\log\sqrt{(-1)^qg}=\frac{-1}{m+1}\hat{\Gamma}{}^{k}{}_{kn},
\end{equation}
for an $m$-dimensional manifold. Furthermore, we define
\begin{equation}\label{alpha-beta-definition}
    \alpha_n:=\frac{-1}{m+1}\Gamma^k{}_{kn},\quad\quad\quad\quad \beta_n:=\frac{-1}{m+1}\Gamma^k{}_{nk}.
\end{equation}
The decomposed connection traces in Eqs. \eqref{affine-conn-traces} may then be concisely written as
\begin{equation}\label{alpha-beta-definition-tors/nonmet}
   \alpha_n=g_n-\frac{1}{m+1}Q_n,\quad\quad\quad\quad\beta_m=g_m-\frac{1}{m+1}(Q_m-T_m).
\end{equation}

Lastly, we define two deviations of connection traces from Levi-Civita. The first,
\begin{equation}\label{cent-def-1}
    \mbox{\textcent}_n:=g_n-\alpha_n=\frac{1}{m+1}Q_n,
\end{equation}
measures the deviation of the connection's internal natural trace from Levi-Civita and is simply proportional to the non-metricity co-vector. In the generalization of the Projective Gauge Theory of Gravity, which this document proposes, $\mbox{\textcent}$ will be identified as a set of fields parameterizing the $m$-dimensional coset space $\mathbb{R}^m_*$. Although we have not yet identified a fundamental use for it, the second definition,
\begin{equation}
    \$_m:=g_m-\beta_m=\frac{1}{m+1}(Q_m-T_m)=\mbox{\textcent}_m-\frac{1}{m+1}T_m,
\end{equation}
measures the deviation of the connection's external natural trace from Levi-Civita. From these expressions, it is clear that the vanishing of the torsion vector is necessary for the hopeful-looking equation
\begin{equation}
    \$_m=\mbox{\textcent}_m.
\end{equation}
The combination
\begin{equation}\label{beta-minus-alpha}
    \beta_n-\alpha_n=\frac{-1}{m+1}\Gamma^k{}_{[nk]}=\frac{-1}{m+1}T^k{}_{nk}=:\frac{-1}{m+1}T_{n}
\end{equation}
is proportional to the natural trace of torsion. The vanishing of the torsion vector is thus necessary and sufficient for the equality $\alpha_n=\beta_n$. 

From Eq. \eqref{gamma-conn-trans}, we find that under a transformation of coordinates $\{x^m\}\rightarrow\{x^{n'}\}$, each connection trace $g_n$, $\alpha_n$, and $\beta_n$ transforms identically:
\begin{equation}\label{intro-conn-trans-lil-j}
    \alpha_{n'}=(\alpha_m+j_m)\frac{\partial x^m}{\partial x^{n'}},
\end{equation}
where
\begin{equation}
    j_m:=\frac{-1}{m+1}\frac{\partial x^{n'}}{\partial x^l}\frac{\partial^2 x^l}{\partial x^m\partial x^{n'}}=\frac{-1}{m+1}\partial_m\log\left|\frac{\partial x^l}{\partial x^{n'}}\right|.
\end{equation}
Thus, under general coordinate transformations, combinations such as
\begin{equation}
    g_n-\beta_n,\quad\quad g_n-\alpha_n,\quad\quad \beta_n-\alpha_n
\end{equation}
contain conspiring factors of $j_n$, ensuring covariance. For example, the torsion vector transforms covariantly, since
\begin{equation}
     \frac{-1}{m+1}T_{n'}=(\beta_m-\alpha_m)\frac{\partial x^m}{\partial x^{n'}}.
\end{equation}
For further discussion on the geometric interpretation of both torsion and non-metricity, see \cite{geometric-meaning-of-TandQ, Hehl}.

\subsection{Curvature Tensor}

In a general $(L_m,g)$, the commutator of covariant derivatives acting on arbitrary rank-$(i,j)^w$ fields produces
\begin{equation}\label{commutator-of-cov-der}
\begin{split}
    [\nabla_m,\nabla_n]V^{p_1\dots p_i}_{\quad q_1\dots q_j}&=\sum_{n=1}^iR^{p_n}{}_{k[mn]}V^{p_1\dots p_{n-1}k p_{n+1}\dots p_i}_{\quad q_1\dots q_j}-\sum_{n=1}^jR^{k}{}_{q_n[mn]}V^{p_1\dots p_i}_{\quad q_1\dots q_{n-1}k q_{n+1}\dots q_j}\\
    &\quad-T^q{}_{mn}\nabla_qV^{p_1\dots p_i}_{\quad q_1\dots q_j}-w\check{\check{R}}_{[mn]}V^{p_1\dots p_i}_{\quad q_1\dots q_j},
    \end{split}
\end{equation}
where Eq. \eqref{cov-der-def-general} was used to derive this expression. Notably, no antisymmetrization brackets appear on the third term due to the definition in Eq. \eqref{torsion-def-gen}. 

In Eq. \eqref{commutator-of-cov-der} we encounter the \textit{Curvature Tensor},
\begin{equation}\label{riemann-curvature}
    R^p{}_{q[mn]}:=\partial_m\Gamma^p{}_{qn}-\partial_n\Gamma^p{}_{qm}+\Gamma^p{}_{km}\Gamma^k{}_{qn}-\Gamma^p{}_{kn}\Gamma^k{}_{qm}.
\end{equation}
Curvature is the third and final geometric deformation of a spacetime manifold and serves as a measure of the \textit{rotation} of a vector under parallel transport, i.e., the \textit{curvature} of $\mathcal{M}$. This object is naturally antisymmetric in its last two (form) indices $m,n$. The third term in Eq. \eqref{commutator-of-cov-der} represents a directional derivative of $V$ along the torsion, while the last term,
\begin{equation}
    \check{\check{R}}_{[mn]}:=R^p{}_{p[mn]}=\partial_{m}\Gamma^k{}_{kn}-\partial_n\Gamma^{k}{}_{km},
\end{equation}
defines the naturally antisymmetric \textit{Homothetic Curvature Tensor}. The homothetic curvature is a natural trace of the curvature tensor over the ``internal" indices. This object is nonzero in the presence of a naturally traced non-metricity (Weyl co-vector), as will be shown later. Its appearance in the commutator of covariant derivatives occurs only when acting on tensor densities of weight $w$. From Eq. \eqref{alpha-beta-definition}, the homothetic curvature may be expressed as
\begin{equation}\label{homo-alpha-intro}
    \check{\check{R}}_{[mn]}=-(m+1)\partial_{[m}\alpha_{n]}.
\end{equation}

Another important trace of the curvature tensor, which does not explicitly appear in the commutator of covariant derivatives is the (non-Riemannian) \textit{Ricci Curvature Tensor}, defined as the external natural trace of the curvature,
\begin{equation}
    R_{qn}:=R^p{}_{q[pn]}.
\end{equation}
Explicitly, in terms of the affine connection, $R_{qn}$ has the form
\begin{equation}
    R_{qn}=\partial_m\Gamma^m{}_{qn}-\partial_n\Gamma^m{}_{qm}+\Gamma^m{}_{km}\Gamma^k{}_{qn}-\Gamma^m{}_{kn}\Gamma^k{}_{qm}.
\end{equation}
A relation may be found between the (non-Riemannian) Ricci tensor and the homothetic curvature tensor. In general, the antisymmetric part of the Ricci tensor satisfies
\begin{equation}
    \begin{split}
        R_{[qn]}&=\partial_m\Gamma^m{}_{[qn]}-\partial_{[n}\Gamma^m{}_{q]m}+\Gamma^m{}_{km}\Gamma^k{}_{[qn]}-\Gamma^m{}_{k[n}\Gamma^k{}_{q]m}\\
        &=\partial_mT^m{}_{qn}+(m+1)\partial_{[n}\beta_{q]}+\Gamma^m{}_{km}T^k{}_{qn}-\Gamma^m{}_{k[n}\Gamma^k{}_{q]m}\\
        &=\nabla_mT^m{}_{qn}-(m+1)\partial_{[q}\beta_{n]}.
    \end{split}
\end{equation}
From Eq. \eqref{beta-minus-alpha}, the vanishing of torsion implies $\alpha_n=\beta_n$ and, therefore,
\begin{equation}
    T^m{}_{pq}=0\quad\quad\Rightarrow\quad\quad R_{mn}=R_{(mn)}+\check{\check{R}}_{[mn]}.
\end{equation}
Thus, the vanishing of both torsion and homothetic curvature is necessary for a symmetric (non-Riemannian) Ricci curvature tensor. 

With a metric available, there exists a third, semi-independent contraction of the curvature tensor. The \textit{Co-Ricci Curvature Tensor} is defined as the only non-vanishing unnatural trace of the curvature tensor. In other words,
\begin{equation}
    \check{R}^p{}_n:=R^p{}_{q[mn]}g^{qm}.
\end{equation}
The reason for the ``semi-" qualifier regarding its independence is that the trace of $\check{R}$ is not independent. Indeed,
\begin{equation}
\begin{split}
    \check{R}&=R^p{}_{q[mn]}g^{qm}\delta^n{}_p\\
    &=-R^p{}_{q[nm]}g^{qm}\delta^n{}_p\\
    &=-R_{qm}g^{qm}\\
    &=-R,
    \end{split}
\end{equation}
which is proportional to the trace of the (non-Riemannian) Ricci tensor, known as the (non-Riemannian) \textit{Ricci scalar}. Thus, in a general $(L_m,g)$, there are three independent contractions of the curvature tensor.

\subsection{Curvature Decomposition}

The decomposition of the connection into its Levi-Civita and distortion parts induces a corresponding decomposition of the curvature tensor. From Eqs. \ref{riemann-curvature} and \ref{connection-partial-decomposition},
\begin{equation}
    R^p{}_{q[mn]}=\hat{R}^p{}_{q[mn]}+H^p{}_{q[mn]},
\end{equation}
where
\begin{equation}\label{hat-riemann-curvature}
    \hat{R}^p{}_{q[mn]}:=\partial_m\hat{\Gamma}^p{}_{qn}-\partial_n\hat{\Gamma}^p{}_{qm}+\hat{\Gamma}^p{}_{km}\hat{\Gamma}^k{}_{qn}-\hat{\Gamma}^p{}_{kn}\hat{\Gamma}^k{}_{qm}
\end{equation}
is the \textit{Riemann Curvature Tensor} constructed from the Levi-Civita connection, and
\begin{equation}
    H^p{}_{q[mn]}:=\hat{\nabla}_mN^p{}_{qn}-\hat{\nabla}_nN^p{}_{qm}+N^p{}_{km}N^k{}_{qn}-N^p{}_{kn}N^{k}{}_{qm}
\end{equation}
is a curvature-like object composed of torsion and non-metricity (contorsion and disformation). Unless explicitly stated otherwise, we will refer to the various curvature traces without the ``non-Riemannian'' qualifier. 

Revisiting the various traces, we obtain a decomposition of the Ricci tensor,
\begin{equation}
    R_{qn}=\hat{R}_{qn}+H_{qn};
\end{equation}
the homothetic curvature,
\begin{equation}
    \check{\check{R}}_{mn}=\check{\check{H}}_{mn};
\end{equation}
and the co-Ricci tensor,
\begin{equation}
    \check{R}^p{}_n=\check{H}^p{}_n.
\end{equation}

The Levi-Civita Ricci tensor $\hat{R}_{qn}$ is naturally symmetric, since $\hat{\Gamma}^k{}_{mn}$ is both torsion-free and its natural trace is proportional to the derivative of the metric, Eq. \eqref{lil-g-intro-def}. This latter fact ensures the vanishing of the Levi-Civita homothetic curvature. The vanishing of the Levi-Civita co-Ricci tensor follows from the symmetries of $\hat{R}_{pqmn}$, in particular,
\begin{equation}
    \hat{R}_{pqmn}=\hat{R}_{mnpq}.
\end{equation}
To show this, we first compute
\begin{equation}
    [\hat{\nabla}_m,\hat{\nabla}_n]V_b=-\hat{R}^a{}_{b[mn]}V_a=-\hat{R}_{ab[mn]}V^a.
\end{equation}
Then, using the compatibility of $\hat{\nabla}_m$ and $g_{mn}$:
\begin{equation}
    \begin{split}
        [\hat{\nabla}_m,\hat{\nabla}_n]V_b&=[\hat{\nabla}_m,\hat{\nabla}_n](g_{ab}V^a)\\
        &=g_{cb}[\hat{\nabla}_m,\hat{\nabla}_n]V^c\\
        &=g_{cb}\hat{R}^c{}_{a[mn]}V^a\\
        &=\hat{R}_{ba[mn]}V^a.
    \end{split}
\end{equation}
Subtracting the two results yields
\begin{equation}\label{LC-R-symmetry}
    \hat{R}_{(ab)[mn]}=0.
\end{equation}
The remainder of the proof requires the first of the following \textit{Bianchi identities} for the Levi-Civita curvature:
\begin{equation}\label{LC-bianchi}
    (1):\;\hat{R}^p{}_{[qmn]}=0,\quad\quad\quad\quad (2):\;\hat{\nabla}_{[r}\hat{R}^p{}_{qmn]}=0.
\end{equation}
Using the first Bianchi identity and the symmetry condition in Eq. \eqref{LC-R-symmetry},
\begin{equation}
        0=(\hat{R}_{pq[mn]}+\hat{R}_{pn[qm]}+\hat{R}_{pm[nq]})-(\hat{R}_{mn[pq]}+\hat{R}_{mq[np]}+\hat{R}_{mp[qn]}).
\end{equation}
Rearranging, we obtain
\begin{equation}
    \begin{split}
        \hat{R}_{pq[mn]}-\hat{R}_{mn[pq]}&=-\hat{R}_{pn[qm]}-
        \hat{R}_{pm[nq]}+\hat{R}_{mq[np]}+\hat{R}_{mp[qn]}\\
        &=-\hat{R}_{pn[qm]}+\hat{R}_{mq[np]}.
    \end{split}
\end{equation}
By relabeling indices in the above relation, Eq. \eqref{LC-R-symmetry} may be used to rearrange the indices to their initial order as
\begin{equation}
    \begin{split}
        \hat{R}_{qp[nm]}-\hat{R}_{nm[qp]}&=-\hat{R}_{qm[pn]}+\hat{R}_{np[mq]}\\
        &=-\hat{R}_{mq[np]}+\hat{R}_{pn[qm]}\\
        &=-(\hat{R}_{pq[mn]}-\hat{R}_{mn[pq]}).
    \end{split}
\end{equation}
Thus, we conclude that
\begin{equation}
    \hat{R}_{pq[mn]}=\hat{R}_{mn[pq]},
\end{equation}
or, more explicitly,
\begin{equation}\label{LC-R-index-sym}
    \hat{R}_{[pq][mn]}=\hat{R}_{[mn][pq]},
\end{equation}
as claimed.

From these properties, the Levi-Civita co-Ricci tensor is found to vanish, since
\begin{equation}
    \begin{split}
        \hat{\check{R}}_{ln}&=\frac{1}{2}(\hat{\check{R}}_{ln}+\hat{\check{R}}_{nl})+\frac{1}{2}(\hat{\check{R}}_{ln}-\hat{\check{R}}_{nl})\\
        &=\frac{-1}{2}(\hat{R}_{lk[nm]}+\hat{R}_{nk[lm]})g^{km}+\frac{1}{2}(\hat{R}_{lk[mn]}-\hat{R}_{nk[ml]})g^{km}\\
        &=\frac{-1}{2}(\hat{R}_{lk[nm]}+\hat{R}_{lm[nk]})g^{km}+\frac{1}{2}(\hat{R}_{lk[mn]}-\hat{R}_{ml[nk]})g^{km}\\
        &=\frac{1}{2}(\hat{R}_{lk[mn]}-\hat{R}_{lm[kn]})g^{km}+\frac{1}{2}(\hat{R}_{lk[mn]}-\hat{R}_{lm[kn]})g^{km}\\
        &=0.
    \end{split}
\end{equation}
In the above, we used the definition of the co-Ricci tensor with indices $2$ and $4$ contracted in the symmetric part and indices $2$ and $3$ contracted in the antisymmetric part. 

The symmetric part of the curvature tensor may be found by acting on the metric with the commutator of covariant derivatives,
\begin{equation}
    [\nabla_m,\nabla_n]g_{pq}=-R^r{}_{p[mn]}g_{rq}-R^r{}_{q[mn]}g_{pr}- T^r{}_{mn}\nabla_rg_{pq}.
\end{equation}
Using the definition of non-metricity and rearranging, one finds that
\begin{equation}\label{sym-curvature}
  R_{(qp)[mn]}=H_{(qp)[mn]}= \nabla_{[m}Q_{n]pq}+ T^r{}_{mn}Q_{rpq}.
\end{equation}
Thus, a vanishing non-metricity is sufficient to ensure $R_{(pq)[mn]}=0$. Furthermore, the homothetic curvature may be expressed entirely in terms of the Weyl co-vector as
\begin{equation}\label{homothetic-Q}
    \begin{split}
        \check{\check{R}}_{mn}&=-(m+1)\partial_{[m}\alpha_{n]}\\
        &=-(m+1)(\partial_{[m}g_{n]}-\frac{1}{m+1}\partial_{[m}Q_{n]})\\
        &=\partial_{[m}Q_{n]}.
    \end{split}
\end{equation}
Equivalently, we obtain
\begin{equation}
     \check{\check{R}}_{mn}=(m+1)\partial_{[m}\mbox{\textcent}_{n]}.
\end{equation}
Thus, only the natural trace of non-metricity gives rise to a nonzero homothetic curvature. As we will find in the generalization of the Projective Gauge Theory of Gravitation, where $\mbox{\textcent}$ is given the status of a coset parameter, the homothetic curvature is \textit{pure gauge}. By this, we mean that one may freely set the (projectively invariant) homothetic curvature to vanish by an appropriate choice of gauge.

For further details on the complete decomposition of the curvature tensor into irreducible parts, see \cite{Hehl}.

\subsection{Projective Transformations}

A \textit{projective transformation} is a particular geometric deformation of the affine connection defined as
\begin{equation}\label{gamma-proj-trans}
    \delta_{p}:\Gamma^l{}_{mn}\rightarrow\Gamma^l{}_{mn}+\delta^l{}_m\xi_n,
\end{equation}
for some set of fields $\xi_n(x)$. Although generalizations of projective transformations have been explored (see, e.g., \cite{generalized-proj-trans}), we confine our attention to the standard implementation above. Subsequently, we will slightly deviate to discuss the more fundamentally motivated \textit{symmetric projective transformations}. 

Under a projective transformation, the curvature tensor transforms as
\begin{equation}
    \delta_{p}:R^p{}_{q[mn]}\rightarrow R^p{}_{q[mn]}+\delta^p{}_q\partial_{[m}\xi_{n]}.
\end{equation}
Thus, the various curvature traces transform under projective transformations of the $m$-dimensional affine connection as
\begin{equation}
   \delta_{p}:\begin{cases} R_{qn}\rightarrow R_{qn}+\partial_{[q}\xi_{n]},\\
    \check{R}^p{}_n\rightarrow \check{R}^p{}_n+g^{pm}\partial_{[m}\xi_{n]},\\
\check{\check{R}}_{[mn]}\rightarrow\check{\check{R}}_{[mn]}+m\partial_{[m}\xi_{n]}.
\end{cases}
\end{equation}
Taking the trace of the first relation above, we find that the Ricci scalar is invariant under the projective transformation of Eq. \eqref{gamma-proj-trans},
\begin{equation}\label{proj-inv-R}
    \delta_{p}:R\rightarrow R.
\end{equation}

From Eq. \eqref{non-metricity-def-gen}, the non-metricity transforms under a projective deformation of the affine connection as
\begin{equation}
    \delta_{p}:Q_{lmn}\rightarrow Q_{lmn}+2\xi_lg_{mn}.
\end{equation}
This provides the first explicit insight into the fundamental nature of projective deformations. Notably, by using the decomposition of the non-metricity in Eq. \eqref{non-metricity-decomp}, the transformation may be rewritten as
\begin{equation}\label{Q-weyl-shift}
    \delta_{p}:Q_{lmn}\rightarrow \frac{2}{m}(Q_l+m\xi_l)g_{mn}+q_{lmn}.
\end{equation}
Thus, a projective deformation of the affine connection contributes a linear shift of the Weyl co-vector. Geometrically, projective transformations add an extra contribution to the change in vector lengths under parallel transport. 

Less interesting is the resulting projective deformation of the torsion, which transforms as
\begin{equation}
    \delta_{p}:T^l{}_{mn}\rightarrow T^l{}_{mn}+\delta^l{}_{[m}\xi_{n]}.
\end{equation}
As will be developed later in this document, a particular class of projective transformations serves as a symmetry of the geodesic equation—provided that one also reparameterizes the path. When one only has access to a Levi-Civita connection, a projective transformation induces torsional deformations:
\begin{equation}\label{proj-torsion-mode}
    \hat{T}{}^l{}_{mn}=\delta^l{}_{[m}\xi_{n]}.
\end{equation}
Effectively, this introduces a more general geometry with $T\neq0$. Since this is not well-motivated, we instead turn to symmetric projective transformations.

\subsection{Symmetric Projective Transformations}
\label{sec:sym-proj-trans}

A \textit{symmetric projective transformation} is a projective deformation of the affine connection defined as
\begin{equation}
\delta_{pp}:\Gamma^l{}_{mn}\rightarrow\Gamma^l{}_{mn}+\delta^l{}_m\xi_n+\delta^l{}_n\xi_m,
\end{equation}
for some set of fields $\xi_n(x)$. This is simply the symmetrization of the transformation discussed in the previous subsection. A fundamental motivation for choosing to work with symmetric projective transformations over standard projective transformations is the following. According to \cite{generalized-proj-trans} and references therein, consider two $m$-dimensional Riemannian manifolds $L_m$ and $L'_m$, such that a diffeomorphism $f:\mathcal{M}\rightarrow\mathcal{M}'$ exists, for which every geodesic $\sigma: I\subset\mathbb{R}\rightarrow\mathcal{M}$ is mapped to another geodesic $f(\sigma)$ in $\mathcal{M}'$. The necessary and sufficient condition for this statement to hold is
\begin{equation}
    \nabla'_qg_{mn}=2\xi_qg'_{mn}+\xi_ng'_{mq}+\xi_mg'_{qn}.
\end{equation}
In other words,
\begin{equation}
    (\Gamma{}')^q{}_{mn}-\Gamma^q{}_{mn}=\delta^q_m\xi_n+\delta^q{}_n\xi_m,
\end{equation}
where
\begin{equation}\label{equiv-metrics}
    \xi_n=\frac{-1}{m+1}\partial_n\log\left(\sqrt{\frac{g'}{g}}\right)
\end{equation}
is the logarithmic derivative of the ratio of metric determinants. Therefore, symmetric projective transformations are more fundamental, as they arise directly from diffeomorphisms or coordinate transformations of geodesics.

Under a symmetric projective transformation, the curvature tensor transforms as
\begin{equation}
    \delta_{pp}:R^p{}_{q[mn]}\rightarrow R^p{}_{q[mn]}+\delta^p{}_q\nabla_{[m}\xi_{n]}-\delta^p{}_{[m}\nabla_{n]}\xi_q+\delta^p{}_{[m}\xi_{n]}\xi_q-T^l{}_{mn}(\delta^p{}_q\xi_l+\delta^p{}_l\xi_q).
\end{equation}
Although the first covariant derivative term above cancels one of the torsion terms, the same does not hold for the second covariant derivative term. Therefore, in the presence of a nonzero torsion, a symmetric projective transformation induces a mixing with torsion. The various curvature traces transform under these symmetric projective deformations as
\begin{equation}
    \delta_{pp}:\begin{cases}R_{qn}\rightarrow R_{qn}+\nabla_{[q}\xi_{n]}-(m-1)(\nabla_n\xi_q-\xi_n\xi_q)-T^l{}_{mn}(\delta^m{}_q\xi_l+\delta^m{}_l\xi_q),\\
    \check{R}^p{}_n\rightarrow \check{R}^p{}_n+g^{pm}(\nabla_m\xi_n-\xi_m\xi_n)-2g^{pm}(\nabla_n\xi_m-\xi_n\xi_m)\\
    \quad\quad\quad\quad+\delta^p{}_n(\nabla^q\xi_q-\xi^2)-T^l{}_{mn}(g^{pm}\xi_l+\xi^m\delta^p{}_l),\\
    \check{\check{R}}_{[mn]}\rightarrow\check{\check{R}}_{[mn]}+(m+1)\left(\nabla_{[m}\xi_{n]}-T^q{}_{mn}\xi_q\right).
    \end{cases}
\end{equation}
Here, we introduce the concise notation
\begin{equation}
    \nabla^q\xi_q:=g^{pq}\nabla_p\xi_q.
\end{equation}

From these expressions, we see that only the homothetic curvature remains independent of the mixing between the projective parameters $\xi$ and torsion. This is to be expected, as homothetic curvature is naturally independent of torsion. The Ricci scalar, however, is no longer invariant. Under symmetric projective transformations of the affine connection, the Ricci scalar transforms as
\begin{equation}\label{proj-noninv-R}
    \delta_{pp}:R\rightarrow R-(m-1)\left(\nabla^n\xi_n-\xi^2\right)+T^n\xi_n.
\end{equation}
Invariance is only regained when both the torsion vanishes and the projective parameters $\xi_n$ satisfy
\begin{equation}
    \nabla^n\xi_n-\xi^2=0.
\end{equation}

Under symmetric projective transformations of the affine connection, the non-metricity transforms as
\begin{equation}
    \delta_{pp}:Q_{lmn}\rightarrow Q_{lmn}+2g_{mn}\xi_l+g_{l(m}\xi_{n)}.
\end{equation}
As is evident from the antisymmetry of torsion, a symmetric projective transformation leaves the torsion unchanged,
\begin{equation}
    \delta_{pp}:T^l{}_{mn}\rightarrow T^l{}_{mn}.
\end{equation}
This will be a key property in the General Projective Gauge Theory of Gravity. The gauge theory of Thomas-Whitehead projective connections \cite{gen-struc}, for example, considers only torsion-free connections. However, it is formulated as a gauge theory of gravity constructed from \textit{symmetric} projective transformations and thus is fundamentally based on diffeomorphisms. Consequently, including torsion in this framework simply amounts to incorporating a most natural invariant tensor.

\section{General Relativity}
\label{sec:intro-gen-rel}

Gravitational theories are theories of spacetime geometry that result from the extremization of an action functional. In its most basic form, these functionals depend only on the metric and describe the geometry of a $V_m$:
\begin{equation}
    S[g]=\int_{\mathcal{M}}\mathcal{L}(g)=\int_{\mathcal{M}}d^mx\sqrt{-g} L(g).
\end{equation}
In the above, $\mathcal{L}(g)$ and $L(g)$ are, respectively, the Lagrangian $m$-form density and Lagrangian function of the metric field, or more precisely, the \textit{gravitational potentials} $g_{mn}$. A $V_m$ is simply an $L_m$ with vanishing torsion and non-metricity but nonzero curvature. Variations of $S[g]$, when set to vanish, provide the field equations for the gravitational system, describing the spatio-temporal dynamics of the gravitational potentials. The most immediate extension of $S[g]$ is to $S[g,\Gamma]$, where a dependence on the affine connection, independent of $g_{mn}$, is included, i.e., $\Gamma\neq\hat{\Gamma}$. These extensions are left for the next section. 

Before reviewing some historically standard choices for $S[g]$, we discuss the variations that lead to the geometry-governing field equations. We then introduce the Einstein-Hilbert action, which describes the theory of General Relativity, along with its two most common modifications: a cosmological constant (dark energy) and matter.

\subsection{Field Variations}

In a $V_m$, the gravitational potentials $g_{mn}$ are the only dynamical field variables. In this section, we compute the variation of each geometric quantity with respect to the dynamical field variables. Much of the content in this subsection may be found in any standard text on General Relativity, for example \cite{carroll}.

\subsubsection*{Metric Determinant}

The explicit variation of the metric determinant in $m$ dimensions is
\begin{equation}
    \begin{split}
        \frac{\delta\sqrt{-g}}{\delta g^{pq}}&=\frac{-1}{2\sqrt{-g}}\frac{\delta |g|}{\delta g^{pq}}\\
        &=\frac{-1}{2\sqrt{-g}}\frac{\delta }{\delta g^{pq}}\left(\frac{(-1)^q}{m!}\hat{\epsilon}^{a_1\dots a_m}\hat{\epsilon}^{b_1\dots b_m}g_{a_1b_1}\dots g_{a_mb_m}\right)\\
        &=\frac{1}{2\sqrt{-g}}\frac{(-1)^{q+1}}{(m-1)!}\hat{\epsilon}^{a_1\dots a_{m-1}a_m}\hat{\epsilon}^{b_1\dots b_{m-1} b_m}g_{a_1b_1}\dots g_{a_{m-1}b_{m-1}}\frac{\delta }{\delta g^{pq}}\left( g_{a_mb_m}\right)\\
        &=\frac{1}{2\sqrt{-g}}\frac{(-1)^{q+1}}{(m-1)!}\hat{\epsilon}^{a_1\dots a_{m-1}a_m}\left((-1)^q|g|\hat{\epsilon}_{a_1\dots a_{m-1} b}g^{bb_m}\right)\frac{\delta }{\delta g^{pq}}\left( g_{a_mb_m}\right)\\
        &=\frac{\sqrt{-g}}{2}\frac{(-1)^{2q+1}}{(m-1)!}\left(\hat{\epsilon}^{a_1\dots a_{m-1}a_m}\hat{\epsilon}_{a_1\dots a_{m-1} b}\right)g^{bb_m}\frac{\delta }{\delta g^{pq}}\left( g_{a_mb_m}\right)\\
        &=\frac{\sqrt{-g}}{2}\frac{-1}{(m-1)!}\left((-1)^q (m-1)!\delta^{a_m}{}_b\right)g^{bb_m}\frac{\delta }{\delta g^{pq}}\left( g_{a_mb_m}\right)\\
        &=\frac{\sqrt{-g}}{2}(-1)^{q+1}g^{a_mb_m}\frac{\delta }{\delta g^{pq}}\left( g_{a_mb_m}\right).
    \end{split}
\end{equation}
Since $g_{mp}g^{pq}=\delta^q{}_n$, we find
\begin{equation}
    (\delta g_{mp})g^{pq}+g_{mp}(\delta g^{pq})=0 \quad\rightarrow\quad \delta g_{mn}=-g_{nq}g_{mp}(\delta g^{pq}).
\end{equation}
Therefore, in an $m=4$-dimensional spacetime of split-signature $(p,q)=(1,3)$,
\begin{equation}
     \frac{1}{\sqrt{-g}}\frac{\delta\sqrt{-g}}{\delta g^{pq}}=\frac{-1}{2}g_{pq}.
\end{equation}

\subsubsection*{Levi-Civita Connection}

The variation of the Levi-Civita Connection is
\begin{equation}
    \begin{split}
        \delta\hat{\Gamma}^l{}_{mn}&=\delta\left(\frac{1}{2}g^{lr}\left(\partial_mg_{nr}+\partial_ng_{rm}-\partial_rg_{mn}\right)\right)\\
        &=\frac{1}{2}(\delta g^{lr})\left(\partial_mg_{nr}+\partial_ng_{rm}-\partial_rg_{mn}\right)+\frac{1}{2} g^{lr}\left(\partial_m\delta g_{nr}+\partial_n\delta g_{rm}-\partial_r\delta g_{mn}\right)\\
        &=\frac{-1}{2}g^{lp}g^{rq}(\delta g_{pq})\left(\partial_mg_{nr}+\partial_ng_{rm}-\partial_rg_{mn}\right)\\
        &\quad+\frac{1}{2} g^{lr}\left(\partial_m\delta g_{nr}+\partial_n\delta g_{rm}-\partial_r\delta g_{mn}\right)\\
        &=-g^{lp}(\delta g_{pq})\hat{\Gamma}^q{}_{mn}+\frac{1}{2} g^{lr}\left(\partial_m\delta g_{nr}+\partial_n\delta g_{rm}-\partial_r\delta g_{mn}\right)\\
        &=\frac{1}{2} g^{lr}\left(\partial_m\delta g_{nr}+\partial_n\delta g_{rm}-\partial_r\delta g_{mn}-2\hat{\Gamma}^q{}_{mn}\delta g_{rq}\right)\\
        &=\frac{1}{2} g^{lr}\left(\hat{\nabla}_m\delta g_{nr}+\hat{\nabla}_n\delta g_{rm}-\hat{\nabla}_r\delta g_{mn}\right),
        \end{split}
\end{equation}
where the appropriate Levi-Civita connections were added and subtracted to arrive at the final covariant expression. 

\subsubsection*{Levi-Civita Curvature}

The variation of the Levi-Civita curvature tensor is then
\begin{equation}
    \begin{split}
        \delta\hat{R}^p{}_{q[mn]}&=\delta\left(\partial_m\hat{\Gamma}^p{}_{qn}-\partial_n\hat{\Gamma}^p{}_{qm}+\hat{\Gamma}^p{}_{km}\hat{\Gamma}^k{}_{qn}-\hat{\Gamma}^p{}_{kn}\hat{\Gamma}^k{}_{qm}\right)\\
        &=\left(\partial_m\delta\hat{\Gamma}^p{}_{qn}-\partial_n\delta\hat{\Gamma}^p{}_{qm}+(\delta\hat{\Gamma}^p{}_{km})\hat{\Gamma}^k{}_{qn}\right)\\
        &\quad+\left(\hat{\Gamma}^p{}_{km}(\delta\hat{\Gamma}^k{}_{qn})-(\delta\hat{\Gamma}^p{}_{kn})\hat{\Gamma}^k{}_{qm}-\hat{\Gamma}^p{}_{kn}(\delta\hat{\Gamma}^k{}_{qm})\right)\\
        &=\hat{\nabla}_m\delta\hat{\Gamma}^p{}_{qn}-\hat{\nabla}_n\delta\hat{\Gamma}^p{}_{qm},
    \end{split}
\end{equation}
since $\delta\Gamma$ is the difference of two connections and, therefore, a tensor. The variation of the Levi-Civita Ricci tensor then follows:
\begin{equation}
\begin{split}
    \delta\hat{R}_{qn}&=\delta(\hat{R}^p{}_{qmn}\delta^m{}_p)\\
    &=\delta^m{}_p(\hat{\nabla}_m\delta\hat{\Gamma}^p{}_{qn}-\hat{\nabla}_n\delta\hat{\Gamma}^p{}_{qm})\\
    &=\hat{\nabla}_m\delta\hat{\Gamma}^m{}_{qn}-\hat{\nabla}_n\delta\hat{\Gamma}^m{}_{qm}.
\end{split}
\end{equation}
Contracting with a metric to produce the Levi-Civita Ricci scalar is where the explicit metric dependence is introduced. The variation of the Levi-Civita Ricci scalar may be written as
\begin{equation}
    \begin{split}
        \delta \hat{R}&=\delta(\hat{R}_{qn}g^{qn})\\
        &=(\delta\hat{R}_{qn})g^{qn}+\hat{R}_{qn}\delta g^{qn}\\
        &=\hat{\nabla}_m(g^{qn}\delta\hat{\Gamma}^m{}_{qn}-g^{qm}\delta\hat{\Gamma}^n{}_{qn})+\hat{R}_{qn} \delta g^{qn}.
    \end{split}
\end{equation}
Since these variations will be integrated over $\mathcal{M}$, any total derivative, such as the one in the above expression, may be omitted. This fact is a direct result of the generalized Stokes' Theorem \cite{carroll},
\begin{equation}\label{stokes}
    \int_{\mathcal{M}}d^mx\sqrt{-g}\;\nabla_kX^k=\int_{\partial\mathcal{M}}d^{m-1}y\sqrt{-h}\;n_kX^k,
\end{equation}
in tandem with the standard assumption that all field variations vanish at the boundary $\partial\mathcal{M}$. In the above, $n_i$ is a unit vector normal to the hypersurface $\partial\mathcal{M}$. Additionally, $\{y\}$ coordinates the hypersurface, for which $h$ is the metric.

\subsection{Einstein-Hilbert}

The Einstein-Hilbert action of a $V_m$ defines General Relativity in $m$ dimensions and has the form \cite{nakahara}
\begin{equation}
    S_{EH}[g]=\frac{1}{2\kappa}\int_{\mathcal{M}}d^mx \mathcal{L}_{EH}(g_{mn})=\frac{1}{2\kappa}\int_{\mathcal{M}} d^mx\sqrt{-g}\;\hat{R},
\end{equation}
where $\kappa:=8\pi G/c^4$ is the gravitational coupling constant, and $G=6.67\times 10^{-11}\frac{\text{m}^3}{\text{kg}\cdot\text{s}^2}$ is Newton's constant \cite{carroll}. This action is a functional of the metric alone since it is constructed from the Levi-Civita connection. Furthermore, according to the relation in Eq. \eqref{proj-inv-R}, the Einstein-Hilbert action is invariant under projective deformations of the connection:
\begin{equation}
    \delta_pS_{EH}[g]=0.
\end{equation}
However, as mentioned previously, the Levi-Civita connection is symmetric. If this symmetry is to be retained, then it is the more fundamentally motivated symmetric projective transformations that are of interest. From Eq. \eqref{proj-noninv-R}, one finds that $S_{EH}$ is no longer invariant when symmetric projective transformations are used:
\begin{equation}
    \begin{split}
        \delta_{pp}S_{EH}[g]=\frac{-(m-1)}{2\kappa}\int_{\mathcal{M}}d^mx\sqrt{-g}\;\left(\hat{\nabla}_n\xi^n-\xi^2\right).
    \end{split}
\end{equation}

As a functional of the metric, field variations produce
\begin{equation}
    \delta S_{EH}[g]=\frac{1}{2\kappa}\int_\mathcal{M} d^mx\sqrt{-g}\;\hat{\mathbb{G}}_{mn}\delta g^{mn},
\end{equation}
where
\begin{equation}
    \hat{\mathbb{G}}_{mn}:=\frac{1}{\sqrt{-g}}\frac{\delta\mathcal{L}_{EH}}{\delta g^{mn}}
\end{equation}
is the \textit{Einstein Tensor} of a $V_m$. The metric field equation is then
\begin{equation}
    \hat{\mathbb{G}}_{mn}=0.
\end{equation}
Explicitly, the metric field equations arise as
\begin{equation}
\begin{split}
    \delta S_{EH}[g]&=\frac{1}{2\kappa}\int_{\mathcal{M}} d^mx\delta(\sqrt{-g}\hat{R})\\
    &=\frac{1}{2\kappa}\int_{\mathcal{M}} d^mx(\sqrt{-g}\delta\hat{R}+\hat{R}\delta\sqrt{-g})\\
    &=\frac{1}{2\kappa}\int_{\mathcal{M}} d^mx\sqrt{-g}\left(\hat{\nabla}_m(g^{q[n}\delta\hat{\Gamma}^{m]}{}_{qn})+\hat{R}_{mn}\delta g^{mn}-\frac{1}{2}\hat{R}g_{mn}\delta g^{mn}\right)\\
    &=\frac{1}{2\kappa}\int_{\mathcal{M}} d^mx\sqrt{-g}\left(\hat{R}_{mn}-\frac{1}{2}\hat{R}g_{mn}\right)\delta g^{mn},
    \end{split}
\end{equation}
where the generalized Stokes' Theorem, Eq. \eqref{stokes}, was used to arrive at the final expression.

Compiling the above statements, the Einstein field equations for the gravitational potentials $g_{mn}$ take the explicit form
\begin{equation}
    \hat{\mathbb{G}}_{mn}=\hat{R}_{(mn)}-\frac{1}{2}\hat{R}g_{mn}=0.
\end{equation}
This equation describes a curved spacetime devoid of any matter or geometric deformations. Interestingly, due to the index symmetry of $\hat{R}_{(mn)}$, and $\hat{R}$'s invariance under projective transformations, the gravitational field equations inherit this invariance:
\begin{equation}
    \delta_{p}\hat{\mathbb{G}}_{mn}=0.
\end{equation}
Therefore, both the action and the field equations are invariant under projective transformations. However, neither statement is true for symmetric projective transformations:
\begin{equation}
    \delta_{pp}\hat{\mathbb{G}}_{mn}\neq0.
\end{equation}
Explicitly,
\begin{equation}
     \frac{-1}{m-1}\delta_{pp}\hat{\mathbb{G}}_{mn}=\left(\hat{\nabla}_{(m}\xi_{n)}-\xi_{(m}\xi_{n)}\right)-\frac{1}{2}g_{mn}\left(\hat{\nabla}_q\xi^q-\xi^2\right).
\end{equation}

\subsection{Modified Einstein-Hilbert}

There are two standard adaptations for the Einstein-Hilbert action of General Relativity in a $V_m$. The first considers adding a matter action $S_M[g,\psi]$, for some set of matter fields $\psi(x)$, which are independent of the gravitational potentials $g_{mn}$:
\begin{equation}
    S[g,\psi]=S_{EH}[g]+S_{M}[g,\psi].
\end{equation}
This addition alters the metric field equations by providing a \textit{source} for the gravitational field. This is due, at the very least, to the presence of the volume element used for integration. Let
\begin{equation}
    \mathbb{M}_{mn}:=\frac{-2}{\sqrt{-g}}\frac{\delta\mathcal{L}_M(g,\psi)}{\delta g^{mn}}
\end{equation}
denote the \textit{energy momentum tensor} of the matter fields $\psi$. The fact that this object acts as a source in the metric field equations can be seen by explicit variation:
\begin{equation}
\begin{split}
    \delta S_M&=\int_\mathcal{M}d^mx\delta\mathcal{L}_M(g,\psi)\\
    &=\int_\mathcal{M}d^mx\delta (\sqrt{-g}L_M(g,\psi))\\
    &=\int_\mathcal{M}d^mx(\sqrt{-g}\delta L_M(g,\psi)+L_M(g,\psi)\delta \sqrt{-g})\\
    &=\int_\mathcal{M}d^mx\left(\sqrt{-g}\frac{\delta L_M(g,\psi)}{\delta g^{mn}}-\frac{\sqrt{-g}}{2}L_M(g,\psi)g_{mn}\right)\delta g^{mn}.
    \end{split}
\end{equation}
Therefore,
\begin{equation}
    \mathbb{M}_{mn}=L_M(g,\psi)g_{mn}-2\frac{\delta L_M(g,\psi)}{\delta g^{mn}}.
\end{equation}
Accounting for the gravitational coupling constant $\kappa$, the metric field equations are modified to
\begin{equation}
    \hat{\mathbb{G}}_{mn}-\kappa\mathbb{M}_{mn}=0\quad\quad\rightarrow\quad\quad \hat{R}_{mn}-\frac{1}{2}\hat{R}g_{mn}-\kappa\mathbb{M}_{mn}=0.
\end{equation}

The second standard adaptation to Einstein's General Theory of Relativity is the addition of a cosmological constant,
\begin{equation}
    S[g,\psi]=S_{EH}[g]+S_{M}[g,\psi]+S_{\Lambda}[g],
\end{equation}
where
\begin{equation}
    S_{\Lambda}[g]:=\frac{1}{2\kappa}\int_{\mathcal{M}}d^mx\sqrt{-g}\;(-2\Lambda).
\end{equation}
The cosmological constant represents the \textit{dark energy} associated with the expansion or contraction of the spacetime, depending on $\text{sgn}(\Lambda)$. Variation of $S_{\Lambda}[g]$ does not introduce a distinct tensorial quantity because $\Lambda$ is simply a constant, and the only contributing variation is from the volume element. Therefore,
\begin{equation}
    \delta S_{\Lambda}[g]=\frac{1}{2\kappa}\int_{\mathcal{M}}d^mx(-2\Lambda)\delta\sqrt{-g}=\frac{1}{2\kappa}\int_{\mathcal{M}}d^mx\sqrt{-g}\;(\Lambda g_{mn})\delta g^{mn}.
\end{equation}
The Einstein field equations are thus modified to include matter and permit an expanding or contracting universe:
\begin{equation}
    \hat{R}_{mn}-\frac{1}{2}\hat{R}g_{mn}+\Lambda g_{mn}=\kappa\mathbb{M}_{mn}.
\end{equation}

Beyond these modifications to the Einstein-Hilbert action, there are many. Of particular interest are the \textit{Euler} and \textit{Pontrjagin} densities. However, as will be shown, for a $V_4$, these are both topological (total derivatives) for the Levi-Civita connection and do not alter the field equations. Thus, permitting a non-Levi-Civita connection—i.e., introducing non-metricity and torsion—is natural when seeking extensions beyond a General Relativistic description of an expanding or contracting matter-filled universe. The most intuitive approach to handling such a generalization is in the language of differential forms and their exterior calculus. To some, this is known as the \textit{Palatini formalism}, to which we turn next.

\section{Exterior Palatini Formalism}

There exist other extensions to General Relativity; however, their appearance is mostly irrelevant when the connection is Levi-Civita. When this condition is removed, it is most convenient to speak in the language of differential forms. The Palatini formalism here refers to obtaining a Levi-Civita connection and both vanishing torsion and non-metricity as a result of the field equations \cite{palatini}. For a more mathematically rigorous approach, see \cite{bundle-palatini}. There is, however, interest in whether or not the exterior formalism loses geometric information \cite{non-unique-levi-civita,inequivalent-palatini-MAG,borunda}. This section introduces the exterior formalism in General Relativity and re-develops the content of Sec. \ref{sec:intro-gen-rel} in this language. Additionally, we review the modifications to General Relativity that were not discussed previously. We then develop and organize a formal program for field variations, mostly for use in later comparison with the General Projective Gauge Theory of Gravitation. Much of this review section may be found in \cite{Hehl}.

\subsection{Fundamentals}

To develop the exterior formalism for gravitational theories, it is convenient to first transform the connection coefficients $\Gamma$ to an arbitrary basis via
\begin{equation}\label{frame-conn-gen-def}
\omega^a{}_{bm}:=e^a{}_l\left(\Gamma^l{}_{nm}+\delta^l{}_n\partial_m\right)(e^{-1})^n{}_b,
\end{equation}
where $\omega^a{}_{bm}$ are the connection coefficients in the new basis. Connecting to the basis differentials $dx^m$, we construct the \textit{connection $1$-form} as
\begin{equation}
    \omega^a{}_{b}:=\omega^a{}_{bm}dx^m.
    \end{equation}
Indices $a,b,\dots=1,2\dots,m$ from the beginning of the Latin alphabet refer to the new basis, while indices $k,l,\dots=1,2,\dots,m$ from the middle of the Latin alphabet refer to the spacetime manifold $\mathcal{M}$. In particular, we will soon be concerned with attributing the new set of indices to a group manifold. 

As discussed previously, antisymmetric products are formed with a \textit{wedge} between forms of various degrees. For example, the wedge product of the $p$-form $A$ and the $q$-form $B$ satisfies
\begin{equation}
    A\wedge B=(-1)^{pq}B\wedge A.
\end{equation}
The exterior covariant derivative acting on a tensor-valued $k$-form $A^{a\dots}{}_{b\dots}$ is defined as
\begin{equation}
    DA^{a\dots}{}_{b\dots}:=dA^{a\dots}{}_{b\dots}+\sum_{c_i}\omega^a{}_c\wedge A^{c\dots}{}_{b\dots}-\sum_{c_i}\omega^c{}_b\wedge A^{a\dots}{}_{c\dots}.
\end{equation}
The second covariant exterior derivative of a tensor-valued $k$-form $A^{a\dots}{}_{b\dots}$ has the convenient expression
\begin{equation}\label{second-cov-deriv-ext}
    DDA^{a\dots}{}_{b\dots}:=\sum_{c_i}R^a{}_c\wedge A^{c\dots}{}_{b\dots}-\sum_{c_i}R^c{}_b\wedge A^{a\dots}{}_{c\dots},
\end{equation}
where $R^a{}_b$ is the curvature $2$-form, which will be defined shortly. The exterior derivative of the wedge product of the $p$-form $A$ and the $q$-form $B$ then follows,
\begin{equation}
    D(A\wedge B)=DA\wedge B+(-1)^pA\wedge DB.
\end{equation}
Since $\omega^a{}_b$ is not a tensor-valued form, expressions such as $D\omega^a{}_b$, using the above prescription, are incorrect. However, the combination
\begin{equation}
    D\omega^a{}_b:=d\omega^a{}_b+\omega^a{}_c\wedge\omega^c{}_b
\end{equation}
is a tensor-valued form and thus has geometric significance. At the expense of potential confusion, we use $D\omega^a{}_b$ to mean the above expression.

The \textit{curvature $2$-form}, the \textit{torsion $2$-form} and the \textit{non-metricity $1$-form} associated with a general $L_m$ may now be viewed as the field strengths of the gravitational field variables $\mathcal{F}_G=\{g_{ab},e^a,\omega^a{}_b\}$, where
\begin{equation}
    g_{ab}:=g_{mn}(e^{-1})^m{}_a(e^{-1})^n{}_b.
\end{equation}
These field strengths are, respectively,
\begin{equation}
    R^a{}_b:=D\omega^a{}_b,\quad\quad T^a:=De^a,\quad\quad Q_{ab}:=-Dg_{ab}.
\end{equation}
The components of these geometric objects may be extracted to give
\begin{equation}
    R^a{}_b=\frac{1}{2}R^a{}_{b[mn]}dx^m\wedge dx^n,\quad\quad T^a=\frac{1}{2}T^a{}_{[mn]}dx^m\wedge dx^n,\quad\quad Q_{ab}=Q_{mab}dx^m.
\end{equation}

Decomposing the connection $1$-form $\omega$ follows from the decomposition of the affine connection $\Gamma$. Substituting into Eq. \eqref{frame-conn-gen-def}, the decomposition of Eq. \eqref{connection-partial-decomposition}, we find
\begin{equation}
    \omega^a{}_{b}=\hat{\omega}{}^a{}_b+N{}^a{}_b,
\end{equation}
where
\begin{equation}
    N^a{}_b:=N^a{}_{bm}dx^m=e^a{}_lN^l{}_{nm}(e^{-1})^n{}_bdx^m
\end{equation}
is the \textit{distortion $1$-form}, and
\begin{equation}
    \hat{\omega}{}^a{}_b:=e^a{}_l\left(\hat{\Gamma}{}^l{}_{nm}+\delta^l{}_n\partial_m\right)(e^{-1})^n{}_bdx^m
\end{equation}
is the Levi-Civita connection in the arbitrary basis provided by $\{e_m\}$. Reflecting this in the curvature $2$-form gives
\begin{equation}
    R^a{}_b=\hat{R}{}^a{}_b+H^a{}_b,
\end{equation}
where
\begin{equation}
    H^a{}_b=\hat{D}N^a{}_b+N^a{}_c\wedge N^c{}_b
\end{equation}
is the field strength of the combined geometric deformations: contorsion and disformation. The components of $H$ are accessed via
\begin{equation}
    H^a{}_b:=\frac{1}{2}H^a{}_{b[mn]}dx^m\wedge dx^n=\frac{1}{2}e^a{}_lH^l{}_{k[mn]}(e^{-1})^k{}_bdx^m\wedge dx^n.
\end{equation}

\subsection{Exterior Variations}
\label{sec:exterior-variation-MAG}

In this section, we vary the action describing a general $L_m$ coupled to matter, given by
\begin{equation}
    S=S[g,e,\omega,\psi].
\end{equation}
This is a functional of both the gravitational field variables $\mathcal{F}_G=\{g_{ab},e^a,\omega^a{}_b\}$ and the matter field variables $\mathcal{F}_M=\{\psi\}$. For generality, we consider the matter fields as $p$-forms. The application to spin-$1/2$ fields follows from taking $p=0$. In terms of the $m$-form Lagrangian density, a general variation produces
\begin{equation}
    \delta S=\int_{\mathcal{M}}\delta \mathcal{L},
\end{equation}
where
\begin{equation}\label{Lm-variation}
    \delta\mathcal{L}=\frac{1}{2}\delta g_{ab}\underline{\mathbb{F}}^{ab}+\delta e^a\wedge\mathbb{F}_a+\delta\omega{}^a{}_b\wedge\mathbb{F}^b{}_a+\delta\psi\wedge\mathbb{F}+d\mathbb{B},
\end{equation}
as shown in \cite{Hehl}. Since we are working with exterior forms, care must be taken when performing functional variation. In particular, we choose to factor all variations to the \textit{left}. 

In the functional variation of Eq. \eqref{Lm-variation}, we encounter the boundary term
\begin{equation}
    \mathbb{B}:=-\delta g_{ab}\wedge\frac{\partial \mathcal{L}}{\partial Q_{ab}}+\delta e^a\wedge\frac{\partial \mathcal{L}}{\partial T^a}+\delta\omega^a{}_b\wedge\frac{\partial\mathcal{L}}{\partial R^a{}_b}+\delta\psi\wedge\frac{\partial\mathcal{L}}{\partial D\psi}.
\end{equation}
This appears as a total derivative and does not alter the field equations. This follows from the standard assumption that field variations vanish on the boundary $\partial\mathcal{M}$. The $p$-form matter field variation,
\begin{equation}
    \mathbb{F}:=\frac{\delta\mathcal{L}}{\delta\psi}=\frac{\partial\mathcal{L}}{\partial\psi}-(-1)^pD\frac{\partial\mathcal{L}}{\partial(D\psi)},
\end{equation}
produces the field equation for $\psi$. For a gauge invariant $\mathcal{L}$, this variational derivative coincides with the $GL(m,\mathbb{R})$-covariant variational derivative \cite{Hehl}. Variation of the connection produces
\begin{equation}
    \mathbb{F}^b{}_a:=\frac{\delta\mathcal{L}}{\delta\omega^a{}_b}=D\frac{\partial\mathcal{L}}{\partial R^a{}_b}+e^b\wedge\frac{\partial\mathcal{L}}{\partial T^a}+2g_{ac}\frac{\partial\mathcal{L}}{\partial Q_{cb}}+\psi\rho(\bm{L}{}^b{}_a)\wedge\frac{\partial\mathcal{L}}{\partial(D\psi)},
\end{equation}
with $\rho(\bm{L}^b{}_a)$ the representation of the group generators appropriate for the $p$-form matter fields $\psi$. Variation of the co-frame yields
\begin{equation}
    \mathbb{F}_a:=\frac{\delta\mathcal{L}}{\delta e^a}=\frac{\partial \mathcal{L}}{\partial e^a}+D\frac{\partial\mathcal{L}}{\partial T^a},
\end{equation}
and variation of the metric yields
\begin{equation}
    \underline{\mathbb{F}}{}^{ab}:=2\frac{\delta\mathcal{L}}{\delta g_{ab}}=2\frac{\partial\mathcal{L}}{\partial g_{ab}}+2D\frac{\partial\mathcal{L}}{\partial Q_{ab}}.
\end{equation}

Assuming $\mathcal{L}$ decomposes nicely into into matter and gravitational sectors, $\mathcal{L}=\mathcal{L}_M+\mathcal{L}_G$, the field variations will split correspondingly: $\mathbb{F}_i=\mathbb{M}_i+\mathbb{G}_i$. The segregation of the matter and gravitational sectors, reflected in the field equations, permits the definition of both material and gravitational currents and momenta. The gravitational gauge field momenta are the $(m-1)$-form
\begin{equation}
    \underline{\mathbb{H}}^{ab}:=-2\frac{\partial\mathcal{L}_G}{\partial Q_{ab}},
\end{equation}
and the $(m-2)$-forms
\begin{equation}
    \mathbb{H}_a:=-\frac{\partial\mathcal{L}_G}{\partial T^a},\quad\quad\quad \mathbb{H}^b{}_a:=-\frac{\partial\mathcal{L}_G}{\partial R^a{}_b}.
\end{equation}
Furthermore, in the gravitational sector, the gravitational currents contain the metrical energy momentum $m$-form
\begin{equation}\label{metric-EM-MAG}
    \underline{\mathbb{E}}{}^{ab}:=2\frac{\partial\mathcal{L}_G}{\partial g_{ab}},
\end{equation}
the canonical energy momentum $(m-1)$-form
\begin{equation}\label{canonical-EM_MAG}
    \mathbb{E}_a:=\frac{\partial \mathcal{L}}{\partial e^a},
\end{equation}
and the hyper-momentum $(m-1)$-form
\begin{equation}\label{hypermom-MAG}
    \mathbb{E}^b{}_a:=-e^b\wedge\mathbb{H}_a-g_{ac}\underline{\mathbb{H}}^{bc}.
\end{equation}

Similarly, in the matter sector, we have, respectively, the metrical energy-momentum current, the canonical energy-momentum current, and the hyper-momentum current:
\begin{equation}\label{matter-currents-MAG}
    \underline{\mathbb{M}}^{ab}:=2\frac{\delta\mathcal{L}_M}{\delta g^{ab}},\quad\quad\quad \mathbb{M}_a:=\frac{\delta\mathcal{L}_M}{\delta e^a},\quad\quad\quad \mathbb{M}^b{}_a:=\frac{\delta\mathcal{L}_M}{\delta\omega^a{}_b}.
\end{equation}
Lastly, since the gravitational sector is typically taken independent of $\psi$, we have $\mathbb{F}=\mathbb{M}$. With these definitions stated, the collective set of field equations, $\mathbb{F}_i=0$, may be written in the concise form:
\begin{equation}\label{general-MAG-equations}
\begin{split}
    \delta\omega^a{}_b:\quad 0&=-D\mathbb{H}^b{}_a+\mathbb{E}^b{}_a+\mathbb{M}^b{}_a,\\
    \delta e^a:\quad 0&=-D\mathbb{H}_a+\mathbb{E}_a+\mathbb{M}_a,\\
    \delta g_{ab}:\quad 0&=-D\underline{\mathbb{H}}^{ab}+\underline{\mathbb{E}}^{ab}+\underline{\mathbb{M}}{}^{ab},\\
    \delta\psi:\quad 0&=\mathbb{M}.
    \end{split}
\end{equation}
The visual symmetry of the first three expressions above is no accident. Appealing to this symmetry would lead one to consider a type of dynamical ``metric," which has the character of a co-vector-valued $1$-form. This would yield a more pleasing visual symmetry among these expressions and is, in fact, what will be discussed in the projective generalization. Furthermore, it was shown in \cite{Hehl} that either $\underline{\mathbb{F}}{}^{ab}$ or $\mathbb{F}_a$ is redundant. This provides further evidence for abandoning this asymmetric treatment of the gravitational field variables in favor of a more symmetric treatment. In other words, abandon the standard notion of treating the metric as a gravitational field variable in favor of a true gauge potential, which only descends to a metric.

\subsection{Palatini}

Briefly, we recount some of the action functionals of Sec. \ref{sec:intro-gen-rel}, now in exterior form. The transitions to exterior form are worked out explicitly in Appendix \hyperref[app-B1:abs-ind-exterior-form]{B.1}. The \textit{Palatini action} is simply the Einstein-Hilbert action in exterior form:
\begin{equation}
    S_{EH}=\frac{1}{2\kappa}\int_{\mathcal{M}}  d^4x\sqrt{-g}R=\frac{1}{2\kappa}\int_{\mathcal{M}} R^{ab}\wedge *(e\wedge e)_{ba}.
\end{equation}
In this form, the integrand is manifestly coordinate-invariant due to the use of exterior forms. This is one of the main conveniences provided by the exterior formalism. Writing the indices within the Palatini action as we have done above is the more natural option, as will be discussed later. For reasons that will also be clarified in a later section, we include the cosmological constant with the standard Palatini action. The cosmological constant in exterior form is found by a similar process, resulting in
\begin{equation}
        S_{\Lambda}=\frac{1}{2\kappa}\int_{\mathcal{M}}d^mx\sqrt{-g}(-2\Lambda)=\frac{1}{2\kappa}\int_{\mathcal{M}}\frac{\Lambda}{6} (e\wedge e)^{ab}\wedge *(e\wedge e)_{ba}.
\end{equation}
Writing the matter action in exterior form, however, would require knowing the specific form of matter considered. A more extended discussion of this is left for Part III, where we discuss projective spinor fields.

Using the field variation technology developed in the previous subsection, the total variation of $S=S_{EH} +S_{\Lambda}+S_M$ provides
\begin{equation}
\begin{split}
  \delta e^a&:\quad\;2\kappa\mathbb{M}_a=*(R_{ba}+\frac{2\Lambda}{3} e_b\wedge e_a)\wedge e^b,\\
  \delta\omega^a{}_b&:\quad2\kappa\mathbb{M}^b{}_a=D*(e\wedge e)^b{}_a,\\
  \delta g_{ab}&:\quad 2\kappa\underline{\mathbb{M}}{}^{ab}=-R^{cb}\wedge*(e\wedge e)^a{}_c,\\
  \delta\psi&:\quad \quad\quad\mathbb{M}=0.
  \end{split}
\end{equation}
The first $m$ of these $\frac{3m(m+1)+2}{2}$ field equations corresponds to the Einstein field equations with a cosmological constant in the presence of matter. The next $m^2$ equations govern the connection, from which one would extract the Levi-Civita connection as a solution. Although we will not discuss these any further, the Levi-Civita connection \textit{must} be a solution for the theory to be physically viable---provided one adheres to the idea that General Relativity is the correct low-energy description of spacetime geometry. The next $\frac{m(m+1)}{2}$ equations determine the metric. In many of the models where the the matter fields are Lorentz spinors, a Dirac-type action typically does not depend explicitly on the metric. Therefore, $\underline{\mathbb{M}}^{ab}=0$.

\subsection{Modified Palatini}
\label{sec:modified-palatini}

There exist more natural terms that may be added to the total action. To introduce these, we restrict our attention to $m=4$ dimensions. These terms were not discussed previously since, in a $V_4$, they are topological for a Levi-Civita connection. Furthermore, they are most naturally discussed in the language of exterior forms. 

The \textit{Euler} term in $4$-dimensions is topological for a Levi-Civita connection and, thus, a total derivative related to the Euler characteristic of $\mathcal{M}$. Calculated explicitly in Appendix \hyperref[app-B1:abs-ind-exterior-form]{B.1}, for a general $L_4$, the Euler term is no longer topological and has the form
\begin{equation}\label{MAG-euler}
        \mathscr{E}:=R^{ab}\wedge*R_{ba}=-d^4x|e|\left(R^{ab}{}_{[mn]} R^{mn}{}_{[ab]}-\Delta R^{mn}\Delta R_{nm}+R^2\right),
\end{equation}
where
\begin{equation}
    \Delta R^m{}_n:=R^{am}{}_{an}-\check{R}{}^{ma}{}_{an}
\end{equation}
is the difference between the Ricci and co-Ricci tensors. It simple to see, without calculation, that the Euler term in a general $L_4$ is not a total derivative. This follows from the necessity of introducing explicit metrics in its definition, which are required to contract indices after taking the internal dual. Therefore, at the very least, non-metricity will appear in addition to the total derivative.

When the symmetric part of the curvature $2$-form vanishes, i.e., when non-metricity vanishes, one finds
\begin{equation}
    R_{(ab)[mn]}=0\quad\quad\Rightarrow \quad\quad\Delta R^{mn}\Delta R_{nm}=4R^{mn}R_{nm}.
\end{equation}
Furthermore, the vanishing of torsion implies
\begin{equation}
    R_{[mn]}=0\quad\quad\Rightarrow \quad\quad4R^{mn}R_{nm}=4R^{mn}R_{mn}.
\end{equation}
Therefore, only when both torsion and non-metricity vanish does one arrive at the usual expression, the \textit{Gauss-Bonnet} term,
\begin{equation}\label{LC-GB}
    R^{ab}{}_{[mn]} R^{mn}{}_{[ab]}-\Delta R^{mn}\Delta R_{nm}+R^2\;\;\overset{Q=T=0}{\longrightarrow} \;\;\hat{R}{}^{abmn}\hat{R}_{abmn}-4\hat{R}{}^{mn} \hat{R}_{mn}+\hat{R}{}^2.
\end{equation}
It would therefore be improper to begin with the expression on the right-hand side when either torsion or non-metricity is present, as done in \cite{gen-struc}. In this limit, the $3$-form whose exterior derivative yields the right-hand side of Eq. \eqref{LC-GB} is the \textit{dual Chern-Simons $3$-form},
\begin{equation}
    \hat{\mathscr{C}}_{*}:=\hat{\epsilon}_{abcd}(\hat{\omega}{}^{ab}\wedge \hat{R}{}^{cd}-\frac{1}{3}\hat{\omega}{}^{ab}\wedge\hat{\omega}{}^c{}_e\wedge\hat{\omega}{}^{ed}).
\end{equation}
This explicitly expresses the topological nature of $\hat{\mathscr{E}}$, since
\begin{equation}
    \hat{\mathscr{E}}=d\mathscr{C}_*.
\end{equation}

Another term available in a general $L_4$ is the topological \textit{Pontrjagin} density. Also calculated explicitly in Appendix \hyperref[app-B1:abs-ind-exterior-form]{B.1}, the Pontrjagin density has the form \cite{mielke-geometro,pontryagin}
\begin{equation}
    \mathscr{P}:=R^{a}{}_b\wedge R^b{}_a=d^4x|e|R^a{}_{bmn}R^b{}_{apq}\epsilon^{mnpq},
\end{equation}
where $\epsilon^{mnpq}$, without a hat, is the Levi-Civita tensor density. A thorough discussion of $\epsilon$, $\hat{\epsilon}$, and the internal $*$-operation may be found in Appendix \hyperref[app-A:orientation]{A.2}. Adding the Pontrjagin density to the total action does not alter the field equations, regardless of whether the connection is Levi-Civita. This is because its variation serves to produce the Bianchi identities. In the language of differential forms, the Bianchi identities listed in Eqs. \eqref{LC-bianchi} generalize in an $L_4$ to \cite{Hehl,chern-bianchi}:
\begin{equation}
(0)\;\; DQ_{ab}=2R_{(ab)},\quad\quad\quad(1)\;\; DT^a=R^a{}_b\wedge\vartheta^b,\quad\quad\quad(2)\;\; DR^a{}_b=0.
\end{equation}
Notice the appearance of a new, independent Bianchi identity, $(0)$. In the Levi-Civita limit, $(0)$ and $(1)$ degenerate to the same statement, i.e., $(1)$ of Eq. \eqref{LC-bianchi}. Due to $\mathscr{P}$'s natural metric-independence, it may be written as a total derivative,
\begin{equation}
    \mathscr{P}=d\mathscr{C},
\end{equation}
where $\mathscr{C}$ is the Chern-Simons $3$-form
\begin{equation}
    \mathscr{C}:=\omega^a{}_b\wedge R^b{}_a-\frac{1}{3}\omega^a{}_b\wedge\omega^b{}_c\wedge\omega^c{}_a.
\end{equation}
Notice that the dual of $\mathscr{C}$ is equivalent to $\hat{\mathscr{C}}_{*}$ for a Levi-Civita connection, i.e., when $\mathscr{C}\rightarrow\hat{\mathscr{C}}$.

The remaining terms that often appear in gravitational theories are the \textit{Holst} \cite{holst2,holst4} and \textit{Nieh-Yan} terms \cite{nieh,nieh-yan-OG-1,generalized-nieh-yan,torsion-nieh-yan-anomaly}, the latter being of deep interest in the General Projective Gauge Gravitational Theory. These terms are not independent, as they are related in spacetimes with vanishing non-metricity by a total derivative \cite{holst-nieh-yan,holst}. The more inclusive term, the \textit{Nieh-Yan density}, is defined as
\begin{equation}
    \mathscr{N}:=R^a{}_b\wedge e^b\wedge e_a+T^a\wedge T_a,
\end{equation}
while the \textit{Holst term} is defined as
\begin{equation}
    \mathscr{H}:=R^a{}_b\wedge e^b\wedge e_a.
\end{equation}
For vanishing non-metricity, one can easily show that
\begin{equation}\label{nieh-yan-exterior-gen}
    \mathscr{N}=d(g_{ab}e^a\wedge T^b).
\end{equation}
One important feature of $\mathscr{N}$ is its role in contributing to a chiral anomaly \cite{torsion-nieh-yan-anomaly,chandia-zinelli}. However, this contribution has been debated \cite{anomaly-torsion2,mielke-anomaly3}. As we will see in the next section, both $\mathscr{P}$ and $\mathscr{N}$ may be seen as resulting from a single, higher-dimensional Pontrjagin density, though the ability to write $\mathscr{N}$ as in Eq. \eqref{nieh-yan-exterior-gen} remains. Subsequently, in the General Projective Gauge Gravitational Theory, we propose an even more fundamental version of the Pontrjagin density—one containing both $\mathscr{P}$ and $\mathscr{N}$, for which $\mathscr{N}$ may be written as an exterior derivative, independent of a vanishing non-metricity. 

This exhausts the list of fundamental terms most commonly encountered in the literature on extensions to the theory of General Relativity. In the next section, we develop the fundamentals of the Metric-Affine gauge theory framework and introduce the M\"{o}bius representation.

\section{Metric-Affine Gravity}

The Metric-Affine theory of gravitation (MAG) is the most general gauge theory of gravity which minimally extends Einstein's General Relativity with geometric deformations. MAG theories are gauge theories of the affine group, which permit a removal of the dependence of the connection on the metric and allow both torsion and non-metricity. This section focuses on the foundations of affine geometry and the corresponding gauge theory. The purpose of this section is to facilitate later comparison with the general projective framework. Therefore, we stop short of discussing specific models for MAG theories. Much of this section follows the standard approach to building MAG theories, as detailed in \cite{Hehl}.

There exist many variants of the Metric-Affine Gravitational theories. The distinguishing features are conditions on the fundamental geometric deformations of the theory, with Special Relativity being the most restrictive and MAG the least restrictive. A table of $m$-dimensional spacetime geometries and their corresponding gravitational theories is copied here, with slight modification, from \cite{spacetime-table}:
\begin{table}[h!]
    \centering
    \begin{tabular}{|c|l|c|c|c|l|} \hline  
  \textbf{Symbol}&\textbf{Spacetime}& $\bm{R}$ & $\bm{T}$ & $\bm{Q}$  &\textbf{Formulation of Gravity} \\\hline \hline
           $M_m$&Minkowski &$=0$&   $=0$&$=0$ &Special Relativity\\ \hline  
           $V_m$&Riemann (Metric) &$\neq0$& 
     $=0$&$=0$ &General Relativity\\ \hline  
  $T_m$&Weitzenb\"{o}ck& $=0$& $\neq0$&$=0$ &(Metric) Teleparallel\\ \hline 
 $S_m$& Symmetric Teleparallel & $=0$& $=0$&$\neq0$ &Symmetric Teleparallel\\ \hline 
 $Z_m$& General Teleparallel & $=0$& $\neq0$&$\neq0$ &General Teleparallel\\ \hline  
  $U_m$&Riemann-Cartan& $\neq0$& $\neq0$&$=0$ &Einstein-Cartan\\ \hline 
 $W_m$& Weyl& $\neq0$& $=0$&$\neq0$ &Weyl\\ \hline
 $L_m$& Metric-Affine & $\neq0$& $\neq0$&$\neq0$ &Metric-Affine\\\hline \end{tabular}
    \caption{Metric spacetime geometries.}
    \label{spacetime-table}
\end{table}

There certainly exist many other variations of these basic units. For example, somewhere between $U_m$ and $W_m$ is the \textit{Weyl-Cartan Geometry}, identified by $R\neq0$, $T\neq0$, and $Q_{lmn}=2Q_lg_{mn}$. Additionally, there is the Generalized Weyl spacetime, where torsion has only a vector's worth of freedom, similar to the projective torsion mode found earlier in Eq. \eqref{proj-torsion-mode}. However, there are arguments against generalized Weyl spacetimes due to the second clock effect \cite{nonmetricity-gravity-weyl-problem}. This issue is present even in MAG theories \cite{second-clock}. The $Z_m,\;U_m,\;W_m$ spacetimes comprise the so-called \textit{Geometrical Trinity of Gravity} \cite{trinity-1,trinity-gravity-capozziello}. See \cite{trinity-gravity-application} for some applications and \cite{trinity-non-rel} for the non-relativistic limit of the Geometrical Trinity. Interestingly, as pointed out in \cite{non-trinity}, obstacles to the trinity via point particle geodesics may be attributed to point particles charged under the shear defects associated with traceless non-metricity. As we will see in the General Projective Gauge Theory of Gravity, it is the projective Schouten form $\overline{\mathcal{P}}_{\underline{b}}$ that is responsible for such shear defects.

\subsection{Affine Group}

Consider the $m$-dimensional real affine group, $\text{Aff}(m,\mathbb{R})$. This is the Lie group with the semi-direct product decomposition
\begin{equation}\label{affine-group-decomposition}
    \text{Aff}(m,\mathbb{R}):=T(\mathbb{R}^m)\rtimes GL(m,\mathbb{R}),
\end{equation}
where $T(\mathbb{R}^m)$ is the group of translations of $\mathbb{R}^m$. Since this is isomorphic to the vector space $\mathbb{R}^m$ itself, we simply write $\mathbb{R}^m$. This $(m+m^2)$-dimensional group of transformations acts on an affine vector $x=(x^a)$ as
\begin{equation}\label{basic-affine-trans}
    x\rightarrow x'=Gx+\tau,
\end{equation}
where $G=\{G^a{}_b\}\in GL(m,\mathbb{R})$, $\tau=\{\tau^a\}\in\mathbb{R}^m$, and $a,b,\dots=1,\dots,m$. The semi-direct product structure is reflected in the composition of consecutive affine transformations. The composition law is given simply by group multiplication. An affine transformation $\begin{pmatrix}
    G'',&\tau''
\end{pmatrix}$ composed of two consecutive affine transformations—first with the pair $\begin{pmatrix}
    G',&\tau'
\end{pmatrix}$, then the pair $\begin{pmatrix}
    G,&\tau
\end{pmatrix}$—produces
\begin{equation}
    \begin{pmatrix}
        G'',&\tau''
    \end{pmatrix}=\begin{pmatrix}
        G,&\tau
    \end{pmatrix}\circ\begin{pmatrix}
        G',&\tau'
    \end{pmatrix}=\begin{pmatrix}
        GG',&G\tau'+\tau
    \end{pmatrix}.
\end{equation}
This complicated composition behavior is easily managed by working instead with the \textit{M\"{o}bius} representation, which will be defined shortly.

The Lie algebra, $\mathfrak{aff}(m,\mathbb{R})$, corresponding to the group decomposition of $\text{Aff}(m,\mathbb{R})$ in Eq. \eqref{affine-group-decomposition} is
\begin{equation}
    \mathfrak{aff}(m,\mathbb{R})=\mathfrak{t}(m,\mathbb{R})\oplus\mathfrak{gl}(m,\mathbb{R}).
\end{equation}
The direct product structure in the algebra follows from the semi-direct product structure of the group. The sub-algebra $\mathfrak{t}(m,\mathbb{R})$, associated with the group $T(m,\mathbb{R})$, is spanned by $\{\bm{P}_a\}$, the set of $m$ generators of $m$-dimensional translations. The sub-algebra $\mathfrak{gl}(m,\mathbb{R})$, associated with the group $GL(m,\mathbb{R})$, is spanned by the set of $m^2$ generators $\{\underline{\bm{L}}^a{}_b\}$ of $m$-dimensional linear transformations. These $m(m+1)$ generators satisfy the Lie bracket structure:
\begin{equation}\label{affine-algebra}
    \begin{split}[\underline{\bm{L}}^a{}_b,\underline{\bm{L}}^c{}_d]&=\delta^a{}_d\underline{\bm{L}}^c{}_b-\delta^c{}_b\underline{\bm{L}}^a{}_d,\\
    [\underline{\bm{L}}^a{}_b,\bm{P}_c]&=\delta^a{}_c\bm{P}_b,\\
    [\bm{P}_a,\bm{P}_b]&=0.
    \end{split}
\end{equation}

Restricting to $GL^+(m,\mathbb{R})$, the linear transformations with positive determinant, there is the isomorphic decomposition
\begin{equation}
    GL^+(m,\mathbb{R})\cong \mathbb{R}^+\times SL(m,\mathbb{R}),
\end{equation}
where $\mathbb{R}^+$ corresponds to the Abelian subgroup of $GL(m,\mathbb{R})$ consisting of elements with positive determinant, and $SL(m,\mathbb{R})$ consists of those elements of $GL(m,\mathbb{R})$ with positive unit determinant. Had the restriction to positive determinant been omitted, one would be required to consider temporal reflections $T$ with $\det(T)=-1$. The isomorphism is then modified to $GL(m,\mathbb{R})\cong\mathbb{R}^+\times \left[T\ltimes SL(m,\mathbb{R})\right]$. In either case, there is an induced splitting of the generators $\underline{\bm{L}}{}^a{}_b$ into the traceless linear transformations $\bm{L}{}^a{}_b$ and \textit{dilations} $\underline{\bm{D}}:=\underline{\bm{L}}{}^a{}_a$, given by
\begin{equation}
    \underline{\bm{L}}{}^a{}_b=\bm{L}{}^a{}_b+\frac{1}{m}\delta^a{}_b\underline{\bm{D}}.
\end{equation}
This splitting does \textit{not} require the introduction of a metric structure. From the first two Lie brackets in Eq. \eqref{affine-algebra}, we find the following commutation relations:
\begin{equation}
\begin{split}
    [\underline{\bm{D}},\underline{\bm{D}}]&=0,\\
    [\underline{\bm{D}},\bm{P}_a]&=\bm{P}_a,\\
    [\underline{\bm{D}},\bm{L}{}^a{}_b]&=0.
    \end{split}
\end{equation}

When the affine space carries a metric $g_{ab}$, a finer splitting of $\bm{L}{}^a{}_b$ is induced:
\begin{equation}
    \underline{\bm{L}}{}_{ab}:=g_{ac}\underline{\bm{L}}{}^c{}_b=\bm{A}_{ab}+\bm{S}_{ab}+\frac{1}{m}g_{ab}\underline{\bm{D}}.
\end{equation}
In the above, $\bm{A}_{ab}:=\bm{L}_{[ab]}\in\mathfrak{so}(m,\mathbb{R})$ generates the $m$-dimensional \textit{rotations} of the Lorentz group $SO(m,\mathbb{R})$, while $\bm{S}_{ab}:=\bm{L}_{(ab)}\in \mathfrak{s}(m,\mathbb{R})$ generates the $m$-dimensional \textit{shear} transformations associated with the \textit{set} of $m\times m$ symmetric traceless matrices, $\text{Sym}(\frac{m(m+1)}{2},\mathbb{R})$. The simple real Lie algebra $\mathfrak{sl}(m,\mathbb{R})$ of $SL(m,\mathbb{R})$ reflects this splitting via the decomposition
\begin{equation}
    \mathfrak{sl}(m,\mathbb{R})=\mathfrak{so}(m,\mathbb{R})\oplus\mathfrak{s}(m,\mathbb{R}).
\end{equation}
This decomposition consists of the maximal compact sub-algebra $\mathfrak{so}(m,\mathbb{R})$, the algebra of the $m$-dimensional Lorentz group, and the non-compact part $\mathfrak{s}(m,\mathbb{R})$, associated with the set of symmetric traceless matrices. This may be understood simply as the decomposition of a traceless $m\times m$ matrix into its (anti)-symmetric parts.

Following \cite{metric-affine-algebra}, with slight departures in notation, the Lie algebra $\mathfrak{aff}(m,\mathbb{R})$ admits a \textit{M\"{o}bius} representation $\rho_M$ in the form of $(m+1)\times (m+1)$ matrices. This follows from the fact that $\text{Aff}(m,\mathbb{R})$ is an $m$-dimensional matrix group. For an element $\alpha\in\mathfrak{aff}(m,\mathbb{R})$ with parameters $\alpha=\alpha(g,\tau)$, given by
\begin{equation}
    \alpha=g^c{}_d\underline{\bm{L}}{}^d{}_c+\tau^c\bm{P}_c,
\end{equation}
the representation $\rho_M(\alpha)$ reads
\begin{equation}
    \tilde{\alpha}:=\rho_M(g^c{}_d\underline{\bm{L}}{}^d{}_c+\tau^c\bm{P}_c)=\begin{pmatrix}
        g^c{}_d\rho_{m\times m}(\underline{\bm{L}}{}^d{}_c)&\tau^c\rho_m(\bm{P}_c)\\0&0
    \end{pmatrix}\equiv\begin{pmatrix}
        g^c{}_d\tilde{\underline{\bm{L}}}{}^d{}_c&\tau^c\tilde{\bm{e}}_c\\0&0
    \end{pmatrix}.
\end{equation}
In the above, $\rho_{m\times m}:\mathfrak{gl}(m,\mathbb{R})\subset\mathfrak{aff}(m,\mathbb{R})\rightarrow M_{m\times m}(\mathbb{R})$ is simply a map which takes the $m\times m$ matrix $\underline{\bm{L}}{}^d{}_c$ to itself, and $\rho_m:\mathfrak{t}(m,\mathbb{R})\subset\mathfrak{aff}(m,\mathbb{R})\rightarrow\mathbb{R}^m$ is a map which takes $\bm{P}_c$ to the basis vector $\tilde{\bm{e}}_c\in\mathbb{R}^m$. Explicitly, 
\begin{equation}\label{self-relations}
\begin{split}
    \left(\rho_{m\times m}(\underline{\bm{L}}{}^d{}_c)\right)^b{}_a&=(\tilde{\underline{\bm{L}}}{}^d{}_c)^b{}_a=\delta^d{}_a\delta^b{}_c,\\
    \left(\rho_m(\bm{P}_c)\right)^a&=(\tilde{\bm{e}}_c)^a=\delta^a{}_c.
    \end{split}
\end{equation}
Using the Lie brackets of Eq. \eqref{affine-algebra}, it is simple to show that $\rho_M$ is a linear representation of $\mathfrak{aff}(m,\mathbb{R})$, since, for any $\alpha,\beta\in\mathfrak{aff}(m,\mathbb{R})$, one has
\begin{equation}
    \rho_M([\alpha,\beta])=[\rho_M(\alpha),\rho_M(\beta)]=[\tilde{\alpha},\tilde{\beta}].
\end{equation}

Since $\mathbb{R}^m$ is the defining module of $GL(m,\mathbb{R})$, the Lie algebra $\mathfrak{gl}(m,\mathbb{R})$ has a natural action on $\mathbb{R}^m$ as a matrix. Therefore, $\mathbb{R}^m$ is the carrier space of the fundamental representation of $\mathfrak{gl}(m,\mathbb{R})$. In other words,
\begin{equation}
    \left(\rho_{m\times m}(\underline{\bm{L}}{}^d{}_c)\right)\tilde{\bm{e}}_a=\delta^d{}_a\tilde{\bm{e}}_c.
\end{equation}
Additionally, the Lie algebra of translations $\mathfrak{t}(m,\mathbb{R})$ provides a carrier space for a representation of $\mathfrak{gl}(m,\mathbb{R})$, given by
\begin{equation}
    \rho_{\mathfrak{t}(m,\mathbb{R})}(\underline{\bm{L}}{}^d{}_c)\bm{P}_b:=[\underline{\bm{L}}{}^d{}_c,\bm{P}_b]=\delta^d{}_b\bm{P}_c.
\end{equation}
With respect to $\mathfrak{gl}(m,\mathbb{R})$ transformations, this relation provides the isomorphism between $\mathbb{R}^m$ and $\mathfrak{t}(m,\mathbb{R})$, thus justifying the interchangeability of the former with the latter. 

Concisely, the M\"{o}bius representation of $\mathfrak{aff}(m,\mathbb{R})$ is
\begin{equation}
    \mathfrak{aff}(m,\mathbb{R})\simeq \left\{\left.\begin{pmatrix}g^c{}_d&\tau^c\\0&0\end{pmatrix}\right|\; g^c{}_d\in \mathfrak{gl}(m,\mathbb{R}),\tau^c\in\mathbb{R}^m\right\}.
\end{equation}
The M\"{o}bius representation of the affine group $\text{Aff}(m,\mathbb{R})$ is easily found using the exponential parameterization of group elements. Consider an affine transformation $A\in\text{Aff}(m,\mathbb{R})$ with
\begin{equation}
    A(g,\tau)=e^{\tau}e^g,
\end{equation}
where $e^{\tau}\in T(m,\mathbb{R})$ and $e^g\equiv G\in GL(m,\mathbb{R})$, with $\tau=\tau^c\bm{P}_c\in\mathbb{R}^m$ and $g=g^c{}_d\underline{\bm{L}}{}^d{}_c\in\mathfrak{gl}(m,\mathbb{R})$, respectively. The M\"{o}bius representation $\tilde{A}$ of $A$ is
\begin{equation}
    \tilde{A}:=\rho_M(A)=\rho_M(e^{\tau})\rho_M(e^g)=e^{\rho_M(\tau)}e^{\rho_M(g)}=\begin{pmatrix}
        \tilde{G}&\tilde{\tau}\\0&1
    \end{pmatrix},
\end{equation}
since
\begin{equation}
    e^{\rho_M(\tau)}:=\sum^{\infty}_{n=1}\frac{1}{n!}\begin{pmatrix}
        0&\tilde{\tau}\\0&0
    \end{pmatrix}^n=\begin{pmatrix}
        \bm{1}_{m}&\tilde{\tau}\\0&1
    \end{pmatrix},
\end{equation}
\begin{equation}\label{exponential-sum-group}
    e^{\rho_M(g)}:=\sum^{\infty}_{n=1}\frac{1}{n!}\begin{pmatrix}
        \tilde{g}&0\\0&0
    \end{pmatrix}^n=\begin{pmatrix}
        e^{\tilde{g}}&0\\0&1\end{pmatrix}=\begin{pmatrix}
        \tilde{G}&0\\0&1\end{pmatrix},
\end{equation}
with $\bm{1}_m$ representing the $m$-dimensional identity matrix. The inverse of $\tilde{A}$ is found by noting 
\begin{equation}
    A(g,\tau)^{-1}=e^{-g}e^{-\tau},
\end{equation}
and therefore,
\begin{equation}
    \tilde{A}{}^{-1}=\rho_M(A^{-1})=\begin{pmatrix}
        \tilde{G}{}^{-1}&-\tilde{G}{}^{-1}\tilde{\tau}\\0&1
    \end{pmatrix}.
\end{equation}

The M\"{o}bius representation of the affine group is thus
\begin{equation}
    \text{Aff}(m,\mathbb{R})\simeq \left\{\left.\tilde{A}=\begin{pmatrix}G&\tau\\0&1\end{pmatrix}\in GL(m+1,\mathbb{R})\right|\; G\in GL(m,\mathbb{R}),\tau\in\mathbb{R}^m\right\}.
\end{equation}
Without risking confusion, the same symbol, $\text{Aff}(m,\mathbb{R})$, is used to denote both representations of the affine group. Due to the relations in Eqs. \eqref{self-relations}, we omit the tildes on the components of $\tilde{A}$. The set of matrices $\tilde{A}=\{\tilde{A}{}^B{}_C\}$, with $A,B,\dots=1,2,\dots,m+1$, forms a subgroup of $GL(m+1,\mathbb{R})$ that leaves invariant the $m$-dimensional hyperplane
\begin{equation}
    \tilde{\mathbb{R}}{}^m:=\left\{\tilde{X}=\begin{pmatrix}x\\1\end{pmatrix}\in\mathbb{R}^{m+1}\right\}.
\end{equation}
The invariance is easily seen by executing the affine transformation
\begin{equation}
\tilde{A}:\tilde{X}\rightarrow\tilde{X}'=\tilde{A}\tilde{X}=\begin{pmatrix}
        G&\tau\\0&1
    \end{pmatrix}\begin{pmatrix}
        x\\1
    \end{pmatrix}=\begin{pmatrix}
       G x+\tau\\1
    \end{pmatrix},
\end{equation}
which further recovers the $m$-dimensional affine transformation encountered in Eq. \eqref{basic-affine-trans}. 

The algebra associated with this group of transformations is the algebra of the corresponding subgroup, $\text{Aff}(m,\mathbb{R})\subset GL(m+1,\mathbb{R})$. The algebra $\mathfrak{gl}(m+1,\mathbb{R})$ of $GL(m+1,\mathbb{R})$ is generated by the set $\{\underline{\bm{L}}{}^A{}_B\}$, satisfying the commutation relations
\begin{equation}\label{MAG-mobius-algebra}
    [\underline{\bm{L}}{}^A{}_B,\underline{\bm{L}}{}^C{}_D]=\delta^A{}_D\underline{\bm{L}}{}^C{}_B-\delta^B{}_C\underline{\bm{L}}{}^A{}_D.
\end{equation}
The affine subalgebra $\mathfrak{aff}(m,\mathbb{R})$ may be found within the above algebra by forming the identifications:
\begin{equation}\label{generator-restrictions}
    \underline{\bm{L}}{}^a{}_b=\underline{\bm{L}}{}^a{}_b,\quad\quad \underline{\bm{L}}{}^{*}{}_b=\bm{P}_b,\quad\quad \underline{\bm{L}}{}^{a}{}_{*}=0,\quad\quad \underline{\bm{L}}{}^{*}{}_{*}=0,
\end{equation}
where we use $*$ to denote the $(m+1)^{th}$ index value. 

A Killing metric may be constructed by extracting the structure constants from the above set of commutators \cite{mielke-geometro}. Explicitly, we write Eq. \eqref{MAG-mobius-algebra} as
\begin{equation}
    [\underline{\bm{L}}{}^A{}_B,\underline{\bm{L}}{}^C{}_D]=f^{ACF}{}_{DEB}\underline{\bm{L}}{}^E{}_F,
\end{equation}
where
\begin{equation}
    f^{ACF}{}_{DEB}:=\delta^A{}_D\delta^C{}_E\delta^F{}_B-\delta^B{}_C\delta^A{}_E\delta^F{}_D
\end{equation}
are the structure constants of the Lie algebra. The Killing metric is then found by tracing the product of two generators $\underline{\bm{L}}{}^A{}_B$. This is simply the ``square" of the structure constants,
\begin{equation}
    \kappa^{AC}{}_{BD}:=\frac{1}{2(m+1)}\text{Tr}\left(\underline{\bm{L}}{}^A{}_B\underline{\bm{L}}{}^C{}_D\right)=\delta^A{}_D\delta^C{}_B-\frac{1}{m+1}\delta^A{}_B\delta^C{}_D.
\end{equation}
Notice that this has the effect of \textit{projecting out the trace} of whatever object it acts on. Using this Killing metric for $\mathfrak{aff}(m,\mathbb{R})$-valued fields requires that the relations in Eqs. \eqref{generator-restrictions} are taken into account.

\subsection{Gauging the Affine Group}

To gauge the affine group, we simply upgrade the transformations to have spacetime dependence, and the group becomes infinite-dimensional \cite{Hehl}. To distinguish this group from the finite-dimensional structure group $\text{Aff}(m,\mathbb{R})$, we write
\begin{equation}
\mathcal{A}(m,\mathbb{R}):=\left\{\left.\tilde{A}=\begin{pmatrix}
    G(x)&\tau(x)\\0&1
\end{pmatrix}\right|\;G(x)\in\mathcal{GL}(m,\mathbb{R}),\tau(x)\in\mathcal{T}(m,\mathbb{R})\right\}.
\end{equation}

The general affine $\mathfrak{aff}(m,\mathbb{R})$-valued connection is defined as
\begin{equation}
    \Omega:=\begin{pmatrix}
        \omega&\vartheta\\0&0
    \end{pmatrix}=\begin{pmatrix}
        \omega^a{}_b\underline{\bm{L}}{}^b{}_a&\frac{1}{l_0}\vartheta^a\bm{P}_a\\0&0
    \end{pmatrix},
\end{equation}
where
\begin{equation}
    \Omega=\Omega^A{}_B\underline{\bm{L}}{}^{B}{}_A=\Omega^A{}_{Bm}\underline{\bm{L}}{}^{B}{}_Adx^m.
\end{equation}
A length parameter $[l_0]=L$ is introduced so that $[\vartheta]=1$. The connection $\Omega$ transforms under the local gauge action of $\tilde{A}(x)$ as
\begin{equation}
\tilde{A}^{-1}(x):\Omega\;\rightarrow\;\Omega'=\tilde{A}^{-1}(x)(\Omega+d)\tilde{A}(x).
\end{equation}
In components, with indices suppressed, we find
\begin{equation}
    \tilde{A}{}^{-1}(x):\omega\;\rightarrow\;\omega'=G^{-1}(x)(\omega+d)G(x),
\end{equation}
and
\begin{equation}\label{MAG-frame-trans}
    \tilde{A}^{-1}(x):\vartheta\;\rightarrow\;\vartheta'=G^{-1}(x)(\vartheta+D\tau(x)),
\end{equation}
where
\begin{equation}
    D\tau(x):=d\tau(x)+\omega\tau(x)
\end{equation}
is the covariant derivative of the translational parameter $\tau(x)$. This transformation behavior of $\vartheta$ does not permit its identification as the standard co-frame $e$, since $e$ is required to transform as a proper gauge vector. In other words,
\begin{equation}
    \vartheta^a=e^a\;\;\Rightarrow\;\; \vartheta'=G^{-1}(x)\vartheta.
\end{equation}

The curvature associated with $\Omega$ is
\begin{equation}
    K=d\Omega+\Omega\wedge\Omega.
\end{equation}
The components of $K$ may be accessed in the same manner as with $\Omega$,
\begin{equation}
    K=\frac{1}{2!}K^A{}_{B[mn]}\underline{\bm{L}}{}^B{}_Adx^m\wedge dx^n.
\end{equation}
The components are calculated to have the form
\begin{equation}\label{MAG-K-matrix}
    K^A{}_B=\begin{pmatrix}
       R^a{}_b&\frac{1}{l_0}T^a\\0&0
    \end{pmatrix},
\end{equation}
where
\begin{equation}
R^a{}_b=d\Omega^a{}_b+\Omega^a{}_c\wedge\Omega^c{}_b=\frac{1}{2}R^a{}_{b[mn]}dx^m\wedge dx^n
\end{equation}
is the $\mathfrak{gl}(m,\mathbb{R})$-valued curvature $2$-form, and
\begin{equation}
T^a:=D\vartheta^a=d\vartheta^a+\Omega^a{}_c\wedge\vartheta^c=\frac{1}{2}T^a{}_{[mn]}dx^m\wedge dx^n
\end{equation}
is the $\mathbb{R}^m$-valued curvature $2$-form. Interestingly, when the non-metricity vanishes and $R_{(ab)}=0$, the gauge theory reduces to that of the (anti)-de Sitter group, and
\begin{equation}
    K^A{}_B\rightarrow\begin{pmatrix}
       R^a{}_b+\frac{1}{l^2_0}\vartheta^a\wedge\vartheta_b&\frac{1}{l_0}T^a\\0&0
    \end{pmatrix},
\end{equation}
\cite{stereo-desitter,coset-vector,kk-ds-poincare,ads-higgs-gravity,Wise,pseudo-proj}. The new addition typically results in a cosmological constant in the dynamical theory. In the General Projective Gauge Gravitational Theory, this term will become metric-independent, and composed of \textit{two independent fields}. 

The curvature $K$ has the local gauge transformation behavior
\begin{equation}
    \tilde{A}^{-1}(x):K\;\rightarrow\;K'=\tilde{A}{}^{-1}(x)K\tilde{A}(x).
\end{equation}
From the definition of $K$, Eq. \eqref{MAG-K-matrix}, the component transformations are found to have the form
\begin{equation}
    \tilde{A}{}^{-1}(x):R\;\rightarrow\;R'=G^{-1}(x)RG(x),
\end{equation}
and
\begin{equation}
    \tilde{A}{}^{-1}(x):T\;\rightarrow\;T'=G^{-1}(x)\left(T+R\tau(x)\right).
\end{equation}
Exposing indices, for example, the last relation implies
\begin{equation}
    \tilde{A}{}^{-1}(x):T^a\;\rightarrow\;(T')^a=(G^{-1}(x))^a{}_b\left(T^b+R^b{}_c\tau^c(x)\right).
\end{equation}
This expression shows that $T^a$ is not yet identifiable with torsion since torsion is a proper vector-valued $2$-form, while $T^a$ above does not have this behavior under the local gauge action of $\tilde{A}{}^{-1}(x)$.

To deal with this misbehavior under local gauge transformations, we may follow the initial insight of Trautman \cite{trautman}. Introduce an $\mathbb{R}^m$-valued $0$-form $\xi=\xi^a\bm{P}_a$ via
\begin{equation}\label{traut-higgs-xi}
    \tilde{\xi}{}^A=\begin{pmatrix}
        \xi^a\\1
    \end{pmatrix}.
\end{equation}
This field is required to satisfy the local gauge transformation behavior
\begin{equation}
    \tilde{A}{}^{-1}(x):\tilde{\xi}\;\rightarrow\;\tilde{\xi}'=\tilde{A}{}^{-1}(x)\tilde{\xi}.
\end{equation}
In components, this reads
\begin{equation}
    \tilde{A}{}^{-1}(x):\xi\;\rightarrow\;\xi'=G^{-1}(x)(\xi-\tau(x)).
\end{equation}
Defining a new translational connection $e^a$ via
\begin{equation}
    \vartheta^a=e^a-D\xi^a,
\end{equation}
we may attribute the $D\tau$ term in Eq. \ref{MAG-frame-trans} as resulting from the transformation of $D\xi$. This permits the identification of $e^a$ as a proper co-frame. The parameters $\xi^a$ go by many names: Poincar\'{e} coordinates, Cartan radius vector, generalized Higgs field, Goldstone coordinates, and many others. The last two monikers follow from $\xi$'s role in ``hiding" the local translational symmetry, as well as it taking values in the coset space $\mathbb{R}^m\simeq \text{Aff}(m,\mathbb{R})/GL(m,\mathbb{R})$. By requiring $D\xi=0$, both $\omega$ and $\vartheta$ are considered \textit{soldered} to $\mathcal{M}$, and the translational symmetry is \textit{spontaneously broken}. Further requiring $\xi=0$, results in $\Omega$ assuming the status of a \textit{Cartan Connection} on the bundle of linear frames \cite{Hehl}.

We will encounter the projective generalization of such a field through the techniques provided by the \textit{nonlinear realization of symmetry groups}. For this reason, we do not dive any deeper into its role in the context of Metric-Affine theory.

\subsection{Action Functionals}

Action functionals, or rather Lagrangian $m$-form densities, are plentiful in Metric-Affine gauge theories of gravity. For this reason, we do not delve into the infinitude of options. We do, however, discuss the Chern-Simons form of the Metric-Affine Gravity theory. This choice is made simply for comparison with later investigations into the General Projective Gauge Theory of Gravitation. To facilitate understanding in the following, we restrict to $m=4$ dimensions and choose a split-signature metric with $(p,q)=(1,3)$. We again follow the instruction of \cite{Hehl}.

In the topological sector, one may form the various $GL(4,\mathbb{R})$ \textit{Chern-Simons $3$-forms} (CS). The first in the set of CS $3$-forms is entirely independent of a metric,
\begin{equation}
\begin{split}
    \mathscr{C}_{RR}&:=-\frac{1}{2}\left(\omega^a{}_b\wedge R^b{}_a-\frac{1}{3}\omega^a{}_b\wedge\omega^b{}_c\wedge\omega^c{}_a\right)\\
    &=-\frac{1}{2}\left(\omega^a{}_b\wedge d\omega^b{}_a+\frac{2}{3}\omega^a{}_b\wedge\omega^b{}_c\wedge\omega^c{}_a\right).
    \end{split}
\end{equation}
The factors of $1/2$ are included to remain consistent with \cite{Hehl}. This may be split into an $SL(4,\mathbb{R})$ part and a dilational part, denoted $\mathscr{C}_{\underline{RR}}$ and $\mathscr{C}_{\text{tr}R\;\text{tr}R}$, respectively:
\begin{equation}
    \mathscr{C}_{RR}=\mathscr{C}_{\underline{RR}}+\frac{1}{4}\mathscr{C}_{\text{tr}R\;\text{tr}R},
\end{equation}
where
\begin{equation}
    \mathscr{C}_{\underline{RR}}:=-\frac{1}{2}\left(\underline{\omega}^a{}_b\wedge\underline{ R}^b{}_a-\frac{1}{3}\underline{\omega}^a{}_b\wedge\underline{\omega}^b{}_c\wedge\underline{\omega}^c{}_a\right),
\end{equation}
and
\begin{equation}
    \mathscr{C}_{\text{tr}R\;\text{tr}R}:=-\frac{1}{2}\omega\wedge R.
\end{equation}
There does exists another, \textit{metric-dependent} CS $3$-form. This is the \textit{translational} CS $3$-form, defined as
\begin{equation}\label{translational-CS-MAG}
    \mathscr{C}_{TT}:=\frac{1}{2l_0^2}g_{ab}\vartheta^a\wedge T^b=\frac{1}{2l_0^2}g_{ab}\vartheta^a\wedge \underline{T}^b=\mathscr{C}_{\underline{TT}}.
\end{equation}
One may form a topological action with exterior derivatives of the CS $3$-forms. These are easily computed:
\begin{equation}
    \mathscr{P}_{TT}:=d\mathscr{C}_{TT}=\frac{1}{2l_0^2}\left[g_{ab}(\underline{T}^a\wedge\underline{T}^b+\underline{R}^a{}_c\wedge\vartheta^c\wedge\vartheta^b)-\underline{Q}{}_{ab}\wedge\vartheta^a\wedge\underline{T}^b\right],
\end{equation}
\begin{equation}
    \mathscr{P}_{RR}:=d\mathscr{C}_{RR}=-\frac{1}{2}R^a{}_{b}\wedge R^b{}_a=-\frac{1}{2}(\underline{R}^a{}_b\wedge\underline{R}^b{}_a+\frac{1}{4}R\wedge R),
\end{equation}
\begin{equation}
\mathscr{P}_{\text{tr}R\;\text{tr}R}:=d\mathscr{C}_{\text{tr}R\;\text{tr}R}=-\frac{1}{2}R\wedge R.
\end{equation}
The last expression above is contained in the preceding expression, which follows from separating out the volume preserving aspects. By forming the combination of $4$-forms,
\begin{equation}
    \mathscr{P}:=\mathscr{P}_{TT}+\mathscr{P}_{RR},
\end{equation}
one may utilize $\mathscr{P}$ to construct the topological action
\begin{equation}\label{boundary-pont}
S_{\mathscr{P}}:=\int_{\mathcal{M}}\mathscr{P}=\int_{\mathcal{M}}d\mathscr{C}=\int_{\partial\mathcal{M}}\mathscr{C},
\end{equation}
where
\begin{equation}
    d\mathscr{C}:=d(\mathscr{C}_{TT}+\mathscr{C}_{RR}).
\end{equation}
Eq. \eqref{boundary-pont} is simply an action functional constructed from the Pontrjagin density. Adding $S_{\mathscr{P}}$ to any other action does not alter the field equations, since the field variations of $S_{\mathscr{P}}$ are satisfied trivially. However, the field equations do, in fact, provide some topological information. In particular, they serve to produce the \textit{Generalized Bianchi identities} encountered previously \cite{chern-bianchi}:
\begin{equation}\label{MAG-bianchi-exterior}
   (0)\quad DQ_{ab}=2R^c{}_{(a}g_{b)c},\quad\quad(1)\;DT^a=R^a{}_b\wedge\vartheta^b,\quad \quad(2)\quad DR^a{}_b=0.
\end{equation}
Notice that these expressions are trivially satisfied as \textit{identities}. For example, from the definition of $T^a$, we find
\begin{equation}
    DT^a=DD\vartheta^a=R^a{}_b\wedge\vartheta^b,
\end{equation}
where Eq. \eqref{second-cov-deriv-ext} was also used.

Up to this point, we have discussed Einstein's General Theory of Relativity and some of its basic modifications and extensions. In particular, we have focused on the exterior form of the Metric-Affine Gravitational Theory in the M\"{o}bius representation. The reason for this is to facilitate later comparison with the General Projective Gauge Gravitational Theory. For the remainder of PART I, we detour to motivate the projective geometric aspects of spacetime and, therefore, of gravitational theories. This task is accomplished through the lens of the Thomas-Whitehead projective gauge theory of gravitation, which this document generalizes. We begin this journey with the string-theoretic motivation for considering projective geometry as the fundamental geometry of nature. 
\section{Thomas-Whitehead Gauge Theory of Gravity}

Projective geometry, and in particular, projective connections, have been studied by many throughout the last century \cite{roberts,roberts-2,roberts-3,projective-relativity,3d-proj-con,matveev2,tanaka,nurowski2, cartan-OG-proj-con}, to name a few. More recently, projective geometry has gained new footing in relation to classical theories of gauged gravity \cite{inflation,gen-struc,dark-e,graded-extension-TW}. 

This section provides a brief introduction to the Thomas-Whitehead gauge theory of gravity and the string-theoretic insights that motivated its construction. The Thomas-Whitehead (TW) gauge theory of gravity further serves as motivation for the General Projective Gauge Gravitational Theory, which this document proposes. The connection of the proposed generalization will be shown to reduce exactly to the TW connection in the limit of vanishing spacetime torsion—i.e., the TW (spin) connection is the General Projective gauge connection in the broken translation phase. Much of the TW review follows the construction of \cite{gen-struc}.

\subsection{String Theory}
\label{sec:string}

Here, we provide the one-dimensional string-theoretic motivation for the $m=4$-dimensional Thomas-Whitehead projective gauge theory of gravity. The particular algebraic structure was investigated in \cite{vince-coadjoint1,vince-coadjoint2} using methods developed in \cite{stringtheory1,stringtheory2,stringtheory3,stringtheory4,stringtheory5,stringtheory6,stringtheory7,stringtheory8}. 

Let $L_N\in\mathfrak{vir}$ and $J^\alpha_N\in\mathfrak{km}$ be the generators of the Virasoro and Kac-Moody algebras, respectively. The Lie algebra corresponding to the semi-direct product $\mathfrak{vir}\ltimes\mathfrak{km}$ is given by:
\begin{equation}
\begin{split}
    [L_N,L_M]&=(N-M)L_{N+M}+cN^3\delta_{N+M,0},\\
[J^\alpha_N,J^\beta_M]&=if^{\alpha\beta\gamma}J^\gamma_{N+M}+Nk\delta_{N+M,0}\delta^{\alpha\beta},\\
    [L_N,J^\alpha_M]&=-MJ^\alpha_{N+M}.
\end{split}
\end{equation}
One may choose to represent the (non-central) generators with the typical parametrization of the circle:
\begin{equation}
\begin{split}
    L_N&=ie^{iN\theta}\partial_\theta=\xi^N(\theta)\partial_\theta,\\
    J^\alpha_N&=\tau^\alpha e^{iN\theta}\partial_\theta.
\end{split}
\end{equation}
These are easily shown to satisfy the above commutation relations. Since we are interested in the centrally extended generators, the basis that realizes the entire centrally extended algebra is the triple $(L_M, J^\alpha_N,\rho)$, where $\rho$ is defined via the elements $(g,\rho)$ of the Virasoro algebra,
\begin{equation}
    (g,\rho)=g(\theta)\frac{d}{d\theta}-ia\rho.
\end{equation}
In the above, $0\leq\theta\leq2\pi$ parameterizes the circle, $a\in\mathbb{R}$, and $\rho$ is the central element.

For two vector fields $\xi$ and $\eta$, with central extensions $a$ and $b$ in the Virasoro sector, the Lie algebra is the pairing
\begin{equation}
    [(\xi,a),(\eta,b)]=\left(\xi\circ\eta,((\xi,\eta))_0\right),
\end{equation}
where the composition is defined via the Lie derivative of vector fields,
\begin{equation}
    \xi\circ\eta=\xi^a\partial_a\eta^b-\eta^a\partial_a\xi^b.
\end{equation}
The central extension is given by the covariantized Gelfand-Fuchs $2$-cocycle \cite{fuchs},
\begin{equation}
((\xi,\eta))_0=\frac{c}{2\pi}\int(\xi\eta''')d\theta=\frac{c}{2\pi}\int\xi^a\nabla_a(g^{bc}\nabla_b\nabla_c\eta^m)g_{mn}d\theta^n,
\end{equation}
where the metric tensor appearing above is $1$-dimensional. The expression above implies an invariant pairing between $\xi$ and $\eta'''$, and exemplifies invariant pairings between vector fields and quadratic differentials:
\begin{equation}
    \langle (\xi,a)|(\mathcal{D},c)\rangle=\int(\xi \mathcal{D})d\theta+ac=\int(\xi^i \mathcal{D}_{ij})d\theta^j+ac.
\end{equation}
The Gelfand-Fuchs two-cocycle is a pairing between a vector $\xi$ and a $1$-cocycle $\eta$. The latter is a projective transformation, taking a vector field to a quadratic differential,
\begin{equation}
    \eta\partial_\theta\rightarrow\eta'''d\theta^2=(\nabla_a g^{mn}\nabla_m\nabla_n\eta^bg_{bc})d\theta^a d\theta^c.
\end{equation}

To define the co-adjoint representation, one requires invariance of the pairing under the action of another centrally extended algebra element, i.e.,
\begin{equation}
    (\eta,d)*\langle(\xi,a)|(\mathcal{D},c)\rangle=0.
\end{equation}
From this, the co-adjoint representation of the Virasoro group may be stated as \cite{stringtheory1,stringtheory3}
\begin{equation}
    ad^*_{(\eta,d)}(\mathcal{D},c)=(\eta \mathcal{D}'+2\eta'\mathcal{D}-c\eta''',0).
\end{equation}
One then sees the Gelfand-Fuchs $2$-cocycle as residing in the pure gauge sector $\left(D=(0,c)\right)$ of the space of co-adjoint elements of a more general invariant $2$-cocycle relative to $D=(\mathcal{D},c)$,
\begin{equation}
    (\eta,\xi)_{(\mathcal{D},c)}=\frac{c}{2\pi}\int(\xi\eta'''-\eta'''\eta)dx+\frac{1}{2\pi}\int(\xi\eta'-\xi'\eta)\mathcal{D}dx.
\end{equation}
Furthermore, there is a one-to-one correspondence between the action of centrally extended co-adjoint Virasoro elements and the action of Sturm-Liouville operators on vector fields
\begin{equation}
    (\mathcal{D},c)\Leftrightarrow -2c\frac{d^2}{dx^2}+\mathcal{D}(x),
\end{equation}
with weight $c$ and potential $\mathcal{D}(x)$. For the homogeneously transforming co-adjoint elements, i.e., tensors, the Sturm-Liouville potential transforms inhomogeneously. The inhomogeneous contribution is exactly the Schwarzian derivative, given by the second summand in
\begin{equation}\label{diff-scwarz}
    \mathcal{D}'(x')=\mathcal{D}(x)\left(f'(x)\right)^2+\frac{c}{2}\left(\frac{f'''(x)}{f(x)}-\frac{3}{2}\left(\frac{f''(x)}{f'(x)}\right)^2\right).
\end{equation}

Returning to the semi-direct product group, the (non-central) co-adjoint elements may be realized as
\begin{equation}
\begin{split}
    \tilde{L}_N&=e^N_{ab}dx^adx^b=-ie^{-iN\theta}d\theta^2,\\
\tilde{J}^\alpha_N&=A^{N,\alpha}_adx^a=\tau^ae^{-iN\theta}d\theta.
\end{split}
\end{equation}
The pairing between arbitrary co-adjoint and adjoint elements is then
\begin{equation}
    \langle(L_A,J^\beta_B,\rho),(\tilde{L}_N,\tilde{J}^\alpha_M,\tilde{\mu})\rangle=\delta_{N,A}+\delta^{\alpha\beta}\delta_{M,B}+\rho\tilde{\mu},
\end{equation}
for arbitrary (co)-adjoint elements expressed as linear combinations of the basis elements. For $\mathfrak{vir}$, these elements are formed from \cite{stringtheory2,stringtheory3}
\begin{equation}
    \xi^j=\sum^\infty_{N=-\infty}e^j_N\xi^N,\quad\quad\quad \mathcal{D}=\sum^\infty_{N=-\infty}\mathcal{D}^n\tilde{L}_n.
\end{equation}
These have the invariant pairing
\begin{equation}
    \langle \xi,\mathcal{D}\rangle=\int\xi^i\mathcal{D}_{ij}dx^j.
\end{equation}
Similarly, for $\mathfrak{km}$, the (co)-adjoint elements are constructed from
\begin{equation}
\Lambda^I_J=\sum^q_{\alpha=1}\sum^\infty_{N=-\infty}\Lambda^N_\alpha(J^\alpha_N)^I_J,\quad\quad\quad A=\sum^M_{\alpha=1}\sum^\infty_{N=-\infty}\Lambda^n_\alpha\tilde{J}^\alpha_n,
\end{equation}
the invariant pairing of which is
\begin{equation}
    \langle\Lambda,A\rangle=\int Tr(\Lambda A_i)dx^i.
\end{equation}
Then, for an arbitrary adjoint element $(\xi(\theta),\Lambda(\theta),a)$, with vector field $\xi$, gauge parameter $\Lambda$, and central element $a$, along with the arbitrary co-adjoint element, we have
\begin{equation}
    (\xi(\theta),\Lambda(\theta),a)*(\mathcal{D}(\theta),A(\theta),\tilde{\mu})=(\delta \mathcal{D}(\theta),\delta A(\theta),0).
\end{equation}
This is the most general form of the co-adjoint action, formed by linear combinations of the action for basis vectors. The terms on the right-hand side of the equality are easily computed \cite{BF-vince}:
\begin{equation}\label{diff-string-trans-1}
\begin{split}
    \delta A&=\underbrace{A'\xi+\xi'A}_{\text{coord. trans.}}-\underbrace{[\Lambda A-A\Lambda]+k\tilde{\mu}\Lambda'}_{\text{gauge trans.}},\\
    &\text{ }\\
    \delta \mathcal{D}&=\underbrace{2\xi'\mathcal{D}+\xi \mathcal{D}'+\frac{c\tilde{\mu}}{2\pi}\xi'''}_{\text{coord. trans.}}-\underbrace{Tr(A\Lambda')}_{\text{gauge trans.}}.
\end{split}
\end{equation}

The transformation $\delta A$ may be seen, in higher dimensions, to represent a Yang-Mills potential. The part corresponding to transformations of coordinates is easily recognized as a transformation of the spatial component of a $1$-form. The gauge transformation may then be identified with a Gauss Law-constraint. The transformation of the Diffeomorphism field $\mathcal{D}$ has a much less obvious interpretation. Its odd behavior under coordinate transformations reveals that the lift to higher dimensions will not produce a tensor, as discussed in \cite{vince2d-4d}, but rather may be identified with a component of the Thomas-Whitehead (TW) connection. 

As we will find in the next few subsections, the coordinate transformation behavior of the $1$-dimensional Diffeomorphism field, when lifted to $m$ dimensions, is
\begin{equation}\label{diff-string-trans-2}
    \mathcal{D}_{p'q'}=\frac{\partial x^m}{\partial x^{p'}}\frac{\partial x^n}{\partial x^{q'}}\mathcal{D}_{mn}+\frac{1}{m+1}\left(\frac{\partial^2 \log |J|}{\partial x^{p'}\partial x^{q'}}-\frac{\partial \log |J|}{\partial x^{r'}}\Pi^{r'}{}_{p'q'}-\frac{1}{m+1}\frac{\partial \log |J|}{\partial x^{p'}}\frac{\partial \log |J|}{\partial x^{q'}}\right).
\end{equation}
In the above, $|J|:=\det (J^{m'}{}_n)$ is the determinant of the Jacobian of the transformation, and $\Pi$, defined shortly, is the fundamental projective invariant. The transformation of $\mathcal{D}$, although not obvious from the above expression, is nothing but the transformation of the components of a co-vector-valued one-form. The correspondence between the $m$-dimensional and $1$-dimensional transformations can be found in \cite{gen-struc}. 

In the next section, we follow the standard development of the higher dimensional classical gravitational theory associated with this string-theoretic construction. In particular, we will find that $\mathcal{D}$, along with $\Pi$, comprise the projective components of an affine connection over a specific $m+1$-dimensional space, the volume bundle.

\subsection{Projective Gauge Symmetry}

Consider a parameterized curve $c(\sigma)$ with tangent vector $\xi^m=\tfrac{dx^m}{d\sigma}$. For $\xi^m$ to be considered geodesic, the way in which it changes when parallel transported along $c(\sigma)$ must be proportional to itself,
\begin{equation}\label{geo1}
    \frac{D\xi^m}{d\sigma}=f(\sigma)\xi^m.
\end{equation}
In this expression, $\frac{D}{d\sigma}$ is the intrinsic derivative operator, which acts on vector fields $v^a$ as
\begin{equation}
    \frac{Dv^m}{d\sigma}=\frac{dv^m}{d\sigma}+\Gamma^m{}_{pq}v^p\xi^q.
\end{equation}
Expanding Eq. \eqref{geo1} leads to the geodesic equation (with pseudo-force),
\begin{equation}
    \frac{d^2x^m}{d\sigma^2}+\Gamma^m{}_{pq}\frac{dx^p}{d\sigma}\frac{dx^q}{d\sigma}=f(\sigma)\frac{dx^m}{d\sigma}.
\end{equation}

One typically wishes to parameterize curves, when possible, with respect to an affine parameter. An affine parameter must always be specified with respect to some connection. Suppose $u$ is such an affine parameter with respect to the connection $\nabla$. This implies
\begin{equation}
    \frac{dx^m}{du}\nabla_mu=1.
\end{equation}
A suitable choice for the function $u(\sigma)$ will render the geodesic equation \textit{geodetic},
\begin{equation}\label{geo2}
    \frac{d^2x^m}{du^2}+\Gamma^m{}_{pq}\frac{dx^p}{du}\frac{dx^q}{du}=0.
\end{equation}
One may also consider this process in reverse, albeit via some other means. Suppose one began with the geodetic equation, Eq. \eqref{geo2}. Then, consider a connection $\breve{\Gamma}$, related to $\Gamma$ via a \textit{symmetric} projective transformation by some field $k_p$,
\begin{equation}\label{proj-trans}
\Gamma^m{}_{pq}=\breve{\Gamma}^m{}_{pq}+\delta^m{}_pk_q+\delta^m{}_qk_p.
\end{equation}
The geodetic equation becomes
\begin{equation}
    \frac{d^2x^m}{du^2}+\Gamma^m{}_{pq}\frac{dx^p}{du}\frac{dx^q}{du}=\left(-2k_p\frac{dx^p}{du}\right)\frac{dx^m}{du}.
\end{equation}
The pseudo-force—the scalar function in parenthesis—may be removed by a suitable reparameterization, effectively reversing the steps above. This establishes the equivalence of symmetric projective transformations and geodesic curve reparametrizations. Since both connections admit the same geodesic curves, they are said to be projectively equivalent, $\breve{\Gamma}^m{}_{pq}\sim\Gamma^m{}_{pq}$. The respective connections then belong to the same projective equivalence class, $\{\Gamma,\breve{\Gamma}\}\in[\nabla]$. 

Recalling the statements of Sec. \ref{sec:sym-proj-trans}, this interplay between reparameterizations and symmetric projective transformations is intimately tied to coordinate transformations. Therefore, a gauge theory built on projective transformations is a gauge theory built on diffeomorphisms \cite{gauge-from-diff-polyakov}. In original work by T.Y. Thomas \cite{thomas-1, thomas-2}, the beginning of such a gauge theory of projectively equivalent connections was established. Thomas' construction begins with the definition of the \textit{fundamental projective invariant} $\Pi^a{}_{bc}$, given as
\begin{equation}
    \Pi^m{}_{pq}:=\Gamma^m{}_{pq}-\tfrac{1}{m+1}\left(\delta^m{}_{p}\Gamma^n{}_{nq}+\delta^m{}_{q}\Gamma^n{}_{np}\right).
\end{equation}
This connection is, by construction, invariant under the projective transformation of Eq. \eqref{proj-trans} and, moreover, is traceless. Using the language developed previously, $\Pi$ is traceless with respect to both natural traces, 
\begin{equation}
    \Pi^m{}_{mq}=\Pi^m{}_{pm}=0.
\end{equation}

The naive approach would be to construct a projectively invariant geodetic equation and investigate particle dynamics. However, a closer look at $\Pi^m{}_{pq}$ reveals it is not a connection in the conventional sense. This is made explicit by considering how it behaves under a general coordinate transformation $\{x^n\}\rightarrow \{x^{m'}(x^n)\}$,
\begin{equation}\label{pi-trans}
    \Pi{}^{m'}{}_{p'q'}=J^{m'}{}_n\left(\Pi^n{}_{rs}J{}^r{}_{p'}J{}^s{}_{q'}+\frac{\partial J{}^n{}_{q'}}{\partial x^{p'}}\right)-j_{p'}\delta^{m'}{}_{q'}-j_{q'}\delta^{m'}{}_{p'}.
\end{equation}
In the above, 
\begin{equation}
    J^{m'}{}_n:=\frac{\partial x^{m'}}{\partial x^n},\quad\quad\quad j_n:=-\frac{1}{m+1}\frac{\partial \log|J|}{\partial x^n}=-\frac{1}{m+1}J^p{}_{q'}\partial_n J^{q'}{}_p,
\end{equation}
have been used to denote, respectively, the Jacobian of the transformation and its logarithmic derivative. The conventional definition $|J|:=\det(J^{m'}{}_n)$ is also utilized. The latter summand in Eq. \eqref{pi-trans} is the term that spoils covariance. However, for special (volume preserving) transformations with $|J|=1$, it is evident that $\Pi^m{}_{pq}$ behaves properly as a connection under general coordinate transformations. The covariance-spoiling term in Eq. \eqref{pi-trans} hints at the deep interplay between gauge and coordinate transformations, the former appearing only when $|J|\neq1$. 

The presence of $|J|$ in the transformation behavior of the connection implies a relationship between the structure of the gauge group and spacetime volumes. This relationship is made concrete via Thomas' construction of a fiber bundle, which is sometimes called the \textit{Thomas Cone}. Here, however, it will be referred to concretely as the volume bundle of the manifold, $V\mathcal{M}$. 

\subsection{Volume Bundle}

To describe the volume bundle, we follow the works of \cite{proj-conn} and \cite{roberts}. Consider the volume bundle of an $m$-dimensional spacetime manifold $\mathcal{M}$. This is simply the bundle of volume forms over $\mathcal{M}$, denoted $V\mathcal{M}$. Consider $\mathcal{E}=T^*\mathcal{M}-\{0\}$, the slit cotangent bundle of $\mathcal{M}$. To construct $V\mathcal{M}$, one takes the $m^{th}$ exterior derivative of $\mathcal{E}$ and enforces the equivalence relation $-\epsilon \sim \epsilon$ on forms $\epsilon\in\mathcal{E}$. Choosing $\{x^n\}=\{ x^1,...,x^{m}\}$ as a basis for $\mathcal{M}$, sections of $V\mathcal{M}$ may then be expressed as
\begin{equation}
    \pm v(x)dx^1\wedge...\wedge dx^{m},
\end{equation}
where $v:\mathcal{M}\rightarrow \mathbb{R}$ is a nowhere vanishing smooth function on $\mathcal{M}$. By construction, it is easily observed that $V\mathcal{M}$ is a line bundle over $\mathcal{M}$ with an $\mathbb{R}^+$-valued fiber. As a manifold, $V\mathcal{M}$ is of dimension $m+1$. Coordinates on $V\mathcal{M}$ are $\{x^M\}:=\{x^m,\;\lambda\}$, where
\begin{equation}
    \lambda=|v|^{\tfrac{1}{m+1}}
\end{equation}
is chosen to represent the $\mathbb{R}^+$-valued fiber coordinate, in accordance with \cite{proj-conn}. 

The fundamental projective invariant may then be lifted to the $\mathcal{M}$-components of an affine connection on $V\mathcal{M}$. The torsion-free \textit{Thomas-Whitehead} (TW) connection is this affine connection on $V\mathcal{M}$, with components
\begin{equation}\label{gamma-tilde}
\begin{split}
    \tilde{\Gamma}^m{}_{pq}&=\Pi^m{}_{pq},\\
    \tilde{\Gamma}^\lambda{}_{pq}&=\lambda\mathcal{D}_{pq},\\
    \tilde{\Gamma}^m{}_{p\lambda}&=\tilde{\Gamma}^m{}_{\lambda p}=\frac{1}{\lambda}\delta^m{}_p.
\end{split}
\end{equation}
The \textit{Diffeomorphism field} $\mathcal{D}_{pq}$ is a rank-$(0,2)^0$ symmetric (non-tensor) field on $\mathcal{M}$, and $\Pi^m{}_{pq}$ is defined in terms of \textit{any} representative member of the projective equivalence class $\Gamma\in[\nabla]$. However, as shown by Thomas \cite{thomas-1,thomas-2}, $\Pi^m{}_{pq}$ may be considered a ``connection" in its own right. The Diffeomorphism field, motivated in the previous section, generalizes the original work of Thomas, as well as the more recent investigations of TW connections \cite{proj-conn,roberts,proj-vs-metric,2-frame}. In those works, the Diffeomorphism field is not independent but rather may be expressed in terms of the equi-projective curvature tensor—the curvature attributed to $\Pi$. In the present case, however, this field is assumed independently dynamical. For later comparison with the generalization of this theory, we present $\tilde{\Gamma}$ in matrix form. The components of this matrix are generally organized as
\begin{equation}
    \tilde{\Gamma}{}^P{}_{QM}=\begin{pmatrix}
        \begin{pmatrix}
            \tilde{\Gamma}{}^p{}_{qm}&\tilde{\Gamma}{}^p{}_{\lambda m}\\\tilde{\Gamma}{}^{\lambda}{}_{qm}&\tilde{\Gamma}{}^{\lambda}{}_{\lambda m}
        \end{pmatrix},\;\begin{pmatrix}
            \tilde{\Gamma}{}^p{}_{q\lambda}&\tilde{\Gamma}{}^p{}_{\lambda \lambda}\\\tilde{\Gamma}{}^{\lambda}{}_{q\lambda}&\tilde{\Gamma}{}^{\lambda}{}_{\lambda \lambda}
        \end{pmatrix}
    \end{pmatrix}.
\end{equation}
Specifically, the TW affine connection has the matrix representation
\begin{equation}
    \tilde{\Gamma}{}^P{}_{QM}=\begin{pmatrix}
        \begin{pmatrix}
            \Pi{}^p{}_{qm}&\frac{1}{\lambda}\delta^p{}_m\\\lambda\mathcal{D}_{qm}&0
        \end{pmatrix},\;\begin{pmatrix}
          \frac{1}{\lambda}\delta^p{}_m&0\\0&0
        \end{pmatrix}
    \end{pmatrix}.
\end{equation}

Now consider a coordinate transformation on $V\mathcal{M}$, such that
\begin{equation}
    \{x^M\}\rightarrow\{x^{M'}\}=\{x^{m'}(x^n),\;\lambda'=\lambda|J|^{\frac{-1}{m+1}}\},
\end{equation}
with $|J|:=\text{det}(J^{m'}{}_n)$. Notice that a mixing of $\{\lambda\}$ into $\{x^m\}$ is not permitted, but a mixing of $\{x^m\}$ into $\{\lambda\}$ is. This is exemplified by the explicit component matrix of the Jacobian of transformation on $V\mathcal{M}$,
\begin{equation}\label{jacobian-old}
    J^{M'}{}_N=\begin{pmatrix}
        J^{m'}{}_n&0\\x^{*'}j_n&|J^{m'}{}_n|^{\frac{-1}{m+1}}
    \end{pmatrix}.
\end{equation}
Since $\tilde{\Gamma}$ is required to be an affine connection on $V\mathcal{M}$, it must transform as such. In other words,
\begin{equation}
\tilde{\Gamma}^M{}_{PQ}\rightarrow\tilde{\Gamma}^{M'}{}_{P'Q'}=J^{M'}{}_N\left(\tilde{\Gamma}^N{}_{RS}J^R{}_{P'}J^S{}_{Q'}+\frac{\partial J^N{}_{P'}}{\partial x^{Q'}}\right).
\end{equation}
From this transformation law, one may expose the corresponding transformation behavior of $\mathcal{D}_{pq}$, 
\begin{equation}
\mathcal{D}_{pq}\rightarrow\mathcal{D}_{p'q'}=\left(\mathcal{D}_{mn}-\partial_nj_m+\Pi^l{}_{mn}j_l-j_mj_n\right)J^m{}_{p'}J^n{}_{q'}.
\end{equation}
This is nothing but the higher-dimensional analogue of the transformation of the Virasoro co-adjoint element—the second expression in Eqs. \eqref{diff-string-trans-1}, lifted and expanded in Eq. \eqref{diff-string-trans-2}. As will be discussed later, the coordinate transformation behavior of $\mathcal{D}_{pq}$ may be identified with the local action of $GL(m,\mathbb{R})$, the $m$-dimensional general linear group.

\subsection{Geodesics}

With the TW connection so defined, the geodesic equation is revisited. Consider a geodetic tangent vector $\tilde{\xi}^M=\frac{dx^M}{du}$ on $V\mathcal{M}$. The parameter $u$ is therefore affine with respect to $\tilde{\nabla}$, and
\begin{equation}
    \tilde{\xi}^M\tilde{\nabla}_M\tilde{\xi}^N=0.
\end{equation}
This results in the following set of generally covariant and projectively invariant ``geodesic" equations on $\mathcal{M}$:
\begin{equation}\label{geo-diff}
\begin{split}
    \frac{d^2x^m}{du^2}+\Pi^m{}_{pq}\frac{dx^p}{du}\frac{dx^q}{du}&=-2\frac{1}{\lambda}\frac{d\lambda}{du}\frac{dx^m}{du},\\
    \frac{d^2\lambda}{du^2}+\lambda \mathcal{D}_{pq}\frac{dx^p}{du}\frac{dx^q}{du}&=0.
\end{split}
\end{equation}
An affine parameter for $\tilde{\Gamma}$ is not necessarily an affine parameter for $\Pi$. Suppose there exists a parameter $\tau$ that is affine with respect to $\Pi$. Let $u\rightarrow \tau(u)$, such that
\begin{equation}
    \frac{d^2\tau}{du^2}=-2\frac{1}{\lambda}\frac{d\lambda}{du}\frac{d\tau}{du}.
\end{equation}
This particular choice for $\tau(u)$ renders the first of Eqs. \eqref{geo-diff} geodetic and implies that
\begin{equation}
    \frac{d^2\lambda}{du^2}=\frac{\lambda}{4}\cdot\frac{3(\tfrac{d^2\tau}{du^2})^2-2(\tfrac{d^3\tau}{du^3})\tfrac{d\tau}{du}}{(\tfrac{d\tau}{du})^2}.
\end{equation}
Substituting this relation into the second expression in Eq. \eqref{geo-diff}, one finds that
\begin{equation}\label{schwarzian}
    \mathcal{D}_{pq}\frac{dx^p}{du}\frac{dx^q}{du}=\frac{1}{2}\cdot\frac{(\tfrac{d^3\tau}{du^3})\tfrac{d\tau}{du}-\frac{3}{2}(\tfrac{d^2\tau}{du^2})^2}{(\tfrac{d\tau}{du})^2}\equiv\frac{1}{2}S(\tau:u).
\end{equation}
The $S(\tau:u)$ appearing above is the Schwarzian derivative of $\tau$ with respect to $u$. Notice that this is just Eq. \eqref{diff-scwarz} for the transformation of the Sturm-Liouville potential to a system where it vanishes. 

When the Diffeomorphism field vanishes, one may consider the base manifold $\mathcal{M}$ to be flat projective space. This produces the typical linear fractional transformations as the permitted \textit{reparameterizations}. There exists a reparameterization $u\rightarrow\tau(u)$ to an affine parameter of $\Pi$, which is of particular importance. In the limit the Diffeomorphism field becomes a constant pure trace field, $\mathcal{D}_{pq}=\frac{-1}{\lambda_0^2}g_{pq}$, with $[\lambda_0]=L$, the affine parameter for $\tilde{\Gamma}$ on $V\mathcal{M}$ is related to an affine parameter for $\Pi$ on $\mathcal{M}$, via
\begin{equation}
\tau(u)=\lambda_0\tan^{-1}\left(\lambda_0a_1(u+a_2)\right)+a_3,
\end{equation}
where the $a_i$ are constants. This is found from choosing a frame where $\{x^1\}=\tau$ and solving Eq. \eqref{schwarzian} for $\tau(u)$. For $u\simeq\epsilon$ or $\lambda_0\rightarrow \epsilon$, with $\epsilon <<1$, there exists a region where $\tau(u)$ is linear,
\begin{equation}
    \tau(u)\simeq b_1u+b_2,
\end{equation}
with $b_i=b_i(\lambda_0,a_i)$. Therefore, a sufficiently large value ($\lambda_0<<1$) defining the vacuum state of the Diffeomorphism field produces a linear relation between the affine parameters of $\tilde{\Gamma}$ and $\Pi$. This state will be of particular interest in later sections of this document. 

Before continuing, it must be noted that there currently exists no known point particle action whose variation results in Eqs. \eqref{geo-diff}.

\subsection{Fundamental Projective Geometric Fields}
\label{sec:fundamental-proj-fields-TW}

The volume bundle $V\mathcal{M}$ possesses a fundamental projective vector field. In a particular gauge, this may be expressed as $\Upsilon\equiv\lambda\partial_\lambda$. Explicitly,
\begin{equation}\label{OG-upsilon-TW}
    \Upsilon^M=\begin{pmatrix}
        0^m\\\lambda
    \end{pmatrix},
\end{equation}
where $0^m:=(0,\;0,\dots,0)^T$ denotes the $m$-dimensional zero vector. The fundamental projective vector field satisfies the compatibility relation
\begin{equation}\label{compatibility}
    \tilde{\nabla}_M\Upsilon^N=\delta^N{}_M,
\end{equation}
and therefore also satisfies the geodesic equation with unit proportionality,
\begin{equation}
    \Upsilon^M\tilde{\nabla}_M\Upsilon^N=\Upsilon^N.
\end{equation}
The fundamental projective vector field may be used to lift vector fields $\chi$ from $\mathcal{M}$ to $V\mathcal{M}$. The lift $\tilde{\chi}$ is accomplished by writing
\begin{equation}
\begin{split}
    \tilde{\chi}&=\tilde{\chi}{}^M\partial_M\\
    &=\chi^m\partial_m-(\lambda\chi^m\kappa_m)\partial_\lambda\\
    &=\chi^m\partial_m-(\chi^m\kappa_m)\Upsilon.
\end{split}
\end{equation}
The field $\kappa_m$ appearing above is required to transform with the addition of $j_m$ under changes of coordinates, similar to the connection traces in, for example Eq. \eqref{intro-conn-trans-lil-j},
\begin{equation}
    \kappa_{m'}=\left(\kappa_n+j_n\right)J^n{}_{m'}.
\end{equation}
The components of $\tilde{\chi}$ are therefore,
\begin{equation}
    \tilde{\chi}{}^M=\begin{pmatrix}
        \chi^m\\-\lambda\chi^m\kappa_m
    \end{pmatrix}.
\end{equation}
More generally, this may be written as
\begin{equation}
    \tilde{\chi}{}^M=\begin{pmatrix}
        \chi^m_{\|}\\\lambda(\chi_{\bot}-\kappa_m\chi^m_{\|})
    \end{pmatrix},
\end{equation}
for components parallel ($\chi_{\|}$) and perpendicular ($\chi_{\bot}$) to $\mathcal{M}$. The latter are unrelated to vectors on $\mathcal{M}$. One may therefore say that in this particular gauge, $\Upsilon$ is purely transverse to $\mathcal{M}$, i.e., $\Upsilon_{\|}=0$. 

A similar construction exists for lifting one-forms $\eta_m$ from $\mathcal{M}$. This results in the components
\begin{equation}
    \tilde{\eta}_M=\begin{pmatrix}
        \eta_m+\kappa_m&\frac{1}{\lambda}
    \end{pmatrix}.
\end{equation}
The result of lifting geometric fields to $V\mathcal{M}$ is not unique due to the arbitrariness of one's choice of $\kappa_m$. It will be shown later in this document that this arbitrariness arises from one's choice of field parameterizing a particular coset space associated with what will be called \textit{pseudo-translations}, or the $\mathbb{R}^m_{*}$ coset. Additionally, we will consider $\Upsilon$ as the \textit{Generalized Projective Higgs Field} associated with the general projective geometry, analogous to the MAG field $\tilde{\xi}$ in Eq. \eqref{traut-higgs-xi}. Notice that the natural inner product on $V\mathcal{M}$ satisfies, for example,
\begin{equation}
    \tilde{\eta}_M\tilde{\chi}{}^M=\eta_m\chi^m_{\|}+\chi_{\bot}.
\end{equation}
In much of \cite{gen-struc}, lifted vectors are taken to have $\chi_{\bot}=0$ in order to preserve inner products. In other words, scalars on $V\mathcal{M}$ project to scalars on $\mathcal{M}$. In the General Projective Gauge Gravitational Theory proposed in this document, we will, however, adopt the view that $\chi_{\bot}=0$ is not permissible.

Assume the existence of a pseudo-Riemannian metric ${g_{mn}}$ defined on $\mathcal{M}$, and fix a convention by choosing $\kappa_m\equiv g_m$. The object $g_m$, initially defined in Eq. \eqref{lil-g-intro-def}, may also be defined in any of the following ways:
\begin{equation}
\begin{split}\label{little-g}
    g_n&=\tfrac{-1}{m+1}\hat{\Gamma}^k{}_{kn}\\
    &=\tfrac{-1}{m+1}\partial_n\log\sqrt{|g|}\\
    &=\tfrac{-1}{2(m+1)}g^{pq}\partial_ng_{pq}\\
    &=\tfrac{-1}{2(m+1)}g^{pq}\nabla^{\Pi}_ng_{pq}\\
    &=\tfrac{1}{m+1}\Tr(|e|^{-1}\partial_n|e|).
\end{split}
\end{equation}
These expressions result from $\Tr(A^{-1}dA)=d\log|A|$, the tracelessness of $\Pi^m{}_{pq}$, and the relationship between general and flat metrics via the inverse or co-frame fields: $g_{mn}=e^{\underline{a}}{}_{m}e^{\underline{b}}{}_{n}\eta_{\underline{ab}}$. The underlined indices denote objects which reside in the flat Minkowski tangent spaces, i.e., they are \textit{Lorentz} indices. Interestingly, $g_n$ may be identified with the Weyl co-vector associated with $\Pi$.

The choice of $\kappa_m\equiv g_m$ makes most natural, the definition of a solder metric $G_{MN}$ on $V\mathcal{M}$, given by
\begin{equation}\label{big-g}
    G_{MN}=\begin{pmatrix}
        g_{mn}-\lambda_0^2g_mg_n&-\frac{\lambda_0^2}{\lambda}g_m\\
        \frac{-\lambda_0^2}{\lambda}g_n&-\frac{\lambda_0^2}{\lambda^2}
    \end{pmatrix},
\end{equation}
which contains only the degrees of freedom inherent to the metric of $\mathcal{M}$. The constant $[\lambda_0]=L$ is introduced to ensure that $G_{MN}$ is dimensionless. The inverse metric is easily found to have the form
\begin{equation}\label{big-g-inv}
    G^{MN}=\begin{pmatrix}
        g^{mn}&-\lambda g^{mp}g_p\\-\lambda g^{qn}g_q&\frac{\lambda^2}{\lambda_0^2}(-1+\lambda_0^2g^{pq}g_pg_q)
    \end{pmatrix}.
\end{equation}
Since $G_{MN}$ contains only metric degrees of freedom, it is invariant under symmetric projective transformations of the connection. Under a transformation of coordinates on $V\mathcal{M}$, the metric $G_{MN}$ furnishes the correct behavior as an $(0,2)^0$-tensor field. In addition, the volume-form of $V\mathcal{M}$ remains invariant under changes of coordinates,
\begin{equation}
    \sqrt{|G(x)|}dx^M=\sqrt{|G'(x')|}d^{M'}.
\end{equation}
From the expression in Eq. \eqref{big-g}, the metric determinants of $\mathcal{M}$ and $V\mathcal{M}$ are related via
\begin{equation}
    |G|=|g|\frac{\lambda_0^2}{\lambda^2}.
\end{equation}
As we will see, the above factor of $\lambda_0/\lambda$, which results from taking the square root, will contribute a logarithmic rescaling of coupling constants in the dynamical theory once integrated out.

\subsection{Projective Spin Connection}

With the $V\mathcal{M}$ metric defined, a type of co-frame may be extracted \cite{gen-struc}. By introducing a flat metric,
\begin{equation}
    \eta_{\underline{AB}}=\text{diag}(1,-1,-1,-1,-1),
\end{equation}
and supposing that $G_{MN}$ may be expressed as
\begin{equation}
    G_{MN}=\eta_{\underline{AB}}\tilde{e}{}^{\underline{A}}{}_M\tilde{e}{}^{\underline{B}}{}_N,
\end{equation}
one may solve these relations for $\tilde{e}$. Doing so provides
\begin{equation}\label{TW-frames}
    \tilde{e}{}^{\underline{A}}{}_M=\begin{pmatrix}
        e^{\underline{a}}{}_m&0\\\lambda_0g_m&\frac{\lambda_0}{\lambda}
    \end{pmatrix},\quad\quad 
    (\tilde{e}{}^{-1})^M{}_{\underline{A}}=\begin{pmatrix}
        (e^{-1})^m{}_{\underline{a}}&0\\-\lambda g_m(e^{-1})^m{}_{\underline{a}}&\frac{\lambda}{\lambda_0}
    \end{pmatrix},
\end{equation}
where we have included $\tilde{e}{}^{-1}$ for completeness. These matrices hint at the existence of a \textit{more general} projective theory of gravitation. Indeed, it is exactly these matrices which inspired the question: What happens to the Thomas-Whitehead theory when the vanishing components of $\tilde{e}$ are nonzero? As we will find, for a genuine projective gauge theory, the associated \textit{projective} frames must account for the equivalence class required by the projective structure. From the components in Eqs. \eqref{TW-frames}, we find that, in general, $\tilde{e}$ does not account for such equivalence class. 

Although it will not be used explicitly in the discussion of TW gravity, the \textit{TW spin connection} may now be formed from $\tilde{e}$, $\tilde{e}{}^{-1},$ and $\tilde{\Gamma}$. The TW spin connection is defined as the parallel transport of the basis-forms $\tilde{e}{}^{-1}$ with respect to $\tilde{\Gamma}$ as
\begin{equation}
    \tilde{\omega}{}^{\underline{A}}{}_{\underline{B}M}=\tilde{e}{}^{\underline{A}}{}_N\tilde{\nabla}_M(\tilde{e}{}^{-1})^N{}_{\underline{B}}.
\end{equation}
The most intuitive way to view the components of $\tilde{\omega}$ is to first contract with a basis form $dx^M$ to create the connection $1$-form $\tilde{\omega}{}^{\underline{A}}{}_{\underline{B}}:=\tilde{\omega}{}^{\underline{A}}{}_{\underline{B}M}dx^M$. The reason for this is that when presented as a list of components, such as in \cite{gen-struc,heavy-lifting}, components of $\tilde{\omega}$ which belong to the same fundamental geometric object are obscured and treated as separate. We thus provide the intuitive form of $\tilde{\omega}$ as
\begin{equation}\label{tw-omega}
    \tilde{\omega}{}^{\underline{A}}{}_{\underline{B}}=\tilde{g}\delta^{\underline{A}}{}_{\underline{B}}+\begin{pmatrix}
        \overline{\omega}{}^{\underline{a}}{}_{\underline{b}}&\frac{1}{\lambda_0}e^{\underline{a}}\\\lambda_0\overline{\mathcal{P}}_{\underline{b}}&0
    \end{pmatrix}.
\end{equation}
In the above, we find the $1$-form version of the manifestly covariant and projectively invariant \textit{Projective Schouten tensor} $\overline{\mathcal{P}}_{mn}$,
\begin{equation}\label{diff-schout-tw-spin}
\overline{\mathcal{P}}_{\underline{b}}:=\overline{\mathcal{P}}_{mn}(e^{-1})^m{}_{\underline{b}}dx^n=\left(\mathcal{D}_{mn}-\partial_ng_m+g_l\Pi^l{}_{mn}-g_mg_n\right)(e^{-1})^m{}_{\underline{b}}dx^n,
\end{equation}
\cite{covariant-tw}. This object will be discussed in more depth in the next subsection and throughout the remainder of this document. We also encounter the manifestly covariant and projectively invariant lift of $g_m$ to $V\mathcal{M}$,
\begin{equation}
    \tilde{g}:=\tilde{g}_Mdx^M=\begin{pmatrix}
        g_m,&\frac{1}{\lambda}
    \end{pmatrix}.
\end{equation}
Lastly, we find the $m$-dimensional spin connection coefficients, which are also manifestly covariant and projectively invariant,
\begin{equation}\label{barred-omega-TW}
    \begin{split}
        \overline{\omega}{}^{\underline{a}}{}_{\underline{b}}&:=\left(e^{\underline{a}}{}_{m}\overline{\nabla}_n(e^{-1})^m{}_{\underline{b}}\right)dx^n\\
        &=e^{\underline{a}}{}_{m}\left(\partial_n(e^{-1})^m{}_{\underline{b}}+\overline{\Gamma}{}^m{}_{ln}(e^{-1})^l{}_{\underline{b}}\right)dx^n,
    \end{split}
\end{equation}
where
\begin{equation}\label{barred-gamma-TW}
    \overline{\Gamma}{}^{m}{}_{ln}:=\Pi^m{}_{ln}-\delta^m{}_lg_n-\delta^m{}_ng_l
\end{equation}
is the covariant and projectively invariant spacetime connection of $\mathcal{M}$. Again, these will not be used explicitly in the review of TW gravity, but we display these relations for later comparison with the generalization of this projective theory.

Written this way, Eq. \eqref{tw-omega} makes explicitly clear what fundamental geometric object gives $\tilde{\omega}$ a non-vanishing trace, i.e., a non-vanishing Weyl co-vector. In the torsional generalization of TW theory proposed in this document, in order to construct a \textit{truly projective} theory, the spin connection $\tilde{\omega}$ is required to be traceless. This follows from the definition of projective geometric fields residing in a formal projective space. Essentially, the determinant function from which $\tilde{g}$ is constructed multiplies the projective geometric fields, creating the equivalence class necessary for their definition as projective fields in a projective space. This removes $\tilde{g}$ from $\tilde{\omega}$, forcing the latter to be traceless. As we will see, this requirement forces every object with group indices to contain an overall multiplicative factor of said determinant function, raised to a power determined by the number of upper and lower group indices, as well as the dimension of the space. All of this will be made explicit in Part II.

\subsection{Thomas-Whitehead Gravity}

We now review the conventional approach to the Thomas-Whitehead gauge theory of gravity. This discussion primarily follows the approach and notations of \cite{gen-struc,dark-e,inflation}, with slight modifications in the presentation to facilitate later comparison with the torsional generalization. The curvature $2$-form is first constructed and used to build an action functional. The action is then reduced by imposing certain simplifying assumptions, and the ensuing implications are briefly discussed.

\subsubsection{Curvature Tensor}

The torsion-free TW connection on $V\mathcal{M}$ may be used to define the curvature tensor via the action of the commutator of covariant derivatives. For example,
\begin{equation}
    [\tilde{\nabla}_P,\tilde{\nabla}_Q]V^M=\mathcal{K}^M{}_{NPQ}V^N.
\end{equation}
The convention chosen here is
\begin{equation}
\mathcal{K}^M{}_{NPQ}=\partial_P\tilde{\Gamma}^M{}_{NQ}-\partial_Q\tilde{\Gamma}^M{}_{NP}+\tilde{\Gamma}^M{}_{LP}\tilde{\Gamma}^L{}_{NQ}-\tilde{\Gamma}^M{}_{LQ}\tilde{\Gamma}^L{}_{NP}.
\end{equation}
Using the components of $\tilde{\Gamma}$, defined in Eq. \eqref{gamma-tilde}, $\mathcal{K}^M{}_{NPQ}$ may be easily evaluated to have the components
\begin{equation}\label{equi-projective}
    \mathcal{K}^M{}_{NPQ}=\begin{pmatrix}
        \mathcal{R}^m{}_{npq}+\delta^m{}_{[p}\mathcal{D}_{q]n}&0\\\lambda\nabla^\Pi_{[p}\mathcal{D}_{q]n}&0
    \end{pmatrix}.
\end{equation}
In this expression, $\mathcal{R}^m{}_{npq}=\mathcal{R}^m{}_{npq}(\Pi)$ is the equi-projective curvature of Thomas \cite{thomas-1,thomas-2}, and $\nabla^\Pi$ is the covariant derivative with respect to the fundamental projective invariant $\Pi$. 

Although $\mathcal{K}^M{}_{NPQ}$ is projectively invariant and transforms as a $(1,3)^0$-tensor on $V\mathcal{M}$, the projectively invariant objects $\Pi^m{}_{pq}$ and $\mathcal{D}_{pq}$ do not transform as a connection and tensor, respectively. It will therefore be convenient to introduce a new geometric object that transforms appropriately. The dynamical projective Schouten tensor $\mathcal{P}_{mn}$ is related to the Diffeomorphism field via
\begin{equation}\label{P-def}
    \mathcal{P}_{mn}=\mathcal{D}_{mn}-\partial_n\alpha_m+\Gamma^l{}_{mn}\alpha_l+\alpha_m\alpha_n,
\end{equation}
where
\begin{equation}
    \alpha_k:=\frac{-1}{m+1}\Gamma^n{}_{nk}
\end{equation}
is the ``internal" natural trace of the underlying connection on the $m$-dimensional manifold $\mathcal{M}$, Eq. \eqref{alpha-beta-definition}. This new object carries the same information as $\mathcal{D}_{mn}$ and transforms homogeneously under changes of coordinates, i.e., as a proper $(0,2)^0$-tensor. The trade-off for manifest tensorality is that $\mathcal{P}_{mn}$, unlike $\mathcal{D}_{mn}$, is not projectively invariant. Under the projective transformation of Eq. \eqref{proj-trans}, $\mathcal{P}_{mn}$ changes according to
\begin{equation}\label{P-proj-trans}
\delta_{pp}:\mathcal{P}_{mn}\rightarrow\mathcal{P}_{mn}+\nabla_nk_m-k_mk_n.
\end{equation}

Substituting out the Diffeomorphism field in favor of the dynamical projective Schouten tensor, one finds the curvature components take the form
\begin{equation}
    \mathcal{K}^M{}_{NPQ}=\begin{pmatrix}
        R^m{}_{npq}+\delta^m{}_{[p}\mathcal{P}_{q]n}-\delta^m{}_n\mathcal{P}_{[pq]}&0\\\lambda\left(\nabla_{[p}\mathcal{P}_{q]n}-\alpha_l\mathcal{K}^l{}_{npq}\right)&0
    \end{pmatrix}.
\end{equation}
Here, $R^m{}_{npq}=R^m{}_{npq}(\Gamma)$ is the curvature of the underlying spacetime connection, and the covariant derivative is built from the same linear affine connection $\Gamma$. Either of the equivalent presentations of $\mathcal{K}^M{}_{NPQ}$ remains generally covariant and projectively invariant, since each component contains the necessary compensating terms. The antisymmetric part of $\mathcal{P}_{mn}$ may easily be found from Eq. \eqref{P-def}. Since $\mathcal{D}_{mn}$ and $\Gamma^l{}_{mn}$ are, by construction, symmetric, one finds
\begin{equation}\label{P-asym}
    \mathcal{P}_{[mn]}=-\partial_{[m}\alpha_{n]}.
\end{equation}
This suggests that when $\alpha$ is exact, $\mathcal{P}_{mn}$ is symmetric. A most restrictive condition for which $\alpha$ is exact is when the connection is Levi-Civita, $\Gamma=\hat{\Gamma}$, resulting in $\alpha_n=g_n$.

For completeness, one may further define an object that is both homogeneously transforming and projectively invariant. This object only \textit{appears} to be gauge-related to the Diffeomorphism field and projective Schouten tensor. However, as discussed in \cite{2-frame}, the relation to $\mathcal{D}_{mn}$ is no gauge transformation at all. The object to which we refer will be encountered extensively in the generalized setting and has the form
\begin{equation}
\begin{split}
\overline{\mathcal{P}}_{mn}&=\mathcal{D}_{mn}-\partial_ng_m+\Pi^l{}_{mn}g_l-g_mg_n\\
    &=\mathcal{P}_{mn}-\partial_n\mbox{\textcent}_m+\Gamma^l{}_{mn}\mbox{\textcent}_l-\mbox{\textcent}_m\mbox{\textcent}_n,
\end{split}
\end{equation}
with $\mbox{\textcent}$ defined in Eq. \eqref{cent-def-1}. From the above statement, we find that when the connection is Levi-Civita, $\alpha_n=g_n$ and
\begin{equation}
    \overline{\mathcal{P}}_{mn}=\mathcal{P}_{mn}.
\end{equation}

Contractions of the curvature relevant to conventional TW theory are the Ricci tensor,
\begin{equation}
    \mathcal{K}_{NQ}=\delta^P{}_M\mathcal{K}^M{}_{NPQ}=\delta^n{}_N\delta^q{}_Q\left(R_{nq}+m\mathcal{P}_{qn}-\mathcal{P}_{nq}\right),
\end{equation}
and the Ricci scalar,
\begin{equation}
    \mathcal{K}=\mathcal{K}_{NQ}G^{NQ}=R+(m-1)\mathcal{P}.
\end{equation}
In the above, $R=R_{mn}g^{mn}$ and $\mathcal{P}=\mathcal{P}_{mn}g^{mn}$ denote metric, or unnatural traces. For notational convenience, one may also introduce a field strength-like object for $\mathcal{P}_{mn}$, the so-called \textit{projective Cotton-York tensor},
\begin{equation}
K_{npq}=g_M\mathcal{K}^M{}_{npq}=\nabla_{[p}\mathcal{P}_{q]n}+\mbox{\textcent}_m\mathcal{K}^m{}_{npq}.
\end{equation}
This will appear explicitly in only the TW action, which we construct in the following subsection.

\subsubsection{Projective Action}

The action functional defining the theory of Thomas-Whitehead dynamical projective connections has been chosen as \cite{gen-struc}
\begin{equation}
    S_{TW}=S_{PEH}+S_{PGB}.
\end{equation}
The first term in the sum is an Einstein-Hilbert-like action constructed from the projective Ricci scalar $\mathcal{K}$ as
\begin{equation}
    S_{PEH}=\frac{-1}{2\tilde{\kappa}_0\lambda_0}\int d^{m+1}x\sqrt{|G|}\mathcal{K}.
\end{equation}
The last term in $S_{TW}$ is a modification of the usual $4$-dimensional Gauss-Bonnet term, extended to an arbitrary $(m+1)$-dimensional $V\mathcal{M}$. Specifically,
\begin{equation}
    S_{PGB}=\frac{\tilde{J}_0c}{\lambda_0}\int d^{m+1}x\sqrt{|G|}\left(\mathcal{K}^M{}_{NPQ}\mathcal{K}_M{}^{NPQ}-4\mathcal{K}^{NP}\mathcal{K}_{NP}+\mathcal{K}^2\right).
\end{equation}
The constants $\tilde{J}_0$ and $\tilde{\kappa}_0$ are related to the angular momentum of the universe and Newton's constant, respectively \cite{dark-e}. Recalling the discussion in Sec. \ref{sec:modified-palatini}, it is clear that $S_{PGB}$ \textit{cannot} be written in exterior form. Since the exterior formalism is used in the generalization of this projective gravitational theory, $S_{TW}$ will be abandoned. Additional factors of $\lambda^{-1}_0$ have been inserted to ensure the expression has the correct units of action. 

The $S_{PGB}$ part of the total action may be expanded to give
\begin{equation}\label{TW-action-no-expand}
\begin{split}
    S_{PGB}&=\frac{J_0c}{\lambda_0}\int d^mx\sqrt{|g|}\left(\mathcal{K}^m{}_{npq}\mathcal{K}_m{}^{npq}-4\mathcal{K}_{np}\mathcal{K}^{np}+\mathcal{K}^2-\lambda_0^2\mathcal{K}_{npq}\mathcal{K}^{npq}\right).
\end{split}
\end{equation}
This expression shows that $S_{PGB}$ may be split into an $m$-dimensional Gauss-Bonnet-like term and an additional ``curvature-squared"-like term for the field strength of $\mathcal{P}_{mn}$. All the $\lambda$-dependence in $S_{TW}$ conveniently factors out as an overall multiplicative factor, which can be integrated out. The coupling constants are then rescaled as a result of integrating over $\lambda$:
\begin{equation}\label{grav-constants-rescale}
    J_0:= \tilde{J}_0\int^{\lambda_f}_{\lambda_i}\frac{d\lambda}{\lambda}= \tilde{J}_0\log\left(\frac{\lambda_f}{\lambda_i}\right),\quad\quad\quad\quad\frac{1}{\kappa_0}:=\frac{1}{\tilde{\kappa}_0}\int^{\lambda_f}_{\lambda_i}\frac{d\lambda}{\lambda}=\frac{\log\left(\frac{\lambda_f}{\lambda_i}\right)}{\tilde{\kappa_0}}.
\end{equation}
This implies that combinations of the form $J_0\kappa_0$ are insensitive to the rescaling which results from integration over $\lambda$.

Expanding the total action reveals
\begin{equation}
\begin{split}
    S_{TW}&=\frac{J_0c}{\lambda_0}S^m_{GB}-\frac{1}{2\kappa_0}\int d^mx \left(R+m_1\mathcal{P} \right)\\
    &\quad+2\lambda_0^2J_0c\int d^mx\left((\nabla_q\mathcal{P}_{mn})^2-(\nabla^q\mathcal{P}^{mn})(\nabla_n\mathcal{P}_{qm})\right)\\
    &\quad-J_0c\int d^mx\left(m_1^2\mathcal{P}^2+2m_1R\mathcal{P}-4m_{3/2}m_1\mathcal{P}_{pq}\mathcal{P}^{pq}-8m_{3/2}R^{pq}\mathcal{P}_{pq}\right).
\end{split}
\end{equation}
Here, $m_k\equiv m-k$ is a numerical factor dependent on the dimension $m$. The term $S^m_{GB}$ takes the form of the standard Gauss-Bonnet term, Eq. \eqref{LC-GB}, even though a Levi-Civita connection has yet to be specified. For $m=4$, this term is proportional to the Euler characteristic of the manifold. As discussed in Sec. \ref{sec:modified-palatini}, in $m=4$-dimensions equipped with a Levi-Civita connection, it can be written as a total derivative and thus does not effect the equations of motion. Following \cite{gen-struc}, we choose $m=4$ and omit this term in what follows. In the expanded form of $S_{TW}$, we observe the emergence of the standard Einstein-Hilbert action, along with kinetic and potential terms for the projective Schouten tensor (Diffeomorphism field) $\mathcal{P}_{mn}$, as well as interactions between $\mathcal{P}_{mn}$ and the curvature. Additionally, the term linear in $\mathcal{P}$ suggests the dynamical generation of a cosmological constant---a dynamical source for dark energy.

\subsubsection{Some Analysis}

According to \cite{einstein-metrics,BGG}, when $\hat{\nabla}\in[\nabla]$, meaning the Levi-Civita connection is a member of the projective equivalence class, the projective Schouten tensor reduces to a constant multiple of the metric. In other words, when the spacetime is considered \textit{Einstein}, the projective Schouten tensor is proportional to the metric. As demonstrated in \cite{dark-e}, one may shift $\mathcal{P}_{mn}$ by a multiple of the metric to effectively expand about this Levi-Civita, or Einstein state.

Impose the Levi-Civita condition $\nabla_lg_{mn}=0$ at the level of the action. Then, consider a decomposition of the dynamical aspects of $\mathcal{P}_{mn}$ into trace and traceless parts:
\begin{equation}
    \mathcal{P}_{mn}=\frac{1}{\lambda_0}\phi g_{mn}+W_{mn},\quad\quad\quad\quad W_{mn}g^{mn}=0.
\end{equation}
The constant factor of $[\lambda_0^{-1}]=M$ has been introduced to ensure $[\phi]=M$. While the assumption of linear trace ($\phi^1$) is maintained here, the generalized setting will favor a quadratic trace ($\phi^2$) as the more dimensionally-natural choice. However, the particular choice of $\lambda_0^{-1}$ allows for the identification of a canonical kinetic term for $\phi$. Implementing this decomposition in the action gives
\begin{equation}\label{expanded-TW-action}
\begin{split}
    S_{TW}&=\frac{-1}{2\kappa_0}\int d^4x\sqrt{|g|}\left((1+\frac{r_0^2}{\lambda_0}\phi)\hat{R}\right)+S_{W,\phi}\\
    &\quad+12J_0c\int d^4x\sqrt{|g|}\left(\frac{1}{2}(\partial_m\phi)^2-\frac{4}{r^2_0\lambda_0}\phi-\frac{2}{\lambda_0^2}\phi^2\right),
\end{split}
\end{equation}
where the rescaling-invariant combination $r^2_0=8J_0c\kappa_0$, with $[r^2_0]=L$, has been introduced. The term $S_{W,\phi}$ encompasses all contributions involving $W_{mn}$, which are ignored under the current assumptions. The second line clearly contains a canonical kinetic and potential term for the trace degree of freedom, $\phi$. The presence of a term linear in $\phi$ suggests a field redefinition. Define the potential energy function
\begin{equation}\label{P-potential}
    V(\phi):=\frac{4}{r^2_0\lambda_0}\phi+\frac{2}{\lambda_0^2}\phi^2.
\end{equation}
Solving for the extremum, $\frac{dV}{d\phi}|_{\phi_0}=0$, we find that
\begin{equation}
    \phi_0=-\frac{\lambda_0}{r^2_0}.
\end{equation}
The factor of $\lambda_0$ is a relic of the assumed form for the trace of $\mathcal{P}_{mn}$. Expanding about this \textit{negative} vacuum state, with $\phi(x)=\Phi(x)+\phi_0$, the action simplifies to
\begin{equation}
\begin{split}
    S_{TW}&=\frac{-1}{2\kappa_0}\int d^4x\sqrt{|g|}\left(\frac{r^2_0}{\lambda_0}\Phi\hat{R}-2\Lambda\right)\\
    &+12J_0c\int d^4x\sqrt{|g|}\left(\frac{1}{2}(\partial_m\Phi)^2-\frac{M_\Phi^2}{2}\Phi^2\right),
    \end{split}
\end{equation}
where
\begin{equation}
    \Lambda=\frac{3}{r^2_0}=\frac{3}{8J_0c\kappa_0},\quad\quad\quad\quad\quad M_\Phi=\frac{2}{\lambda_0}.
\end{equation}
These results match those obtained in \cite{dark-e} for the $\mathcal{P}$-induced cosmological constant $\Lambda$. However, in this context, they are interpreted as an expansion of $\phi$ about its potential minimum. The value of $\Lambda$ also agrees with the solutions to the the field equations in \cite{gen-struc}, which are omitted here for brevity. It is worth noting that additional contributions to $\Lambda$ may exist, which would render this result incomplete.

A mass term has emerged for the scalar mode $\Phi$, distinct from the cosmological constant $\Lambda$. Similar dimensional reduction schemes, driven by a scalar field acquiring a nonzero vacuum value, have been investigated previously (e.g.,\cite{dim-redux}). In \cite{inflation}, $\phi$ is treated as a non-Higgs inflaton-like scalar. The term \textit{non-Higgs} here refers to $\phi$ having no direct connection to the Standard Model Higgs field. However, its role as \textit{a} Higgs field remains an open possibility. Using a slow roll approximation and fitting to Planck, BICEP2, and Keck Array data, a range of possible values for $\lambda_0$ was determined,
\begin{equation}
    10^3<\lambda_0M_p<10^6,
\end{equation}
where $M_p\simeq1.22 \times 10^{19}\;\text{GeV}$ is the Planck mass. This suggests that $M_\Phi$ is on the order of
\begin{equation}
    10^{13}\;\text{GeV}<M_\Phi< 10^{16}\;\text{GeV}.
\end{equation}
For later convenience, we also convert this range for $\lambda_0$ into a metrical length scale, providing
\begin{equation}
    10^{-31}\;\text{m}<\lambda_0<10^{-28}\;\text{m}.
\end{equation}

The ideas discussed thus far will be reformulated and generalized in Part II of this document. The forthcoming perspective will align more closely with that of Cartan and Veblen, while still incorporating the fundamental insights of T.Y. Thomas. The proposed generalization allows for torsional deformations of $\mathcal{M}$, effectively extending the underlying $GL(m,\mathbb{R})$ transformations of $\mathcal{M}$ to full $\text{Aff}(m,\mathbb{R})$ transformations, thereby completing the projective linear group. 

\pagebreak
\thispagestyle{empty}
\phantomsection
\addcontentsline{toc}{section}{\hspace{-1.5em}\textbf{PART II: PROJECTIVE GRAVITY}}
\begingroup
\renewcommand{\addcontentsline}[3]{}
\vspace*{5cm}
\noindent
\makebox[\textwidth]{\Huge \textbf{PART II:}}\vspace{+3em}
\makebox[\textwidth]{\Huge \textbf{PROJECTIVE GRAVITY}}
\vfill
\label{part-II}
\endgroup

\pagebreak

\section{Introduction}
\label{sec:PII-intro}

Projective gravitational theories have existed since the inception of general relativity itself. From Veblen and friends \cite{projective-relativity,veblen-generalization-quad-form,det-power}, Schouten and van Dantzig \cite{schouten,schouten-2,vandantzig}, Pauli \cite{pauli-5homo-1,pauli-5homo-2,pauli-mayer-1,pauli-mayer-2}, and Lessner \cite{projective-gravity-lessner}, to Yano and Ohgane \cite{projective-gravity-old}, Cartan \cite{cartan-OG-proj-con}, Ludwig \cite{proj-rel-ludwig}, Evans and Sen \cite{homo-map}, Hoffmann \cite{hoffmann}, and Gibbons and Warnick \cite{projective-dark-e}. More generally, interest in the field of projective geometry has grown in recent years, evidenced by the development of projective models of consciousness \cite{projective-consciousness}, and the rewriting of Special Relativity in the language of projective geometry \cite{projective-special-relativity}. There have also been attempts at using projective spaces as a means of unification, such as the Projective Unified Field theory suggested by Schmutzer \cite{puft} (for an overview of unification schemes, see \cite{unified-gravity-review,unified-history,unified-history-2}. Higher dimensional conformal unified descriptions of gravity have also been of interest, such as in \cite{conformal-unified-gravity}. These conformal theories, such as in holography \cite{weyl-holography}, differ from the projective setting, in that the latter includes (symmetric) shear defects.

Some gauge gravitational theories appear very closely related to the projective setting when attempting to extend the (Anti)-de Sitter and Poincar\'{e} models, for example, \cite{almost-proj-ds}. Furthermore, the Kaluza-Klein-type model of Ehresmann connections studied in \cite{kk-ds-poincare} attempts to generalize from the Poincar\'{e} and (Anti)-de Sitter gauge models by introducing a variable length scale. The square of this field is utilized to relax into a cosmological constant. The projective formalism naturally contains an analogous field, the projective Schouten tensor, and is central to our investigation.

In Part II of this document, we propose yet another variant: The General Projective Gauge Theory of Gravity. Within this construction, there appear possible limiting situations for which every formulation of gravity outlined in Table \ref{spacetime-table} has a projective extension. Such an identification, however, appears to require that one abandon the standard concept of a gravitational potential, i.e., spacetime metric tensor $g_{mn}$, in favor of a dynamical projective linear gauge connection-form. A particular component of this connection, the projective Schouten form, may be viewed as dynamically providing access to a conformal class of spacetime metrics, offering a symmetrical treatment of all fundamental gravitational fields. For this reason, the fundamental field strength of the gravitational potential is rather attributed to the projective Cotton-York $2$-form, and the non-metricity as a related, secondary concept. These ideas appear to be shared by \cite{wilczek}, where, in the early universe, there does not exist any well-defined notion of spatial and temporal distances. This point of view rather emphasizes the role played by spacetime volumes, wherein the condensation of a preferred spacetime volume results in well-defined spacetime length intervals. We thus suggest a new, \textit{projective} table of spacetimes to be investigated:
\begin{table}[ht!]
    \centering
    \begin{tabular}{|c|l|c|c|c|}\hline  
  \textbf{Symbol}&\textbf{Spacetime}& $\bm{\mathcal{R}}$& $\bm{\mathcal{T}}$& $\bm{\mathcal{S}}$\\\hline \hline
           $PM_m$&$\mathfrak{p}$-Minkowski&$=0$&   $=0$&$=0$ \\ \hline  
           $PV_m$&$\mathfrak{p}$-Riemann&$\neq0$& 
     $=0$&$=0$ \\ \hline  
  $PT_m$&$\mathfrak{p}$-Weitzenb\"{o}ck& $=0$& $\neq0$&$=0$ \\ \hline 
 $PS_m$& $\mathfrak{p}$-Symmetric Teleparallel& $=0$& $=0$&$\neq0$ \\ \hline 
 $PZ_m$& $\mathfrak{p}$-General Teleparallel& $=0$& $\neq0$&$\neq0$ \\ \hline  
  $PU_m$&$\mathfrak{p}$-Riemann-Cartan& $\neq0$& $\neq0$&$=0$ \\ \hline 
 $PW_m$& $\mathfrak{p}$-Weyl& $\neq0$& $=0$&$\neq0$ \\ \hline
 $PL_m$& $\mathfrak{p}$-Metric-Affine& $\neq0$& $\neq0$&$\neq0$ \\\hline \end{tabular}
    \caption{Projective spacetime geometries.}
    \label{proj-spacetime-table}
\end{table}

Our main focus will be the Projective, or $\mathfrak{p}$-Metric-Affine Gravitational theory, associated with an affine-projective spacetime geometry $PL_m$. As pointed out in \cite{Hehl}, the proper affine group, on which the Metric-Affine theory is built, cannot be obtained by means of contracting the projective linear group. For this reason, we will instead refer to $PL_m$ as the \textit{General Projective Gravitational Theory}.

Generalizing the gauge theory side of the projective gravitational theory discussed in \cite{gen-struc} to include torsion is of interest for a multitude of reasons. In the non-projective setting, for example, the coupling of torsion to spinors has been shown to result in big-bounce cosmologies \cite{torsion-spin-big-bounce}. Additionally, torsion has been utilized in attempts to explain inflation and dark energy \cite{poincare-inflaton,torsion-dynamical-dark-e}. Constructing a gauge theory that supports both dynamical torsion and dynamical non-metricity, treated on equal footing, may aid in understanding their observable effects. The observable effects of torsion were investigated in \cite{observable-effects-torsion}, and the observable effects of non-metricity, which are of most interest in the context of the general projective theory, were studied in \cite{observable-nonmetricity}. In \cite{aspects1}, these observable effects were attributable to a dynamical matter field, which, in the projective setting, appears to have an analogous field---the projective Schouten tensor.

Part II is organized as follows. We begin with a review of projective spaces, taking as a starting point the concept of a curve or material path. Associated with these projective spaces of paths are affine spaces, whose elements are of a particular type. We then develop the group of transformations acting on elements of the projective spaces and discuss the geometrical meaning of each constituent transformation through the affine lens. We then develop the associated projective algebra. This is shown to permit a contraction, which results in an affine-adjacent algebra. The volume bundle is reconstructed, wherein the local linear transformations of $\mathcal{M}$ are generalized to affine transformations. We then develop a map between the projective space and the volume bundle, and derive conditions which result in the associated projective structure. As a result, a scalar field's worth of redundancy arises and is shown to induce a non-vanishing projective Schouten tensor. The corresponding $(m+1)$-dimensional affine connection is identified as the novel \textit{projective symmetric teleparallel connection}.

Removing the inertial frame condition, we introduce a torsional generalization of the projective linear gauge connection. Following the prescription of treating the metric tensor as secondary, we review and utilize the technique of nonlinearly realizing the local projective symmetry group over the Lorentz subgroup. This is also known as the coset construction. This process provides access to a flat reference space in which we may construct inner products and, therefore, non-topological actions. We discuss the application of this technique to the general projective gauge connection, its curvature, and the Goldstone-Lorentz metric. The fundamental projective fields are then constructed and reinterpreted as generalized Higgs fields. Successive derivatives of the generalized Higgs fields are shown to produce the frame field and connection of the volume bundle. 

We then redevelop the field variation technology relevant to the General Projective Gravitational Theory and interpret some of the variational objects encountered. From there, two basic action functionals are constructed in order to describe the dynamics associated with a projective gravitational system: the projective Pontrjagin and projective Lovelock actions. The former is shown to contain a metric-independent, projectively invariant generalization of the Nieh-Yan form and, upon variation, produces the generalized projectively invariant Bianchi identities. Features of the non-topological projective Lovelock action are discussed, and field variations found for two distinct lines of reasoning: both dynamical and non-dynamical (pure gauge) generalized Higgs vector. We briefly investigate the solution space of both conceptual sectors. In the exterior formalism, it is found that a degenerate \textit{projective} frame is unavoidable, signaling a change in topology.

We note that the present construction is distinct from the torsional projective geometries studied in \cite{kobayashi-proj-conn} and \cite{heavy-lifting}, the latter being a proposal for a type of torsional generalization of Thomas-Whitehead projective connections. The distinction is that here we construct a \textit{gauge theory} of gravity by following the standard gauge prescription, assigning an algebra-valued connection one-form to each generator of the local gauge group. The present construction and \cite{heavy-lifting}, for example, may agree in some aspects, but are nonetheless foundationally distinct.

\section{Projective Geometry}
\label{sec:proj-geo}

\subsection{Homogeneous Coordinates and Projective Space}
\label{subsec:homo-coord}

This subsection is a review of homogeneous coordinates and projective spaces following \cite{proj-geo-n-dim}, modified with information from \cite{nakahara}. A thorough review of projective differential geometry is provided in \cite{proj-geo-old-and-new}. Following the introductory material in Sec. \ref{sec:intro-fundamentals}, consider an $m$-dimensional spacetime manifold $\mathcal{M}$, with coordinates $\{y^m\}$. A map $c(\tau):(a,b)\rightarrow\mathcal{M}$, with $\tau\in(a,b)$ an open interval, defines a parameterized curve in $\mathcal{M}$. Let $f:\mathcal{M}\rightarrow\mathbb{R}$ be a smooth invertible function. The directional derivative of $f(c(\tau))$ at $\tau=0$,
\begin{equation}
    \left.\frac{d(f(\tau))}{d\tau}\right|_{\tau=0},
\end{equation}
defines a vector at the point $c(0)\in\mathcal{M}$. If two curves $c_1(\tau)$ and $c_2(\tau)$ exist such that $c_1(0)=c_2(0)$ gives the same point $p_{\mathcal{M}}$ of $\mathcal{M}$, and their directional derivatives produce the same vector, then $c_1(t)$ is equivalent to $c_2(\tau)$, i.e., $c_1(\tau)\sim c_2(\tau)$. A tangent vector is then defined as the equivalence class of curves, $[c(t)]$. The collection of equivalence classes of curves at $p_{\mathcal{M}}$, i.e., the set of all vectors at that point, forms an $m$-dimensional vector space $T_p\mathcal{M}$, the tangent space of $\mathcal{M}$ at the point $p_{\mathcal{M}}$. A basis for $T_p\mathcal{M}$ is provided by the set of holonomic vectors $\{\partial_m\}$, with
\begin{equation}
    \partial_m:=\frac{\partial}{\partial y^m}.
\end{equation}
Let the vector $\{x\}:=\{\frac{dy^m}{d\tau}\partial_m\}$ denote an element of $T_p\mathcal{M}$, with components $\{x^m\}\equiv\{\frac{dy^m}{d\tau}\}$ in the coordinate basis. This provides the interpretation that elements of $T_p\mathcal{M}$ may be viewed as representing parameterized displacements from the point $p_{\mathcal{M}}$. However, by releasing from $T_p\mathcal{M}$ its point of contact with $\mathcal{M}$, i.e., exclude vanishing displacements from $p_{\mathcal{M}}$, we may view the resulting space as the $m$-dimensional affine space
\begin{equation}\label{affine-space-def}
    \mathcal{A}_m:=\biggl\{\{x\}\in T_p\mathcal{M}\;\Bigg| \;\{x\}\neq0\biggr\}
\end{equation}
consisting of the nonzero \textit{points} $\{x\}$. Points have no addition operation, but may be translated by vectors, relative to other points. Since there is no distinguished origin in $\mathcal{A}_m$, only relative displacements are accessible. For ease, we will often use the vector $\{x\}$ and its components $\{x^m\}$ interchangeably.

Consider a point $p\in\mathcal{A}_m$ of an $m$-dimensional real affine space $\mathcal{A}_m$, with coordinates $p=\{x^m\}$ and $m,n,\dots=1,2,\dots,m$. Consider also an ordered $(m+1)$-tuple of real numbers of the form $P=\{X^M\}$, with $M,N,\dots=1,2,\dots,m+1$, and for which any one of the $\{X^i\}\neq0$, with $i\in\{1,2,\dots,m+1\}$. We here fix a convention by arbitrarily assigning $\{X^{m+1}\}\neq0$. The points $p$ and $P$ are said to be in correspondence, $P\equiv p$, when
\begin{equation}\label{gen-homo-to-inhomo}
    \{x^1\}=\left\{\frac{X^1}{X^{m+1}}\right\}, \quad\{x^2\}=\left\{\frac{X^2}{X^{m+1}}\right\},\quad\dots,\quad\{x^m\}=\left\{\frac{X^m}{X^{m+1}}\right\}.
\end{equation}
Therefore, to each $P$ there corresponds one and only one point of $\mathcal{A}_m$, with coordinates $\{x^m\}\equiv\{\frac{X^m}{X^{m+1}}\}$. However, to each $p$ there corresponds an infinity of $(m+1)$-tuples $P$. To see this, consider two $(m+1)$-tuples, $P$ and $\bar{P}$, with $\bar{P}$ given as
\begin{equation}
    \bar{P}:=\{\bar{X}{}^M\}=\{\lambda X^M\},
\end{equation}
for some $\lambda\in\mathbb{R}-\{0\}$. Then,
\begin{equation}
    \left\{\frac{X^i}{X^{m+1}}\right\}= \left\{\frac{\lambda X^i}{\lambda X^{m+1}}\right\}= \left\{\frac{\bar{X}{}^i}{\bar{X}{}^{m+1}}\right\},\quad\quad \forall \;i\in\{1,2,\dots,m\},
\end{equation}
denotes the same point of $\mathcal{A}_m$. For such points, one may solve the above expression for $\bar{X}{}^i$ to find
\begin{equation}
     \left\{\bar{X}{}^i\right\}= \left\{\frac{\bar{X}{}^{m+1}}{X^{m+1}}X^i\right\}.
\end{equation}
This provides a possible choice for $\lambda$,
\begin{equation}
    \lambda=\frac{\bar{X}{}^{m+1}}{X^{m+1}}.
\end{equation}
Thus, two $(m+1)$-tuples $P$ and $\bar{P}$, with $\{X^{m+1}\}\neq0$ and $\{\bar{X}{}^{m+1}\}\neq0$, are found to correspond to the same point $p$, only when there exists a $\lambda\in\mathbb{R}-\{0\}$ such that $\{\bar{X}{}^M\}=\{\lambda X^M\}$. This statement is nothing but that of an equivalence relation on points $P$. In other words,
\begin{equation}
    \{X^M\}\sim \lambda \cdot\{X^M\},\quad\quad\lambda\in\mathbb{R}-\{0\}.
\end{equation}
This removes one degree of freedom from the system, ensuring that only $m$ relations of the form Eq. \eqref{gen-homo-to-inhomo} are meaningful.

All $(m+1)$-tuples associated with the point $p\in\mathcal{A}_m$ may thus be collectively expressed as
\begin{equation}
    P=\{X^M\}=\begin{Bmatrix}\lambda x^m\\\lambda \end{Bmatrix},
\end{equation}
with $\lambda\neq0$. Since the $(m+1)$-tuple $P$ determines uniquely the corresponding point $p\in\mathcal{A}_m$, we may consider $P$ as the coordinates of $p$ in a different space. For $p$ and $P$ which correspond in the manner of Eq. \eqref{gen-homo-to-inhomo}, the coordinates $P$ are called the \textit{homogeneous coordinates} of $p$, while the coordinates of $p$ themselves are called the \textit{inhomogeneous coordinates} of $p$. 

The $(m+1)^{\text{th}}$ homogeneous coordinate must be non-vanishing. When this coordinate does vanish, we find a description of the points at infinity in $\mathcal{A}_m$. Consider the fixed point $p_0\in\mathcal{A}_m$, for which at least any one of its components $\{X^i_0\}\neq0$. Let
\begin{equation}
    q:=\kappa p_0,\quad\quad\kappa\in\mathbb{R}^+,
\end{equation}
be some point along the line segment containing $p_0$. For every value of $\kappa\in(0,\;\infty)$, the collection of points $\{q_{\kappa}\}$ traces a ray extending exclusively from the ``origin" to infinity, passing through $p_0$. Homogeneous coordinates for the point $q$ may be taken as
\begin{equation}
    Q:=\begin{Bmatrix}
        p_0\\\frac{1}{\kappa}
    \end{Bmatrix},
\end{equation}
since division by the $(m+1)^{\text{th}}$ coordinate, $1/\kappa$, produces $q$. The point with homogeneous coordinates
\begin{equation}
    \lim_{\kappa\rightarrow\infty}Q=\begin{Bmatrix}
        p_0\\0
    \end{Bmatrix},
\end{equation}
is called a \textit{point at infinity}. By repeating the above arguments for all fixed points $p_0\in\mathcal{A}_m$, one obtains the collection of points at infinity. We denote this collection of points as
\begin{equation}\label{infinity-surface}
    \mathcal{H}_{\infty}:=\biggl\{P\;\bigg|\;\{X^i\}\neq0,\;\{X^{m+1}\}=0\;\biggr\},
\end{equation}
and refer to $\mathcal{H}_{\infty}$ as the \textit{hypersurface at infinity}. Considering the case where $p_0\notin \mathcal{A}_m$, with $p_0=\{0^m\}$ and
\begin{equation}
    \{0^m\}:=\begin{Bmatrix}
        0\\\vdots\\0
    \end{Bmatrix},
\end{equation}
leads one to an identification of the origin with points at infinity. For this reason, the homogeneous coordinates given by
\begin{equation}
    \{0^M\}:=\begin{Bmatrix}
        0^m\\0
    \end{Bmatrix},
\end{equation}
describe neither a point of $\mathcal{A}_m$ nor of $\mathcal{H}_{\infty}$. Therefore, $\{0^M\}$ is excluded entirely. Amending $\mathcal{A}_m$ with the collection of points at infinity produces the $m$-dimensional \textit{projective space}
\begin{equation}
    \mathcal{P}_m:=\mathcal{A}_m\cup\mathcal{H}_{\infty}.
\end{equation}
The points of the affine space $p\in\mathcal{A}_m$ may thus be described as those points $P\in\mathcal{P}_m$ with $\{X^{m+1}\}\neq0$ and any one of the $\{X^i\}\neq0$. 

Associated with $\mathcal{P}_{m}$ is an $(m+1)$-dimensional vector space $\mathcal{V}_{m+1}$. Drawing this association permits the use of standard vector calculus on the associates of points $P$. A point $P\in\mathcal{P}_m$ and a vector $V\in\mathcal{V}_{m+1}$ are said to correspond if and only if
\begin{equation}\label{proj-aff-vec-corresp}
    P\equiv V\quad\Leftrightarrow\quad P=\{X^M\},\;\;\;V=(X^M).
\end{equation}
In other words, the components $(X^1,X^2,\dots, X^{m+1})^T$ of the \textit{vector} $V$ are the homogeneous coordinates $\{X^M\}$ of the \textit{point} $P$. The only vector $V$ with no corresponding point $P$ is the zero vector. From the properties of $\mathcal{P}_m$, one finds that the vectors of $\mathcal{V}_{m+1}$ also satisfy the equivalence relation: If $(X)=\lambda\cdot(Y)$, with $\lambda\neq0$, then $(X)\sim(Y)$. Therefore, to each $\{X\}$ there corresponds all non-vanishing vectors of the form $\lambda\cdot (X)$. This sets up a one-to-one correspondence between points of $\mathcal{P}_m$ and $1$-dimensional linear vector spaces of $\mathcal{V}_{m+1}$. We believe this to be the connection between the $1$-dimensional string theoretic concepts of Sec. \ref{sec:string} and the higher-dimensional classical projective theory. 

To summarize, we have accomplished the following. From a set of $m$-dimensional coordinates $\{x\}$ of a point $p$ of an affine space $\mathcal{A}_m$, an $m$-dimensional projective space $\mathcal{P}_m$ was constructed by adjoining to $\mathcal{A}_m$ the collection of points at infinity. The image of $\{x\}$ in $\mathcal{P}_m$ is a line of equivalent points $\{X\}$. To make use of $\mathcal{P}_m$, an $(m+1)$-dimensional vector space $\mathcal{V}_{m+1}$ is formed, such that the image of $\{X\}$ in $\mathcal{V}_{m+1}$ is a line of equivalent points $(X)$. In other words, a nonzero $(m+1)$-dimensional vector whose components are homogeneous coordinates in $\mathcal{P}_m$. 

In order to apply this concept of projective space to the General Projective Gauge Theory of Gravity, we must construct the affine space $\mathcal{A}_m$ from \textit{unparameterized displacements}. Essentially, this requires the projective space $\mathcal{P}_m$ to be constructed from $T^*_p\mathcal{M}$, rather than from $T_p\mathcal{M}$. Therefore, elements of the $m$-dimensional affine space of interest, $\mathcal{A}^*_m$, are the non-vanishing unparameterized displacements $\{x^m\}=\{dy^m\}$.

\subsection{Projective Linear Group}
\label{subsec:pgl}

Consider a vector $V=(X)$ of the $(m+1)$-dimensional vector space $\mathcal{V}_{m+1}$. Let $y:=\{x^M\}=\{x^1,\dots, x^{m+1}\}$ denote a set of coordinates for $\mathcal{V}_{m+1}$ and let
\begin{equation}
    e_M:=\begin{pmatrix}
        e_1,&\dots,&e_{m+1}
    \end{pmatrix}
\end{equation}
be a set of non-holonomic basis vectors spanning $\mathcal{V}_{m+1}$. The coordinate-dependent components of $V$ may be accessed by connecting to this basis, i.e.,
\begin{equation}
    V=V^Me_M,\quad\quad\quad V^M=\begin{pmatrix}
        X^1\\\vdots\\X^{m+1}
    \end{pmatrix}.
\end{equation}
Consider a transformation of coordinates $\{x\}\rightarrow\{x'\}$ which maps the non-holonomic basis to a holonomic one, given by
\begin{equation}
    e_{M'}=e_N(G^{-1})^N{}_{M'},\quad\quad\quad e_{M'}:=\partial_{M'}.
\end{equation}
This induces a linear transformation of the components of the vector $V$,
\begin{equation}
    V^{M'}=G^{M'}{}_NV^N.
\end{equation}
At a point of $\mathcal{V}_{m+1}$, the $(m+1)\times (m+1)$ matrix $G^{M'}{}_N$ is identified as an element of the $(m+1)$-dimensional group of invertible matrices with real entries. This is the $(m+1)$-dimensional General Linear Group, denoted $GL(m+1,\mathbb{R})$. Assuming this to hold for all points, we may write
\begin{equation}
    G:=\{G^{M'}{}_N\}\in GL(m+1,\mathbb{R}).
\end{equation}

The projective linear group $PGL(m,\mathbb{R})$ is defined from $GL(m+1,\mathbb{R})$ as the quotient
\begin{equation}
    PGL(m,\mathbb{R})\cong GL(m+1,\mathbb{R})/\mathbb{R}^{\times},
\end{equation}
with $\mathbb{R}^{\times}$ representing nonzero multiples of the identity. Restricting to the parts positively connected to the identity, $\mathbb{R}^{\times}\rightarrow \mathbb{R}^{+}$, there exists a decomposition,
\begin{equation}\label{gl-proj-decomp}
    GL^+(m+1,\mathbb{R})\cong SL(m+1,\mathbb{R})\times \mathbb{R}^+,
\end{equation}
with $SL(m+1,\mathbb{R})$ consisting of those elements of $GL^+(m+1,\mathbb{R})$ with positive unit determinant. Without the positivity restriction, the reflections, i.e., discrete transformations, must be accounted for \cite{Hehl}. We restrict all considerations to $GL^{+}(m+1,\mathbb{R})$ and omit the $(+)$-superscript for notational convenience. We therefore have the isomorphism
\begin{equation}
    SL(m+1,\mathbb{R})\cong PGL(m,\mathbb{R}).
\end{equation}
The projective linear group may be decomposed into the semi-direct product
\begin{equation}\label{semi-direct}
    PGL(m,\mathbb{R})\cong\mathbb{R}^m\rtimes GL(m,\mathbb{R})\ltimes \mathbb{R}^m_*,
\end{equation}
\cite{kobayashi,kobayashi-proj-conn}. These correspond, respectively, to $m$-dimensional (internal/vertical) translations, general linear transformations and what we will refer to as pseudo-translations. In other words, 
\begin{equation}
    PGL(m,\mathbb{R})\cong\text{Aff}(m,\mathbb{R})\ltimes \mathbb{R}^m_*,
\end{equation}
with $\text{Aff}(m,\mathbb{R})$ initially defined in Eq. \eqref{affine-group-decomposition}.

An element $S\in SL(m+1,\mathbb{R})\cong PGL(m,\mathbb{R})$ may be constructed from an element $G\in GL(m+1,\mathbb{R})$ in a manner similar to the $2$-jet construction of \cite{2-frame}. For convenience, we restrict attention to those $G\in GL(m+1,\mathbb{R})$ which have components of the form
\begin{equation}
    G^{M'}{}_N=\begin{pmatrix}
        A^{m'}{}_n&\frac{1}{\lambda}b^{m'}\\\lambda c_n&d
    \end{pmatrix}.
\end{equation}
As will be discussed shortly, this is simply a restriction to those transformations $\{x'\}$ which are homogeneous of degree-$1$. Recall the correspondence $\{X\}\equiv (X)$ of Eq. \eqref{proj-aff-vec-corresp}, and the map
\begin{equation}\label{x-X-map}
    \{x^m\}=\left\{\lambda_0\frac{X^m}{X^{m+1}}\right\}
\end{equation}
of Eq. \eqref{gen-homo-to-inhomo}, where a constant $[\lambda_0]=L$ has been inserted for proper physical coordinate dimensions. A similar map is used in the momentum space description of (A)-dS gauge gravitational theories; the so-called \textit{Beltrami coordinates} \cite{beltrami}. From $(X')=(G\cdot X)$, we deduce the corresponding action of $G\in GL(m+1,\mathbb{R})$ on $\{x^m\}$:
\begin{equation}\label{x-X-relation}
    \{x^{m'}\}=\left\{\lambda_0\frac{X^{m'}}{X^{(m+1)'}}\right\}\equiv\left(\lambda_0\frac{X^{m'}}{X^{(m+1)'}}\right).
\end{equation}

Consider the second order expansion of $\{x^{m'}\}$ in $\{x^m\}$, about the point $x_0\equiv\{x_0\}=0$:
\begin{equation}
    x^m\rightarrow x^{m'}_{(2)}=\left.x^{m'}\right|_{x=0}+x^n\left.\frac{\partial x^{m'}}{\partial x^{n}}\right|_{x=0}+x^mx^n\left.\frac{\partial^2 x^{m'}}{\partial x^m\partial x^n}\right|_{x=0}.
\end{equation}
In the above, we have omitted explicit coordinate brackets for notational convenience. Using the \textit{vector} expression in the right-hand side equality of Eq. \eqref{x-X-relation}, we find the partial derivatives to have the form:
\begin{equation}\label{s-der-expressions}
\begin{split}
    s^{m'}&:=\left.x^{m'}\right|_{x=0}=\frac{\lambda_0}{\lambda d}b^{m'},\\
    s^{m'}{}_n&:=\left.\frac{\partial x^{m'}}{\partial x^{n}}\right|_{x=0}=\frac{1}{d}(A^{m'}{}_n-\frac{1}{d}b^{m'}c_n),\\
    s^{m'}{}_{mn}&:=\left.\frac{\partial^2 x^{m'}}{\partial x^m\partial x^n}\right|_{x=0}=-\frac{\lambda}{\lambda_0d}(s^{m'}{}_mc_n+s^{m'}{}_nc_m).
    \end{split}
\end{equation}
Inverting $s^{m'}{}_n$, the expression for $s^{m'}{}_{mn}$ may be solved to find
\begin{equation}
    \lambda c_m=\frac{-\lambda_0d}{m+1}(s^{-1})^n{}_{m'}s^{m'}{}_{mn}=d\lambda_0s_m.
\end{equation}
By using the definitions in Eqs. \eqref{s-der-expressions}, the right-hand equality in the above expression may be written in a much more familiar form,
\begin{equation}\label{log-der-s-trans}
    s_m=\partial_m\log|s^{m'}{}_n|^{\frac{-1}{m+1}}.
\end{equation}
This permits, in addition to Eqs. \eqref{s-der-expressions}, the ability to express $G\in GL(m+1,\mathbb{R})$ in terms of the new set of transformations, $\{s^{m'},s^{m'}{}_n,s_n\}$, obtained by second order expansion about the origin,
\begin{equation}
    G^{M'}{}_N=:d\cdot\hat{S}^{M'}{}_N=d\cdot\begin{pmatrix}
        s^{m'}{}_n+s^{m'}s_n&\frac{1}{\lambda_0}s^{m'}\\\lambda_0s_n&1
    \end{pmatrix}.
\end{equation}
However, the $d$ still remains. To solve for $d$ , we take the determinant of the above expression \cite{det-power},
\begin{equation}
    d=|G|^{\frac{1}{m+1}}|\hat{S}|^{\frac{-1}{m+1}}=|G|^{\frac{1}{m+1}}|s|^{\frac{-1}{m+1}},
\end{equation}
where, for example, $|s|\equiv|s^{m'}{}_n|:=\det(s^{m'}{}_n)$. We thus arrive at
\begin{equation}\label{gl-S-trans}
    G^{M'}{}_N=|G|^{\frac{1}{m+1}}S^{M'}{}_N,
\end{equation}
with
\begin{equation}
    S^{M'}{}_N=|s|^{\frac{-1}{m+1}}\begin{pmatrix}
        s^{m'}{}_n+s^{m'}s_n&\frac{1}{\lambda_0}s^{m'}\\\lambda_0s_n&1
    \end{pmatrix}.
\end{equation}
One may easily check that the matrix $S$ satisfies $|S|=1$, and is indeed an element $S\in SL(m+1,\mathbb{R})$. That $S$ is also an element $S\in PGL(m,\mathbb{R})$ is confirmed by noting that $S$ may be decomposed into independent transformations, in exact accordance with the semi-direct product structure outlined in Eq. \eqref{semi-direct}. In other words,
\begin{equation}
    S^{M'}{}_N=|s|^{\frac{-1}{m+1}}\begin{pmatrix}
      \delta^{m'}{}_{p'}&\frac{1}{\lambda_0}s^{m'}\\0&1
    \end{pmatrix}\begin{pmatrix}
        s^{p'}{}_q&0\\0&1
    \end{pmatrix}\begin{pmatrix}
       \delta^q{}_n&0\\\lambda_0s_n&1
    \end{pmatrix}.
\end{equation}

We thus find the procedure for mapping elements $G\in GL(m+1,\mathbb{R})$ to elements $S\in PGL(m+1,\mathbb{R})\equiv GL(m+1,\mathbb{R})/\mathbb{R}^{\times}$ of the homogeneous space---simply divide out from $G$, the determinant $|G|$ raised to the $(\frac{-1}{m+1})^{\text{th}}$ power. This conveniently provides a means for mapping transformations of \textit{vectors} in $\mathcal{V}_{m+1}$ to transformations of \textit{points} in $\mathcal{P}_m$ and, therefore, transformations of unparameterized displacements in $T^*_p\mathcal{M}$. We note that, formally, the equivalence relation must be used in $\mathcal{V}_{m+1}$ to promote the correspondence to equality, i.e., $\{X\}=|G|^{\frac{-1}{m+1}}(X)$. However, the $\{x\}$ are insensitive to this distinction. 

To reel in the abstract mathematics, we discuss the physical interpretation of the component transformations contained in $S$. First, notice that in the transition $G\rightarrow S$ we have enacted the replacement $\lambda\rightarrow \lambda_0$. Recall that $\{x\}=\{dy\}$ represents unparameterized displacements. Using the form of $G\in GL(m+1,\mathbb{R})$ from Eq. \eqref{gl-S-trans}, we transform the components $(X)$ of the vector $V\in\mathcal{V}_{m+1}$ to find
\begin{equation}
    V^{M'}=G^{M'}{}_NV^N=|G|^{\frac{1}{m+1}}\begin{pmatrix}
        s^{m'}{}_nX^n+s^{m'}(\frac{1}{\lambda_0}X^{m+1}+s_nX^n)\\\lambda_0(\frac{1}{\lambda_0}X^{m+1}+s_nX^n)
    \end{pmatrix}.
\end{equation}
We then use the equivalence relation inherited from the identification $\{X\}\equiv(X)$ to write 
\begin{equation}
    V^{M'}\sim|G|^{\frac{-1}{m+1}}V^{M'}.
\end{equation}
In other words, the equivalence relation provides
\begin{equation}
    V^{M'}= S^{M'}{}_NV^N,
\end{equation}
where one must keep in mind that this expression was only obtained through the use of the equivalence relation. Using Eq. \eqref{x-X-relation}, we enact the transition to the transformed $\{x^{m'}\}$ to find
\begin{equation}\label{x-trans-gen-result}
    \{x^{m'}\}=\left\{\frac{s^{m'}{}_nx^n+s^{m'}(1+x^ns_n)}{1+x^ns_n}\right\}.
\end{equation}

From Eq. \eqref{x-trans-gen-result}, we extract a physical interpretation of the independent constituent transformations. First, upon fixing $\{x^n\}=\{x^n_0\}$, one finds
\begin{equation}
    \{x^{m'}\}=\{z_0^{m'}+s^{m'}\},
\end{equation}
where $\{z^{m'}_0\}=\{s^{m'}{}_nx^n/(1+x^ns_n)\}$. Thus, the set of transformations for which $s^{m'}=0$ preserves points of the $m$-dimensional affine space. In particular, there is a fixed point of contact between $\mathcal{P}_m$ and the homogeneous space. This preservation of the contact point permits a foliation of the space along surfaces of constant $\{x^{m+1}\}$ and is where the theory of Thomas-Whitehead connections \cite{gen-struc}, and a torsional version \cite{heavy-lifting} reside. In the context of Cartan geometry, this condition is known as \textit{rolling without slipping}. The rolling refers to the interpretation that the model homogeneous space $GL(m+1,\mathbb{R})/\mathbb{R}^{\times}$ is used to probe the geometry of the locally isomorphic projective space $\mathcal{P}_m$ over $\mathcal{M}$ \cite{into-cartan}. The condition of no slipping then refers to a static relation between the two points of contact, i.e., between the two spaces. In other words, the two points may be identified. 

Second, the surface at infinity in $\mathcal{P}_m$, defined by Eq. \eqref{infinity-surface}, may be arrived at by setting $\{X^{m+1}\}=\{0\}$. This is best understood when viewed in the transformed coordinates, since by setting $\{X^{(m+1)'}\}=\{0\}$, we may simply read off the result from Eq. \eqref{x-trans-gen-result}. Thus, the collection of points which are \textit{inaccessible} to elements of $\mathcal{A}_m$ in the transformed system, are defined by the relation
\begin{equation}
    1+x^ns_n=0.
\end{equation}
Therefore, the collection of $s_n$ satisfying $x^ns_n=-1$ are transformations which are inaccessible in $\mathcal{A}_m$. This provides the following interpretation. Transformations for which $s_n=0$, taking $\{x^m\}$ to $\{x^{m'}\}=\{s^{m'}{}_nx^n+s^{m'}\}$, are those transformations which preserve the \textit{affine structure} of $\mathcal{A}_m$. This is the main reason for referring to the general projective construction of the present as a type of \textit{Projective Metric-Affine} theory. Described in tangible geometrical terms that parallel the $s^{m'}=0$ case, transformations $s_n$ for which $x^ns_n=-1$, preserve the surface at infinity. Furthermore, from Eq. \eqref{log-der-s-trans}, we find these are exactly the volume-preserving transformations of $\mathcal{A}_m$.

Lastly, the transformations $s^{m'}{}_n$ are identified simply as linear transformations of $\mathcal{A}_m$. In this document, we will be most interested in those transformations for which $s_{m'n}=-s_{nm'}$ is antisymmetric. This condition provides access to an $m$-dimensional (projective) Minkowski space $PM_m$, since the transformations preserving $|x|^2=x^mx_m$, i.e., the isometries of $PM_m$, are of the form $\{x^{m'}\}=\{a^{m'}{}_nx^n+s^{m'}\}$, with $a_{m'n}:=s_{[m'n]}$ a Lorentz transformation.

\subsection{Projective Linear Algebra}
\label{subsec:pgl-algebra}

As vector spaces, the Lie algebra corresponding to $SL(m+1,\mathbb{R})\cong PGL(m,\mathbb{R})$ permits the direct-sum decomposition
\begin{equation}
    \mathfrak{sl}(m+1,\mathbb{R})\cong \mathfrak{t}(m,\mathbb{R})\oplus\mathfrak{gl}(m,\mathbb{R})\oplus\mathfrak{t}^*(m,\mathbb{R}_*),
\end{equation}
which follows from the semi-direct product structure in the group decomposition, Eq. \eqref{semi-direct}. Again, the explicit $(+)$-superscript is omitted from $\mathfrak{sl}(m+1,\mathbb{R})$. The Lie algebra $\mathfrak{gl}(m+1,\mathbb{R})$ of $GL(m+1,\mathbb{R})$ is defined as
\begin{equation}\label{gl-commutators}
    [\underline{\bm{L}}{}^{A}{}_{B},\underline{\bm{L}}{}^{C}{}_{D}]=\delta^{A}{}_{D}\underline{\bm{L}}{}^{C}{}_{B}-\delta^{C}{}_{B}\underline{\bm{L}}{}^{A}{}_{D},
\end{equation}
where $\underline{\bm{L}}$ are the Lie algebra generators. Separating the trace from traceless parts of the generators produces
\begin{equation}
\begin{split}
     \bm{L}^{A}{}_{B}&:=\underline{\bm{L}}{}^{A}{}_{B}-\tfrac{1}{m+1}\delta^{A}{}_{B}\underline{\bm{L}}{}^{C}{}_{C}\in\mathfrak{sl}(m+1,\mathbb{R}),\\
     \underline{\bm{D}}&:=\underline{\bm{L}}{}^{C}{}_{C}\in\mathfrak{t}^{+}(1,\mathbb{R}^{+}).
     \end{split}
     \end{equation}
The former contains the generators associated with the $m$-dimensional projective linear group, while the latter generates $(m+1)$-dimensional positive dilations, represented by the $|G|$-factor of the previous section. Recall that this was quotiented away in order to provide a projective structure.

To further separate the $m$-dimensional generators of (pseudo)-translations and $m$-dimensional linear transformations from $\bm{L}\in\mathfrak{sl}(m+1,\mathbb{R})$, the generic relation between (in)-homogeneous coordinates, Eq. \eqref{x-X-map}, is used. For convenience, we choose the coordinate basis for the linear generators and translate between the two coordinate sets. Additionally, we let $*$ denote the $(m+1)^{\text{th}}$ index and replace the constant $\lambda_0\rightarrow x^*_0$ for notational convenience. For translations, we find that
\begin{equation}
    \bm{L}^{*}{}_{b}=X^{*}\frac{\partial}{\partial X^{b}}=X^{*}\frac{\partial x^{a}}{\partial X^{b}}\frac{\partial}{\partial x^{a}}=x^*_0\frac{\partial}{\partial x^{b}},
\end{equation}
and for pseudo-translations, we find that
\begin{equation}
    \bm{L}^{a}{}_{*}=X^{a}\frac{\partial}{\partial X^{*}}=X^{a}\frac{\partial x^{b}}{\partial X^{*}}\frac{\partial}{\partial x^{b}}=\frac{-1}{x^*_0}x^{a}x^{b}\frac{\partial}{\partial x^{b}}.
\end{equation}
We therefore define
\begin{equation}\label{pp*-generators}
\begin{split}
    \bm{P}_{b}&:=\frac{1}{x^*_0}\bm{L}^{*}{}_{b}\in\mathfrak{t}(m,\mathbb{R}),\\
    \bm{P}{}^{a}_*&:=-x^*_0\bm{L}^{a}{}_{*}\in\mathfrak{t}^*(m,\mathbb{R}_*)
    \end{split}
\end{equation}
as the translation and pseudo-translation generators, respectively. Notice the independence of the result on the particular form of $X^{*}$. Lastly, since $\bm{L}{}^{A}{}_B$ generates $\mathfrak{sl}(m+1,\mathbb{R})$, they are naturally traceless,
\begin{equation}
    \bm{L}{}^A{}_A=0\;\;\Rightarrow\;\; \bm{L}{}^a{}_a=-\bm{L}{}^{*}{}_{*}.
\end{equation}

Using the definitions of $\bm{P},\;\bm{P}_*,\;\bm{L}$, and ${\underline{\bm{D}}}$, as well as the traceless property of the $\mathfrak{sl}(m+1,\mathbb{R})$ generators, the projective decomposition of the Lie algebra $\mathfrak{gl}^+(m+1,\mathbb{R})$ may now be stated:
\begin{equation}\label{pgl+D-algebra-commutators}
\begin{split}
    [\bm{L}^{a}{}_{b},\bm{L}^{c}{}_{d}]&=\delta^{a}{}_{d}\bm{L}^{c}{}_{b}-\delta^{c}{}_{b}\bm{L}^{a}{}_{d},\\
       [\bm{P}^{a}_*,\bm{P}_{b}]&=\bm{L}^{a}{}_{b}+\delta^{a}{}_{b}\bm{L}^{c}{}_{c},\\
    [\bm{P}_{a},\bm{L}^{b}{}_{c}]&=-\delta^{b}{}_{a}\bm{P}_{c},\\
    [\bm{P}^{a}_*,\bm{L}^{b}{}_{c}]&=\delta^{a}{}_{c}\bm{P}^{b}_*,
\end{split}
\quad\quad\begin{split}
    [\bm{P}^{a}_*,\bm{P}^{b}_*]&=0,\\
    [\bm{P}_{a},\bm{P}_{b}]&=0,\\
    [\bm{L}^{A}{}_{B},\underline{\bm{D}}]&=0,\\
    [\underline{\bm{D}},\underline{\bm{D}}]&=0.
\end{split}
\end{equation}
The algebra $\mathfrak{pgl}(m,\mathbb{R})$ of $PGL(m,\mathbb{R})$ is obtained by removing $\underline{\bm{D}}$ from the above set of commutators. From these relations, it appears that $x^*_0$ cannot play the role of contraction parameter \cite{contraction-og-iw,graded-contract}, since it is absent from all commutators. It must be noted that redefining the (pseudo)-translational generators with their $x^*_0$ factors moved to the opposite side of each expression in Eqs. \eqref{pp*-generators} results in the same conclusion. Furthermore, these relations make explicit the $\mathbb{Z}_2$-graded structure of $\mathfrak{pgl}(m,\mathbb{R})$, since if $\bm{P}\subset\mathfrak{g}_{-1}$, $\bm{L}\subset\mathfrak{g}_{0}$, and $\bm{P}_*\subset\mathfrak{g}_{1}$, we have \cite{2-frame}
\begin{equation}\label{graded-algebra}
[\mathfrak{g}_{m},\mathfrak{g}_{n}]\subset\mathfrak{g}_{m+n},
\end{equation}
expressing this as a \textit{reductive} or \textit{reducible algebra} \cite{into-cartan}.

When a metric is available, a finer decomposition of the algebra is available. Since we are most interested in $\mathfrak{pgl}(m,\mathbb{R})$, we omit all contributions from $\underline{\bm{D}}$ in the following discussion. For ease, we consider the available metric to be that of an $(m+1)$-dimensional Minkowski space $M_{m+1}$ given by $\eta_{\underline{AB}}=\text{diag}(1,1,,\dots,-1,-1,\eta_0)$, with $p$ positive entries and $q$ negative entries, and $p+q=m$. For now, we leave the $(m+1)^{\text{th}}$ slot arbitrary, denoted by the constant $\eta_0=\pm1$. The indices are taken underlined to distinguish them from the previous discussion. The decomposition of $\bm{L}$ takes the form
\begin{equation}
\bm{L}_{\underline{AB}}=\bm{S}_{\underline{AB}}+\bm{A}_{\underline{AB}},
\end{equation}
where $\bm{S}_{\underline{AB}}:=\frac{1}{2}\bm{L}_{(\underline{AB})}$ and $\bm{A}_{\underline{AB}}:=\frac{1}{2}\bm{L}_{[\underline{AB}]}$ are the symmetric and antisymmetric parts of $\bm{L}$, respectively. The $\bm{S}$ generate the \textit{set} of $(m+1)$-dimensional real symmetric matrices, $\text{Sym}(\tfrac{(m+1)(m+2)}{2},\mathbb{R})$, while the $\bm{A}$ generate the $(m+1)$-dimensional real special orthogonal group $SO(m+1,\mathbb{R})$. 

Upon lowering an index, the translation generator picks up a factor of $\eta_0$,
\begin{equation}\label{p-generator-lower-index}
    \bm{L}_{\underline{*b}}:=\eta_{\underline{*A}}\bm{L}^{\underline{A}}{}_{\underline{b}}=\eta_0x^*_0\bm{P}_b.
\end{equation}
No such amendment occurs for
\begin{equation}
    \bm{L}_{\underline{a*}}:=\eta_{\underline{aB}}\bm{L}^{\underline{B}}{}_{\underline{*}}=\frac{-1}{x^*_0}\bm{P}^*_a.
\end{equation}
Let
\begin{equation}
    \bm{S}_{\underline{a*}}=\frac{1}{2}\bm{L}_{(\underline{a*})}:=\frac{1}{x^*_0}\bm{M}^{+}_{\underline{a}}\quad\text{and}\quad\bm{A}_{\underline{a*}}=\frac{1}{2}\bm{L}_{[\underline{a*}]}:=\frac{1}{x^*_0}\bm{M}^{-}_{\underline{a}}.
\end{equation}
Explicitly, using the coordinate map between (in)-homogeneous coordinates, these generators have the form
\begin{equation}\label{M-generator-def}
\bm{M}^{\pm}_{\underline{a}}=\frac{-1}{2}\left(\bm{P}^*_{\underline{a}}\mp\eta_0(x^*_0)^2\bm{P}_{\underline{a}}\right).
\end{equation}
Reflecting this decomposition in $\mathfrak{pgl}(m,\mathbb{R})$, the relations in Eqs. \eqref{pgl+D-algebra-commutators} become
\begin{equation}\label{pgl-lorentz-algebra}
\begin{split}
    [\bm{S}_{\underline{ab}},\bm{S}_{\underline{cd}}]&=-\eta_{\underline{ca}}\bm{A}_{\underline{bd}}-\eta_{\underline{cb}}\bm{A}_{\underline{ad}}+\eta_{\underline{ad}}\bm{A}_{\underline{cb}}+\eta_{\underline{bd}}\bm{A}_{\underline{ca}},\\
[\bm{A}_{\underline{ab}},\bm{A}_{\underline{cd}}]&=\eta_{\underline{ca}}\bm{A}_{\underline{bd}}-\eta_{\underline{cb}}\bm{A}_{\underline{ad}}+\eta_{\underline{ad}}\bm{A}_{\underline{cb}}-\eta_{\underline{bd}}\bm{A}_{\underline{ca}},\\
[\bm{A}_{\underline{ab}},\bm{S}_{\underline{cd}}]&=\eta_{\underline{ca}}\bm{S}_{\underline{bd}}-\eta_{\underline{cb}}\bm{S}_{\underline{ad}}+\eta_{\underline{ad}}\bm{S}_{\underline{cb}}-\eta_{\underline{bd}}\bm{S}_{\underline{ca}},\\
[\bm{M}^{\pm}_{\underline{a}},\bm{A}_{\underline{bc}}]&=\eta_{\underline{ac}}\bm{M}^{\pm}_{\underline{b}}-\eta_{\underline{ab}}\bm{M}^{\pm}_{\underline{c}},\\
[\bm{M}^{\pm}_{\underline{a}},\bm{S}_{\underline{bc}}]&=\eta_{\underline{ac}}\bm{M}^{\mp}_{\underline{b}}+\eta_{\underline{ab}}\bm{M}^{\mp}_{\underline{c}},\\
[\bm{M}^{\pm}_{\underline{a}},\bm{M}^{\pm}_{\underline{b}}]&=\mp\eta_0(x^*_0)^2\bm{A}_{\underline{ab}},\\
[\bm{M}^{\pm}_{\underline{a}},\bm{M}^{\mp}_{\underline{b}}]&=\pm\eta_0(x^*_0)^2(\bm{S}_{\underline{ab}}+\eta_{\underline{ab}}\bm{S}).
\end{split}
\end{equation}
From the definition in Eq. \eqref{M-generator-def}, and the above commutation relations, we recognize $\bm{M}^{\pm}_{\underline{a}}$ as the vector operator used in \cite{world-spinors} for the construction of world spinors. Moreover, we find the emergence of $x^*_0$'s potential role as a contraction parameter. This comes only at the expense of introducing a metric. If one considers the limit $(x^*_0)^2\rightarrow 0$, or imposes a Grassmann-valued constraint $(x^*_0)^2=0$, then
\begin{equation}
    \bm{M}_{\underline{a}}:=\bm{M}{}^{\pm}_{\underline{a}}=\bm{M}{}^{\mp}_{\underline{a}},
\end{equation}
i.e., the two sets of now Abelian ``translations" become equivalent. 

Another possible way to view this contraction is not directly through the constant contraction parameter $x^*_0$, but rather through the initial projective vector $X$. Since
\begin{equation}
X^{\underline{A}}=\begin{pmatrix}X^{\underline{a}}\\X^{\underline{*}} \end{pmatrix}=\frac{X^{\underline{*}}}{x^*_0} \begin{pmatrix}x^{\underline{a}}\\x^*_0\end{pmatrix}=:\frac{X^{\underline{*}}}{x^*_0} x^{\underline{A}},
\end{equation}
we have $x^{\underline{A}}x_{\underline{A}}=\eta_{\underline{AB}}x^{\underline{A}}x^{\underline{B}}=x^{\underline{a}}x_{\underline{a}}+\eta_0(x^{*}_0)^2$. Then, imposing
\begin{equation}\label{algebra-condition}
    x^{\underline{a}}\bm{M}{}^{-}_{\underline{a}}f(x)=0,\quad\forall f(x),
\end{equation}
one finds that either $f(x)\in\mathcal{H}_0(x^{\underline{a}})$ is a degree-$0$ homogeneous function, $x^{\underline{a}}\partial_{\underline{a}}f(x)=0$, or $x^{\underline{a}}x_{\underline{a}}+\eta_0(x^*_0)^2=0$, restricting the length of $x^{\underline{a}}$. Since this is intended to hold for all $f(x)$, we have the satisfaction of the latter expression. While this does not contract the algebra in the usual sense, it does effectively ``break" the generator in the sense that its action on anything will return zero. Both interpretations, $(x^*_0)^2\rightarrow 0$ and $x^{\underline{a}}x_{\underline{a}}+\eta_0(x^*_0)^2=0$, become compatible when one further imposes the \textit{Lorentz-invariant} constraint $x^{\underline{a}}x_{\underline{a}}=0$. 

In any case, when $(x^*_0)^2\rightarrow 0$ is imposed, one forces the vanishing of the second summand in Eq. \eqref{M-generator-def}, effectively breaking the \textit{translational} symmetry as opposed to the pseudo-translational symmetry. To make explicit the identification of the contracted algebra, consider recombining $\bm{S}_{\underline{ab}}$ and $\bm{A}_{\underline{ab}}$ to form the $m$-dimensional linear generators $\bm{L}_{\underline{ab}}=\bm{S}_{\underline{ab}}+\bm{A}_{\underline{ab}}$. In this contraction limit, the resulting algebra is
\begin{equation}\label{pgl-lorentz-algebra-contracted}
    \begin{split}[\bm{L}_{\underline{ab}},\bm{L}_{\underline{cd}}]&=\eta_{\underline{ad}}\bm{L}_{\underline{cb}}-\eta_{\underline{cb}}\bm{L}_{\underline{ad}},\\
[\bm{M}_{\underline{a}},\bm{L}_{\underline{bc}}]&=\eta_{\underline{ac}}\bm{M}_{\underline{b}},\\
    [\bm{M}_{\underline{a}},\bm{M}_{\underline{b}}]&=0.
    \end{split}
\end{equation}
The above algebra only \textit{resembles} that of $\mathfrak{aff}(m,\mathbb{R})$, but is indeed \textit{not} the algebra of the affine group. The reason for this lies in the middle commutator, which would be required to read $[\bm{M}_{\underline{a}},\bm{L}_{\underline{bc}}]=-\eta_{\underline{ab}}\bm{M}_{\underline{c}}$ for such an identification to exist. We will therefore refer to the above contraction as resulting in the algebra, $\mathfrak{p}\mathfrak{aff}(m,\mathbb{R})$, of the \textit{projective affine group} $\mathfrak{p}\text{Aff}(m,\mathbb{R})$. For some applications of the In\"{o}nu-Wigner contraction to other common algebras, see \cite{ds-gauge1}.

\subsection{Volume Bundle}
\label{subsec:vol-bundle}

We follow the constructions of \cite{proj-conn,roberts} and deviate where necessary, in order to release the constraint of fixed contact point. In other words, we wish to permit slipping when rolling the model homogeneous space along the projective space associated with the volume bundle. Consider the bundle $T^*\mathcal{M}$ of co-tangent spaces of the $m$-dimensional spacetime manifold $\mathcal{M}$. If $\{x^m\}$ are a set of coordinates for $\mathcal{M}$, then the natural or coordinate basis of $T^*\mathcal{M}$ is $\{dx^m\}$, i.e., unparameterized displacements. The $m^{\text{th}}$ exterior product is constructed as
\begin{equation}
    {\bigwedge}{}^{\!m}(T^*\mathcal{M})=\biggl\{\alpha dx^1\wedge\dots\wedge dx^m\;\bigg|\;\alpha\in\mathbb{R}\biggr\}.
\end{equation}
Furthermore, define the bundle of non-vanishing elements $\mathcal{V}\in\mathcal{E}(T^*\mathcal{M})$ as
\begin{equation}
    \mathcal{E}(T^*\mathcal{M})=\Lambda^m(T^*\mathcal{M})-\{0\}.
\end{equation}
The object $\mathcal{V}$ may then be considered a volume element. The set of pairs $[\pm\mathcal{V}]$ of elements forms what is called the volume bundle of $\mathcal{M}$, denoted $V\mathcal{M}$. Define the bundle projection $\nu:\mathcal{N}\rightarrow\mathcal{M}$ by $\nu[\pm\mathcal{V}]=p$ when $\pm\mathcal{V}\in\mathcal{E}(T^*_p\mathcal{M})$, for any point $p\in\mathcal{M}$. Choose the fiber coordinate on the unit-dimensional fiber of $\nu$ as $|v|$, where 
\begin{equation}
    \mathcal{V}=v(\mathcal{V})\left(dx^1\wedge\dots\wedge dx^m\right)|_p
\end{equation}
is satisfied for any $\mathcal{V}$. For convenience, we take instead
\begin{equation}
    x^*:=x^*_0|v|^{\frac{1}{m+1}},
\end{equation}
with $[x^*]=L$ as the $\mathbb{R}^+$-valued fiber coordinate and $[x^*_0]=L$ a positive constant. The absolute value is simply an imposed equivalence relation on forms in $\mathcal{E}(T^*\mathcal{M})$, and is imposed to avoid orientation reversals. Essentially, this is related to the earlier restriction to $GL^+(m+1,\mathbb{R})$. Interestingly, $\mathbb{R}^+$-extending gauge groups in a manner similar to this allows one to gain insight on Higgs fields and symmetry breaking \cite{R+extended-higgs}. 

Consider $\{x^M\}=\{x^m,x^*\}$ as the coordinates of the $(m+1)$-dimensional volume bundle $V\mathcal{M}$. In the previous construction of $V\mathcal{M}$, a diffeomorphism $\{x^m\}\rightarrow \{x^{m'}\}$ of $\mathcal{M}$, with Jacobian of transformation $J^{m'}{}_n$, induces a transformation of the fiber $\{x^*\}\rightarrow \{x^{*'}\}$, where
\begin{equation}
    x^{*'}=x^*|J^{m'}{}_n|^{\frac{-1}{m+1}}.
\end{equation}
Presently, we generalize this transformation behavior to include those transformations for which $\{x^{m'}\}=\{x^{m'}(x,x^*)\}$. In other words, we permit a mixing of $\{x^*\}$ into $\{x^m\}$. On $V\mathcal{M}$, the Jacobian of transformation then has the matrix form
\begin{equation}\label{jacobian-new}
    J^{M'}{}_N=\begin{pmatrix}
        J^{m'}{}_n&\frac{1}{x^*}k^{m'}\\x^{*'}j_n&|J^{m'}{}_n|^{\frac{-1}{m+1}}
    \end{pmatrix},
\end{equation}
where
\begin{equation}\label{shift-vector}
    j_n:=\frac{\partial\log|J^{m'}{}_n|^{\frac{-1}{m+1}}}{\partial x^n},\quad\quad\quad k^{m'}:=x^*\frac{\partial x^{m'}}{\partial x^*}.
\end{equation}
The translation vector $k^{m'}$ simply measures the ``vertical" change of the new coordinate set and vanishes when $\{x^{m'}\}\neq \{x^{m'}(x^*)\}$. A foliation of \textit{spacetime} sheets then exists along surfaces of constant $\{x^*\}=\{x^*_0\}$ when $k^{m'}=0$. At any point $P\in V\mathcal{M}$, the transformation matrix is an element of the positively oriented linear group $J\in GL^{+}(m+1,\mathbb{R})$. We assume this to extend to all points $P$ of $V\mathcal{M}$. 

Essentially, the Jacobian in Eq. \eqref{jacobian-new} generalizes the previous construction, Eq. \eqref{jacobian-old}, by extending the original $m$-dimensional linear transformations $J^{m'}{}_n$ to \textit{affine} transformations via the additional shift of $k^{m'}$. Notice also that, from the definition of this shift vector in Eq. \eqref{shift-vector}, it may be viewed as measuring the change in the new $m$-coordinates $\{x^{m'}\}$ along the fiber direction $\{x^*\}$. Interestingly, the power to which $\{x^*\}$ is raised in each of the block components of \eqref{jacobian-new} appears to correspond to the grading of the algebra, Eq. \eqref{graded-algebra}.

In the generalization process, we are interested in working with a non-coordinate basis, denoted $\breve{e}{}^a=\breve{e}{}^a{}_ndx^n$. The tautological $m$-form of \cite{proj-conn}, in the basis $\breve{e}{}^a$, is then
\begin{equation}
\begin{split}
    \mathcal{T}&:=\left(\frac{x^*}{x^*_0}\right)^{m+1}dx^1\wedge\dots\wedge dx^m\\
    &=\left(\frac{x^*}{x^*_0}\right)^{m+1}\left(\frac{1}{m!}\hat{\epsilon}_{n_1\dots n_m}dx^{n_1}\wedge\dots\wedge dx^{n_m}\right)\\
    &=\left(\frac{x^*}{x^*_0}|\breve{e}{}|^{\frac{-1}{m+1}}\right)^{m+1}\left(\breve{e}{}^{1}\wedge\dots\wedge  \breve{e}{}^{m}\right)\\
    &=\breve{\mathfrak{p}}{}^{m+1}\breve{e}{}^{1}\wedge\dots\wedge \breve{e}{}^{m},
    \end{split}
\end{equation}
where $|\breve{e}{}|:=|\breve{e}{}^a{}_m|$. This expression is utilized only as a convenient way to motivate and introduce the projective factor
\begin{equation}
    \breve{\mathfrak{p}}:=\frac{x^*}{x^*_0}|\breve{e}{}^a{}_m|^{\frac{-1}{m+1}}.
\end{equation}
The projective factor $\breve{\mathfrak{p}}$ is invariant with respect to $m$-dimensional \textit{coordinate} transformations $J$, but is not invariant with respect to \textit{gauge} transformations $S$, as will be discussed shortly. The volume bundle will be taken to play the role of the auxiliary space $\mathcal{V}_{m+1}$ discussed in Sec. \ref{sec:proj-geo}.

\subsection{Affine-to-Projective Map}
\label{subsec:coord-map}

In this section, we develop the map between the coordinates of $V\mathcal{M}$ and the coordinates of the associated projective space $P\mathcal{M}$, identified with the homogeneous space $GL^+(m+1,\mathbb{R})/\mathbb{R}^+$. This process is analogous to the construction outlined in Sec. \ref{sec:proj-geo}.

Let $\{x^N\}$ denote the set of coordinates of a point $q\in V\mathcal{M}$, and let $\{X^A\}$ denote the coordinates of the associated point $Q\in P\mathcal{M}$. Indices from the middle of the Latin alphabet, $N=1,2,\dots,m+1$, refer to $V\mathcal{M}$, while indices from the beginning of the Latin alphabet, $A=1,2,\dots,m+1$, have the same range and refer to $P\mathcal{M}$. We single out the fiber coordinate contained in $\{x^N\}$ and denote its index with $*$. The remaining $m$ coordinates will often be denoted with no index, i.e., $\{x^N\}\equiv\{x,x^*\}$.

The map $\mathcal{F}$, given by
\begin{equation}\label{F-map}
    \mathcal{F}:\{x^N\}\rightarrow\{\mathcal{F}^A(x^N)\}\equiv\{X^A(x^N)\},
\end{equation}
which takes the coordinates of $V\mathcal{M}$ to the coordinates of $P\mathcal{M}$ is assumed to have the general form
\begin{equation}\label{x-X-map-specific}
    \{X^A(x^N)\}=\left\{\mathfrak{q}(x,x^*) y^a(x),\; x^*_0\mathfrak{q}^*(x,x^*)\right\}.
\end{equation}
The dimensionless functions $\mathfrak{q}\neq \mathfrak{q}{}^{*}$ are considered here as degree-$1$ homogeneous functions of their argument, $\mathfrak{q},\mathfrak{q}^*\in\mathcal{H}_1(x,x^*)$. Homogeneous functions $\mathcal{H}_{\mathfrak{w}}$ of degree $\mathfrak{w}\neq 1$ are considered in \cite{schouten}. The imposition of homogeneity on $\{X^A\}$, combined with the assumption $\mathfrak{q},\mathfrak{q}^*\in\mathcal{H}_1(x,x^*)$, requires that the functions $y^a(x)$ be inhomogeneous. The dimensionful factor $[x^*_0]=L$ is included to retain proper physical dimensions. The above map appears analogous to the generalized homogeneous coordinates discussed in \cite{BGG-2} for exponential $\mathfrak{q}=\mathfrak{q}^*$.

An $\mathcal{H}_{\mathfrak{w}}(X)$ function $f(X)$ satisfies
\begin{equation}
f(\mathfrak{r}X)=\mathfrak{r}^{\mathfrak{w}}f(X),
\end{equation}
and, equivalently, Euler's condition of homogeneity states that \cite{schouten}
\begin{equation}
    X[f(X)]=X\cdot\partial f(X)=\mathfrak{w}f(X),\quad\quad \partial:=\frac{\partial}{\partial X}.
\end{equation}
As can easily be shown using the above relations, the partial derivative of an $\mathcal{H}_{\mathfrak{w}}$ function is an $\mathcal{H}_{\mathfrak{w}-1}$ function. 

The functions $y^a(x)$ may be identified with the unparameterized displacements about a point in $\mathcal{M}$, i.e., the coordinate differentials $dx^m$ \cite{projective-relativity}. These are accessed via the following inhomogeneous map \cite{schouten}, 
\begin{equation}
    dx^a=x^*_0\frac{X^a}{X^*}=\frac{\mathfrak{q}}{\mathfrak{q}^*}y^a(x).
\end{equation}
This is simply a particular instantiation of Eq. \eqref{x-X-map}. 

A projective frame associated with $\mathcal{F}$ may be constructed at each point $Q\in P\mathcal{M}$ via the exterior differential of the homogeneous coordinate functions $\{X^A\}$,
\begin{equation}
    \mathcal{E}^A:=dX^A.
\end{equation}
The matrix of components may be accessed by pulling back to $V\mathcal{M}$,
\begin{equation}
    \mathcal{E}^A=\mathcal{E}^A{}_Ndx^N,
\end{equation}
where
\begin{equation}
    \mathcal{E}{}^A{}_N:=\frac{\partial X^A}{\partial x^N}
\end{equation}
is the component matrix, with inverse given as $(\mathcal{E}^{-1})^N{}_A:=\frac{\partial x^N}{\partial X^A}$, such that
\begin{equation}
\mathcal{E}^A{}_N(\mathcal{E}^{-1})^N{}_B=\delta^A{}_B,\quad\quad\quad(\mathcal{E}^{-1})^N{}_A\mathcal{E}^A{}_M=\delta^N{}_M.
\end{equation}

The matrix $\mathcal{E}$ must contain either homogeneous factor $\mathfrak{q}^{(*)}(x,x^*)$ proportional to the identity. For the purpose of investigating the $m$-dimensional physics near $x^*=x^*_0$, we choose $\mathfrak{q}^*(x,x^*)$. Using Eq. \eqref{x-X-map-specific}, the linear matrix representation of $\mathcal{E}$ is easily calculated to have the form
\begin{equation}\label{curly-solder-matrix}
    \mathcal{E}^{A}{}_N=\mathfrak{q}^*\begin{pmatrix}
\breve{e}{}^{a}{}_n+\breve{y}{}^{a}\breve{e}^*_n&\breve{y}{}^{a}\breve{e}^*_*+\breve{y}{}^{a}_*\\x^*_0\breve{e}^*_n&x^*_0\breve{e}^{*}_{*}\end{pmatrix},
\end{equation}
with
\begin{equation}
    \breve{y}{}^a:=\frac{\mathfrak{q}}{\mathfrak{q}{}^*}y^a,\quad\;\;\breve{e}{}^a{}_n:=\frac{\partial \breve{y}{}^a}{\partial x^n},\;\;\quad
    \breve{y}{}^a_*:=\frac{\partial \breve{y}{}^a}{\partial x^*},\;\;\quad \breve{e}^*_N=\begin{pmatrix}
        \frac{\partial \log\mathfrak{q}{}^*}{\partial x^n},\; \frac{\partial \log\mathfrak{q}{}^*}{\partial x^*}
    \end{pmatrix}.
\end{equation}
Although $y^a=y^a(x)$, scaling with the ratio $\mathfrak{q}/\mathfrak{q}^*$ leads to $\breve{y}{}^a=\breve{y}^a(x,x^*)$. The purpose of defining $\breve{y}{}^a$ in this way is as follows. For $\mathcal{E}$ to represent a genuine projective frame, we require the point-wise identification
\begin{equation}
    \mathcal{E}^A{}_N\in PGL(m,\mathbb{R})\cong SL(m+1,\mathbb{R}).
\end{equation}
More precisely, we require $\mathcal{E}$ to represent the map
\begin{equation}\label{mathcal-E-map}
    \mathcal{E}:PGL(m,\mathbb{R})\rightarrow GL(m+1,\mathbb{R}).
\end{equation}
For the following discussion, it will be more convenient to instead work with $\mathcal{E}{}^{-1}$, which obviously represents the inverse of Eq. \eqref{mathcal-E-map},
\begin{equation}
    \mathcal{E}{}^{-1}: GL(m+1,\mathbb{R})\rightarrow PGL(m,\mathbb{R}).
\end{equation}
Since $PGL(m,\mathbb{R})$ may be identified as a homogeneous space, we have the equivalent expression,
\begin{equation}
    \mathcal{E}{}^{-1}: GL(m+1,\mathbb{R})/\mathbb{R}^{+}\rightarrow GL(m+1,\mathbb{R}).
\end{equation}
To ensure the validity of Eq. \eqref{mathcal-E-map}, we consider the generic matrix representation of $\mathcal{E}^{-1}$,
\begin{equation}
\mathcal{E}^{-1}=\begin{pmatrix}M&V\\C&s\end{pmatrix},
\end{equation}
with $M=\{M^n{}_a\}\in GL(m,\mathbb{R})$, $V=\{V^n\}\in\mathbb{R}^m$, $C=\{C_a\}\in \mathbb{R}^m_*$, and $s\in\mathbb{R}^+$, and require that $\mathcal{E}^{-1}$ permit a factorization of the form
\begin{equation}\label{matrix-factors}
    (\mathcal{E}{}^{-1})^M{}_A=\begin{pmatrix}s'&0\\0&s'\end{pmatrix}\begin{pmatrix}1&V'\\0&1\end{pmatrix}\begin{pmatrix}M'&0\\0&1\end{pmatrix}\begin{pmatrix}1&0\\C'&1\end{pmatrix}.
\end{equation}
In general, $s', V', M', C'$ are each functions of $s,V,M,C$. When Eq. \eqref{matrix-factors} is satisfied, the inverse matrix $\mathcal{E}$ is given by Eq. \eqref{inverse-proj-matrix-gen},
\begin{equation}
    \mathcal{E}=\begin{pmatrix}
M^{-1}+\frac{1}{u}\left(M^{-1}V\right)\left(CM^{-1}\right)&\frac{-1}{u}M^{-1}V\\\frac{-1}{u}CM^{-1}&\frac{1}{u}\end{pmatrix},
\end{equation}
where $u=s-CM^{-1}V$. Identifying these components with those of Eq.~\eqref{curly-solder-matrix}, the satisfaction of this factorization of matrices requires that $\breve{y}{}^a_*=0$. Therefore,
\begin{equation}
    \frac{\mathfrak{q}(x,x^*)}{\mathfrak{q}{}^{*}(x,x^*)}=\varphi(x).
\end{equation}
In other words, the ratio of $\mathcal{H}_1$ factors $\mathfrak{q}^{(*)}$ is independent of the fiber coordinate $\{x^*\}$, and provides access to a spacetime-dependent scalar field $\varphi(x)$. Had $\breve{y}{}^a$ not been introduced in this way, only $\varphi=1$ permits a factorization of the form Eq.~\eqref{matrix-factors}. We may satisfy the above conditions by simply choosing as the $\mathcal{H}_1$ factor $\mathfrak{q}^*$ the previously defined projective factor $\breve{\mathfrak{p}}$,
\begin{equation}
    \mathfrak{q}{}^{*}(x,x^*):=\breve{\mathfrak{p}}(x,x^*)=\frac{x^*}{x^*_0}|\breve{e}{}^a{}_n|^{\frac{-1}{m+1}}.
\end{equation}
These statements appear to imply that $\mathcal{E}$ may be identified with a generalized \textit{projective $2$-frame} \cite{2-frame}. For more information on projective $2$-frames and their relationship to general coordinate transformations, i.e., Polyakov's so-called ``diffeomorphisms from gauge transformations," see \cite{gauge-from-diff-polyakov,polyakov-soldering}, and for the closely related idea of gauging the diffeomorphism group, see \cite{gauge-diffeomorphism,gauge-diff-lord-1,gauge-diff-lord-2,gauge-diff-lord-3}.

Following from Eq. \eqref{mathcal-E-map}, the projective $2$-frame $\mathcal{E}$ is seen to transform nonlinearly as
\begin{equation}
    \mathcal{E}'=S\mathcal{E}J^{-1},
\end{equation}
with $S\in SL(m+1,\mathbb{R})\simeq PGL(m,\mathbb{R})$ and $J\in GL(m+1,\mathbb{R})$. A $GL(m+1,\mathbb{R})$ vector $V$ is mapped to a $PGL(m,\mathbb{R})$ vector via the projective $2$-frame,
\begin{equation}
    V^A=\mathcal{E}{}^A{}_MV^M.
\end{equation}
A projective vector $V^A\in\mathcal{H}_1(x,x^*)$ will be called a $\breve{\mathfrak{p}}$-vector. In general, we will refer to $\mathcal{H}_{\mathfrak{w}}$ objects as $\breve{\mathfrak{p}}{}^{\mathfrak{w}}$-objects. As a further example, a rank-$(r,s)$ $\breve{\mathfrak{p}}{}^{\mathfrak{w}}$-tensor, 
\begin{equation}
    T^{A_1\dots A_r}{}_{B_1\dots A_s}:=\breve{\mathfrak{p}}{}^{\mathfrak{w}}t^{A_1\dots A_r}{}_{B_1\dots A_s},
\end{equation}
has $\mathfrak{w}=r-s$. We also introduce here the notation of using upper-case symbols to denote $\breve{\mathfrak{p}}{}^{\mathfrak{w}}$-objects and the corresponding lower-case symbol to denote its non-$\breve{\mathfrak{p}}{}^{\mathfrak{w}}$ parts. Notice that the above expression is nothing but a particular instantiation of a rank-$(r,s)^{\mathfrak{w}}$ tensor density of weight $\mathfrak{w}$.

As a final note, when a metric is available in $P\mathcal{M}$, it may be pulled back to $V\mathcal{M}$ via the general projective $2$-frames $\mathcal{E}$. The image of this metric in $V\mathcal{M}$ contains the scalar $\varphi(x)$ in a manner very similar to the Brans-Dicke scalar field \cite{witten}. For example, consider the $\{X^A\}$ as inertial coordinates and introduce a flat metric $\breve{\eta}_{AB}$ as
\begin{equation}
    \breve{\eta}_{AB}:=\breve{\mathfrak{p}}{}^{-2}\begin{pmatrix}
        \eta_{ab}&0\\0&\eta_0
    \end{pmatrix},
\end{equation}
with $\eta_{ab}=\text{diag}(1,-1,-1,\dots,-1)$ and $\eta_0=\pm1$. The exact form of this flat metric will be discussed in more depth in Sec. \ref{subsec:goldstone-metric}. Then, by forming
\begin{equation}
    \breve{G}_{MN}:=\breve{\eta}_{AB}\mathcal{E}^A{}_M\mathcal{E}^B{}_N,
\end{equation}
we find
\begin{equation}
\breve{G}_{MN}=\begin{pmatrix}\breve{g}_{mn}+\breve{y}_{{(m}}\breve{e}^*_{n)}+\breve{Y}{}^2\breve{e}^*_m\breve{e}^*_n&\frac{1}{x^*}\left(\breve{y}_n+\breve{Y}{}^2\breve{e}^*_n\right)\\\frac{1}{x^*}\left(\breve{y}_m+\breve{Y}{}^2\breve{e}^*_m\right)&\left(\frac{\breve{Y}}{x^*}\right)^2\end{pmatrix},
\end{equation}
where
\begin{equation}
    \breve{Y}{}^A:=\begin{pmatrix}
        \breve{y}{}^{a}\\x^*_0
    \end{pmatrix},\quad\quad\quad \breve{g}_{mn}:=\eta_{ab}\breve{e}{}^a{}_m\breve{e}{}^b{}_n.
\end{equation}
Exposing the scalar field $\varphi$, and evaluating at $y^a=0$, reveals
\begin{equation}
\breve{G}_{MN}=\begin{pmatrix}\varphi^2g_{mn}+\eta_0(x^*_0)^2g_mg_n&\eta_0\frac{(x^*_0)^2}{x^*}g_n\\\eta_0\frac{(x^*_0)^2}{x^*}g_m&\eta_0\left(\frac{x^*_0}{x^*}\right)^2\end{pmatrix},
\end{equation}
where $ g_m:=\partial_m\log\left|\frac{\partial y^a}{\partial x^n}\right|^{\frac{-1}{m+1}}$. One may further impose a vanishing $g_m$ by restricting to constant volumes, $|\frac{\partial y^a}{\partial x^n}|=\text{const.}$, in which case
\begin{equation}\label{breve-metric-twice-reduced}
\breve{G}_{MN}=\begin{pmatrix}\varphi^2g_{mn}&0\\0&\eta_0\left(\frac{x^*_0}{x^*}\right)^2\end{pmatrix}.
\end{equation}
Thus, $\varphi(x)$ may be viewed as providing a conformal class of $m$-dimensional metrics. We  now transition to developing the projective symmetric teleparallel connection.

\subsection{Projective Symmetric Teleparallel Connection}
\label{subsec:proj-st-conn}

Symmetric Teleparallel Connections \cite{symmetric-tel-connections-1,symmetric-tel-connections-2,teleparallel-palatini}, Symmetric teleparallel Equivalents of General Relativity (STEGR) \cite{og-stegr,symmetric-tel-gravity,STEGR-1,STEGR-2,teleparallel-3+1}, and teleparallel geometries in general \cite{teleparallel-general,teleparallel-intro,teleparallel-intro-2} are of recent interest. Symmetric teleparallel theories are, essentially, gravitational models of (symmetric teleparallel) spacetimes $S_m$ with vanishing torsion and curvature. The Einstein-Hilbert action may then be written in terms of the non-metricity, to which all gravitational effects are attributed. In an $S_m$, the vanishing of torsion and curvature determines the spacetime connection strictly in terms of the solder forms (frames), resulting in the \textit{Symmetric Teleparallel Connection}. Here, we develop the \textit{Projective Symmetric Teleparallel Connection} of a projective symmetric teleparallel spacetime $PS_m$. The projective symmetric teleparallel connection is intimately related with the entire group of general coordinate transformations, generated by the combined set of $m$-dimensional conformal and special linear transformations \cite{ogievetsky2,Ogievetsky3}.

Consider a geodesic vector $V$ on $P\mathcal{M}$ in an inertial frame,
\begin{equation}
    V^A\partial_AV^B=0.
\end{equation}
Using the projective $2$-frames of Eq. \eqref{curly-solder-matrix}, this geodesic equation may be pulled back to $V\mathcal{M}$, producing
\begin{equation}
    V^M\mathcal{E}^A{}_N\left(\partial_MV^N+\tilde{\Gamma}{}^N{}_{LM}V^L\right)=0,
\end{equation}
where $V^M:=(\mathcal{E}^{-1})^M{}_AV^A$ is the in-homogeneous image of $V^A$ in $V\mathcal{M}$. Mapping out of the inertial frame produces the \textit{Projective Symmetric Teleparallel Connection Coefficients} of $V\mathcal{M}$,
\begin{equation}\label{sym-tel-conn}
    \tilde{\Gamma}{}^N{}_{LM}:=(\mathcal{E}^{-1}){}^N{}_B\partial_M\mathcal{E}^B{}_L.
\end{equation}
Taking, for example, $V^N=\frac{dx^N}{d\tau}$ for some parameter $\tau$ gives the geodesic equation in standard form,
\begin{equation}\label{PST-conn-geodesic}
    \frac{d^2x^N}{d\tau^2}+\tilde{\Gamma}{}^N{}_{LM}\frac{dx^L}{d\tau}\frac{dx^M}{d\tau}=0.
\end{equation}
From this, we define the \textit{covariant derivative}
\begin{equation}
    \tilde{\nabla}_M:=\partial_M+\tilde{\Gamma}_M,
\end{equation}
and cast the geodesic equation in manifestly covariant form
\begin{equation}
    V^M\tilde{\nabla}_MV^N=0.
\end{equation}

The linear matrix, or M\"{o}bius representation will be used extensively throughout this document. We therefore represent $\tilde{\Gamma}$ as a matrix-valued co-vector, i.e., $ \tilde{\Gamma}{}^L{}_{NM}=\begin{pmatrix}\tilde{\Gamma}{}^L{}_{Nm},&\tilde{\Gamma}{}^L{}_{N*}\end{pmatrix}$, where the co-vector character resides in the form index. Explicitly, the general component-form is taken to be organized as
\begin{equation}
    \begin{split}
        \tilde{\Gamma}{}^L{}_{NM}=\begin{pmatrix}\begin{pmatrix}
            \tilde{\Gamma}{}^l{}_{nm}&\tilde{\Gamma}{}^l{}_{*m}\\\tilde{\Gamma}{}^*{}_{nm}&\tilde{\Gamma}{}^*{}_{*m}
        \end{pmatrix},&
        \begin{pmatrix}
            \tilde{\Gamma}{}^l{}_{n*}&\tilde{\Gamma}{}^l{}_{**}\\\tilde{\Gamma}{}^*{}_{n*}&\tilde{\Gamma}{}^*{}_{**}
        \end{pmatrix}\end{pmatrix}.
    \end{split}
\end{equation}

To compute the connection coefficients $\tilde{\Gamma}$, mapped from the projective inertial $m$-frame $\breve{e}$, we note that in addition to the various conditions of the previous subsection, we also have
\begin{equation}
\partial_*e^*_n=\partial_ne^*_*=\partial_*(\tfrac{1}{\mathfrak{q}^*}\mathcal{E}^a{}_n)=0.
\end{equation}
The connection coefficients, in general, are computed to be
\begin{equation}
    \tilde{\Gamma}{}^L{}_{NM}=\begin{pmatrix}
\begin{pmatrix}\breve{\Pi}^l{}_{nm}&\delta^l{}_me^*_*\\\frac{1}{e^*_*}\mathcal{D}_{bm}&0
        \end{pmatrix},\;\begin{pmatrix}
            \delta^l{}_ne^*_*&0 \\0 &e^*_*+e^*_{**}
        \end{pmatrix}
    \end{pmatrix}.
\end{equation}
In this expression, we have introduced the $m$-dimensional projective connection symbols,
\begin{equation}\label{pstegr-pi}
    \breve{\Pi}{}^l{}_{nm}:=(\breve{e}{}^{-1})^l{}_a\partial_m\breve{e}^a{}_n+\delta^l{}_ne^*_m+\delta^l{}_me^*_n,
\end{equation}
the Diffeomorphism field,
\begin{equation}
    \mathcal{D}_{nm}:=\partial_{m}e^*_n -e^*_l\Pi^l{}_{nm}+e^*_me^*_n,
\end{equation}
and the convenient notation
\begin{equation}
    e^*_{**}:=\partial_*\log e^*_*=\partial_*\log\left(\partial_*\log\mathfrak{q}^{*}\right).
\end{equation}
Choosing $\mathfrak{q}^*=\breve{\mathfrak{p}}$ as in the previous subsection simply amounts to $e^*_*=(x^*)^{-1}$ and $e^*_*+e^*_{**}=0$. 

By construction, $\tilde{\Gamma}$ is symmetric and, therefore, has a vanishing object of anholonimity \cite{anholonimity},
\begin{equation}
    C^B{}_{ML}:=\partial_{[M}\mathcal{E}^B{}_{L]}=0.
\end{equation}
Thus, $\tilde{\Gamma}$ also has a vanishing torsion and is symmetric in its lower indices. Additionally, the curvature associated with $\tilde{\Gamma}$ vanishes, as can easily be seen from Eq.~\eqref{sym-tel-conn}. These two properties define, in general, a \textit{Symmetric Teleparallel Connection}. 

The above connection, where $\varphi(x)$ is hidden or taken to be unity, is common in the literature on (flat) projective connections. Retaining the $\mathfrak{q}^*=\breve{\mathfrak{p}}$ choice and exposing the scalar field $\varphi(x)$ reveals
\begin{equation}\label{compatible-conf-proj-conn}
    \tilde{\Gamma}{}^L{}_{NM}=\begin{pmatrix}
        \begin{pmatrix}\Pi^l{}_{nm}-y^l\mathcal{P}_{nm}&\frac{1}{x^*}\delta^l{}_m\\x^*\mathcal{D}_{nm}&0
        \end{pmatrix},\;\begin{pmatrix}
      \frac{1}{x^*}\delta^l{}_n&0 \\0&0
        \end{pmatrix}
    \end{pmatrix}.
\end{equation}
In this version of $\tilde{\Gamma}$, we find the $m$-dimensional projective connection
\begin{equation}
    \Pi^l{}_{nm}:=(e^{-1})^l{}_a\partial_me^a{}_n+\delta^a{}_b\alpha_m+\delta^a{}_m\alpha_n,
\end{equation}
with $\alpha_m:=e^*_m=\partial_m\log|e^a{}_n|^{\frac{-1}{m+1}}$, and the Diffeomorphism field,
\begin{equation}
\mathcal{D}_{nm}=\mathcal{P}_{nm}+\partial_m\alpha_n-\alpha_l\Pi^l{}_{nm}+\alpha_m\alpha_n.
\end{equation}
With the scalar field $\varphi(x)$ unpacked, we also encounter the projective Schouten tensor,
\begin{equation}
    \mathcal{P}_{nm}=-(\tfrac{1+\alpha\cdot y}{1+\phi\cdot y})(\partial_m\phi_n-\phi_l\Pi^l{}_{nm}-\phi_m\phi_n+\mathcal{E}_m\phi_n+\mathcal{E}_n\phi_m),
\end{equation}
where $\phi_n:=\partial_n\log\varphi$. If $\varphi=\varphi(y(x))$ is rather taken to depend on $\{x\}$ via $\{y(x)\}$, then the above collapses to
\begin{equation}\label{P-metric}
   \mathcal{P}_{nm}=(\tfrac{1+\alpha\cdot y}{1+\phi\cdot y})e^a{}_me^b{}_n(\varphi\partial_a\partial_b\varphi^{-1}).
\end{equation}
The connection of Eq.~\eqref{compatible-conf-proj-conn} may be seen to represent a compatible projective/conformal structure, typically studied in the so-called Unimodular Conformal/Projective Relativity (UCPR) theory \cite{compatible-proj-conf-struc-1,compatible-proj-conf-struc-3,compatible-proj-conf-struc-4}. In \cite{compatible-proj-conf-struc-2,ehlers,ehlers-2}, it was argued that the fundamental structure of spacetime be such compatibility, and a follow-up paper discusses the prediction, measurement, and observation in UCPR, \cite{ucpr-measurement}. Interestingly, they introduce in the context of UCPR, a concept called \textit{retrodiction}, which is quite reminiscient of the assembly index in assembly theory, \cite{assembly-theory}, which investigates an emergent description for the origins of biological life.

Identifying the connection of Eq.~\eqref{compatible-conf-proj-conn} as representing a compatible projective/conformal structure requires that $\mathcal{P}_{nm}$ be viewed as descending to a metric. Additionally, up to the rational pre-factor, it may be possible to identify $\varphi^{-1/2}$ with a pre-Finsler function \cite{pre-finsler-func}, as this connection appears to be a particular instantiation of a Berwald connection or, more specifically, a Berwald-Thomas-Whitehead connection \cite{berwald-proj-trans}. These concepts find their footing in the study of projective and more general path geometries \cite{berwald,douglas,haantjes,thomas-3,projective-radius-vector} (see \cite{path-geometry-grassman} for their almost Grassmann structures).

Explicitly, consider Eq. \eqref{P-metric}, for ease, in the particular scenario where $\alpha\cdot y=\phi\cdot y=0$, and the tangent indices are Lorentz, i.e., $a,b\rightarrow\underline{a},\underline{b}$ forcing $\{y\}$ into a set of $m$-dimensional inertial coordinates. In light of the above discussion, the projective Schouten tensor may then be viewed as providing a pair of conformally related metrics,
\begin{equation}\label{P-metric-star}
    g_{mn}:=\mathcal{P}_{mn}=\varphi g^*_{mn},
\end{equation}
where
\begin{equation}
    g^*_{mn}:=\eta^*_{\underline{ab}}e^{\underline{a}}{}_me^{\underline{b}}{}_n,
\end{equation}
and
\begin{equation}
    \eta^*_{\underline{ab}}:=\frac{\partial^2\varphi^{-1}}{\partial y^{\underline{a}}\partial y^{\underline{b}}},
\end{equation}
provided $\varphi\neq0$ and $\partial_{\underline{a}}\partial_{\underline{b}}\varphi^{-1}(y)\neq 0$. With these conditions satisfied, $\mathcal{P}_{mn}$ is a symmetric, non-degenerate bilinear form and satisfies the properties required of a (pseudo)-Riemannian metric.

\subsection{Geodesics}
\label{subsec:proj-st-geodesic}

The scalar field $\varphi(x)=\mathfrak{q}(x,x^*)/\mathfrak{q}^*(x,x^*)$ may be shown to provide a parameterization of geodesics via the Schwarzian derivative. Others have studied the Schwarzian derivative as the Euler-Lagrange equation for a system with vanishing Diffeomorphism field \cite{schwarzian-euler-lagrange}. For notational ease, we write $\mathfrak{q}_*\equiv\mathfrak{q}^*$ for the remainder of this section. 

Consider $\mathfrak{q}(x,x^*)$ and $\mathfrak{q}_*(x,x^*)$ as two independent solutions to the $1$-dimensional Sturm-Liouville equation \cite{dynamical-proj-curv},
\begin{equation}\label{specific-sturm}
    \hat{\bm{L}}\mathfrak{q}_{(*)}(x,x^*):=\left(-2c\frac{d^2}{dx^2}+\mathcal{D}(x)\right)\mathfrak{q}_{(*)}(x,x^*)=0.
\end{equation}
The above follows a one-to-one correspondence with centrally extended coadjoint elements of the Virasoro algebra,
\begin{equation}
    (\mathcal{D},c)\Leftrightarrow -2c\frac{d^2}{dx^2}+\mathcal{D}(x).
\end{equation}
Consider the first derivative of $\varphi(x,x^*)$,
\begin{equation}
    \varphi'(x,x^*):=\frac{d\varphi(x,x^*)}{dx}=\frac{\mathfrak{q}'\mathfrak{q}_*-\mathfrak{q}\mathfrak{q}_*'}{\mathfrak{q}_*^2}.
\end{equation}
The numerator is simply the Wronskian of the two solutions and guarantees $\varphi'(x,x^*)\neq0$. The second derivative provides
\begin{equation}
\varphi''(x,x^*)=-2\frac{\mathfrak{q}'\mathfrak{q}_*'}{\mathfrak{q}_*^2}+2\frac{\mathfrak{q}(\mathfrak{q}_*')^2}{\mathfrak{q}_*^3}+\frac{\mathfrak{q}''}{\mathfrak{q}_*}-\frac{\mathfrak{q}\mathfrak{q}_*''}{\mathfrak{q}_*^2}.
\end{equation}
Using the assumption that $\mathfrak{q}_*$ is a solution to the Sturm-Liouville system, Eq. \eqref{specific-sturm}, the second derivative may be written as
\begin{equation}
\varphi''(x,x^*)=-2\frac{\mathfrak{q}'\mathfrak{q}_*'}{\mathfrak{q}_*^2}+2\frac{\mathfrak{q}(\mathfrak{q}_*')^2}{\mathfrak{q}_*^3}.
\end{equation}
Lastly, the third derivative of $\varphi(x,x^*)$ with respect to $x$ yields
\begin{equation}\label{third-der-phi}
\varphi'''(x,x^*)=\frac{2(\mathfrak{q}_*\mathfrak{q}'-\mathfrak{q}\mathfrak{q}_*')(\mathfrak{q}_*^2\mathcal{D}+3c(\mathfrak{q}_*')^2)}{c\mathfrak{q}_*^4},
\end{equation}
where Eq. \eqref{specific-sturm} was used to introduce $\mathcal{D}(x)$ into the expression. Solving Eq. \eqref{third-der-phi} for $\mathcal{D}(x)$ results in the Schwarzian derivative of the scalar field $\varphi(x)=\mathfrak{q}(x,x^*)/\mathfrak{q}^*(x,x^*)$,
\begin{equation}\label{diff-schwarzian-phi}
    \mathcal{D}(x)=\frac{c}{2}\left(\frac{\varphi'''}{\varphi'}-\frac{3}{2}\left(\frac{\varphi''}{\varphi'}\right)^2\right)=\frac{c}{2}S(\varphi(x):x).
\end{equation}
We therefore find $\log\varphi$ is analogous to the ratio of two geodesically equivalent metric determinants, as considered in \cite{diff-scalar-ratio}. Moreover, this is \textit{exactly} the projective parameters in Eq. \eqref{equiv-metrics} resulting from the restricted class of diffeomorphisms that map geodesics to geodesics. According to \cite{no-2-metric-with-same-diff-field}, we may conclude that no two metrics have the same $\varphi$.

In order to connect Eq. \eqref{diff-schwarzian-phi} to reparameterizations, we turn to the geodesic equation with respect to the connection of Eq. \eqref{compatible-conf-proj-conn}. Expanding the sums in Eq. \eqref{PST-conn-geodesic} and separating the $x^*$ equation yields
\begin{equation}\label{general-pi-geo}
    \frac{d^2 x^n}{d\tau^2}+\Pi^n{}_{lm}\frac{dx^l}{d\tau}\frac{dx^m}{d\tau}=\frac{-2}{x^*}\frac{dx^*}{d\tau}\frac{dx^n}{d\tau}+y^n\mathcal{P}_{lm}\frac{dx^l}{d\tau}\frac{dx^m}{d\tau},
\end{equation}
and
\begin{equation}\label{general-diff-geo}
    \frac{d^2x^*}{d\tau^2}+x^*\mathcal{D}_{nm}\frac{dx^n}{d\tau}\frac{dx^m}{d\tau}=0.
\end{equation}
Compare with Eqs. \eqref{geo-diff} for the geodesics of Thomas-Whitehead theory. Recall that $\tau$ is an affine parameter with respect to $\tilde{\Gamma}$, which is not necessarily also affine with respect to $\Pi$. Suppose there exists a parameter $\pi$ which is affine with respect to $\Pi$,
\begin{equation}
     \frac{d^2 x^n}{d\pi^2}+\Pi^n{}_{lm}\frac{dx^l}{d\pi}\frac{dx^m}{d\pi}=0.
\end{equation}
Then, consider a reparameterization $\tau\rightarrow \pi(\tau)$. The first set of geodesic equations in Eqs. \eqref{general-pi-geo} leads to
\begin{equation}
    \left(\frac{d^2\pi}{d\tau^2}+\frac{2}{x^*}\frac{dx^*}{d\tau}\frac{d\tau}{d\pi}\right)\frac{dx^n}{d\pi}=y^n\mathcal{P}_{lm}\frac{dx^l}{d\tau}\frac{dx^m}{d\tau}.
\end{equation}
This differs from the geodesic equation of TW theory by the nonzero term on the right-hand side, which is equivalent to
\begin{equation}
    y^n\mathcal{P}_{lm}\frac{dx^l}{d\tau}\frac{dx^m}{d\tau}\equiv \varphi y^n\left(\frac{ds_*}{d\tau}\right)^2,
\end{equation}
where $ds^2_*=g^*_{mn}dx^mdx^n$. Consider the simple scenario $y=0$. Then, the parameterization satisfies
\begin{equation}
     \frac{d^2\pi}{d\tau^2}=\frac{-2}{x^*}\frac{dx^*}{d\tau}\frac{d\tau}{d\pi}.
 \end{equation}
Making use of Eq. \eqref{general-diff-geo}, we find
\begin{equation}\label{diff-schwarzian-pi-tau}
     \mathcal{D}_{nm}\frac{dx^n}{d\tau}\frac{dx^m}{d\tau}=\frac{1}{2}\frac{(\frac{d^3\pi}{d\tau^3})\frac{d\pi}{d\tau}-\frac{3}{2}(\frac{d^2\pi}{d\tau^2})^2}{(\frac{d\pi}{d\tau})^2}\equiv S(\pi:\tau).
 \end{equation}
This again shows the role played by $\mathcal{D}$ in modulating the space of permissible reparameterizations. The $y\neq0$ scenario is left for future investigation. 

Reducing Eq. \eqref{diff-schwarzian-pi-tau} to $m=1$ dimension, we may identify a relationship between the affine parameter for $\Pi$ and the scalar field $\varphi$. Comparing with Eq. \eqref{diff-schwarzian-phi} implies a charge $c=1$, and 
\begin{equation}
    \frac{d\pi}{d\tau}\equiv\frac{d\varphi}{dx}.
\end{equation}
Further investigation of this relationship is left for future research.

\section{Nonlinear Realization of Projective Linear Group}
\label{sec:nlr-pgl}

In this section we remove the inertial condition on the $\{X^A\}$ coordinates, permitting the existence of a non-inertial framed connection $\Omega$ of an arbitrary frame. This generalization of Eq. \eqref{sym-tel-conn} complicates the relationship between the theory written on $V\mathcal{M}$ and the theory written on $P\mathcal{M}$. This is due to the presence of a non-trivial torsion, induced by the fundamental projective vector field. For this reason, and many others, we nonlinearly realize the Projective Linear Group over the Lorentz stability subgroup. Complementary to the method of coadjoint orbits \cite{nonlinear-orbit,stringtheory3} and the reverse of the dressing field method \cite{2-frame,2-frame-dressing-field-nonprojective}, this particular nonlinear realization process provides \textit{access to} a set of $m$-dimensional inertial coordinates and, therefore, access to a flat metric. Moreover, this process elucidates the physical and gauge content of the theory by reorganizing the available degrees of freedom. 

This section is structured as follows. We begin with a short review of the nonlinear realization technique for gauge symmetries, followed by a thorough discussion of the coset fields parameterizing the projective coset space of interest. We then develop and discuss the flat projective metric, which will used extensively throughout the remainder of this document. This section closes with an exhaustive construction of the nonlinearly realized projective gauge connection-form and the associated curvature-form.

For more information on the nonlinear realization technique and its application to various gauge groups, see \cite{isham-salam-strathdee,salam-nonlinear,ivanenko-cosets-bundle-gravity,kharuk-coset-1,ogievetsky1,diffeomorphism-coset,goon,goon-2,ogievetsky-proceedings,schwarzian-inverse-higgs,Leclerc,AdS-cosets,ads-higgs-gravity,without-metric}. For the relation between the nonlinear realization and orbit methods, see \cite{nonlinear-orbit}, and for the proposed ``improved" technique of composite bundles, see \cite{composite-bundle,composite-bundle-1,composite-bundles-gl4}.

\subsection{Nonlinear Realization Technique}
\label{subsec:nlr}

The nonlinear realization technique or coset construction offers a systematic method to handle scenarios in which a spacetime symmetry group $G$ is spontaneously or explicitly broken down to a closed subgroup $H$. Initially, the full system admits symmetries under all transformations contained in $G$. After symmetry breaking, only those transformations in $H\subset G$ remain linearly realized, preserving a chosen geometric or physical structure. Any further reduction stages (e.g. $H_1 \rightarrow H_2$) likewise lead to additional cosets, each describing newly broken directions.

Following \cite{higgs-mech-for-grav}, let $G$ be a Lie group and $H\subset G$ a closed subgroup. One forms the \textit{homogeneous} or \textit{coset space} $G/H$ from the left cosets of $H$ as
\begin{equation}
    G/H=\{gH,g\in G\}.
\end{equation}
This space describes the degrees of freedom associated with the broken generators of $G$. Each coset $g_iH$ is disjoint from the others, and one can view
\begin{equation}
    G=\{H\cup g_1H\cup g_2H\cup... \},
\end{equation}
in which $g_1\notin H,\;g_2\notin\{H,g_1H\}$, etc., as providing a covering of $G$. In this sense, each coset $g_iH$ may be viewed as a ``copy” of $H$, with each copy moved around by $g_i$. This is the essence of the complementarity between the orbit technique and the nonlinear realization technique.

A convenient way to organize these cosets is to pick a \textit{coset representative},
\begin{equation}
    \sigma(p_i)\in G,
\end{equation}
where $\{p_i\}$ is a set of coordinates labeling points in $G/H$. This choice effectively identifies each point in the coset space with a unique element of $G$, since one may interpret $\sigma(p_i)\in G$ as a ``point” in the coset. In the gravitational context, this may be interpreted as a (local) embedding of the spacetime coordinates and other parameters, all contained in $\{p_i\}$, into the group. In a global setting, these coordinates typically represent the Goldstone modes, or in general, the ``nonlinear” fields that ``hide" transformation misbehavior when the system in question is acted on by the full group $G$. In other words, the coordinates $\{p_i\}$ are the compensating fields used to ensure fields transform linearly with respect to the unbroken subgroup $H$. Because $H$ remains unbroken, it is identified with the local (or ``residual”) symmetry still present in the reduced system. In a local setting, the coordinates $\{p_i\}$ in $G/H$ capture the ``would-be Goldstone modes," or in general, the ``nonlinear” fields that describe fluctuations away from an $H$-invariant configuration when acted on by the full group $G$. However, both the local and global setting follow identical procedures, only departing from one another in the physical interpretation of the coset coordinates.

From a geometric viewpoint, $G$ then serves as a principal fiber bundle over $G/H$, with structure group $H$. Concretely, one has:
\begin{itemize}
    \item A \textbf{base space}: the coset $G/H$, parameterized by $\{p_i\}$.
    \item A \textbf{fiber}: isomorphic to $H$, since each coset $g_iH$ is shifted by elements of $H$.
    \item A \textbf{projection} $\sigma^{-1}:G\rightarrow G/H$, taking each $g\in G$ to its corresponding coset.
\end{itemize}
Because $H$ is the unbroken subgroup, it acts as the “local symmetry” preserving a chosen reference configuration or vacuum. The coset space, $G/H$, collects all the directions in $G$ that are ``broken” and thus parameterized by the additional fields.

A crucial feature of the \textit{nonlinear realization} process is that the coset coordinates $\{p_i\}$ transform non-trivially under $G$. Specifically, if $g\in G$ acts on $\sigma(p_i)$ by left multiplication, one obtains
\begin{equation}\label{gen-non-lin-trans-law}
g\sigma(p_i)=\sigma(p_i')h(p_i,g),\;\;\quad h\in H.
\end{equation}
Equivalently, with arguments suppressed,
\begin{equation}
    \sigma'=g\sigma h^{-1}.
\end{equation}
The new parameters $\{p'_i\}$ thus do not transform under a linear representation of $G$; rather they are related nonlinearly to the old parameters, governed by an element $h(p_i,g)$ depending on both $\{p_i\}$ and $g$. This ensures that the coset degrees of freedom do not sit in an ordinary (linear) representation of $G$, but rather mix among themselves via elements of $H$.

When the symmetry is successively broken in, for example, $n$ stages, a composite coset may be constructed from the coset of each stage and the applicability of the above program remains. To show this, consider the $n$-stage process $H_0\rightarrow \dots\rightarrow H_n$, with each of the $n$-stages identified by its stability subgroup $H_i$. Let $\sigma_i^{-1}:H_{i-1}\rightarrow H_{i-1}/H_i$, with $i\in\mathbb{N}^{+}$ and $i\leq n$, represent each coset element. We then identify $H_0$ with $G$, so that $\sigma_1^{-1}:G\rightarrow G/H_1$. These coset elements each satisfy the nonlinear transformation law, Eq. \eqref{gen-non-lin-trans-law},
\begin{equation}
    \sigma'_i=h_{i-1}\sigma_{i} h^{-1}_{i},\;\;\quad h_{i}\in H_i.
\end{equation}
We construct the composite projection, $\sigma^{-1}:H_0\rightarrow H_0/H_n\equiv G/H_n$, from the $\sigma_i$ as
\begin{equation}
    \sigma:=\prod_{i=1}^n\sigma_i.
\end{equation}
The composite element then satisfies
\begin{equation}
    \begin{split}
\sigma'&=\prod_{i=1}^n\sigma_i'\\
&=\sigma_1'\sigma_2'\dots\sigma_n'\\
&=h_0\sigma_1h_1^{-1}h_1\sigma_2h_2^{-1}\dots h_{n-1}\sigma_nh_n\\
&=h_0\sigma_1\sigma_2\dots\sigma_nh_n\\
&=h_0\sigma h_n\\
&\equiv g\sigma h_n,
    \end{split}
\end{equation}
proving the claim.

A key ingredient in the coset method for global symmetries  is the (un-gauged) \textit{Maurer–Cartan form}, defined as
\begin{equation}
    \omega:=\sigma^{-1}d\sigma,
\end{equation}
where $d$ is the exterior derivative on the manifold spanned by the fields $\{p_i\}$. Because $\sigma\in G$, the one-form $\omega$ takes values in the Lie algebra $\mathfrak{g}$ of $G$. The Maurer-Cartan form simply organizes how the coset parameters $\{p_i\}$ transform globally. For $\mathfrak{h}\subset\mathfrak{g}$ a Lie subalgebra of $\mathfrak{g}$, with $\mathfrak{h}$ the Lie algebra of the unbroken symmetry group $H$, $\mathfrak{g}$ may be decomposed with respect to $\mathfrak{h}$ and its complementary subspace $\mathfrak{k}$, representing the space spanned by the broken generators. The complementarity simply means $\mathfrak{h}\cap\mathfrak{k}=\{0\}$. This decomposition takes the form of a direct sum of vector spaces,
\begin{equation}
\mathfrak{g}=\mathfrak{h}\oplus\mathfrak{k}.
\end{equation}
Accordingly, the Maurer-Cartan form $\omega$ decomposes into
\begin{equation}
\omega=\omega_{\mathfrak{h}}+\omega_{\mathfrak{k}},
\end{equation}
enabling one to track separately how the broken and unbroken sectors transform. Under the action of $g\in G$, the Maurer–Cartan form transforms in a way that preserves its group-theoretic structure, but typically acquires an inhomogeneous shift in the $\mathfrak{h}$-valued part. In many physical applications, $\omega_{\mathfrak{k}}$ corresponds to ``would-be Goldstone modes” for the broken symmetry transformations. Under the left global action of $g\in G$ and the induced local $h\in H$, the Maurer-Cartan form $\omega$ transforms in a predictable way that allows the construction of invariant Lagrangians or other geometric objects through appropriate contractions, wedges, and/or traces. For example, the (gauged) Wess-Zumino terms investigated in \cite{brauner-gauged-WZW}, and references therein. Explicitly, the transformation of the Maurer-Cartan form is easily computed as
\begin{equation}
    g:\omega\rightarrow\omega'=(\sigma^{-1})'d\sigma'=h(d+\omega)h^{-1}.
\end{equation}

When we \textit{gauge} these symmetries by promoting, for example, all of the transformations in $G$ to local ones, i.e., $g(x)\in G$, one introduces a gauge connection $\Omega$ for the relevant local symmetry generators. In that case, we follow the standard \textit{minimal coupling} prescription, replacing the exterior derivative with the covariant derivative,
\begin{equation}
    d\rightarrow D:=d+\Omega.
\end{equation}
Covariance then requires that $\Omega$ satisfy the transformation behavior 
\begin{equation}
    g(x):\Omega\rightarrow \Omega'=g(d+\Omega)g^{-1}.
\end{equation}
The Maurer–Cartan form then generalizes to a \textit{gauged} or \textit{covariant} Maurer–Cartan form, which we denote as $\tilde{\Omega}$:
\begin{equation}
\tilde{\Omega}:=\sigma^{-1}D\sigma=\sigma^{-1}(d+\Omega)\sigma.
\end{equation}
Under local gauge transformations $g(x)\in G$, both $\sigma$ and $\Omega$ adjust so that $\tilde{\Omega}$ transforms covariantly as
\begin{equation}
    g(x):\tilde{\Omega}\rightarrow\tilde{\Omega}'=h(d+\tilde{\Omega})h^{-1}
\end{equation}
This is essential if we wish to interpret part of $G$ as a local spacetime symmetry—commonly the subgroup $H$, but in some cases, a larger portion of $G$. The \textit{geometrical} fields then appear within $\tilde{\Omega}$, with components associated to different generators of $\mathfrak\{g\}$. 

As we will find in the remainder of this section, when applied to the projective linear group, the geometrical fields contained in $\tilde{\Omega}$ consist of the projectively invariant Lorentz spin connection, non-metricity, co-frame, and Schouten form. Each of these are associated with, respectively, local Lorentz transformations, local shears, local translations, and local projective transformations (pseudo-translations). We now turn to the construction of such a projective coset element.

\subsection{Coset Element}
\label{subsec:coset}

Consider $S(x)\in SL(m+1,\mathbb{R})\cong PGL(m,\mathbb{R})$, where $S\equiv\{S^{A'}{}_B\}$ is the matrix
\begin{equation}\label{SL-matrix}
    S^{A'}{}_B=|s|^{\frac{-1}{m+1}}\begin{pmatrix}
        s^{a'}{}_b-t^{a'}u_b&\frac{1}{x^*_0}t^{a'}\\-x^*_0u_b&1
    \end{pmatrix}.
\end{equation}
The inverse, $S^{-1}(x)$, is easily found to have the form
\begin{equation}\label{SL-matrix-inv}
    (S^{-1})^{A}{}_{B'}=|s|^{\frac{1}{m+1}}\begin{pmatrix}
       (s^{-1})^a{}_{b'}&\frac{-1}{x^*_0}(s^{-1})^a{}_{b'}t^{b'}\\x^*_0u_a(s^{-1})^a{}_{b'}&1-u\cdot s^{-1}\cdot t
    \end{pmatrix}.
\end{equation}
In the above, $|s|:=|s^{a'}{}_b|$ denotes the determinant of $s$. Furthermore, we consider
\begin{equation}\label{u-trans-parameter-def}
    u_b=\partial_b\log|s|^{\frac{1}{m+1}}.
\end{equation}
The gauge transformation matrices, $S$ and $S^{-1}$, are easily verified as elements of $SL(m+1,\mathbb{R})$. Using Eq. \eqref{det-formula} to form the determinant of a $PGL(m,\mathbb{R})$ matrix, one finds that $|S|=|S^{-1}|=1$.

Let $G\equiv SL(m+1,\mathbb{R})\cong PGL(m,\mathbb{R})$ and $H\equiv SO(m,\mathbb{R})$, and consider the coset projection $\sigma^{-1}:G\rightarrow G/H$. We define
\begin{equation}\label{coset-element}
    \sigma^A{}_{\underline{B}}:=|r|^{\frac{-1}{m+1}}\begin{pmatrix}
        r^a{}_{\underline{b}}&\frac{1}{x^*_0}r^a{}_{\underline{b}}\xi^{\underline{b}}\\
        -x^*_0\change_ar^a{}_{\underline{b}}&1-\change\cdot r\cdot\xi
    \end{pmatrix}
\end{equation}
as the matrix form of the coset element, and
\begin{equation}\label{coset-element-inv}
    (\sigma^{-1}){}^{\underline{B}}{}_A:=|r|^{\frac{1}{m+1}}\begin{pmatrix}
        (r^{-1}){}^{\underline{b}}{}_a-\xi^{\underline{a}}\change_b&\frac{-1}{x^*_0}\xi^{\underline{a}}\\x^*_0\change_a&1
    \end{pmatrix}
\end{equation}
as its inverse, with $|r|:=|r^a{}_{\underline{b}}|$. The coset parameter known as the ``reducing matrix" \cite{Leclerc}, $r^a{}_{\underline{b}}$, is required to be symmetric since it parameterizes the $\tfrac{m(m+1)}{2}$-dimensional coset space $\text{Sym}(\tfrac{m(m+1)}{2})$. The coset parameters $\xi$ and $\change$ parameterize, respectively, the $m$-dimensional coset spaces $\mathbb{R}^m$ and $\mathbb{R}^m_*$. The former may be identified with the $m$-dimensional manifold $\mathcal{M}$ once the symmetry has been broken and $\xi$ provides access to coordinates \cite{higgs-mech-for-grav}. The collection $\{(r^{-1})^{\underline{a}}{}_b,\;\xi^{\underline{a}},\;\change_b\}$ thus parameterizes the $\frac{m(m+5)}{2}$-dimensional coset space
\begin{equation}
    SL(m+1,\mathbb{R})/SO(m,\mathbb{R}).
\end{equation}

The coset parameters naturally satisfy the nonlinear transformation law. Specifying the general definition in Eq. \eqref{gen-non-lin-trans-law} to the present, we find that
\begin{equation}
    \sigma^{A'}{}_{\underline{B}'}=S^{A'}{}_C\sigma^C{}_{\underline{D}}(\Lambda^{-1})^{\underline{D}}{}_{\underline{B}'},
\end{equation}
where
\begin{equation}
    \Lambda^{\underline{A}'}{}_{\underline{B}}:=\begin{pmatrix}
        \lambda^{\underline{a}'}{}_{\underline{b}}&0\\0&1
    \end{pmatrix},\quad\quad\quad |\lambda^{\underline{a}'}{}_{\underline{b}}|=1,
\end{equation}
is the M\"{o}bius representation of an $m$-dimensional Lorentz transformation. We will often refer to these as projective Lorentz transformations. The exact form of this Lorentz transformation appears in the literature on non-projective (anti)-de Sitter gauge theories \cite{de-sitter-fermion-2}. The inverse coset element transforms as
\begin{equation}
    (\sigma^{-1}){}^{\underline{B}'}{}_{A'}=\Lambda^{\underline{B}'}{}_{\underline{C}}(\sigma^{-1}){}^{\underline{C}}{}_{D}(S^{-1}){}^{D}{}_{A'}.
\end{equation}

The components of the transformed coset element $\sigma^{A'}{}_{\underline{B}'}$ are, explicitly,
\begin{equation}
    |rs|^{\frac{-1}{m+1}}\begin{pmatrix}
       \left(s^{a'}{}_cr^c{}_{\underline{d}}-t^{a'}(\change_b+u_b)r^b{}_{\underline{d}}\right)\overline{\lambda}{}^{\underline{d}}{}_{\underline{b}'}&\frac{1}{x^*_0}\left(s^{a'}{}_cr^c{}_{\underline{b}}\xi^{\underline{b}}+t^{a'}-t^{a'}(\change_c+u_c) r^c{}_{\underline{b}}\xi^{\underline{b}})\right)\\-x^*_0\left(\change_c+u_c\right)r^c{}_{\underline{d}}\overline{\lambda}{}^{\underline{d}}{}_{\underline{b}'}
       &\left(1-(\change_c+u_c)r^c{}_{\underline{b}}\xi^{\underline{b}}\right)
    \end{pmatrix}.
\end{equation}
From this, we find the transformation of the overall determinant factor is as expected:
\begin{equation}\label{det-trans-r}
    |r'|^{\frac{-1}{m+1}}=|r|^{\frac{-1}{m+1}}|s|^{\frac{-1}{m+1}}=|s^{a'}{}_br^b{}_{\underline{c}}|^{\frac{-1}{m+1}}.
\end{equation}
Taking the determinant of the transformed $r$-components, i.e., $\sigma^{a'}{}_{\underline{b}'}$, and noting that for Lorentz transformations $|(\lambda^{-1})^{\underline{a}}{}_{\underline{b}}'|=1$, we find that
\begin{equation}
        |r'|=|s||r||\delta^b{}_c-(s^{-1}){}^b{}_{d'}t^{d'}(\change_c+u_c)|.
\end{equation}
For this to agree with Eq. \eqref{det-trans-r}, it must be that
\begin{equation}
    |\delta^b{}_c-(s^{-1}){}^b{}_{d'}t^{d'}(\change_c+u_c)|=1.
\end{equation}
The satisfaction of this expression imposes restrictions on the coset parameters. To find such restrictions, we utilize the components of $(\sigma^{-1})'$. Explicitly, the transformed components of $\sigma^{-1}$ are:
\begin{equation}\label{inv-coset-element-trans}
    \begin{split}
          (\sigma^{-1}){}^{\underline{a}'}{}_{b'}&=|rs|^{\frac{1}{m+1}}\lambda^{\underline{a}'}{}_{\underline{c}} \left((r^{-1}){}^{\underline{c}}{}_{d}-\xi^{\underline{c}}(\change_d+u_d)\right)(s^{-1}){}^{d}{}_{b'},\\
            (\sigma^{-1}){}^{\underline{a}'}{}_{*'}&=\tfrac{-1}{x^*_0}|rs|^{\frac{1}{m+1}}\lambda^{\underline{a}'}{}_{\underline{b}}\left(\xi^{\underline{b}}+t^{\underline{b}}-\xi^{\underline{b}}(\change_c+u_c)(s^{-1}){}^{c}{}_{d'}t^{d'}\right),\\
              (\sigma^{-1}){}^{\underline{*}'}{}_{b'}&=x^*_0|rs|^{\frac{1}{m+1}}(\change_{c}+u_c)(s^{-1}){}^{c}{}_{b'},\\
                (\sigma^{-1}){}^{\underline{*}'}{}_{*'}&=|rs|^{\frac{1}{m+1}}\left(1-(\change_{c}+u_c)(s^{-1}){}^c{}_{b'}t^{b'}\right),\\
    \end{split}
\end{equation}
where $t^{\underline{b}}\equiv(r^{-1}){}^{\underline{b}}{}_{c}(s^{-1}){}^{c}{}_{d'}t^{d'}$. According to Eq. \eqref{det-trans-r}, we must require the transformation of the scalar component, 
\begin{equation}
    (\sigma^{-1}){}^{\underline{*}}{}_{*}=|r|^{\frac{1}{m+1}},
\end{equation}
takes the form
\begin{equation}
    (\sigma^{-1}){}^{\underline{*}'}{}_{*'}=|rs|^{\frac{1}{m+1}}.
\end{equation}
Reflecting this in the last expression in Eqs. \eqref{inv-coset-element-trans} yields $1=1-(\change_{c}+u_c)t^{c}$, which may be resolved to find
\begin{equation}\label{cent-t-orth}
    (\change_{c}+u_c)(s^{-1}){}^c{}_{b'}t^{b'}=\change_{b'}t^{b'}=0.
\end{equation}
We may now prove that $|\delta^b{}_c-(s^{-1}){}^b{}_{d'}t^{d'}(\change_c+u_c)|=1$. For convenience, we define $X^b{}_c:=(s^{-1}){}^b{}_{d'}t^{d'}(\change_c+u_c)$ and $Y^b{}_c:=\delta^b{}_c-X^b{}_c$, and consider an $m=4$-dimensional manifold of split-signature $(p,q)=(1,3)$. The determinant of interest may then be expressed as
\begin{equation}\label{Y-det}
       Y=\frac{-1}{4!}\hat{\epsilon}_{abcd}\hat{\epsilon}^{efgh}Y^a{}_eY^b{}_fY^c{}_gY^d{}_h.
\end{equation}
See Appendix \hyperref[app-A:orientation]{A.2} for further information on expressing determinants in this manner. Executing the sums in Eq. \eqref{Y-det} and defining $X:=X^a{}_a$, one finds, in addition to $1$, a sum of terms each contributing a $\text{tr}(X\dots X)$ containing $\leq 4$ factors of $X$. Each of these terms vanishes due to Eq.~\eqref{cent-t-orth}. A third order term, for example, has the form $X^a{}_bX^b{}_cX^c{}_a=(t^{d'}\change_{d'})^3=0$. Therefore, $|\delta^b{}_c-X^b{}_c|=1$, as claimed.

Collecting the results, the transformation of each component contained in both $\sigma$ and $\sigma^{-1}$ are:
\begin{equation}
    \begin{split}
    \change_{b'}&=(\change_c+u_c)(s^{-1}){}^c{}_{b'},\\
    \xi^{\underline{a}'}&=\lambda^{\underline{a}'}{}_{\underline{b}}(\xi^{\underline{b}}+(r^{-1}){}^{\underline{b}}{}_{c}(s^{-1}){}^{c}{}_{d'}t^{d'})=\lambda^{\underline{a}'}{}_{\underline{b}}\xi^{\underline{b}}+t^{\underline{a}'},\\
    r^{a'}{}_{\underline{b}'}&=s^{a'}{}_c(\delta^c{}_b-(s^{-1}){}^c{}_{d'}t^{d'}\change_{e'}s^{e'}{}_b)r^b{}_{\underline{d}}\overline{\lambda}{}^{\underline{d}}{}_{\underline{b}'},\\
    (r^{-1}){}^{\underline{a}'}{}_{b'}&=\lambda^{\underline{a}'}{}_{\underline{b}}(r^{-1}){}^{\underline{b}}{}_c(s^{-1}){}^c{}_{d'}(\delta^{d'}{}_{b'}+t^{d'}\change_{b'}),\\
        (\change_ar^{a}{}_{\underline{b}'})&=\change_{a'}r^{a'}{}_{\underline{b}}=(\change_a+u_a)r^a{}_{\underline{d}}\overline{\lambda}{}^{\underline{d}}{}_{\underline{b}'},\\
        (\change\cdot r\cdot \xi)'&=\change_{c'}r^{c'}{}_{\underline{b}'}\xi^{\underline{b}'}=(\change_c+u_c)r^c{}_{\underline{b}}\xi^{\underline{b}},\\
        (r^{a'}{}_{\underline{b}}\xi^{\underline{b}})&=r^{a'}{}_{\underline{b}'}\xi^{\underline{b}'}=(\delta^{a'}{}_{b'}-t^{a'}\change_{b'})s^{b'}{}_cr^c{}_{\underline{b}}\xi^{\underline{b}}+t^{a'}.
    \end{split}
\end{equation}
The orthogonality relation between $t^{a'}$ and $\change_{a'}$ ensures that $r^{a}{}_{\underline{b}}$ and $(r^{-1}){}^{\underline{b}}{}_{c}$ remain inverses of each other after the transformation, i.e.,
\begin{equation}
        r^{a'}{}_{\underline{b}'}(r^{-1}){}^{\underline{b}'}{}_{c'}=\delta^{a'}{}_{c'},\quad\quad  (r^{-1}){}^{\underline{b}'}{}_{c'}r^{c'}{}_{\underline{a}'}=\delta^{\underline{b}'}{}_{\underline{a}'}
\end{equation}
Notice that when the vertical translation symmetry is broken, $\xi^{\underline{a}}=0$ and $t^{a'}=0$, for example, one has the familiar transformation behavior for the reducing matrix:
\begin{equation}
\begin{split}
    r^{a'}{}_{\underline{b}'}&=s^{a'}{}_br^b{}_{\underline{d}}\overline{\lambda}{}^{\underline{d}}{}_{\underline{b}'},\\
    (r^{-1}){}^{\underline{a}'}{}_{b'}&=\lambda^{\underline{a}'}{}_{\underline{b}}(r^{-1}){}^{\underline{b}}{}_c(s^{-1}){}^c{}_{b'}.
    \end{split}
\end{equation}

\subsection{Goldstone Metric}
\label{subsec:goldstone-metric}

We introduce an $SL(m+1,\mathbb{R})$ metric $H_{AB}$, which is required to satisfy the transformation law
\begin{equation}
        S(x):H\rightarrow H_{A'B'}=H_{CD}(S^{-1}){}^C{}_{A'}(S^{-1}){}^D{}_{B'},
\end{equation}
with $S$ and $S^{-1}$ given in Eqs. \eqref{SL-matrix}, \eqref{SL-matrix-inv}. To satisfy this transformation behavior, $H_{AB}$ must be of the form
\begin{equation}
    H_{AB}=\mathfrak{p}{}^{-2}\begin{pmatrix}H_{ab}&H_{a*}\\H_{*b}&H_{**}\end{pmatrix},
\end{equation}
where we take
\begin{equation}\label{p-factor-1}
    \mathfrak{p}:=\frac{x^*}{x^*_0}|\vartheta^a{}_m|^{\frac{-1}{m+1}},
\end{equation}
for some $\vartheta$ with the appropriate transformation behavior. Following our naming convention, $H_{AB}$ is the $\mathfrak{p}^{-2}$-metric. Additionally, we will often refer to $H_{AB}$ as the \textit{Goldstone metric}, for reasons that will be clear soonlyish. Nonlinearly realizing the symmetries with $\sigma$ produces what we will call the $\bar{\mathfrak{p}}^{-2}$-metric:
\begin{equation}
\tilde{\eta}_{\underline{AB}}=H_{AB}\sigma^A{}_{\underline{A}}\sigma^B{}_{\underline{B}}.
\end{equation}
We will continue to use a tilde to denote fields resulting from the nonlinear realization process.

Under the action of a local gauge transformation, we find
\begin{equation}
    \begin{split}
        S(x):\tilde{\eta}_{\underline{AB}}\rightarrow \tilde{\eta}_{\underline{A}'\underline{B}'}&= H_{AB}\sigma^A{}_{\underline{A}'}\sigma^B{}_{\underline{B}'}\\
        &=H_{AB}\sigma^A{}_{\underline{C}}\sigma^B{}_{\underline{D}}(\Lambda^{-1})^{\underline{C}}{}_{\underline{A}'}(\Lambda^{-1})^{\underline{D}}{}_{\underline{B}'}\\
        &=\tilde{\eta}_{\underline{CD}}(\Lambda^{-1})^{\underline{C}}{}_{\underline{A}'}(\Lambda^{-1})^{\underline{D}}{}_{\underline{B}'}.
    \end{split}
\end{equation}
In other words, $\tilde{\eta}_{\underline{AB}}$ is the $(m+1)$-dimensional flat projective metric. The form of $\tilde{\eta}_{\underline{AB}}$ as a $\bar{\mathfrak{p}}{}^{-2}$-tensor is
\begin{equation}\label{eta-tilde}
\tilde{\eta}_{\underline{AB}}=\bar{\mathfrak{p}}^{-2}\eta_{\underline{AB}},
\end{equation}
where
\begin{equation}
    \eta_{\underline{AB}}=\text{diag}\begin{pmatrix}
      1,&-1,&-1,&\dots,&\eta_0
    \end{pmatrix}
\end{equation}
is the standard $SO(p,m+1-p)$ metric in the mostly minus convention. The sign of the $\eta_{\underline{**}}$ component, $\eta_0:=\text{sgn}(\eta_{\underline{**}})=\pm1$, determines its (A)-dS status, with lower (upper) sign corresponding to (Anti)-de Sitter. In $m+1$ dimensions the determinant of $\eta_{\underline{AB}}$, denoted $|\eta|$, is
\begin{equation}
    |\eta|=\begin{cases}+\eta_0\quad \Leftrightarrow\quad m=2k-1,\quad k\in\mathbb{R}^+\\
    -\eta_0\quad \Leftrightarrow\quad m=2k,\quad \quad \;\;\;k\in\mathbb{R}^+.
    \end{cases}
\end{equation}
In Eq.~\eqref{eta-tilde}, we encounter the \textit{coordinate}- and \textit{gauge}-invariant combination
\begin{equation} \bar{\mathfrak{p}}:=\mathfrak{p}|r^a{}_{\underline{b}}|{}^{\frac{1}{m+1}}=\frac{x^*}{x^*_0}|(r^{-1})^{\underline{a}}{}_{b}\vartheta^b{}_n|{}^{\frac{-1}{m+1}}.
\end{equation}
The gauge invariance follows from $|\lambda|=1$, and the coordinate invariance follows from the opposing transformation behavior of $\{x^*\}$ and the spacetime ``$n$" index in the determinant factor.

The components of the Goldstone metric $H_{AB}$ may be resolved from the definition of $\tilde{\eta}_{\underline{AB}}$,
\begin{equation}\label{H-metric}
    H_{AB}=\mathfrak{p}^{-2}\begin{pmatrix}
        h_{ab}+\tilde{\xi}{}^2\change_a\change_b-\change_{(a}\xi_{b)}&\tfrac{1}{x^*_0}(\tilde{\xi}{}^2\change_a-\xi_a)\\\tfrac{1}{x^*_0}(\tilde{\xi}{}^2\change_b-\xi_b)&\tfrac{1}{(x^*_0)^2}\tilde{\xi}{}^2
    \end{pmatrix},
\end{equation}
with $\xi_a:=h_{ab}\xi^b=h_{ab}r^b{}_{\underline{c}}\xi^{\underline{c}}$ and
\begin{equation}
    h_{ab}:=(r^{-1}){}^{\underline{c}}{}_{a}(r^{-1}){}^{\underline{d}}{}_{b}\eta_{\underline{cd}}.
\end{equation}
Additionally, we have introduced the Goldstone $m+1$-vector \cite{coset-vector},
\begin{equation}
    \tilde{\xi}{}^{\underline{A}}:=\mathfrak{p}\begin{pmatrix}
        -\xi^{\underline{a}}\\x^*_0
    \end{pmatrix},
\end{equation}
whose square is $\tilde{\xi}{}^2=\xi^{2}+\eta_0(x^*_0)^2$. Compare with the $\tilde{\xi}$ discussed in the context of Metric-Affine gauge theories in Eq. \eqref{traut-higgs-xi}. The minus sign appearing in the first $m$-components of $\tilde{\xi}$ permits the identification
\begin{equation}
    \tilde{\xi}{}^{\underline{A}}\bumpeq\tilde{\Upsilon}{}^{\underline{A}},
\end{equation}
where $\tilde{\Upsilon}$ is the $\bar{\mathfrak{p}}$-Generalized Higgs Vector, and the equality with a ``bump-up" is used to denote expressions which are true for $y^{\underline{a}}=0$. Both of these statements will be developed in Sec. \ref{sec:gen-higgs-field}.

The inverse Goldstone metric $H^{AB}$ may be solved for by constructing the $\bar{\mathfrak{p}}{}^{2}$-metric,
\begin{equation}
    \tilde{\eta}{}^{\underline{AB}}:=(\sigma^{-1})^{\underline{A}}{}_A(\sigma^{-1})^{\underline{B}}{}_BH^{AB}.
\end{equation}
Thus, the inverse Goldstone metric $H^{AB}$ has the explicit matrix form
\begin{equation}
    H^{AB}=\mathfrak{p}^2\begin{pmatrix}
        h^{ab}+\frac{\eta_0}{(x^*_0)^2}\xi^a\xi^b&-x^*_0(\change^a-\tfrac{\eta_0}{(x^*_0)^2}\xi^a)\\-x^*_0(\change^b-\tfrac{\eta_0}{(x^*_0)^2}\xi^b)&\eta_0+(x^*_0)^2\change^2
    \end{pmatrix},
\end{equation}
with $\change^2:=\change^a\change_a=h^{ab}\change_a\change_b$ and
\begin{equation}
    h^{ab}:=r^a{}_{\underline{c}}r^b{}_{\underline{c}}\eta^{\underline{cd}}.
\end{equation}
The reason for calling $H_{AB}$ the \textit{Goldstone metric} should now be clear. Essentially, all of the degrees of freedom contained in $H_{AB}$ come from the Goldstone fields or coset parameters $\{(r^{-1})^{\underline{a}}{}_b,\;\xi^{\underline{a}},\;\change_b\}$.

\subsection{\texorpdfstring{$\bar{\mathfrak{p}}$}{\bar{\mathfrak{p}}}-Connection}
\label{subsec:nlr-proj-conn}

Similar to the gravitational gauge theories of \cite{SL5-gravity-1} and \cite{mielke}, we define the $\mathfrak{sl}(m+1,\mathbb{R})$-valued connection $1$-form $\Omega$ to have the components
\begin{equation}\label{big-omega-gen}
    \Omega=\Omega^A{}_B\bm{L}{}^B{}_A=\varpi^a{}_b\bm{L}{}^b{}_a+\vartheta^a\bm{L}{}^*{}_a+\mathcal{P}_b\bm{L}{}^b{}_*+\varpi^*_*\bm{L}{}^*{}_*.
\end{equation}
In the above, $\mathcal{P}$ is the $\mathbb{R}^m_*$-valued \textit{projective Schouten} connection $1$-form and $\vartheta$ is the $\mathbb{R}^m$-valued \textit{projective translational} or \textit{co-frame} connection $1$-form used in the projective factor $\mathfrak{p}$, (e.g. Eq. \eqref{p-factor-1}). The purpose of initially identifying $\vartheta^a$ in this manner is to significantly simplify calculations. Removing this assumption and permitting their independence is left as a future area of research. 

As an $\mathfrak{sl}(m+1,\mathbb{R})$-valued connection, $\Omega^A{}_B$ is traceless. Therefore,
\begin{equation}
    \varpi^*_*=-\delta^a{}_b\varpi^b{}_a.
\end{equation}
Using the traceless property of the generators, $\bm{L}^A{}_A=\bm{L}^a{}_a+\bm{L}^*{}_*=0$, we are permitted to move $\varpi^*_*$ to the $m\times m$-block of $\Omega$. This is, essentially, the homomorphism between the connection on the principle bundle and the connection on the projective $2$-frame bundle discussed in \cite{2-frame}. We therefore define the $m$-dimensional component connection $\omega^a{}_b$ on the projective $2$-frame bundle as
\begin{equation}\label{minus-little-trace}
    \omega^a{}_b:=\varpi^a{}_b-\delta^a{}_b\varpi^*_*.
\end{equation}
For $\Omega$ to remain traceless, we must require that
\begin{equation}
    \omega^a{}_a=(m+1)\varpi^a{}_a.
\end{equation}
As a linear matrix, i.e., the M\"{o}bius representation, Eq. \eqref{big-omega-gen} may be expressed as
\begin{equation}
    \Omega^A{}_B=\begin{pmatrix}
        \omega^a{}_b&\frac{1}{x^*_0}\vartheta^a\\x^*_0\mathcal{P}_b&0
    \end{pmatrix},
\end{equation}
where we use the same symbol $\Omega$ to denote the principle and $2$-frame connections. The factors of $x^*_0$ result from the the transformation of generators outlined in Eq. \eqref{pp*-generators}. 

Under local gauge transformations, the $\mathfrak{sl}(m+1,\mathbb{R})$-valued connection $\Omega$ becomes
\begin{equation}
    S(x):\Omega\rightarrow\Omega'=S(\Omega+d)S^{-1}.
\end{equation}
In components, this transformation is expressed as
\begin{equation}
    \Omega^{A'}{}_{B'}=S^{A'}{}_{C}(\Omega^C{}_D+\delta^C{}_Dd)(S^{-1}){}^D{}_{B'}.
\end{equation}
Such transformations will necessarily take $\Omega^*{}_*=0\rightarrow\Omega^{*'}{}_{*'}\neq0$. Therefore, following any operation, we subtract $\delta^{a'}{}_{b'}\Omega^{*'}{}_{*'}$ from $\Omega^{a'}{}_{b'}$ as was done in Eq. \eqref{minus-little-trace}. Executing the $SL(m+1,\mathbb{R})$ transformation yields, for the connection components:
\begin{equation}\label{gauge-trans-of-PGL-conn}
    \begin{split}
        \omega^{a'}{}_{b'}&=s^{a'}{}_{c}(\omega^c{}_d+\delta^c{}_dd)(s^{-1}){}^d{}_{b'}+s^{a'}{}_{c}\vartheta^cu_d(s^{-1})^d{}_{b'}+\delta^{a'}{}_{b'}u_b\vartheta^b+t^{a'}\mathcal{P}_{b'}+\delta^{a'}{}_{b'}\mathcal{P}_{c'}t^{c'},\\
        \mathcal{P}_{b'}&=(\mathcal{P}_d+du_d-u_c\omega^c{}_d-u_{c}\vartheta^cu_d)(s^{-1})^d{}_{b'},\\
        \vartheta^{a'}&=s^{a'}{}_{c}\vartheta^c-dt^{a'}-\omega^{a'}{}_{b'}t^{b'}+t^{a'}t^{c'}\mathcal{P}_{c'}.
    \end{split}
\end{equation}
In the above, $\Omega^{*'}{}_{*'}=-\mathcal{P}_{b'}t^{b'}$ was subtracted from $\Omega^{a'}{}_{b'}$. These transformations constitute the general local \textit{projective transformations}. 

The nonlinear connection is then defined as
\begin{equation}
    \tilde{\Omega}=\sigma^{-1}(\Omega+d)\sigma.
\end{equation}
This nonlinear connection transforms under the local gauge action of $S(x)\in SL(m+1,\mathbb{R})$ as
\begin{equation}
    S(x):\tilde{\Omega}\rightarrow\tilde{\Omega}{}'=\sigma^{-1}{}'(\Omega'+d)\sigma'=\Lambda(\tilde{\Omega}+d)\Lambda^{-1}.
\end{equation}
Therefore, $\tilde{\Omega}$ is a nonlinear connection transforming under the projective Lorentz group. Using the coset element of Eqs. \eqref{coset-element}, \eqref{coset-element-inv}, the nonlinear projective connection has the M\"{o}bius representation
\begin{equation}\label{nonlinear-omega-tilde}
    \tilde{\Omega}{}^{\underline{A}}{}_{\underline{B}}=\begin{pmatrix}
        \overline{\omega}{}^{\underline{a}}{}_{\underline{b}}&\frac{1}{x^*_0}\overline{\vartheta}{}^{\underline{a}}\\x^*_0\overline{\mathcal{P}}_{\underline{b}}&0
    \end{pmatrix},
\end{equation}
where
\begin{equation}
    \begin{split}
    \overline{\omega}{}^{\underline{a}}{}_{\underline{b}}&=(r^{-1}){}^{\underline{a}}{}_{b}(\omega^b{}_c+\delta^b{}_cd)r^c{}_{\underline{b}}-(r^{-1}){}^{\underline{a}}{}_{b}\vartheta^b\change_cr^c{}_{\underline{b}}-\delta^{\underline{a}}{}_{\underline{b}}\change_c\vartheta^c-\xi^{\underline{a}}\overline{\mathcal{P}}_{\underline{b}}-\delta^{\underline{a}}{}_{\underline{b}}\overline{\mathcal{P}}\cdot\xi,\\
    \overline{\mathcal{P}}_{\underline{b}}&=(\mathcal{P}_c-d\change_c+\change_b\omega^b{}_c-\change_b\vartheta^b\change_c)r^c{}_{\underline{b}},\\
    \overline{\vartheta}{}^{\underline{a}}&=(r^{-1}){}^{\underline{a}}{}_{b}\vartheta^b+d\xi^{\underline{a}}+\overline{\omega}{}^{\underline{a}}{}_{\underline{b}}\xi^{\underline{b}}+\xi^{\underline{a}}\overline{\mathcal{P}}\cdot\xi.
    \end{split}
\end{equation}
In calculating $\tilde{\Omega}$, the component $\tilde{\Omega}{}^{\underline{*}}{}_{\underline{*}}=\overline{\mathcal{P}}\cdot\xi:=\overline{\mathcal{P}}_{\underline{b}}\xi^{\underline{b}}$ was subtracted from $\tilde{\Omega}{}^{\underline{a}}{}_{\underline{b}}$.

Interestingly, had the roles of $\sigma$ and $(\sigma^{-1})$ been interchanged (along with the gauge transformation matrices $S$ and $(S^{-1})$, i.e., $ \{\sigma,\;S\}\leftrightarrow\{(\sigma^{-1}),\;(S^{-1})\}$), one would find that $\vartheta^a$ and $\mathcal{P}_b$ exchange their behavior. Explicitly, 
\begin{equation}
    \begin{split}
        \overline{\omega}{}^{\underline{a}}{}_{\underline{b}}&=(r^{-1}){}^{\underline{a}}{}_c(\omega^{c}{}_{d}+\delta^c{}_dd)r^d{}_{\underline{b}}-\overline{\vartheta}{}^{\underline{a}}\change_cr^c{}_{\underline{b}}-\delta^{\underline{a}}{}_{\underline{b}}\change_b\overline{\vartheta}{}^b-\xi^{\underline{a}}\mathcal{P}_cr^c{}_{\underline{b}}-\delta^{\underline{a}}{}_{\underline{b}}\mathcal{P}\cdot r\cdot\xi,\\
\overline{\mathcal{P}}_{\underline{b}}&=\mathcal{P}_{\underline{b}}-d\change_{\underline{b}}+\change_{\underline{c}}\overline{\omega}{}^{\underline{c}}{}_{\underline{b}}+\change_{\underline{b}}\change\cdot\overline{\vartheta},\\
  \overline{\vartheta}{}^{\underline{a}}&=(r^{-1}){}^{\underline{a}}{}_{b}\left(\vartheta^b+d\xi^b+\omega^{b}{}_{c}\xi^c-\xi^b\mathcal{P}\cdot r\cdot\xi\right).
    \end{split}
\end{equation}
However, this does \textit{not} generally produce the $V\mathcal{M}$-connection of interest, $\tilde{\Gamma}^L{}_{NM}$.

Since the base $\mathfrak{sl}(m+1,\mathbb{R})$-valued connection $\Omega$ is traceless, so too is the nonlinear connection,
\begin{equation}
    \tilde{\Omega}{}^{\underline{A}}{}_{\underline{A}}=0.
\end{equation}
For the above property to hold true, one must require $\overline{\omega}{}^{\underline{a}}{}_{\underline{a}}=0$. Enforcing this condition leads to
\begin{equation}
    \omega^a{}_a=(m+1)\varpi^a{}_a=d\log|r^{-1}|+(m+1)(\change\cdot\vartheta+\overline{\mathcal{P}}\cdot\xi).
\end{equation}
The traceless property of the base connection $\Omega$ further requires $\omega^a{}_a=\varpi^a{}_a=0$, which results in
\begin{equation}
    d\log|r|^{\frac{1}{m+1}}=\change\cdot\vartheta+\overline{\mathcal{P}}\cdot\xi.
\end{equation}
In accordance with \cite{proj-conn}, we assume $\overline{\mathcal{P}}\cdot\xi=0$. Therefore, 
\begin{equation}\label{proj-inv-higgs-app}
    \change_b\vartheta^b=d\log|r|^{\frac{1}{m+1}}.
\end{equation}
This is simply a rearranging of the gauge degrees of freedom and follows from the \textit{Inverse Higgs Theorem} of \cite{ogievetsky1}. In \cite{schwarzian-inverse-higgs}, a variation of the Inverse Higgs Theorem was applied to the $SL(2,\mathbb{R})\times\mathbb{R}$ connection in order to recover the Schwarzian derivative in terms of Maurer-Cartan forms. Presently, due to the last commutation relation in Eqs. \eqref{pgl-lorentz-algebra}, the theorem states that the coset fields $\change$ and $r$ parameterizing, respectively, $\mathfrak{t}^*(m,\mathbb{R})$ and $\text{Sym}(\tfrac{m(m+1)}{2})$ may be solved for in terms of each other and their derivatives. This is precisely the statement of Eq. \eqref{proj-inv-higgs-app}. Further application of this theorem may be found in \cite{inverse-higgs-application-3}, and conditions for applicability are further discussed in \cite{inverse-higgs-application-1}, and alternative interpretations in \cite{inverse-higgs-application-2}. Further investigation of this particular reorganization of degrees of freedom is of utmost importance and is left for a future area of research.

The statement $\overline{\mathcal{P}}\cdot\xi=0$ further permits a convenient and manifestly gauge-covariant expression for the nonlinear co-frame,
\begin{equation}
    \overline{\vartheta}{}^{\underline{a}}=(r^{-1}){}^{\underline{a}}{}_{b}\vartheta^b+\overline{D}\xi^{\underline{a}},
\end{equation}
where
\begin{equation}
\overline{D}\xi^{\underline{a}}:=d\xi^{\underline{a}}+\overline{\omega}{}^{\underline{a}}{}_{\underline{b}}\xi^{\underline{b}}
\end{equation}
is the nonlinear projective $SO(m,\mathbb{R})$-covariant derivative. 

Attempting to write the nonlinear projective Schouten form in a similar fashion reveals a series of terms with increasing powers of $\change\cdot\xi$. Since $\change\cdot t=0$, we make the minimal assumption $\change\cdot \xi=0$. Therefore,
\begin{equation}\label{p-bar-aint-shit}
        \overline{\mathcal{P}}_{\underline{b}}=\mathcal{P}_{\underline{b}}-d\change_{\underline{b}}+\change_{\underline{a}}\overline{\omega}{}^{\underline{a}}{}_{\underline{b}}+\change_b\vartheta^b\change_{\underline{b}}.
\end{equation}
Since $\change_b$ is constrained via Eq. \eqref{proj-inv-higgs-app}, its nonlinear covariant derivative $\overline{D}$ requires a compensating factor for gauge-covariance. This compensating factor is precisely the reason for $\change_b\vartheta^b\change_{\underline{b}}$ appearing in Eq. \eqref{p-bar-aint-shit}. Thus, 
\begin{equation}
\overline{\mathcal{P}}_{\underline{b}}=\mathcal{P}_{\underline{b}}-\overline{D}\change_{\underline{b}}+\change_b\vartheta^b\change_{\underline{b}}.
\end{equation}

Lastly, we note that when one imposes
\begin{equation}
\xi^{\underline{a}}=0,\quad\quad\quad\overline{D}\xi^{\underline{a}}=0,\quad\quad\quad\change_{b}=0,\quad\quad\quad\overline{D}\change_b=0,
\end{equation}
by a convenient choice of gauge for these coset parameters, then the nonlinear connection $\tilde{\Omega}$ reduces to a \textit{Cartan Connection} \cite{Hehl}. Furthermore, imposing only the first two conditions on $\xi$ reduces the translational connection-form $\overline{\vartheta}$ to the \textit{solder form}, providing a fixed point of contact (a soldering) between the two spaces. These gauge choices will be discussed in more depth in Sec. \ref{sec:gen-higgs-field}, where we construct the generalized projective Higgs fields.

\subsection{\texorpdfstring{$\bar{\mathfrak{p}}$}{\bar{\mathfrak{p}}}-Curvature}
\label{subsec:nlr-proj-curv}

From the $\mathfrak{sl}(m+1,\mathbb{R})$-valued connection $\Omega$, we define the $\mathfrak{sl}(m+1,\mathbb{R})$-valued curvature $2$-form as
\begin{equation}
    \mathcal{K}:=d\Omega+[\Omega,\;\Omega].
\end{equation}
Exposing both the manifold and Lie algebra indices yields
\begin{equation}
    \mathcal{K}=\mathcal{K}^A{}_{B}\bm{L}{}^B{}_A=\frac{1}{2!}\mathcal{K}^A{}_{B[MN]}\bm{L}{}^B{}_Adx^M\wedge dx^N.
\end{equation}
In terms of $\Omega$, this is simply
\begin{equation}\mathcal{K}=\left(d\Omega^A{}_B+\Omega^A{}_C\wedge\Omega^C{}_B\right)\bm{L}{}^B{}_A,
\end{equation}
where $\Omega^A{}_C\wedge\Omega^C{}_B= \Omega^A{}_C \Omega^C{}_B-\Omega^C{}_B\Omega^A{}_C$ follows from the structure constants extracted from Eqs. \eqref{pgl+D-algebra-commutators}. Since $\Omega$ is independent of $x^*$ and, additionally, $\Omega^A{}_{B*}=0$ for all $A$ and $B$, we have that
\begin{equation}
    \Omega^A{}_C\wedge\Omega^C{}_B= \frac{1}{2!}\Omega^A{}_{C[m} \Omega^C{}_{|B|n]}dx^m\wedge dx^n
\end{equation}
in the coordinate basis. The curvature naturally inherits this independence,
\begin{equation}
    \mathcal{K}=\mathcal{K}^A{}_{B}\bm{L}{}^B{}_A=\frac{1}{2!}\mathcal{K}^A{}_{B[mn]}\bm{L}{}^B{}_Adx^m\wedge dx^n.
\end{equation}

We may express the components of $\mathcal{K}$ in the M\"{o}bius representation as
\begin{equation}
    \mathcal{K}^A{}_B=\begin{pmatrix}
    k^a{}_b&\frac{1}{x^*_0}\mathcal{T}^a\\
    x^*_0\mathcal{S}_b&k^*_*
    \end{pmatrix}.
\end{equation}
Utilizing the traceless property of the generators, we again make the replacement $k^a{}_b\rightarrow K^a{}_b:=k^a{}_b-\delta^a{}_bk^*_*$, as was done with the connection, $\Omega$. Explicitly, the components of $\mathcal{K}$ are found to be:
\begin{equation}
\begin{split}
K^a{}_b&:=\mathcal{R}^a{}_b+\vartheta^a\wedge\mathcal{P}_b+\delta^a{}_b\vartheta^c\wedge\mathcal{P}_c,\\
    \mathcal{T}^a&:=D\vartheta^a=d\vartheta^a+\omega^a{}_c\wedge \vartheta^c,\\
\mathcal{S}_b&:=D\mathcal{P}_b=d\mathcal{P}_b+\mathcal{P}_a\wedge\omega^a{}_b,
    \end{split}
\end{equation}
where $k^*_*=\mathcal{P}_c\wedge \vartheta^c$. In the above,
\begin{equation}
    \mathcal{R}^a{}_b:=d\omega^a{}_b+\omega^a{}_c\wedge\omega^c{}_b
\end{equation}
is the field strength of the $\mathfrak{gl}(m,\mathbb{R})$-valued connection. In the coordinate basis, for example, the components of $\mathcal{R}{}^a{}_b$ are accessed via
\begin{equation}
    \mathcal{R}{}^a{}_b:=\frac{1}{2!}\mathcal{R}{}^a{}_{b[mn]}dx^m\wedge dx^n.
\end{equation}
The $\mathbb{R}^m$-valued component of the curvature $2$-form is simply the field strength of the $\mathbb{R}^m$-valued connection $1$-form $\vartheta^a$, i.e., the torsion $2$-form. In the coordinate basis,
\begin{equation}
    \mathcal{T}^a:=\frac{1}{2!}\mathcal{T}^a{}_{[mn]}dx^m\wedge dx^n,
\end{equation}
where $\mathcal{T}^a{}_{[mn]}$ is the torsion tensor.

The $\mathbb{R}^m_*$-valued component of the curvature $2$-form is simply the field strength of the $\mathbb{R}^m_*$-valued connection $1$-form $\mathcal{P}_b$. This is the projective Cotton-York $2$-form in the presence of torsion. In the coordinate basis, we find the projective Cotton-York tensor, $\mathcal{S}_{b[mn]}$, accessed via
\begin{equation}
    \mathcal{S}_b:=\frac{1}{2!}\mathcal{S}_{b[mn]}dx^m\wedge dx^n.
\end{equation}

Since $\mathcal{K}$ is the curvature of an $\mathfrak{sl}(m+1,\mathbb{R})$-valued connection, it is required traceless. This provides
\begin{equation}
    \check{\check{\mathcal{R}}}=-(m+1)\vartheta^c\wedge \mathcal{P}_c.
\end{equation}
Therefore, the homothetic curvature, $\check{\check{\mathcal{R}}}:=\mathcal{R}^a{}_a$, associated to $\omega$ is proportional to $\vartheta^c\wedge \mathcal{P}_c$, in agreement with \cite{covariant-tw}. If $\vartheta$ had the required transformation properties to be considered a \textit{true} co-frame, this would be the antisymmetric part of $\mathcal{P}$ with respect to that basis. As shown in Eq. \eqref{homothetic-Q}, the homothetic curvature is nothing but the exterior differential of the Weyl co-vector. Therefore, 
\begin{equation}\label{antisym-P-dQ}
    \vartheta^c\wedge\mathcal{P}_c=\frac{-1}{m+1}dQ,
\end{equation}
where $Q=Q_mdx^m$ is the Weyl $1$-form associated with $\omega$. These considerations imply that $\mathcal{K}$ has a vanishing homothetic curvature and, therefore, a vanishing Weyl $1$-form. However, the interior does not share this property, as exemplified by Eq. \eqref{antisym-P-dQ}.

The curvature $2$-form $\mathcal{K}$ transforms under the local gauge action of $SL(m+1,\mathbb{R})$ as
\begin{equation}
    S(x):\mathcal{K}\rightarrow\mathcal{K}'=S\mathcal{K}S^{-1}.
\end{equation}
Exposing indices,
\begin{equation}
  \mathcal{K}^{A'}{}_{B'}=S^{A'}{}_{C}\mathcal{K}^C{}_D(S^{-1})^D{}_{B'}.
\end{equation}
Using the form of the gauge transformation matrices $S$ and $S^{-1}$ from Eqs. \eqref{SL-matrix}, \eqref{SL-matrix-inv}, we find the transformation of the component field strengths:
\begin{equation}
    \begin{split}
    \mathcal{K}{}^{a'}{}_{b'}&=s^{a'}{}_{c}(K^c{}_d+\delta^{c}{}_{d}\mathcal{T}^bu_b+\mathcal{T}^cu_d+\delta^{c}{}_{d}t^{d'}\mathcal{S}_{d'}+t^{c}\mathcal{S}_{d'}s{}^{d'}{}_d)(s^{-1})^d{}_{b'},\\
        \mathcal{S}_{b'}&=\left(\mathcal{S}_c-u_b(K^b{}_c+\mathcal{T}^bu_c)\right)(s^{-1})^c{}_{b'},\\
        \mathcal{T}^{a'}&=s^{a'}{}_{b}\left(\mathcal{T}^b-(s^{-1})^{b}{}_{d'}(\mathcal{K}{}^{d'}{}_{b'}-\delta^{d'}{}_{b'}t^{c'}\mathcal{S}_{c'})t^{b'}\right),
    \end{split}
\end{equation}
where $t^a:=(s^{-1})^a{}_{b'}t^{b'}$. Although $\mathcal{K}^A{}_B$ gauge transforms homogeneously, the components of $\mathcal{K}^A{}_B$ do not individually share this property, since a homogeneous transformation of the components would simply be the first term in each of the expressions above. To force the components of $\mathcal{K}$ to behave properly, we may nonlinearly realize the symmetries. 

The nonlinear curvature $2$-form is constructed from the nonlinear connection as
\begin{equation}
    \tilde{\mathcal{K}}=d\tilde{\Omega}+[\tilde{\Omega},\;\tilde{\Omega}].
\end{equation}
Exposing indices yields
\begin{equation}
    \tilde{\mathcal{K}}{}^{\underline{A}}{}_{\underline{B}}=d\tilde{\Omega}{}^{\underline{A}}{}_{\underline{B}}+\tilde{\Omega}{}^{\underline{A}}{}_{\underline{C}}\wedge \tilde{\Omega}{}^{\underline{C}}{}_{\underline{B}}.
\end{equation}
Furthermore, $\tilde{\mathcal{K}}{}^{\underline{A}}{}_{\underline{B}}$ inherits from $\tilde{\Omega}$ the independence on $x^*$,
\begin{equation}
\tilde{\mathcal{K}}{}^{\underline{A}}{}_{\underline{B}}=\frac{1}{2!}\tilde{\mathcal{K}}{}^{\underline{A}}{}_{\underline{B}mn}dx^m\wedge dx^n.
\end{equation}
The nonlinear curvature is easily computed and has the M\"{o}bius representation
\begin{equation}
    \tilde{\mathcal{K}}{}^{\underline{A}}{}_{\underline{B}}=\begin{pmatrix}
        \overline{\mathcal{R}}{}^{\underline{a}}{}_{\underline{b}}+\overline{\vartheta}{}^{\underline{a}}\wedge\overline{\mathcal{P}}_{\underline{b}}+\delta^{\underline{a}}{}_{\underline{b}}\overline{\vartheta}{}^{\underline{c}}\wedge\overline{\mathcal{P}}_{\underline{c}}&\frac{1}{x^*_0}\overline{\mathcal{T}}{}^{\underline{a}}\\x^*_0\overline{\mathcal{S}}_{\underline{b}}&0
    \end{pmatrix}.
\end{equation}
The $m$-dimensional nonlinear curvature $2$-form $\overline{\mathcal{R}}{}^{\underline{a}}{}_{\underline{b}}$ is defined as
\begin{equation}
    \overline{\mathcal{R}}{}^{\underline{a}}{}_{\underline{b}}:=d\overline{\omega}{}^{\underline{a}}{}_{\underline{b}}+\overline{\omega}{}^{\underline{a}}{}_{\underline{c}}\wedge\overline{\omega}{}^{\underline{c}}{}_{\underline{b}}.
\end{equation}
The nonlinear torsion $2$-form or co-frame field strength is defined as
\begin{equation}\label{barred-torsion}
    \overline{\mathcal{T}}{}^{\underline{a}}:=\overline{D}\overline{\vartheta}{}^{\underline{a}}=d\overline{\vartheta}{}^{\underline{a}}+\overline{\omega}{}^{\underline{a}}{}_{\underline{b}}\wedge\overline{\vartheta}{}^{\underline{b}},
\end{equation}
and the nonlinear projective Cotton-York $2$-form or projective Schouten field strength is defined as
\begin{equation}
\overline{\mathcal{S}}_{\underline{b}}:=\overline{D}\overline{\mathcal{P}}_{\underline{b}}=d\overline{\mathcal{P}}_{\underline{b}}+\overline{\mathcal{P}}_{\underline{a}}\wedge\overline{\omega}{}^{\underline{a}}{}_{\underline{b}}.
\end{equation}

For the nonlinear curvature $\tilde{\mathcal{K}}$ to retain a vanishing homothetic curvature, one must have
\begin{equation}\label{r-bar-trace}
        \overline{\mathcal{R}}{}^{\underline{a}}{}_{\underline{a}}=-(m+1)\overline{\vartheta}{}^{\underline{a}}\wedge\overline{\mathcal{P}}_{\underline{a}}.
\end{equation}
However, since $\overline{\omega}{}^{\underline{a}}{}_{\underline{b}}$ is traceless and wedge products are antisymmetric, $\overline{\mathcal{R}}{}^{\underline{a}}{}_{\underline{a}}=0$. Furthermore, the transformation properties of $\overline{\vartheta}$ permit its use as a proper co-frame, and thus, may be used as a basis-form. Therefore,
\begin{equation}\label{anti-sym-p-bar}
    \overline{\vartheta}{}^{\underline{a}}\wedge\overline{\mathcal{P}}_{\underline{a}}=\overline{\mathcal{P}}_{[\underline{ab}]}\overline{\vartheta}{}^{\underline{a}}\wedge\overline{\vartheta}{}^{\underline{b}}=0.
\end{equation}
In other words, $\overline{\mathcal{P}}_{\underline{ab}}$ is symmetric. If $\overline{\mathcal{R}}{}^{\underline{a}}{}_{\underline{b}mn}(\overline{\vartheta}{}^{-1})^{m}{}_{\underline{a}}=0$ as well, then $\overline{\mathcal{R}}{}^{\underline{a}}{}_{\underline{b}}$ may be identified with the projective Weyl tensor \cite{proj-vs-metric}. Comparing Eq. \eqref{anti-sym-p-bar} with Eq. $4.1$ of \cite{p-bar-metric}, it is possible to identify the antisymmetric part of $\overline{\mathcal{P}}_{\underline{ab}}$ with a symplectic form through the lens of the co-tractor bundle construction. Interestingly, the present application of the nonlinear realization framework appears then to force a vanishing of this symplectic form. Moreover, Eq. \eqref{r-bar-trace} says that this symplectic form may be identified with the homothetic curvature. It may thus be possible to view the symmetric and antisymmetric parts of the pseudo-translational connection-form, $\overline{\mathcal{P}}_{\underline{ab}}$, of a projective gauge structure as providing, respectively, a metric tensor and a symplectic form. These statements require a significant amount of mathematics to be made concrete, and will be prioritized in future investigations.

It will be convenient for later discussions to have available the (anti)-symmetric parts of $\tilde{\mathcal{K}}$. Lowering one index with $\tilde{\eta}$ provides the general organization of the components
\begin{equation}
\tilde{\mathcal{K}}_{\underline{AB}}:=\tilde{\eta}_{\underline{AC}}\tilde{\mathcal{K}}{}^{\underline{C}}{}_{\underline{B}}=\begin{pmatrix}
\tilde{\mathcal{K}}_{\underline{ab}}&\tilde{\mathcal{K}}_{\underline{a*}}\\\tilde{\mathcal{K}}_{\underline{*b}}&\tilde{\mathcal{K}}_{\underline{**}}
    \end{pmatrix}.
\end{equation}
Taking the antisymmetric parts (no factor of $1/2$) provides
\begin{equation}\label{antisymmetric-K}
\tilde{\mathcal{K}}_{[\underline{AB}]}=\bar{\mathfrak{p}}{}^{-2}\begin{pmatrix}
\overline{\mathcal{R}}_{[\underline{ab}]}+\overline{\vartheta}_{[\underline{a}}\wedge\overline{\mathcal{P}}_{\underline{b}]}&-\eta_0x^*_0\left(\overline{\mathcal{S}}_{\underline{a}}-\frac{\eta_0}{(x^*_0)^2}\eta_{\underline{ac}}\overline{\mathcal{T}}{}^{\underline{c}}\right)\\\eta_0x^*_0\left(\overline{\mathcal{S}}_{\underline{b}}-\frac{\eta_0}{(x^*_0)^2}\eta_{\underline{bc}}\overline{\mathcal{T}}{}^{\underline{c}}\right)&0
    \end{pmatrix},
\end{equation}
while the symmetric parts, related to non-metricity, yield
\begin{equation}\label{symmetric-K}
\tilde{\mathcal{K}}_{(\underline{AB})}=\bar{\mathfrak{p}}{}^{-2}\begin{pmatrix}
\overline{\mathcal{R}}_{(\underline{ab})}+\overline{\vartheta}_{(\underline{a}}\wedge\overline{\mathcal{P}}_{\underline{b})}&\eta_0x^*_0\left(\overline{\mathcal{S}}_{\underline{a}}+\frac{\eta_0}{(x^*_0)^2}\eta_{\underline{ac}}\overline{\mathcal{T}}{}^{\underline{c}}\right)\\\eta_0x^*_0\left(\overline{\mathcal{S}}_{\underline{b}}+\frac{\eta_0}{(x^*_0)^2}\eta_{\underline{bc}}\overline{\mathcal{T}}{}^{\underline{c}}\right)&0
    \end{pmatrix}.
\end{equation}
Had we instead raised the indices, this would simply move around factors of $\eta_0$. The combination
\begin{equation}
    \overline{\mathcal{S}}{}^{\pm}_{\underline{b}}:=\overline{\mathcal{S}}_{\underline{b}}\pm\frac{\eta_0}{(x^*_0)^2}\eta_{\underline{bc}}\overline{\mathcal{T}}{}^{\underline{c}},
\end{equation}
up to a yet-to-be-discussed non-metricity, may be identified as the field strength of a shifted projective Schouten form, i.e., the gauge-connection associated with the generators in Eqs. \eqref{M-generator-def}. In particular, for
\begin{equation}
    \overline{\mathcal{P}}{}^{\pm}_{\underline{b}}:=\overline{\mathcal{P}}_{\underline{b}}\pm\frac{\eta_0}{(x^*_0)^2}\eta_{\underline{ab}}\overline{\vartheta}{}^{\underline{a}},
\end{equation}
the above field strength represents
\begin{equation}
    \overline{\mathcal{S}}{}^{\pm}_{\underline{b}}=\overline{D}\overline{\mathcal{P}}{}^{\pm}_{\underline{b}}\mp\frac{\eta_0}{(x^*_0)^2}\overline{Q}_{\underline{ab}}\wedge\overline{\vartheta}{}^{\underline{a}}.
\end{equation}
In the Poincar\'{e} gauge gravity of \cite{poincare-cotton-york}, a projectively invariant combination of torsion and non-metricity parallels the above $\overline{\mathcal{S}}{}^{\pm}$ for a Metric-Affine framework, and may be considered in exact analogy when the projective Schouten form is constant. In particular, consider the decomposed curvature in Eqs. \eqref{symmetric-K} and \eqref{antisymmetric-K}, and further, assume the projective Schouten form takes on the negative solution
\begin{equation}
    \overline{\mathcal{P}}_{\underline{b}}=\frac{-\eta_0}{(x^*_0)^2}\eta_{\underline{ab}}\overline{\vartheta}{}^{\underline{a}}.
\end{equation}
Then, 
\begin{equation}
    \overline{\mathcal{S}}{}^{+}_{\underline{b}}=\frac{-\eta_0}{(x^*_0)^2}\overline{Q}_{\underline{ab}}\wedge\overline{\vartheta}{}^{\underline{a}},
\end{equation}
\begin{equation}
    \overline{\mathcal{S}}{}^{-}_{\underline{b}}=\frac{\eta_0}{(x^*_0)^2}\overline{Q}_{\underline{ab}}\wedge\overline{\vartheta}{}^{\underline{a}}-\frac{2\eta_0}{(x^*_0)^2}\eta_{\underline{ab}}\overline{\mathcal{T}}{}^{\underline{a}},
\end{equation}
showing that $\overline{\mathcal{S}}{}^{+}$ may be identified with the non-metricity in this limit.

\section{Generalized Higgs Fields}
\label{sec:gen-higgs-field}

In this section we develop the general projective analogues of the Generalized Higgs fields \cite{trautman}. These are very closely related to projective tractors \cite{einstein-metrics,tractors-1,tractors-2,BGG}. In particular, it appears that these $\mathfrak{p}{}^{\mathfrak{w}}$-fields are intimately related to sections of the tractor bundle. The standard vectorial Generalized Higgs field was briefly discussed in the context of Metric-Affine gauge theories in \cite{Hehl} and, further, in the projective setting by Whitehead \cite{projective-radius-vector}. In general, these projective fields arise in the context of studying projective normal coordinates \cite{proj-normal-coord-chern,proj-norm-coord-path-geometry}. 

The generalized projective Higgs vector is shown to exactly reproduce the fundamental projective vector discussed in Sec. \ref{sec:fundamental-proj-fields-TW}, once a particular gauge is specified. The first gauge-covariant derivative of this field gives rise to a generalized projective $2$-frame, such as those utilized in \cite{2-frame}. The second gauge-covariant derivative then defines the affine connection on $V\mathcal{M}$. Therefore, the entire General Projective Gauge Gravitational Theory may be thought of as constructed from this single vector. 

From the general projective $2$-frames, we develop the notion of a $\bar{\mathfrak{p}}^{-(m+1)}$-orientation symbol, and discuss the ensuing internal duality operation. With these objects defined, we detour to develop and discuss the $\bar{\mathfrak{p}}{}^{-2}$-non-metricity.

Lastly, without an inherent metric structure---aside from the one provided by the nonlinear realization process---a generalized projective Higgs co-vector is introduced. This object is not as well-understood, but its first gauge-covariant derivative seems to make appearances in non-projective settings, such as the bi-metric theory considered in \cite{bi-frame}. The second gauge-covariant derivative of the generalized projective Higgs co-vector leads to what appears to be a genuine Higgs-metric, whose $m\times m$-block components are identified with the diffeomorphism field $\mathcal{D}$.

\subsection{Generalized \texorpdfstring{$\bar{\mathfrak{p}}$}{\bar{\mathfrak{p}}}-Higgs Vector}
\label{subsec:gen-higgs-v}

We define an $\mathfrak{sl}(m+1,\mathbb{R})\cong\mathfrak{pgl}(m,\mathbb{R})$-valued vector $Y$ by requiring it transform under the local gauge action of $S\in SL(m+1,\mathbb{R})\cong PGL(m,\mathbb{R})$ according to
\begin{equation}\label{gen-higgs-vec-trans-S}
    S(x):\;Y^A\rightarrow Y^{A'}=S^{A'}{}_BY^B,
\end{equation}
with $S$ given in Eq. \eqref{SL-matrix}. For the satisfaction of Eq.  \eqref{gen-higgs-vec-trans-S}, we assume $Y$ takes the general form
\begin{equation}\label{gen-higgs-vec-gen-def}
    Y^A=\mathfrak{p}\begin{pmatrix}
        y^a\\x^*_0(q+\alpha\cdot y)
    \end{pmatrix},
\end{equation}
where $q\in\mathbb{R}^+$ is a dimensionless constant, $\mathfrak{p}=\frac{x^*}{x^*_0}|\vartheta^a{}_m|^{\frac{-1}{m+1}}$ is the projective factor, and $\alpha\in\mathbb{R}^m_*$, with 
\begin{equation}\label{alpha-def-higgs}
\alpha_b:=\partial_b\log|\vartheta^a{}_m|^{\frac{-1}{m+1}}.
\end{equation}
As a means of simplification, we fix here and throughout, $q=1$. The components of $Y$ are found to transform as
\begin{equation}
   Y^{A'}=S^{A'}{}_BY^B=\mathfrak{p}|s|^{\frac{-1}{m+1}}\begin{pmatrix}
       s^{a'}{}_by^b+t^{a'}(1+\alpha'\cdot y)\\x^*_0(1+\alpha'\cdot y)
   \end{pmatrix}.
\end{equation}
With respect to these local gauge transformations, the gauge co-vector $\alpha_a$ admits the transformation behavior
\begin{equation}
    \alpha_{b'}=(\alpha_b-u_b)(s^{-1}){}^{b}{}_{c'},
\end{equation}
which, in light of the definitions made in Eqs. \eqref{u-trans-parameter-def} and \eqref{alpha-def-higgs}, is to be expected. Furthermore, $q=1$ is left invariant and
\begin{equation}
    y^{a'}=s^{a'}{}_by^b+\tau^{a'}(y)
\end{equation}
is found to transform as an affine gauge vector with a $y$-dependent translation provided by $\tau^{a'}=t^{a'}(1+\alpha'\cdot y)$.

To access vectors which coherently project to the $m$-dimensional spacetime manifold $\mathcal{M}$, we simply follow the prescription for transforming a homogeneous (projective) vector into its inhomogeneous image. This task is accomplished by dividing out the $*$-component of $Y$, as was done in Eq. \eqref{x-X-map}. Similarly, in retention of proper physical dimension, we scale by the dimensionful constant $[x^*_0]=L$, 
\begin{equation}
    z^a:=x^*_0\frac{Y^a}{Y^*}=\frac{y^a}{1+\alpha\cdot y}.
\end{equation}
The local gauge transformations $S$ act on $z^a$ via fractional linear transformations,
\begin{equation}
    z^{a'}=x^*_0\frac{Y^{a'}}{Y^{*'}}=\frac{s^{a'}{}_by^b}{1+\alpha'\cdot y}+t^{a'}.
\end{equation}
This is simply a particular instantiation of the transformation encountered in Eq. \eqref{x-trans-gen-result}. Had we instead taken $q=0$, as was done in \cite{gen-struc,heavy-lifting}, we would be required to exclude $\alpha=0$ from any consideration. This follows from the fact that the $q=\alpha=0$ scenario lives on the hypersurface at infinity, Eq. \eqref{infinity-surface}, which is necessarily inaccessible to the spacetime. We may thus consider this particular scenario as \textit{un-physical} and, therefore, leave further investigation to the mathematicians. 

In the method of nonlinear realizations, all fields that couple to the connection must also have their symmetries realized nonlinearly. The nonlinear projective Higgs vector is constructed in the standard fashion as
\begin{equation}
    \tilde{\Upsilon}:=\sigma^{-1}Y.
\end{equation}
Explicitly, the components are found to have the form
\begin{equation}\label{upsilon-def}
    \tilde{\Upsilon}{}^{\underline{A}}=\bar{\mathfrak{p}}\begin{pmatrix}
        y^{\underline{a}}-\xi^{\underline{a}}(1+(\change+\alpha)\cdot y)\\x^*_0(1+(\change+\alpha)\cdot y)
    \end{pmatrix}.
\end{equation}
Following our previously developed notation, we write
\begin{equation}
     \tilde{\Upsilon}{}^{\underline{A}}=\bar{\mathfrak{p}}\tilde{\upsilon}{}^{\underline{A}},
\end{equation}
where the lower-case vector $\tilde{\upsilon}$ represents all parts of $\tilde{\Upsilon}$ which are not proportional to the identity. The transformation behavior of $\tilde{\Upsilon}$ under the local gauge action of $S\in SL(m+1,\mathbb{R})$ is just as expected:
\begin{equation}
    \begin{split}
        S(x):\;\tilde{\Upsilon}\rightarrow\tilde{\Upsilon}'&=(\sigma^{-1}Y)'\\
        &=(\Lambda\sigma^{-1}S^{-1})(SY)\\
        &=\Lambda\sigma^{-1}Y\\
        &=\Lambda\tilde{\Upsilon}.
    \end{split}
\end{equation}
In other words, $\tilde{\Upsilon}$ transforms nonlinearly as a projective Lorentz vector. 

The $m$ gauge degrees of freedom provided by the coset coordinates $\xi$ permit a choice of gauge wherein
\begin{equation}
    \xi^{\underline{a}}=\frac{y^{\underline{a}}}{1+(\change+\alpha)\cdot y}.
\end{equation}
Inverting this choice produces
\begin{equation}
    y^{\underline{a}}=\frac{\xi^{\underline{a}}}{1-(\change+\alpha)\cdot r\cdot\xi}.
\end{equation}
To retain consistency with the minimal assumption $\change\cdot \xi=0$, we further assume $\alpha\cdot \xi=0$. Therefore, this particular choice of gauge forces the generalized Higgs $\bar{\mathfrak{p}}$-vector to assume the form
\begin{equation}\label{upsilon-APV}
    \tilde{\Upsilon}{}^{\underline{A}}\overset{\circ}{=}\bar{\mathfrak{p}}\begin{pmatrix}
        0^{\underline{a}}\\x^*_0
    \end{pmatrix},
\end{equation}
where $0^{\underline{a}}:=(0,\;0,\dots,0)^T$ is used to denote the $m$-dimensional zero vector. We refer to this particular gauge as the \textit{almost-physical} \textit{vector-gauge} (APV-gauge) and use $\overset{\circ}{=}$ to denote expressions valid in this gauge. Comparing Eq. \eqref{upsilon-APV} above to Eq. \eqref{OG-upsilon-TW} in the context of Thomas-Whitehead theory, we identify $\tilde{\Upsilon}$ with the fundamental projective vector field associated with a projective structure. Additionally, we identify the APV-gauge with the gauge choice resulting in Eq. \eqref{OG-upsilon-TW}, its image in $V\mathcal{M}$. This provides further evidence for the claim that the present construction describes a generalization of Thomas-Whitehead theories.

The length of $\tilde{\Upsilon}$ with respect to the $\bar{\mathfrak{p}}{}^{-2}$-metric $\tilde{\eta}$ is easily calculated in the APV-gauge,
\begin{equation}\label{upsilon-fixed-length}
    \tilde{\Upsilon}{}^2:=\tilde{\eta}_{\underline{AB}}\tilde{\Upsilon}{}^{\underline{A}}\tilde{\Upsilon}{}^{\underline{B}}\overset{\circ}{=}\eta_0(x^*_0)^2.
\end{equation}
The timelike or spacelike nature of $\tilde{\Upsilon}$ in the APV-gauge entirely determined by the timelike or spacelike nature of the $x^*$-coordinate, i.e., $\eta_0=\pm 1$. In either situation, we find the length of $\tilde{\Upsilon}$ may be fixed by a choice of gauge. Recalling the discussion following Eq. \eqref{algebra-condition}, we may consider the generalized Higgs $\bar{\mathfrak{p}}$-vector $\tilde{\Upsilon}$ as a symmetry breaking vector field, and the APV-gauge as the Lorentz-invariant constraint required for a coherent group contraction---paired with the $x^*_0\rightarrow0$ limit. Analogous vector fields have been studied in the context of non-projective (A)-dS gauge theories, for example, \cite{into-cartan,stelle-west,dynamical-stelle-west-1,5d-LL-total-der-action,topological-gravity,MM-conf-flat,Wise}. In Sec. \ref{sec:dynamical-theory}, where the dynamical projective gravitational theory is constructed, we show that Eq. \eqref{upsilon-fixed-length} may be arrived at as a result of the field equations for $\tilde{\Upsilon}$. Essentially, we show that the APV-gauge may be determined dynamically. 

\subsubsection{First Derivative}
\label{subsubsec:gen-higgs-v-1d}

The nonlinear projective gauge-covariant derivative acts on the generalized projective Higgs vector $\tilde{\Upsilon}$ as
\begin{equation}
    \tilde{D}\tilde{\Upsilon}{}^{\underline{A}}=d\tilde{\Upsilon}{}^{\underline{A}}+\tilde{\Omega}{}^{\underline{A}}{}_{\underline{B}}\tilde{\Upsilon}{}^{\underline{B}}.
\end{equation}
In the APV-gauge, the first $m(m+1)$ components of the vector-valued $1$-form $\tilde{D}\tilde{\Upsilon}$ become
\begin{equation}\label{cov-d-upsilon}
    \begin{split}
        \tilde{D}\tilde{\Upsilon}{}^{\underline{a}}&=d\tilde{\Upsilon}{}^{\underline{a}}+\tilde{\Omega}{}^{\underline{a}}{}_{\underline{b}}\tilde{\Upsilon}{}^{\underline{b}}+\tilde{\Omega}{}^{\underline{a}}{}_{\underline{*}}\tilde{\Upsilon}{}^{\underline{*}}\\
        &=\bar{\mathfrak{p}}(\frac{\tilde{\upsilon}{}^{\underline{*}}}{x^*_0}\overline{\vartheta}{}^{\underline{a}}+\tilde{\upsilon}{}^{\underline{a}}\tilde{g}+\overline{D}\tilde{\upsilon}{}^{\underline{a}})\\
        &\overset{\circ}{=}\bar{\mathfrak{p}}\overline{\vartheta}{}^{\underline{a}},
    \end{split}
\end{equation}
where 
\begin{equation}\label{little-g-tilde-def}
    \tilde{g}:=d\log\bar{\mathfrak{p}}=\begin{pmatrix}
        g,&\frac{1}{x^*}
    \end{pmatrix}=\begin{pmatrix}
        \change+\alpha,&\frac{1}{x^*}
    \end{pmatrix}
\end{equation}
is the only $x^*$-containing term in the penultimate line of Eq. \eqref{cov-d-upsilon}. The remaining $m+1$ components of $\tilde{D}\tilde{\Upsilon}$ are found similarly,
\begin{equation}
    \begin{split}
        \tilde{D}\tilde{\Upsilon}{}^{\underline{*}}&=d\tilde{\Upsilon}{}^{\underline{*}}+\tilde{\Omega}{}^{\underline{*}}{}_{\underline{b}}\tilde{\Upsilon}{}^{\underline{b}}+\tilde{\Omega}{}^{\underline{*}}{}_{\underline{*}}\tilde{\Upsilon}{}^{\underline{*}}\\
        &=\bar{\mathfrak{p}}(\tilde{\upsilon}{}^{\underline{*}}\tilde{g}+x^*_0\overline{\mathcal{P}}_{\underline{b}}\tilde{\upsilon}{}^{\underline{b}})\\
        &\overset{\circ}{=}\bar{\mathfrak{p}}x^*_0\tilde{g}.
    \end{split}
\end{equation}
Due to the explicit appearance of $\overline{\vartheta}$ in the final equality of Eq. \eqref{cov-d-upsilon}, we define the \textit{general projective $2$-frame} as
\begin{equation}\label{big-E-def}
    \tilde{E}{}^{\underline{A}}:=\tilde{D}\tilde{\Upsilon}{}^{\underline{A}}.
\end{equation}
This is the generalization of $\tilde{e}$ in the Thomas-Whitehead theory of \cite{gen-struc,heavy-lifting}, and the generalization of the projective $2$-frames used in \cite{2-frame}. Exposing the $V\mathcal{M}$ index, the \textit{$\bar{\mathfrak{p}}$-co-frame} has the matrix components given by
\begin{equation}
    \tilde{E}{}^{\underline{A}}{}_{M}=\bar{\mathfrak{p}}\begin{pmatrix}\frac{\tilde{\upsilon}{}^{\underline{*}}}{x^*_0}\overline{\vartheta}{}^{\underline{a}}{}_m+\tilde{\upsilon}{}^{\underline{a}}g_m+\overline{D}_m\tilde{\upsilon}{}^{\underline{a}}&\frac{1}{x^*}\tilde{\upsilon}{}^{\underline{a}}\\\tilde{\upsilon}{}^{\underline{*}}g_m+x^*_0\overline{\mathcal{P}}_{\underline{b}m}\tilde{\upsilon}{}^{\underline{b}}&\frac{\tilde{\upsilon}{}^{\underline{*}}}{x^*}
    \end{pmatrix}.
\end{equation}
These $\tilde{E}{}^{\underline{A}}$ may be viewed as a particular instantiation of the annihilating $1$-form used to define a projective Ehresmann connection \cite{density-algebra-2}. For more information on the algebra of densities and their use in projective geometry, see \cite{density-algebra-1,tensor-density,density-algebra-3,density-algebra-4}. In the APV-gauge,
\begin{equation}
     \tilde{E}{}^{\underline{A}}{}_{M}\overset{\circ}{=}\bar{\mathfrak{p}}\begin{pmatrix}
         \overline{\vartheta}{}^{\underline{a}}{}_m&0\\x^*_0g_m&\frac{x^*_0}{x^*}
     \end{pmatrix}.
\end{equation}
Following our prescription, we define
\begin{equation}\label{big-E-little-e}
    \tilde{E}{}^{\underline{A}}{}_M:=\bar{\mathfrak{p}}\tilde{e}{}^{\underline{A}}{}_M,
\end{equation}
so that $\tilde{e}$ is simply the $\bar{\mathfrak{p}}{}^0$-port of $\tilde{E}$. We may therefore say that in the APV-gauge, $\tilde{e}$ in Eq. \eqref{big-E-little-e} may be identified with the $\tilde{e}$ of TW theory in Eq. \eqref{TW-frames} \cite{gen-struc,heavy-lifting}. 

Notice that when the additional constraint of
\begin{equation}\label{cov-der-xi}
    \overline{D}\xi^{\underline{a}}=0
\end{equation}
is imposed on the projective ``radius vector" $\xi$ \cite{affine-cartan-book}, then
\begin{equation}
    \overline{\vartheta}{}^{\underline{a}}=(r^{-1})^{\underline{a}}{}_b\vartheta^{b}=:e^{\underline{a}}
\end{equation}
is an ordinary $m$-dimensional Lorentz co-frame, $e$. When both the APV-gauge and Eq.~\eqref{cov-der-xi} are chosen, equalities will be expressed with a filled in circle. For example,
\begin{equation}
    \tilde{E}{}^{\underline{A}}{}_{M}\overset{\blcirc}{=}\bar{\mathfrak{p}}\begin{pmatrix}e^{\underline{a}}{}_m&0\\x^*_0g_m&\frac{x^*_0}{x^*}
    \end{pmatrix}.
\end{equation}
The combination of these two choices will then be referred to as the \textit{physical vector-gauge} (PV-gauge). 

When $\tilde{E}$ is non-degenerate, the inverse exists and may be found by requiring the duality relations be satisfied:
\begin{equation}
\tilde{E}{}^{\underline{A}}{}_{M}(\tilde{E}{}^{-1})^M{}_{\underline{B}}=\delta^{\underline{A}}{}_{\underline{B}},
\end{equation}
and
\begin{equation}
    (\tilde{E}{}^{-1})^M{}_{\underline{A}}\tilde{E}{}^{\underline{A}}{}_{N}=\delta^M{}_N.
\end{equation}
Assuming the inverse to exist, one finds
\begin{equation}
    (\tilde{E}{}^{-1})^M{}_{\underline{A}}\overset{\circ}{=}\bar{\mathfrak{p}}^{-1}\begin{pmatrix}
        (\overline{\vartheta}{}^{-1})^m{}_{\underline{a}}&0\\-x^*g_m(\overline{\vartheta}{}^{-1})^m{}_{\underline{a}}&\frac{x^*}{x^*_0}
    \end{pmatrix}
\end{equation}
in the APV-gauge.

In any gauge, the determinant of the the $\bar{\mathfrak{p}}$-co-frame has its $x^*$-dependence separated out. From Eq. \eqref{det-formula}, we find
\begin{equation}\label{det-big-E-gen}
|\tilde{E}|:=|\tilde{E}{}^{\underline{A}}{}_M|=\bar{\mathfrak{p}}{}^{m+1}\frac{\tilde{\upsilon}{}^{\underline{*}}}{x^*}\left|\frac{\tilde{\upsilon}{}^{\underline{*}}}{x^*_0}\overline{\vartheta}{}^{\underline{a}}{}_m+\overline{D}_m\tilde{\upsilon}{}^{\underline{a}}-\frac{x^*_0}{\tilde{\upsilon}{}^{\underline{*}}}\overline{\mathcal{P}}_{\underline{b}m}\tilde{\upsilon}{}^{\underline{b}}\tilde{\upsilon}{}^{\underline{a}}\right|.
\end{equation}
Using Eq. \eqref{big-E-little-e}, the above may be written more concisely as
\begin{equation}
|\tilde{E}|=\bar{\mathfrak{p}}{}^{m+1}|\tilde{e}|.
\end{equation}
In the APV-gauge, $|\tilde{E}|$ reduces significantly to
\begin{equation}
|\tilde{E}|\overset{\circ}{=}\frac{x^*_0}{x^*}\bar{\mathfrak{p}}{}^{m+1}|\overline{\vartheta}{}^{\underline{a}}{}_m|,
\end{equation}
and in the PV-gauge, we recover
\begin{equation}
    |\tilde{E}|\overset{\blcirc}{=}\frac{x^*_0}{x^*}\bar{\mathfrak{p}}{}^{m+1}|e|,
\end{equation}
which, when expanded, may be identified with the density comprising the natural volume form on $V\mathcal{M}$ \cite{proj-conn}.

Using the $\bar{\mathfrak{p}}$-co-frame $\tilde{E}$, we may construct a Higgs-metric on $V\mathcal{M}$, distinct from the Goldstone metric $H$, as
\begin{equation}
    \tilde{G}_{MN}:=\tilde{\eta}_{\underline{AB}}(\tilde{D}_M\tilde{\Upsilon}{}^{\underline{A}})(\tilde{D}_N\tilde{\Upsilon}{}^{\underline{B}}).
\end{equation}
In the APV-gauge, the matrix components are found to be
\begin{equation}\label{higgs-metric-EE}
    \tilde{G}_{MN}\overset{\circ}{=}\begin{pmatrix}
       \overline{g}_{mn}+\eta_0(x^*_0)^2g_mg_n&\eta_0\frac{(x^*_0)^2}{x^*}g_n\\\eta_0\frac{(x^*_0)^2}{x^*}g_m&\eta_0\left(\frac{x^*_0}{x^*}\right)^2
    \end{pmatrix},
\end{equation}
where $\overline{g}_{mn}:= \eta_{\underline{ab}}\overline{\vartheta}{}^{\underline{a}}{}_m\overline{\vartheta}{}^{\underline{b}}{}_n$. For $\eta_0=-1$, this $\tilde{G}_{MN}$ is exactly the generalization of the metric defined in Eq. \eqref{big-g} in the context of Thomas-Whitehead theory. For more information on these types of Higgs metrics, see \cite{Leclerc,composite-bundle,gl4-symmetry-breaking,composite-bundles-gl4,higgs-metric}.

\subsubsection{Second Derivative}
\label{subsubsec:gen-higgs-v-2d}

When pulled back to $V\mathcal{M}$, the second derivative of the generalized projective Higgs vector $\tilde{\Upsilon}$ defines the connection $\tilde{\Gamma}$ on the volume bundle,
\begin{equation}\label{gen-higgs-connection}
    \tilde{\Gamma}{}^L{}_{NM}:=(\tilde{E}^{-1})^L{}_{\underline{A}}\tilde{D}_M\tilde{E}{}^{\underline{A}}{}_N.
\end{equation}
The definition of the $\bar{\mathfrak{p}}$-co-frame $\tilde{E}$, Eq. \eqref{big-E-def}, permits a decomposition of the gauge-covariant derivative operators in Eq. \eqref{gen-higgs-connection} into their (anti)-symmetric parts,
\begin{equation}
    \tilde{D}_M\tilde{D}_N\tilde{\Upsilon}{}^{\underline{A}}=\tfrac{1}{2}[\tilde{D}_M,\tilde{D}_N]\tilde{\Upsilon}{}^{\underline{A}}+\tfrac{1}{2}\{\tilde{D}_M,\tilde{D}_N\}\tilde{\Upsilon}{}^{\underline{A}}.
\end{equation}
In the language of Clifford's \textit{geometric algebra} \cite{geometric-algebra}, 
\begin{equation}
    \tilde{D}\tilde{D}\tilde{\Upsilon}{}^{\underline{A}}=\left(\tilde{D}\cdot \tilde{D}+\tilde{D}\wedge\tilde{D}\right)\tilde{\Upsilon}{}^{\underline{A}}.
\end{equation}
The antisymmetric part may be taken to define the nonlinear curvature, since
\begin{equation}
    [\tilde{D}_M,\;\tilde{D}_N]\tilde{\Upsilon}{}^{\underline{A}}=\tilde{\mathcal{K}}{}^{\underline{A}}{}_{\underline{B}[MN]}\tilde{\Upsilon}{}^{\underline{B}}.
\end{equation}
From the definition of $\tilde{\Gamma}$ in Eq.~\eqref{gen-higgs-connection}, this simply provides the torsion,
\begin{equation}\label{K-tilde-Upsilon-torsion}
    \tilde{\mathcal{K}}{}^{\underline{A}}{}_{\underline{B}}\tilde{\Upsilon}{}^{\underline{B}}=\tilde{D}\wedge\tilde{D}\tilde{\Upsilon}{}^{\underline{A}}=\tilde{D}\tilde{E}{}^{\underline{A}}
    =\tilde{\mathcal{T}}{}^{\underline{A}},
\end{equation}
where we use the short-hand notation $\tilde{D}\tilde{E}\equiv\tilde{D}\wedge\tilde{E}$. In the APV-gauge, this reduces precisely to the nonlinear torsion $2$-form defined in Eq. \eqref{barred-torsion},
\begin{equation}\label{APV-torsion}
    \tilde{\mathcal{T}}{}^{\underline{A}}=\tilde{\mathcal{K}}{}^{\underline{A}}{}_{\underline{B}}\tilde{\Upsilon}{}^{\underline{B}}=\begin{pmatrix}
        \tilde{\mathcal{K}}{}^{\underline{a}}{}_{\underline{b}}\tilde{\Upsilon}{}^{\underline{b}}+\tilde{\mathcal{K}}{}^{\underline{a}}{}_{\underline{*}}\tilde{\Upsilon}{}^{\underline{*}}\\
         \tilde{\mathcal{K}}{}^{\underline{*}}{}_{\underline{b}}\tilde{\Upsilon}{}^{\underline{b}}
    \end{pmatrix}\overset{\circ}{=}\bar{\mathfrak{p}}\begin{pmatrix}
       \overline{\mathcal{T}}{}^{\underline{a}}\\0
    \end{pmatrix}.
\end{equation}
Connecting this back to $V\mathcal{M}$ yields
\begin{equation}
    (\tilde{E}^{-1})^L{}_{\underline{A}} \tilde{\mathcal{K}}{}^{\underline{A}}{}_{\underline{B}}\tilde{\Upsilon}{}^{\underline{B}}\overset{\circ}{=}\begin{pmatrix}
         (\overline{\vartheta}{}^{-1})^l{}_{\underline{a}} \overline{\mathcal{T}}{}^{\underline{a}}\\
          -x^*g_m(\overline{\vartheta}{}^{-1})^m{}_{\underline{a}}\overline{\mathcal{T}}{}^{\underline{a}}
    \end{pmatrix}.
\end{equation}
The symmetric part thus defines a torsion-free linear connection via
\begin{equation}
\tilde{E}{}^{\underline{A}}{}_L\tilde{\Gamma}{}^L{}_{(NM)}=\{\tilde{D}_M,\;\tilde{D}_N\}\tilde{\Upsilon}{}^{\underline{A}}=\tilde{D}_{(M}\tilde{E}{}^{\underline{A}}{}_{N)}.
\end{equation}

We display explicitly the computation of only the $\mathcal{M}$-components of $\tilde{\Gamma}{}^L{}_{NM}$. For calculational convenience we impose the APV-gauge, then restrict further to the PV-gauge. In the end, we separate the (anti)-symmetric parts and discuss the result:
\begin{equation}\label{gamma-tilde-calculation}
    \begin{split}
        \tilde{\Gamma}{}^{l}{}_{nm}&=(\tilde{E}{}^{-1})^l{}_{\underline{A}}\tilde{D}_{m}\tilde{E}{}^{\underline{A}}{}_{n}\\
        &\overset{\circ}{=}(\tilde{E}{}^{-1})^l{}_{\underline{a}}(\partial_{m}\tilde{E}{}^{\underline{a}}{}_{n}+\tilde{\Omega}{}^{\underline{a}}{}_{\underline{b}m}\tilde{E}{}^{\underline{b}}{}_{n}+\tilde{\Omega}{}^{\underline{a}}{}_{\underline{*}m}\tilde{E}{}^{\underline{*}}{}_{n})\\
        &\overset{\circ}{=}(\overline{\vartheta}{}^{-1})^l{}_{\underline{a}}(\partial_{m}\overline{\vartheta}{}^{\underline{a}}{}_{n}+\overline{\omega}{}^{\underline{a}}{}_{\underline{b}m}\overline{\vartheta}{}^{\underline{b}}{}_{n}+\overline{\vartheta}{}^{\underline{a}}{}_{n}g_{m}+\overline{\vartheta}{}^{\underline{a}}{}_{m}g_n)\\
        &\overset{\circ}{=}(\overline{\vartheta}{}^{-1})^l{}_{\underline{a}}(r^{-1}){}^{\underline{a}}{}_{c}\left(\delta^b{}_c\partial_{m}+\omega^b{}_{cm}-\delta^b{}_c\change_d\vartheta^d{}_m-\vartheta^b{}_m\change_c\right)r^c{}_{\underline{b}}\overline{\vartheta}{}^{\underline{b}}{}_{n}\\
        &\quad-(\overline{\vartheta}{}^{-1})^l{}_{\underline{a}}\xi^{\underline{a}}\overline{\mathcal{P}}_{\underline{b}m}\overline{\vartheta}{}^{\underline{b}}{}_{n}+\delta^l{}_mg_{n}+\delta^l{}_ng_m\\
        &\overset{\blcirc}{=}(e^{-1})^l{}_{\underline{a}}\left(\omega^{\underline{a}}{}_{\underline{b}m}+\delta^{\underline{a}}{}_{\underline{b}}\partial_m\right)e^{\underline{b}}{}_{n}+\delta^l{}_n\alpha_m+\delta^l{}_m\alpha_n-\xi^l\overline{\mathcal{P}}_{am}e^a{}_{n},
    \end{split}
\end{equation}
with $\xi^l=(e^{-1})^l{}_{\underline{a}}\xi^{\underline{a}}$ and
\begin{equation}
    \omega^{\underline{a}}{}_{\underline{b}m}:=(r^{-1})^{\underline{a}}{}_b\left(\omega^b{}_{cm}+\delta^b{}_c\partial_m\right)r^c{}_{\underline{b}}.
\end{equation}
Upon symmetrizing the third line in the above calculation, the result may be identified as a generalized relation between the ``barred" and ``un-barred" connections of \cite{covariant-tw}, since
\begin{equation}
    \Pi^{l}{}_{nm}\overset{\circ}{:=}(\overline{\vartheta}{}^{-1})^l{}_{\underline{a}}(\overline{\omega}{}^{\underline{a}}{}_{\underline{b}(m}+\delta^{\underline{a}}{}_{\underline{b}}\partial_{(m})\overline{\vartheta}{}^{\underline{b}}{}_{n)}+\delta^l{}_{(m}g_{n)}.
\end{equation}
Compare with Eqs. \eqref{barred-omega-TW} and \eqref{barred-gamma-TW} in the Thomas-Whitehead theory, and Eq. \eqref{pstegr-pi} in the projective symmetric teleparallel theory. Further restricting to the PV-gauge, we recover the standard projective connection initially proposed by Thomas \cite{thomas-1},
\begin{equation}
    \Pi^l{}_{nm}\overset{\blcirc}{:=}\Gamma^l{}_{(nm)}+\delta^l{}_{(n}\alpha_{m)},
\end{equation}
where
\begin{equation}
     \Gamma^l{}_{nm}\overset{\blcirc}{:=}(e^{-1})^l{}_{\underline{a}}\left(\omega^{\underline{a}}{}_{\underline{b}m}+\delta^{\underline{a}}{}_{\underline{b}}\partial_m\right)e^{\underline{b}}{}_{n}
\end{equation}
is the linear connection on $\mathcal{M}$. Although not obvious, to conclude that $\Pi^l{}_{nm}$ \textit{is} the standard projective connection, we must require that
\begin{equation}
    \partial_m(\vartheta^{-1})^m{}_b\overset{\blcirc}{=}(\vartheta^{-1})^m{}_a\omega^a{}_{bm}
\end{equation}
is satisfied to ensure both the internal and external traces of $\Pi$ vanish,
\begin{equation}
    \Pi^l_{lm}\overset{\blcirc}{=}\Pi^l{}_{ml}\overset{\blcirc}{=}0.
\end{equation}
Additionally, from the definition of $\overline{\mathcal{T}}$ in Eq. \eqref{barred-torsion}, we have
\begin{equation}
    \overline{\mathcal{T}}{}^l{}_{mn}:=(\overline{\vartheta}{}^{-1})^l{}_{\underline{a}}\overline{\mathcal{T}}{}^{\underline{a}}{}_{mn}\overset{\blcirc}{=}(e^{-1})^l{}_{\underline{a}}\partial_{[m}e^{\underline{a}}{}_{n]}+(e^{-1})^l{}_{\underline{a}}\omega^{\underline{a}}{}_{\underline{b}[m}e^{\underline{b}}{}_{n]}=\mathcal{T}{}^l{}_{mn}.
\end{equation}
And lastly, the $\xi$ term in Eq. \eqref{gamma-tilde-calculation} may be expressed as
\begin{equation}
\xi^l\overline{\mathcal{P}}_{nm}=\xi^l\overline{\mathcal{P}}_{(nm)}\overset{\blcirc}{=}\tfrac{y^l}{1+g\cdot y}(\mathcal{P}_{nm}-\partial_m\change_n+\change_l\Gamma^l{}_{nm}-\change_n\change_m).
\end{equation}
Therefore, the $\mathcal{M}$-components of $\tilde{\Gamma}$ in the PV-gauge result in
\begin{equation}
    \tilde{\Gamma}{}^{l}{}_{nm}\overset{\blcirc}{=}\Pi^l{}_{nm}+\mathcal{T}{}^l{}_{nm}-y^l\overline{\mathcal{P}}_{nm}.
\end{equation}

The component $\tilde{\Gamma}{}^{*}{}_{nm}$, not computed explicitly, takes the form
\begin{equation}
    \tilde{\Gamma}{}^{*}{}_{nm}\overset{\blcirc}{=}x^*\mathcal{D}_{nm},
\end{equation}
where
\begin{equation}\label{diff-field-gen-proj}
    \mathcal{D}_{nm}:=\mathcal{P}_{nm}+\partial_{m}\alpha_n-\alpha_l\Gamma^l{}_{nm}-\alpha_n\alpha_m
\end{equation}
is the \textit{Diffeomorphism field} encountered previously, see Eq. \eqref{P-def}. Notice that $\mathcal{D}_{nm}$ has been given an antisymmetric part. This is precisely the sought-after extension of \cite{gen-struc}. The antisymmetric part of $\mathcal{D}_{nm}$ has a natural interpretation in the PV-gauge given by
\begin{equation}
    \begin{split}
        \mathcal{D}_{[nm]}&\overset{\blcirc}{=}\mathcal{P}_{[nm]}+\partial_{[m}\alpha_{n]}-\alpha_l\Gamma^l{}_{[nm]}\\
        &=\tfrac{-1}{m+1}\check{\check{\mathcal{R}}}_{[nm]}-\tfrac{1}{m+1}\check{\check{\mathcal{R}}}_{[mn]}-\alpha_l\mathcal{T}^l{}_{nm}\\
        &=\alpha_l\mathcal{T}^l{}_{mn},
    \end{split}
\end{equation}
where Eqs. \eqref{homo-alpha-intro} and \eqref{r-bar-trace} were utilized to arrive at the final result. We thus find that in the PV-gauge, the antisymmetric part of the generalized Diffeomorphism field is nothing but the torsion tensor in the direction of $\alpha$. This provides another layer of certainty that the present generalization of Thomas-Whitehead connections correctly reduces to the standard notion of Thomas-Whitehead connections for vanishing torsion in the PV-gauge. In order to maintain direct comparison throughout, we use $\mathcal{D}_{nm}:=\mathcal{D}_{(nm)}$ to represent the symmetric part and always explicitly separate the torsion unless stated otherwise.

All together, the $V\mathcal{M}$-connection is computed to have the components
\begin{equation}\label{final-gamma-tilde-PV}
    \tilde{\Gamma}{}^L{}_{NM}\overset{\blcirc}{=}\begin{pmatrix}
        \begin{pmatrix}
            \Pi^l{}_{nm}+\mathcal{T}{}^l{}_{nm}-y^l\overline{\mathcal{P}}_{nm}&\frac{1}{x^*}\delta^l{}_m\\x^*\mathcal{D}_{nm}-x^*\alpha_l\mathcal{T}{}^{l}{}_{nm}&0
        \end{pmatrix},\;\begin{pmatrix}
            \frac{1}{x^*}\delta^l{}_n&0\\0&0
        \end{pmatrix}
    \end{pmatrix}.
\end{equation}
Compare with the projective symmetric teleparallel connection in Eq. \eqref{compatible-conf-proj-conn}. As a future direction of research, it would be interesting to develop a concrete relation between the connection in Eq. \eqref{final-gamma-tilde-PV} above and the Berwald-Thomas-Whitehead connections discussed in \cite{berwald,berwald-proj-trans,proj-conn}. In particular, through the study of sprays and their relationship to the nonlinear connections \cite{nonlinear-spray-connections}.

\subsection{\texorpdfstring{$\bar{\mathfrak{p}}$}{\bar{\mathfrak{p}}}-Orientation Symbol}
\label{subsec:nl-orient-sym}

In order to form a dynamical theory for the General Projective Gauge Gravitational Theory, we seek a definition for the internal projective dual operation. This will provide a means of forming a projective Euler density, as was accomplished in the Metric-Affine gauge setting in Eq. \eqref{MAG-euler}. However, physically meaningful expressions must be independent of the equivalence class provided by $\bar{\mathfrak{p}}$, i.e., only those expressions which are of type-$\bar{\mathfrak{p}}{}^0$ permit physical viability. Interestingly, the nonlinear projective or $\bar{\mathfrak{p}}{}^{\mathfrak{w}}$-orientation symbol cannot be constructed by the prescription followed thus far. This fact follows from the role played by the orientation symbol in forming determinants, Appendix \hyperref[app-A:orientation]{A.2}, in combination with the unit-volume property of the coset projection, $|\sigma|=|\sigma^{-1}|=1$. For example, a $\mathfrak{p}^{-(m+1)}$-orientation symbol $\hat{\eta}_{A_1\dots A_{m+1}}$ transforms under the local gauge action of $S^{-1}\in SL(m+1,\mathbb{R})$, with $S^{-1}$ given in Eq.~\eqref{SL-matrix-inv}, as
\begin{equation}
S^{-1}(x):\hat{\eta}_{A_1\dots A_{m+1}}\rightarrow \hat{\eta}_{A_1'\dots A_{m+1}'}
=|S^{-1}|\hat{\eta}_{A_1'\dots A_{m+1}'}=\hat{\eta}_{A_1'\dots A_{m+1}'}.
\end{equation}
Clearly, $\hat{\eta}_{A_1\dots A_{m+1}}$ is invariant as a result of $|S|=|S^{-1}|=1$. Applying the nonlinear realization process with the coset element $\sigma$ leads to
\begin{equation}
    \hat{\eta}_{B_1\dots B_{m+1}}\sigma^{B_1}{}_{\underline{A}_1}\dots\sigma^{B_{m+1}}{}_{\underline{A}_{m+1}}=|\sigma|\hat{\eta}_{\underline{A}_1\dots\underline{A}_{m+1}}=\hat{\eta}_{\underline{A}_1\dots\underline{A}_{m+1}},
\end{equation}
which has only the effect of changing the index reference space. Thus, the standard procedure will not produce the orientation symbol we require.

We therefore begin by requiring the invariant nonlinear projective volume-form, $\hat{\tilde{\epsilon}}=* 1$, be defined as
\begin{equation}\label{invariant-proj-vol}
    \hat{\tilde{\epsilon}}:=\frac{1}{(m+1)!}\hat{\tilde{\epsilon}}_{\underline{A}_1\dots\underline{A}_{m+1}}\tilde{E}{}^{\underline{A}_1}\wedge\dots\wedge\tilde{E}{}^{\underline{A}_{m+1}}.
\end{equation}
The ``hat" is here used to distinguish between the projective orientation \textit{symbol}, $\hat{\tilde{\epsilon}}_{\underline{A}_1\dots\underline{A}_{m+1}}$, and the projective orientation \textit{tensor density}, $\tilde{\epsilon}_{\underline{A}_1\dots\underline{A}_{m+1}}$. Invariance in this particular nonlinear projective context loses some specificity as a result of the gauge- and coordinate-invariance of the projective factor $\bar{\mathfrak{p}}$, as well as the preservation of volumes inherent to the chosen stability subgroup. However, Eq. \eqref{invariant-proj-vol} is indeed the correct definition of the invariant nonlinear projective volume-form, as can easily be confirmed by reversing the steps of the nonlinear realization process, i.e., ``removing tildes" and replacing $\underline{A},\underline{B},\dots$ with $A,B,\dots$. We therefore use the term invariance unequivocally. 

Essentially, invariance requires the nonlinear projective orientation symbol $\hat{\tilde{\epsilon}}_{\underline{A}_1\dots\underline{A}_{m+1}}$ to be identified with the $\bar{\mathfrak{p}}{}^{-(m+1)}$-orientation symbol,
\begin{equation}
\hat{\tilde{\epsilon}}_{\underline{A}_1\dots\underline{A}_{m+1}}:=\bar{\mathfrak{p}}{}^{-(m+1)}\hat{\epsilon}_{\underline{A}_1\dots\underline{A}_{m+1}}.
\end{equation}
In the above expression, $\hat{\epsilon}_{\underline{A}_1\dots\underline{A}_{m+1}}$ is the $(m+1)$-dimensional completely antisymmetric ``Levi-Civita" symbol, identified by its transformation behavior under the projective Lorentz group. From the above definitions, it is clear that a positively oriented volume form follows from a restriction to positive projective factors:
\begin{equation}
    \bar{\mathfrak{p}}>0.
\end{equation}
Using the properties outlined in Appendix \hyperref[app-A:orientation]{A.2}, we show that $\hat{\tilde{\epsilon}}$ is indeed the invariant volume-form. For a spacetime manifold $\mathcal{M}$ of split-signature $(p,q)$, we find
\begin{equation}
    \begin{split}
        \hat{\tilde{\epsilon}}&=\frac{1}{(m+1)!}\hat{\tilde{\epsilon}}_{\underline{A}_1\dots\underline{A}_{m+1}}\tilde{E}{}^{\underline{A}_1}\wedge\dots\wedge\tilde{E}{}^{\underline{A}_{m+1}}\\
        &=\frac{(-1)^q\eta_0}{(m+1)!}|\tilde{E}|\hat{\tilde{\epsilon}}_{N_1\dots N_{m+1}}dx^{N_1}\wedge\dots\wedge dx^{N_{m+1}}\\
        &=\frac{(-1)^q\eta_0}{(m+1)!}\bar{\mathfrak{p}}^{m+1}|\tilde{e}|\hat{\tilde{\epsilon}}_{N_1\dots N_{m+1}}dx^{N_1}\wedge\dots\wedge dx^{N_{m+1}}\\
        &=\frac{(-1)^q\eta_0}{(m+1)!}|\tilde{e}|\hat{\epsilon}_{N_1\dots N_{m+1}}dx^{N_1}\wedge\dots\wedge dx^{N_{m+1}}\\
        &=(-1)^q\eta_0|\tilde{e}|d^{m+1}x,
    \end{split}
\end{equation}
where $|\tilde{E}|\equiv|\tilde{E}{}^{\underline{A}}{}_M|$ and $|\tilde{e}|\equiv|\tilde{e}{}^{\underline{A}}{}_M|$. Due to the semi-direct product structure of $SL(m+1,\mathbb{R})\cong PGL(m,\mathbb{R})$ and the particular $x^*$-dependence displayed by $\tilde{e}^{\underline{A}}{}_M$, the determinant $|\tilde{e}|$ has its $x^*$-dependence projected out as an overall factor, see Eqs. \eqref{det-big-E-gen} and \eqref{big-E-little-e}. Therefore, integration with respect to the invariant projective volume-form in Eq. \eqref{invariant-proj-vol} properly recovers the overall rescaling factor, $\int dx^*/x^*$, encountered in Eq. \eqref{grav-constants-rescale} for the Thomas-Whitehead gravitational theory.

The gauge-covariant derivative of $\hat{\tilde{\epsilon}}_{\underline{A}_1\dots\underline{A}_{m+1}}$ conveniently yields a familiar quantity. To calculate this, we note that when any two indices of $\hat{\tilde{\epsilon}}_{\underline{A}_1\dots\underline{A}_{m+1}}$ are equal, the result vanishes due to the inherent antisymmetry. Therefore, it is sufficient to consider
\begin{equation}
    \begin{split}
\tilde{D}\hat{\tilde{\epsilon}}_{\underline{12\dots m+1}}&=d\tilde{\epsilon}_{\underline{12\dots m+1}}-\tilde{\Omega}{}^{\underline{A}}{}_{\underline{1}}\tilde{\epsilon}_{\underline{A2\dots m+1}}-\tilde{\Omega}{}^{\underline{A}}{}_{\underline{2}}\tilde{\epsilon}_{\underline{1A\dots m+1}}-\dots-\tilde{\Omega}{}^{\underline{A}}{}_{\underline{m+1}}\tilde{\epsilon}_{\underline{12\dots A}}\\
&=d\hat{\tilde{\epsilon}}_{\underline{12\dots m+1}}-\tilde{\Omega}{}^{\underline{A}}{}_{\underline{A}}\hat{\tilde{\epsilon}}_{\underline{12\dots m+1}}\\
&=d\hat{\tilde{\epsilon}}_{\underline{12\dots m+1}}\\
&=\hat{\tilde{\epsilon}}_{\underline{12\dots m+1}}d\log\bar{\mathfrak{p}}^{-(m+1)}.
    \end{split}
\end{equation}
Recalling Eq. \eqref{little-g-tilde-def}, we arrive at the general expression
\begin{equation}\label{cov-der-nlp-orientation-symbol}
\tilde{D}\hat{\tilde{\epsilon}}_{\underline{A}_1\dots\underline{A}_{m+1}}=-(m+1)\tilde{g}\hat{\tilde{\epsilon}}_{\underline{A}_1\dots\underline{A}_{m+1}}.
\end{equation}
Compare with the analogous expression for Metric-Affine geometries, Eq. $3.8.5$ of \cite{Hehl}. Obviously, the gauge-covariant derivative of the invariant nonlinear projective volume-form vanishes, $\tilde{D}\hat{\tilde{\epsilon}}=0$, as a result of producing an $m+2$-form.

The $\bar{\mathfrak{p}}{}^{-(m+1)}$-orientation symbol, along with the $\bar{\mathfrak{p}}{}^{-2}$-metric, allows for the construction of a type of \textit{internal} Hodge ``dual" operation. This operation is defined to act on the internal \textit{algebra}-valued indices, rather than the form indices. For this reason, the internal dual operation is defined to map a $\bar{\mathfrak{p}}{}^{\mathfrak{w}}$-$k$-form to a $\bar{\mathfrak{p}}{}^{\mathfrak{w}}$-$(m+1-k)$-form via the Lie-algebra valued orientation symbol $\hat{\tilde{\epsilon}}_{\underline{A}_1\dots\underline{A}_{m+1}}$. In other words, we define this operation in a manner that preserves projective weights. As a specific example, one that will be encountered most frequently, consider the $\bar{\mathfrak{p}}{}^{0}$-$2$-form $\tilde{T}{}^{\underline{A}}{}_{\underline{B}}$ in $m+1=5$ dimensions. The internal dual operation, which we will casually denote by the left action of $*$, yields the $\bar{\mathfrak{p}}{}^{0}$-$3$-form
\begin{equation}\label{star-K-gen}
    *\tilde{T}{}^{\underline{A}}{}_{\underline{B}}
    =\tilde{\eta}{}^{\underline{AF}}\hat{\tilde{\epsilon}}_{\underline{FBCDE}}\tilde{T}{}^{\underline{C}}{}_{\underline{G}}\tilde{\eta}{}^{\underline{GD}}\wedge\tilde{E}{}^{\underline{E}}.
\end{equation}

It will be convenient to have available the (internal) projective analogues of the so-called $\eta$-bases \cite{Hehl}. To construct these, we first introduce the internal product, which we denote by $\rfloor$. Using the invariant projective volume-form, the internal product is defined to have the operation
\begin{equation}\label{internal-product}
    \hat{\tilde{\epsilon}}_{\underline{A}_1}:=(\tilde{E}{}^{-1})_{\underline{A}_1}\rfloor\hat{\tilde{\epsilon}}=\frac{1}{m!}\hat{\tilde{\epsilon}}_{\underline{A}_1\underline{A}_2\dots\underline{A}_{m+1}}\tilde{E}{}^{\underline{A}_2}\wedge\dots\wedge\tilde{E}{}^{\underline{A}_{m+1}}=*(\tilde{E})_{\underline{A}}.
\end{equation}
Taking successive internal products of the previous result, terminating with the $\bar{\mathfrak{p}}{}^{-(m+1)}$-orientation symbol, we thus form the projective analogues of the $\eta$-bases---the $\hat{\tilde{\epsilon}}$-bases:
\begin{equation}
\begin{split}
    \hat{\tilde{\epsilon}}&:=* 1=\tfrac{1}{(m+1)!}\hat{\tilde{\epsilon}}_{\underline{A}{}_1\dots\underline{A}{}_{m+1}}\tilde{E}{}^{\underline{A}{}_1}\wedge\dots\wedge\tilde{E}{}^{\underline{A}_{m+1}},\\
    \hat{\tilde{\epsilon}}_{\underline{A}{}_1}&:=*(\tilde{E})_{\underline{A}_1}=\tfrac{1}{m!}\hat{\tilde{\epsilon}}_{\underline{A}{}_1\dots\underline{A}{}_{m+1}}\tilde{E}{}^{\underline{A}{}_2}\wedge\dots\wedge\tilde{E}{}^{\underline{A}_{m+1}},\\
    &\vdots\\
    \hat{\tilde{\epsilon}}_{\underline{A}{}_1\dots\underline{A}{}_m}&:=*(\tilde{E}\wedge\dots\wedge\tilde{E})_{\underline{A}{}_1\dots\underline{A}{}_m}=\hat{\tilde{\epsilon}}_{\underline{A}{}_1\dots\underline{A}{}_{m+1}}\tilde{E}{}^{\underline{A}{}_{m+1}},\\
    \hat{\tilde{\epsilon}}_{\underline{A}{}_1\dots\underline{A}{}_{m+1}}&:=*(\tilde{E}\wedge\dots\wedge\tilde{E})_{\underline{A}{}_1\dots\underline{A}{}_{m+1}}.
    \end{split}
\end{equation}
Acting on the $\hat{\tilde{\epsilon}}$-bases with the gauge-covariant derivative provides, in total,
\begin{equation}
    \begin{split}
    \tilde{D}\hat{\tilde{\epsilon}}&=0,\\
        \tilde{D}\hat{\tilde{\epsilon}}_{\underline{A}_1}&=-(m+1)\tilde{g}\wedge\hat{\tilde{\epsilon}}_{\underline{A}_1}+\tilde{\mathcal{T}}{}^{\underline{B}}\wedge\hat{\tilde{\epsilon}}_{\underline{A}_1\underline{B}},\\
        \tilde{D}\hat{\tilde{\epsilon}}_{\underline{A}_1\underline{A}_2}&=-(m+1)\tilde{g}\wedge\hat{\tilde{\epsilon}}_{\underline{A}_1\underline{A}_2}+\tilde{\mathcal{T}}{}^{\underline{B}}\wedge\hat{\tilde{\epsilon}}_{\underline{A}_1\underline{A}_2\underline{B}},\\
         &\vdots\\
         \tilde{D}\hat{\tilde{\epsilon}}_{\underline{A}_1\dots\underline{A}_{m}}&=-(m+1)\tilde{g}\hat{\tilde{\epsilon}}_{\underline{A}_1\dots\underline{A}_{m+1}}+\tilde{\mathcal{T}}{}^{\underline{B}}\hat{\tilde{\epsilon}}_{\underline{A}_1\dots\underline{A}_m\underline{B}},\\
         \tilde{D}\hat{\tilde{\epsilon}}_{\underline{A}_1\dots\underline{A}_{m+1}}&=-(m+1)\tilde{g}\hat{\tilde{\epsilon}}_{\underline{A}_1\dots\underline{A}_{m+1}}.
    \end{split}
\end{equation}
These results again correspond exactly to the analogous non-projective $\eta$-basis relations in \cite{Hehl}.

\subsection{\texorpdfstring{$\bar{\mathfrak{p}}$}{\bar{\mathfrak{p}}}-Non-Metricity}
\label{subsec:nl-non-metricity}

In this section, we seek to isolate the (anti)-symmetric parts of the nonlinear connection $\tilde{\Omega}$. By lowering (raising) an index with the $\bar{\mathfrak{p}}{}^{-2(+2)}$-metric $\tilde{\eta}$, a projective weight of $\mathfrak{w}=-2(+2)$ will be introduced. The $\bar{\mathfrak{p}}{}^{-2}$-connection with lowered indices is
\begin{equation}
        \tilde{\Omega}_{\underline{AB}}:=\tilde{\eta}_{\underline{AC}}\tilde{\Omega}{}^{\underline{C}}{}_{\underline{B}}=\bar{\mathfrak{p}}^{-2}\tilde{\omega}_{\underline{AB}},
\end{equation}
where, following our convention,
\begin{equation}
\tilde{\omega}_{\underline{AB}}:=\begin{pmatrix}\overline{\omega}_{\underline{ab}}&\frac{1}{x^*_0}\eta_{\underline{ab}}\overline{\vartheta}{}^{\underline{b}}\\x^*_0\eta_0\overline{\mathcal{P}}_{\underline{b}}&0
    \end{pmatrix}.
\end{equation}
Define the antisymmetric parts of the $\bar{\mathfrak{p}}{}^{-2}$-connection $\tilde{\Omega}$ as
\begin{equation}
 \tilde{\Omega}{}^{-}_{\underline{AB}}:=\tilde{\Omega}_{[\underline{AB}]}=\begin{pmatrix}
\tilde{\Omega}_{[\underline{ab}]}&\tilde{\Omega}_{[\underline{a*}]}\\\tilde{\Omega}_{[\underline{*b}]}&0
    \end{pmatrix},
\end{equation}
where no factor of $1/2$ is present in this definition. Explicitly, the components are computed to have the form
\begin{equation}
    \tilde{\Omega}{}^{-}_{\underline{AB}}=\bar{\mathfrak{p}}^{-2}\begin{pmatrix}
        \hat{\overline{\omega}}_{\underline{ab}}&-x^*_0\eta_0\overline{\mathcal{P}}{}^{-}_{\underline{a}}\\x^*_0\eta_0\overline{\mathcal{P}}{}^{-}_{\underline{b}}&0
    \end{pmatrix},
\end{equation}
where
\begin{equation}\label{p-bar-plus-minus}
    \overline{\mathcal{P}}{}^{\pm}_{\underline{b}}:= \overline{\mathcal{P}}_{\underline{b}}\pm\frac{\eta_0}{(x^*_0)^2}\eta_{\underline{ab}}\overline{\vartheta}{}^{\underline{a}}
\end{equation}
is the nonlinear projective Schouten $1$-form with its zero shifted to $\overline{\mathcal{P}}{}_{\underline{b}}=\mp\frac{\eta_0}{(x^*_0)^2}\eta_{\underline{ab}}\overline{\vartheta}{}^{\underline{a}}$. Essentially, Eq. \eqref{p-bar-plus-minus} represents the connection-forms associated with the generators $\bm{M}{}^{\pm}_{\underline{b}}$ defined in Eq. \eqref{M-generator-def}. The connection-form
\begin{equation}
    \hat{\overline{\omega}}_{\underline{ab}}:=\overline{\omega}_{[\underline{ab}]}
\end{equation}
may be identified with the $m$-dimensional compatible Lorentz connection. Following our convention established in Eq.\eqref{connection-partial-decomposition}, a ``hat" is used to denote the Lorentz connection. Its status as the compatible Lorentz connection follows from its antisymmetry and its local gauge transformation behavior under the Lorentz group. Furthermore, $\hat{\overline{\omega}}$ only contains $\frac{m(m-1)}{2}$ gauge degrees of freedom.

Define the symmetric part of the nonlinear connection as
\begin{equation}
    \tilde{\Omega}{}^{+}_{\underline{AB}}:=\tilde{\Omega}_{(\underline{AB})}=\begin{pmatrix}
\tilde{\Omega}_{(\underline{ab})}&\tilde{\Omega}_{(\underline{a*})}\\\tilde{\Omega}_{(\underline{*b})}&0
    \end{pmatrix},
\end{equation}
where no factor of $1/2$ is present in this definition. Explicitly,
\begin{equation}
    \tilde{\Omega}{}^{+}_{\underline{AB}}=\bar{\mathfrak{p}}^{-2}\begin{pmatrix}
        \overline{Q}{}_{\underline{ab}}&x^*_0\eta_0\overline{\mathcal{P}}{}^{+}_{\underline{a}}\\x^*_0\eta_0\overline{\mathcal{P}}{}^{+}_{\underline{b}}&0
    \end{pmatrix}.
\end{equation}
In the above, we have introduced the nonlinear projective non-metricity,
\begin{equation}\label{barred-Q}
    \overline{Q}{}_{\underline{ab}}:=\overline{\omega}_{(\underline{ab})}.
\end{equation}
This object is naturally traceless, symmetric, and projectively invariant. Therefore, $\overline{Q}_{\underline{ab}}$ is responsible for modulating the shear deformations associated with $\overline{\omega}$. In connection with the Thomas-Whitehead theory, using $\overline{Q}_{\underline{ab}}$ to form the disformation tensor, Eq. \eqref{disformation-general}, leads to the so-called \textit{Palatini field}, $\tilde{C}$, used in \cite{gen-struc,covariant-tw,heavy-lifting}. We note that Eq. \eqref{barred-Q} may be written in the equivalent form
\begin{equation}
    \overline{Q}{}_{\underline{ab}}=-\overline{D}\eta_{\underline{ab}}.
\end{equation}

The conventional relation, here between the $\bar{\mathfrak{p}}{}^{-2}$-connection $\tilde{\Omega}{}^{+}$ and the $\bar{\mathfrak{p}}{}^{-2}$-non-metricity $\tilde{Q}$, may be found as follows. The nonlinear gauge-covariant derivative of the $\bar{\mathfrak{p}}{}^{-2}$-metric provides,
\begin{equation}\label{gen-proj-non-met}
\begin{split}
    \tilde{Q}_{\underline{AB}}:=-\tilde{D}\tilde{\eta}_{\underline{AB}}&=-d\tilde{\eta}_{\underline{AB}}+\tilde{\eta}_{\underline{CB}}\tilde{\Omega}{}^{\underline{C}}{}_{\underline{A}}+\tilde{\eta}_{\underline{AC}}\tilde{\Omega}{}^{\underline{C}}{}_{\underline{B}}\\
    &= -\tilde{\eta}_{\underline{AB}}d\log\left(\bar{\mathfrak{p}}^{-2}\right)+\tilde{\Omega}{}^{+}_{\underline{AB}}\\
    &= 2\tilde{g}\tilde{\eta}_{\underline{AB}}+\tilde{\Omega}{}^{+}_{\underline{AB}}.
    \end{split}
\end{equation}
Therefore, the symmetric part of the connection is contained in the $\bar{\mathfrak{p}}{}^{-2}$-non-metricity in the standard manner. We may thus identify $\tilde{g}$ as the $\bar{\mathfrak{p}}{}^0$-Weyl $1$-form. It is interesting to note the departure from the commonly encountered relation in gauge gravitational theories, where the Weyl $1$-form is contained within $\tilde{\Omega}{}^{+}$ and only appears upon decomposing $\tilde{Q}$ \cite{Hehl}. This is yet another one of the few main differences between the present work and Thomas-Whitehead gravitational theory put forth in \cite{gen-struc}, since there, the trace of the spin connection is proportional to $\tilde{g}$, see Eq. \eqref{tw-omega}. This occurs as a result of the projective factors being absorbed by the connection, rather than the more natural, projective, approach in which the projective factor appears in the geometric fields on which the connection acts. This appears as the more natural option, since without this factor, the geometric fields do not satisfy the equivalence relation required for their projective geometric description. Moreover, due to the traceless property of $\tilde{\Omega}{}^{+}$, we may further attribute additional shearing defects to $\overline{\mathcal{P}}{}^{+}$. Therefore, when $\overline{\omega}_{\underline{ab}}$ is compatible with $\eta_{\underline{ab}}$, i.e., $\overline{Q}{}_{\underline{ab}}=0$, all of the shear defects result from a non-vanishing $\overline{\mathcal{P}}{}^{+}$.

There are two independent traces of the $\bar{\mathfrak{p}}{}^{-2}$-non-metricity: \textit{natural} and \textit{unnatural}. The natural trace is independent of the connection $\tilde{\Omega}$, since
\begin{equation}
    \tilde{Q}{}^{\underline{B}}{}_{\underline{B}}:=\tilde{\eta}^{\underline{AB}}\tilde{Q}_{\underline{AB}}=2(m+1)\tilde{g}.
\end{equation}
The above relation confirms $\tilde{g}$ as the $\bar{\mathfrak{p}}{}^0$-Weyl $1$-form responsible for the rescaling effects or dilations of length under parallel transport. The unnatural trace requires first connecting to $V\mathcal{M}$,
\begin{equation}
    \tilde{Q}_{M\underline{AB}}:=\tilde{Q}_{M\underline{AB}}dx^M,
\end{equation}
where we deviate from our convention by placing the form index closest to the base character $(Q)$. Taking the unnatural trace, for example, over $M$ and $\underline{A}$, we find that
\begin{equation}\label{unnatural-trace-Q-tilde}
\begin{split}
    \tilde{Q}_{\underline{A}}{}^{\underline{A}}{}_{\underline{B}}&:=\tilde{\eta}^{\underline{CA}}(\tilde{E}{}^{-1})^M{}_{\underline{C}}\tilde{Q}_{M\underline{AB}}\\&=(\tilde{E}{}^{-1})^M{}_{\underline{C}}\tilde{\eta}^{\underline{CA}}\left(2\tilde{g}_M\tilde{\eta}_{\underline{AB}}+\tilde{\Omega}{}^{+}_{\underline{AB}M}\right)\\
    &=(\tilde{E}{}^{-1})^M{}_{\underline{C}}\left(2\tilde{g}_M\delta^{\underline{C}}{}_{\underline{B}}+(\tilde{\Omega}{}^{+})^{\underline{C}}{}_{\underline{B}M}\right)\\
    &=2\tilde{g}_M(\tilde{E}{}^{-1})^M{}_{\underline{B}}+(\tilde{\Omega}{}^{+})^{\underline{C}}{}_{\underline{B}M}(\tilde{E}{}^{-1})^M{}_{\underline{C}}\\
    &\overset{\circ}{=}\bar{\mathfrak{p}}^{-1}\begin{pmatrix}
       \overline{Q}{}_{\underline{b}},&\frac{2}{x^*_0}+x^*_0\eta_0\overline{\mathcal{P}}{}^{+}
    \end{pmatrix},
    \end{split}
\end{equation}
since $\tilde{g}_M(\tilde{E}{}^{-1})^M{}_{\underline{b}}\overset{\circ}{=}0$. In the above expression, we encounter the unnatural trace of the $m$-dimensional nonlinear non-metricity of Eq. \eqref{barred-Q},
\begin{equation}
      \overline{Q}{}_{\underline{b}}:=\overline{Q}{}_{m\underline{ab}}\eta^{\underline{ac}}(\overline{\vartheta}{}^{-1})^m{}_{\underline{c}},
\end{equation}
and the only trace of the nonlinear (symmetric) projective Schouten \textit{tensor},
\begin{equation}
    \overline{\mathcal{P}}{}^{+}:=\overline{\mathcal{P}}{}^{+}_{\underline{a}m}\eta^{\underline{ac}}(\overline{\vartheta}{}^{-1})^m{}_{\underline{c}}.
\end{equation}
The resulting expression in Eq. \eqref{unnatural-trace-Q-tilde} will play an important role when discussing projective matter fields. We thus leave extended discussion of Eq. \eqref{unnatural-trace-Q-tilde} for Secs. \ref{sec:spinor-action} and \ref{sec:tw-gen-proj}, and simply comment that the dynamical aspects of this object have been considered in the context of Metric-Affine geometries, Lorentz violation, and black holes \cite{torsion-matter2,second-clock,observable-nonmetricity}.

\subsection{Generalized \texorpdfstring{$\bar{\mathfrak{p}}$}{\bar{\mathfrak{p}}}-Higgs Co-Vector}
\label{subsec:gen-higgs-cv}

The $SL(m+1,\mathbb{R})\cong PGL(m,\mathbb{R})$ gauge group permits, in addition to the generalized projective Higgs vector field $\tilde{\Upsilon}$, a generalized projective Higgs co-vector field. An $\mathfrak{sl}(m+1,\mathbb{R})\cong\mathfrak{pgl}(m,\mathbb{R})$-valued co-vector $A$ is required to transform under the local gauge action of $S\in SL(m+1,\mathbb{R})\cong PGL(m,\mathbb{R})$ as
\begin{equation}\label{A-trans}
    S(x):\; A_B\rightarrow A_{B'}=A_C(S^{-1}){}^C{}_{B'},
\end{equation}
with $(S^{-1}){}^A{}_{B'}$ given in Eq.~\eqref{SL-matrix-inv}. In order to explicitly define the components of $A$, we form the metric-independent, invariant pair with unit norm,
\begin{equation}\label{AY1}
    A\cdot Y=1.
\end{equation}
Had we considered $q\neq1$ following Eq. \eqref{gen-higgs-vec-gen-def}, the above expression would simply read $A\cdot Y=q$. From Eqs. \eqref{AY1} and \eqref{gen-higgs-vec-gen-def}, we may easily calculate the components of $A$, for which we find
\begin{equation}
    A_B=\mathfrak{p}^{-1}\begin{pmatrix}
        -\alpha_b,&\frac{1}{x^*_0}
    \end{pmatrix}.
\end{equation}
According to Eq. \eqref{A-trans}, the components of $A$ are found to transform as
\begin{equation}
   A_{B'}=\mathfrak{p}^{-1}|s|^{\frac{1}{m+1}}\begin{pmatrix}
       -\alpha_{b'},&\frac{1}{x^*_0}(1+\alpha'\cdot t)
   \end{pmatrix},
\end{equation}
with
\begin{equation}
    \alpha_{b'}=(\alpha_c-u_c)(s^{-1}){}^c{}_{b'}.
\end{equation}
Due to our earlier minimal assumption $\alpha\cdot\xi=0$, we further assume $\alpha\cdot t=0$. This provides the orthogonality relation
\begin{equation}
   \alpha'\cdot t=(\alpha-u)\cdot s^{-1}\cdot t=0.
\end{equation}
Under local gauge transformations, the projective Higgs co-vector thus behaves as
\begin{equation}
   A_{B'}=\mathfrak{p}^{-1}|s|^{\frac{1}{m+1}}\begin{pmatrix}
       -(\alpha_c-u_c)(s^{-1}){}^c{}_{b'},&\frac{1}{x^*_0}
   \end{pmatrix}.
\end{equation}
 
With the right action of $\sigma$, we form the nonlinear projective Higgs co-vector $\tilde{\Theta}$ from $A$ as
\begin{equation}
    \tilde{\Theta}:=A\sigma.
\end{equation}
This object transforms linearly as a projective Lorentz co-vector under local gauge transformations $S(x)$, since
\begin{equation}
\begin{split}
    S(x):\;\tilde{\Theta}\rightarrow\tilde{\Theta}'&=(A\sigma)'\\
    &=(AS^{-1})(S\sigma\Lambda^{-1})\\
    &=A\sigma\Lambda^{-1}\\
    &=\tilde{\Theta}\Lambda^{-1}.
    \end{split}
\end{equation}
The nonlinear projective Higgs co-vector has the explicit form
\begin{equation}
\tilde{\Theta}_{\underline{A}}=A_B\sigma^B{}_{\underline{A}}=\bar{\mathfrak{p}}^{-1}\begin{pmatrix}
      -g_br^b{}_{\underline{a}},&\frac{1}{x^*_0}
    \end{pmatrix},
\end{equation}
since
\begin{equation}\label{little-g-underlined}
    g_br^b{}_{\underline{a}}=g_m(\overline{\vartheta}{}^{-1})^m{}_{\underline{a}}=(\alpha_b+\change_b)r^b{}_{\underline{a}}.
\end{equation}
Following our convention, we further define the $\bar{\mathfrak{p}}{}^0$-part, $\tilde{\theta}_{\underline{A}}$, via
\begin{equation}
    \tilde{\Theta}_{\underline{A}}:=\bar{\mathfrak{p}}{}^{-1}\tilde{\theta}_{\underline{A}}.
\end{equation}

Analogous to the APV-gauge defined in Eq. \eqref{upsilon-APV}, we define the \textit{almost-physical co-vector-gauge} (APC-gauge) as the choice of gauge wherein $\tilde{\theta}_{\underline{a}}=0$. This choice leads to $\alpha=-\change$, or equivalently,
\begin{equation}
    g_br^b{}_{\underline{a}}=0.
\end{equation}
Expressions that hold true in the APC-gauge will be denoted with a lower circle, 
\begin{equation}
    \tilde{\Theta}_{\underline{A}}\underset{\circ}{=}\bar{\mathfrak{p}}^{-1}\begin{pmatrix}
    0_{\underline{a}},&\frac{1}{x^*_0}
    \end{pmatrix},
\end{equation}
where $ 0_{\underline{a}}:=(0,\;0,\dots,0)$ is the $m$-dimensional zero-co-vector.

\subsubsection{First Derivative}
\label{subsubsec:gen-higgs-cv-1d}

The nonlinear projective gauge-covariant derivative acts on the generalized projective Higgs co-vector $\tilde{\Theta}$ as
\begin{equation}\label{first-cov-der-higg-covec}
    \tilde{D}\tilde{\Theta}_{\underline{B}}=d\tilde{\Theta}_{\underline{B}}-\tilde{\Theta}_{\underline{A}}\tilde{\Omega}{}^{\underline{A}}{}_{\underline{B}}.
\end{equation}
The components of this expression are easily calculated. The first $m(m+1)$ components are found to have the form
\begin{equation}
\begin{split}
\tilde{D}\tilde{\Theta}_{\underline{b}}&=d\tilde{\Theta}_{\underline{b}}-\tilde{\Theta}_{\underline{a}}\tilde{\Omega}{}^{\underline{a}}{}_{\underline{b}}-\tilde{\Theta}_{\underline{*}}\tilde{\Omega}{}^{\underline{*}}{}_{\underline{b}}\\
&=-\bar{\mathfrak{p}}^{-1}(\overline{\mathcal{P}}_{\underline{b}}+dg_{\underline{b}}-g_{\underline{a}}\overline{\omega}{}^{\underline{a}}{}_{\underline{b}}-g_{\underline{b}}\tilde{g})\\
    &=-\bar{\mathfrak{p}}^{-1}\begin{pmatrix}\mathcal{D}_{\underline{b}},&\frac{-1}{x^*}g_{\underline{b}}\end{pmatrix}\\
    &\underset{\circ}{=}-\bar{\mathfrak{p}}^{-1}\begin{pmatrix}\mathcal{D}_{\underline{b}},&0\end{pmatrix},
    \end{split}
\end{equation}
where we have defined
\begin{equation}
\mathcal{D}_{\underline{b}}:=\overline{\mathcal{P}}_{\underline{b}}+dg_{\underline{b}}-g_{\underline{a}}\overline{\omega}{}^{\underline{a}}{}_{\underline{b}}-g_{\underline{b}}g
\end{equation}
as the Diffeomorphism $1$-form, and use the short-hand notation $g_{\underline{b}}:=g_br^b{}_{\underline{a}}$. Compare with Eqs. \eqref{diff-schout-tw-spin} and \eqref{diff-field-gen-proj}. Additionally, in the APC-gauge, we have the equality
\begin{equation}
    \mathcal{D}_{\underline{b}}\underset{\circ}{=}\overline{\mathcal{P}}_{\underline{b}}.
\end{equation}
The remaining $m+1$ components of Eq. \eqref{first-cov-der-higg-covec} are found similarly,
\begin{equation}
\begin{split}
\tilde{D}\tilde{\Theta}_{\underline{*}}&=d\tilde{\Theta}_{\underline{*}}-\tilde{\Theta}_{\underline{a}}\tilde{\Omega}{}^{\underline{a}}{}_{\underline{*}}\\
    &=\tfrac{-1}{x^*_0}\bar{\mathfrak{p}}^{-1}(\tilde{g}-g_{\underline{a}}\overline{\vartheta}{}^{\underline{a}})\\
    &=\tfrac{-1}{x^*_0}\bar{\mathfrak{p}}^{-1}\begin{pmatrix}
        0,&\frac{1}{x^*}
    \end{pmatrix}.
    \end{split}
\end{equation}

Analogous to the projective $2$-frames constructed from $\tilde{\Upsilon}$, Eq. \eqref{big-E-def}, we define the dimensionless $\bar{\mathfrak{p}}{}^{-1}$-co-vector-valued $1$-form via
\begin{equation}
    \tilde{\mathcal{G}}_{\underline{B}M}:=(x^*_0)^2\tilde{D}_M\tilde{\Theta}_{\underline{B}}.
\end{equation}
The components of $\tilde{\mathcal{G}}_{\underline{B}M}$ are
\begin{equation}
    \tilde{\mathcal{G}}_{\underline{B}M}=\begin{pmatrix}
        \tilde{\mathcal{G}}_{\underline{b}m}&\tilde{\mathcal{G}}_{\underline{b}*}\\\tilde{\mathcal{G}}_{\underline{*}m}&\tilde{\mathcal{G}}_{\underline{*}*}
    \end{pmatrix}=\bar{\mathfrak{p}}^{-1}\begin{pmatrix}
      -( x^*_0)^2\mathcal{D}_{\underline{b}m}&\frac{(x^*_0)^2}{x^*}g_{\underline{b}}\\0&-\frac{x^*_0}{x^*}
    \end{pmatrix}.
\end{equation}
Although $\tilde{\mathcal{G}}_{M\underline{B}}$ is presented as a matrix for visual purposes, one must bear in mind it is rather a Lie algebra co-vector-valued $1$-form. In the APC-gauge, it simplifies to
\begin{equation}
    \tilde{\mathcal{G}}_{\underline{B}M}\underset{\circ}{=}\bar{\mathfrak{p}}^{-1}\begin{pmatrix}
       -(x^*_0)^2\overline{\mathcal{P}}_{\underline{b}m}&0\\0&-\left(\frac{x^*_0}{x^*}\right)^2
    \end{pmatrix}.
\end{equation}

It appears that one may construct a dimensionless, dynamical $\bar{\mathfrak{p}}{}^0$-``metric" $\tilde{\mathcal{G}}_{MN}$ from both the $\bar{\mathfrak{p}}$- and $\bar{\mathfrak{p}}{}^{-1}$-Higgs fields in a manner similar to what is done in bi-metric gravitational theories \cite{bi-frame},
\begin{equation}
    \tilde{\mathcal{G}}_{MN}:=\tilde{E}{}^{\underline{A}}{}_{M}\tilde{\mathcal{G}}_{\underline{A}N}.
\end{equation}
Interestingly, this expression does not require the existence of a metric structure, but rather only a projective structure. In the almost physical, AP-gauge (APV$+$APC), we find
\begin{equation}
    \tilde{\mathcal{G}}_{MN}\underset{\circ}{\overset{\circ}{=}}\begin{pmatrix}
        -(x^*_0)^2\overline{\mathcal{P}}_{mn}&0\\0&-\left(\frac{x^*_0}{x^*}\right)^2
    \end{pmatrix},
\end{equation}
where $\overline{\mathcal{P}}_{nm}\equiv\overline{\mathcal{P}}_{\underline{b}m}\overline{\vartheta}{}^{\underline{b}}{}_n$. This Higgs metric $\tilde{\mathcal{G}}$ is naturally symmetric, however, the consideration of $\tilde{\mathcal{G}}$ as a metric also requires the inverse to exist. For this to occur, a dynamical theory of $\overline{\mathcal{P}}_{mn}$ must be constructed, wherein the degenerate solution, $\overline{\mathcal{P}}_{mn}=0$ is not permitted. A more natural interpretation is not that $\overline{\mathcal{P}}_{mn}$ \textit{is} the spacetime metric, but rather $\overline{\mathcal{P}}_{mn}$ \textit{provides} the spacetime metric. This subtle distinction relieves one of then considering only those models for which $\overline{\mathcal{P}}_{mn}=0$ is not a solution. Recalling Sec. \ref{subsec:coord-map} of the introduction, when $\mathcal{E}$ maps vectors in $V\mathcal{M}$ to $\mathfrak{p}$-vectors in $P\mathcal{M}$, then $\mathcal{P}_{mn}= e^a{}_me^b{}_n(\varphi\partial_a\partial_b\varphi^{-1})$. The factor in parenthesis may be identified as a Hessian metric. In particular, it takes the form of a centro-affine metric describing the geometry of $m$-dimensional hypersurfaces invariant under $GL(m+1,\mathbb{R})$ \cite{hessian-metric-1}. This projective Schouten tensor is also used in the study of statistical manifolds \cite{hessian-metric-stat-man}. We will therefore often consider the ``pure-trace" projective Schouten form
\begin{equation}
    \overline{\mathcal{P}}_{\underline{b}}=-\chi^2\eta_{\underline{ab}}\overline{\vartheta}{}^{\underline{a}},
\end{equation}
with $\chi=\chi(\varphi(x))$.

For completeness, we display the algebra-valued $\bar{\mathfrak{p}}{}^{-2}$-Higgs metric, defined as
\begin{equation}
\tilde{\mathcal{G}}_{\underline{AB}}:=\tilde{\mathcal{G}}_{\underline{A}N}(\tilde{E}{}^{-1})^N{}_{\underline{B}}.
\end{equation}
The components of which are
\begin{equation}
\tilde{\mathcal{G}}_{\underline{AB}}\overset{\circ}{=}\bar{\mathfrak{p}}{}^{-2}\begin{pmatrix}
        -(x^*_0)^2\mathcal{D}_{\underline{ab}}-(x^*_0)^2g_{\underline{a}}g_{\underline{b}}&x^*_0g_{\underline{b}}\\x^*_0g_{\underline{a}}&-1
    \end{pmatrix},
\end{equation}
where $\mathcal{D}_{\underline{ab}}\equiv \mathcal{D}_{\underline{a}m}(\overline{\vartheta}{}^{-1})^m{}_{\underline{b}}$.

The utilization of $\overline{\mathcal{P}}_{mn}$ as a metric is not a new idea. Others have explicitly done so for both projective and conformal structures \cite{diff-metric,p-bar-metric,p-bar-metric-2}. This idea has even been investigated for the closely related \textit{conformal} Schouten tensor \cite{p-bar-metric-so5}. There, the so-called Deser-Van Nieuwenhuizen condition is that the Schouten tensor is symmetric in the basis provided by the co-frame gauge connection-form, which in turn, is shown to ensure a well-defined inverse. Therefore, there is certainly reason to believe that the present $\overline{\mathcal{P}}_{mn}$ may be viewed as providing $\mathcal{M}$ with a metric. However, much work is to be completed before this statement is to be considered in any official capacity.

\subsubsection{Second Derivative}
\label{subsubsec:gen-higgs-cv-2d}

The second derivative of the $\bar{\mathfrak{p}}{}^{-1}$-Higgs co-vector $\tilde{\Theta}$ defines a $\bar{\mathfrak{p}}{}^0$-non-metricity-type object,
\begin{equation}
    \tilde{\mathcal{Q}}_{MNL}:=-(\tilde{D}_M\tilde{D}_N\tilde{\Theta}_{\underline{B}})\tilde{D}_L\tilde{\Upsilon}{}^{\underline{B}}.
\end{equation}
Using the convenient definitions developed thus far, this may be expressed as
\begin{equation}
    \tilde{\mathcal{Q}}_{MNL}=-(\tilde{D}_M\tilde{\mathcal{G}}_{\underline{B}N})\tilde{E}{}^{\underline{B}}{}_{L}.
\end{equation}
The components of $\tilde{\mathcal{Q}}_{MNL}$ are easily calculated in the APV-gauge. Using the template,
\begin{equation}
    \tilde{\mathcal{Q}}_{MNL}=\left\{\begin{pmatrix}
         \tilde{\mathcal{Q}}_{mnl}& \tilde{\mathcal{Q}}_{m*l}\\ \tilde{\mathcal{Q}}_{*nl}& \tilde{\mathcal{Q}}_{**l}
    \end{pmatrix},\;\begin{pmatrix}
         \tilde{\mathcal{Q}}_{mn*}& \tilde{\mathcal{Q}}_{m**}\\ \tilde{\mathcal{Q}}_{*n*}& \tilde{\mathcal{Q}}_{***}
    \end{pmatrix}\right\},
\end{equation}
we find
\begin{equation}
    \tilde{\mathcal{Q}}_{MNL}\overset{\circ}{=}\left\{\begin{pmatrix}
         (x^*_0)^2(\overline{\vartheta}{}^{\underline{b}}{}_l\overline{D}_m\mathcal{D}_{\underline{b}n}-g_{(m}\mathcal{D}_{l)n})&\frac{-(x^*_0)^2}{x^*}\mathcal{D}_{lm}\\ \frac{-(x^*_0)^2}{x^*}\mathcal{D}_{ln}&0
    \end{pmatrix},\;\begin{pmatrix}
         \frac{-(x^*_0)^2}{x^*}\mathcal{D}_{mn}& 0\\ 0&\frac{-2}{x^*}\left(\frac{x^*_0}{x^*}\right)^2
    \end{pmatrix}\right\},
\end{equation}
where $\overline{D}_m\mathcal{D}_{\underline{b}n}:=\partial_m\mathcal{D}_{\underline{b}n}-\mathcal{D}_{\underline{a}n}\overline{\omega}{}^{\underline{a}}{}_{\underline{b}m}$. For brevity, all appearances of $\mathcal{D}$ above contain their antisymmetric parts.

Restricting further to the AP-gauge provides
\begin{equation}
    \tilde{\mathcal{Q}}_{MNL}\underset{\circ}{\overset{\circ}{=}}\left\{\begin{pmatrix}
         (x^*_0)^2\overline{\mathcal{S}}_{lmn}&\frac{-(x^*_0)^2}{x^*}\overline{\mathcal{P}}_{lm}\\ \frac{-(x^*_0)^2}{x^*}\overline{\mathcal{P}}_{ln}&0
    \end{pmatrix},\begin{pmatrix}
         \frac{-(x^*_0)^2}{x^*}\overline{\mathcal{P}}_{mn}& 0\\ 0&\frac{-2}{x^*}\left(\frac{x^*_0}{x^*}\right)^2
    \end{pmatrix}\right\},
\end{equation}
where $\overline{\mathcal{S}}_{lmn}=\overline{\vartheta}{}^{\underline{b}}{}_l\overline{D}_m\overline{\mathcal{P}}_{\underline{b}n}$ contains the field strength of $\overline{\mathcal{P}}$. We use ``contains" due to the inclusion of the symmetric part, $\overline{\mathcal{S}}_{l(mn)}$. We reserve ``field strength" for genuine $2$-forms, i.e., $\overline{\mathcal{S}}_{l[mn]}$. Making this distinction, we find that the only non-vanishing components of the antisymmetric part of $\tilde{\mathcal{Q}}$ are the field strength,
\begin{equation}
    \tilde{\mathcal{Q}}_{[mn]l}\underset{\circ}{\overset{\circ}{=}}(x^*_0)^2\overline{\mathcal{S}}_{l[mn]}.
\end{equation}
Although $\tilde{\mathcal{Q}}$ does not exactly match the form of $\tilde{Q}$ given in Eq. \eqref{gen-proj-non-met}, that certainly does not imply there is no relation to uncover between them. We leave further investigation of $\tilde{\mathcal{Q}}$ and its relationship to the $\bar{\mathfrak{p}}{}^{-2}$-non-metricity for future investigation.

\section{The Dynamical Theory}
\label{sec:dynamical-theory}

In this final section of Part II, we develop the dynamical model of the General Projective Gauge Gravitational Theory. We begin with deriving the field theoretic machinery necessary to describe the dynamics. The field variations are found in exterior form for general actions, including matter, and are constructed in a manner most useful for future canonical quantization. Furthermore, we outline the relationships between the projective setting and the Metric-Affine setting.

There are two action functionals to which we apply the field variation machinery in $m+1=5$ dimensions. The first is a general projective extension of the Pontrjagin density. This is found to contain a projectively invariant version of the ordinary Pontrjagin density, as well as a new, \textit{metric-independent} and projectively invariant topological term. The latter is shown to be related to the Nieh-Yan form once a conformal relation is imposed between the (pseudo)-translational connection-forms. From the general expression, the associated Chern-Simons forms and the generalized projective Bianchi identities are derived, and some interesting properties discussed for both. We further show how these are related to the Metric-Affine counterparts.

The second action functional discussed is a general projective Lovelock theory, the definition of which requires the $\bar{\mathfrak{p}}$-Higgs vector. We discuss the form of the action in depth. In particular, we show the abstract-index form, and discuss its relationship with the Thomas-Whitehead theory. Additionally, we show how the general projective Lovelock model is related to the Macdowell-Mansouri and Stelle-West models, as well as what conditions must be imposed to arrive at a projective Lovelock-Chern-Simons theory. In finding the field equations, we follow two modes of reasoning: dynamical and non-dynamical $\tilde{\Upsilon}$. In the dynamical case, we show that the APV-gauge choice may be obtained as a solution of the field equations, which then exactly reduces the entire set to the non-dynamical case. We find non-vanishing torsion solutions that appear to yield vanishing volumes.

\subsection{Field Variations}
\label{subsec:field-variations}

We wish to investigate dynamical models describing both the General Projective Gauge Theory of Gravity and Matter. In general, the actions considered herein are functionals of the form
\begin{equation}
    S=\int_{V\mathcal{M}}\mathcal{L}(\tilde{\mathcal{K}},\tilde{\Upsilon},\tilde{D}\tilde{\Upsilon},\tilde{\Psi},\tilde{D}\tilde{\Psi}),
\end{equation}
where $\mathcal{L}$ is the $(m+1)$-form Lagrangian density. The arguments of $\mathcal{L}$ are fixed by the requirement that $\mathcal{L}$ is invariant under local gauge transformations, since it may then only depend on $\tilde{\Omega}$ via the gauge-covariant derivative $\tilde{D}$ \cite{Hehl}. The matter fields $\tilde{\Psi}$ will be treated generally as $p$-forms for the purpose of general variation. In the particular case studied in Part III of this document, where the matter fields are taken to be projective spinors, $p=0$. It is conceivable to allow for dependence on higher derivatives, such as $\tilde{D}\tilde{D}\tilde{\Upsilon}$. However, any power of exterior gauge-covariant derivatives of $\tilde{\Upsilon}$ are reducible in terms of $\tilde{\mathcal{K}}$ and $\tilde{E}$, and are therefore, not considered independent. Such higher-derivative contributions were considered by Lovelock in \cite{lovelock-1}. We do not use a tilde on the symbols representing the action $S$ or the Lagrangian density $\mathcal{L}$ to signify their independence of the nonlinear realization process, which follows from the requirement that $\mathcal{L}$ be of type-$\bar{\mathfrak{p}}{}^0$. Furthermore, we assume the total action separates into two sectors, $S=S_G+S_M$, with $S_G$ the gravitational sector and $S_M$ the matter sector. It then follows that each sector is composed from their respective $(m+1)$-form Lagrangian densities, which we take to have the functional dependencies:
\begin{subequations}
 \begin{equation}
\mathcal{L}_G=\mathcal{L}_G(\tilde{\mathcal{K}},\tilde{\Upsilon},\tilde{D}\tilde{\Upsilon}),
 \end{equation}
 \begin{equation}
\mathcal{L}_M=\mathcal{L}_M(\tilde{\Upsilon},\tilde{D}\tilde{\Upsilon},\tilde{\Psi},\tilde{D}\tilde{\Psi}).
 \end{equation}
 \end{subequations}
 
 We stick to the convention of pulling all variations to the left, as was done in Sec. \ref{sec:exterior-variation-MAG} in the context of the Metric-Affine gauge framework. The total functional variation $\delta$ of the composite action $S$ is then
\begin{equation}
\begin{split}
    \delta S&=\int_{V\mathcal{M}}\delta\tilde{\mathcal{K}}{}^{\underline{A}}{}_{\underline{B}}\wedge\frac{\partial\mathcal{L}}{\partial\tilde{\mathcal{K}}{}^{\underline{A}}{}_{\underline{B}}}\\
    &\quad\quad\quad+\delta\tilde{\Upsilon}{}^{\underline{A}}\wedge\frac{\partial\mathcal{L}}{\partial\tilde{\Upsilon}{}^{\underline{A}}}+\delta(\tilde{D}\tilde{\Upsilon}{}^{\underline{A}})\wedge\frac{\partial\mathcal{L}}{\partial(\tilde{D}\tilde{\Upsilon}{}^{\underline{A}})}\\
    &\quad\quad\quad+\delta\tilde{\Psi}\wedge\frac{\partial\mathcal{L}}{\partial\tilde{\Psi}}+\delta(\tilde{D}\tilde{\Psi})\wedge\frac{\partial\mathcal{L}}{\partial(\tilde{D}\tilde{\Psi})}.
    \end{split}
\end{equation}
We consider the connection $\tilde{\Omega}$ as the fundamental field variable and, therefore, write the curvature variation in terms of the variation of the fundamental field variable as
\begin{equation}
    \begin{split}
        \delta\tilde{\mathcal{K}}{}^{\underline{A}}{}_{\underline{B}}\wedge\frac{\partial\mathcal{L}}{\partial\tilde{\mathcal{K}}{}^{\underline{A}}{}_{\underline{B}}}
        &=\delta\tilde{\Omega}{}^{\underline{A}}{}_{\underline{B}}\wedge \tilde{D}\frac{\partial\mathcal{L}}{\partial\tilde{\mathcal{K}}{}^{\underline{A}}{}_{\underline{B}}},
    \end{split}
\end{equation}
plus the omitted boundary term $d(\delta\tilde{\Omega}{}^{\underline{A}}{}_{\underline{B}}\wedge\frac{\partial\mathcal{L}}{\partial\tilde{\mathcal{K}}{}^{\underline{A}}{}_{\underline{B}}})$. We then write the total functional variation of the composite action concisely as
\begin{equation}\label{proj-field-variation-gen}
    \delta\mathcal{S}=\int_{V\mathcal{M}} \delta\tilde{\Omega}{}^{\underline{A}}{}_{\underline{B}}\wedge\tilde{\mathbb{F}}{}^{\underline{B}}{}_{\underline{A}}+\delta\Psi\wedge\tilde{\mathbb{F}}+\delta\tilde{\Upsilon}{}^{\underline{A}}\wedge\tilde{\mathbb{F}}_{\underline{A}}+\int_{\partial V\mathcal{M}}d\mathbb{B},
\end{equation}
where $\mathbb{B}$ denotes the collection of boundary terms
\begin{equation}
\begin{split}
    \mathbb{B}&=\delta\tilde{\Omega}{}^{\underline{A}}{}_{\underline{B}}\wedge\frac{\partial\mathcal{L}}{\partial\tilde{\mathcal{K}}{}^{\underline{A}}{}_{\underline{B}}}+\delta\tilde{\Upsilon}{}^{\underline{A}}\wedge\frac{\partial\mathcal{L}}{\partial(\tilde{D}\tilde{\Upsilon}{}^{\underline{A}})}+\delta\Psi\wedge\frac{\partial\mathcal{L}}{\partial(\tilde{D}\Psi)}.
    \end{split}
\end{equation}
For manifolds with boundary, these terms should be retained. Variation of $\mathbb{B}$, restricted to the space of compact solutions, allows one to probe the symplectic structure \cite{instanton-angle}. Investigating specific implementations of $\mathbb{B}$ in the general projective setting may thus prove fruitful in understanding the role played by the antisymmetric part of the projective Schouten form \cite{p-bar-metric}. For simplicity, we omit the boundary contributions and leave all investigation of $\mathbb{B}$ for future work. 

The total variation in Eq. \eqref{proj-field-variation-gen} provides the complete set of field equations:
\begin{equation}
\begin{split}
    \tilde{\mathbb{F}}{}^{\underline{B}}{}_{\underline{A}}&:=\tilde{\mathbb{G}}{}^{\underline{B}}{}_{\underline{A}}+\tilde{\mathbb{M}}{}^{\underline{B}}{}_{\underline{A}}=0,\\
\tilde{\mathbb{F}}_{\underline{A}}&:=\tilde{\mathbb{G}}_{\underline{A}}+\tilde{\mathbb{M}}_{\underline{A}}=0,\\
\tilde{\mathbb{F}}&=\tilde{\mathbb{M}}:=\frac{\delta\mathcal{L}_M}{\delta\tilde{\Psi}}=\frac{\partial\mathcal{L}_M}{\partial\Psi}-(-1)^{p} \tilde{D}(\frac{\partial\mathcal{L}_M}{\partial(\tilde{D}\Psi)})=0,
    \end{split}
\end{equation}
where we have split contributions from the gravitational, $\tilde{\mathbb{G}}$, and matter, $\tilde{\mathbb{M}}$, sectors. The gravitational contributions to the connection and $\bar{\mathfrak{p}}$-Higgs vector field equations are, respectively,
\begin{equation}
    \tilde{\mathbb{G}}{}^{\underline{B}}{}_{\underline{A}}:=\frac{\delta\mathcal{L}_G}{\delta\tilde{\Omega}{}^{\underline{A}}{}_{\underline{B}}}=-\tilde{D}\tilde{\mathbb{H}}{}^{\underline{B}}{}_{\underline{A}}+\tilde{\Upsilon}{}^{\underline{B}}\tilde{\mathbb{X}}_{\underline{A}},\quad\quad\quad\tilde{\mathbb{G}}_{\underline{A}}:=\frac{\delta\mathcal{L}_G}{\delta\tilde{\Upsilon}{}^{\underline{A}}}=-\tilde{D}\tilde{\mathbb{X}}_{\underline{A}}+\tilde{\mathbb{Y}}_{\underline{A}}.
\end{equation}
In these expressions, we introduce the general projective gravitational gauge field momenta or excitations, defined as
\begin{equation}
    \tilde{\mathbb{H}}{}^{\underline{B}}{}_{\underline{A}}:=-\frac{\partial\mathcal{L}_G}{\partial\tilde{\mathcal{K}}{}^{\underline{A}}{}_{\underline{B}}},
\end{equation}
and the $\bar{\mathfrak{p}}$-Higgs vector field momentum and potential as
\begin{equation}
\tilde{\mathbb{X}}_{\underline{A}}:=-\frac{\partial\mathcal{L}_G}{\partial(\tilde{D}\tilde{\Upsilon}{}^{\underline{A}})},\quad\quad\quad \tilde{\mathbb{Y}}_{\underline{A}}:=\frac{\partial\mathcal{L}_G}{\partial\tilde{\Upsilon}{}^{\underline{A}}}.
\end{equation}
In the matter sector $\tilde{\mathbb{M}}$, we find the total general projective matter currents
\begin{equation}\label{proj-matter-currents}
    \tilde{\mathbb{M}}{}^{\underline{B}}{}_{\underline{A}}:=\frac{\delta\mathcal{L}_M}{\delta\tilde{\Omega}{}^{\underline{A}}{}_{\underline{B}}}=-\tilde{D} \tilde{\mathbb{I}}{}^{\underline{B}}{}_{\underline{A}}+\tilde{\Upsilon}{}^{\underline{B}}\tilde{\mathbb{U}}_{\underline{A}}+\tilde{\mathbb{J}}{}^{\underline{B}}{}_{\underline{A}},
\end{equation}
and the $\bar{\mathfrak{p}}$-Higgs vector field contributions to the matter sector,
\begin{equation}    \tilde{\mathbb{M}}_{\underline{A}}:=\frac{\delta\mathcal{L}_M}{\delta\tilde{\Upsilon}{}^{\underline{A}}}=-\tilde{D}\tilde{\mathbb{U}}_{\underline{A}}+\tilde{\mathbb{V}}_{\underline{A}}.
\end{equation}
In these expressions, we introduce the definitions
\begin{equation}
    \tilde{\mathbb{I}}{}^{\underline{B}}{}_{\underline{A}}:=-\frac{\partial\mathcal{L}_M}{\partial\tilde{\mathcal{K}}{}^{\underline{A}}{}_{\underline{B}}},\quad\quad\quad\tilde{\mathbb{V}}_{\underline{A}}:=\frac{\partial\mathcal{L}_M}{\partial\tilde{\Upsilon}{}^{\underline{A}}},\quad  \quad\quad\tilde{\mathbb{U}}_{\underline{A}}:=-\frac{\partial\mathcal{L}_M}{\partial(\tilde{D}\tilde{\Upsilon}{}^{\underline{A}})},
\end{equation}
as well as the projective extensions of the ``ordinary" matter currents
\begin{equation}\label{proj-ext-ordinary-matter-current}
    \tilde{\mathbb{J}}{}^{\underline{B}}{}_{\underline{A}}:=\rho(\bm{L}{}^{\underline{B}}{}_{\underline{A}})\Psi\wedge\frac{\partial\mathcal{L}_M}{\partial(\tilde{D}\Psi)}.
\end{equation}
Since we will be omitting the boundary term $\mathbb{B}$, the actions considered will not depend explicitly on $\tilde{\Upsilon}$. The models considered herein thus have $\tilde{\mathbb{V}}_{\underline{A}}=\tilde{\mathbb{Y}}_{\underline{A}}=0$, providing a total gauge-covariant derivative field equation to be satisfied by $\tilde{\Upsilon}$, 
\begin{equation}
    \tilde{\mathbb{F}}_{\underline{A}}=-\tilde{D}(\tilde{\mathbb{X}}_{\underline{A}}+\tilde{\mathbb{U}}_{\underline{A}}).
\end{equation}
The total projective matter currents $\tilde{\mathbb{M}}{}^{\underline{B}}{}_{\underline{A}}$ are easily seen to contain the projective analogues of the standard MAG matter currents defined in Eqs. \eqref{matter-currents-MAG},
\begin{equation}
    \tilde{\mathbb{M}}{}^{\underline{B}}{}_{\underline{A}}=\begin{pmatrix}\frac{\delta\mathcal{L}_M}{\delta\overline{\omega}{}^{\underline{a}}{}_{\underline{b}}}&\frac{1}{x^*_0}\frac{\delta\mathcal{L}_M}{\delta\overline{\mathcal{P}}_{\underline{b}}}\\x^*_0\frac{\delta\mathcal{L}_M}{\delta\overline{\vartheta}{}^{\underline{a}}}&\frac{\delta\mathcal{L}_M}{\delta\overline{\omega}{}^{\underline{a}}{}_{\underline{a}}}\end{pmatrix}.
\end{equation}
In other words, we identify the total projective matter current components with
\begin{equation}
    \tilde{\mathbb{M}}{}^{\underline{B}}{}_{\underline{A}}=\begin{pmatrix}
        \text{Hyper-momentum}&\text{``metric"\;Energy-Momentum}\\
        \text{Canonical Energy-Momentum}&\text{Tr(Hyper-momentum)}
    \end{pmatrix}.
\end{equation}
Notice that this naming convention requires one to view $\frac{\delta\mathcal{L}_M}{\delta\overline{\mathcal{P}}_{\underline{b}}}$ as descending to the standard metrical energy-momentum current. 

To separate the gravitational momenta from currents for $\tilde{\Omega}$, we isolate the independent connection-forms contained within the variation. This is accomplished by simply summing over the $\underline{*}$-components as
\begin{equation}\label{connection-variation}
\begin{split}
    \delta\tilde{\Omega}{}^{\underline{A}}{}_{\underline{B}}\wedge\tilde{\mathbb{G}}{}^{\underline{B}}{}_{\underline{A}}&=\delta\tilde{\Omega}{}^{\underline{a}}{}_{\underline{b}}\wedge\tilde{\mathbb{G}}{}^{\underline{b}}{}_{\underline{a}}+\delta\tilde{\Omega}{}^{\underline{*}}{}_{\underline{*}}\wedge\tilde{\mathbb{G}}{}^{\underline{*}}{}_{\underline{*}}+\delta\tilde{\Omega}{}^{\underline{a}}{}_{\underline{*}}\wedge\tilde{\mathbb{G}}{}^{\underline{*}}{}_{\underline{a}}+\delta\tilde{\Omega}{}^{\underline{*}}{}_{\underline{b}}\wedge\tilde{\mathbb{G}}{}^{\underline{b}}{}_{\underline{*}}.
    \end{split}
\end{equation}
The partial derivatives with respect to the curvature contained in $\tilde{\mathbb{G}}{}^{\underline{B}}{}_{\underline{A}}$ must be expanded and written in terms of the component field strengths. The final variation appearing in Eq.~\eqref{connection-variation} may be expressed as
\begin{equation}
\delta\tilde{\Omega}{}^{\underline{*}}{}_{\underline{b}}\wedge\tilde{\mathbb{G}}{}^{\underline{b}}{}_{\underline{*}}=\delta\overline{\mathcal{P}}_{\underline{b}}\wedge\left(-\overline{D}\overline{\mathbb{H}}{}^{\underline{b}}+\overline{\mathbb{E}}{}^{\underline{b}}+x^*_0\tilde{\Upsilon}{}^{\underline{b}}\tilde{\mathbb{X}}_{\underline{*}}\right),
\end{equation}
where
\begin{equation}
    \overline{\mathbb{H}}{}^{\underline{b}}:=-\frac{\partial\mathcal{L}_G}{\partial\overline{\mathcal{S}}_{\underline{b}}},\quad\quad\quad\overline{\mathbb{E}}{}^{\underline{b}}:=\frac{\partial\mathcal{L}_G}{\partial\overline{\mathcal{P}}_{\underline{b}}}
\end{equation}
are, respectively, the gauge field momenta of the projective Schouten connection-form and the ``metrical" energy momentum. To arrive at this result, we utilized the identity
\begin{equation}
        \delta^{\underline{b}}{}_{\underline{c}}=\frac{\partial\overline{\mathcal{P}}_{\underline{c}}}{\partial\overline{\mathcal{P}}_{\underline{b}}}=\frac{\partial\tilde{\mathcal{K}}{}^{\underline{d}}{}_{\underline{e}}}{\partial\overline{\mathcal{P}}_{\underline{b}}}\wedge\frac{\partial\overline{\mathcal{P}}_{\underline{c}}}{\partial\tilde{\mathcal{K}}{}^{\underline{d}}{}_{\underline{e}}}=\overline{\vartheta}{}^{\underline{d}}\wedge\frac{\partial\overline{\mathcal{P}}_{\underline{c}}}{\partial\tilde{\mathcal{K}}{}^{\underline{d}}{}_{\underline{b}}},
\end{equation}
along with the traceless property of $\tilde{\mathcal{K}}$. We note that when $\overline{\mathcal{P}}$ has a non-vanishing antisymmetric part and $\tilde{\mathcal{K}}$ a non-vanishing trace, the above identity contains an additional term on the right-hand side. The vectorial variation in Eq.~\eqref{connection-variation} may be expressed as
\begin{equation}
        \delta\tilde{\Omega}{}^{\underline{a}}{}_{\underline{*}}\wedge\tilde{\mathbb{G}}{}^{\underline{*}}{}_{\underline{a}}=\delta\overline{\vartheta}{}^{\underline{a}}\wedge\left(-\overline{D}\overline{\mathbb{H}}_{\underline{a}}+\overline{\mathbb{E}}_{\underline{a}}+\tfrac{1}{x^*_0}\tilde{\Upsilon}{}^{\underline{*}}\tilde{\mathbb{X}}_{\underline{a}}\right),
\end{equation}
where
\begin{equation}
    \overline{\mathbb{H}}_{\underline{a}}:=-\frac{\partial\mathcal{L}_G}{\partial\overline{\mathcal{T}}{}^{\underline{a}}},\quad\quad\quad\overline{\mathbb{E}}_{\underline{a}}:=\frac{\partial\mathcal{L}_G}{\partial\overline{\vartheta}{}^{\underline{a}}}
\end{equation}
are the projective co-frame gauge field momentum and canonical energy momentum, respectively. To arrive at this result a similar identity was utilized,
\begin{equation}
    \delta^{\underline{c}}{}_{\underline{a}}=\frac{\partial\overline{\vartheta}{}^{\underline{c}}}{\partial\overline{\vartheta}{}^{\underline{a}}}=\frac{\partial\tilde{\mathcal{K}}{}^{\underline{d}}{}_{\underline{b}}}{\partial\overline{\vartheta}{}^{\underline{a}}}\wedge\frac{\partial\overline{\vartheta}{}^{\underline{c}}}{\partial\tilde{\mathcal{K}}{}^{\underline{d}}{}_{\underline{b}}}=\overline{\mathcal{P}}_{\underline{b}}\wedge\frac{\partial\overline{\vartheta}{}^{\underline{c}}}{\partial\tilde{\mathcal{K}}{}^{\underline{a}}{}_{\underline{b}}},
\end{equation}
along with the traceless property of $\tilde{\mathcal{K}}$. Again, this identity is modified in the presence of a non-vanishing antisymmetric part in $\overline{\mathcal{P}}$ and a non-vanishing trace in $\tilde{\mathcal{K}}$. Lastly, the first variation in Eq.~\eqref{connection-variation} may be expressed as
\begin{equation}
    \delta\tilde{\Omega}{}^{\underline{a}}{}_{\underline{b}}\wedge\tilde{\mathbb{G}}{}^{\underline{b}}{}_{\underline{a}}=\delta\overline{\omega}{}^{\underline{a}}{}_{\underline{b}}\wedge\left(-  \overline{D}\overline{\mathbb{H}}{}^{\underline{b}}{}_{\underline{a}}+\overline{\mathbb{E}}{}^{\underline{b}}{}_{\underline{a}}+\tilde{\Upsilon}{}^{\underline{b}}\tilde{\mathbb{X}}_{\underline{a}}\right),
\end{equation}
where
\begin{equation}
    \overline{\mathbb{H}}{}^{\underline{b}}{}_{\underline{a}}:=-\frac{\partial\mathcal{L}_G}{\partial\overline{\mathcal{R}}{}^{\underline{a}}{}_{\underline{b}}},\quad\quad\quad \quad\overline{\mathbb{E}}{}^{\underline{b}}{}_{\underline{a}}:=-\overline{\vartheta}{}^{\underline{b}}\wedge\overline{\mathbb{H}}_{\underline{a}}-\overline{\mathbb{H}}{}^{\underline{b}}\wedge\overline{\mathcal{P}}_{\underline{a}} ,
\end{equation}
are the gauge field momentum of the projective connection and the projective hyper-momentum, respectively. To arrive at this result, we note that
\begin{equation}
    \frac{\partial\mathcal{L}}{\partial\tilde{\mathcal{K}}{}^{\underline{a}}{}_{\underline{b}}}=\frac{\partial\overline{\mathcal{R}}{}^{\underline{c}}{}_{\underline{d}}}{\partial\tilde{\mathcal{K}}{}^{\underline{a}}{}_{\underline{b}}}\frac{\partial\mathcal{L}}{\partial\overline{\mathcal{R}}{}^{\underline{c}}{}_{\underline{d}}}=\frac{\partial\mathcal{L}}{\partial\overline{\mathcal{R}}{}^{\underline{a}}{}_{\underline{b}}}.
\end{equation}
We therefore express the components of the gravitational contribution to the $\tilde{\Omega}$ field equations as
\begin{equation}
    \tilde{\mathbb{G}}{}^{\underline{B}}{}_{\underline{A}}=\begin{pmatrix}
        -\overline{D}\overline{\mathbb{H}}{}^{\underline{b}}{}_{\underline{a}}+\overline{\mathbb{E}}{}^{\underline{b}}{}_{\underline{a}}+\tilde{\Upsilon}{}^{\underline{b}}\tilde{\mathbb{X}}_{\underline{a}}&-\overline{D}\overline{\mathbb{H}}{}^{\underline{b}}+\overline{\mathbb{E}}{}^{\underline{b}}+x^*_0\tilde{\Upsilon}{}^{\underline{b}}\tilde{\mathbb{X}}_{\underline{*}}\\
        -\overline{D}\overline{\mathbb{H}}_{\underline{a}}+\overline{\mathbb{E}}_{\underline{a}}+\tfrac{1}{x^*_0}\tilde{\Upsilon}{}^{\underline{*}}\tilde{\mathbb{X}}_{\underline{a}}& \overline{\mathbb{E}}{}^{\underline{a}}{}_{\underline{a}}+\tilde{\Upsilon}{}^{\underline{*}}\tilde{\mathbb{X}}_{\underline{*}}
    \end{pmatrix},
\end{equation}
where the traceless property of $\overline{\mathcal{R}}$ was utilized. Up to the presence of $\tilde{\mathbb{X}}$ in each component, we find that $ \tilde{\mathbb{G}}{}^{\underline{B}}{}_{\underline{A}}$ contains all the usual MAG momenta and currents, provided one entertains the idea that $\overline{\mathcal{P}}$ provides access to a dynamical metric. 

To summarize, the field equations of the combined general projective gravitational and material theories are
\begin{equation}\label{gen-proj-field-eq}
    \begin{split}
        \delta\overline{\omega}{}^{\underline{a}}{}_{\underline{b}}:\quad 0&=  -\overline{D}\overline{\mathbb{H}}{}^{\underline{b}}{}_{\underline{a}}+\overline{\mathbb{E}}{}^{\underline{b}}{}_{\underline{a}}+\tilde{\Upsilon}{}^{\underline{b}}\tilde{\mathbb{X}}_{\underline{a}}+\tilde{\mathbb{M}}{}^{\underline{b}}{}_{\underline{a}},\\
        \delta\overline{\vartheta}{}^{\underline{a}}:\quad 0&=-\overline{D}\overline{\mathbb{H}}_{\underline{a}}+\overline{\mathbb{E}}_{\underline{a}}+\tfrac{1}{x^*_0}\tilde{\Upsilon}{}^{\underline{*}}\tilde{\mathbb{X}}_{\underline{a}}+\tilde{\mathbb{M}}{}^{\underline{*}}{}_{\underline{a}},\\
        \delta\overline{\mathcal{P}}_{\underline{a}}:\quad 0&=-\overline{D}\overline{\mathbb{H}}{}^{\underline{b}}+\overline{\mathbb{E}}{}^{\underline{b}}+x^*_0\tilde{\Upsilon}{}^{\underline{b}}\tilde{\mathbb{X}}_{\underline{*}}+\tilde{\mathbb{M}}{}^{\underline{b}}{}_{\underline{*}},\\
        \delta\Psi:\quad 0&=\tilde{\mathbb{M}},
    \end{split}
\end{equation}
along with the auxiliary set of field equations to be satisfied by a dynamical generalized projective Higgs vector $\tilde{\Upsilon}$. In general, 
\begin{equation}\label{upsilon-field-eq-gen}
   \delta\tilde{\Upsilon}:\quad 0=-\tilde{D}\tilde{\mathbb{X}}_{\underline{A}}+\tilde{\mathbb{Y}}_{\underline{A}}+\tilde{\mathbb{M}}_{\underline{A}}.
\end{equation}
For a non-dynamical $\tilde{\Upsilon}$, Eq. \eqref{upsilon-field-eq-gen} is trivial and $\tilde{\mathbb{X}}$ vanishes from all variations in Eqs. \eqref{gen-proj-field-eq}. In this scenario, the complete set of general projective field equations exactly parallels the MAG field equations listed in Eqs. \eqref{general-MAG-equations}, provided $\overline{\mathcal{P}}$ is considered in place of a dynamical metric. In what follows, we will investigate both situations---dynamical and non-dynamical $\tilde{\Upsilon}$---showing that a fixed length may be arrived at as a result of the field equations, thus providing a dynamical interpretation for the validity of the PV-gauge choice. Before discussing the dynamical theory, we turn to the projective Pontrjagin from, wherein a new topological term is presented.

\subsection{\texorpdfstring{$\bar{\mathfrak{p}}$}{\bar{\mathfrak{p}}}-Pontrjagin}
\label{subsec:pontrjagin}

The naturally \textit{metric-free} $PGL(m,\mathbb{R})\cong SL(m+1,\mathbb{R})$ Pontrjagin form, denoted $\mathscr{P}_{\bar{\mathfrak{p}}}$, may be written for any $m=4k$ with $k\in\mathbb{Z}^+$, and has the interpretation of giving the difference between the number of self- and anti-self-dual harmonic (projective) connections on $\mathcal{M}$ \cite{anomaly-torsion}. Interest in the Pontrjagin form and the closely related, \textit{metric-dependent} Nieh-Yan form stems from its contribution to the chiral anomaly, though the latter's relevance has been debated \cite{chandia-zinelli,torsion-nieh-yan-anomaly,mielke-anomaly2,mielke-anomaly3,anomaly-torsion2,anomaly-torsion}. Generalizations of the Nieh-Yan form for the $5$-dimensional de Sitter group were considered in \cite{generalized-nieh-yan}, and the general affine group in \cite{pontryagin-5d-brst}. Projective geometry, as we will see, provides a \textit{metric-independent} generalization which descends to the standard Nieh-Yan form in the limit of constant $\overline{\mathcal{P}}$ and vanishing non-metricity $\overline{Q}$, and may thus provide novel contributions to the chiral anomaly. Moreover, the nonlinear realization formalism renders the generalized Nieh-Yan form \textit{projectively invariant}.

The projective Pontrjagin form $\mathscr{P}_{\bar{\mathfrak{p}}}$ is found by first constructing the Killing metric from the structure constants of the algebra \cite{mielke-geometro},
\begin{equation}
    [\bm{L}{}^{\underline{A}}{}_{\underline{B}},\bm{L}{}^{\underline{A}}{}_{\underline{B}}]=f^{\underline{AC},\underline{E}}{}_{\underline{BD},\underline{F}}\bm{L}{}^{\underline{F}}{}_{\underline{E}},
\end{equation}
as
\begin{equation}\label{SL-killing-metric}
    \kappa^{\underline{AC}}{}_{\underline{BD}}:=\tfrac{1}{2(m+1)}\text{Tr}\left(\bm{L}{}^{\underline{A}}{}_{\underline{B}}\bm{L}{}^{\underline{C}}{}_{\underline{D}}\right)=\delta^{\underline{A}}{}_{\underline{D}}\delta^{\underline{C}}{}_{\underline{B}}-\tfrac{1}{(m+1)}\delta^{\underline{A}}{}_{\underline{B}}\delta^{\underline{C}}{}_{\underline{D}}.
\end{equation}
The non-vanishing components of $f^{\underline{AC},\underline{E}}{}_{\underline{BD},\underline{F}}$ are easily extracted from the commutators in Eqs. \eqref{pgl+D-algebra-commutators}. The normalization in Eq. \eqref{SL-killing-metric} is chosen to make explicit the Killing metric's effect of projecting out traces. Restricting to $m+1=5$ dimensions, the projective Pontrjagin $4$-form is defined as
\begin{equation}
    S_{\bar{\mathscr{P}}}:=\frac{1}{8\pi^2}\int_{\mathcal{M}_{4}}\mathscr{P}_{\bar{\mathfrak{p}}}=\frac{1}{8\pi^2}\int_{\mathcal{M}_{4}}\text{Tr}_{\kappa}(\tilde{\mathcal{K}}\wedge\tilde{\mathcal{K}}),
\end{equation}
with the standard $m=4$ normalization \cite{chandia-zinelli}. Expanding $\mathscr{P}_{\bar{\mathfrak{p}}}$ with respect to the Killing metric in Eq. \eqref{SL-killing-metric} yields
\begin{equation}\label{pontrjagin-KK}
    \mathscr{P}_{\bar{\mathfrak{p}}}=\text{Tr}_{\kappa}(\tilde{\mathcal{K}}\wedge\tilde{\mathcal{K}})=\kappa^{\underline{AC}}{}_{\underline{BD}}\tilde{\mathcal{K}}{}^{\underline{B}}{}_{\underline{A}}\wedge\tilde{\mathcal{K}}{}^{\underline{D}}{}_{\underline{C}}=\tilde{\mathcal{K}}{}^{\underline{A}}{}_{\underline{B}}\wedge\tilde{\mathcal{K}}{}^{\underline{B}}{}_{\underline{A}}.
\end{equation}
Upon restricting to antisymmetric $\tilde{\mathcal{K}}$, Eq. \eqref{pontrjagin-KK} reduces to a Pontrjagin form for a type of projective (Anti)-de Sitter group \cite{generalized-nieh-yan,anomaly-torsion}.

Summing over the $\underline{*}$-components in the right-hand side of Eq. \eqref{pontrjagin-KK} reveals
\begin{equation}
    \tilde{\mathcal{K}}{}^{\underline{A}}{}_{\underline{B}}\wedge\tilde{\mathcal{K}}{}^{\underline{B}}{}_{\underline{A}}
    =\overline{\mathcal{R}}{}^{\underline{a}}{}_{\underline{b}}\wedge\overline{\mathcal{R}}{}^{\underline{b}}{}_{\underline{a}}+2\overline{\mathcal{R}}{}^{\underline{a}}{}_{\underline{b}}\wedge\overline{\vartheta}{}^{\underline{b}}\wedge\overline{\mathcal{P}}_{\underline{a}}+2\overline{\mathcal{T}}{}^{\underline{b}}\wedge\overline{\mathcal{S}}_{\underline{b}},
\end{equation}
in limiting agreement with \cite{Hehl,mielke}. The last two terms in the expression above may be combined to yield
\begin{equation}\label{proj-nieh-yan-1}
\begin{split}
\mathscr{N}_{\bar{\mathfrak{p}}}:&=2\overline{\mathcal{R}}{}^{\underline{a}}{}_{\underline{b}}\wedge\overline{\vartheta}{}^{\underline{b}}\wedge\overline{\mathcal{P}}_{\underline{a}}+2\overline{\mathcal{T}}{}^{\underline{a}}\wedge\overline{\mathcal{S}}_{\underline{a}}\\
&=(\overline{\mathcal{R}}{}^{\underline{a}}{}_{\underline{b}}\wedge\overline{\vartheta}{}^{\underline{b}}\wedge\overline{\mathcal{P}}_{\underline{a}}+\overline{\mathcal{T}}{}^{\underline{a}}\wedge\overline{\mathcal{S}}_{\underline{a}})+(\overline{\mathcal{R}}{}^{\underline{a}}{}_{\underline{b}}\wedge\overline{\vartheta}{}^{\underline{b}}\wedge\overline{\mathcal{P}}_{\underline{a}}+\overline{\mathcal{T}}{}^{\underline{a}}\wedge\overline{\mathcal{S}}_{\underline{a}})\\
&=(\overline{D}{}^2\overline{\vartheta}{}^{\underline{a}}\wedge\overline{\mathcal{P}}_{\underline{a}}+\overline{\mathcal{T}}{}^{\underline{a}}\wedge\overline{D}\overline{\mathcal{P}}_{\underline{a}})+(\overline{D}{}^2\overline{\mathcal{P}}_{\underline{a}}\wedge\overline{\vartheta}{}^{\underline{a}}+\overline{D}\overline{\vartheta}{}^{\underline{a}}\wedge\overline{\mathcal{S}}_{\underline{a}})\\
&=(\overline{D}\overline{\mathcal{T}}{}^{\underline{a}}\wedge\overline{\mathcal{P}}_{\underline{a}}+\overline{\mathcal{T}}{}^{\underline{a}}\wedge\overline{D}\overline{\mathcal{P}}_{\underline{a}})+(\overline{D}\overline{\mathcal{S}}_{\underline{a}}\wedge\overline{\vartheta}{}^{\underline{a}}+\overline{\mathcal{S}}_{\underline{a}}\wedge\overline{D}\overline{\vartheta}{}^{\underline{a}})\\
&=\overline{D}(\overline{\mathcal{T}}{}^{\underline{a}}\wedge\overline{\mathcal{P}}_{\underline{a}})+\overline{D}(\overline{\mathcal{S}}_{\underline{a}}\wedge\overline{\vartheta}{}^{\underline{a}})\\
&=\overline{D}(\overline{\mathcal{T}}{}^{\underline{a}}\wedge\overline{\mathcal{P}}_{\underline{a}}+\overline{\mathcal{S}}_{\underline{a}}\wedge\overline{\vartheta}{}^{\underline{a}}).
    \end{split}
\end{equation}
From the form of $\mathscr{N}_{\bar{\mathfrak{p}}}$, we may identify it as the metric-independent, projectively invariant generalization of the Nieh-Yan form: the \textit{$\bar{\mathfrak{p}}$-Nieh-Yan form}. This identification follows from the limit where $\overline{\mathcal{P}}_{\underline{a}}=-l^{-2}_0\overline{\vartheta}{}^{\underline{b}}\eta_{\underline{ab}}$ is proportional to $\overline{\vartheta}{}^{\underline{a}}$, with $[l_0]=L$ some constant, since then
\begin{equation}
\begin{split}
    \mathscr{N}_{\bar{\mathfrak{p}}}&\rightarrow -\frac{1}{l_0^2}\overline{D}(\overline{\mathcal{T}}{}^{\underline{a}}\wedge\eta_{\underline{ab}}\overline{\vartheta}{}^{\underline{b}}-\overline{Q}_{\underline{ab}}\wedge\overline{\vartheta}{}^{\underline{b}}\wedge\overline{\vartheta}{}^{\underline{a}}+\overline{\mathcal{T}}{}^{\underline{a}}\wedge\eta_{\underline{ab}}\overline{\vartheta}{}^{\underline{b}})\\
    &=-\frac{2}{l_0^2}\overline{D}(\overline{\mathcal{T}}{}^{\underline{a}}\wedge\overline{\vartheta}_{\underline{a}})\\
    &=-\frac{2}{l_0^2}\mathscr{N},
    \end{split}
\end{equation}
where $\mathscr{N}$ is the standard Nieh-Yan form \cite{nieh,nieh-yan-OG-1,chandia-zinelli}, and $\overline{Q}_{[\underline{ab}]}=0$. Following \cite{chandia-zinelli}, we may view $\mathscr{N}_{\bar{\mathfrak{p}}}$ as providing the difference of two topological invariants, since
\begin{equation}\label{topological-proj-statement}
    \mathscr{N}_{\bar{\mathfrak{p}}}=\frac{1}{2}(\mathscr{P}_{\bar{\mathfrak{p}}}-\mathscr{P})=\frac{1}{2}\left(\tilde{\mathcal{K}}{}^{\underline{A}}{}_{\underline{B}}\wedge\tilde{\mathcal{K}}{}^{\underline{B}}{}_{\underline{A}}
    -\overline{\mathcal{R}}{}^{\underline{a}}{}_{\underline{b}}\wedge\overline{\mathcal{R}}{}^{\underline{b}}{}_{\underline{a}}\right).
\end{equation}
Effectively, due to the employment of the nonlinear realization process, Eq. \eqref{topological-proj-statement} is the projective analogue of Eq. $13$ in \cite{chandia-zinelli}, and therefore, $\mathscr{N}_{\bar{\mathfrak{p}}}$ is indeed a \textit{new} topological invariant. In particular, had we not made use of the nonlinear realization process, $\mathscr{N}_{\mathfrak{p}}$ would be identified as the difference of $SL(5,\mathbb{R})$ and $SL(4,\mathbb{R})$ Pontrjagin densities. This new metric-independent and projectively invariant topological term may prove to be a fruitful area of investigation outside of the present context, for example, in condensed matter systems \cite{nieh-yan-cond-matt-1,nieh-yan-cond-matt-2}

We may relate $\mathscr{P}_{\bar{\mathfrak{p}}}$ to the projective Chern-Simons $3$-form $\mathscr{C}_{\bar{\mathfrak{p}}}$ via the exterior derivative
\begin{equation}
    \mathscr{P}_{\bar{\mathfrak{p}}}=d\mathscr{C}_{\bar{\mathfrak{p}}},
\end{equation}
where
\begin{equation}
    \mathscr{C}_{\bar{\mathfrak{p}}}=\tilde{\Omega}{}^{\underline{A}}{}_{\underline{B}}\wedge\tilde{\mathcal{K}}{}^{\underline{B}}{}_{\underline{A}}-\tfrac{1}{3}\tilde{\Omega}{}^{\underline{A}}{}_{\underline{B}}\wedge\tilde{\Omega}{}^{\underline{B}}{}_{\underline{C}}\wedge\tilde{\Omega}{}^{\underline{C}}{}_{\underline{A}}.
\end{equation}
This further splits into the $3$-forms $\mathscr{C}_{\bar{\mathfrak{p}}}= \mathscr{C}_{\bar{\mathfrak{p}}\mathcal{R}\mathcal{R}}+ \mathscr{C}_{\bar{\mathfrak{p}}\mathcal{S}\mathcal{T}}$, given by
\begin{equation}
    \mathscr{C}_{\bar{\mathfrak{p}}\mathcal{R}\mathcal{R}}=\overline{\omega}{}^{\underline{a}}{}_{\underline{b}}\wedge\overline{\mathcal{R}}{}^{\underline{b}}{}_{\underline{a}}-\tfrac{1}{3}\overline{\omega}{}^{\underline{a}}{}_{\underline{b}}\wedge\overline{\omega}{}^{\underline{b}}{}_{\underline{c}}\wedge\overline{\omega}{}^{\underline{c}}{}_{\underline{a}},
\end{equation}
\begin{equation}
\mathscr{C}_{\bar{\mathfrak{p}}\mathcal{ST}}=d\mathscr{N}_{\bar{\mathfrak{p}}}=\overline{\mathcal{P}}_{\underline{a}}\wedge\overline{\mathcal{T}}{}^{\underline{a}}+\overline{\vartheta}{}^{\underline{a}}\wedge\overline{\mathcal{S}}_{\underline{a}}.
\end{equation}
Here, no analogue of the dilational Chern-Simons form, $\mathscr{C}_{\text{Tr}{\mathcal{R}}\text{Tr}{\mathcal{R}}}$, appears as a result of the traceless property of $\tilde{\mathcal{K}}$. From the field variation machinery developed in Sec. \ref{subsec:field-variations}, the \textit{boundary} variation of $\mathscr{C}_{\bar{\mathfrak{p}}}$ provides the maximally symmetric solution
\begin{equation}
    \tilde{\mathcal{K}}{}^{\underline{A}}{}_{\underline{B}}=0.
\end{equation}
This follows from the independent variations producing the beautiful set of relations:
\begin{equation}
    \frac{\delta\mathscr{C}_{\bar{\mathfrak{p}}\mathcal{ST}}}{\delta\overline{\vartheta}{}^{\underline{b}}}=-\overline{\mathcal{S}}_{\underline{b}},\quad\quad\quad\frac{\delta\mathscr{C}_{\bar{\mathfrak{p}}\mathcal{ST}}}{\delta\overline{\mathcal{P}_{\underline{a}}}}=-\overline{\mathcal{T}}{}^{\underline{a}},\quad\quad\quad \frac{\delta\mathscr{C}_{\bar{\mathfrak{p}}\mathcal{ST}}}{\delta\overline{\omega}{}^{\underline{b}}{}_{\underline{a}}}=-\overline{\vartheta}{}^{\underline{a}}\wedge\mathcal{P}_{\underline{b}},
\end{equation}
and
\begin{equation}
    \frac{\delta\mathscr{C}_{\bar{\mathfrak{p}}\mathcal{R}\mathcal{R}}}{\delta\overline{\omega}{}^{\underline{b}}{}_{\underline{a}}}=-\overline{\mathcal{R}}{}^{\underline{a}}{}_{\underline{b}}.
\end{equation}
Compare with Eqs. $3.9.10-3.9.13$ of \cite{Hehl} for the MAG Chern-Simons variations.

In the limit where $\overline{\mathcal{P}}$ is pure trace, i.e., conformally related to the translational connection-form via $\overline{\mathcal{P}}_{\underline{a}}=-\chi^2\eta_{\underline{ab}}\overline{\vartheta}{}^{\underline{b}}$, with $[\chi(x)]=M$, it appears possible to relate $\chi$ to the Barbero-Immirzi \textit{field} \cite{instanton-angle,BI-inflaton}. However, its appearance is not in the usual manner, as it only couples to the partially reduced Nieh-Yan term and not the $m$-dimensional Pontrjagin term. This, however, was the case discussed in \cite{BI-matter-interaction,immirzi} and references therein, where the Barbero-Immirzi \textit{field} removes divergences from the gravitational action. In this limit, we find
\begin{equation}\label{specific-nieh-yan-gen}
    \mathscr{P}_{\bar{\mathfrak{p}}}\rightarrow\overline{\mathcal{R}}{}^{\underline{a}}{}_{\underline{b}}\wedge\overline{\mathcal{R}}{}^{\underline{b}}{}_{\underline{a}}-2\chi^2\left(\overline{\mathcal{R}}{}^{\underline{a}}{}_{\underline{b}}\wedge\overline{\vartheta}{}^{\underline{b}}\wedge\overline{\vartheta}_{\underline{a}}+\overline{\mathcal{T}}_{\underline{a}}\wedge\overline{\mathcal{T}}{}^{\underline{a}}+\overline{q}_{\underline{ab}}\wedge\overline{\vartheta}{}^{\underline{b}}\wedge\overline{\mathcal{T}}{}^{\underline{a}}\right).
\end{equation}
Recalling $\overline{Q}{}^{\underline{a}}{}_{\underline{a}}=0$, two terms leading to the above expression combined to form the \textit{trace-full} non-metricity-like object
\begin{equation}
\overline{q}_{\underline{ab}}:=2\eta_{\underline{ab}}d\log\chi-\overline{Q}_{\underline{ab}}.
\end{equation}
The term in parenthesis in Eq. \eqref{specific-nieh-yan-gen} may be viewed as a particular instantiation of the \textit{generalized Nieh-Yan form} proposed in \cite{general-nieh-yan} for general Metric-Affine geometries. This follows from the fact that the first two terms in parenthesis combine to form $\mathscr{N}$, while the last term couples to a non-metricity $\overline{q}$, whose Weyl form is provided by the logarithmic exterior derivative of the projective Schouten scalar $\chi$. 

Based on the previous discussion, we find that in this conformal limit, the entire second summand in Eq. \eqref{specific-nieh-yan-gen} has an associated \textit{conformal-translational} Chern-Simons $3$-form, defined by
\begin{equation}
    d\mathscr{C}_{\bar{\mathfrak{p}}\mathcal{TT}}=-\chi^2\left(\overline{\mathcal{R}}{}^{\underline{a}}{}_{\underline{b}}\wedge\overline{\vartheta}{}^{\underline{b}}\wedge\overline{\vartheta}_{\underline{a}}+\overline{\mathcal{T}}_{\underline{a}}\wedge\overline{\mathcal{T}}{}^{\underline{a}}+\overline{q}_{\underline{ab}}\wedge\overline{\vartheta}{}^{\underline{b}}\wedge\overline{\mathcal{T}}{}^{\underline{a}}\right),
\end{equation}
where
\begin{equation}
\mathscr{C}_{\bar{\mathfrak{p}}\mathcal{TT}}:=-\chi^2\mathscr{C}_{\mathcal{TT}}=-\chi^2\eta_{\underline{ab}}\overline{\vartheta}{}^{\underline{a}}\wedge\overline{\mathcal{T}}{}^{\underline{b}}.
\end{equation}
In this limit, the total Chern-Simons $3$-form $\mathscr{C}_{\bar{\mathfrak{p}}}$ may be written in terms of a projectively invariant translational Chern-Simons $3$-form $\mathscr{C}_{\mathcal{TT}}$. Up to a difference in normalization, this is an exact analogue to the translational Chern-Simons $3$-form encountered in Metric-Affine geometries, see Eq. \eqref{translational-CS-MAG}. Explicitly,
\begin{equation}
\mathscr{C}_{\bar{\mathfrak{p}}}\rightarrow\mathscr{C}_{\bar{\mathfrak{p}}\mathcal{R}\mathcal{R}}+2\chi^2\mathscr{C}_{\mathcal{TT}},
\end{equation}
where, up to the presence of the scalar field $-\chi^2$, is in agreement with \cite{mielke}. According to \cite{nieh-yan-cond-matt-1,nieh-yan-cond-matt-2} and references therein, it may be possible to identify the scalar degree of freedom contained in $\overline{\mathcal{P}}_{\underline{b}}$ as providing an UV cutoff scale. It would certainly be interesting, then, to relate the additional degrees of freedom contained in $\overline{\mathcal{P}}_{\underline{b}}$ to the anomalous thermal contributions investigated in these references.

Variation of $S_{\bar{\mathscr{P}}}$ with respect to the gravitational field variables is trivially satisfied as the General Projective Bianchi identities. These have been studied, in part, in the (not projectively invariant) torsion-free sector \cite{proj-vs-metric}. The General Projective Bianchi identities may be concisely written as the single equality
\begin{equation}\label{pontrjagin-field-eq}
    \tilde{D}\tilde{\mathcal{K}}{}^{\underline{A}}{}_{\underline{B}}=0.
\end{equation}
The component equations are then the \textit{General Projective Bianchi Identities}:
\begin{equation}\label{General-Projective-Bianchi-Identities-02}
(0)\;\;\overline{D}\overline{\mathcal{S}}_{\underline{b}}+\overline{\mathcal{R}}{}^{\underline{a}}{}_{\underline{b}}\wedge\overline{\mathcal{P}}_{\underline{a}}=0,\quad\quad (1)\;\;\overline{D}\overline{\mathcal{T}}{}^{\underline{a}}-\overline{\mathcal{R}}{}^{\underline{a}}{}_{\underline{b}}\wedge\overline{\vartheta}{}^{\underline{b}}=0,\quad\quad (2)\;\;\overline{D}\overline{\mathcal{R}}{}^{\underline{a}}{}_{\underline{b}}=0,
\end{equation}
and
\begin{equation}\label{General-Projective-Bianchi-Identities-3}
(3)\;\;\overline{\mathcal{T}}{}^{\underline{a}}\wedge\overline{\mathcal{P}}_{\underline{a}}-\overline{\mathcal{S}}_{\underline{a}}\wedge\overline{\vartheta}{}^{\underline{a}}=0.
\end{equation}
Curiously, the nonlinear realization process has rendered Eq. $(3)$ trivial. This follows from the ability to identify $\overline{\vartheta}$ with the standard co-frame, which, when combined with the traceless nature of the curvatures, provides the symmetry of $\overline{\mathcal{P}}$, i.e., $\overline{\vartheta}{}^{\underline{a}}\wedge\overline{\mathcal{P}}_{\underline{a}}=0$. Explicitly, taking the gauge-covariant derivative of this wedge product provides
\begin{equation}\label{nieh-yan-difference}
     \overline{D}(\overline{\vartheta}{}^{\underline{a}}\wedge\overline{\mathcal{P}}_{\underline{a}})=\overline{\mathcal{T}}{}^{\underline{a}}\wedge\overline{\mathcal{P}}_{\underline{a}}-\overline{\mathcal{S}}_{\underline{a}}\wedge\overline{\vartheta}{}^{\underline{a}}=0.
 \end{equation}
Therefore, the General Projective Bianchi Identities properly yield $(m+1)^2-1=24$ equations. Notice that if the symmetry condition were not present, identity $(3)$ is satisfied whenever $\overline{\mathcal{P}}$ is conformally related to the co-frame, due to the antisymmetry of the wedge product. Prior to applying the nonlinear realization process, $\vartheta^{a}\wedge\mathcal{P}_{a}\neq0$. In this case, identity $(3)$ must be subtracted from identity $(2)$ in order to retain the proper number of equations. Interestingly, identity $(3)$ permits two further equivalent expressions for $\mathscr{N}_{\bar{\mathfrak{p}}}$:
\begin{equation}
    \mathscr{N}_{\bar{\mathfrak{p}}}\overset{(3)}{\equiv}2\overline{D}(\overline{\mathcal{S}}_{\underline{a}}\wedge\overline{\vartheta}{}^{\underline{a}})\overset{(3)}{\equiv}2\overline{D}(\overline{\mathcal{T}}{}^{\underline{a}}\wedge\overline{\mathcal{P}}_{\underline{a}}).
\end{equation}

One may obtain the familiar non-metricity Bianchi identity, $(0)$ in Eq. \eqref{MAG-bianchi-exterior}, by taking either of the (anti)-symmetric parts of Eq. \eqref{pontrjagin-field-eq},
\begin{equation}
\begin{split}
    \tilde{D}\tilde{\mathcal{K}}_{(\underline{*b})}&:=\tilde{\eta}_{\underline{A}(\underline{*}}\tilde{D}\tilde{\mathcal{K}}{}^{\underline{A}}{}_{\underline{b})}\\
    &=\eta_0x^*_0\bar{\mathfrak{p}}{}^{-2}\overline{D}\left(\overline{\mathcal{S}}_{\underline{b}}+\tfrac{\eta_0}{(x^*_0)^2}\eta_{\underline{ab}}\overline{\mathcal{T}}{}^{\underline{a}}+\overline{Q}_{\underline{ab}}\wedge\overline{\mathcal{T}}{}^{\underline{a}}\right)\\
    &\quad+\eta_0x^*_0\bar{\mathfrak{p}}{}^{-2}\left(\overline{\mathcal{R}}{}^{\underline{a}}{}_{\underline{b}}\wedge\overline{\mathcal{P}}_{\underline{a}}-\eta_{\underline{ab}}\overline{\mathcal{R}}{}^{\underline{a}}{}_{\underline{c}}\wedge\underline{\overline{\vartheta}}{}^{\underline{c}}\right)\\
    &=\eta_0x^*_0\bar{\mathfrak{p}}{}^{-2}\left(\overline{D}\overline{\mathcal{S}}{}^{+}_{\underline{b}}+\overline{\mathcal{R}}{}^{\underline{a}}{}_{\underline{b}}\wedge\overline{\mathcal{P}}{}^{+}_{\underline{a}}+\overline{D}\overline{Q}_{\underline{ab}}\wedge\underline{\overline{\vartheta}}{}^{\underline{a}}-\overline{\mathcal{R}}_{(\underline{ab})}\wedge\underline{\overline{\vartheta}}{}^{\underline{a}}\right).
    \end{split}
\end{equation}
In the expression above, we have made use of the shifted projective Schouten-form $\overline{\mathcal{P}}{}^{\pm}_{\underline{b}}=\overline{\mathcal{P}}_{\underline{b}}\pm\eta_{\underline{ab}}\underline{\overline{\vartheta}}{}^{\underline{a}}$, and let $\underline{\overline{\vartheta}}{}^{\underline{a}}:=\frac{\eta_0}{(x^*_0)^2}\overline{\vartheta}{}^{\underline{a}}$ denote the signature-dependent rescaled co-frame. Furthermore, we have introduced the field strength of the shifted projective Schouten-form,
\begin{equation}
    \overline{\mathcal{S}}{}^{+}_{\underline{b}}=\overline{D}\overline{\mathcal{P}}{}^{+}_{\underline{b}}.
\end{equation}
Since 
\begin{equation}
    \overline{D}\overline{\mathcal{S}}{}^{+}_{\underline{b}}+\overline{\mathcal{R}}{}^{\underline{a}}{}_{\underline{b}}\wedge\overline{\mathcal{P}}{}^{+}_{\underline{a}}=0
\end{equation}
is trivially satisfied as a result of Eqs. \eqref{General-Projective-Bianchi-Identities-02}, we arrive at the final result
\begin{equation}\label{proj-non-met-bianchi}
    \overline{\mathcal{R}}_{(\underline{ab})}\wedge\overline{\vartheta}{}^{\underline{a}}=\overline{D}\overline{Q}_{\underline{ab}}\wedge\overline{\vartheta}{}^{\underline{a}},
\end{equation}
where overall factors of $\frac{\eta_0}{(x^*_0)^2}$ have been canceled. This is the typical form of the non-metricity Bianchi identity in exterior form \cite{chern-bianchi}. Having taken instead the antisymmetric part, one obtains the same result after utilizing the trivial identity $
    \overline{D}\overline{\mathcal{S}}{}^{-}_{\underline{b}}+\overline{\mathcal{R}}{}^{\underline{a}}{}_{\underline{b}}\wedge\overline{\mathcal{P}}{}^{-}_{\underline{a}}=0$. It appears, however, that the projective geometric framework permits a second, independent identity, in analogy to Eq. \eqref{proj-non-met-bianchi}. This follows from repeating the above steps, but rather with indices raised,
\begin{equation}
        \tilde{D}\tilde{\mathcal{K}}{}^{(\underline{*b})}=\tilde{\eta}{}^{\underline{bc}}\tilde{D}\tilde{\mathcal{K}}{}^{\underline{*}}{}_{\underline{c}}+\tilde{\eta}{}^{\underline{**}}\tilde{D}\tilde{\mathcal{K}}{}^{\underline{b}}{}_{\underline{*}}.
    \end{equation}
This is found to produce
\begin{equation}
        \overline{\mathcal{R}}{}^{(\underline{ab})}\wedge\overline{\mathcal{P}}_{\underline{b}}=-\overline{D}\overline{Q}{}^{\underline{ab}}\wedge\overline{\mathcal{P}}_{\underline{b}},
    \end{equation}
which only descends to Eq. \eqref{proj-non-met-bianchi} for a non-vanishing $\overline{\mathcal{P}}$ conformally related to $\overline{\vartheta}$.

The trivial identity $(3)$ permits three equivalent expressions for $\mathscr{C}_{\bar{\mathfrak{p}}\mathcal{S}\mathcal{T}}$,
\begin{equation}
     \begin{split}
\mathscr{C}_{\bar{\mathfrak{p}}\mathcal{S}\mathcal{T}}&=\overline{\mathcal{P}}_{\underline{a}}\wedge\overline{\mathcal{T}}{}^{\underline{a}}+\overline{\vartheta}{}^{\underline{a}}\wedge\overline{\mathcal{S}}_{\underline{a}}\\
&=2\overline{\vartheta}{}^{\underline{a}}\wedge\overline{\mathcal{S}}_{\underline{a}}\\
&=2\overline{\mathcal{P}}_{\underline{a}}\wedge\overline{\mathcal{T}}{}^{\underline{a}}.
     \end{split}
 \end{equation}
From this, a vanishing projective torsion $\overline{\mathcal{T}}$, analogous to acquiring Weyl nullity \cite{nullity}, is sufficient for both $\mathscr{C}_{\bar{\mathfrak{p}}}=\mathscr{C}_{\bar{\mathfrak{p}}\mathcal{R}\mathcal{R}}$ and, from Eqs. $(1)$ and $(3)$,
\begin{equation}
    \overline{\mathcal{R}}{}^{\underline{a}}{}_{[\underline{bcd}]}=0,\quad\quad \overline{\mathcal{S}}_{[\underline{abc}]}=0.
\end{equation}
Furthermore, in the vanishing torsion regime, there is a complete reduction of $\mathscr{P}_{\bar{\mathfrak{p}}}$ to the projectively invariant $\mathscr{P}$ of $4$-dimensions,
\begin{equation}
     \tilde{\mathcal{K}}{}^{\underline{A}}{}_{\underline{B}}\wedge\tilde{\mathcal{K}}{}^{\underline{B}}{}_{\underline{A}}
    =\overline{\mathcal{R}}{}^{\underline{a}}{}_{\underline{b}}\wedge\overline{\mathcal{R}}{}^{\underline{b}}{}_{\underline{a}}.
\end{equation}
When the curvature $\overline{\mathcal{R}}$ is proportional to the metric, the projective Schouten field strength vanishes \cite{einstein-metrics}. Its vanishing is sufficient for both $\mathscr{C}_{\bar{\mathfrak{p}}}=\mathscr{C}_{\bar{\mathfrak{p}}\mathcal{R}\mathcal{R}}$ and
\begin{equation}
    \overline{\mathcal{R}}{}^{\underline{a}}{}_{\underline{b}}\wedge\overline{\mathcal{P}}_{\underline{a}}=0,\quad\quad \overline{\mathcal{P}}_{\underline{a}}\wedge\overline{\mathcal{T}}{}^{\underline{a}}=0.
\end{equation}
In this conformal limit, with $\chi^2\neq0$, these expressions descend to
\begin{equation}
    \overline{\mathcal{R}}_{[\underline{a}|\underline{b}|\underline{cd}]}=0,\quad\quad \overline{\mathcal{T}}_{[\underline{abc}]}=0.
\end{equation}
Additionally, a vanishing $\overline{\mathcal{S}}$ also produces
\begin{equation}
    \tilde{\mathcal{K}}{}^{\underline{A}}{}_{\underline{B}}\wedge\tilde{\mathcal{K}}{}^{\underline{B}}{}_{\underline{A}}=\overline{\mathcal{R}}{}^{\underline{a}}{}_{\underline{b}}\wedge\overline{\mathcal{R}}{}^{\underline{b}}{}_{\underline{a}}.
\end{equation}
Thus, a vanishing $\overline{\mathcal{S}}$ also forces a vanishing $\mathscr{N}_{\bar{\mathfrak{p}}}$. This provides a further complication to the matters regarding the status of the Nieh-Yan form and its contribution to the chiral anomaly. In particular, the nonlinearly realized projective setting seems to imply the Nieh-Yan form vanishes whenever either $\overline{\mathcal{T}}{}^{\underline{a}}=0$ or $\overline{\mathcal{S}}_{\underline{a}}=0$, and thus $\mathscr{N}$ cannot contribute to anything in either of these sectors. Therefore, in the present context, \textit{there exists no projectively invariant torsional topological invariant independent of the projective Schouten form}. Given these statements, it may be more appropriate to speak only of $\overline{\mathcal{S}}{}^{\pm}$, as opposed to $\overline{\mathcal{S}}$ and $\overline{\mathcal{T}}$. The implications of these considerations on the chiral anomaly are very much apart of future investigations, and an explicit calculation of such is necessary before drawing any definite conclusion. This will be discussed further in Sec. \ref{sec:proj-matter-D2}.

\subsection{\texorpdfstring{$\bar{\mathfrak{p}}$}{\bar{\mathfrak{p}}}-Lovelock}
\label{subsec:lovelock}

In this section, we focus on constructing the dynamical aspects of the General Projective Gauge Theory of Gravitation. In particular, we form a projectively invariant Lovelock theory. More information on Lovelock theories can be found in \cite{lovelock-2,lovelock-1,love1,love2,love3,zumino}. To simplify calculations, we restrict attention to the physically meaningful $m+1=5$-dimensional scenario. We discuss properties of this model in both exterior and abstract-index form, and show how its related to the Macdowell-Mansouri and Stelle-West formulations of gravity \cite{MM-OG,stelle-west}. The field variations of the general theory are found, and some matter-free solutions investigated in the PV-gauge. Inspired by \cite{MM-OG,stelle-west}, and the more recent applications \cite{dynamical-stelle-west-1,MM-conf-flat,into-cartan,pseudo-proj,holst}, we show that the $m+1=5$-dimensional projective Lovelock gravitational theory allows for the PV-gauge choice to be viewed as resulting from the fields equations. Additionally, it is shown that many solutions lead to a degenerate co-frame. These were discussed in \cite{pseudo-proj} in the context of Macdowell-Mansouri (A)-dS gravity.

The projective Lovelock action in $m+1=5$ dimensions is 
\begin{equation}
    S_{\bar{\mathfrak{p}}L}=\frac{1}{x^*_0}\int_{V\mathcal{M}}\tilde{\alpha}_2\tilde{\mathcal{K}}{}^{\underline{AB}}\wedge *\tilde{\mathcal{K}}_{\underline{BA}}+\frac{\tilde{\alpha}_1}{3}\tilde{\mathcal{B}}{}^{\underline{AB}}\wedge*\tilde{\mathcal{K}}_{\underline{BA}}+\frac{\tilde{\alpha}_0}{5}\tilde{\mathcal{B}}{}^{\underline{AB}}\wedge*\tilde{\mathcal{B}}_{\underline{BA}},
\end{equation}
where
\begin{equation}
    \tilde{\mathcal{B}}{}^{\underline{AB}}:=\frac{1}{2!}\tilde{E}{}^{\underline{A}}\wedge\tilde{E}{}^{\underline{B}}=\frac{1}{2}\tilde{D}\tilde{\Upsilon}{}^{\underline{A}}\wedge\tilde{D}\tilde{\Upsilon}{}^{\underline{B}},
\end{equation}
\begin{equation}
*\tilde{\mathcal{B}}_{\underline{AB}}\equiv\hat{\tilde{\epsilon}}_{\underline{AB}}=\frac{1}{3}\hat{\tilde{\epsilon}}_{\underline{ABCDE}}\tilde{\mathcal{B}}{}^{\underline{CD}}\wedge\tilde{E}{}^{\underline{E}},
\end{equation}
and $*\tilde{\mathcal{K}}$ has the form of Eq. \eqref{star-K-gen}. The symbol $\tilde{\mathcal{B}}$ has been chosen to represent $\tilde{E}\wedge\tilde{E}$ for the simple purpose of eliciting parallels with the $BF$-formalism for constructing gravitational theories \cite{mielke,BF-symmetry-break,BF-gravity,BF-vince}. The coefficients introduced above are given the dimensions
\begin{equation}
    [\tilde{\alpha}_2]=L^{-1},\quad\quad [\tilde{\alpha}_1]=L^{-3},\quad\quad [\tilde{\alpha}_0]=L^{-5},
\end{equation}
along with the overall factor $[x^*_0]=L$, such that $[S_{\bar{\mathfrak{p}}L}]=1$. The numerical factors in $S_{\bar{\mathfrak{p}}L}$ are chosen similar to those used in \cite{love2,5d-de-sitter-constants-2} for $5$-dimensional (A)-dS Lovelock gravitational theories, as opposed to the standard Lovelock coefficients \cite{transgression}. A similar non-projective model of $m+1=5$-dimensional gravity, \cite{gb-constant-types}, explores various choices for the $\alpha_i$ coefficients and the resulting implications on the underlying spacetime manifold $\mathcal{M}$. 

Had our gauge group generators been taken antisymmetric, we would ``have a Lagrangian without having a Lagrangian" provided by transgression \cite{chamseddine,transgression}. For an application, see \cite{holographic-5d-chern-simon,holographic-5d-chern-simon-2}. Therefore, if the present model were reduced to the (Anti)-de Sitter group, the projective Lovelock action would be a total derivative \cite{5d-LL-total-der-action}. When this holds, the theory is identified with a projective Lovelock-Chern-Simons theory. The obstruction to this identification is attributed to the necessary metric-dependence and the unavoidable incompatibility of the connection. According to \cite{traceless-nonmetricity-gb-nontopological}, it is specifically the traceless parts of non-metricity which determine this. Interestingly, this implies that it is also the non-trivial character of $\overline{\mathcal{P}}$ which is responsible for the obstruction to the action being written as a total derivative.

The $m+1=5$-dimensional projective Lovelock action $S_{\bar{\mathfrak{p}}L}$ may be expressed in abstract-index form. This transition between formalisms is explicitly worked out in Appendix \hyperref[app-B2:exterior-abs-ind-form-proj-grav]{B.2}. Up to the signature-dependent factor $(-1)^{q+1}\eta_0$, the associated Lagrangian density $\mathcal{L}_{\bar{\mathfrak{p}}L}$ is
\begin{equation}\label{proj-LL-lagrangian}
\mathcal{L}_{\bar{\mathfrak{p}}L}=d^5x|\tilde{e}|\left(\tilde{\alpha}_2\left(\tilde{\mathcal{K}}^{\underline{AB}}{}_{[\underline{CD}]}\tilde{\mathcal{K}}^{\underline{CD}}{}_{[\underline{AB}]}-\Delta\tilde{\mathcal{K}}^{\underline{A}}{}_{\underline{B}}\Delta\tilde{\mathcal{K}}^{\underline{B}}{}_{\underline{A}}+\tilde{\mathcal{K}}^2\right)+\tilde{\alpha}_1\tilde{\mathcal{K}}+2\tilde{\alpha}_0\right),
\end{equation}
where
\begin{equation}
    \Delta\tilde{\mathcal{K}}^{\underline{A}}{}_{\underline{B}}:=\tilde{\mathcal{K}}{}^{[\underline{CA}]}{}_{[\underline{CB}]}=\tilde{\mathcal{K}}{}^{\underline{CA}}{}_{[\underline{CB}]}-\tilde{\mathcal{K}}{}^{\underline{AC}}{}_{[\underline{CB}]}=\tilde{\mathcal{K}}{}^{\underline{A}}{}_{\underline{B}}-\check{\tilde{\mathcal{K}}}{}^{\underline{A}}{}_{\underline{B}}
\end{equation}
is the difference between the projective Ricci and co-Ricci curvature tensors. The first term in $\mathcal{L}_{\bar{\mathfrak{p}}L}$ differs from the quadratic term used in the Thomas-Whitehead theory of \cite{gen-struc,heavy-lifting,covariant-tw}, but becomes equivalent when the curvature is antisymmetric in both sets of indices and symmetric under interchanging the sets. Recalling the discussion leading to Eq. \eqref{LC-R-index-sym}, the Thomas-Whitehead action is seen to coincide with the projective Lovelock action when both the torsion $\tilde{\mathcal{T}}$ and non-metricity $\tilde{Q}$ vanish. Note that although $\tilde{\mathcal{K}}{}^{\underline{A}}{}_{\underline{B}M*}=0$ for all values of the index $M$, this is not true of $\tilde{\mathcal{K}}{}^{\underline{A}}{}_{\underline{BC*}}$. This is where the couplings to $\overline{\mathcal{S}}$ reside, providing dynamics for $\overline{\mathcal{P}}$. When a foliation is permitted along slices of constant $x^*$, i.e., $\tilde{E}{}^{\underline{a}}{}_*=0$, these terms miraculously drop out of the action. This, however, is precisely the statement of the APV-gauge. Thus, from these statements alone, we conclude that $S_{\bar{\mathfrak{p}}L}$ provides \textit{no dynamics} for $\overline{\mathcal{P}}$ in the APV-gauge. 

We also note that the first term in Eq. \eqref{proj-LL-lagrangian} is generally different from the Kretschmann scalar, which would read
\begin{equation}
    \tilde{\mathcal{K}}{}^{\underline{AB}}{}_{\underline{CD}}\tilde{\mathcal{K}}_{\underline{AB}}{}^{\underline{CD}}.
\end{equation}
This is the term utilized in the ``Projective Gauss-Bonnet" action of \cite{gen-struc,heavy-lifting,covariant-tw}. It has been shown that the Kretschmann scalar may be set to vanish by a convenient choice of parameters in the projective transformation of the connection \cite{projective-kretchman}. Presently, however, this cannot be accomplished, since the connection is invariant under projective transformations. 

In the APV-gauge, $S_{\bar{\mathfrak{p}}L}$ takes the abstract-index form
\begin{equation}\label{proj-LL-abs-ind-APV-prior}
\begin{split}
    S_{\bar{\mathfrak{p}}L}\overset{\circ}{=}(-1)^{q+1}\eta_0\int\frac{dx^*}{x^*}\int_{\mathcal{M}}&d^4x|\overline{\vartheta}|\left(\tilde{\alpha}_2(\overline{\mathcal{R}}{}^{\underline{ab}}{}_{[\underline{cd}]}\overline{\mathcal{R}}{}^{\underline{cd}}{}_{[\underline{ab}]}-\Delta\overline{\mathcal{R}}{}^{\underline{a}}{}_{\underline{b}}\Delta\overline{\mathcal{R}}{}^{\underline{b}}{}_{\underline{a}}+\overline{\mathcal{R}}{}^2)\right.\\
&\quad\quad\quad+2\tilde{\alpha}_2(\overline{\mathcal{R}}\overline{\mathcal{P}}-\Delta\overline{\mathcal{R}}{}^{\underline{ab}}\overline{\mathcal{P}}_{\underline{ab}}-\overline{\mathcal{P}}{}^{\underline{ab}}\overline{\mathcal{P}}_{\underline{ab}}+\overline{\mathcal{P}}{}^{2})\\
&\quad\quad\quad+\left.\tilde{\alpha}_1(\overline{\mathcal{R}}+3\overline{\mathcal{P}})+2\tilde{\alpha}_0\right),
    \end{split}
\end{equation}
and we find that the $\tilde{\alpha}_i$ coefficients may be rescaled by integrating over the $x^*$-direction,
\begin{equation}
    \alpha_i:=\tilde{\alpha}_i\int^{x^*_f}_{x^*_i}\frac{x^*_0}{x^*}dx^*=\tilde{\alpha}_i\log\left(\frac{x^*_f}{x^*_i}\right).
\end{equation}
This rescaling behavior is the same as that observed in Eqs. \eqref{grav-constants-rescale}, and permits rescaling invariance for combinations such as $\tilde{\alpha}_i/\tilde{\alpha}_j$. We note that, upon rescaling the coefficients in Eq. \eqref{proj-LL-abs-ind-APV-prior}, the action reduces purely to a $4$-dimensional theory. This may then be recast in exterior form as
\begin{equation}\label{4-dim-pLL}
\begin{split}
    S_{\bar{\mathfrak{p}}L}\overset{\circ}{=}\eta_0\int_{\mathcal{M}}&\alpha_2\left(\overline{\mathcal{R}}{}^{\underline{a}}{}_{\underline{b}}\wedge\star\overline{\mathcal{R}}{}^{\underline{b}}{}_{\underline{a}}+2\overline{\mathcal{R}}{}^{\underline{ab}}\wedge\star(\overline{\vartheta}\wedge\overline{\mathcal{P}})_{\underline{ba}}+\overline{\vartheta}{}^{\underline{a}}\wedge\overline{\mathcal{P}}{}^{\underline{b}}\wedge\star(\overline{\vartheta}\wedge\overline{\mathcal{P}})_{\underline{ba}}\right)\\
    &+\alpha_1\left(\overline{\mathcal{R}}{}^{\underline{ab}}\wedge\star(\overline{\vartheta}\wedge\overline{\vartheta})_{\underline{ba}}+\overline{\vartheta}{}^{\underline{a}}\wedge\overline{\mathcal{P}}{}^{\underline{b}}\wedge\star(\overline{\vartheta}\wedge\overline{\vartheta})_{\underline{ba}}\right)\\
    &+\frac{\alpha_0}{6}\left(\overline{\vartheta}{}^{\underline{a}}\wedge\overline{\vartheta}{}^{\underline{b}}\wedge\star(\overline{\vartheta}\wedge\overline{\vartheta})_{\underline{ba}}\right),
    \end{split}
\end{equation}
where
\begin{equation}
\star(\overline{\vartheta}\wedge\overline{\vartheta})_{\underline{ba}}:=\frac{1}{2!}\hat{\epsilon}_{\underline{bacd}}\overline{\vartheta}{}^{\underline{c}}\wedge\overline{\vartheta}{}^{\underline{d}},\quad\quad\quad\star(\overline{\vartheta}\wedge\overline{\mathcal{P}})_{\underline{ba}}:=\hat{\epsilon}_{\underline{bacd}}\overline{\vartheta}{}^{\underline{c}}\wedge\overline{\mathcal{P}}{}^{\underline{d}}.
\end{equation}

In the limit where $\overline{\mathcal{P}}_{\underline{a}}=\mathcal{P}^2_0\eta_{\underline{ab}}\overline{\vartheta}{}^{\underline{b}}$ and the connection is compatible, $ S_{\bar{\mathfrak{p}}L}$ reduces significantly and has the abstract-index form
\begin{equation}\label{proj-LL-abs-ind-APV}
\begin{split}
    S_{\bar{\mathfrak{p}}L}\overset{\circ}{=}\alpha_2S_{\overline{\mathscr{E}}}+\alpha_1\int_{\mathcal{M}}&d^4x|\overline{\vartheta}|\left(\overline{\mathcal{R}}+\frac{4\alpha_2}{\alpha_1}\overline{\mathcal{R}}\mathcal{P}^2_0+\frac{24\alpha_2}{\alpha_1}(\mathcal{P}^2_0+\beta^{+})(\mathcal{P}^2_0+\beta^{-})\right),
    \end{split}
\end{equation}
where the overall factor of $(-1)^{q+1}\eta_0$ was discarded, and
\begin{equation}
    S_{\overline{\mathscr{E}}}:=\int_{\mathcal{M}}d^4x|\overline{\vartheta}|\mathscr{E}_{\bar{\mathfrak{p}}}
\end{equation}
is the projectively invariant Euler or Gauss-Bonnet term,
\begin{equation}
    \mathscr{E}_{\bar{\mathfrak{p}}}:=\overline{\mathcal{R}}{}^{\underline{ab}}{}_{[\underline{cd}]}\overline{\mathcal{R}}{}^{\underline{cd}}{}_{[\underline{ab}]}-4\overline{\mathcal{R}}{}^{\underline{a}}{}_{\underline{b}}\overline{\mathcal{R}}{}^{\underline{b}}{}_{\underline{a}}+\overline{\mathcal{R}}{}^2.
\end{equation}
Since the connection is taken compatible in this limit, i.e., $\overline{\mathcal{T}}=\overline{Q}=0$, the Euler term $\mathscr{E}_{\bar{\mathfrak{p}}}$ is topological and does not alter the field equations. In Eq. \eqref{proj-LL-abs-ind-APV}, there appears a factored potential function with rescaling invariant roots given by
\begin{equation}
    \mathcal{P}_0^2=\frac{-\alpha_1}{4\alpha_2}\left(1\pm\sqrt{1-\frac{\alpha_2}{3\alpha_1}\frac{4\alpha_0}{\alpha_1}}\right).
\end{equation}
As noted in \cite{love2,5d-de-sitter-constants-2}, when the discriminant vanishes, this partially reduced theory results in a projective Lovelock-Chern-Simons theory. In this limit, Eq. \eqref{proj-LL-abs-ind-APV} contains a Higgs-type potential for the scalar part of $\overline{\mathcal{P}}$, similar to what is done in \cite{dim-red-higgs-pot}, where Higgs potentials are constructed via dimensional reduction. Additionally, if one were to choose such a vanishing discriminant, and evaluate $S_{\bar{\mathfrak{p}}L}$ on the nonzero $\mathcal{P}_0$ solution, then $L_{\bar{\mathfrak{p}}L}=0$. We note that if the interaction term, $\overline{\mathcal{R}}\mathcal{P}^2_0$, was considered part of the potential, then $\beta^{\pm}$ is amended to
\begin{equation}
    \beta^{\pm}\rightarrow\frac{1}{12}\left((\overline{\mathcal{R}}+\frac{3\alpha_1}{\alpha_2})\pm\sqrt{(\overline{\mathcal{R}}+\frac{3\alpha_1}{\alpha_2})^2-\frac{12\alpha_0}{\alpha_2}}\right).
\end{equation}
In this case, having either $\mathcal{P}^2_0=-\beta^{\pm}$ results in $L_{\bar{\mathfrak{p}}L}\propto \overline{\mathcal{R}}$, rather than zero. Although this may seem unnatural, it is in fact more consistent with the usual interpretation of $\overline{\mathcal{P}}$ and $\overline{\mathcal{R}}$ being related when the connection is torsion-free and compatible \cite{proj-conn}.

Before analyzing the dynamical theory, we show how $\mathcal{L}_{\bar{\mathfrak{p}}L}$ is related to the Macdowell-Mansouri- \cite{MM-OG}, and Stelle-West-type constructions \cite{stelle-west}. This relationship follows from removing the gauge-covariant derivatives contained in each $\tilde{E}$. The Macdowell-Mansouri (MM) and Stelle-West (SW) models typically appear in gauge gravitational theories when there exists a generalized Higgs vector \cite{Wise,MM-conf-flat}, and in some cases, when a dynamical cosmological constant is sought \cite{dynamical-stelle-west-1}.

We consider first, removing the gauge-covariant derivative of $\tilde{\Upsilon}$ in the $\tilde{\alpha}_2$ term. Define the MM-dual as
\begin{equation}
\circledast\tilde{\mathcal{K}}_{\underline{AB}}:=\hat{\tilde{\epsilon}}_{\underline{ABCDE}}\tilde{\mathcal{K}}{}^{\underline{CD}}\tilde{\Upsilon}{}^{\underline{E}}.
\end{equation}
We then find
\begin{equation}
    \tilde{\mathcal{K}}{}^{\underline{AB}}\wedge*\tilde{\mathcal{K}}_{\underline{BA}}=\tilde{D}\left(\tilde{\mathcal{K}}{}^{\underline{AB}}\wedge\circledast\tilde{\mathcal{K}}_{\underline{BA}}\right)+\left(5\tilde{g}\tilde{\eta}{}^{\underline{BD}}-2\tilde{Q}{}^{\underline{BD}}\right)\wedge\tilde{\mathcal{K}}{}^{\underline{A}}{}_{\underline{D}}\wedge\circledast\tilde{\mathcal{K}}_{\underline{BA}},
\end{equation}
which differs from the MM construction by the projective Weyl and non-metricity forms. Their vanishing permits the use of the covariant Stokes' Theorem, Eq. \eqref{stokes}, resulting in the MM model,
\begin{equation}
    \int_{V\mathcal{M}}\tilde{\mathcal{K}}{}^{\underline{AB}}\wedge*\tilde{\mathcal{K}}_{\underline{BA}}=\int_{\partial V\mathcal{M}}\tilde{\mathcal{K}}{}^{\underline{AB}}\wedge\circledast\tilde{\mathcal{K}}_{\underline{BA}}.
\end{equation}

The remaining two terms of the action may be expressed similarly. Taking
\begin{equation}
\circledast\tilde{\mathcal{B}}_{\underline{AB}}:=\hat{\tilde{\epsilon}}_{\underline{ABCDE}}\tilde{\mathcal{B}}{}^{\underline{CD}}\tilde{\Upsilon}{}^{\underline{E}},
\end{equation}
we find that for the $\tilde{\alpha}_1$ term
\begin{equation}
\begin{split}
        \tilde{\mathcal{B}}{}^{\underline{AB}}\wedge*\tilde{\mathcal{K}}_{\underline{BA}}&=\tilde{D}\left(\tilde{\mathcal{B}}{}^{\underline{AB}}\wedge\circledast\tilde{\mathcal{K}}_{\underline{BA}}\right)+\left(5\tilde{g}\tilde{\eta}{}^{\underline{BC}}-\tilde{Q}{}^{\underline{BC}}\right)\wedge\tilde{\mathcal{K}}{}^{\underline{A}}{}_{\underline{C}}\wedge\circledast\tilde{\mathcal{B}}_{\underline{BA}}\\
        &\quad-\tilde{\mathcal{T}}{}^{\underline{A}}\wedge\tilde{E}{}^{\underline{B}}\wedge\circledast\tilde{\mathcal{K}}_{\underline{BA}},
        \end{split}
\end{equation}
and for the $\tilde{\alpha}_0$ term
\begin{equation}
        3\tilde{\mathcal{B}}{}^{\underline{AB}}\wedge*\tilde{\mathcal{B}}_{\underline{BA}}=\tilde{D}\left(\tilde{\mathcal{B}}{}^{\underline{AB}}\wedge\circledast\tilde{\mathcal{B}}_{\underline{BA}}\right)+\left(5\tilde{g}\wedge\tilde{\mathcal{B}}{}^{\underline{AB}}-2\tilde{\mathcal{T}}{}^{\underline{A}}\wedge\tilde{E}{}^{\underline{B}}\right)\wedge\circledast\tilde{\mathcal{B}}_{\underline{BA}}.
\end{equation}
The factor of $3$ is simply an artifact of the difference in dual definitions. A Projective Macdowell-Mansouri or Stelle-West-type construction is not utilized in this document due to the availability of $V\mathcal{M}$, which provides all the tools necessary to coherently construct an $m+1=5$-dimensional theory. In particular, $V\mathcal{M}$ provides the ability to define non-trivial $(m+1)$-forms as a result of the existence of a genuine coordinate, $x^*$.

We study the dynamics resulting from $S_{\bar{\mathfrak{p}}L}$ via two different approaches. In the first approach, we do not treat $\tilde{\Upsilon}$ as a dynamical field, and consider its descent to rigidity purely as a choice of gauge. In the second approach, we permit $\tilde{\Upsilon}$ to be a dynamical field, and seek the same choice of gauge as a result of the field equations. Both of these scenarios have been investigated for an analogous vector field in non-projective gauge gravitational theories \cite{cartan-gravity-1,cartan-gravity-2}. 

In general, the field equations for $S_{\bar{\mathfrak{p}}L}$ are readily found using the field variation machinery developed in Sec. \ref{subsec:field-variations}. The General Projective Gauge Gravitational field equations are
\begin{equation}\label{proj-LL-field-eq-gen}
    -\tilde{\mathbb{M}}{}^{\underline{B}}{}_{\underline{A}}=-\tilde{D}\tilde{\mathbb{H}}{}^{\underline{B}}{}_{\underline{A}}+\tilde{\Upsilon}{}^{\underline{B}}\tilde{\mathbb{X}}_{\underline{A}},\quad\quad\quad-\tilde{\mathbb{M}}_{\underline{A}}=\tilde{D}\tilde{\mathbb{X}}_{\underline{A}}.
\end{equation}
Explicitly,
\begin{equation}\label{proj-LL-field-eq-omega}
\begin{split}
    -\tilde{\mathbb{M}}{}^{\underline{B}}{}_{\underline{A}}&=2\tilde{\alpha}_2\tilde{D}*\left(\tilde{\mathcal{K}}{}^{\underline{B}}{}_{\underline{A}}+\frac{\tilde{\alpha}_1}{2\tilde{\alpha}_2}\tilde{\mathcal{B}}{}^{\underline{B}}{}_{\underline{A}}\right)\\
    &\quad-2\tilde{\alpha}_2\tilde{\Upsilon}{}^{\underline{B}}\hat{\tilde{\epsilon}}_{\underline{ACDEF}}\left(\frac{1}{2}\tilde{\mathcal{K}}{}^{\underline{CD}}\wedge\tilde{\mathcal{K}}{}^{\underline{EF}}+\frac{\tilde{\alpha}_1}{2\tilde{\alpha}_2}\tilde{\mathcal{B}}{}^{\underline{CD}}\wedge\tilde{\mathcal{K}}{}^{\underline{EF}}+\frac{\tilde{\alpha}_0}{6\tilde{\alpha}_2}\tilde{\mathcal{B}}{}^{\underline{CD}}\wedge\tilde{\mathcal{B}}{}^{\underline{EF}}\right),
\end{split}
\end{equation}
\begin{equation}\label{proj-LL-field-eq-upsilon}
-\tilde{\mathbb{M}}_{\underline{A}}=2\tilde{\alpha}_2\tilde{D}\left(\hat{\tilde{\epsilon}}_{\underline{ACDEF}}\left(\frac{1}{2}\tilde{\mathcal{K}}{}^{\underline{CD}}\wedge\tilde{\mathcal{K}}{}^{\underline{EF}}+\frac{\tilde{\alpha}_1}{2\tilde{\alpha}_2}\tilde{\mathcal{B}}{}^{\underline{CD}}\wedge\tilde{\mathcal{K}}{}^{\underline{EF}}+\frac{\tilde{\alpha}_0}{6\tilde{\alpha}_2}\tilde{\mathcal{B}}{}^{\underline{CD}}\wedge\tilde{\mathcal{B}}{}^{\underline{EF}}\right)\right).
\end{equation}

\subsubsection{Non-Dynamical \texorpdfstring{$\tilde{\Upsilon}$}{\tilde{\Upsilon}}}

In the non-dynamical $\tilde{\Upsilon}$ approach, Eq. \eqref{proj-LL-field-eq-upsilon} is absent and the connection field equation, Eq. \eqref{proj-LL-field-eq-omega}, reduces to
\begin{equation}\label{non-dyn-upsilon-max-sym-expanded}
    0=2\tilde{\alpha}_2\tilde{D}*\left(\tilde{\mathcal{K}}{}^{\underline{B}}{}_{\underline{A}}+\frac{\tilde{\alpha}_1}{2\tilde{\alpha}_2}\tilde{\mathcal{B}}{}^{\underline{B}}{}_{\underline{A}}\right)
\end{equation}
in a universe devoid of matter. Distributing the covariant derivative, canceling overall factors, and using the definitions developed thus far provides
\begin{equation}\label{non-dyn-upsilon-1}
\begin{split}
    0&=\left(\tilde{Q}{}^{\underline{BG}}-5\tilde{g}\tilde{\eta}{}^{\underline{BG}}\right)\wedge*\left(\tilde{\mathcal{K}}_{\underline{GA}}+\frac{\tilde{\alpha}_1}{2\tilde{\alpha}_2}\tilde{\mathcal{B}}_{\underline{GA}}\right)\\
    &\quad\quad+\hat{\tilde{\epsilon}}{}^{\underline{B}}{}_{\underline{ACDE}}\left(\left(\tilde{\mathcal{K}}{}^{\underline{CD}}+\frac{\tilde{\alpha}_1}{2\tilde{\alpha}_2}\tilde{\mathcal{B}}{}^{\underline{CD}}\right)\wedge\tilde{\mathcal{T}}{}^{\underline{E}}+\tilde{Q}{}^{\underline{FD}}\wedge\tilde{\mathcal{K}}{}^{\underline{C}}{}_{\underline{F}}\wedge\tilde{E}{}^{\underline{E}}\right).
    \end{split}
\end{equation}
In order to arrive at this result, the projective Bianchi identity, Eq. \eqref{pontrjagin-field-eq}, was also used. In this form, a particular solution is obvious and is given by
\begin{equation}\label{Q-T-solution}
    \tilde{\mathcal{T}}{}^{\underline{A}}=\tilde{g}\wedge\tilde{E}{}^{\underline{A}},\quad\quad\quad \tilde{Q}{}^{\underline{AB}}=2\tilde{g}\tilde{\eta}{}^{\underline{AB}}.
\end{equation}
This particular torsion solution parallels the non-vanishing torsion solution found in \cite{torsion-dynamical-dark-e}. From the above solution for $\tilde{\mathcal{T}}$, a vanishing $\tilde{\mathcal{T}}$ implies that $\tilde{E}$ is linearly dependent on $\tilde{g}$, which may be assumed to take the form
\begin{equation}\label{tilde-e-g-upsilon}
\tilde{E}{}^{\underline{A}}=\tilde{g}\tilde{\Upsilon}{}^{\underline{A}}.
\end{equation}
The above solution for $\tilde{Q}$ implies, in addition to $\overline{Q}_{\underline{ab}}=0$,
\begin{equation}
    \overline{\mathcal{P}}{}^{+}_{\underline{b}}=0.
\end{equation}
However, one may only obtain a completely vanishing $\tilde{Q}$ by imposing $\tilde{g}=0$. This may be accomplished by choosing the APC-gauge, along with taking the $x^*\rightarrow\infty$ limit. 

Since all terms in Eq. \eqref{non-dyn-upsilon-1} cancel once Eqs. \eqref{Q-T-solution} are imposed, there appears to be no restriction on $\tilde{\mathcal{K}}{}^{\underline{a}}{}_{\underline{b}}$. For this reason, we investigate a few other particular solutions via alternative routes. The simplest solution permitted by these $24$ field equations that we will investigate is the \textit{maximally symmetric} solution, $\tilde{\mathcal{K}}{}^{\underline{B}}{}_{\underline{A}}=0$. Additionally, we investigate the solution with only $\overline{Q}=\overline{\mathcal{T}}=\overline{\mathcal{S}}=0$ imposed. This is found to result in a Projective General Relativity with cosmological constant. Many of the other available solutions, outlined in Table \ref{proj-spacetime-table}, are left for future areas of research.

\subsubsection{Non-Dynamical \texorpdfstring{$\tilde{\Upsilon}$}{\tilde{\Upsilon}}: Maximally Symmetric}

Consider the maximally symmetric solution given by
\begin{equation}\label{max-sym-sol-non-dyn-upsilon}
    \tilde{\mathcal{K}}{}^{\underline{A}}{}_{\underline{B}}=0.
\end{equation}
Within the confines of this maximally symmetric solution, we seek to determine what conditions are required of the field variables and their field strengths. Obviously, Eq. \eqref{max-sym-sol-non-dyn-upsilon} requires that $\overline{\mathcal{T}}=\overline{\mathcal{S}}=0$, leading to constant $\overline{\mathcal{P}}$ and constant $\overline{\vartheta}$ solutions. As discussed previously, in its descent to rigidity, $\overline{\mathcal{P}}$ is generally assumed to become conformally related to the co-frame connection, $\overline{\mathcal{P}}_{\underline{a}}\propto\eta_{\underline{ab}}\overline{\vartheta}{}^{\underline{b}}$. For a constant $\overline{\mathcal{P}}$, we simply take a constant of proportionality, denoted here as $\mathcal{P}^2_0$, with $[\mathcal{P}_0]=M$. The remaining components of the curvature $\tilde{\mathcal{K}}{}^{\underline{a}}{}_{\underline{b}}$ must then satisfy
\begin{equation}
\begin{split}
\tilde{\mathcal{K}}{}^{(\underline{ab})}&=0\quad\Rightarrow\quad\overline{\mathcal{R}}{}^{(\underline{ab})}=0,\\
\tilde{\mathcal{K}}{}^{[\underline{ab}]}&=0\quad\Rightarrow\quad\overline{\mathcal{R}}{}^{[\underline{ab}]}=-\mathcal{P}^2_0\overline{\vartheta}{}^{\underline{a}}\wedge\overline{\vartheta}{}^{\underline{b}}.
\end{split}
\end{equation}
The first expression above follows from the anti-commutativity of the wedge product. Recalling Eq. \eqref{proj-non-met-bianchi}, for the symmetric part of $\overline{\mathcal{R}}$ to vanish, it is sufficient to have $\overline{Q}_{\underline{ab}}=0$. The condition on the antisymmetric part of $\overline{\mathcal{R}}$ states that $\mathcal{M}$ is a(n) (anti)-de Sitter spacetime, depending on the sign of $\mathcal{P}^2_0$. 

We now utilize the field equations to find the obvious value for $\mathcal{P}^2_0$. Setting $\tilde{\mathcal{K}}{}^{\underline{B}}{}_{\underline{A}}=0$ in Eq. \eqref{non-dyn-upsilon-1} and canceling overall factors,
\begin{equation}
    0=\frac{1}{3}\left(\tilde{Q}{}^{\underline{BG}}-5\tilde{g}\tilde{\eta}{}^{\underline{BG}}\right)\wedge\hat{\tilde{\epsilon}}_{\underline{GACDE}}\tilde{\mathcal{B}}{}^{\underline{CD}}\wedge\tilde{E}{}^{\underline{E}}+\hat{\tilde{\epsilon}}{}^{\underline{B}}{}_{\underline{ACDE}}\tilde{\mathcal{B}}{}^{\underline{CD}}\wedge\tilde{\mathcal{T}}{}^{\underline{E}}.
\end{equation}
Since $\tilde{\Upsilon}$ is taken non-dynamical, we may freely choose to work in the APV-gauge. Additionally, we must follow the procedure employed for both the connection and curvature. Since the $\mathfrak{sl}(m+1,\mathbb{R})$ generators are traceless, we are required to subtract $\tilde{\mathbb{G}}{}^{\underline{*}}{}_{\underline{*}}$ from $\tilde{\mathbb{G}}{}^{\underline{b}}{}_{\underline{a}}$ in the collective field equations $\tilde{\mathbb{G}}{}^{\underline{B}}{}_{\underline{A}}$. This action properly provides $(m+1)^2-1=24$ equations. However, the manner in which we investigate solutions for all cases considered, this will not be necessary. We therefore display all component equations separately.

Separating the independent variations and imposing the maximally symmetric solution, the field equations reduce to:
\begin{equation}\label{component-field-eq-1}
\begin{split}
    \tilde{\mathbb{G}}{}^{\underline{*}}{}_{\underline{a}}:\quad 0&\overset{\circ}{=}\tilde{\alpha}_1\eta_0\bar{\mathfrak{p}}{}^2\hat{\tilde{\epsilon}}_{\underline{acde*}}\tilde{\mathcal{B}}{}^{\underline{cd}}\wedge\left(\tilde{\mathcal{T}}{}^{\underline{e}}+\tfrac{(x^*_0)^2}{\eta_0}\bar{\mathfrak{p}}\tilde{g}\wedge\overline{\mathcal{P}}{}^{\underline{e}}\right),\\
    \tilde{\mathbb{G}}{}^{\underline{b}}{}_{\underline{*}}:\quad 0&\overset{\circ}{=}\tilde{\alpha}_1\hat{\epsilon}_{\underline{acde*}}\tilde{\mathcal{B}}{}^{\underline{cd}}\wedge\left(\tfrac{1}{3}\bar{\mathfrak{p}}{}^{2}\overline{Q}{}^{\underline{ab}}\wedge\tilde{E}{}^{\underline{e}}-\tilde{\eta}{}^{\underline{ab}}(\tilde{\mathcal{T}}{}^{\underline{e}}-\tilde{g}\wedge\tilde{E}{}^{\underline{e}})\right),\\
    \tilde{\mathbb{G}}{}^{\underline{b}}{}_{\underline{a}}:\quad 0&\overset{\circ}{=}x^*_0\tilde{\alpha}_1\hat{\tilde{\epsilon}}_{\underline{acde*}}\left(\bar{\mathfrak{p}}{}^3\tilde{g}\wedge\overline{Q}{}^{\underline{be}}\wedge\tilde{\mathcal{B}}{}^{\underline{cd}}+\tfrac{1}{3}\tilde{\eta}{}^{\underline{bf}}\overline{\mathcal{P}}{}^{+}_{\underline{f}}\wedge\tilde{\mathcal{B}}{}^{\underline{cd}}\wedge\tilde{E}{}^{\underline{e}}-\bar{\mathfrak{p}}\tilde{\eta}{}^{\underline{bc}}\tilde{g}\wedge\tilde{E}{}^{\underline{d}}\wedge\tilde{\mathcal{T}}{}^{\underline{e}}\right),\\
    \tilde{\mathbb{G}}{}^{\underline{*}}{}_{\underline{*}}:\quad 0&\overset{\circ}{=}\tfrac{-x^*_0}{3}\overline{\mathcal{P}}{}^{+}_{\underline{a}}\hat{\tilde{\epsilon}}{}^{\underline{a}}{}_{\underline{bcd*}}\tilde{\mathcal{B}}{}^{\underline{bc}}\wedge\tilde{E}{}^{\underline{d}}.
    \end{split}
\end{equation}
Although the above is not presented in a form where the APV-gauge choice is obvious, we utilized its consequences to arrive at these expressions. The purpose of presenting the field equations in this way is to aid in understanding the role of $\bar{\mathfrak{p}}$ in modulating the difference between the $(m+1)$-dimensional $\tilde{\mathcal{T}}$ and its $m$-dimensional components, $\overline{\mathcal{T}}$. 

For Eqs. \eqref{component-field-eq-1} to be satisfied, it is sufficient for the terms in parenthesis to individually vanish. Since the symmetric part of $\overline{\mathcal{R}}$ vanishes for $\overline{Q}=0$, the projective Schouten and co-frame field equations, $\tilde{\mathbb{G}}{}^{\underline{b}}{}_{\underline{*}}$ and $\tilde{\mathbb{G}}{}^{\underline{*}}{}_{\underline{a}}$, respectively, provide
\begin{equation}\label{max-sym-conn-field-eq-torsion-sol}
   \tilde{\mathcal{T}}{}^{\underline{e}}\overset{\circ}{=}\tilde{g}\wedge\tilde{E}{}^{\underline{e}}, \quad\quad\quad \tilde{\mathcal{T}}{}^{\underline{e}}\overset{\circ}{=}\frac{-(x^*_0)^2}{\eta_0}\bar{\mathfrak{p}}\tilde{g}\wedge\overline{\mathcal{P}}{}^{\underline{e}}.
\end{equation}
Setting these expressions equal to one another, we find that
\begin{equation}
    \overline{\mathcal{P}}{}^{+}_{\underline{a}}\overset{\circ}{=}0\quad\Rightarrow\quad\mathcal{P}_{0}=\frac{-\eta_0}{(x^*_0)^2}.
\end{equation}
This value should appear quite obvious, since it is $\overline{\mathcal{P}}{}^{+}$ that appears in the symmetric part of $\tilde{\Omega}$, i.e., the shear defects present in the $\bar{\mathfrak{p}}{}^{-2}$-non-metricity $\tilde{Q}$. For this particular value of $\mathcal{P}_0$, we find that 
\begin{equation}\label{max-sym-conn-field-eq}
    \tilde{\mathbb{G}}{}^{\underline{b}}{}_{\underline{a}}-\delta^{\underline{b}}{}_{\underline{a}}\tilde{\mathbb{G}}{}^{\underline{*}}{}_{\underline{*}}=0\quad\Rightarrow\quad \bar{\mathfrak{p}}\tilde{\eta}{}^{\underline{bc}}\tilde{g}\wedge\tilde{E}{}^{\underline{d}}\wedge\tilde{\mathcal{T}}{}^{\underline{e}}=0.
\end{equation}
This condition is easily seen to be satisfied upon substituting either of the expressions in Eqs. \eqref{max-sym-conn-field-eq-torsion-sol}, and using the antisymmetry of the wedge product. We thus find that Eq. \eqref{non-dyn-upsilon-max-sym-expanded} permits the maximally symmetric solution, with the $m$-dimensional curvature satisfying
\begin{equation}
    \overline{\mathcal{R}}{}^{[\underline{ab}]}=-\frac{1}{(x^*_0)^4}\overline{\vartheta}{}^{\underline{a}}\wedge\overline{\vartheta}{}^{\underline{b}}.
\end{equation}

Recall that $\tilde{\mathcal{T}}$ is defined as the second exterior gauge-covariant derivative of $\tilde{\Upsilon}$. The satisfaction of the $\tilde{\mathbb{G}}{}^{\underline{b}}{}_{\underline{*}}$ field equation reduces this second order relation to first order, since
\begin{equation}
    \tilde{\mathbb{G}}{}^{\underline{*}}{}_{\underline{a}}=0\quad\Rightarrow \quad\tilde{D}\tilde{D}\tilde{\Upsilon}{}^{\underline{A}}=\tilde{\mathcal{K}}{}^{\underline{A}}{}_{\underline{B}}\tilde{\Upsilon}{}^{\underline{B}}=\tilde{\mathcal{T}}{}^{\underline{A}}\overset{\circ}{=}\tilde{g}\wedge\tilde{E}{}^{\underline{A}}=\tilde{g}\wedge\tilde{D}\tilde{\Upsilon}{}^{\underline{A}}.
\end{equation}
Recalling that $\tilde{E}{}^{\underline{A}}=\bar{\mathfrak{p}}\tilde{e}{}^{\underline{A}}$, we find that $\tilde{\mathcal{T}}$ may generally be expressed as
\begin{equation}
    \tilde{\mathcal{T}}{}^{\underline{A}}=\tilde{D}\tilde{E}{}^{\underline{A}}=\tilde{g}\wedge\tilde{E}{}^{\underline{A}}+\bar{\mathfrak{p}}\tilde{D}\tilde{e}{}^{\underline{A}}.
\end{equation}
Therefore, the maximally symmetric solution imposes the covariant constancy not of $\tilde{E}{}^{\underline{A}}$, but of $\tilde{e}{}^{\underline{A}}$,
\begin{equation}\label{little-e-der-vanish}
    \tilde{D}\tilde{e}{}^{\underline{A}}=0.
\end{equation}
In the APV-gauge, the components of $\tilde{D}\tilde{e}$ take the form
\begin{equation}
    \tilde{D}\tilde{e}{}^{\underline{A}}\overset{\circ}{=}\begin{pmatrix}
        \overline{\mathcal{T}}{}^{\underline{a}}+\overline{\vartheta}{}^{\underline{a}}\wedge\tilde{g}\\0
    \end{pmatrix}.
\end{equation}
Thus, if Eq. \eqref{little-e-der-vanish} is to be satisfied, it is sufficient to impose
\begin{equation}
     \overline{\mathcal{T}}{}^{\underline{a}}=\tilde{g}\wedge\overline{\vartheta}{}^{\underline{a}}.
\end{equation}
However, the solution $\overline{\mathcal{T}}{}^{\underline{a}}=0$ may only be satisfied for vanishing $\overline{\vartheta}$. We note that the imposition of Eq. \eqref{tilde-e-g-upsilon} yields this same result. All hope is not lost, since $\overline{\vartheta}=0$ does not necessarily imply a vanishing spacetime metric or spacetime volume. Explicitly, imposing $\overline{\vartheta}=0$ simply results in the identification $\vartheta=\overline{D}\tilde{\upsilon}$. In other words, the co-frame may be written as the gauge-covariant derivative of a vector field. In light of these solutions, it therefore seems that $\overline{D}\tilde{\upsilon}=0$ is not admissible by any choice of gauge. Degenerate co-frames have been studied, for example, in \cite{MAG-degenerate-coframe-1,degenerate-coframe-2}. This degeneracy leads to a vanishing volume and has been shown to signal topology changes, see \cite{degenerate-coframe-topology-change,degenerate-frame-topology-change} and references therein. We leave extensive investigation of the $\overline{\vartheta}=0$ solutions and their implications, in particular, on the topology of the underlying spacetime manifold $\mathcal{M}$ as a futures area of research.

\subsubsection{Non-Dynamical \texorpdfstring{$\tilde{\Upsilon}$}{\tilde{\Upsilon}}: Projective General Relativity}

We seek to recover General Relativity with cosmological constant from the general projective Lovelock theory, by first imposing only $\overline{Q}=\overline{\mathcal{S}}=\overline{\mathcal{T}}=0$. We again choose to work in the APV-gauge. The condition of vanishing $\overline{\mathcal{S}}$ implies that $\overline{\mathcal{P}}$ is constant, for which we take
\begin{equation}
\overline{\mathcal{P}}_{\underline{a}}=\mathcal{P}^2_0\eta_{\underline{ab}}\overline{\vartheta}{}^{\underline{b}}.
\end{equation}
The field equations, Eq. \eqref{non-dyn-upsilon-max-sym-expanded}, then separate into the component equations:
\begin{equation}
\begin{split}
    \tilde{\mathbb{G}}{}^{\underline{*}}{}_{\underline{a}}:\quad 0&\overset{\circ}{=}2\tilde{\alpha}_2(x^*_0)^2\mathcal{P}^2_0\tilde{g}\wedge\hat{\epsilon}_{\underline{acde*}}\left(\overline{\mathcal{R}}{}^{\underline{cd}}+(\mathcal{P}^2_0+\tfrac{\alpha_1}{4\alpha_2})\overline{\vartheta}{}^{\underline{c}}\wedge\overline{\vartheta}{}^{\underline{d}}\right)\wedge\overline{\vartheta}{}^{\underline{e}},\\
    \tilde{\mathbb{G}}{}^{\underline{b}}{}_{\underline{*}}:\quad 0&\overset{\circ}{=}2\tilde{\alpha}_2\tilde{g}\wedge\hat{\epsilon}{}^{\underline{b}}{}_{\underline{cde*}}\left(\overline{\mathcal{R}}{}^{\underline{cd}}+(\mathcal{P}^2_0+\tfrac{\alpha_1}{4\alpha_2})\overline{\vartheta}{}^{\underline{c}}\wedge\overline{\vartheta}{}^{\underline{d}}\right)\wedge\overline{\vartheta}{}^{\underline{e}},\\
    \tilde{\mathbb{G}}{}^{\underline{b}}{}_{\underline{a}}:\quad 0&\overset{\circ}{=}2\tilde{\alpha}_2x^*_0(\mathcal{P}^2_0+\tfrac{\eta_0}{(x^*_0)^2})\hat{\epsilon}_{\underline{acde*}}\left(\overline{\mathcal{R}}{}^{\underline{cd}}+(\mathcal{P}^2_0+\tfrac{\alpha_1}{12\alpha_2})\overline{\vartheta}{}^{\underline{c}}\wedge\overline{\vartheta}{}^{\underline{d}}\right)\wedge\overline{\vartheta}{}^{\underline{b}}\wedge\overline{\vartheta}{}^{\underline{e}},\\
    \tilde{\mathbb{G}}{}^{\underline{*}}{}_{\underline{*}}:\quad 0&\overset{\circ}{=}2\tilde{\alpha}_2x^*_0(\mathcal{P}^2_0+\tfrac{\eta_0}{(x^*_0)^2})\hat{\epsilon}_{\underline{acde*}}\left(\overline{\mathcal{R}}{}^{\underline{cd}}+(\mathcal{P}^2_0+\tfrac{\alpha_1}{12\alpha_2})\overline{\vartheta}{}^{\underline{c}}\wedge\overline{\vartheta}{}^{\underline{d}}\right)\wedge\overline{\vartheta}{}^{\underline{a}}\wedge\overline{\vartheta}{}^{\underline{e}}.
    \end{split}
\end{equation}
Up to the overall factor of $(x^*_0)^2\mathcal{P}^2_0$, the co-frame and projective Schouten field equations, $\tilde{\mathbb{G}}{}^{\underline{*}}{}_{\underline{a}}$ and $\tilde{\mathbb{G}}{}^{\underline{b}}{}_{\underline{*}}$, respectively, are identical. The $m$-dimensional connection field equation $\tilde{\mathbb{G}}{}^{\underline{b}}{}_{\underline{a}}$ is related to $\tilde{\mathbb{G}}{}^{\underline{*}}{}_{\underline{*}}$ via the trace operation. Therefore, the two independent field equations to solve are
\begin{equation}
    \begin{split}
        \tilde{\mathbb{G}}{}^{\underline{b}}{}_{\underline{*}}:\quad 0&\overset{\circ}{=}2\tilde{\alpha}_2\tilde{g}\wedge\hat{\epsilon}{}^{\underline{b}}{}_{\underline{cde*}}\left(\overline{\mathcal{R}}{}^{\underline{cd}}+(\mathcal{P}^2_0+\tfrac{\alpha_1}{4\alpha_2})\overline{\vartheta}{}^{\underline{c}}\wedge\overline{\vartheta}{}^{\underline{d}}\right)\wedge\overline{\vartheta}{}^{\underline{e}},\\
    \tilde{\mathbb{G}}{}^{\underline{b}}{}_{\underline{a}}:\quad 0&\overset{\circ}{=}2\tilde{\alpha}_2x^*_0(\mathcal{P}^2_0+\tfrac{\eta_0}{(x^*_0)^2})\hat{\epsilon}_{\underline{acde*}}\left(\overline{\mathcal{R}}{}^{\underline{cd}}+(\mathcal{P}^2_0+\tfrac{\alpha_1}{12\alpha_2})\overline{\vartheta}{}^{\underline{c}}\wedge\overline{\vartheta}{}^{\underline{d}}\right)\wedge\overline{\vartheta}{}^{\underline{b}}\wedge\overline{\vartheta}{}^{\underline{e}}.
    \end{split}
\end{equation}
The first set of equations, $\tilde{\mathbb{G}}{}^{\underline{b}}{}_{\underline{*}}$, provides the general projective Einstein field equations. For the expression to vanish, it is sufficient for
\begin{equation}\label{R-sol-gen-1}
    \overline{\mathcal{R}}{}^{\underline{cd}}=-(\mathcal{P}^2_0+\frac{\alpha_1}{4\alpha_2})\overline{\vartheta}{}^{\underline{c}}\wedge\overline{\vartheta}{}^{\underline{d}}.
\end{equation}
This yields a(n) (Anti)-de Sitter spacetime, depending on the sign of $(\mathcal{P}^2_0+\frac{\alpha_1}{4\alpha_2})$. Substituting this expression into $\tilde{\mathbb{G}}{}^{\underline{b}}{}_{\underline{a}}$, the $m$-dimensional projective connection field equation becomes
\begin{equation}
    \tilde{\mathbb{G}}{}^{\underline{b}}{}_{\underline{a}}:\quad 0\overset{\circ}{=}\frac{-\tilde{\alpha}_1x^*_0}{3}(\mathcal{P}^2_0+\frac{\eta_0}{(x^*_0)^2})\hat{\epsilon}_{\underline{acde*}}\overline{\vartheta}{}^{\underline{b}}\wedge\overline{\vartheta}{}^{\underline{c}}\wedge\overline{\vartheta}{}^{\underline{d}}\wedge\overline{\vartheta}{}^{\underline{e}}.
\end{equation}
This expression is solved for 
\begin{equation}
    \mathcal{P}^2_0=-\frac{\eta_0}{(x^*_0)^2}.
\end{equation}
We therefore find a cosmological constant with contributions from both $\overline{\mathcal{P}}$ and $\tilde{E}$:
\begin{equation}\label{R-sol-1}
    \overline{\mathcal{R}}{}^{\underline{cd}}=(\frac{\eta_0}{(x^*_0)^2}-\frac{\alpha_1}{4\alpha_2})\overline{\vartheta}{}^{\underline{c}}\wedge\overline{\vartheta}{}^{\underline{d}}.
\end{equation}

We relate $\alpha_1$ and $\alpha_2$ to the coefficients of \cite{covariant-tw,gen-struc} for the Thomas-Whitehead theory of gravity, see Eqs. \eqref{grav-constants-rescale}. Deviating slightly, we make the identifications:
\begin{equation}
    \alpha_1:=\frac{\eta_0}{2\kappa_0},\quad\quad\quad\alpha_2:=J_0c,\quad\quad\quad\frac{\eta_0}{r^2_0}:=\frac{\alpha_1}{4\alpha_2}=\frac{\eta_0}{8J_0c\kappa_0}.
\end{equation}
Reflecting these identifications in Eq. \eqref{R-sol-1} provides
\begin{equation}
    \overline{\mathcal{R}}{}^{\underline{cd}}=\eta_0(\frac{1}{(x^*_0)^2}-\frac{1}{r_0^2})\overline{\vartheta}{}^{\underline{c}}\wedge\overline{\vartheta}{}^{\underline{d}}.
\end{equation}
Letting $\Lambda:=\eta_0(\tfrac{1}{(x^*_0)^2}-\tfrac{1}{r_0^2})$ denote the cosmological constant, the projective Minkowski space $PM_4$ may be accessed when
\begin{equation}
    \Lambda=0\;\;\Rightarrow\;\;\frac{1}{(x^*_0)^2}=\frac{1}{r^2_0}.
\end{equation}
This relation provides the first explicit expression for $x^*_0$ in terms of known ($\kappa_0$) and measurable ($J_0$) \cite{inflation} quantities. Thus, from the perspective of $PM_4$, the general projective Lovelock theory has one less free parameter than the Thomas-Whitehead theory.

Interestingly, both the sign of $\eta_0$ \textit{and} the difference $r_0-x^*_0$ combine to determine whether $\Lambda<0$ or $\Lambda>0$. This differs from ordinary (anti)-de Sitter gauge theories, wherein only $\eta_0$ determines this. Additionally, since $x^*_0$ represents a fundamental length scale set by the projective dimension $x^*$, and assumed by $\overline{\mathcal{P}}$, the expanding universe may be viewed as a result of $r_0$'s deviation from this fundamental value. Recalling that ratios of the $\tilde{\alpha}_i$ are independent of the rescaling that results from integration over $x^*$, the deviation of $r_0$ from $x_0$ inherits this independence. Thus, any additional deviation is modulated entirely by $\overline{\mathcal{P}}$. Similar types of varying cosmological ``constants" can typically be found in non-projective (Anti)-de Sitter models \cite{dS-variable-cosmo-const,dS-variable-cosmo-const-2}.

For completeness, we note that had we began with the action in the form of Eq. \eqref{4-dim-pLL}, the field equations would read:
\begin{equation}
\begin{split}
    \tilde{\mathbb{G}}{}^{\underline{*}}{}_{\underline{a}}:\quad0&\overset{\circ}{=}\hat{\epsilon}_{\underline{bacd}}\left(\overline{\mathcal{R}}{}^{\underline{ab}}\wedge(\overline{\mathcal{P}}{}^{\underline{d}}+\tfrac{\alpha_1}{2\alpha_2}\overline{\vartheta}{}^{\underline{d}})+(\overline{\mathcal{P}}{}^{\underline{a}}+\gamma_{+}\overline{\vartheta}{}^{\underline{a}})\wedge(\overline{\mathcal{P}}{}^{\underline{b}}+\gamma_{-}\overline{\vartheta}{}^{\underline{b}})\wedge\overline{\vartheta}{}^{\underline{d}}\right),\\
    \tilde{\mathbb{G}}{}^{\underline{b}}{}_{\underline{*}}:\quad0&\overset{\circ}{=}\hat{\epsilon}{}^{\underline{b}}{}_{\underline{acd}}\left(\overline{\mathcal{R}}{}^{\underline{ac}}+\overline{\vartheta}{}^{\underline{a}}\wedge(\overline{\mathcal{P}}{}^{\underline{c}}+\tfrac{\alpha_1}{4\alpha_2}\overline{\vartheta}{}^{\underline{c}})\right)\wedge\overline{\vartheta}{}^{\underline{d}},\\
    \tilde{\mathbb{G}}{}^{\underline{b}}{}_{\underline{a}}:\quad0&\overset{\circ}{=}\overline{D}\left(\hat{\epsilon}{}^{\underline{b}}{}_{\underline{acd}}\left(\overline{\mathcal{R}}{}^{\underline{cd}}+\overline{\vartheta}{}^{\underline{c}}\wedge(\overline{\mathcal{P}}{}^{\underline{d}}+\tfrac{\alpha_1}{4\alpha_2}\overline{\vartheta}{}^{\underline{d}})\right)\right),
\end{split}
\end{equation}
where
\begin{equation}
    \gamma_{\pm}:=\frac{3\alpha_1}{8\alpha_2}\left(1\mp\sqrt{1-\frac{8}{9}\frac{\alpha_2}{3\alpha_1}\frac{4\alpha_0}{\alpha_1}}\right).
\end{equation}
These equations obviously result in a slightly different conclusion. The reason for this is due, in part, to the now present $\alpha_0$ contribution. In particular, choosing the gauge at the level of the action, Eq. \eqref{4-dim-pLL}, renders the $\tilde{\alpha}_0$ term as explicitly dependent on $\overline{\vartheta}$. 

The projective Lovelock-Chern-Simons theory is found to result from the choice of coefficients,
\begin{equation}
    \gamma_{+}=\frac{\alpha_1}{2\alpha_2},\quad\quad\quad \gamma_{-}=\frac{\alpha_1}{4\alpha_2}.
\end{equation}
For constant $\overline{\mathcal{P}}\propto\mathcal{P}^2_0$, the connection and co-frame field equations, $\tilde{\mathbb{G}}{}^{\underline{*}}{}_{\underline{a}}$ and $\tilde{\mathbb{G}}{}^{\underline{b}}{}_{\underline{a}}$, respectively, are trivially satisfied by
\begin{equation}
    \overline{\mathcal{R}}{}^{\underline{ab}}=-(\mathcal{P}^2_0+\frac{\alpha_1}{4\alpha_2})\overline{\vartheta}{}^{\underline{a}}\wedge\overline{\vartheta}{}^{\underline{b}},
\end{equation}
just as in Eq. \eqref{R-sol-gen-1}. Reflecting this solution for $\overline{\mathcal{R}}$ in the $\overline{\mathcal{P}}$ field equations $\tilde{\mathbb{G}}{}^{\underline{*}}{}_{\underline{a}}$ leads to the algebraic relation between coefficients 
\begin{equation}
    \frac{4\alpha_0}{\alpha_1}=\frac{3\alpha_1}{\alpha_2}.
\end{equation}
Thus, when one chooses the APV-gauge at the level of the action, the general projective Lovelock theory is solved by the choice of coefficients which result in the projective Lovelock-Chern-Simons theory.

\subsubsection{Dynamical \texorpdfstring{$\tilde{\Upsilon}$}{\tilde{\Upsilon}}}
\label{sec:dynamical-upsilon}

We now seek to describe the APV-gauge choice as a result of the field equations. In this dynamical $\tilde{\Upsilon}$ approach, both sets of equations remain, and in a universe devoid of matter, Eqs. \eqref{proj-LL-field-eq-gen} become
\begin{equation}\label{dynamical-upsilon-field-eq}
    0=-\tilde{D}\tilde{\mathbb{H}}{}^{\underline{B}}{}_{\underline{A}}+\tilde{\Upsilon}{}^{\underline{B}}\tilde{\mathbb{X}}_{\underline{A}},\quad\quad\quad0=\tilde{D}\tilde{\mathbb{X}}_{\underline{A}}.
\end{equation}
Explicitly, these expressions read:
\begin{equation}
\begin{split}
    0&=\left(\tilde{Q}{}^{\underline{BG}}-5\tilde{g}\tilde{\eta}{}^{\underline{BG}}\right)\wedge*\left(\tilde{\mathcal{K}}_{\underline{GA}}+\tfrac{\tilde{\alpha}_1}{2\tilde{\alpha}_2}\tilde{\mathcal{B}}_{\underline{GA}}\right)\\
    &\quad\quad+\hat{\tilde{\epsilon}}{}^{\underline{B}}{}_{\underline{ACDE}}\left(\left(\tilde{\mathcal{K}}{}^{\underline{CD}}+\tfrac{\tilde{\alpha}_1}{2\tilde{\alpha}_2}\tilde{\mathcal{B}}{}^{\underline{CD}}\right)\wedge\tilde{\mathcal{T}}{}^{\underline{E}}+\tilde{Q}{}^{\underline{FD}}\wedge\tilde{\mathcal{K}}{}^{\underline{C}}{}_{\underline{F}}\wedge\tilde{E}{}^{\underline{E}}\right)\\
    &\quad-\tilde{\Upsilon}{}^{\underline{B}}\hat{\tilde{\epsilon}}_{\underline{ACDEF}}\left(\tfrac{1}{2}\tilde{\mathcal{K}}{}^{\underline{CD}}\wedge\tilde{\mathcal{K}}{}^{\underline{EF}}+\tfrac{\tilde{\alpha}_1}{2\tilde{\alpha}_2}\tilde{\mathcal{B}}{}^{\underline{CD}}\wedge\tilde{\mathcal{K}}{}^{\underline{EF}}+\tfrac{\tilde{\alpha}_0}{6\tilde{\alpha}_2}\tilde{\mathcal{B}}{}^{\underline{CD}}\wedge\tilde{\mathcal{B}}{}^{\underline{EF}}\right),
    \end{split}
\end{equation}
\begin{equation}
    0=\tilde{D}\left(\hat{\tilde{\epsilon}}_{\underline{ACDEF}}\left(\tfrac{1}{2}\tilde{\mathcal{K}}{}^{\underline{CD}}\wedge\tilde{\mathcal{K}}{}^{\underline{EF}}+\tfrac{\tilde{\alpha}_1}{2\tilde{\alpha}_2}\tilde{\mathcal{B}}{}^{\underline{CD}}\wedge\tilde{\mathcal{K}}{}^{\underline{EF}}+\tfrac{\tilde{\alpha}_0}{6\tilde{\alpha}_2}\tilde{\mathcal{B}}{}^{\underline{CD}}\wedge\tilde{\mathcal{B}}{}^{\underline{EF}}\right)\right).
\end{equation}
These field equations may be solved simultaneously by forming the scalar expression
\begin{equation}
    \tilde{\Upsilon}{}^{\underline{A}}\tilde{\mathbb{G}}_{\underline{A}}+\beta\tilde{\Upsilon}_{\underline{B}}\tilde{\mathbb{G}}{}^{\underline{B}}{}_{\underline{A}}\wedge\tilde{E}{}^{\underline{A}}=0,
\end{equation}
for some initially arbitrary constant parameter $\beta$. We expand this expression and move the gauge-covariant derivatives around to find
\begin{equation}
    \begin{split}
        0&=\tilde{\Upsilon}{}^{\underline{A}}\tilde{\mathbb{G}}_{\underline{A}}+\beta\tilde{\Upsilon}_{\underline{B}}\tilde{\mathbb{G}}{}^{\underline{B}}{}_{\underline{A}}\wedge\tilde{E}{}^{\underline{A}}\\
        &=\tilde{\Upsilon}{}^{\underline{A}}\tilde{D}\tilde{\mathbb{X}}_{\underline{A}}+\beta\tilde{\Upsilon}_{\underline{B}}(-\tilde{D}\tilde{\mathbb{H}}{}^{\underline{B}}{}_{\underline{A}}+\tilde{\Upsilon}{}^{\underline{B}}\tilde{\mathbb{X}}_{\underline{A}})\wedge\tilde{E}{}^{\underline{A}}\\
        &=\tilde{D}(\tilde{\Upsilon}{}^{\underline{A}}\tilde{\mathbb{X}}_{\underline{A}})-\tilde{\mathbb{X}}_{\underline{A}}\wedge\tilde{E}{}^{\underline{A}}-\beta\tilde{\Upsilon}_{\underline{B}}\tilde{D}\tilde{\mathbb{H}}{}^{\underline{B}}{}_{\underline{A}}\wedge\tilde{E}{}^{\underline{A}}+\beta\tilde{\Upsilon}{}^{2}\tilde{\mathbb{X}}_{\underline{A}}\wedge\tilde{E}{}^{\underline{A}}\\
        &=\beta\left(\tilde{\Upsilon}{}^{2}-\beta^{-1}\right)\tilde{\mathbb{X}}_{\underline{A}}\wedge\tilde{E}{}^{\underline{A}}-\beta\tilde{\Upsilon}_{\underline{B}}\tilde{D}\tilde{\mathbb{H}}{}^{\underline{B}}{}_{\underline{A}}\wedge\tilde{E}{}^{\underline{A}}.
    \end{split}
\end{equation}
The last line above follows from viewing the trace of the $\tilde{\Omega}$ field equation,
\begin{equation}
     0=-\tilde{D}\tilde{\mathbb{H}}{}^{\underline{A}}{}_{\underline{A}}+\tilde{\Upsilon}{}^{\underline{A}}\tilde{\mathbb{X}}_{\underline{A}}.
\end{equation}
From Eq. \eqref{non-dyn-upsilon-1}, we easily find that $\tilde{D}\tilde{\mathbb{H}}{}^{\underline{A}}{}_{\underline{A}}=0$, and therefore,
\begin{equation}
    \tilde{\Upsilon}{}^{\underline{A}}\tilde{\mathbb{X}}_{\underline{A}}=0\quad\Rightarrow\quad\tilde{D}(\tilde{\Upsilon}{}^{\underline{A}}\tilde{\mathbb{X}}_{\underline{A}})=0.
\end{equation}

We thus find that a solution to Eqs. \eqref{dynamical-upsilon-field-eq} is provided by
\begin{equation}
    \tilde{\Upsilon}{}^{2}=\beta^{-1},\quad\quad\quad \tilde{\Upsilon}_{\underline{B}}\tilde{D}\tilde{\mathbb{H}}{}^{\underline{B}}{}_{\underline{A}}\wedge\tilde{E}{}^{\underline{A}}=0.
\end{equation}
For the latter relation to vanish, it sufficient to have $\tilde{D}\tilde{\mathbb{H}}{}^{\underline{B}}{}_{\underline{A}}=0$. The combination of $\tilde{\Upsilon}{}^{2}=\beta^{-1}$ and $\tilde{D}\tilde{\mathbb{H}}{}^{\underline{B}}{}_{\underline{A}}=0$ exactly reduces the system of equations in Eqs. \eqref{dynamical-upsilon-field-eq} to the non-dynamical $\tilde{\Upsilon}$ scenarios discussed previously. A Lorentz invariant condition for which $\tilde{\Upsilon}{}^{2}=\beta^{-1}$ is satisfied is given simply by the APV-gauge,
\begin{equation}
    \tilde{\Upsilon}{}^{2}\overset{\circ}{=}\eta_0(x^*_0)^2\quad\Rightarrow\quad \beta^{-1}=\eta_0(x^*_0)^2.
\end{equation}
Since $\eta_0=\pm1$, one may take the equivalent expression,
\begin{equation}
    \beta=\frac{\eta_0}{(x^*_0)^2}.
\end{equation}

We have thus shown that the generalized projective Higgs vector field $\tilde{\Upsilon}$ may be considered a genuinely dynamical field. For the general projective Lovelock theory, the collective set of field equations for $\tilde{\Upsilon}$ and $\tilde{\Omega}$ permit the APV-gauge choice as resulting from the field equations. Once chosen, the model may be seen to reduce to the non-dynamical $\tilde{\Upsilon}$ sector. However, in the presence of projective matter, i.e., $\tilde{\mathbb{M}}_i\neq0$, this relationship is unlikely to retain its precision, if at all. We therefore focus the remainder of this document on the development of projective matter fields.

\pagebreak
\thispagestyle{empty}
\phantomsection
\addcontentsline{toc}{section}{\hspace{-1.5em}\textbf{PART III: PROJECTIVE MATTER}}
\begingroup
\renewcommand{\addcontentsline}[3]{}
\vspace*{5cm}
\noindent
\makebox[\textwidth]{\Huge \textbf{PART III:}}\vspace{+3em}
\makebox[\textwidth]{\Huge \textbf{PROJECTIVE MATTER}}
\vfill
\label{part-III}
\endgroup

\pagebreak

\section{Introduction}

Projective spinors have long been studied in various contexts, each leading to slightly different results and interpretations. Some notable approaches to the theory of projective spinors are those of Veblen and friends \cite{projective-spinors-veblen-2comp,projective-spinors-veblen-4comp,projective-spinors-veblen-differentiation,projective-spinors-veblen-intro,projective-spinors-veblen-dirac-eq}, Pauli \cite{pauli-mayer-2,pauli-5homo-1,pauli-5homo-2,pauli-mayer-1}, Cartan \cite{cartan-spinors}, Lee \cite{projective-spinors-lee}, and Frescura \cite{projective-spinors-frescura}. Many of these are nicely compiled into a detailed review by Zund \cite{projective-spinors-zund}. Interest in projective theories of matter spans beyond the spinors themselves, including attempts to describe all the Standard Model fermions in a complex projective framework \cite{complex-proj-fermions}. 

Closely related to projective spinors are those of Kaluza-Klein theories \cite{5d-spinors}, (Anti)-de Sitter theories \cite{de-sitter-fermion,de-sitter-fermion-2}, and the so-called \textit{world spinors} associated with Metric-Affine gravitational theories \cite{GL-spinor4,world-spinors,world-spinors-2,sl5-spin,sl5-spin-2,without-metric}. World spinors are infinite dimensional representations of the double covering of the $4$-dimensional special linear group, and follow from nonlinearly realizing the double cover cover of the homogeneous (holonomic) diffeomorphism group over the double cover of the special linear group. This internal special linear group has even gained attention in the context of the entire Standard Model spectrum \cite{sl4-standard-model}. For these reasons, the connection between world spinors and the following construction is of deep interest, and will be a large part of future investigation.

In Part III of this document, we begin with a construction of the $\mathfrak{p}^{\mathfrak{w}}$-gamma matrices satisfying the Clifford algebra associated with the $SL(m+1,\mathbb{R})\cong PGL(m,\mathbb{R})$ Goldstone metric $H$. We then nonlinearly realize the symmetry group to form the $\bar{\mathfrak{p}}{}^{\mathfrak{w}}$-gamma matrices and discuss some of their useful and salient features. We then define $\mathfrak{p}^{\mathfrak{w}}$-spinor fields according to their behavior under the appropriate spin representation of the local $SL(m+1,\mathbb{R})$ gauge transformations. From these, we derive the adjoint $\mathfrak{p}{}^{\mathfrak{w}}$-spinor fields from a general, spacetime-dependent $\mathfrak{p}{}^{-\mathfrak{r}}$-spinor metric. These objects are then subjected to the nonlinear realization procedure, resulting in (adjoint) $\bar{\mathfrak{p}}{}^{\mathfrak{w}}$-spinors and a $\bar{\mathfrak{p}}{}^{-\mathfrak{r}}$-spinor metric. We construct a $\mathfrak{p}^{\mathfrak{w}}$-spinor gauge-covariant derivative and derive its action on the relevant fields. This is then extended to a $\bar{\mathfrak{p}}{}^{\mathfrak{w}}$-spinor gauge-covariant derivative. A covariant Dirac-type action is chosen, such that it is consistent with the projective Lovelock action of the gravitational sector. Requiring reality of the action forces the construction of an explicitly self-adjoint projective Dirac-type operator. The field equations of this system are found and some properties investigated. In particular, we find an induced chiral mass, and we find that reality of the action prevents an interaction with the projective Schouten form. 

Part III concludes with a short detailed discussion of the current density formed from $\bar{\mathfrak{p}}{}^{\mathfrak{w}}$-spinors, and the square of the self-adjoint operator for use in future investigations of the chiral anomaly. We then consider the projective matter sector of the coupled projective gravity/spinor theory. This includes finding the field variations for the gravitational contributions to the matter sector, and the matter contributions to the gravitational sector. The fully coupled theory is left for future investigation. Lastly, we discuss the Thomas-Whitehead-Dirac action \cite{gen-struc} applied to the present formalism. 

In all of Part III, we suppress projective spinor indices. Additionally, we restrict all discussion of projective spinors to even $m=2k$ dimensions, with $k\in\mathbb{N}$, so that a notion of chirality exists. Furthermore, we take the $m=2k$ dimensions to be of split-signature $(p,q)$ with $p$ positive (timelike) entries and $q$ negative (spacelike) entries, and consider only odd $p=2n+1$ with $n\in\mathbb{N}_0$. This restricts the number of spacelike entries to also be odd, $q=2r-1$ with $r=k-n\in\mathbb{N}$. 

\section{Gamma Matrices}
\label{sec:gamma-matrix}

In this section, we develop the $\mathfrak{p}$-gamma matrices for use in discussing dynamical $\mathfrak{p}$-spinor fields. We then nonlinear realize the symmetry group over the projective Lorentz subgroup to form the $\bar{\mathfrak{p}}$-gamma matrices. Since the $\bar{\mathfrak{p}}$-gamma matrices will essentially contain the ordinary flat gamma matrices over Minkowski space, we first briefly review the latter. Much of this review content can be found in any standard reference on quantum field theory \cite{peskin}, group theory \cite{lorentz-group}, the Dirac equation \cite{curved-dirac}, or supersymmetry \cite{muller-supersymmetry}. For ease, we restrict the introductory section to $m=4$ dimensions with split-signature $(p,q)=(1,3)$.

\subsection{Lorentz Gamma Matrices}

The ordinary flat gamma matrices $\gamma^{\underline{a}}$ of an $m=4$-dimensional Minkowski space $M_4$ satisfy the Clifford algebra $CL(M_4)$ given by
\begin{equation}\label{CL(M4)}
\{\gamma^{\underline{a}},\gamma^{\underline{b}}\}=2\eta^{\underline{ab}}\bm{1}_m,
\end{equation}
where
\begin{equation}
    \{\gamma^{\underline{a}},\gamma^{\underline{b}}\}:=\gamma^{\underline{a}}\gamma^{\underline{b}}+\gamma^{\underline{b}}\gamma^{\underline{a}}
\end{equation}
is the anti-commutator. A particularly important property of $\gamma^{\underline{a}}$ that will require modification in the projective generalization is the Hermitian adjoint, defined as the conjugate-transpose over the spin indices. This may be accomplished via
\begin{equation}\label{hermitian-flat-gamma-4}
(\gamma^{\underline{a}})^{\dagger}=\gamma^{\underline{0}}\gamma^{\underline{a}}\gamma^{\underline{0}}.
\end{equation}
In general, this expression depends on $\eta^{\underline{00}}$, since
\begin{equation}
(\gamma^{\underline{0}})^{\dagger}=\eta^{\underline{00}}\gamma^{\underline{0}},\quad\quad\quad(\gamma^{\underline{i}})^{\dagger}=-\eta^{\underline{00}}\gamma^{\underline{i}},
\end{equation}
where $\underline{i}=\underline{1},\underline{2},\underline{3}$. For our particular choice of signature $(p,q)=(1,3)$,
\begin{equation}
(\gamma^{\underline{0}})^{\dagger}=\gamma^{\underline{0}},\quad\quad\quad(\gamma^{\underline{i}})^{\dagger}=-\gamma^{\underline{i}}.
\end{equation}
As we will find, Eq. \eqref{hermitian-flat-gamma-4} will require modification in the projective generalization so that it does not alter projective weights. From Eq. \eqref{CL(M4)}, we find these matrices satisfy
\begin{equation}\label{flat-gamma-square}
\gamma^{\underline{a}}\gamma_{\underline{a}}=\eta_{\underline{ab}}\gamma^{\underline{a}}\gamma^{\underline{b}}=m=4.
\end{equation}

Under local Lorentz (gauge) transformations $\lambda(x)\equiv\{\lambda^{\underline{a}'}{}_{\underline{b}}\}\in SO(1,3)$, the gamma matrices transform as Lorentz vectors with an additional change of spin-basis, given by the similarity transformation
\begin{equation}
\rho(\lambda^{-1})\gamma^{\underline{a}'}\rho(\lambda)=\lambda^{\underline{a}'}{}_{\underline{b}}\gamma^{\underline{b}},
\end{equation}
where $\rho(\lambda)$ is the appropriate spinor representation of $\lambda$. The spinor representation $\rho(\lambda)$ may be arrived via the exponential map of the Lorentz generators. The Lorentz generators
\begin{equation}\label{lorentz-sigma-4d}
    \sigma^{\underline{ab}}:=\frac{i}{2}[\gamma^{\underline{a}},\gamma^{\underline{b}}]
\end{equation}
form a representation of the Lorentz group and therefore, satisfy the (Lorentz) $SO(p,q)$ algebra $\mathfrak{so}(p,q)$,
\begin{equation}\label{sigma-sigma-4d}
[\sigma^{\underline{ab}},\sigma^{\underline{cd}}]=-2i\left(\eta^{\underline{ac}}\sigma^{\underline{bd}}-\eta^{\underline{bc}}\sigma^{\underline{ad}}+\eta^{\underline{ad}}\sigma^{\underline{cb}}-\eta^{\underline{bd}}\sigma^{\underline{ca}}\right),
\end{equation}
where the factor of $2$ results from the normalization chosen in Eq. \eqref{lorentz-sigma-4d}. The spinor representation of a Lorentz transformation is then given by the exponential map
\begin{equation}
    \rho(\lambda)=\exp\left(\frac{i}{4}\alpha_{\underline{ab}}\sigma^{\underline{ab}}\right),
\end{equation}
with $\alpha(x)=\{\alpha_{[\underline{ab}]}\}$ a set of $\frac{m(m-1)}{2}=6$ spacetime-dependent parameters.

The $m=2k$-dimensional chiral gamma matrix of $M_m$ is defined as
\begin{equation}
    \gamma^{\underline{m+1}}:=\frac{i^{\lfloor\frac{m-2}{2}\rfloor}}{m!}\hat{\epsilon}_{\underline{a}{}_1\dots \underline{a}{}_m}\gamma^{\underline{a}{}_1}\dots\gamma^{\underline{a}{}_m}.
\end{equation}
In odd dimensions, this expression is reducible entirely in terms of the $\gamma^{\underline{a}}$ and does not constitute an independent generator. For the physical case of interest, $m=4$, we have
\begin{equation}\label{4d-gamma-5}
    \gamma^{\underline{5}}:=\frac{i}{4!}\hat{\epsilon}_{\underline{abcd}}\gamma^{\underline{a}}\gamma^{\underline{b}}\gamma^{\underline{c}}\gamma^{\underline{d}},
\end{equation}
with $\hat{\epsilon}_{\underline{abcd}}=+1$. In other words,
\begin{equation}
\gamma^{\underline{5}}=i\gamma^{\underline{0}}\gamma^{\underline{1}}\gamma^{\underline{2}}\gamma^{\underline{3}}.
\end{equation}
In any representation, the flat chiral gamma matrix $\gamma^{\underline{m+1}}$ naturally satisfies
\begin{equation}
    (\gamma^{\underline{m+1}})^2=\bm{1}_m,\quad\quad\quad\{\gamma^{\underline{m+1}},\gamma^{\underline{a}}\}=\bm{0}_m,\quad\quad\quad (\gamma^{\underline{m+1}})^{\dagger}=\gamma^{\underline{m+1}}.
\end{equation}

Having $\gamma^{\underline{5}}$ available provides a means of generating the Poincar\'{e} group. Let
\begin{equation}
    \pi^{\underline{a}}_{\pm}:=(1\pm\gamma^{\underline{5}})\gamma^{\underline{a}},
\end{equation}
which is reminiscient of Eq. \eqref{M-generator-def}. Using the information above, it is simple to show that
\begin{equation}
[\pi^{\underline{a}}_{\pm},\pi^{\underline{b}}_{\pm}]=0,\quad\quad\quad [\pi^{\underline{a}}_{\pm},\sigma^{\underline{bc}}]=-2i(\eta^{\underline{ac}}\pi^{\underline{b}}_{\pm}-\eta^{\underline{ab}}\pi^{\underline{c}}_{\pm}).
\end{equation}
Combined with Eq. \eqref{sigma-sigma-4d}, we find a representation of the algebra associated with the Poincar\'{e} group. When discussing fermionic fields in (anti)-de Sitter gauge theories, the $\pi_{\pm}^{\underline{a}}$ and $\sigma^{\underline{ab}}$ generators are combined into a collective object via the M\"{o}bius representation, providing the generators of the (anti)-de Sitter group \cite{de-sitter-fermion,de-sitter-fermion-2}. This set of gamma matrix generators may be enlarged even further to the conformal group \cite{gamma-conformal-algebra,gamma-algebras} , which is very closely related to the projective linear group. In particular, the algebra of the conformal group may be found by taking the decomposed algebra of the projective linear group in Eq. \eqref{pgl-lorentz-algebra}, and removing all appearances of the symmetric generators, while retaining its trace. However, due to Eq. \eqref{CL(M4)}, the symmetric-traceless generators contained in the $\mathfrak{pgl}(m,\mathbb{R})$ algebra do not appear to have a corresponding representation in terms of $\gamma$. Though not discussed here, the symmetry group of the present document may be viewed as a particular instantiation of \textit{disformal symmetries} \cite{disformal-1}, since these arise from extending the conformal geometry to include traceless non-metricity or shear degrees of freedom \cite{disformal-2}.

Having these basic objects and their defining characteristics available, they will serve as the foundation on which the projective gamma matrices are constructed.

\subsection{\texorpdfstring{$\mathfrak{p}$}{\mathfrak{p}}-Gamma Matrices}
\label{subsec:proj-gamma-matrix}

We define the $\mathfrak{p}$-gamma matrices in accordance with the Clifford algebra of the homogeneous space
\begin{equation}
    CL(P\mathcal{M}):=CL(GL^{+}(m+1,\mathbb{R})/\mathbb{R}^{+})\cong CL(SL(m+1,\mathbb{R})).
\end{equation}
Utilizing the inverse $SL(m+1,\mathbb{R})$ Goldstone metric $H^{AB}$ of Eq.~\eqref{H-metric}, the $CL(P\mathcal{M})$ algebra is simply
\begin{equation}\label{sl5-clifford}
    \{\Gamma^A,\;\Gamma^B\}=2H^{AB}\bm{1}_{m}.
\end{equation}
Note that this algebra is still proportional to the $m\times m$ identity matrix $\bm{1}_m$. From the $CL(P\mathcal{M})$ algebra, we find that the $\mathfrak{p}$-gamma matrices satisfy a relation analogous to Eq. \eqref{flat-gamma-square},
\begin{equation}
    \Gamma^A\Gamma_A=H_{AB}\Gamma^A\Gamma^B=m+1.
\end{equation}
These $\mathfrak{p}$-gamma matrices satisfy the $SL(m+1,\mathbb{R})$ transformation law
\begin{equation}\label{sl-gamma-trans}
    \rho(S^{-1})\Gamma^{A'}\rho(S)=S^{A'}{}_B\Gamma^B
\end{equation}
for the appropriate spinor representation $\rho$ of $S\in SL(m+1,\mathbb{R})$. The right-hand side states that the index transforms as a proper $\mathfrak{sl}(m+1,\mathbb{R})$-vector, while the left-hand-side states that a similarity transformation, or change of spin-basis must compensate.

To find the explicit form of $\Gamma^A$, we assign the generic components
\begin{equation}
    \Gamma^A:=\mathfrak{p}\begin{pmatrix}
        \Gamma^a\\\Gamma^*
    \end{pmatrix},
\end{equation}
where the projective weight $\mathfrak{w}=1$ follows from there being only one upper (vector) index. The $CL(P\mathcal{M})$ relations of Eq.~\eqref{sl5-clifford} then provide:
\begin{equation}
\begin{split}
    \{\Gamma^a,\;\Gamma^b\}&=2\mathfrak{p}^2(h^{ab}+\frac{\eta_0}{(x^*_0)^2}\xi^a\xi^b)\bm{1}_{m},\\
    \{\Gamma^*,\;\Gamma^b\}&=-2x^*_0\mathfrak{p}^2(\change^b-\frac{\eta_0}{(x^*_0)^2}\xi^b)\bm{1}_{m}\\
    \{\Gamma^*,\;\Gamma^*\}&=2(\Gamma^{*})^2=2\mathfrak{p}^2(\eta_0+(x^*_0)^2\change^2)\bm{1}_{m}.
    \end{split}
\end{equation}
These may be solved to find explicit expressions for the components of $\Gamma^A$,
\begin{equation}\label{sl5-gamma-def}
    \Gamma^A=\mathfrak{p}\begin{pmatrix}
        \gamma^a\pm\frac{\sqrt{\eta_0}}{x^*_0}\xi^a\gamma^{\underline{m+1}}\\\pm\sqrt{\eta_0}\gamma^{\underline{m+1}}-x^*_0\change_b\gamma^b
    \end{pmatrix}.
\end{equation}
In this expression,
\begin{equation}
    \gamma^a:=r^a{}_{\underline{b}}\gamma^{\underline{b}}
\end{equation}
is defined in terms of the flat $\gamma^{\underline{a}}$ satisfying \eqref{CL(M4)}, and thus satisfy their own Clifford algebra with respect to $h^{ab}$,
\begin{equation}
    \{\gamma^a,\gamma^b\}=2h^{ab}\bm{1}_m.
\end{equation}
The $(\pm)$ appearing in Eq. \eqref{sl5-gamma-def} results from taking the square root of $\eta_0$ and is left arbitrary for generality. The $\mathfrak{p}^{-1}$-gamma matrices are found simply by lowering the vector index with $H_{AB}$,
\begin{equation}
    \Gamma_A=H_{AB}\Gamma^B=\mathfrak{p}^{-1}\begin{pmatrix}
        \gamma_a-\change_a(\xi^b\gamma_b\mp\frac{x^*_0}{\sqrt{\eta_0}}\gamma^{\underline{5}}),&
        -\frac{1}{x^*_0}(\xi^b\gamma_b\mp\frac{x^*_0}{\sqrt{\eta_0}}\gamma^{\underline{5}})
    \end{pmatrix}
\end{equation}

\subsection{\texorpdfstring{$\bar{\mathfrak{p}}$}{\bar{\mathfrak{p}}}-Gamma Matrices}
\label{subsec:nlp-gamma-matrix}

We may now construct the $\bar{\mathfrak{p}}$-counterparts of $\Gamma^A$ by applying the inverse coset element to the $\mathfrak{sl}(m+1,\mathbb{R})$-vector index. However, the transformation behavior of $\Gamma^A$ given by Eq. \eqref{sl-gamma-trans} requires an additional change of spin-basis,
\begin{equation}\label{pre-p-bar-gamma}
   \rho_{\mathfrak{w}}(\sigma)\Gamma^{\underline{A}}\rho_{\mathfrak{w}}(\sigma^{-1})=(\sigma^{-1}){}^{\underline{A}}{}_B\Gamma^B.
\end{equation}
Since the left-hand side represents a change of spin-basis, and we require all considerations invariant under such changes of spin-basis, we simply define the above as the nonlinearly realized $\bar{\mathfrak{p}}$-gamma matrix,
\begin{equation}\label{p-bar-gamma}
   \tilde{\Gamma}{}^{\underline{A}}:=(\sigma^{-1}){}^{\underline{A}}{}_B\Gamma^B =\bar{\mathfrak{p}}\begin{pmatrix}
        \gamma^{\underline{a}}\\\pm\sqrt{\eta_0}\gamma^{\underline{m+1}}
    \end{pmatrix}.
\end{equation}

These nonlinearly realized gamma matrices now satisfy the Clifford algebra of the projective Lorentz group, which may easily be related to the (anti)-de Sitter group, since
\begin{equation}\label{CL5}
    \{\tilde{\Gamma}{}^{\underline{A}},\;\tilde{\Gamma}{}^{\underline{B}}\}=2\tilde{\eta}{}^{\underline{AB}}\bm{1}_{m}.
\end{equation}
From these relations, we once again find $\tilde{\Gamma}_{\underline{A}}\tilde{\Gamma}{}^{\underline{A}}=m+1$, where
\begin{equation}
    \tilde{\Gamma}_{\underline{A}}:=\tilde{\eta}_{\underline{AB}}\tilde{\Gamma}{}^{\underline{B}}=\bar{\mathfrak{p}}{}^{-1}\begin{pmatrix}
        \gamma_{\underline{a}},&\pm\frac{1}{\sqrt{\eta_0}}\gamma^{\underline{m+1}}
    \end{pmatrix},
\end{equation}
and $\gamma_{\underline{a}}:=\eta_{\underline{ab}}\gamma^{\underline{b}}$. For convenience, we further define the $\bar{\mathfrak{p}}{}^0$-gamma matrices $\tilde{\gamma}{}^{\underline{A}}$ and $\tilde{\gamma}_{\underline{A}}$, following our prescription, as
\begin{equation}
\tilde{\Gamma}{}^{\underline{A}}:=\bar{\mathfrak{p}}\tilde{\gamma}{}^{\underline{A}},\quad\quad\quad\tilde{\Gamma}_{\underline{A}}:=\bar{\mathfrak{p}}{}^{-1}\tilde{\gamma}_{\underline{A}}.
\end{equation}

Pulling back the $\bar{\mathfrak{p}}$-gamma matrices to $V\mathcal{M}$ produces, in the APV-gauge, 
\begin{equation}
    \tilde{\Gamma}{}^M:=(\tilde{E}{}^{-1})^M{}_{\underline{A}}\tilde{\Gamma}{}^{\underline{A}}\overset{\circ}{=}\begin{pmatrix}
        \gamma^{m}\\\frac{x^*}{x^*_0}(\pm\sqrt{\eta_0}\gamma^{\underline{m+1}}-x^*_0g\cdot\gamma)
    \end{pmatrix},
\end{equation}
where
\begin{equation}
    \gamma^m\overset{\circ}{:=}(\overline{\vartheta}{}^{-1})^m{}_{\underline{a}}\gamma^{\underline{a}},
\end{equation}
and
\begin{equation}
    g\cdot\gamma\overset{\circ}{:=}g_m(\overline{\vartheta}{}^{-1})^m{}_{\underline{a}}\gamma^{\underline{a}}.
\end{equation}
These satisfy the Clifford algebra $CL(V\mathcal{M})$, given by
\begin{equation}
    \{\tilde{\Gamma}{}^{M},\tilde{\Gamma}{}^N\}=2G^{MN}\bm{1}_m,
\end{equation}
with $G^{MN}$ the inverse of the Higgs-metric in Eq. \eqref{higgs-metric-EE}. We therefore find that the $\tilde{\Gamma}{}^M$ reduce in the APV-gauge precisely to the gamma matrices of Thomas-Whitehead theory \cite{gen-struc}. The \textit{key difference} is that here, we treat the flat tangent spaces as projective, resulting in factors of $\mathfrak{p}$.

The $\bar{\mathfrak{p}}{}^2$-Lorentz ((A)-dS) generator is, up to a complex half, the antisymmetric combination
\begin{equation}\label{big-flat-sigma-def}
\tilde{\Sigma}{}^{\underline{AB}}:=\frac{i}{2}[\tilde{\Gamma}{}^{\underline{A}},\tilde{\Gamma}{}^{\underline{B}}]=\bar{\mathfrak{p}}{}^2\tilde{\sigma}{}^{\underline{AB}},
\end{equation}
where
\begin{equation}
    \tilde{\sigma}{}^{\underline{AB}}:=\frac{i}{2}[\tilde{\gamma}{}^{\underline{A}},\tilde{\gamma}{}^{\underline{B}}]
\end{equation}
is the ordinary $(m+1)$-dimensional extension of the Lorentz generator in Eq. \eqref{lorentz-sigma-4d}. 

Following \cite{clifforms}, it will be convenient to have defined the gamma matrix $1$-form---the so-called $\tilde{\Gamma}$-basis or \textit{Clifform} basis,
\begin{equation}
    \tilde{\Gamma}:=\tilde{\Gamma}_{\underline{A}}\tilde{E}{}^{\underline{A}}.
\end{equation}
These are genuine projective scalars ($\bar{\mathfrak{p}}$-independent). Although these will not be used extensively, their introduction will provide convenience in some areas. In particular, the dual operator acting on the $\tilde{\Gamma}$-basis is defined here as acting on the $\bar{\mathfrak{p}}$-co-frame $\tilde{E}{}^{\underline{A}}$,
\begin{equation}
\begin{split}
    *\tilde{\Gamma}&=\tilde{\Gamma}{}^{\underline{A}}*(\tilde{E})_{\underline{A}}\\
    &=\tilde{\Gamma}{}^{\underline{A}}\hat{\tilde{\epsilon}}_{\underline{A}}\\
    &=\tilde{\Gamma}_{\underline{A}}*\tilde{E}{}^{\underline{A}}\\
    &=\frac{1}{m!}\tilde{\Gamma}{}^{\underline{A}_1}\hat{\tilde{\epsilon}}_{\underline{A}_1\underline{A}_2\dots\underline{A}_{m+1}}\tilde{E}{}^{\underline{A}_2}\wedge\dots\wedge\tilde{E}{}^{\underline{A}_{m+1}}.
    \end{split}
\end{equation}
In this basis, the generator in Eq. \eqref{big-flat-sigma-def} becomes the $2$-form
\begin{equation}
    \tilde{\Sigma}:=\frac{i}{2}\tilde{\Gamma}\wedge\tilde{\Gamma}.
\end{equation}

It will also be convenient to have available the Hermitian conjugate of the dual gamma matrix. For this, we note that $\hat{\tilde{\epsilon}}_{\underline{A}_1\underline{A}_2\dots\underline{A}_{m+1}}$ is simply a set of real constants proportional to the spinor identity $\bm{1}_m$, and is thus unaffected by such an operation. However, the wedge product of real $\bar{\mathfrak{p}}$-co-frames will reverse order. We therefore find
\begin{equation}
\begin{split}
    (*\tilde{\Gamma}){}^{\dagger}&=(\frac{1}{m!}\tilde{\Gamma}{}^{\underline{A}_1}\hat{\tilde{\epsilon}}_{\underline{A}_1\underline{A}_2\dots\underline{A}_{m+1}}\tilde{E}{}^{\underline{A}_2}\wedge\dots\wedge\tilde{E}{}^{\underline{A}_{m+1}})^{\dagger}\\
    &=\frac{1}{m!}(\tilde{E}{}^{\underline{A}_2}\wedge\dots\wedge\tilde{E}{}^{\underline{A}_{m+1}})^{\dagger}(\hat{\tilde{\epsilon}}_{\underline{A}_1\underline{A}_2\dots\underline{A}_{m+1}})^{\dagger}(\tilde{\Gamma}{}^{\underline{A}_1})^{\dagger}\\
    &=\frac{1}{m!}(\tilde{\Gamma}{}^{\underline{A}_1})^{\dagger}\hat{\tilde{\epsilon}}_{\underline{A}_1\underline{A}_2\dots\underline{A}_{m+1}}\tilde{E}{}^{\underline{A}_{m+1}}\wedge\dots\wedge\tilde{E}{}^{\underline{A}_2}\\
    &=\frac{1}{m!}(-1)^{\mho}(\tilde{\Gamma}{}^{\underline{A}_1})^{\dagger}\hat{\tilde{\epsilon}}_{\underline{A}_1\underline{A}_2\dots\underline{A}_{m+1}}\tilde{E}{}^{\underline{A}_{2}}\wedge\dots\wedge\tilde{E}{}^{\underline{A}_{m+1}}\\
     &=(-1)^{\mho}(\tilde{\Gamma}{}^{\underline{A}})^{\dagger}\hat{\tilde{\epsilon}}_{\underline{A}},
    \end{split}
\end{equation}
where, in $m=2k$,
\begin{equation}\label{upsidedown-omega}
    \mho:=\sum_{j=1}^{m-1}j=\frac{m(m-1)}{2}=k(k-1)
\end{equation}
accounts for the parity factors resulting from the anti-commutativity of the wedge product. For the physical dimension of interest, $m+1=5$, we have $\mho=2$ and
\begin{equation}\label{pre-damma-flat-tilde-gamma}
     (*\tilde{\Gamma}){}^{\dagger}=(\tilde{\Gamma}{}^{\underline{A}})^{\dagger}\hat{\tilde{\epsilon}}_{\underline{A}}.
\end{equation}
Since we have not yet developed any notion of conjugation for $\tilde{\Gamma}{}^{\underline{A}}$, Eq. \eqref{pre-damma-flat-tilde-gamma} is as far as we can go. This notion will be developed in Sec. \ref{subsec:proj-spinors}.

Lastly, we note that in $m+1=5$ dimensions,
\begin{equation}
    \pm\sqrt{\eta_0}=\frac{i}{5!}\hat{\tilde{\epsilon}}_{\underline{ABCDE}}\tilde{\Gamma}{}^{\underline{A}}\tilde{\Gamma}{}^{\underline{B}}\tilde{\Gamma}{}^{\underline{C}}\tilde{\Gamma}{}^{\underline{D}}\tilde{\Gamma}{}^{\underline{E}}.
\end{equation}
This is exemplary of the fact that one cannot define any notion of chirality in $5$-dimensions, see Eq. \eqref{4d-gamma-5} for reference. However, the value of $\eta_0=\pm1$ may, from the above identity, provide one information about the orientation of the flat projective tangent spaces. From this identity, we may also peel off one factor of $\tilde{\Gamma}$ to obtain an alternative expression for $\tilde{\Gamma}$,
\begin{equation}
    \tilde{\Gamma}_{\underline{A}}=\pm\frac{i}{\sqrt{\eta_0}}\frac{1}{4!}\hat{\tilde{\epsilon}}_{\underline{ABCDE}}\tilde{\Gamma}{}^{\underline{B}}\tilde{\Gamma}{}^{\underline{C}}\tilde{\Gamma}{}^{\underline{D}}\tilde{\Gamma}{}^{\underline{E}}.
\end{equation}
This relation simply provides a convenient means of transitioning expressions between the $\tilde{\Gamma}$-basis exterior form and the typical abstract-index form. As a simple consistency check, the above relation provides
\begin{equation}
    \tilde{\Gamma}_{\underline{*}}=\pm\frac{i}{\sqrt{\eta_0}}\frac{1}{4!}\hat{\tilde{\epsilon}}_{\underline{*bcde}}\tilde{\Gamma}{}^{\underline{b}}\tilde{\Gamma}{}^{\underline{c}}\tilde{\Gamma}{}^{\underline{d}}\tilde{\Gamma}{}^{\underline{e}}=\bar{\mathfrak{p}}{}^{-1}(\frac{\pm1}{\sqrt{\eta_0}}\gamma^{\underline{m+1}}),
\end{equation}
as expected.

The generators of $PGL(m,\mathbb{R})\cong SL(m+1,\mathbb{R})$ are not easily represented by these gamma matrices. As noted previously, omitting the $m$-dimensional shear generators, these may be shown to generate the $m$-dimensional conformal group \cite{gamma-conformal-algebra}. The complication arises from the necessity of a vector operator, of which $PGL(m,\mathbb{R})$ has two, defined in Eq \eqref{M-generator-def}. Finding a solution to this issue is not undertaken in this document, but will likely be found in combining the ideas of \cite{Ogievetsky3} and \cite{sl5-spin-2,world-spinors,world-spinors-2}. Understanding the role of the $\bar{\mathfrak{p}}$-gamma matrices in generating the projective linear group will be a part of future investigation.

\subsubsection{Clifford Basis and Representation}
\label{subsubsec:clifford-basis}

We attempt to provide a derivation of the weighted projective spinor representation utilized in \cite{gen-struc,heavy-lifting} and maintained in this document. In this section only, we include the combinatorial factors in all index (anti)-symmetrization brackets for brevity. 

In an $m$-dimensional vector space, the basis vectors $\{e^a\}$ satisfy the Clifford algebra \cite{clifford-basis-expansion}
\begin{equation}
    e^a\cdot e^b:=\frac{1}{2}\left(e^ae^b+e^be^a\right)=\frac{1}{2}\left\{e^a,e^b\right\}=e^{(a}e^{b)}=g^{ab}\bm{1}_m,
\end{equation}
with $a,b=1,2,\dots,m$, and where $g^{ab}$ is the metric of the space and $\bm{1}_m$ the $m\times m$ identity. The dot operation used is shorthand notation for the inner product. The Clifford algebra is the symmetric part of the \textit{geometric product} \cite{geometric-algebra}
\begin{equation}
    e^ae^b=e^a\cdot e^b+e^a\wedge e^b,
\end{equation}
where the antisymmetric part of the geometric product is given by the outer product
\begin{equation}
    e^a\wedge e^b:=\frac{1}{2}\left(e^ae^b-e^be^a\right)=\frac{1}{2}\left[e^a,e^b\right]=e^{[a}e^{b]}.
\end{equation}
The inner product produces a scalar, while the outer product produces a bi-vector. In general, an $r$-vector may be composed from the outer product of $r$ basis vectors as
\begin{equation}
    e^{a_1}\wedge\dots\wedge e^{a_r}=\frac{1}{r!}\left[e^{a_1},\dots,e^{a_r}\right]=e^{[a_1},\dots,e^{a_r]},
\end{equation}
with the last equality denoting the complete antisymmetrization of the indices (See Appendix \hyperref[app-A:orientation]{A.2}). An arbitrary element $W$ of the Clifford algebra may be expanded in terms of the $r$-vectors as
\begin{equation}
    W=\sum_{n=0}^{m}\wedge^ne^{a_{n}}W_{[a_{n}]},
\end{equation}
for some generally complex set of coefficients $W_{[a_{n}]}$, which are completely antisymmetric for $n>2$. As an explicit example, consider $m=4$ dimensions. Then $W$ has the form
\begin{equation}
    W=W_0\bm{1}_m+W_{a}e^a+W_{[ab]}e^a\wedge e^b+W_{[abc]}e^a\wedge e^b\wedge e^c+W_{[abcd]}e^a\wedge e^b\wedge e^c\wedge e^d.
\end{equation}

In a $m=4$-dimensional flat space, we may take $\{e^a\}\rightarrow\{\gamma^{\underline{a}}\}$. The algebra element $W$ may then be written as
\begin{equation}\label{C-gamma-expansion}
W=W_0\bm{1}_m+W_{\underline{a}}\gamma^{\underline{a}}+W_{[\underline{ab}]}\gamma^{\underline{a}}\wedge \gamma^{\underline{b}}+W_{[\underline{abc}]}\gamma^{\underline{a}}\wedge \gamma^{\underline{b}}\wedge \gamma^{\underline{c}}+W_{[\underline{abcd}]}\gamma^{\underline{a}}\wedge \gamma^{\underline{b}}\wedge \gamma^{\underline{c}}\wedge \gamma^{\underline{d}}.
\end{equation}
From Eq. \eqref{4d-gamma-5}, we may write
\begin{equation}
    i\hat{\epsilon}{}^{\underline{abcd}}\gamma^{\underline{5}}=\gamma^{\underline{a}}\wedge \gamma^{\underline{b}}\wedge \gamma^{\underline{c}}\wedge \gamma^{\underline{d}},
\end{equation}
and
\begin{equation}
    i\hat{\epsilon}{}^{\underline{abcd}}\gamma^{\underline{5}}\gamma_{\underline{a}}=\gamma^{\underline{b}}\wedge \gamma^{\underline{c}}\wedge \gamma^{\underline{d}}.
\end{equation}
Then, introducing also the Lorentz generators from Eq. \eqref{lorentz-sigma-4d}, we may rewrite Eq. \eqref{C-gamma-expansion} as
\begin{equation}\label{expanded-clifford-C}
W=W_0\bm{1}_m+W_{\underline{a}}\gamma^{\underline{a}}+iW_{\underline{ab}}\sigma^{\underline{ab}}+i\hat{W}_{\underline{a}}\gamma^{\underline{a}}\gamma^{\underline{5}}+i\hat{W}\gamma^{\underline{5}},
\end{equation}
where, for notational convenience:
\begin{equation}
 W_{\underline{ab}}:=-W_{[\underline{ab}]},\quad \quad\hat{W}_{\underline{d}}:=W_{[\underline{abc}]}\hat{\epsilon}{}^{\underline{abc}}{}_{\underline{d}},\quad\quad\hat{W}:=W_{\underline{abcd}}\hat{\epsilon}{}^{\underline{abcd}}.
\end{equation}
We therefore find that the collection of $16$ linearly-independent objects
\begin{equation}
    B_{\Gamma}=\left\{\eta^{\underline{ab}}\bm{1}_m,\gamma^{\underline{a}},i\sigma^{\underline{ab}},i\gamma^{\underline{a}}\gamma^{\underline{5}},i\gamma^{\underline{5}}\right\}
\end{equation}
forms a basis for the space of $4\times4$ complex matrices. In other words, any $4\times 4$ complex matrix $W$ can be written as a linear combination in the basis provided by $B_{\Gamma}$.

The basis $B_{\Gamma}$ may be repackaged concisely in terms of the nonlinear projective gamma matrices, $\tilde{\Gamma}$. In other words, the repackaging provides a projective reinterpretation. In order to permit some generality in the choice of signature, the factors of $i$ accompanying each $\gamma^{\underline{5}}$ are replaced with factors of $\pm\sqrt{\eta_0}$. Since $\tilde{\Gamma}{}^{\underline{A}}$ simply adjoins $\pm\sqrt{\eta_0}\gamma^{\underline{5}}$ to $\gamma^{\underline{a}}$, we may repackage $B_{\Gamma}$ by forming
\begin{equation}
    \tilde{\Gamma}{}^{\underline{AB}}=\frac{1}{4}\tilde{\Gamma}{}^{\underline{A}}\tilde{\Gamma}{}^{\underline{B}},
\end{equation}
where we have chosen a normalization factor for later convenience. Since
\begin{equation}
    \tilde{\Gamma}{}^{(\underline{AB})}=\frac{1}{4}\tilde{\eta}{}^{\underline{AB}}\bm{1}_m,\quad\quad\quad \tilde{\Gamma}{}^{[\underline{AB}]}=\frac{-i}{4}\tilde{\Sigma}{}^{\underline{AB}},
\end{equation}
we have the equivalent, repackaged basis provided by $B_{\tilde{\Gamma}}=\{\tilde{\Gamma}{}^{\underline{AB}},\tilde{\Gamma}{}^{\underline{A}}\}$. Writing
\begin{equation}
    \tilde{W}_{\underline{A}}:=\begin{pmatrix}
        \tilde{W}_{\underline{a}},&\hat{\tilde{W}}
\end{pmatrix},\quad\tilde{W}_{\underline{AB}}:=4(\tilde{W}_{(\underline{AB})}+\tilde{W}_{[\underline{AB}]}),\quad\tilde{W}_{[\underline{AB}]}:=\begin{pmatrix}
        \tilde{W}_{[\underline{ab}]}&-\hat{\tilde{W}}_{\underline{b}}\\\hat{\tilde{W}}_{\underline{a}}&0
    \end{pmatrix}
\end{equation}
as the collective set of generally complex coefficients, where the factor of $4$ is an artifact of normalization, we may express the $\bar{\mathfrak{p}}{}^0$ nonlinear projective Clifford algebra element $\tilde{W}$ as
\begin{equation}\label{C}
    \tilde{W}=\tilde{W}_{\underline{A}}\tilde{\Gamma}{}^{\underline{A}}+\tilde{W}_{\underline{AB}}\tilde{\Gamma}{}^{\underline{AB}}.
\end{equation}
When expanded, this produces
\begin{equation}
\tilde{W}=\tilde{W}_0\bm{1}_m+\tilde{W}_{\underline{a}}\tilde{\Gamma}{}^{\underline{a}}+\hat{\tilde{W}}\tilde{\Gamma}{}^{\underline{*}}-i\tilde{W}_{[\underline{ab}]}\tilde{\Sigma}{}^{\underline{ab}}+\hat{\tilde{W}}_{\underline{a}}\tilde{\Gamma}{}^{\underline{a}}\tilde{\Gamma}{}^{\underline{*}},
\end{equation}
where $\tilde{W}_0:=\tilde{W}{}^{\underline{A}}{}_{\underline{A}}$. Since $\tilde{\Gamma}{}^{\underline{A}}$ and $\tilde{\Gamma}{}^{\underline{AB}}$ are of type-$\bar{\mathfrak{p}}{}^1$ and -$\bar{\mathfrak{p}}{}^2$, respectively, the coefficients $\tilde{W}_{\underline{A}}$ and $\tilde{W}_{\underline{AB}}$ are required to be of types $\bar{\mathfrak{p}}{}^{-1}$ and $\bar{\mathfrak{p}}{}^{-2}$, respectively. This is easily confirmed by counting indices and noting their upper and lower placements. This requirement ensures that $\tilde{W}$ is of a single type, i.e., $\bar{\mathfrak{p}}{}^{0}$. Following our conventions, we may therefore express $\tilde{W}$ as
\begin{equation}\label{nonlinear-clifford-C}
    \tilde{W}=\tilde{w}_{\underline{A}}\tilde{\gamma}{}^{\underline{A}}+\tilde{w}_{\underline{AB}}\tilde{\gamma}{}^{\underline{AB}}=\tilde{w}_0\bm{1}_m+\tilde{w}_{\underline{a}}\gamma^{\underline{a}}\pm\sqrt{\eta_0}\;\hat{\tilde{w}}\gamma^{\underline{5}}+i\tilde{w}_{\underline{ab}}\sigma{}^{\underline{ab}}\pm\sqrt{\eta_0}\;\hat{\tilde{w}}_{\underline{a}}\gamma{}^{\underline{a}}\gamma^{\underline{5}},
\end{equation}
exactly reproducing Eq. \eqref{expanded-clifford-C} as claimed.

On $P\mathcal{M}$, we expect weighted (projective) spinor representations $\rho_{\mathfrak{w}}(\tilde{W})$, where $\rho(\dots)$ is the representation and $\mathfrak{w}\in \mathbb{C}$ a generally complex projective weight. We focus on representing the nonlinearly realized objects $\rho_{\mathfrak{w}}(\tilde{W})$, though the following holds just as well for $\rho_{\mathfrak{w}}(W)$. An element of the weighted spin representation may generally be formed by applying the exponential map to a matrix expanded in $B_{\tilde{\Gamma}}$. Applying the exponential map to (minus) Eq. \eqref{nonlinear-clifford-C}, and using the fact that $\tilde{W}$ is of type-$\bar{\mathfrak{p}}{}^0$, we find
\begin{equation}
    \rho_{\mathfrak{w}}(\tilde{W})=\exp\left(-\tilde{w}_{\underline{A}}\tilde{\gamma}{}^{\underline{A}}-\tilde{w}_{\underline{AB}}\tilde{\gamma}{}^{\underline{AB}}\right).
\end{equation}
Since $\tilde{w}_{\underline{AB}}$ is a general complex set of coefficients, we may separate out its trace. Let $\cancel{\tilde{w}}_{\underline{AB}}$ denote the traceless parts of $\tilde{w}_{\underline{AB}}$. Due to the Clifford algebra relations, the symmetric part of the $\tilde{w}_{\underline{AB}}$ coefficients collapses to its trace. As a result, the symmetric traceless parts of $\tilde{w}_{\underline{AB}}$ are inaccessible. Therefore, the traceless parts of $\tilde{w}_{\underline{AB}}$ are also antisymmetric, $\cancel{\tilde{w}}_{\underline{AB}}=\tilde{w}_{[\underline{AB}]}$. Thus, 
\begin{equation}
    \rho_{\mathfrak{w}}(\tilde{W})=\exp\left(-\tilde{w}_{\underline{A}}\tilde{\gamma}{}^{\underline{A}}-\tilde{w}_{[\underline{AB}]}\tilde{\gamma}{}^{\underline{AB}}-\frac{1}{m+1}\tilde{w}_0\eta_{\underline{AB}}\tilde{\gamma}{}^{\underline{AB}}\right),
\end{equation}
where $\tilde{w}_0:=\tilde{w}_{\underline{CD}}\eta{}{}^{\underline{CD}}$. Note that $\eta_{\underline{AB}}$ is used in place of $\tilde{\eta}_{\underline{AB}}$, since we have simplified the expression, canceling all factors of $\bar{\mathfrak{p}}$. The expression $\eta_{\underline{AB}}\tilde{\gamma}{}^{\underline{AB}}$ appearing in $\rho_{\mathfrak{w}}(\tilde{W})$ is proportional to the spinor identity, since
\begin{equation}
    \eta_{\underline{AB}}\tilde{\gamma}{}^{\underline{AB}}=\frac{1}{4}\eta_{\underline{AB}}\tilde{\gamma}{}^{\underline{A}}\tilde{\gamma}{}^{\underline{B}}=\frac{1}{4}\eta_{\underline{AB}}\eta{}^{\underline{AB}}\bm{1}_m=\frac{m+1}{4}\bm{1}_m.
\end{equation}
Therefore, this term commutes with all other basis elements, and we may segregate this exponential as
\begin{equation}
    \rho_{\mathfrak{w}}(\tilde{W})=\exp\left(-\tilde{w}_{\underline{A}}\tilde{\gamma}{}^{\underline{A}}-\tilde{w}_{[\underline{AB}]}\tilde{\gamma}{}^{\underline{AB}}\right)\exp\left(-\frac{1}{4}\frac{1}{m+1}\tilde{w}_0\eta_{\underline{AB}}\eta{}^{\underline{AB}}\bm{1}_m\right).
\end{equation}

As stated previously, the coefficients $\tilde{w}_{\underline{AB}}$ are generally complex. We would like to relate its trace to a real parameter. We therefore consider
\begin{equation}\label{reality-trace}
\tilde{w}_{0}\eta_{\underline{AB}}=\mathfrak{w}\tilde{n}_{\underline{AB}},
\end{equation}
with the symmetric matrix $\tilde{n}_{\underline{AB}}$ having $\tilde{n}{}^{\underline{A}}{}_{\underline{A}}$ real or $i\tilde{n}{}^{\underline{A}}{}_{\underline{A}}$ real, and $\mathfrak{w}\in\mathbb{C}$ some constant coefficient. These conditions may be enforced, for example, by considering $\tilde{w}_0=\frac{1}{2}(\tilde{r}_0+i\tilde{c}_0)$ and $\mathfrak{w}=\frac{1}{2}(\mathfrak{r}+i\mathfrak{c})$. Then, reality of the trace $\tilde{n}\equiv \tilde{n}{}^{\underline{A}}{}_{\underline{A}}$ simply requires that
\begin{equation}
    \text{Im}(\tilde{n})=0\quad\Rightarrow\quad\frac{\tilde{c}_0}{\tilde{r}_0}=\frac{\mathfrak{c}}{\mathfrak{r}},
\end{equation}
while reality of $i\tilde{n}$ requires
\begin{equation}
    \text{Im}(i\tilde{n})=0\quad\Rightarrow\quad\frac{\tilde{c}_0}{\tilde{r}_0}=-\frac{\mathfrak{r}}{\mathfrak{c}}.
\end{equation}
Reflecting Eq. \eqref{reality-trace} in $\rho_{\mathfrak{w}}(\tilde{W})$ and rearranging a little, we find
\begin{equation}
    \rho_{\mathfrak{w}}(\tilde{W})=\exp\left(-\tilde{w}_{\underline{A}}\tilde{\gamma}{}^{\underline{A}}-\tilde{w}_{[\underline{AB}]}\tilde{\gamma}{}^{\underline{AB}}\right)\exp\left(\frac{\tilde{n}}{4}\bm{1}_m\right){}^{-\frac{\mathfrak{w}}{m+1}}.
\end{equation}
Since $\tilde{n}_{\underline{AB}}$ is a matrix of coefficients for an element of the algebra, the exponential of $\tilde{n}_{\underline{AB}}$, connected to the appropriate basis element ($\eta^{\underline{AB}}\bm{1}_m$), is a group element. Let
\begin{equation}
\tilde{Z}:=\exp\left(\frac{\tilde{n}}{4}\bm{1}_m\right)
\end{equation}
denote such a (central) group element. Then
\begin{equation}\label{center-separated-rho}
    \rho_{\mathfrak{w}}(\tilde{W})=\tilde{Z}{}^{-\frac{\mathfrak{w}}{m+1}}\rho(\tilde{w}_{\underline{A}},\tilde{w}_{[\underline{AB}]}),
\end{equation}
where
 \begin{equation}
\rho(\tilde{w}_{\underline{A}},\tilde{w}_{[\underline{AB}]}):=\exp\left(-\tilde{w}_{\underline{A}}\tilde{\gamma}{}^{\underline{A}}-\tilde{w}_{[\underline{AB}]}\tilde{\gamma}{}^{\underline{AB}}\right)
\end{equation}
may be shown to contain an element of the conformal group \cite{gamma-conformal-algebra}. To confirm this, we first expand $\rho(\tilde{w}_{\underline{A}},\tilde{w}_{[\underline{AB}]})$ to
\begin{equation}\label{conformal-rho}
\rho(\tilde{w}_{\underline{A}},\tilde{w}_{[\underline{AB}]})=\exp\left(-\tilde{w}_{\underline{a}}\gamma^{\underline{a}}\mp\sqrt{\eta_0}\;\hat{\tilde{w}}_{\underline{a}}\gamma{}^{\underline{a}}\gamma^{\underline{5}}\mp\sqrt{\eta_0}\;\hat{\tilde{w}}\gamma^{\underline{5}}-i\tilde{w}_{\underline{ab}}\sigma{}^{\underline{ab}}\right).
\end{equation}
Then, by considering
\begin{equation}
    \tilde{w}_{\underline{a}}\equiv\frac{1}{2}\left(\tilde{u}_{\underline{a}}-\tilde{v}_{\underline{a}}\right),\quad\quad\quad\hat{\tilde{w}}_{\underline{a}}\equiv\frac{1}{2}(\tilde{u}_{\underline{a}}+\tilde{v}_{\underline{a}}),
\end{equation}
the first two terms in Eq. \eqref{conformal-rho} combine to form
\begin{equation}
-\tilde{w}_{\underline{a}}\gamma^{\underline{a}}\mp\sqrt{\eta_0}\;\hat{\tilde{w}}_{\underline{a}}\gamma{}^{\underline{a}}\gamma^{\underline{5}}=-\tilde{u}_{\underline{a}}\bm{P}{}^{\underline{a}}-\tilde{v}_{\underline{a}}\bm{K}^{\underline{a}},
\end{equation}
where
\begin{equation}
    \bm{P}{}^{\underline{a}}:=\frac{1}{2}\left(\gamma^{\underline{a}}\pm\sqrt{\eta_0}\gamma^{\underline{a}}\gamma^{\underline{5}}\right),\quad\quad\quad \bm{K}{}^{\underline{a}}:=\frac{-1}{2}\left(\gamma^{\underline{a}}\mp\sqrt{\eta_0}\gamma^{\underline{a}}\gamma^{\underline{5}}\right)
\end{equation}
are, respectively, the generators of translations and special conformal transformations in $4$ dimensions \cite{gamma-conformal-algebra}. When taken together with the generators of $4$-dimensional dilations $\bm{D}=\pm\sqrt{\eta_0}\gamma^{\underline{5}}$ and Lorentz transformations $\bm{A}{}^{\underline{ab}}=\frac{-i}{2}\sigma^{\underline{ab}}$, the collection generates the $4$-dimensional conformal group.

Returning to Eq. \eqref{center-separated-rho}, since
\begin{equation}
    \frac{\tilde{n}}{4}\equiv\frac{1}{4}\tilde{n}_{\underline{AB}}\eta^{\underline{AB}}=\text{tr}\left(\frac{1}{4}\tilde{n}_{\underline{AB}}\right),
\end{equation}
we may utilize the identity $\exp(\text{tr}(X))=\det(\exp(X))$ to write
\begin{equation}
    \tilde{Z}{}^{\frac{-\mathfrak{w}}{m+1}}\equiv\det\left(\exp\left(\frac{1}{4}\tilde{n}_{\underline{AB}}\bm{1}_m\right)\right){}^{\frac{-\mathfrak{w}}{m+1}}=\left|\exp\left(\frac{1}{4}\tilde{n}_{\underline{AB}}\bm{1}_m\right)\right|{}^{\frac{-\mathfrak{w}}{m+1}}=:\mathfrak{z}{}^{\mathfrak{w}}.
\end{equation}
We thus arrive at the final expression for the weighted spinor representation $\rho_{\mathfrak{w}}(\dots)$,
\begin{equation}
    \rho_{\mathfrak{w}}(\tilde{W})=\mathfrak{z}^{\mathfrak{w}}\rho(\tilde{w}_{\underline{A}},\tilde{w}_{[\underline{AB}]}).
\end{equation}
Notice in Eq. \eqref{conformal-rho}, if $\mp\sqrt{\eta_0}\;\hat{\tilde{w}}\gamma^{\underline{5}}$ was included in $\tilde{Z}$, then we would also recover the chiral weight used in \cite{gen-struc}.

As a specific example, consider 
\begin{equation}
    \tilde{n}{}^{\underline{A}}{}_{\underline{B}}=\begin{pmatrix}
        h^{\underline{a}}{}_{\underline{b}}&0\\0&0
    \end{pmatrix},
\end{equation}
with $h_{[\underline{ab}]}=0$ a symmetric $m\times m$ matrix. Following the same process as in Eq. \eqref{exponential-sum-group}, and omitting the explicit spinor identity $\bm{1}_m$, we find
\begin{equation}
\begin{split}
    \mathfrak{z}{}^{\mathfrak{w}}=\tilde{Z}{}^{\frac{-\mathfrak{w}}{m+1}}&=\left|\sum^{\infty}_{n=1}\frac{1}{n!}\begin{pmatrix}
         \frac{1}{4} h^{\underline{a}}{}_{\underline{b}}&0\\0&0
    \end{pmatrix}^n\right|^{\frac{-\mathfrak{w}}{m+1}}\\
    &=\left|\begin{pmatrix}
        \exp\left(\frac{1}{4} h^{\underline{a}}{}_{\underline{b}}\right)&0\\0&1\end{pmatrix}\right|^{\frac{-\mathfrak{w}}{m+1}}\\
        &=:\left|\begin{pmatrix}
        r^{\underline{a}}{}_{\underline{b}}&0\\0&1\end{pmatrix}\right|^{\frac{-\mathfrak{w}}{m+1}}\\
        &=\left|r^{\underline{a}}{}_{\underline{b}}\right|^{\frac{-\mathfrak{w}}{m+1}}.
        \end{split}
\end{equation}
Had we taken the exponential parameterization of the coset element, as was done in \cite{ogievetsky1} for the projective linear group and \cite{higgs-mech-for-grav} for the diffeomorphism group, we would have found that the symmetric coset parameters $r^{\underline{a}}{}_{\underline{b}}$ are identified with the exponential of a set of algebraic symmetric parameters $h^{\underline{a}}{}_{\underline{b}}$. The difference in index types then follows from explicitly computing the transformation of the exponential group element, and the result will necessarily depend on the choice of stability subgroup (see Sec. \ref{sec:nlr-pgl}).  We may therefore consider the (central) group element $\tilde{Z}$ as representing the determinant of the symmetric algebraic parameters utilized in the Clifford basis expansion.

In order to obtain obtain projective factors $\mathfrak{p}$ in the weighted spinor representation, we must require the map in Eq. \eqref{F-map} to have a weighted spinor representation containing
\begin{equation}
\begin{split}
    \mathfrak{z}{}^{\mathfrak{w}}=\tilde{Z}{}^{\frac{-\mathfrak{w}}{m+1}}&=\left|\sum^{\infty}_{n=1}\frac{1}{n!}\begin{pmatrix}f^a{}_m&0\\0&-(m+1)\log\left(\frac{x^*}{x^*_0}\right)\end{pmatrix}\right|^{\frac{-\mathfrak{w}}{m+1}}\\
    &=\left|\begin{pmatrix}
        \exp(f^a{}_m)&0\\0&\left(\frac{x^*}{x^*_0}\right)^{-(m+1)}
    \end{pmatrix}\right|^{\frac{-\mathfrak{w}}{m+1}}\\
    &=|\vartheta^a{}_m|^{\frac{-\mathfrak{w}}{m+1}}\left(\frac{x^*}{x^*_0}\right)^{\mathfrak{w}}\\
    &=\mathfrak{p}^{\mathfrak{w}},
    \end{split}
\end{equation}
where $\vartheta^a{}_m:=\exp(f^a{}_m)$, just as in \cite{ogievetsky1,higgs-mech-for-grav}. The details of where this particular form of matrix comes from is left for future investigation. In the remainder of this document, we assume its validity, as it produces the sought-after generalization of \cite{gen-struc}. 

To recapitulate, the weighted spinor representation produces a factor proportional to the spinor identity, containing the determinant of the group element constructed from the symmetric parameters, raised to the power $-\mathfrak{w}/(m+1)$. We interpret this in the context of the ideas presented hitherto as contributions to the projective factor raised to the power $\mathfrak{w}\in\mathbb{C}$. To fix a convenient notation, we will often write
\begin{equation}\label{factored-p-spinor-rep}
   \rho_{\mathfrak{w}}(\tilde{W})=\rho_{\mathfrak{w}}(\bar{\mathfrak{p}})\rho(\tilde{w})=\bar{\mathfrak{p}}{}^{\mathfrak{w}}\rho(\tilde{w}),
\end{equation}
where it is understood that $\rho(\tilde{w})\equiv\rho(\tilde{w}_{\underline{A}},\tilde{w}_{[\underline{AB}]})$ depends only on the $m$-dimensional pseudoscalar, (pseudo)-vector, and bi-vector (Lorentz) parameters.

\section{Spinors}
\label{sec:spinors}

In this section we develop spinors, their projective extensions and their subsequent nonlinear realization. On $\mathcal{M}$, an ordinary $4$-component Lorentz bi-spinor satisfies the transformation law
\begin{equation}
\lambda(x):\psi(x)\rightarrow\psi'(x')=\rho(\lambda)\psi(\lambda^{-1}x),
\end{equation}
for $\rho(\lambda)$ the appropriate spin representation of the local Lorentz transformation matrix $\lambda\in SO(1,3)$. This transformation property of spinor fields will serve as the foundation on which the projective spinor fields are constructed. We simply replace the group of transformations with the particular gauge group of interest, and derive many of the resulting implications.

\subsection{\texorpdfstring{$\mathfrak{p}$}{\mathfrak{p}}-Spinors}
\label{subsec:proj-spinors}

Since the $\bar{\mathfrak{p}}$-gamma matrices are constructed from the usual flat $\gamma^{\underline{a}}$ and $\gamma^{\underline{5}}$, and further, satisfy Eq. \eqref{CL5}, spinors on $V\mathcal{M}$ will retain their $m$-dimensional character. We thus require the transformation 
\begin{equation}\label{gl-spinors}
      G(x):\breve{\Psi}(x,x^*)\rightarrow \breve{\Psi}'(x,x^*)=\rho(G)\breve{\Psi}(x,x^*),
\end{equation}
for $\rho(G)$ the appropriate spin representation of $G\in GL(m+1,\mathbb{R})$. In the above and throughout, we omit the explicit action on coordinates for brevity. The $\rho(G)$ are known to be infinite dimensional representations \cite{sl5-spin,sl5-spin-2}, and are tied with the basis-complication note of the previous section.

In order to apply the nonlinear framework, these spinor fields must be carried over to $P\mathcal{M}$. Recall that the purpose of mapping to $P\mathcal{M}$ is to separate all dependence on $x^*$ into a factor proportional to the identity. We assume a similar statement to hold for projective spinors, and therefore consider
\begin{equation}\label{p-spinor-def}
    \mathcal{F}:\breve{\Psi}(x,x^*)\rightarrow \Psi_{\mathfrak{w}}(x,x^*):=\rho_{\mathfrak{w}}(\mathfrak{p}(x,x^*))\psi(x),
\end{equation}
where $\rho_{\mathfrak{w}}(\mathfrak{p})$ denotes the appropriate \textit{weighted} spinor representation of $\mathfrak{p}$, and $\mathcal{F}$ is the map defined in Eq. \eqref{F-map}. Assuming the spinor action of this map to take the form of Eq. \eqref{p-spinor-def} accounts for the possibility of exotic projective weights. To distinguish this (potential) subclass of representations, we refer to these as $\mathfrak{p}{}^{\mathfrak{w}}$-spinors. By definition, the spinor $\psi(x)$ is a $\mathfrak{p}^0$-spinor. Related information on weighted spinors or spinor densities may be found in \cite{projective-spinors-veblen-4comp,spinor-density}. For simplicity, we will write $\Psi$ in place of $\Psi_{\mathfrak{w}}$ unless it is unclear from context.

Local $SL(m+1,\mathbb{R})$ gauge transformations now act on $\Psi$ as
\begin{equation}
      S(x):\Psi(x,x^*)\rightarrow \Psi'(x,x^*)=\rho_{\mathfrak{w}}(S)\Psi(x,x^*).
\end{equation}
For this transformation to hold, we appeal to Eq. \eqref{factored-p-spinor-rep} and require a factorization of the form
\begin{equation}
    \rho(S)=\rho_{\mathfrak{w}}(|s|^{\frac{-1}{m+1}})\rho(s,t,u),
\end{equation}
so that
\begin{equation}
    \Psi'(x,x^*)=\rho_{\mathfrak{w}}(\mathfrak{p}|s|^{\frac{-1}{m+1}})\rho(s,t,u)\psi(x),
\end{equation}
where $s,t,u$ are the gauge parameters used in Eqs. \eqref{SL-matrix} and \eqref{SL-matrix-inv}.

The adjoint $\mathfrak{p}$-spinor $\overline{\Psi}(x,x^*)$ is defined in such a way that bilinear combinations transform appropriately. To construct spinor bilinears, we must introduce a non-degenerate, homogeneously gauge-transforming spinor-bilinear form or simply a \textit{spinor metric}, denoted $M$. Let the \textit{spacetime-dependent} spinor metric $M=M(x)$ be the $\mathfrak{p}^\mathfrak{-r}$-spinor-valued $m\times m$ matrix that defines the inner product $(\cdot\Vert\cdot)$ on the $\mathfrak{p}$-spinor representation space,
\begin{equation}\label{spinor-product}
    (\Psi\Vert\Psi):=\Psi^{\dagger}M\Psi=\overline{\Psi}\Psi.
\end{equation}
The inner product $(\cdot\Vert\cdot)$ is defined to have the operation of transposition and complex conjugation, denoted $\dagger$, on the left object, followed by a (spin-index) contraction with the right object via the $\mathfrak{p}^\mathfrak{-r}$-spinor metric $M$. As is obvious, we suppress all spinor-valued indices, and explicit $x$-dependence for ease of readability. The introduction of $M$ provides the definition of the Pauli-adjoint $\mathfrak{p}$-spinor, or simply, the adjoint $\mathfrak{p}$-spinor,
\begin{equation}\label{p-spinor-adjoint}
\overline{\Psi}:=\Psi^{\dagger}M=\psi^{\dagger}[\rho_{\mathfrak{w}}(\mathfrak{p})]{}^{\dagger}M.
\end{equation}
Following the arguments leading to Eq. \eqref{p-spinor-def}, we consider
\begin{equation}
    M(x)=\rho_{-\mathfrak{r}}(\mathfrak{p})M_0(x),
\end{equation}
where $M_0$ is the generally spacetime-dependent $\mathfrak{p}^0$-spinor metric. We deviate from the convention $(M_0\rightarrow m)$ in this definition to avoid confusion with both dimensional factors and masses.

The scalar bilinear $\overline{\Psi}\Psi$, defined in Eq. \eqref{spinor-product}, represents a true \textit{spacetime}-scalar when it is independent of $x^*$. Therefore, $(\Psi\Vert\Psi)$ must be of type-$\mathfrak{p}^0$ if it is to have any physical meaning. In other words, we require
\begin{equation}\label{scalar-bilinear}
(\Psi\Vert\Psi)=\overline{\Psi}\Psi=\overline{\psi}\psi=(\psi\Vert\psi).
\end{equation}
Expanding the left-hand side produces
\begin{equation}
    \begin{split}
\overline{\Psi}\Psi&=\Psi^{\dagger}M\Psi\\
        &=\psi^{\dagger}[\rho_{\mathfrak{w}}(\mathfrak{p})]{}^{\dagger}M\rho_{\mathfrak{w}}(\mathfrak{p})\psi\\
        &=\psi^{\dagger}M_0M_0^{-1}[\rho_{\mathfrak{w}}(\mathfrak{p})]{}^{\dagger}M\rho_{\mathfrak{w}}(\mathfrak{p})\psi\\
        &=\overline{\psi}M_0^{-1}[\rho_{\mathfrak{w}}(\mathfrak{p})]{}^{\dagger}M\rho_{\mathfrak{w}}(\mathfrak{p})\psi,
    \end{split}
\end{equation}
where $\overline{\psi}:=\psi^{\dagger}M_0$ follows from the $\mathfrak{p}^0$-spinor nature of $\psi$. For the scalar relation in Eq.~\eqref{scalar-bilinear} to hold, 
\begin{equation}\label{spin-metric-pre-condition}
    M_{0}^{-1}\left[\rho_{\mathfrak{w}}(\mathfrak{p})\right]{}^{\dagger}M\rho_{\mathfrak{w}}(\mathfrak{p})=1.
\end{equation}
Since we are considering only those representations which permit a factoring of the projective weight, it follows that $\rho_{\mathfrak{a}}(\mathfrak{p})\rho_{\mathfrak{b}}(\mathfrak{p})=\rho_{\mathfrak{a}+\mathfrak{b}}(\mathfrak{p})$. Recalling the discussion in Sec. \ref{subsubsec:clifford-basis}, we suppose the expressions for $\rho_{\mathfrak{a}}(\mathfrak{p})$,
\begin{equation}
    \rho_{\mathfrak{a}}(\mathfrak{p})=\mathfrak{p}{}^{\mathfrak{a}}.
\end{equation}
This choice is similar to the one utilized in \cite{gen-struc,heavy-lifting}, and derived in \cite{spinor-density}. We note that it may be possible to include a chiral weight by amending the above with $\exp(i\mathfrak{v}\gamma^{\underline{5}}\log{\mathfrak{p}})$ \cite{spinor-density}. However, we do not consider such additions in this document. Reflecting these statements in Eq. \eqref{spin-metric-pre-condition} provides
\begin{equation}
\begin{split}
    1&=M_{0}^{-1}\left[\rho_{\mathfrak{w}}(\mathfrak{p})\right]{}^{\dagger}M_0\rho_{\mathfrak{w}-\mathfrak{r}}(\mathfrak{p})\\
    &=M_{0}^{-1}\exp(\mathfrak{w}^*\log\mathfrak{p})M_0\exp((\mathfrak{w}-\mathfrak{r})\log\mathfrak{p})\\
    &=\exp((\mathfrak{w}^*+\mathfrak{w}-\mathfrak{r})\log\mathfrak{p}).
    \end{split}
\end{equation}
Thus,
\begin{equation}\label{weight-condition}
\mathfrak{w}^*+\mathfrak{w}=\mathfrak{r}\quad\Rightarrow\quad\begin{cases}\text{Re}(\mathfrak{r})=2\text{Re}(\mathfrak{w})\\
\text{Im}(\mathfrak{r})=0,
    \end{cases}
\end{equation}
restricting $\mathfrak{r}$ to an $\mathbb{R}$-valued projective weight. However, no restriction is imposed on $\text{Im}(\mathfrak{w})$, the complex part of $\mathfrak{w}$. We thus write the general projective spinor weight as
\begin{equation}
    \mathfrak{w}=\frac{1}{2}(\mathfrak{r}+i\mathfrak{c}).
\end{equation}
These are the basic requirements for $\overline{\Psi}\Psi=\overline{\psi}\psi$ to be a genuine projective scalar, i.e., of type-$\mathfrak{p}^0$. Concisely, the $\mathfrak{p}^{\mathfrak{w}}$-spinor and its adjoint may now be written as
\begin{subequations}
\begin{align}
    \Psi&=\mathfrak{p}^{\frac{1}{2}(\mathfrak{r}+i\mathfrak{c})}\psi\\
    \overline{\Psi}&=\Psi^{\dagger}M=\overline{\psi}\mathfrak{p}^{-\frac{1}{2}(\mathfrak{r}+i\mathfrak{c})}.
\end{align}
\end{subequations}

We require the scalar bilinear to be \textit{real}, or \textit{Hermitian}. Taking the Hermitian conjugate of the scalar bilinear, we find
\begin{equation}
    \begin{split}
        (\overline{\Psi}\Psi)^{\dagger}&=\Psi^{\dagger}\overline{\Psi}{}^{\dagger}\\
        &=\Psi^{\dagger}MM^{-1}(\Psi^{\dagger}M){}^{\dagger}\\
        &=\overline{\Psi}M^{-1}M^{\dagger}\Psi.
    \end{split}
\end{equation}
Therefore, Hermiticity of the scalar bilinear requires Hermiticity of the $\mathfrak{p}^{-\mathfrak{r}}$-spinor metric $M$,
\begin{equation}\label{hermitian-m}
    M^{\dagger}=M.
\end{equation}

Considering now, the $\mathfrak{p}$-vector (vector and pseudo-vector) bilinear,
\begin{equation}
    J^A:=(\Psi\Vert\Gamma^A\Psi)=\overline{\Psi}\Gamma^A\Psi,
\end{equation}
we may derive further properties of the present construction. For example, we require $J^A$ be Hermitian, $(J^A)^{\dagger}=J^A$. For the Hermitian conjugate of $J^A$, we find
\begin{equation}
\begin{split}
    (J^A)^{\dagger}&=(\overline{\Psi}\Gamma^A\Psi)^{\dagger}\\
    &=\Psi^{\dagger}(\Gamma^A)^{\dagger}\overline{\Psi}{}^{\dagger}\\
    &=\Psi^{\dagger}MM^{-1}(\Gamma^A)^{\dagger}(\Psi^{\dagger}M)^{\dagger}\\
    &=\overline{\Psi}M^{-1}(\Gamma^A){}^{\dagger}M^{\dagger}\Psi.
    \end{split}
\end{equation}
Thus, for $J^A$ to be Hermitian, we require
\begin{equation}
    \Gamma^A=M^{-1}(\Gamma^A){}^{\dagger}M^{\dagger}.
\end{equation}
In other words, the combination $M\Gamma^A$ is Hermitian and $M$ is sometimes referred to as the \textit{Hermitianizing Matrix} \cite{hermitianizing-matrix}. Such a statement is nothing new, for example \cite{pauli-5homo-1,pauli-5homo-2}. Solving for $\Gamma^\dagger$ and using the Hermiticity of $M$, we find the projective analogue of the gamma matrix Hermitian conjugate, Eq. \eqref{hermitian-flat-gamma-4},
\begin{equation}\label{general-dagger-gamma}
    (\Gamma^A){}^{\dagger}=M\Gamma^AM^{-1}.
\end{equation}
This relation ensures that
\begin{equation}
    (\Psi\Vert\Gamma^A\Psi)=(\Gamma^A\Psi\Vert\Psi)
\end{equation}
is satisfied.

We may begin to make contact with the more familiar realization of Hermitian conjugation for gamma matrices, Eq. \eqref{hermitian-flat-gamma-4}, as well as the standard choice of spinor metric $M_0\sim \gamma^{\underline{0}}$. Following \cite{spin-metric}, since we are considering even $m=2k$-dimensional spaces of split-signature $(p,q)=(2n+1,2r-1)$, Eq. \eqref{general-dagger-gamma} is generally satisfied for $M$ proportional to the product of all timelike $\Gamma$'s,
\begin{equation}\label{gen-specific-spin-metric}
    M=\beta\Gamma_0\Gamma_1\dots\Gamma_{2n},
\end{equation}
for some proportionality constant $\beta\in\mathbb{C}$. In particular, for Hermitian timelike $\Gamma$ and anti-Hermitian spacelike $\Gamma$, Eq. \eqref{gen-specific-spin-metric} provides
\begin{equation}
    (\Gamma^A){}^{\dagger}=(-1)^{2n}M\Gamma^AM^{-1},
\end{equation}
which is equivalent to Eq. \eqref{general-dagger-gamma} for all $n$. In order to satisfy the Hermiticity requirement in Eq. \eqref{hermitian-m}, we utilize $CL(P\mathcal{M})$ to show
\begin{equation}
    \begin{split}
        M^{\dagger}&=(\beta\Gamma_0\Gamma_1\dots\Gamma_{2n}){}^{\dagger}\\
&=\beta^*\Gamma_{2n}^{\dagger}\dots\Gamma_{1}^{\dagger}\Gamma_{0}^{\dagger}\\
&=(-1)^{k}\beta^*\Gamma_0\Gamma_1\dots\Gamma_{2n}\\
&=(-1)^{k}\beta^*\beta^{-1}M,
    \end{split}
\end{equation}
where
\begin{equation}
    k:=\sum_{j=1}^{2n}j=\frac{2n(2n+1)}{2}
\end{equation}
results from anti-commuting each timelike Hermitian $\Gamma$. For a Hermitian $M$, we must require
\begin{equation}
    \beta=(-1)^k\beta^*=i^{2k}\beta^*.
\end{equation}
Taking $\beta=i^w$, for some constant $w\in\mathbb{R}$, we then find
\begin{equation}
    w=k=n(2n+1).
\end{equation}
Therefore, the spinor metric $M$ may be taken as
\begin{equation}\label{M-def}
    M=i^{n(2n+1)}\Gamma_0\Gamma_1\dots\Gamma_{2n},
\end{equation}
Interestingly, the physical scenario $n=0$ provides $M=\Gamma_0$, restricting $M$ to be a $\mathfrak{p}^{-1}$-spin matrix. Generally, for arbitrary $n$, we find that $M$ is of type-$\mathfrak{p}^{-\mathfrak{r}}$, with $\mathfrak{r}=2n+1$.

Lastly, for the spinor inner product to locally gauge transform homogeneously, the $\mathfrak{p}^{-\mathfrak{r}}$-spinor metric must transform appropriately. To find its transformation behavior, we note that the scalar bilinear must be invariant, 
\begin{equation}\label{spinor-scalar-trans}
    S(x):(\Psi\Vert\Psi)\rightarrow (\Psi\Vert\Psi)'=(\Psi\Vert\Psi).
\end{equation}
Expanding this relation provides the local gauge transformation behavior of $M$,
\begin{equation}
    \begin{split}
    (\Psi\Vert\Psi)'&=\overline{\Psi}{}'M'\Psi'\\
    &=\Psi^{\dagger}\left[\rho_{\mathfrak{w}}(S)\right]{}^{\dagger}M'\rho_{\mathfrak{w}}(S)\Psi.
    \end{split}
\end{equation}
Therefore, the satisfaction of Eq. \eqref{spinor-scalar-trans} requires that $M$ transform under local $SL(m+1,\mathbb{R})$-gauge transformations as
\begin{equation}
   S(x):M\rightarrow M'=\left[\rho_{\mathfrak{w}}(S^{-1})\right]{}^{\dagger}M\rho_{\mathfrak{w}}(S^{-1}),
\end{equation}
where we have taken $\left[\rho_{\mathfrak{w}}(S)\right]{}^{-1}\equiv\rho_{\mathfrak{w}}(S^{-1})$.

\subsection{\texorpdfstring{$\bar{\mathfrak{p}}$}{\bar{\mathfrak{p}}}-Spinors}
\label{subsec:nlp-spinors}

The $\mathfrak{p}^0$-spinor $\psi(x)$ of Eq.~\eqref{p-spinor-def} is best understood through the nonlinear realization of symmetry groups. In order to nonlinearly realize the symmetries over the projective Lorentz group for spinors, we simply follow the prescription and form $\bar{\mathfrak{p}}{}^{\mathfrak{w}}$-spinors via
\begin{equation}\label{p-bar-spinor}
    \tilde{\Psi}(x,x^*):=\rho_{\mathfrak{w}}(\sigma^{-1})\Psi(x,x^*).
\end{equation}
The local gauge transformation behavior is now
\begin{equation}
    S(x): \tilde{\Psi}(x,x^*)\rightarrow \tilde{\Psi}'(x,x^*)=\rho_{\mathfrak{w}}(\Lambda(x) )\tilde{\Psi}(x,x^*),
\end{equation}
with $\rho_{\mathfrak{w}}(\Lambda)$ the appropriate spin representation of $\Lambda\in SO(m+1,\mathbb{R})$. We note that $\rho_{\mathfrak{w}}(\Lambda )=\rho(\Lambda)$, since $|\Lambda|=|\lambda|=1$. For this local gauge transformation behavior we require $\rho(A)\rho(B)=\rho(AB)$, so that no additional (phase) factor is introduced. However, recalling Eq. \eqref{weight-condition}, there is no restriction on a complex part of the projective weight $\mathfrak{w}$. This complex weight may be viewed as resulting from considering a projective representation, i.e., $c\rho(A)\rho(B)=\rho(AB)$, where $c\in\mathbb{C}$ \cite{proj-geo-old-and-new}. Following the aforementioned requirement, we may write
\begin{equation}
    \rho_{\mathfrak{w}}(\sigma^{-1})=\rho_{\mathfrak{w}}(|r|^{\frac{-1}{m+1}})\rho(r,\xi,\change).
\end{equation}
This means the $\bar{\mathfrak{p}}{}^{\mathfrak{w}}$-spinor takes the form
\begin{equation}
    \tilde{\Psi}(x,x^*)=\rho_{\mathfrak{w}}(\bar{\mathfrak{p}})\tilde{\psi}(x),
\end{equation}
and has the local gauge transformation behavior
\begin{equation}
    S(x):\tilde{\Psi}(x,x^*)\rightarrow\tilde{\Psi}'(x,x^*)=\rho_{\mathfrak{w}}(\bar{\mathfrak{p}})\tilde{\psi}'(x)=\rho_{\mathfrak{w}}(\bar{\mathfrak{p}})\rho(\Lambda)\tilde{\psi}(x).
\end{equation}
We thus find that
\begin{equation}
    \tilde{\psi}(x):=\rho(r,\xi,\change)\psi(x)
\end{equation}
transforms as a projective Lorentz spinor under the nonlinear local action of $S\in SL(m+1,\mathbb{R})\cong PGL(m,\mathbb{R})$. This appears to fall right in-line with what is known about the manifield approach to world spinors in the Metric-Affine gauge theories. Amongst the infinitude of manifield spinors of various half-integer spins, the lowest order state \textit{behaves} as an ordinary Lorentz spinor \cite{Hehl}. Here, we find what appears to be the image of this statement for projective spinors, illuminated by the nonlinear symmetry realization approach.

For the $\bar{\mathfrak{p}}{}^{\mathfrak{w}}$-spinors, we gain access to the familiar form of defining a spinor adjoint. Following the discussion of the previous section, Eq. \eqref{p-spinor-adjoint} in particular, we define adjoint $\bar{\mathfrak{p}}{}^{\mathfrak{w}}$-spinors as
\begin{equation}\label{p-bar-adjoint-spinor}
    \tilde{\overline{\Psi}}(x,x^*):=\tilde{\Psi}{}^{\dagger}\tilde{M}.
\end{equation}
Analogous to the reality requirement in Eq. \eqref{scalar-bilinear}, we require
\begin{equation}
(\tilde{\Psi}\Vert\tilde{\Psi})=\tilde{\overline{\Psi}}\tilde{\Psi}=\overline{\Psi}\Psi=(\Psi\Vert\Psi).
\end{equation}
In other words, the scalar bilinear is insensitive to the nonlinear realization process. This condition, along with Eqs. \eqref{p-bar-spinor} and \eqref{p-bar-adjoint-spinor}, provides a relation between the $\mathfrak{p}^{-\mathfrak{r}}$-spinor metric $M$ and the $\bar{\mathfrak{p}}{}^{-\mathfrak{r}}$-spinor metric $\tilde{M}$,
\begin{equation}\label{M-tilde-def}
    \tilde{M}=\left[\rho_{\mathfrak{w}}(\sigma)\right]{}^{\dagger}M\rho_{\mathfrak{w}}(\sigma).
\end{equation}
From the definition of $M$ in Eq. \eqref{M-def}, the inverse of the nonlinear gamma matrix in Eq. \eqref{pre-p-bar-gamma}, and the assumption of a unitary representation $\left[\rho_{\mathfrak{w}}(\sigma)\right]{}^{\dagger}=\rho_{\mathfrak{w}}(\sigma^{-1})$, we find
\begin{equation}
    \begin{split}
        \tilde{M}&=\left[\rho_{\mathfrak{w}}(\sigma)\right]{}^{\dagger}M\rho_{\mathfrak{w}}(\sigma)\\
        &=i^{n(2n+1)}\rho_{\mathfrak{w}}(\sigma^{-1})\Gamma_0\Gamma_1\dots\Gamma_{2n}\rho_{\mathfrak{w}}(\sigma)\\
        &=i^{n(2n+1)}\rho_{\mathfrak{w}}(\sigma^{-1})\Gamma_0\rho_{\mathfrak{w}}(\sigma)\rho_{\mathfrak{w}}(\sigma^{-1})\Gamma_1\rho_{\mathfrak{w}}(\sigma)\dots\rho_{\mathfrak{w}}(\sigma^{-1})\Gamma_{2n}\rho_{\mathfrak{w}}(\sigma)\\
&=i^{n(2n+1)}\tilde{\Gamma}_{\underline{0}}\tilde{\Gamma}_{\underline{1}}\dots\tilde{\Gamma}_{\underline{2n}}.
    \end{split}
\end{equation}
We may thus write the \textit{representation- and dimension-dependent} adjoint as
\begin{equation}
    \tilde{\overline{\Psi}}(x,x^*):=\tilde{\Psi}{}^{\dagger}\tilde{M}=\tilde{\Psi}^{\dagger}\tilde{\Gamma}_{\underline{0}}.
\end{equation}
Notice that one must be mindful of index placement, since using $\tilde{\Gamma}{}^{\underline{0}}$ will introduce a factor of $\bar{\mathfrak{p}}{}^{+1}$, while using $\tilde{\Gamma}_{\underline{0}}$ will introduce a factor of $\bar{\mathfrak{p}}{}^{-1}$. Essentially, from Eq. \eqref{M-def}, this instantiation of spinor adjoint is a particular choice for the spinor metric, and therefore a particular dimension. For if
\begin{equation}
    \tilde{M}\equiv\tilde{\Gamma}_{\underline{0}},
\end{equation}
then $m=4$ with $(p,q)=(1,3)$, and $\tilde{M}$ is a $\bar{\mathfrak{p}}{}^{-1}$-spinor metric. As for the representation dependence, comparing Eqs. \eqref{pre-p-bar-gamma} and \eqref{M-tilde-def}, the choice $\tilde{M}\equiv\tilde{\Gamma}_{\underline{0}}$ restricts the representation to unitary, since it imposes $\left(\rho_{\mathfrak{w}}(\sigma)\right)^{\dagger}=\rho_{\mathfrak{w}}(\sigma^{-1})$. For these choices, we have
\begin{equation}
  \tilde{\overline{\Psi}}(x,x^*)=\tilde{\Psi}{}^{\dagger}\tilde{\Gamma}_{\underline{0}}\bar{\mathfrak{p}}{}^{\frac{1}{2}-i\frac{\mathfrak{c}}{2}}=\tilde{\psi}{}^{\dagger}\tilde{\gamma}_{\underline{0}}\bar{\mathfrak{p}}{}^{-\frac{1}{2}-i\frac{\mathfrak{c}}{2}}=\tilde{\psi}{}^{\dagger}\gamma_{\underline{0}}\bar{\mathfrak{p}}{}^{-\frac{1}{2}-i\frac{\mathfrak{c}}{2}}=\tilde{\overline{\psi}}(x)\bar{\mathfrak{p}}{}^{-\frac{1}{2}-i\frac{\mathfrak{c}}{2}},
\end{equation}
and of course
\begin{equation}
     \tilde{\Psi}(x,x^*)=\bar{\mathfrak{p}}{}^{\frac{1}{2}+i\frac{\mathfrak{c}}{2}}\tilde{\psi}(x).
\end{equation}

In $m=4$ dimensions, the Hermitian conjugation operation of Eq.~\eqref{general-dagger-gamma} now takes the form
\begin{equation}\label{specific-dagger-gamma-0}
    (\tilde{\Gamma}{}^{\underline{A}})^{\dagger}=\tilde{\Gamma}_{\underline{0}}\tilde{\Gamma}{}^{\underline{A}}\tilde{\Gamma}{}^{\underline{0}},
\end{equation}
since $(\tilde{\Gamma}_{\underline{0}})^{-1}=\tilde{\Gamma}{}^{\underline{0}}$. This definition is formally independent of $\eta^{\underline{00}}$, as was required. Somewhat hidden between the lines in these discussions is ones ``freedom" in choosing anti-Hermiticity as opposed to Hermiticity. Since we have restricted considerations to even $m=2k$-dimensional spaces with an odd number of timelike dimensions, our choice of Hermiticity is consistent with the developments thus far. 

From the $\tilde{\Gamma}$-Clifford algebra in Eq. \eqref{CL5} and the Hermitian conjugation operation in Eq.~\eqref{specific-dagger-gamma-0}, we recover the usual (anti)-Hermiticity properties:
\begin{equation}
    (\tilde{\Gamma}{}^{\underline{0}})^{\dagger}=+\tilde{\Gamma}{}^{\underline{0}},\quad\quad\quad(\tilde{\Gamma}{}^{\underline{k}})^{\dagger}=-\tilde{\Gamma}{}^{\underline{k}},\quad\quad\quad (\tilde{\Gamma}{}^{\underline{*}})^{\dagger}=-\tilde{\Gamma}{}^{\underline{*}},
\end{equation}
where $\underline{k}=\underline{1},\underline{2},\underline{3}$. Notice, the choice of $\tilde{M}\equiv\tilde{\Gamma}_{\underline{0}}$ appears to restrict $\eta_0$, since
\begin{equation}
    (\tilde{\Gamma}{}^{\underline{*}})^{\dagger}=\pm\bar{\mathfrak{p}}\sqrt{\eta_0}{}^{\dagger}\gamma^{\underline{m+1}}
\end{equation}
is only minus itself when $\eta_0=-1$. 

Lastly, we note that it may be of interest to define a type of generalized projective chiral, or Weyl spinors. To construct these, following \cite{fermion-upsilon-projector}, one forms a projection operator from the normalized general projective Higgs vector as
\begin{equation}
    \hat{\tilde{\bm{P}}}_{\pm}:=\frac{1}{2}\left(\bm{1}_m\pm\frac{\tilde{\Upsilon}{}^{\underline{A}}\tilde{\Gamma}{}_{\underline{A}}}{|\tilde{\Upsilon}|}\right).
\end{equation}
As can easily be checked, this operator is indeed a projector since it satisfies the required properties:
\begin{equation}
\hat{\tilde{\bm{P}}}_{\pm}\hat{\tilde{\bm{P}}}_{\pm}=\hat{\tilde{\bm{P}}}_{\pm},\quad\quad\quad \hat{\tilde{\bm{P}}}_{\pm}\hat{\tilde{\bm{P}}}_{\mp}=0.
\end{equation}
The generalized chiral $\bar{\mathfrak{p}}{}^{\mathfrak{w}}$-spinors are thus defined as
\begin{equation}
    \tilde{\Psi}_{\pm}:=\hat{\tilde{\bm{P}}}_{\pm}\tilde{\Psi}.
\end{equation}
In the APV-gauge, $\tilde{\Upsilon}$ assumes the form of Eq. \eqref{upsilon-APV} and the generalized chiral projector resolves to
\begin{equation}\label{APV-projector}
    \hat{\tilde{\bm{P}}}_{\pm}\overset{\circ}{=}\frac{1}{2}\left(\bm{1}_m\pm\eta_0\gamma^{\underline{m+1}}\right),
\end{equation}
for $\tilde{\gamma}{}^{\underline{*}}=+\sqrt{\eta_0}\gamma^{\underline{m+1}}$. Choosing the other sign, $\tilde{\gamma}{}^{\underline{*}}=-\sqrt{\eta_0}\gamma^{\underline{m+1}}$, simply results in the replacement $\pm\rightarrow\mp$ in Eq. \eqref{APV-projector}.

Interestingly, the availability of a general projective Higgs co-vector would seem to offer a novel definition for the chiral projection operator. This would be defined as
\begin{equation}
    \hat{\tilde{\bm{Q}}}_{\pm}:=\frac{1}{2}\left(\bm{1}_m\pm\frac{\tilde{\Theta}_{\underline{A}}\tilde{\Gamma}{}^{\underline{A}}}{|\tilde{\Theta}|}\right),
\end{equation}
and may also easily be shown to satisfy
\begin{equation}
\hat{\tilde{\bm{Q}}}_{\pm}\hat{\tilde{\bm{Q}}}_{\pm}=\hat{\tilde{\bm{Q}}}_{\pm},\quad\quad\quad \hat{\tilde{\bm{Q}}}_{\pm}\hat{\tilde{\bm{Q}}}_{\mp}=0.
\end{equation}
Further investigation into this construction and its theoretical implications may be valuable, and is included in the future research program.

\section{Spinor Covariant Derivative}
\label{sec:spinor-der}

In order to construct a dynamical theory for projective spinors, more foundational information must be acquired. We build each of these foundational concepts utilizing the formal procedure to ensure any differences inherent to the present general projective setting are uncovered. We begin this section with the gauge-covariant derivative, deriving its form for the various $\mathfrak{p}{}^{\mathfrak{w}}$-spinor-valued objects on which it acts and confirm its transformation behavior. We then derive the form of the gauge-covariant $\mathfrak{p}{}^{\mathfrak{w}}$-spinor connection and state some fundamental relations it provides. From these basic results, the nonlinear realization procedure is applied, resulting in the gauge-covariant $\bar{\mathfrak{p}}{}^{\mathfrak{w}}$-spinor connection. We then reevaluate some of the essential properties.

\subsection{\texorpdfstring{$\mathfrak{p}$}{\mathfrak{p}}-Spinor Covariant Derivative}
\label{subsec:proj-spinor-cov-der}

To derive the $\mathfrak{p}{}^{\mathfrak{w}}$-spinor gauge-covariant derivative, we simply use the $\mathfrak{p}{}^{\mathfrak{w}}$-spinor bilinears and require their gauge-covariant derivative behave accordingly under local gauge transformations. The scalar bilinear $\overline{\Psi}\Psi$ provides us with a notion of how the $\mathfrak{p}$-spinor gauge-covariant derivative acts on both $\mathfrak{p}$-spinors and their adjoint. Since $\overline{\Psi}\Psi$ is a scalar, we require
\begin{equation}\label{cov-der-spinor-scalar}
    D(\overline{\Psi}\Psi)=d(\overline{\Psi}\Psi).
\end{equation}
For ease, we consider defining its action on $\Psi$ as
\begin{equation}\label{cov-der-psi}
    D\Psi:=d\Psi+\Omega\Psi.
\end{equation}
Expanding the left-hand side of Eq.~\eqref{cov-der-spinor-scalar} and using both the linearity of derivations and the definition in Eq. \eqref{cov-der-psi}, one finds that Eq.~\eqref{cov-der-spinor-scalar} is satisfied for
\begin{equation}\label{cov-der-psi-bar}
    D\overline{\Psi}=d\overline{\Psi}-\overline{\Psi}\Omega.
\end{equation}

Consider now, the $\mathfrak{p}{}^{\mathfrak{w}}$-vector (and pseudo-vector) bilinear $J^A=\overline{\Psi}\Gamma^A\Psi$, which is also a $\mathfrak{p}{}^{\mathfrak{w}}$-spinor-valued scalar. The gauge-covariant derivative of $J^A$ must respect its vectorial nature, providing
\begin{equation}\label{cov-der-vector-bilinear}
    DJ^A=dJ^A+\Omega^A{}_BJ^B.
\end{equation}
Consider, generically,
\begin{equation}
    D\Gamma^A=d\Gamma^A+[X\Gamma]^A,
\end{equation}
for some object $X$. Using the definition of $J^A$ and Eqs. \eqref{cov-der-psi} and \eqref{cov-der-psi-bar}, we expand $DJ^A$ to find
\begin{equation}
    \begin{split}
        DJ^A&=D(\overline{\Psi}\Gamma^A\Psi)\\
        &=(D\overline{\Psi})\Gamma^A\Psi+\overline{\Psi}(D\Gamma^A)\Psi+\overline{\Psi}\Gamma^A(D\Psi)\\
        &=(d\overline{\Psi}-\overline{\Psi}\Omega)\Gamma^A\Psi+\overline{\Psi}(d\Gamma^A+[X\Gamma]^A)\Psi+\overline{\Psi}\Gamma^A(d\Psi+\Omega\Psi)\\
        &=d(\overline{\Psi}\Gamma^A\Psi)+\overline{\Psi}([X\Gamma]^A+[\Gamma^A,\Omega])\Psi.
    \end{split}
\end{equation}
To recover Eq.~\eqref{cov-der-vector-bilinear}, we must require that
\begin{equation}
    [X\Gamma]^A=\Omega^A{}_{B}\Gamma^B+[\Omega,\Gamma^A].
\end{equation}
Therefore, the gauge-covariant derivative of $\Gamma^A$ takes the standard form \cite{curved-dirac}
\begin{equation}\label{gamma-cov-der}
    D\Gamma^A=d\Gamma^A+\Omega^A{}_{B}\Gamma^B+[\Omega,\Gamma^A],
\end{equation}
where 
\begin{equation}\label{spin-conn-def}
    \Omega=\Omega_{AB}\Gamma^{AB}=\frac{1}{4}\Omega_{AB}\Gamma^{A}\Gamma^{B}=\frac{-i}{8}\Omega_{[AB]}\Sigma^{AB}.
\end{equation}
We omit the derivation of this last expression and refer the reader to \cite{curved-dirac}. The last equality in Eq. \eqref{spin-conn-def} follows from the Clifford algebra $CL(P\mathcal{M})$ and the fact that $\Omega$ is a \textit{traceless} connection.

Lastly, the transformation property which leads to $D$'s status as a true gauge-covariant derivative is that it transforms like the object on which it acts. Consider the gauge-transformed covariant derivative of a $\mathfrak{p}$-spinor,
\begin{equation}
    \begin{split}
        (D\Psi)'&=d\Psi'+\Omega'\Psi'\\
        &=d(\rho_{\mathfrak{w}}(S)\Psi)+\Omega'(\rho_{\mathfrak{w}}(S)\Psi)\\
        &=\rho_{\mathfrak{w}}(S)(d\Psi)+(d\rho_{\mathfrak{w}}(S))\Psi+\Omega'\rho_{\mathfrak{w}}(S)\Psi\\
        &=\rho_{\mathfrak{w}}(S)(D\Psi)+(d\rho_{\mathfrak{w}}(S))\Psi+\Omega'\rho_{\mathfrak{w}}(S)\Psi-\Omega\rho_{\mathfrak{w}}(S)\Psi.
    \end{split}
\end{equation}
For $(D\Psi)'=\rho_{\mathfrak{w}}(S)(D\Psi)$, we must require that
\begin{equation}
    \Omega'=\rho_{\mathfrak{w}}(S)(\Omega+d)\rho_{\mathfrak{w}}(S^{-1}).
\end{equation}
In other words, $\Omega$ is the weighted spinor representation of the $SL(m+1,\mathbb{R})\cong PGL(m,\mathbb{R})$ gauge connection-form.

The Hermitian conjugate of the $\mathfrak{p}^{\mathfrak{w}}$-spinor gauge-covariant derivative is easily found. Noting that $M=M^{\dagger}$ and that the traceless $\Omega_{AB}$ is proportional to the $\mathfrak{p}$-spinor identity $\bm{1}_m$,
\begin{equation}\label{proj-spinor-cov-der-herm}
    \begin{split}
        (D\Psi)^{\dagger}&=(d\Psi+\Omega\Psi)^{\dagger}\\
        &=(d\Psi)^{\dagger}+(\frac{1}{4}\Omega_{AB}\Gamma^A\Gamma^B\Psi)^{\dagger}\\
        &=d\Psi^{\dagger}+\frac{1}{4}\Psi^{\dagger}(\Gamma^B)^{\dagger}(\Gamma^A)^{\dagger}\Omega^{\dagger}_{AB}\\
        &=d\Psi^{\dagger}+\frac{1}{4}\Psi^{\dagger}M\Gamma^B\Gamma^AM^{-1}\Omega_{AB}\\
        &=(d\overline{\Psi})M^{-1}+\overline{\Psi}dM^{-1}-\frac{1}{4}\overline{\Psi}\Omega_{AB}\Gamma^A\Gamma^BM^{-1}\\
        &=(D\overline{\Psi})M^{-1}+\overline{\Psi}dM^{-1}.
    \end{split}
\end{equation}
The Hermitian conjugate of the adjoint $\mathfrak{p}^{\mathfrak{w}}$-spinor gauge-covariant derivative follows simply by conjugating Eq. \eqref{proj-spinor-cov-der-herm} once more and solving the expression for the term of interest. Doing so yields
\begin{equation}\label{ajoint-proj-spinor-cov-der-herm}
    (D\overline{\Psi})^{\dagger}=MD\Psi+(d M)\Psi.
\end{equation}
Since Eq. \eqref{gamma-cov-der} implies the gauge-covariant derivative of the $\mathfrak{p}{}^{-\mathfrak{r}}$-spinor metric takes the form
\begin{equation}
    DM=dM+[\Omega,M],
\end{equation}
we find that both Eqs. \eqref{proj-spinor-cov-der-herm} and \eqref{ajoint-proj-spinor-cov-der-herm} lose their gauge-covariance under the action of Hermitian conjugation. Later, we will correct this misbehavior, as well as others that will arise, via the introduction of a self-adjoint operator.

\subsection{\texorpdfstring{$\bar{\mathfrak{p}}$}{\bar{\mathfrak{p}}}-Spinor Covariant Derivative}
\label{subsec:nlp-spinor-cov-der}

We construct the $\bar{\mathfrak{p}}{}^{\mathfrak{w}}$-spinor gauge-covariant derivative by first noting the gauge-covariant derivative is required to transform in the same manner as the object on which it acts. We therefore begin with
\begin{equation}
    \tilde{D}\tilde{\Psi}=\rho_{\mathfrak{w}}(\sigma^{-1})D\Psi.
\end{equation}
Expanding the right-hand side provides
\begin{equation}
\begin{split}
    \rho_{\mathfrak{w}}(\sigma^{-1})D\Psi&=\rho_{\mathfrak{w}}(\sigma^{-1})(d\Psi+\Omega\Psi)\\
    &=(d\rho_{\mathfrak{w}}(\sigma^{-1})\Psi)-(d\rho_{\mathfrak{w}}(\sigma^{-1}))\Psi+\rho_{\mathfrak{w}}(\sigma^{-1})\Omega\Psi\\
    &=d\tilde{\Psi}-(d\rho_{\mathfrak{w}}(\sigma^{-1}))[\rho_{\mathfrak{w}}(\sigma^{-1})]^{-1}\tilde{\Psi}+\rho_{\mathfrak{w}}(\sigma^{-1})\Omega[\rho_{\mathfrak{w}}(\sigma^{-1})]^{-1}\tilde{\Psi}\\
    &=d\tilde{\Psi}+\tilde{\Omega}\tilde{\Psi},
    \end{split}
\end{equation}
where $\tilde{\Omega}$ is related to $\Omega$ via
\begin{equation}\label{almost-omega-tilde}
    \tilde{\Omega}=\rho_{\mathfrak{w}}(\sigma^{-1})(\Omega+d)\rho_{\mathfrak{w}}(\sigma).
\end{equation}
We would like for this nonlinear connection to have the form
\begin{equation}\label{spinor-omega-tilde}
\tilde{\Omega}\equiv\tilde{\Omega}{}^{\underline{A}}{}_{\underline{B}}\rho_{\mathfrak{w}}(\bm{L}{}^{\underline{B}}{}_{\underline{A}})=\tilde{\Omega}_{\underline{AB}}\tilde{\Gamma}{}^{\underline{AB}}=\frac{1}{4}\tilde{\Omega}_{\underline{AB}}\tilde{\Gamma}{}^{\underline{A}}\tilde{\Gamma}{}^{\underline{B}}=\frac{-i}{8}\tilde{\Omega}_{[\underline{AB}]}\tilde{\Sigma}{}^{\underline{AB}},
\end{equation}
with $\tilde{\Omega}$ given in Eq. \eqref{nonlinear-omega-tilde}. To accomplish this, we substitute Eqs. \eqref{spin-conn-def} and \eqref{pre-p-bar-gamma} into Eq. \eqref{almost-omega-tilde} to find
\begin{equation}
    \tilde{\Omega}=\tilde{\Omega}_{\underline{AB}}\tilde{\Gamma}{}^{\underline{AB}}-\frac{1}{4}(\sigma^{-1})^{\underline{A}}{}_Ad\sigma^A{}_{\underline{B}}\tilde{\Gamma}_{\underline{A}}\tilde{\Gamma}{}^{\underline{B}}+\rho_{\mathfrak{w}}(\sigma^{-1})d\rho_{\mathfrak{w}}(\sigma).
\end{equation}
We thus require the representation satisfies
\begin{equation}\label{rep-relation}
    \rho_{\mathfrak{w}}(\sigma^{-1})d\rho_{\mathfrak{w}}(\sigma)=\frac{1}{4}(\sigma^{-1})^{\underline{A}}{}_Ad\sigma^A{}_{\underline{B}}\tilde{\Gamma}_{\underline{A}}\tilde{\Gamma}{}^{\underline{B}}=\frac{-i}{4}(\sigma^{-1})^{\underline{A}}{}_Ad\sigma^A{}_{[\underline{B}}\tilde{\Sigma}_{\underline{A}]}{}^{\underline{B}},
\end{equation}
so that Eq. \eqref{spinor-omega-tilde} is recovered. Note that the vanishing of the symmetric part in Eq. \eqref{rep-relation} is due to $|\sigma|=1$.

The nonlinear gauge-covariant derivative acting on $\bar{\mathfrak{p}}{}^{\mathfrak{w}}$-spinors is thus
\begin{equation}
    \tilde{D}\tilde{\Psi}=d\tilde{\Psi}+\tilde{\Omega}\tilde{\Psi}.
\end{equation}
Applying these arguments to the adjoint $\bar{\mathfrak{p}}{}^{\mathfrak{w}}$-spinors gives the obvious relation
\begin{equation}
    \tilde{D}\tilde{\overline{\Psi}}=d\tilde{\overline{\Psi}}-\tilde{\overline{\Psi}}\tilde{\Omega},
\end{equation}
while applied to $\tilde{\Gamma}$ gives
\begin{equation}\label{nonlinear-cov-der-p-bar-gamma}
    \tilde{D}\tilde{\Gamma}{}^{\underline{A}}=d\tilde{\Gamma}{}^{\underline{A}}+\tilde{\Omega}{}^{\underline{A}}{}_{\underline{B}}\tilde{\Gamma}{}^{\underline{B}}+[\tilde{\Omega},\tilde{\Gamma}{}^{\underline{A}}].
\end{equation}
For each of these, we find the proper behavior under local gauge transformations---nonlinearly with the projective Lorentz group. For example,
\begin{equation}
S(x):\tilde{D}\tilde{\Psi}\rightarrow(\tilde{D}\tilde{\Psi})'=\rho(\Lambda(x))\tilde{D}\tilde{\Psi}.
\end{equation}

We have now developed all the machinery necessary to relate $\tilde{D}\tilde{\Gamma}{}^{\underline{A}}$ to a more familiar geometric object---the $\bar{\mathfrak{p}}{}^{-2}$-non-metricity. Expanding Eq. \eqref{nonlinear-cov-der-p-bar-gamma}, we find
\begin{equation}\label{gamma-non-metricity}
    \begin{split}
        \tilde{D}\tilde{\Gamma}{}^{\underline{A}}&=d\tilde{\Gamma}{}^{\underline{A}}+\tilde{\Omega}{}^{\underline{A}}{}_{\underline{B}}\tilde{\Gamma}{}^{\underline{B}}+[\tilde{\Omega},\tilde{\Gamma}{}^{\underline{A}}]\\
        &=\tilde{g}\tilde{\Gamma}{}^{\underline{A}}+\bar{\mathfrak{p}}d\tilde{\gamma}{}^{\underline{A}}+\frac{1}{2}\tilde{\Omega}{}^{(\underline{AB})}\tilde{\Gamma}_{\underline{B}}+\frac{1}{2}\tilde{\Omega}{}^{[\underline{AB}]}\tilde{\Gamma}_{\underline{B}}+[\tilde{\Omega}{}^{-},\tilde{\Gamma}{}^{\underline{A}}]\\
        &=\frac{1}{2}\left(2\tilde{g}\tilde{\eta}{}^{\underline{AB}}+\tilde{\Omega}{}^{(\underline{AB})}\right)\tilde{\Gamma}_{\underline{B}}+\bar{\mathfrak{p}}\left(d\tilde{\gamma}{}^{\underline{A}}+\frac{1}{2}\tilde{\omega}{}^{[\underline{AB}]}\tilde{\gamma}_{\underline{B}}+[\tilde{\omega}{}^{-},\tilde{\gamma}{}^{\underline{A}}]\right)\\
        &=\frac{1}{2}\tilde{Q}{}^{\underline{AB}}\tilde{\Gamma}_{\underline{B}}+\bar{\mathfrak{p}}\tilde{D}{}^{-}\tilde{\gamma}{}^{\underline{A}}\\
        &=\frac{1}{2}\tilde{\eta}{}^{\underline{AB}}\tilde{Q}_{\underline{BC}}\tilde{\Gamma}{}^{\underline{C}}.
    \end{split}
\end{equation}
This result follows from
\begin{equation}
   \tilde{D}{}^{-}\tilde{\gamma}{}^{\underline{A}}:= d\tilde{\gamma}{}^{\underline{A}}+\frac{1}{2}\tilde{\omega}{}^{[\underline{AB}]}\tilde{\gamma}_{\underline{B}}+[\tilde{\omega}{}^{-},\tilde{\gamma}{}^{\underline{A}}]=0,
\end{equation}
which is the analogous projective statement of covariant conservation of the Lorentz gamma matrices, $\tilde{\gamma}{}^{\underline{A}}=\gamma^{\underline{A}}$ \cite{curved-dirac}. Note that, following our convention, $\tilde{\omega}$ is the $\bar{\mathfrak{p}}{}^{0}$-part of $\tilde{\Omega}_{\underline{AB}}:=\bar{\mathfrak{p}}{}^{-2}\tilde{\omega}_{\underline{AB}}$, and therefore, the above expression is formally independent of $\bar{\mathfrak{p}}$. For the reader's convenience, we show this covariant constancy holds for the $\underline{*}$-component,
\begin{equation}
    \begin{split}
        \tilde{D}{}^{-}\tilde{\gamma}{}^{\underline{*}}&= d\tilde{\gamma}{}^{\underline{*}}+\frac{1}{2}\tilde{\omega}^{[\underline{*B}]}\tilde{\gamma}_{\underline{B}}+[\tilde{\omega},\tilde{\gamma}{}^{\underline{*}}]\\
        &=\pm\sqrt{\eta_0}d\gamma^{\underline{m+1}}+\frac{\eta_0}{2}\tilde{\omega}_{[\underline{*B}]}\tilde{\gamma}{}^{\underline{B}}\pm\frac{\sqrt{\eta_0}}{4}\tilde{\omega}_{\underline{AB}}[\tilde{\gamma}{}^{\underline{A}}\tilde{\gamma}{}^{\underline{B}},\gamma^{\underline{m+1}}]\\
        &=\frac{\eta_0}{2}\tilde{\omega}_{[\underline{*B}]}\tilde{\gamma}{}^{\underline{B}}\pm\frac{\sqrt{\eta_0}}{4}\tilde{\omega}_{\underline{*b}}[\tilde{\gamma}{}^{\underline{*}}\tilde{\gamma}{}^{\underline{b}},\gamma^{\underline{m+1}}]\pm\frac{\sqrt{\eta_0}}{4}\tilde{\omega}_{\underline{b*}}[\tilde{\gamma}{}^{\underline{b}}\tilde{\gamma}{}^{\underline{*}},\gamma^{\underline{m+1}}]\pm\frac{\sqrt{\eta_0}}{4}\tilde{\omega}_{\underline{ab}}[\tilde{\gamma}{}^{\underline{a}}\tilde{\gamma}{}^{\underline{b}},\gamma^{\underline{m+1}}]\\
        &=\frac{\eta_0}{2}\tilde{\omega}_{[\underline{*b}]}\tilde{\gamma}{}^{\underline{b}}\pm\frac{\sqrt{\eta_0}}{4}\tilde{\omega}_{\underline{*b}}[\tilde{\gamma}{}^{\underline{*}}\tilde{\gamma}{}^{\underline{b}},\gamma^{\underline{m+1}}]\pm\frac{\sqrt{\eta_0}}{4}\tilde{\omega}_{\underline{b*}}[\tilde{\gamma}{}^{\underline{b}}\tilde{\gamma}{}^{\underline{*}},\gamma^{\underline{m+1}}]\\
        &=\frac{\eta_0}{2}\tilde{\omega}_{[\underline{*b}]}\tilde{\gamma}{}^{\underline{b}}\mp\frac{\sqrt{\eta_0}}{4}\tilde{\omega}_{[\underline{*b}]}[\tilde{\gamma}{}^{\underline{b}}\tilde{\gamma}{}^{\underline{*}},\gamma^{\underline{m+1}}]\\
        &=\frac{\eta_0}{2}\tilde{\Omega}_{[\underline{*b}]}\tilde{\gamma}{}^{\underline{b}}-\frac{\eta_0}{4}\tilde{\omega}_{[\underline{*b}]}[\tilde{\gamma}{}^{\underline{b}}\tilde{\gamma}{}^{\underline{m+1}},\gamma^{\underline{m+1}}]\\
        &=\frac{\eta_0}{2}\tilde{\omega}_{[\underline{*b}]}\tilde{\gamma}{}^{\underline{b}}-\frac{\eta_0}{2}\tilde{\omega}_{[\underline{*b}]}\tilde{\gamma}{}^{\underline{b}}\\
        &=0.
    \end{split}
\end{equation}

Lastly, the Hermitian conjugate of the $\bar{\mathfrak{p}}{}^{\mathfrak{w}}$-spinor covariant derivative may be obtained from Eq. \eqref{proj-spinor-cov-der-herm} by replacing all fields and operators with their nonlinear extensions,
\begin{equation}
        (\tilde{D}\tilde{\Psi})^{\dagger}=(\tilde{D}\tilde{\overline{\Psi}})\tilde{M}{}^{-1}+\tilde{\overline{\Psi}}d\tilde{M}{}^{-1}.
\end{equation}
The same follows for the adjoint,
\begin{equation}
(\tilde{D}\tilde{\overline{\Psi}})^{\dagger}=\tilde{M}\tilde{D}\tilde{\Psi}+(d \tilde{M})\Psi.
\end{equation}
While not very illuminating at this point, these expressions will be crucial in constructing a proper \textit{Hermitian} action functional describing the spinor dynamics.

\section{Spinor Action}
\label{sec:spinor-action}

We seek a \textit{real}, or \textit{Hermitian}, gauge- and coordinate-invariant Lagrangian $(m+1)$-form density. This task is not as simple in the general projective setting as it is for the ordinary Dirac theory of Lorentz spinors. General coordinate invariance is easily obtained by working in the exterior formalism, since all $V\mathcal{M}$ indices are packaged away. Gauge invariance is also easily accomplished by utilizing the gauge-covariant derivative. However, the requirement of Hermiticity is shown to obstruct the gauge invariance. To reconcile these two properties, we construct an explicitly self-adjoint gauge-covariant derivative operator. 

We first form a Hermitian Lagrangian $(m+1)$-form density for $\mathfrak{p}$-spinors, invariant under the local gauge action of $SL(m+1,\mathbb{R})\cong PGL(m,\mathbb{R})$. Hermiticity of the Lagrangian density requires one to add a Lagrangian $(m+1)$-form density for the adjoint $\mathfrak{p}$-spinors, resulting in an expression which is no longer gauge-invariant. In order to recover gauge invariance, we construct the self-adjoint operator by adding the required terms. We then construct the fully gauge- and coordinate-invariant Hermitian Lagrangian $(m+1)$-form density and discuss some salient features. 

By replacing all objects with their nonlinear extensions (adding a tilde), the Lagrangian density is rendered invariant under local Lorentz transformations. We utilize this nonlinearly realized form of the Lagrangian density to discuss the dynamics. We first investigate the dimensional reduction at the level of the action, and discuss a few key features. We then find the field equations for the system in the absence of the gravitational sector. 

We then detour slightly to discuss the conservation of the projective current density, and begin the investigation of a projective description of the chiral anomaly. This section closes with finding the projective matter contributions to the projective gravitational sector, as well as the projective gravitational contributions to the projective matter sector. Investigation of the fully coupled projective theory is left as a future line of research.

\subsection{\texorpdfstring{$\mathfrak{p}$}{\mathfrak{p}}-Spinor Action}
\label{subsec:proj-spinor-action}

A Hermitian gauge-invariant action may be constructed by first considering the $\mathfrak{p}$-spinor kinetic term. In exterior form, this is
\begin{equation}
    \mathcal{L}_{\Psi}:=\overline{\Psi}*\Gamma\wedge D\Psi.
\end{equation}
Recall the adjoint of the $\mathfrak{p}$-spinor gauge-covariant derivative, Eq. \eqref{proj-spinor-cov-der-herm}. Using this, the Hermitian conjugate of $\mathcal{L}_{\Psi}$ reveals
\begin{equation}\label{L-psi-dagger}
    \begin{split}
        \mathcal{L}_{\Psi}^{\dagger}&=(\overline{\Psi}*\Gamma\wedge D\Psi)^{\dagger}\\
        &=(D\Psi)^{\dagger}\wedge(*\Gamma)^{\dagger} \overline{\Psi}{}^{\dagger}\\
        &=(-1)^{\mho}\left((D\overline{\Psi})M^{-1}+\overline{\Psi}dM^{-1}\right)\wedge M*\Gamma M^{-1}(\Psi^{\dagger}M)^{\dagger}\\
        &=(-1)^{\mho}\left((D\overline{\Psi})-\overline{\Psi}M^{-1}dM\right)\wedge *\Gamma \Psi\\
        &=(-1)^{\mho}\left((D\overline{\Psi})\wedge *\Gamma \Psi-\overline{\Psi}\underline{m}\wedge *\Gamma \Psi\right)\\
        &=(-1)^{\mho}(\overline{\mathcal{L}}_{\Psi}-\mathcal{L}_{\underline{m}\Gamma}),
    \end{split}
\end{equation}
where $\mho$ is given in Eq. \eqref{upsidedown-omega} and $M=M^{\dagger}$ is Hermitian. In the above, we have introduced the following convenient definitions:
\begin{equation}
    \overline{\mathcal{L}}_{\Psi}:=D\overline{\Psi}\wedge*\Gamma \Psi,\quad\quad\quad\mathcal{L}_{\underline{m}\Gamma}:=\overline{\Psi}\underline{m}\wedge *\Gamma \Psi,\quad\quad\quad \underline{m}:=M^{-1}dM.
\end{equation}
We thus find that, although $\mathcal{L}_{\Psi}$ is invariant under transformations of coordinates and gauge, the same is \textit{not} true for its Hermitian conjugate $\mathcal{L}_{\Psi}^{\dagger}$. In order to restore invariance, an extension of the gauge-covariant $D$-operator must be used. For convenience, we further define
\begin{equation}
    \mathcal{L}_{\Gamma}:=\overline{\Psi}(D*\Gamma)\Psi.
\end{equation}

The extended gauge-covariant $D$-operator $\mathfrak{D}$ which restores invariance under the action of Hermitian conjugation is defined as
\begin{equation}\label{frak-D-def}
    \mathfrak{D}:=*\Gamma\wedge D+\frac{1}{2}(-1)^mM^{-1}D(M*\Gamma),
\end{equation}
where the dimension-dependent parity factor ensures consistency in arbitrary-dimensional spacetimes. Our restriction to $m=2k$ gives $(-1)^m=+1$. We note, however, that in odd $m=(2k+1)$-dimensional spacetimes $\mathfrak{D}$ does \textit{not} appear to take the same form as Eq. \eqref{frak-D-def}. Additionally, we note that the form of this operator has been arrived at elsewhere via similar arguments in a non-projective setting \cite{self-adjoint-operator-1,self-adjoint-operator-2}. 

The operator $\mathfrak{D}$ is simply one-half the sum of $\mathcal{L}_{\Psi}$ and the ordinary $D$-operator acting on everything to the right of $(\Psi\dots)$ in the inner product, 
\begin{equation}\label{frak-D-lagrangian}
    \mathfrak{L}_{\Psi}:=(\Psi\Vert\mathfrak{D}\Psi)=\frac{1}{2}\left((\Psi\Vert*\Gamma\wedge D\Psi)+(\Psi D(\Vert*\Gamma\Psi))\right).
\end{equation}
Explicitly, the first term of the right-hand side is
\begin{equation}
    \frac{1}{2}(\Psi\Vert*\Gamma\wedge D\Psi)=\frac{1}{2}\overline{\Psi}*\Gamma\wedge D\Psi,
\end{equation}
while the second term of the right-hand side is
\begin{equation}
    \begin{split}
        \frac{1}{2}(\Psi D(\Vert*\Gamma\Psi))&=\frac{1}{2}\Psi^{\dagger}D(M*\Gamma\Psi)\\
    &=\frac{1}{2}\overline{\Psi}\left(*\Gamma\wedge D\Psi+M^{-1}D(M*\Gamma)\Psi\right).
    \end{split}
\end{equation}
Therefore,
\begin{equation}
    \frac{1}{2}\left((\Psi\Vert*\Gamma\wedge D\Psi)+(\Psi D(\Vert*\Gamma\Psi))\right)=\overline{\Psi}*\Gamma\wedge D\Psi+\frac{1}{2}\overline{\Psi}M^{-1}D(M*\Gamma)\Psi,
\end{equation}
as claimed.

To check that the Hermitian conjugate of $\mathfrak{L}_{\Psi}$ is indeed gauge-invariant, we must find only the Hermitian conjugate of the correction term, since $\mathcal{L}_{\Psi}^{\dagger}$ was found in Eq. \eqref{L-psi-dagger}. We thus calculate
\begin{equation}\label{lagrangian-gamma-adjoint}
    \begin{split}
        (\overline{\Psi}M^{-1}D(M*\Gamma)\Psi)^{\dagger}
        &=(\overline{\Psi}M^{-1}DM\wedge*\Gamma\Psi)^{\dagger}+(\overline{\Psi}D(*\Gamma)\Psi)^{\dagger}\\
        &=(-1)^{\mho}\overline{\Psi}*\Gamma \wedge M^{-1}(DM)^{\dagger}\Psi+\overline{\Psi}M^{-1}(D*\Gamma)^{\dagger}M\Psi\\
        &=(-1)^{\mho}\overline{\Psi}*\Gamma \wedge(\underline{m}+\Omega+\Omega^{\dagger})\Psi\\
        &\quad+(-1)^{\mho}\overline{\Psi}(D(*\Gamma)+\underline{m}\wedge*\Gamma-*\Gamma\wedge\underline{m})\Psi\\
        &=(-1)^{\mho}\overline{\Psi}*\Gamma \wedge(\Omega+\Omega^{\dagger})\Psi+(-1)^{\mho}\overline{\Psi}(D(*\Gamma)+\underline{m}\wedge*\Gamma)\Psi.
    \end{split}
\end{equation}
In order to investigate if we have recovered gauge-invariance, this must be added to $\mathcal{L}_{\Psi}^{\dagger}$. We note that it is in this step one would be led to restricting considerations to even $m=2k$-dimensional spacetimes, resulting in the particular operator defined in Eq. \eqref{frak-D-def}. Adding the above result to $\mathcal{L}_{\Psi}^{\dagger}$ gives
\begin{equation}
        \mathfrak{L}_{\Psi}^{\dagger}=(-1)^{\mho}\left(\overline{\mathcal{L}}_{\Psi}+\frac{1}{2}\mathcal{L}_{\Gamma}-\frac{1}{2}\overline{\Psi}\underline{m}\wedge*\Gamma\Psi+\frac{1}{2}\overline{\Psi}*\Gamma\wedge(\Omega+\Omega^{\dagger})\Psi\right).
\end{equation}
The gauge-invariance of $\mathfrak{L}_{\Psi}^{\dagger}$ requires the latter two terms to combine to form a gauge-covariant derivative. For this to occur, the $*\Gamma$ must appear on the same side of both terms. We therefore need to satisfy
\begin{equation}\label{m-gamma-commute}
\underline{m}\wedge*\Gamma=*\Gamma\wedge\underline{m}.
\end{equation}
This expression is trivially satisfied in $m=2k$ dimensions, since
\begin{equation}
    \underline{m}=M^{-1}dM=(d\log |M|)\bm{1}_m
\end{equation}
is proportional to the $\mathfrak{p}{}^{\mathfrak{w}}$-spinor identity, and the parity factor resulting from the anti-commutativity of wedge products,
\begin{equation}
\underline{m}\wedge*\Gamma=(-1)^m*\Gamma\wedge\underline{m},
\end{equation}
returns $+1$. Furthermore, using $M=\mathfrak{p}^{-\mathfrak{r}}M_0$, the explicit definition of $M$ in Eq. \eqref{M-def}, and the Clifford algebra, we have $(M_0)^2=\pm1$. Then, since
\begin{equation}
\begin{split}
    M^{-1}DM&=M^{-1}dM+M^{-1}[\Omega,M]\\
    &=\underline{m}+M^{-1}\Omega M-\Omega\\
    &=\underline{m}-M^{-2}\Omega^{\dagger} M^2-\Omega\\
    &=\underline{m}-\Omega^{\dagger}-\Omega,
    \end{split}
\end{equation}
we arrive at the final result
\begin{equation}\label{frak-D-dagger}
    \mathfrak{L}_{\Psi}^{\dagger}=(-1)^{\mho}\left(\overline{\mathcal{L}}_{\Psi}+\frac{1}{2}\mathcal{L}_{\Gamma}-\frac{1}{2}\mathcal{L}_{\Gamma m}\right).
\end{equation}
In this expression, we have introduced the definitions
\begin{equation}
    \mathcal{L}_{\Gamma m}:=\overline{\Psi}*\Gamma\wedge m\Psi,\quad\quad\quad m:=M^{-1}DM.
\end{equation}
The gauge invariance is now manifest, since each term of the sum is a $(2k+1)$-form density constructed from gauge-covariant derivatives. Furthermore, recall that in $m=2k$ dimensions, $\mho=2$.

Notice that in Eq. \eqref{frak-D-dagger}, the difference $\mathcal{L}_{\Gamma}-\mathcal{L}_{\Gamma m}$ combines to form the manifestly covariant expression,
\begin{equation}
    \mathcal{L}_{\Gamma}-\mathcal{L}_{\Gamma m}=\overline{\Psi}D(*\Gamma M^{-1})M.
\end{equation}
Thus, we find the Hermitian conjugate of the $\mathfrak{D}$-operator,
\begin{equation}
    \mathfrak{D}{}^{\dagger}:=D\wedge*\Gamma+\frac{1}{2}D(*\Gamma M^{-1})M,
\end{equation}
where it is understood the action of $\mathfrak{D}{}^{\dagger}$ is to the left.

A Hermitian Lagrangian $(2k+1)$-form density may now be constructed by taking one-half the sum of Eqs. \eqref{frak-D-lagrangian} and \eqref{frak-D-dagger}, multiplied by some constant coefficient $\tilde{\alpha}\in\mathbb{C}$, with $[\tilde{\alpha}]=1$. Explicitly, this is
\begin{equation}
    \mathcal{L}_{\mathfrak{D}}:=\frac{1}{2}(\tilde{\alpha}\mathfrak{L}_{\Psi}+\tilde{\alpha}{}^{\dagger}\mathfrak{L}_{\Psi}^{\dagger}).
\end{equation}
Expanding the right-hand side yields the visually pleasing expression,
\begin{equation}
    \mathcal{L}_{\mathfrak{D}}=\frac{\tilde{\alpha}}{2}\left(\overline{\Psi}*\Gamma\wedge D\Psi+\frac{1}{2}\overline{\Psi}M^{-1}D(M*\Gamma)\Psi\right)+\frac{\tilde{\alpha}{}^{\dagger}}{2}\left(D\overline{\Psi}\wedge *\Gamma\Psi+\frac{1}{2}\overline{\Psi}D(*\Gamma M^{-1})M\Psi\right).
\end{equation}
Distributing the gauge-covariant derivatives and combining all like-terms provides the more standard form of the expression,
\begin{equation}\label{expanded-lagrangian-frak-D}
    \mathcal{L}_{\mathfrak{D}}=\frac{\tilde{\alpha}}{2}\overline{\Psi}*\Gamma\wedge D\Psi+\frac{\tilde{\alpha}{}^{\dagger}}{2}D\overline{\Psi}\wedge *\Gamma\Psi+\frac{\tilde{\alpha}+\tilde{\alpha}{}^{\dagger}}{4}\overline{\Psi}D(*\Gamma)\Psi+\frac{\tilde{\alpha}-\tilde{\alpha}{}^{\dagger}}{4}\overline{\Psi}m\wedge*\Gamma\Psi,
\end{equation}
where Eq. \eqref{m-gamma-commute} was utilized. Notice that the coefficients of the latter two terms are formed from the sum and difference of $\tilde{\alpha}$ and $\tilde{\alpha}{}^{\dagger}$. Therefore, the $\mathcal{L}_{\Gamma}$ term is \textit{purely real} and the $\mathcal{L}_{m\Gamma}$ term is \textit{purely imaginary}.

The form of $\mathcal{L}_{\mathfrak{D}}$ in Eq. \eqref{expanded-lagrangian-frak-D} makes explicit the \textit{non-Hermiticity} of the Thomas-Whitehead-Dirac action utilized in \cite{gen-struc}, due to the reality of $\mathcal{L}_{\Gamma}$. A non-Hermitian Lagrangian density leads to complex energies and is therefore discarded on grounds of physical viability. We note that non-Hermitian theories of spinors are plentiful in the literature \cite{non-hermitian1,anti-herm-mass-4,anti-herm-mass,anti-herm-mass-3,non-hermitian2,nonhermitian3,non-herm-interaction}, and may have some physical justification for investigation. The Thomas-Whitehead-Dirac action of \cite{gen-struc} was chosen using the argument that it is related to the standard $\mathcal{L}_{\Psi}$-type Lagrangian via a total derivative. Applying that same line of reasoning to the present, we subtract from $\mathcal{L}_{\mathfrak{D}}$ the total derivative
\begin{equation}
\mathcal{L}_{td}:=\frac{\tilde{\alpha}{}^{\dagger}}{2}D(\overline{\Psi}*\Gamma\Psi),
\end{equation}
to find
\begin{equation}
    \mathcal{L}_{\mathfrak{D}}-\mathcal{L}_{td}=\frac{\tilde{\alpha}-\tilde{\alpha}{}^{\dagger}}{2}(\overline{\Psi}*\Gamma\wedge D\Psi+\frac{1}{2}\overline{\Psi}D(*\Gamma)\Psi+\frac{1}{2}\overline{\Psi}m\wedge*\Gamma\Psi)=\frac{\tilde{\alpha}-\tilde{\alpha}{}^{\dagger}}{2}\mathfrak{L}_{\Psi}.
\end{equation}
Therefore, the present construction implies that $\mathcal{L}_{\mathfrak{D}}$ is related to $\mathfrak{L}_{\Psi}$ via a total derivative, at no expense to the Hermiticity. In light of the discussion in \cite{matter-torsion2}, we have thus constructed an operator which avoids much of the complication associated with the minimal coupling of spinor fields to gravity in the presence of both torsion and non-metricity.

\subsection{\texorpdfstring{$\bar{\mathfrak{p}}$}{\bar{\mathfrak{p}}}-Spinor Action}
\label{subsec:nlp-spinor-action}

The nonlinearly realized Lagrangian $(m+1)$-form density is found by simply replacing all fields and operators with their nonlinearly realized extensions. Since the total Lagrangian $(2k+1)$-form density $\mathcal{L}_{\mathfrak{D}}$ was previously $SL(2k+1,\mathbb{R})\cong PGL(2k,\mathbb{R})$-invariant, the nonlinear realization method, $\mathcal{L}_{\mathfrak{D}}\rightarrow\mathcal{L}_{\tilde{\mathfrak{D}}}$, provides local Lorentz invariance under the local action of $SL(2k+1,\mathbb{R})\cong PGL(2k,\mathbb{R})$. The nonlinearly realized Lagrangian $(2k+1)$-form density $\mathcal{L}_{\tilde{\mathfrak{D}}}$ thus takes the form
\begin{equation}\label{L-D-bar}
\mathcal{L}_{\tilde{\mathfrak{D}}}=\frac{\tilde{\alpha}}{2}\mathcal{L}_{\tilde{\Psi}}+\frac{\tilde{\alpha}{}^{\dagger}}{2}\overline{\mathcal{L}}_{\tilde{\Psi}}+\frac{\tilde{\alpha}+\tilde{\alpha}{}^{\dagger}}{4}\mathcal{L}_{\tilde{\Gamma}}+\frac{\tilde{\alpha}-\tilde{\alpha}{}^{\dagger}}{4}\mathcal{L}_{\tilde{m}\tilde{\Gamma}}.
\end{equation}

The total action $S_{\tilde{\mathfrak{D}}}\equiv S_{\mathfrak{D}}$ is expressed as
\begin{equation}\label{S-D-bar}
    S_{\tilde{\mathfrak{D}}}=\frac{1}{x^*_0}\int_{V\mathcal{M}}\mathcal{L}_{\tilde{\mathfrak{D}}},
\end{equation}
where $[x^*_0]=L$ is included for proper physical dimensions, and the integration is taken over the $(2k+1)$-dimensional volume bundle, $V\mathcal{M}$. 

Before investigating the field equations, we first discuss some interesting properties of $\mathcal{L}_{\tilde{\mathfrak{D}}}$ via dimensional reduction at the level of the action, and further, what happens when the APV-gauge is chosen. The purely projective matter sector, with Hermitian $\mathcal{L}_{\tilde{\mathfrak{D}}}$, is found to provide the standard (projectively invariant) torsion coupling. Interestingly, we also find that there is \textit{no interaction with the projective Schouten form} when reality is enforced. Furthermore, we find a chiral mass induced by the complex part of the projective weight $\mathfrak{w}$. 

The projective current density is then discussed and its conservation shown. This is followed by the beginning of an investigation into a projective description of the chiral anomaly. For this, we simply calculate the square of the self-adjoint projective $\mathfrak{D}$-operator. This section closes with the field equations of the projective matter contribution to the projective gravitational sector, and the associated projective gravitational contributions to the projective matter sector. 

\subsubsection{Action Level Dimensional Reduction}

Using the transition from exterior form to abstract-index notation outlined in Appendix \hyperref[app-B3:exterior-abs-ind-form]{B.3}, the action $S_{\tilde{\mathfrak{D}}}$ in Eq. \eqref{S-D-bar} becomes
\begin{equation}\label{nlp-spinor-action}
\begin{split}
    S_{\tilde{\mathfrak{D}}}&=\frac{(-1)^q\eta_0}{2x^*_0}\int_{V\mathcal{M}}d^{m+1}x|\tilde{e}|\left(\tilde{\alpha}\tilde{\overline{\Psi}}\tilde{\Gamma}{}^M\tilde{D}_M\tilde{\Psi}+\tilde{\alpha}{}^{\dagger}(\tilde{D}_M\tilde{\overline{\Psi}})\tilde{\Gamma}{}^M\tilde{\Psi}\right)\\
    &\quad+\frac{(-1)^q\eta_0}{2x^*_0}\int_{V\mathcal{M}}d^{m+1}x|\tilde{e}|\left(\frac{\tilde{\alpha}+\tilde{\alpha}{}^{\dagger}}{4}\tilde{\overline{\Psi}}\tilde{I}_M\tilde{\Gamma}{}^M\tilde{\Psi}+\frac{\tilde{\alpha}-\tilde{\alpha}{}^{\dagger}}{2}\tilde{\overline{\Psi}}\tilde{m}_M\tilde{\Gamma}{}^M\tilde{\Psi}\right),
    \end{split}
\end{equation}
The last term in Eq. \eqref{nlp-spinor-action} reduces to a familiar object in $m=2k=4$ dimensions, since
\begin{equation}\label{M-g-relation-action}
    \tilde{M}\equiv\tilde{\Gamma}_{\underline{0}}\quad\Rightarrow \quad\tilde{m}=-\tilde{g}.
\end{equation}
As shown in Appendix \hyperref[app-B4:calc-spinor]{B.4}, the \textit{real} projective interaction $\tilde{I}_M$ appearing in Eq. \eqref{nlp-spinor-action} is simply the unnatural trace of the antisymmetric part of the projective distortion tensor,
\begin{equation}\label{distortion-interaction}
    \tilde{I}_M\equiv\tilde{N}_M:=\tilde{N}_{[ML]}{}^L=\tilde{Q}_M-2(m+1)\tilde{g}_M+2\tilde{\mathcal{T}}_M,
\end{equation}
where $\tilde{Q}_M=\tilde{Q}_{N\underline{AB}}(\tilde{E}{}^{-1}){}^{N}{}_{\underline{C}}\tilde{\eta}{}^{\underline{AC}}$ is the unnatural trace of the $\bar{\mathfrak{p}}{}^{-2}$-non-metricity, Eq. \eqref{unnatural-trace-Q-tilde}. Recall the ordinary definition of the distortion tensor, Eq. \eqref{distortion-def}. Moreover, it is shown in Appendix \hyperref[app-B4:calc-spinor]{B.4} that this interaction is equivalent to the commutator of distortion and the dual gamma matrix,
\begin{equation}
    \tilde{D}*\tilde{\Gamma}=\tilde{N}\wedge*\tilde{\Gamma}-*\tilde{\Gamma}\wedge\tilde{N},
\end{equation}
where
\begin{equation}
    \tilde{N}:=\tilde{N}_Mdx^M:=\frac{1}{4}\tilde{N}_{\underline{AB}M}\tilde{\Gamma}{}^{\underline{A}}\tilde{\Gamma}{}^{\underline{B}}dx^M
\end{equation}
is the $\bar{\mathfrak{p}}{}^{-2}$-distortion tensor expressed in the spin-basis $B_{\tilde{\Gamma}}$. The interaction $\tilde{I}_M$ may also be viewed as arising from a $\mathfrak{gl}(m+1,\mathbb{R})$-valued connection, $\Delta$, expressed in the spin-basis $B_{\tilde{\Gamma}}$. For then, if the nonlinear realization process were carried out on $\Delta$,
\begin{equation}
    \tilde{\Delta}_M=\frac{1}{4}\tilde{\Delta}_{\underline{AB}M}\tilde{\Gamma}{}^{\underline{A}}\tilde{\Gamma}{}^{\underline{B}}=\frac{1}{4}\tilde{\Omega}_{\underline{AB}M}\tilde{\Gamma}{}^{\underline{A}}\tilde{\Gamma}{}^{\underline{B}}+\frac{1}{4}\tilde{N}_{\underline{AB}M}\tilde{\Gamma}{}^{\underline{A}}\tilde{\Gamma}{}^{\underline{B}},
\end{equation}
where $\tilde{\Omega}$ is the nonlinear projective connection of Eq. \eqref{spinor-omega-tilde}. The above decomposition follows from splitting the $V\mathcal{M}$-connection $\tilde{\Gamma}$ into its Levi-Civita and distortion parts, $\hat{\tilde{\Gamma}}+\tilde{N}$, and then using the relation between $\tilde{\Omega}$ and $\tilde{\Gamma}$ via $\tilde{E}$, Eq. \eqref{gen-higgs-connection}.

In order to relate $S_{\tilde{\mathfrak{D}}}$ to the standard (projectively invariant) Dirac theory of Lorentz spinors, we work in the APV-gauge and sum over the $*$-indices, effectively providing a dimensional reduction at the level of the action. We then isolate $\tilde{\psi}$ and $\tilde{\overline{\psi}}$, resulting in an action written in terms of spinors transforming with respect to the Lorentz group. Working through each term separately, the first line of Eq. \eqref{nlp-spinor-action} is seen to produce an interaction with $\tilde{g}$, the projective Weyl-form. For the $\bar{\mathfrak{p}}{}^{\mathfrak{w}}$-spinor kinetic term, we find
\begin{equation}\label{spinor-g-der-split}
\tilde{\overline{\Psi}}\tilde{\Gamma}{}^M\tilde{D}_M\tilde{\Psi}=\tilde{\overline{\psi}}\tilde{\Gamma}{}^M\tilde{D}_M\tilde{\psi}+\mathfrak{w}\tilde{\overline{\Psi}}\tilde{g}_M\tilde{\Gamma}{}^M\tilde{\Psi},
\end{equation}
and for the adjoint $\bar{\mathfrak{p}}{}^{\mathfrak{w}}$-spinor kinetic term,
\begin{equation}\label{adj-spinor-g-der-split}
(\tilde{D}_M\tilde{\overline{\Psi}})\tilde{\Gamma}{}^M\tilde{\Psi}=(\tilde{D}_M\tilde{\overline{\psi}})\tilde{\Gamma}{}^M\tilde{\psi}-\mathfrak{w}\tilde{\overline{\Psi}}\tilde{g}_M\tilde{\Gamma}{}^M\tilde{\Psi}.
\end{equation}
Recall that $\tilde{\Omega}_M=\tilde{\Omega}_m\delta^m{}_M$ vanishes whenever the form-index $M=*$. Using this fact and choosing the APV-gauge, the interaction with the projective Weyl co-vector becomes
\begin{equation}\label{lil-g-APV}
    \tilde{g}_M\tilde{\Gamma}{}^M\overset{\circ}{=}\pm\frac{\sqrt{\eta_0}}{x^*_0}\gamma^{\underline{m+1}}.
\end{equation}
Reflecting these statements in the $\bar{\mathfrak{p}}{}^{\mathfrak{w}}$-spinor kinetic term yields
\begin{equation}\label{spinor-g-contribution}
    \tilde{\overline{\Psi}}\tilde{\Gamma}{}^M\tilde{D}_M\tilde{\Psi}\overset{\circ}{=}\tilde{\overline{\psi}}\tilde{\Gamma}{}^m\tilde{D}_m\tilde{\psi}\pm\frac{\mathfrak{w}\sqrt{\eta_0}}{x^*_0}\tilde{\overline{\psi}}\gamma^{\underline{m+1}}\tilde{\psi},
\end{equation}
while doing so for the adjoint $\bar{\mathfrak{p}}{}^{\mathfrak{w}}$-spinor kinetic term provides
\begin{equation}\label{adj-spinor-g-contribution}
   (\tilde{D}_M\tilde{\overline{\Psi}})\tilde{\Gamma}{}^M\tilde{\Psi}\overset{\circ}{=}(\tilde{D}_m\tilde{\overline{\psi}})\tilde{\Gamma}{}^m\tilde{\psi}\mp\frac{\mathfrak{w}\sqrt{\eta_0}}{x^*_0}\tilde{\overline{\psi}}\gamma^{\underline{m+1}}\tilde{\psi}.
\end{equation}
Both kinetic terms may be reduced further, since the interaction with the connection still contains internal $\underline{*}$-indices, i.e.,
\begin{equation}
    \tilde{D}_m=\partial_m\pm\frac{1}{4}\tilde{\Omega}_{\underline{AB}m}\tilde{\Gamma}{}^{\underline{A}}\tilde{\Gamma}{}^{\underline{B}}.
\end{equation}
Summing over these reduces Eq. \eqref{spinor-g-contribution} to
\begin{equation}\label{expanded-spin-kin}
    \tilde{\overline{\Psi}}\tilde{\Gamma}{}^M\tilde{D}_M\tilde{\Psi}\overset{\circ}{=}\tilde{\overline{\psi}}\gamma^m\hat{\overline{D}}_m\tilde{\psi}\mp\frac{x^*_0}{4\sqrt{\eta_0}}\tilde{\overline{\psi}}\gamma^{\underline{m+1}}\overline{\mathcal{P}}{}^{-}\tilde{\psi}\pm\frac{\mathfrak{w}\sqrt{\eta_0}}{x^*_0}\tilde{\overline{\psi}}\gamma^{\underline{m+1}}\tilde{\psi},
\end{equation}
and Eq. \eqref{adj-spinor-g-contribution} to
\begin{equation}\label{adj-expanded-spin-kin}
      (\tilde{D}_M\tilde{\overline{\Psi}})\tilde{\Gamma}{}^M\tilde{\Psi}\overset{\circ}{=}(\hat{\overline{D}}_m\tilde{\overline{\psi}})\gamma^m\tilde{\psi}\mp\frac{x^*_0}{4\sqrt{\eta_0}}\tilde{\overline{\psi}}\gamma^{\underline{m+1}}\overline{\mathcal{P}}{}^{-}\tilde{\psi}\mp\frac{\mathfrak{w}\sqrt{\eta_0}}{x^*_0}\tilde{\overline{\psi}}\gamma^{\underline{m+1}}\tilde{\psi},
\end{equation}
where
\begin{equation}
\hat{\overline{D}}_m\tilde{\psi}:=\partial_m\tilde{\psi}+\hat{\overline{\omega}}_{m}\tilde{\psi}=\partial_m\tilde{\psi}-\frac{i}{8}\hat{\overline{\omega}}_{[\underline{ab}]m}\sigma^{\underline{ab}}\tilde{\psi}
\end{equation}
may be identified with the ordinary (projectively invariant) Lorentz kinetic term. Notice that the $(\pm)$ resulting from Eq. \eqref{p-bar-gamma} is opposite $(\mp)$ between the two $\mathfrak{w}$-terms in Eqs. \eqref{expanded-spin-kin} and \eqref{adj-expanded-spin-kin}. However, the $(\mp)$ is the same in both for the coupling to the trace of the negatively-shifted projective Schouten tensor. This follows from anti-commuting the chiral matrix $\gamma^{\underline{m+1}}$. 

To complete the dimensional reduction at the level of the action, we must reduce the interaction term. This task is easily accomplished in the APV-gauge. From the relation in Eq. \eqref{distortion-interaction}, we find
\begin{equation}\label{summed-proj-interaction}
    \tilde{\overline{\Psi}}\tilde{I}_M\tilde{\Gamma}{}^M\tilde{\Psi}\overset{\circ}{=}\pm\tilde{\overline{\psi}}\left(\frac{x^*_0}{\sqrt{\eta_0}}\overline{\mathcal{P}}{}^{+}\gamma^{\underline{m+1}}-\frac{2m\sqrt{\eta_0}}{x^*_0}\gamma^{\underline{m+1}}\pm2\overline{\mathcal{T}}_m\gamma^m\pm\overline{Q}_m\gamma^m\right)\tilde{\psi},
\end{equation}
where Eqs. \eqref{unnatural-trace-Q-tilde} and \eqref{APV-torsion} were utilized.

Returning to the action, Eq. \eqref{nlp-spinor-action}, and recalling that the choice made in Eq. \eqref{M-g-relation-action} corresponds to $m=2k=4$ dimensions of split-signature $(p,q)=(1,3)$, we substitute all the above expanded expressions and combine like-terms to find
\begin{equation}\label{partially-reduced-proj-spin-action}
    \begin{split}
        S_{\tilde{\mathfrak{D}}}&\overset{\circ}{=}\frac{-\eta_0}{2}\int\frac{dx^*}{x^*}\int_{\mathcal{M}}d^{4}x|\overline{\vartheta}|\left(\tilde{\alpha}\tilde{\overline{\psi}}\gamma^m\hat{\overline{D}}_m\tilde{\psi}+\tilde{\alpha}{}^{\dagger}(\hat{\overline{D}}_m\tilde{\overline{\psi}})\gamma^m\tilde{\psi}+\frac{\tilde{\alpha}+\tilde{\alpha}{}^{\dagger}}{4}\tilde{\overline{\psi}}\gamma^m\overline{N}_m\tilde{\psi}\right)\\
    &\quad\quad\quad\quad\quad\quad\quad\quad\quad\quad\quad\pm\left(\frac{\tilde{\alpha}-\tilde{\alpha}{}^{\dagger}}{2}\frac{2\sqrt{\eta_0}}{x^*_0}(\mathfrak{w}-\frac{1}{2})\tilde{\overline{\psi}}\gamma^{\underline{5}}\tilde{\psi}\right),
    \end{split}
\end{equation}
where
\begin{equation}
    \overline{N}_m:=\overline{Q}_m+2\overline{\mathcal{T}}_m
\end{equation}
is the unnatural trace of the anti-symmetrized projectively invariant $m=4$-dimensional distortion tensor associated with $\omega$. Notice that we have exactly recovered the correct coupling to the non-Levi-Civita parts of the connection, as was expected from the identification in Eq. \eqref{distortion-interaction}. We thus find that dimensional reduction at the level of the action, paired with the requirement of Hermiticity, leads to there being \textit{no} interaction with $\overline{\mathcal{P}}$. This is quite interesting when juxtaposed to the non-Hermitian Thomas-Whitehead-Dirac action considered in \cite{gen-struc}, where the simple replacement of the $\mathcal{L}_{\tilde{\Gamma}}$ coefficient, $\tilde{\alpha}+\tilde{\alpha}{}^{\dagger}\rightarrow \tilde{\alpha}-\tilde{\alpha}{}^{\dagger}$, leads to an interaction with $\overline{\mathcal{P}}$. Since $\overline{\mathcal{P}}$ modulates dark energy, and in some cases interacts with spinors, it was postulated as a dark matter portal. However, as shown here, this portal behavior would require the action have no definite Hermiticity. Obviously, this is of utmost importance for reality, and therefore, is the subject of future investigation.

Considering the standard choice of coefficient, where $\tilde{\alpha}=-\tilde{\alpha}{}^{\dagger}=i\tilde{\beta}$ with $\tilde{\beta}\in\mathbb{R}$, the action reduces tremendously. The most important realization is that this choice of coefficient completely kills off the coupling to distortion, i.e., all the non-Levi-Civita deformations of the connection. Substituting the \textit{dimension-dependent} expression $\mathfrak{w}=\frac{1}{2}(1+i\mathfrak{c})$ for the projective weight, the reduction of $S_{\tilde{\mathfrak{D}}}$ to the physical, projectively invariant Dirac-type theory of Lorentz spinors is
\begin{equation}
        S_{\tilde{\mathfrak{D}}}\overset{\circ}{=}-\eta_0\beta\int_{\mathcal{M}}d^{4}x|\overline{\vartheta}|\left(\frac{i}{2}\tilde{\overline{\psi}}\gamma^m\hat{\overline{D}}_m\tilde{\psi}-\frac{i}{2}(\hat{\overline{D}}_m\tilde{\overline{\psi}})\gamma^m\tilde{\psi}\mp m_{\mathfrak{c}}\tilde{\overline{\psi}}\gamma^{5}\tilde{\psi}\right),
\end{equation}
where
\begin{equation}
   \beta:=\tilde{\beta}\int^{x^*_f}_{x^*_i}\frac{dx^*}{x^*}=\tilde{\beta}\log\left(\frac{x^*_f}{x^*_i}\right)
\end{equation}
is the``renormalized" dimensionless coefficient, and
\begin{equation}
m_{\mathfrak{c}}:=\frac{\mathfrak{c}}{2x^*_0}
\end{equation}
is the induced chiral mass, proportional to the undetermined complex part of the projective weight, $\mathfrak{c}$. Following \cite{heavy-lifting}, we have also made the identification that it is $\sqrt{\eta_0}\gamma^{\underline{5}}$ which is the spacetime chiral gamma matrix. Since, for $\eta_0=-1$, the proper spacetime chiral gamma matrix is $i\gamma^{\underline{5}}$. We therefore make the general statement
\begin{equation}
    \gamma^{5}:=\sqrt{\eta_0}\gamma^{\underline{5}}.
\end{equation}
Lastly, we comment on the renormalized coefficient $\beta$. Since the standard choice of coefficient is $\tilde{\alpha}=-\tilde{\alpha}{}^{\dagger}=i$, leading to $\tilde{\beta}=1$, it would appear that some new constant has emerged. However, one must keep in mind we have been working in natural units, $\hbar=c=1$. Therefore, we may consider the renormalization as being enacted on, say, $\hbar$. The value of $\hbar$ which results from such a process would be the value measured today, $\hbar\approx 6.626\times 10^{-34}\;\text{J}\cdot \text{s}$.

\subsubsection{Field Equations}
\label{sec:proj-matter-field-eq}

The variation of the action, Eq. \eqref{S-D-bar}, with respect to the projective matter field variables $\mathcal{F}_M=\{\tilde{\overline{\Psi}},\tilde{\Psi}\}$ produces
\begin{equation}
    \delta S_{\tilde{\mathfrak{D}}}=\frac{1}{x^*_0}\int_{V\mathcal{M}}\delta\tilde{\overline{\Psi}}\;\tilde{\overline{\mathbb{F}}}+\tilde{\mathbb{F}}\;\delta\tilde{\Psi},
\end{equation}
where
\begin{equation}
    \tilde{\overline{\mathbb{F}}}:=\frac{\delta\mathcal{L}_{\tilde{\mathfrak{D}}}}{\delta\tilde{\overline{\Psi}}}=\frac{\partial\mathcal{L}_{\tilde{\mathfrak{D}}}}{\partial\tilde{\overline{\Psi}}}-\tilde{D}\frac{\partial\mathcal{L}_{\tilde{\mathfrak{D}}}}{\partial(\tilde{D}\tilde{\overline{\Psi}})},
\end{equation}
\begin{equation}
    \tilde{\mathbb{F}}:=\frac{\delta\mathcal{L}_{\tilde{\mathfrak{D}}}}{\delta\tilde{\Psi}}=\frac{\partial\mathcal{L}_{\tilde{\mathfrak{D}}}}{\partial\tilde{\Psi}}-\tilde{D}\frac{\partial\mathcal{L}_{\tilde{\mathfrak{D}}}}{\partial(\tilde{D}\tilde{\Psi})}
\end{equation}
are, respectively, the adjoint $\bar{\mathfrak{p}}{}^{\mathfrak{w}}$-spinor and $\bar{\mathfrak{p}}{}^{\mathfrak{w}}$-spinor field variations. Setting these to vanish and dividing by the overall factor $\frac{\tilde{\alpha}-\tilde{\alpha}{}^{\dagger}}{2}$ produces the beautifully symmetric pair of equations:
\begin{equation}
     \delta\tilde{\overline{\Psi}}:\quad \tilde{\mathfrak{D}}\tilde{\Psi}=0,\quad\quad\quad\text{and}\quad\quad \quad\delta\tilde{\Psi}:\quad\tilde{\mathfrak{D}}{}^{\dagger}\tilde{\overline{\Psi}}=0.
\end{equation}
The simplicity of these expressions results from $\mathcal{L}_{\mathfrak{D}}$ being related to $\mathfrak{L}_{\Psi}$ via a total derivative. Therefore, the field equations are independent of the collective coefficient $\frac{\tilde{\alpha}-\tilde{\alpha}{}^{\dagger}}{2}$ and no new degree of freedom is introduced. Since these field equations are \textit{conjugates of one another}, it is sufficient to view only one. We choose to work with $\tilde{\overline{\mathbb{F}}}$, which expands to
\begin{equation}
    \delta\tilde{\overline{\Psi}}:\quad  *\tilde{\Gamma}\wedge\tilde{D}\tilde{\Psi}+\frac{1}{2}(\tilde{D}*\tilde{\Gamma})\tilde{\Psi}=\frac{-1}{2}\tilde{m}\wedge*\tilde{\Gamma}\tilde{\Psi}.
\end{equation}
The conjugate equation simply replaces $\tilde{m}\rightarrow-\tilde{m}$. In order to connect this expression with the conventional (projectively invariant) Dirac theory of Lorentz spinors, we again express this in abstract-index notation. The details of this transition to abstract-index form may be found in Appendix \hyperref[app-B3:exterior-abs-ind-form]{B.3}. For this, we find
\begin{equation}
    \delta\tilde{\overline{\Psi}}:\quad\tilde{\Gamma}{}^M\tilde{D}_M\tilde{\Psi}+\frac{1}{4}\tilde{\Gamma}{}^M\tilde{I}_M\tilde{\Psi}=\frac{1}{2}\tilde{g}_M\tilde{\Gamma}{}^M\tilde{\Psi},
\end{equation}
where we have canceled overall factors and used $\tilde{m}=-\tilde{g}$. Recall, this latter choice is only compatible with choosing $m=2k=4$ dimensions with a split-signature $(p,q)=(1,3)$.

In order to investigate the space of solutions for the projective matter field variable $\mathcal{F}_M$, we dimensionally reduce by summing over the $*$-indices. This process is similar to what was accomplished previously, now at the level of the field equations. To simplify discussions, we choose to work in the APV-gauge. Recall that the APV-gauge is simply a fixing of the origin or contact point of the model homogeneous space $(G/H)$ with the spacetime $\mathcal{M}$. 

Reflecting these statements in the field equations and using the expressions of the previous subsection for the interactions, Eqs. \eqref{summed-proj-interaction} and \eqref{lil-g-APV}, we find
\begin{equation}
    \delta\tilde{\overline{\Psi}}:\quad i\gamma^m\hat{\overline{D}}_m\tilde{\psi}+\frac{i}{4}\gamma^m\overline{N}_m\tilde{\psi}\overset{\circ}{=}\pm m_{\mathfrak{c}}\gamma^{5}\tilde{\psi},
\end{equation}
and
\begin{equation}
    \delta\tilde{\Psi}:\quad i(\hat{\overline{D}}_m\tilde{\overline{\psi}})\gamma^m+\frac{i}{4}\tilde{\overline{\psi}}\gamma^m\overline{N}_m\overset{\circ}{=}\mp m_{\mathfrak{c}}\tilde{\overline{\psi}}\gamma^{5},
\end{equation}
where the factors of $i$ result from $\mathfrak{w}=\frac{1}{2}(1+i\mathfrak{c})$. Notice the retention of the coupling to the projectively invariant $4$-dimensional distortion (non-metricity and torsion). Due to the nonlinear realization procedure, the antisymmetric connection $\hat{\overline{\omega}}$ contains only $\frac{m(m-1)}{2}=6$ degrees of freedom and, therefore, does not contain any dependence on the torsion or non-metricity $(\overline{N})$. For this reason, it is identified with the projectively invariant Lorentz connection. It is interesting to note that, comparing with \cite{clifforms}, we find that when $\overline{Q}=0$ the interaction with $\overline{N}$ reduces to a coupling to the \textit{torsion vector} and \textit{not} the \textit{axial torsion}. Moreover, we have recovered exactly what has been claimed in Metric-Affine gravitational theories to be impossible---a coupling between spinors and the symmetric traceless parts of the connection, i.e., $\overline{Q}$ \cite{janssen}.

Furthermore, the projective Lorentz spinors were provided the induced chiral mass $m_{\mathfrak{c}}$, dependent on the undetermined complex projective weight $\mathfrak{c}$. These induced chiral masses are real, and therefore, do not lead to any $CP$-violation. It is interesting to note that, in contrast to the Thomas-Whitehead-Dirac theory of \cite{gen-struc}, \textit{no choice of $\mathfrak{w}$ can eliminate $m_{\mathfrak{c}}$}. According to the estimates of \cite{inflation}, summarized briefly in \cite{covariant-tw}, the induced chiral mass is on the order of
\begin{equation}
    m_{\mathfrak{c}}\approx\mathfrak{c}\times 10^{12}\;\text{GeV}.
\end{equation}
Provided that $\mathfrak{c}\sim 1$, this extremely large mass may be useful in a \textit{seesaw mechanism} for the generation of extremely light masses for left-handed neutrinos $(\nu)$, and extremely large masses for sterile right-handed neutrinos $(N)$ \cite{wienberg,nonhermitian-seesaw}. Briefly, if one considers the Lorentz-transforming spinor $\tilde{\psi}$ to be of Majorana type, then $m_{\mathfrak{c}}$ may be identified with the Majorana mass. Assuming this field also couples to the standard model Higgs field, then upon spontaneously breaking the electroweak symmetry a standard Dirac-type mass is generated, $m_D=\frac{Y v_{H}}{2}$, where $v_{H}\approx 246 \;\text{Gev}$ is the Higgs vacuum expectation value and $Y$ is the Yukawa coupling strength. Diagonalizing the mass matrix and performing a field rotation on the spinor to the mass eigen-basis provides two mass eigenvalues. In the limit $m_{\mathfrak{c}}>>m_D$, the two masses are
\begin{equation}
    m_{\nu}\approx\frac{m_D^2}{m_{\mathfrak{c}}},\quad\quad\quad m_{N}\approx m_{\mathfrak{c}}(1+\frac{m^2_{\nu}}{m_{\mathfrak{c}}}).
\end{equation}
Inputting the above approximate values and considering $Y\approx 0.1$ and $\mathfrak{c}\approx1$, these masses are of the order
\begin{equation}
    m_{\nu}\approx 0.1\;\text{eV},\quad\quad\quad m_N\approx 10^{12}\;\text{Gev}.
\end{equation}
This possibility is of deep interest and will play a large part of future research projects.

Since no chiral interaction with the projective Schouten trace is found when Hermiticity is enforced, a modification must be introduced if one wishes to recover such a coupling. One possibility would be to consider a Pauli-type modification, such as the one considered in \cite{pauli-coupling}. Essentially, this exploits the fact that adding a total derivative to $\mathcal{L}_{\tilde{\mathfrak{D}}}$ will not alter the field equations. A convenient choice for this term provides a direct coupling to the antisymmetric part of the curvature tensor (spinor). This, however, may not produce the same type of chiral coupling as in \cite{gen-struc}, since this process would appear to rather descend directly to a chiral coupling to the field strength $\overline{\mathcal{S}}_{\underline{a}}$. Only after integration by parts could one recover a coupling to the projective Schouten tensor, which will likely lead to quite a different overall form than intended. Such a process is certainly of interest, as it may provide access to understanding the dark sector. Investigation into this additional term is left for future research.

\subsubsection{Projective Currents}
\label{sec:proj-matter-D2}

Since $\mathcal{L}_{\tilde{\mathfrak{D}}}$ is initially a massless theory and, according to \cite{massive-anomaly}, the Hermiticity of $\mathcal{L}_{\tilde{\mathfrak{D}}}$ together provide the \textit{projective vector current density}
\begin{equation}
    \tilde{J}:=\tilde{\overline{\Psi}}*\tilde{\Gamma}\tilde{\Psi}.
\end{equation}
The components of $\tilde{J}$ may be extracted via the internal product, defined in Eq. \eqref{internal-product},
\begin{equation}
    \tilde{J}_{\underline{A}}:=(\tilde{E}{}^{-1})_{\underline{A}}\rfloor*\tilde{J}=(-1)^q\eta_0\tilde{\overline{\Psi}}\tilde{\Gamma}_{\underline{A}}\tilde{\Psi}.
\end{equation}
The projective vector current-density, when restricted to the space of solutions, i.e., ``on-shell", is seen to satisfy the covariant conservation law $\tilde{D}\tilde{J}=0$. Notice that $\tilde{J}$ is conserved with respect to $\tilde{D}$ and \textit{not} $\tilde{\mathfrak{D}}$. This may be seen by direct calculation,
\begin{equation}
    \begin{split}
        \tilde{D}\tilde{J}&=\tilde{D}(\tilde{\overline{\Psi}}*\tilde{\Gamma}\tilde{\Psi})\\
        &=\tilde{D}\tilde{\overline{\Psi}}\wedge*\tilde{\Gamma}\tilde{\Psi}+\tilde{\overline{\Psi}}\tilde{D}(*\tilde{\Gamma})\tilde{\Psi}+\tilde{\overline{\Psi}}*\tilde{\Gamma}\wedge\tilde{D}\tilde{\Psi}\\
        &=\left(\tilde{D}\tilde{\overline{\Psi}}\wedge*\tilde{\Gamma}\tilde{\Psi}+\frac{1}{2}\tilde{\overline{\Psi}}\tilde{D}(*\tilde{\Gamma})\tilde{\Psi}\right)+\left(\tilde{\overline{\Psi}}*\tilde{\Gamma}\wedge\tilde{D}\tilde{\Psi}+\frac{1}{2}\tilde{\overline{\Psi}}\tilde{D}(*\tilde{\Gamma})\tilde{\Psi}\right)\\
&=(\tilde{\mathfrak{D}}^{\dagger}\tilde{\overline{\Psi}})\tilde{\Psi}+\tilde{\overline{\Psi}}\tilde{\mathfrak{D}}\tilde{\Psi}\\
&=\tilde{\mathbb{F}}\tilde{\Psi}+\tilde{\overline{\Psi}}\tilde{\overline{\mathbb{F}}}\\
        &=0,
    \end{split}
\end{equation}
when the (purely projective matter) field equations $\tilde{\mathbb{F}}$ and $\tilde{\overline{\mathbb{F}}}$ are satisfied. 

As part of the future research program, constructing an object which descends to the \textit{projective axial vector current density}, if possible, may shed light on a projective interpretation of the chiral anomaly. The naive attempt would be, for example, the $\bar{\mathfrak{p}}$-$4$-form
\begin{equation}
\tilde{A}{}^{\underline{*}}=\tilde{\overline{\Psi}}\tilde{\Gamma}{}^{\underline{*}}*\tilde{\Gamma}\tilde{\Psi},
\end{equation}
since its components,
\begin{equation}
    \tilde{A}{}^{\underline{*}}{}_{\underline{B}}:=(\tilde{E}{}^{-1})_{\underline{B}}\rfloor*\tilde{A}{}^{\underline{*}}=(-1)^q\eta_0\tilde{\overline{\Psi}}\tilde{\Gamma}{}^{\underline{*}}\tilde{\Gamma}_{\underline{B}}\tilde{\Psi},
\end{equation}
are seen to contain the usual axial vector current in the $\underline{b}$-components, and is properly of type-$\bar{\mathfrak{p}}{}^0$, as is required for physicality. Alternatively, following \cite{massive-anomaly}, the natural extension of the $4$-dimensional axial current $3$-form,
\begin{equation}
    \mathcal{A}:=\frac{i}{3!}\overline{\psi}\gamma\wedge\gamma\wedge\gamma\psi=\frac{1}{3}\overline{\psi}\sigma\wedge\gamma\psi,
\end{equation}
to the projectively invariant setting of $2k+1=5$ dimensions would be the $4$-form
\begin{equation}
\tilde{\mathcal{A}}:=\frac{1}{4!}\tilde{\overline{\Psi}}\tilde{\Gamma}\wedge\tilde{\Gamma}\wedge\tilde{\Gamma}\wedge\tilde{\Gamma}\tilde{\Psi}=\frac{-1}{3!}\tilde{\overline{\Psi}}\tilde{\Sigma}\wedge\tilde{\Sigma}\tilde{\Psi}.
\end{equation}
However, the components of this object are just the vector current density,
\begin{equation}
    \tilde{\mathcal{A}}_{\underline{B}}:=(\tilde{E}{}^{-1})_{\underline{B}}\rfloor*\tilde{\mathcal{A}}=(-1)^q\eta_0\tilde{\overline{\Psi}}\tilde{\Gamma}_{\underline{B}}\tilde{\Psi}\equiv \tilde{J}_{\underline{B}}.
\end{equation}
The gauge-covariant derivative of this object, calculated in Appendix \hyperref[app-B3:exterior-abs-ind-form]{B.3}, is found in the APV gauge to have the form
\begin{equation}
    \tilde{D}\tilde{\mathcal{A}}\overset{\circ}{=}\pm i(-1)^q(\frac{1}{\sqrt{\eta_0}}+\sqrt{\eta_0})\tilde{\overline{\Psi}}\gamma^{\underline{a}}\overline{\mathcal{T}}_{\underline{a}}\tilde{\Psi},
\end{equation}
which vanishes for spacelike $x^*$, i.e., $\eta_0=-1$. 

We are thus confronted with the difficulty of describing chiral attributes in $5$ dimensions. Since the descent to $4$ dimensions imparts a chiral mass to the Lorentz-transforming spinors, one may expect a sort of chiral symmetry breaking to have occurred in the process. Much of the investigation into this topic is left for a future research program. However, since this future research likely includes a direct calculation of the potential chiral anomaly present in the theory, as a first step in this direction, we compute the ``square" of the Hermitian operator
\begin{equation}
    \tilde{\mathfrak{D}}=*\tilde{\Gamma}\wedge\tilde{D}+\frac{1}{2}\tilde{M}{}^{-1}\tilde{D}(\tilde{M}*\tilde{\Gamma}).
\end{equation}
The square of this operator is to be used in the Fujikawa method of anomaly calculation \cite{chandia-zinelli}. The components are extracted via the internal dual,
\begin{equation}
    *\tilde{\mathfrak{D}}=(-1)^q\eta_0\tilde{\Gamma}{}^{\underline{A}}(\tilde{D}_{\underline{A}}+\frac{1}{2}\tilde{X}^{+}_{\underline{A}}),
\end{equation}
valid in $2k+1$-dimensions. In the above, we have introduced the convenient notation
\begin{equation}
\tilde{X}^{\pm}_{\underline{A}}:=\tilde{m}_{\underline{A}}\pm\frac{1}{2}\tilde{I}_{\underline{A}}.
\end{equation}
The adjoint operator then has the components
\begin{equation}
    *\tilde{\mathfrak{D}}^{\dagger}=(-1)^q\eta_0\tilde{\Gamma}{}^{\underline{A}}(\tilde{D}_{\underline{A}}-\frac{1}{2}\tilde{X}^{-}_{\underline{A}}).
\end{equation}

The square of the $\mathfrak{D}$-operator,
\begin{equation}
|\tilde{\mathfrak{D}}|^2:=*\tilde{\mathfrak{D}}^{\dagger}*\tilde{\mathfrak{D}}=\tilde{\Gamma}{}^{\underline{A}}(\tilde{D}_{\underline{A}}-\frac{1}{2}\tilde{X}^{-}_{\underline{A}})\tilde{\Gamma}{}^{\underline{B}}(\tilde{D}_{\underline{B}}+\frac{1}{2}\tilde{X}^{+}_{\underline{B}}),
\end{equation}
 is easily found, with a bit of simplification, to have the signature-independent form
\begin{equation}
     |\tilde{\mathfrak{D}}|^2=\tilde{\Gamma}{}^{\underline{A}}\tilde{\Gamma}{}^{\underline{B}}\tilde{D}_{\underline{A}}\tilde{D}_{\underline{B}}+\tilde{Y}{}^{\underline{A}}\tilde{D}_{\underline{A}}+\tilde{Z}.
 \end{equation}
 In this expression, we have introduced the following convenient definitions:
\begin{equation}
     \tilde{\Gamma}{}^{\underline{A}}\tilde{\Gamma}{}^{\underline{B}}\tilde{D}_{\underline{A}}\tilde{D}_{\underline{B}}=\tilde{D}^{\underline{A}}\tilde{D}_{\underline{A}}-\frac{1}{8}\tilde{\Sigma}{}^{{\underline{AB}}}\tilde{\Sigma}{}^{\underline{CD}}\tilde{\mathcal{K}}_{\underline{CD}}{}_{\underline{AB}},
 \end{equation}
\begin{equation}
    \tilde{Y}{}^{\underline{A}}:=\frac{1}{2}\left(\tilde{Q}_{\underline{B}}{}^{\underline{A}}{}_{\underline{C}}\tilde{\Gamma}{}^{\underline{B}}\tilde{\Gamma}{}^{\underline{C}}+\tilde{\Gamma}{}^{\underline{A}}\tilde{\Gamma}{}^{\underline{B}}\tilde{X}{}^{+}_{\underline{B}}-\tilde{\Gamma}{}^{\underline{B}}\tilde{\Gamma}{}^{\underline{A}}\tilde{X}{}^{-}_{\underline{B}}\right),
\end{equation}
\begin{equation}
    \tilde{Z}:=\frac{1}{2}\tilde{\Gamma}{}^{\underline{A}}\tilde{\Gamma}{}^{\underline{B}}\left(\tilde{D}_{\underline{A}}\tilde{X}{}^{+}_{\underline{B}}+\frac{1}{2}\tilde{Q}_{\underline{A}}{}^{\underline{C}}{}_{\underline{B}}\tilde{X}{}^{+}_{\underline{C}}-\frac{1}{2}\tilde{X}{}^{-}_{\underline{A}}\tilde{X}{}^{+}_{\underline{B}}\right).
\end{equation}
The remainder of the calculation is quite involved, and again, is left as a future research program. For more information on the chiral anomaly, see \cite{chandia-zinelli,anomaly-torsion,anomaly-torsion2,torsion-nieh-yan-anomaly,massive-anomaly,mielke-anomaly2,mielke-anomaly3,PV-reg-anomaly-torsion,zumino-anomaly}.

\subsubsection{Gravitational Sector}

For the coupling of $\bar{\mathfrak{p}}{}^{\mathfrak{w}}$-spinors to gravity, recall from Eq. \eqref{proj-matter-currents} that the variation of the projective matter sector, $\mathcal{L}_M\equiv\mathcal{L}_{\tilde{\mathfrak{D}}}$, with respect to the connection $\tilde{\Omega}$ yields
\begin{equation}
    \tilde{\mathbb{M}}{}^{\underline{B}}{}_{\underline{A}}=-\tilde{D}\tilde{\mathbb{I}}{}^{\underline{B}}{}_{\underline{A}}+\tilde{\Upsilon}{}^{\underline{B}}\tilde{\mathbb{U}}_{\underline{A}}+\tilde{\mathbb{J}}{}^{\underline{B}}{}_{\underline{A}}.
\end{equation}
Since for $\mathcal{L}_{\tilde{\mathfrak{D}}}$ there is no explicit coupling to the curvature, $\tilde{\mathbb{I}}{}^{\underline{B}}{}_{\underline{A}}=0$. The case for $\tilde{\mathbb{I}}{}^{\underline{B}}{}_{\underline{A}}\neq0$, discussed at the end of Sec. \ref{sec:proj-matter-field-eq}, would follow from a Pauli-type coupling \cite{pauli-coupling}. Additionally, the explicit Hermiticity of $\mathcal{L}_{\tilde{\mathfrak{D}}}$ modifies the projective extension of the ``ordinary" spin-current density, Eq. \eqref{proj-ext-ordinary-matter-current}, to
\begin{equation}
    \tilde{\mathbb{J}}{}^{\underline{B}}{}_{\underline{A}}=\tilde{\overline{\Psi}}\rho_{\mathfrak{w}}(\bm{L}{}^{\underline{B}}{}_{\underline{A}})\frac{\partial\mathcal{L}_{\tilde{\mathfrak{D}}}}{\partial(\tilde{D}\tilde{\overline{\Psi}})}+\frac{\partial\mathcal{L}_{\tilde{\mathfrak{D}}}}{\partial(\tilde{D}\tilde{\Psi})}\rho_{\mathfrak{w}}(\bm{L}{}^{\underline{B}}{}_{\underline{A}})\tilde{\Psi},
\end{equation}
where
\begin{equation}
    \rho_{\mathfrak{w}}(\bm{L}^{\underline{B}}{}_{\underline{A}}):=\frac{1}{4}\tilde{\Gamma}{}^{\underline{B}}\tilde{\Gamma}_{\underline{A}}
\end{equation}
is our choice of representation, consistent with Eq. \eqref{spinor-omega-tilde}. This projective, or rather, $\bar{\mathfrak{p}}{}^{0}$-spin-current density is found to have the form
\begin{equation}
    \tilde{\mathbb{J}}{}^{\underline{B}}{}_{\underline{A}}=\frac{\tilde{\alpha}{}^{\dagger}}{2}\tilde{\overline{\Psi}}\rho_{\mathfrak{w}}(\bm{L}^{\underline{B}}{}_{\underline{A}})*\tilde{\Gamma}\tilde{\Psi}+(-1)^{1+m}\frac{\tilde{\alpha}}{2}\tilde{\overline{\Psi}}*\tilde{\Gamma}\rho_{\mathfrak{w}}(\bm{L}^{\underline{B}}{}_{\underline{A}})\tilde{\Psi}.
\end{equation}
The $(-1)^{1+m}$ appears as a result of commuting the partial derivative of a spinor with a spinor, and commuting a one-form with an $m=2k$-form. Choosing $\tilde{\alpha}=-\tilde{\alpha}{}^{\dagger}=i\tilde{\beta}$, and using the definition of $\rho_{\mathfrak{w}}(\bm{L}^{\underline{B}}{}_{\underline{A}})$ provides
\begin{equation}
    \tilde{\mathbb{J}}{}^{\underline{B}}{}_{\underline{A}}=\frac{-i\tilde{\beta}}{8}\tilde{\overline{\Psi}}\left\{*\tilde{\Gamma},\tilde{\Gamma}{}^{\underline{B}}\tilde{\Gamma}_{\underline{A}}\right\}\tilde{\Psi}.
\end{equation}
We may view the components of the $\bar{\mathfrak{p}}{}^{0}$-spin-current density by taking the internal dual, paired with an internal product. Explicitly,
\begin{equation}
    \tilde{\mathbb{J}}{}^{\underline{B}}{}_{\underline{A}}{}_{\underline{C}}:=(\tilde{E}{}^{-1})_{\underline{C}}\rfloor*\tilde{\mathbb{J}}{}^{\underline{B}}{}_{\underline{A}}.
\end{equation}
To remain consistent with our conventions, the index arising from the interior product will be placed on the outside. Piece-by-piece, the internal dual of the $\bar{\mathfrak{p}}{}^{0}$-spin-current density is
\begin{equation}
    *\tilde{\mathbb{J}}{}^{\underline{B}}{}_{\underline{A}}=-\frac{i\tilde{\beta}}{8}\tilde{\overline{\Psi}}\left\{**\tilde{\Gamma},\tilde{\Gamma}{}^{\underline{B}}\tilde{\Gamma}_{\underline{A}}\right\}\tilde{\Psi}=(-1)^{q+1}\frac{i\tilde{\beta}\eta_0}{8}\tilde{\overline{\Psi}}\left\{\tilde{\Gamma}_{\underline{D}}\tilde{E}{}^{\underline{D}},\tilde{\Gamma}{}^{\underline{B}}\tilde{\Gamma}_{\underline{A}}\right\}\tilde{\Psi}.
\end{equation}
Then, the internal product gives
\begin{equation}
\begin{split}
    (\tilde{E}{}^{-1})_{\underline{C}}\rfloor*\tilde{\mathbb{J}}{}^{\underline{B}}{}_{\underline{A}}&=(-1)^{q+1}\frac{i\tilde{\beta}\eta_0}{8}\tilde{\overline{\Psi}}\left\{\tilde{\Gamma}_{\underline{D}}\tilde{E}{}^{\underline{D}}{}_M(\tilde{E}{}^{-1})^M{}_{\underline{C}},\tilde{\Gamma}{}^{\underline{B}}\tilde{\Gamma}_{\underline{A}}\right\}\tilde{\Psi}\\
    &=(-1)^{q+1}\frac{i\tilde{\beta}\eta_0}{8}\tilde{\overline{\Psi}}\left\{\tilde{\Gamma}_{\underline{C}},\tilde{\Gamma}{}^{\underline{B}}\tilde{\Gamma}_{\underline{A}}\right\}\tilde{\Psi}.
    \end{split}
\end{equation}
Separating the (anti)-symmetric parts, the final form of the $\bar{\mathfrak{p}}{}^{0}$-spin-current density components is
\begin{equation}
    \tilde{\mathbb{J}}{}^{\underline{B}}{}_{\underline{A}}{}_{\underline{C}}=(-1)^{q+1}\frac{i\tilde{\beta}\eta_0}{4}\tilde{\overline{\Psi}}\delta^{\underline{B}}{}_{\underline{A}}\tilde{\Gamma}_{\underline{C}}\tilde{\Psi}+(-1)^{q+1}\frac{\tilde{\beta}\eta_0}{8}\tilde{\overline{\Psi}}\left\{\tilde{\Sigma}{}^{\underline{B}}{}_{\underline{A}},\tilde{\Gamma}_{\underline{C}}\right\}\tilde{\Psi}.
\end{equation}
It must be noted, however, that $\tilde{\mathbb{J}}$ arises from variation of the connection, which is traceless in the basis $B_{\tilde{\Gamma}}$. Therefore, the first term above is likely to be removed. The second term appearing above is the standard spin-current, extended to the projective sector, in agreement with \cite{spin-metric}. Since we are only considering non-chiral projective weights $\mathfrak{w}$, the above may be written in terms of the projective Lorentz spinors. To further clean up the expression, consider $m=2k=4$ dimensions with a split-signature $(p,q)=(1,3)$, and spacelike $\eta_0$,
\begin{equation}
    \tilde{\mathbb{J}}{}^{\underline{B}}{}_{\underline{A}}{}_{\underline{C}}=\frac{-i\tilde{\beta}}{4}\tilde{\overline{\psi}}\left(\delta^{\underline{B}}{}_{\underline{A}}\tilde{\Gamma}_{\underline{C}}-\frac{i}{2}\left\{\tilde{\Sigma}{}^{\underline{B}}{}_{\underline{A}},\tilde{\Gamma}_{\underline{C}}\right\}\right)\tilde{\psi}.
\end{equation}

The remaining current to calculate is $\tilde{\mathbb{U}}_{\underline{B}}$. For convenience, we recall its definition here as
\begin{equation}
    \tilde{\mathbb{U}}_{\underline{A}}:=-\frac{\partial\mathcal{L}_M}{\partial(\tilde{D}\tilde{\Upsilon}{}^{\underline{A}})}.
\end{equation}
Explicitly, we find
\begin{equation}\label{expanded-spinor-U}
\begin{split}
    \tilde{\mathbb{U}}_{\underline{B}}&=\frac{\tilde{\alpha}}{2}\tilde{\overline{\Psi}}\tilde{\Gamma}{}^{\underline{A}}\hat{\tilde{\epsilon}}_{\underline{AB}}\wedge\tilde{D}\tilde{\Psi}-\frac{\tilde{\alpha}{}^{\dagger}}{2}\tilde{D}\tilde{\overline{\Psi}}\wedge\tilde{\Gamma}{}^{\underline{A}}\hat{\tilde{\epsilon}}_{\underline{AB}}\tilde{\Psi}\\
    &\quad-\frac{\tilde{\alpha}-\tilde{\alpha}{}^{\dagger}}{4}\tilde{\overline{\Psi}}\tilde{m}\wedge\tilde{\Gamma}{}^{\underline{A}}\hat{\tilde{\epsilon}}_{\underline{AB}}\tilde{\Psi}-\frac{\tilde{\alpha}+\tilde{\alpha}{}^{\dagger}}{4}\tilde{\overline{\Psi}}[\tilde{N},\tilde{\Gamma}{}^{\underline{A}}\hat{\epsilon}_{\underline{AB}}]\tilde{\Psi}.
    \end{split}
\end{equation}
Again, choosing the standard coefficient $\tilde{\alpha}=-\tilde{\alpha}{}^{\dagger}=i\tilde{\beta}$ forces the last term to vanish. The components of $\tilde{\mathbb{U}}_{\underline{B}}$ are accessed in the same manner as $\tilde{\mathbb{J}}{}^{\underline{A}}{}_{\underline{B}}$,
\begin{equation}
     \tilde{\mathbb{U}}_{\underline{B}}{}_{\underline{C}}:=(\tilde{E}{}^{-1})_{\underline{C}}\rfloor*\tilde{\mathbb{U}}_{\underline{B}}.
\end{equation}
We first take the internal dual, which provides
\begin{equation}
        *\tilde{\mathbb{U}}_{\underline{B}}=\frac{i\tilde{\beta}}{2}\left(\tilde{\overline{\Psi}}\tilde{\Gamma}{}^{\underline{A}}*(\hat{\tilde{\epsilon}}_{\underline{AB}}\wedge\tilde{D}\tilde{\Psi})+*(\tilde{D}\tilde{\overline{\Psi}}\wedge\hat{\tilde{\epsilon}}_{\underline{AB}})\tilde{\Gamma}{}^{\underline{A}}\tilde{\Psi}-\tilde{\overline{\Psi}}*(\tilde{m}\wedge\hat{\tilde{\epsilon}}_{\underline{AB}})\tilde{\Gamma}{}^{\underline{A}}\tilde{\Psi}\right).
\end{equation}
The first term, in arbitrary $m=2k$ dimensions, becomes
\begin{equation}
    \begin{split}
        \tilde{\overline{\Psi}}\tilde{\Gamma}{}^{\underline{A}_1}*(\hat{\tilde{\epsilon}}_{\underline{A}_1\underline{B}}\wedge\tilde{D}\tilde{\Psi})&=\frac{1}{(m-1)!}\tilde{\overline{\Psi}}\tilde{\Gamma}{}^{\underline{A}_1}\hat{\tilde{\epsilon}}_{\underline{A}_1\underline{B}\underline{A}_3\dots\underline{A}_{m+1}}*(\tilde{E}{}^{\underline{A}_3}\wedge\dots\wedge\tilde{E}{}^{\underline{A}_{m+1}}\wedge\tilde{E}{}^{\underline{A}_2})\tilde{D}_{\underline{A}_2}\tilde{\Psi}\\
        &=\frac{1}{(m-1)!}\tilde{\overline{\Psi}}\tilde{\Gamma}{}^{\underline{A}_1}\hat{\tilde{\epsilon}}_{\underline{A}_1\underline{B}\underline{A}_3\dots\underline{A}_{m+1}}\hat{\tilde{\epsilon}}{}^{\underline{A}_3\dots\underline{A}_{m+1}\underline{A}_2}{}_{\underline{A}}\tilde{E}{}^{\underline{A}}\tilde{D}_{\underline{A}_2}\tilde{\Psi}\\
        &=(-1)^q\eta_0\tilde{\overline{\Psi}}\tilde{\Gamma}{}^{\underline{A}_1}(\delta^{\underline{A}_2}{}_{\underline{A}_1}\tilde{\eta}_{\underline{BA}}-\delta^{\underline{A}_2}{}_{\underline{B}}\tilde{\eta}_{\underline{A}\underline{A}_1})\tilde{E}{}^{\underline{A}}\tilde{D}_{\underline{A}_2}\tilde{\Psi}\\
        &=(-1)^q\eta_0\tilde{\overline{\Psi}}(\tilde{\Gamma}{}^{\underline{A}_2}\tilde{\eta}_{\underline{BA}}-\delta^{\underline{A}_2}{}_{\underline{B}}\tilde{\Gamma}{}^{\underline{A}})\tilde{E}{}^{\underline{A}}\tilde{D}_{\underline{A}_2}\tilde{\Psi}.
    \end{split}
\end{equation}
Then, taking the interior product provides the components
\begin{equation}
    \begin{split}
        (\tilde{E}{}^{-1})_{\underline{C}}\rfloor\tilde{\overline{\Psi}}\tilde{\Gamma}{}^{\underline{A}_1}*(\hat{\tilde{\epsilon}}_{\underline{A}_1\underline{B}}\wedge\tilde{D}\tilde{\Psi})&=(-1)^q\eta_0\tilde{\overline{\Psi}}(\tilde{\Gamma}{}^{\underline{A}_2}\tilde{\eta}_{\underline{BA}}-\delta^{\underline{A}_2}{}_{\underline{B}}\tilde{\Gamma}{}^{\underline{A}})\delta^{\underline{A}}{}_{\underline{C}}\tilde{D}_{\underline{A}_2}\tilde{\Psi}\\
        &=(-1)^q\eta_0(\tilde{\overline{\Psi}}\tilde{\eta}_{\underline{BC}}\tilde{\Gamma}{}^{\underline{A}}\tilde{D}_{\underline{A}}\tilde{\Psi}-\tilde{\overline{\Psi}}\tilde{\Gamma}_{\underline{C}}\tilde{D}_{\underline{B}}\tilde{\Psi}).
    \end{split}
\end{equation}
For the other two terms of $*\tilde{\mathbb{U}}_{\underline{B}}$, they receive an additional factor of $(-1)^{m-1}$ from commuting the indices of the $\bar{\mathfrak{p}}^{-(m+1)}$-orientation symbol. Thus, in $m=2k$ dimensions, both the second and third term acquire a minus sign in the process. The result is
\begin{equation}
    \tilde{\mathbb{U}}_{\underline{B}}{}_{\underline{C}}=(-1)^{q}\frac{i\tilde{\beta}\eta_0}{2}\left(\tilde{\eta}_{\underline{BC}}L_{\tilde{\mathfrak{D}}}-\left(\tilde{\overline{\Psi}}\tilde{\Gamma}_{\underline{C}}\tilde{D}_{\underline{B}}\tilde{\Psi}-(\tilde{D}_{\underline{B}}\tilde{\overline{\Psi}})\tilde{\Gamma}_{\underline{C}}+\tilde{\overline{\Psi}}\tilde{m}_{\underline{B}}\tilde{\Gamma}_{\underline{C}}\tilde{\Psi}\right)\right),
\end{equation}
where $L_{\tilde{\mathfrak{D}}}$ is the $\bar{\mathfrak{p}}{}^{\mathfrak{w}}$-spinor Lagrangian
\begin{equation}
L_{\tilde{\mathfrak{D}}}:=\tilde{\overline{\Psi}}\tilde{\Gamma}{}^{\underline{A}}\tilde{D}_{\underline{A}}\tilde{\Psi}-(\tilde{D}_{\underline{A}}\tilde{\overline{\Psi}})\tilde{\Gamma}{}^{\underline{A}}\tilde{\Psi}+\tilde{\overline{\Psi}}\tilde{m}_{\underline{A}}\tilde{\Gamma}{}^{\underline{A}}\tilde{\Psi}.
\end{equation}
This is simply the Lagrangian associated with the Lagrangian density of Eq. \eqref{L-D-bar}, with the choice $\tilde{\alpha}=-\tilde{\alpha}{}^{\dagger}=i\tilde{\beta}$ substituted.

Therefore, the total contribution of the $\bar{\mathfrak{p}}{}^{\mathfrak{w}}$-spinor to the gravitational sector is
\begin{equation}
\begin{split}
    \tilde{\mathbb{M}}{}^{\underline{B}}{}_{\underline{A}}&=\frac{i\tilde{\beta}}{2}\tilde{\Upsilon}{}^{\underline{B}}\left(\tilde{\overline{\Psi}}\tilde{\Gamma}{}^{\underline{C}}\hat{\tilde{\epsilon}}_{\underline{CA}}\wedge\tilde{D}\tilde{\Psi}+\tilde{D}\tilde{\overline{\Psi}}\wedge\tilde{\Gamma}{}^{\underline{C}}\hat{\tilde{\epsilon}}_{\underline{CA}}\tilde{\Psi}-\tilde{\overline{\Psi}}\tilde{m}\wedge\tilde{\Gamma}{}^{\underline{C}}\hat{\tilde{\epsilon}}_{\underline{CA}}\tilde{\Psi}\right)\\
    &\quad-\frac{i\tilde{\beta}}{8}\tilde{\overline{\Psi}}\left\{*\tilde{\Gamma},\tilde{\Gamma}{}^{\underline{B}}\tilde{\Gamma}_{\underline{A}}\right\}\tilde{\Psi}.
    \end{split}
\end{equation}
In components,
\begin{equation}
\begin{split}
        \tilde{\mathbb{M}}{}^{\underline{B}}{}_{\underline{A}}{}_{\underline{C}}&=\tilde{\beta}{}'\tilde{\Upsilon}{}^{\underline{B}}\left(-\tilde{\eta}_{\underline{AC}}L_{\tilde{\mathfrak{D}}}+\tilde{\overline{\Psi}}\tilde{\Gamma}_{\underline{C}}\tilde{D}_{\underline{A}}\tilde{\Psi}-(\tilde{D}_{\underline{A}}\tilde{\overline{\Psi}})\tilde{\Gamma}_{\underline{C}}+\tilde{\overline{\Psi}}\tilde{m}_{\underline{A}}\tilde{\Gamma}_{\underline{C}}\tilde{\Psi}\right)\\
        &\quad+\frac{\tilde{\beta}{}'}{2}\left(\tilde{\overline{\Psi}}\delta^{\underline{B}}{}_{\underline{A}}\tilde{\Gamma}_{\underline{C}}\tilde{\Psi}-\frac{i}{2}\tilde{\overline{\Psi}}\left\{\tilde{\Sigma}{}^{\underline{B}}{}_{\underline{A}},\tilde{\Gamma}_{\underline{C}}\right\}\tilde{\Psi}\right),
        \end{split}
        \end{equation}
where we have called $\tilde{\beta}{}':=(-1)^{q+1}\frac{i\tilde{\beta}\eta_0}{2}$ for convenience. Additionally, we have the contribution of $\tilde{\mathbb{U}}_{\underline{A}}$ to the generalized projective Higgs vector field equations, $\tilde{\mathbb{F}}_{\underline{A}}=0$, since
\begin{equation}
\tilde{D}\tilde{\mathbb{X}}_{\underline{A}}=-\tilde{D}\tilde{\mathbb{U}}_{\underline{A}}.
\end{equation}
Since $\tilde{D}\tilde{\mathbb{X}}_{\underline{A}}=0$ was utilized in Sec. \ref{sec:dynamical-upsilon} to dynamically arrive at the APV-gauge choice, the non-vanishing expression for $\tilde{\mathbb{U}}_{\underline{A}}$ in Eq. \eqref{expanded-spinor-U} implies that the fully coupled theory of projective gravity and projective matter may not enjoy this possibly. Re-investigating the solutions of the general projective theory with both matter and gravity present is left for the future research program. For the remainder of the discussion of projective matter fields, we investigate the Thomas-Whitehead-Dirac model, as applied to the present $\bar{\mathfrak{p}}{}^{\mathfrak{w}}$-spinor construction.

\section{Thomas-Whitehead \texorpdfstring{$\bar{\mathfrak{\MakeLowercase{p}}}$}{\bar{\mathfrak{\MakeLowercase{p}}}}-Spinor Action}
\label{sec:tw-gen-proj}

For completeness, we review the Thomas-Whitehead-Dirac action of \cite{gen-struc}, as applied to the present, torsion-containing generalization. The choice of \cite{gen-struc} was, in exterior form, the \textit{non-Hermitian} Lagrangian $(m+1)$-form density
\begin{equation}
\tilde{\mathcal{L}}_{TWD}:=\frac{i}{2}\left(\tilde{\overline{\Psi}}*\tilde{\Gamma}\wedge \tilde{D}\tilde{\Psi}-(-1)^m\tilde{D}\tilde{\overline{\Psi}}\wedge*\tilde{\Gamma}\tilde{\Psi}-(-1)^m\tilde{\overline{\Psi}}\tilde{D}(*\tilde{\Gamma})\tilde{\Psi}\right).
\end{equation}
The additional factors of $(-1)^m$ are to account for the possible application to odd $m=2k+1$ dimensions. These factors appear only on the latter terms for the same reason that $(-1)^m$ appeared only in $\mathcal{L}_{\tilde{\Psi}}$ when expressed in abstract-index form. See Appendix \hyperref[app-B3:exterior-abs-ind-form]{B.3}. Recalling Eq. \eqref{lagrangian-gamma-adjoint}, the non-Hermiticity of $\tilde{\mathcal{L}}_{TWD}$ is easily identified by the last term, $i\mathcal{L}_{\tilde{\Gamma}}$, acquiring a minus sign under the action of Hermitian conjugation. As stated previously, $\tilde{\mathcal{L}}_{TWD}$ was chosen because it differs from the standard spinor kinetic term $\mathcal{L}_{\tilde{\Psi}}$ by a total gauge-covariant derivative,
\begin{equation}
\tilde{\mathcal{L}}_{TWD}:=i\tilde{\overline{\Psi}}*\tilde{\Gamma}\wedge\tilde{D}\tilde{\Psi}-\frac{i}{2}(-1)^m\tilde{D}(\tilde{\overline{\Psi}}*\tilde{\Gamma}\tilde{\Psi}),
\end{equation}
valid in any dimension.

Using the expressions outlined in Appendix \hyperref[app-B3:exterior-abs-ind-form]{B.3} and restricting to even $m=2k$ dimensions, $\tilde{\mathcal{L}}_{TWD}$ has the abstract-index form
\begin{equation}
    \tilde{\mathcal{L}}_{TWD}=(-1)^{q}\frac{i\eta_0}{2}d^{m+1}x|\tilde{e}|\left(\tilde{\overline{\Psi}}\tilde{\Gamma}{}^M(\tilde{D}_M\tilde{\Psi})-(\tilde{D}_M\tilde{\overline{\Psi}})\tilde{\Gamma}{}^M\tilde{\Psi}-\frac{1}{2}\tilde{\overline{\Psi}}\tilde{\Gamma}{}^M\tilde{I}_M\tilde{\Psi}\right).
\end{equation}
The (adjoint)-spinor kinetic terms conveniently combine to produce an anti-commutator,
\begin{equation}\label{anti-commutator-TWD}
    \begin{split}
        \tilde{\overline{\Psi}}\tilde{\Gamma}{}^{M}\tilde{D}_M\tilde{\Psi}-\tilde{D}_M\tilde{\overline{\Psi}}\tilde{\Gamma}{}^{M}\tilde{\Psi}&=\tilde{\overline{\Psi}}\tilde{\Gamma}{}^{M}(\partial_M\tilde{\Psi}+\tilde{\Omega}_M\tilde{\Psi})-(\partial_M\tilde{\overline{\Psi}}-\tilde{\overline{\Psi}}\tilde{\Omega}_M)\tilde{\Gamma}{}^{M}\tilde{\Psi}\\
        &=\tilde{\overline{\Psi}}\tilde{\Gamma}{}^{M}\partial_M\tilde{\Psi}-(\partial_M\tilde{\overline{\Psi}})\tilde{\Gamma}{}^{M}\tilde{\Psi}+\tilde{\overline{\Psi}}\{\tilde{\Gamma}{}^{M},\tilde{\Omega}_M\}\tilde{\Psi}.
    \end{split}
\end{equation}
The anti-commutator reduces to an interaction with the projectively invariant Lorentz connection, plus a part which vanishes in the APV-gauge, since it is proportional to the antisymmetric parts of $\overline{\mathcal{P}}_{\underline{ab}}$. To see this, we first decompose the connection into its (anti)-symmetric parts as
\begin{equation}
        \{\tilde{\Gamma}{}^M,\tilde{\Omega}_M\}=\{\tilde{\Gamma}{}^M,\tilde{\Omega}{}^{+}_M\}+\{\tilde{\Gamma}{}^M,\tilde{\Omega}{}^{-}_M\}.
\end{equation}
The first term vanishes due to the traceless property of the nonlinear connection $\tilde{\Omega}$,
\begin{equation}
        \tilde{\Omega}^+_M=\frac{1}{16}\tilde{\Omega}_{(\underline{AB})M}\{\tilde{\Gamma}{}^{\underline{A}},\tilde{\Gamma}{}^{\underline{B}}\}=\frac{1}{8}\tilde{\Omega}_{(\underline{AB})M}\tilde{\eta}{}^{\underline{AB}}=0.
\end{equation}
Recalling that $\tilde{\Omega}{}^{\underline{A}}{}_{\underline{B}*}=0$ for all values of $\underline{A}$ and $\underline{B}$, the anti-commutator with $\tilde{\Omega}{}^{-}$ sums to
\begin{equation}
    \begin{split}
        \{\tilde{\Gamma}{}^M,\tilde{\Omega}{}^{-}_M\}&=\frac{1}{16}\tilde{\Omega}_{[\underline{BC}]M}\{\tilde{\Gamma}{}^{M},[\tilde{\Gamma}{}^{\underline{B}},\tilde{\Gamma}{}^{\underline{C}}]\}\\
        &=\frac{1}{16}\tilde{\Omega}_{[\underline{bc}]m}\{\tilde{\Gamma}{}^{m},[\tilde{\Gamma}{}^{\underline{b}},\tilde{\Gamma}{}^{\underline{c}}]\}+\frac{1}{8}\tilde{\Omega}_{[\underline{b*}]m}\{\tilde{\Gamma}{}^{m},[\tilde{\Gamma}{}^{\underline{b}},\tilde{\Gamma}{}^{\underline{*}}]\}\\
        &=\frac{1}{16}\hat{\overline{\omega}}_{\underline{bc}m}\{\tilde{\Gamma}{}^{m},[\tilde{\gamma}{}^{\underline{b}},\tilde{\gamma}{}^{\underline{c}}]\}-\frac{x^*_0\eta_0}{8}\overline{\mathcal{P}}{}^{-}_{m\underline{b}}\{\tilde{\Gamma}{}^{m},[\tilde{\gamma}{}^{\underline{b}},\tilde{\gamma}{}^{\underline{*}}]\},
    \end{split}
\end{equation}
where the $\bar{\mathfrak{p}}$-factors canceled leaving the $\bar{\mathfrak{p}}{}^0$-matrix $\tilde{\gamma}{}^{\underline{A}}$. The coupling to the projective Schouten form has its symmetric part projected out,
\begin{equation}
       \overline{\mathcal{P}}{}^{-}_{m\underline{b}}\{\tilde{\Gamma}{}^{m},[\tilde{\gamma}{}^{\underline{b}},\tilde{\gamma}{}^{\underline{*}}]\}
       =-i\overline{\mathcal{P}}{}^{-}_{m[\underline{a}}(\tilde{e}{}^{-1})^m{}_{\underline{b}]}\tilde{\sigma}{}^{\underline{ab}}\tilde{\gamma}{}^{\underline{*}}.
\end{equation}
Recall that $\tilde{\mathcal{K}}{}^{\underline{A}}{}_{\underline{A}}=0$ implies $\overline{\mathcal{P}}_{[\underline{ab}]}\overline{\vartheta}{}^{\underline{a}}\wedge\overline{\vartheta}{}^{\underline{b}}=0$. However, 
\begin{equation}
    \overline{\mathcal{P}}{}^{-}_{m[\underline{b}}(\tilde{e}{}^{-1})^m{}_{\underline{a}]}= \overline{\mathcal{P}}_{m[\underline{b}}(\tilde{e}{}^{-1})^m{}_{\underline{a}]}-\frac{\eta_0}{(x^*_0)^2}\eta_{\underline{c}[\underline{b}}\overline{\vartheta}{}^{\underline{c}}{}_m(\tilde{e}{}^{-1})^m{}_{\underline{a}]}
\end{equation}
does not necessarily vanish, since, at the very least, $\overline{\vartheta}{}^{\underline{c}}{}_m(\tilde{e}{}^{-1})^m{}_{\underline{a}}\neq\delta^{\underline{a}}{}_{\underline{c}}$. Only in the APV-gauge does one have
\begin{equation}\label{anti-P}
\overline{\mathcal{P}}{}^{-}_{m[\underline{b}}(\tilde{e}{}^{-1})^m{}_{\underline{a}]}\overset{\circ}{=}0.
\end{equation}
The coupling to the antisymmetric connection $\hat{\overline{\omega}}$ also reduces conveniently in the APV-gauge,
\begin{equation}
       \hat{\overline{\omega}}_{\underline{bc}m}\{\tilde{\Gamma}{}^{m},[\tilde{\gamma}{}^{\underline{b}},\tilde{\gamma}{}^{\underline{c}}]\}\overset{\circ}{=}-2i\hat{\overline{\omega}}_{\underline{bc}m}\{\gamma^m,\sigma^{\underline{bc}}\}.
\end{equation}
Putting the pieces back together, we find
\begin{equation}
        \{\tilde{\Gamma}{}^M,\tilde{\Omega}{}^{-}_M\}=\{\tilde{\Gamma}{}^{m},\hat{\overline{\omega}}_{m}\}\mp\frac{ix^*_0}{\sqrt{\eta_0}}\overline{\mathcal{P}}{}^{-}_{m[\underline{a}}(\tilde{e}{}^{-1})^m{}_{\underline{b}]}\sigma^{\underline{ab}}\gamma^{\underline{5}}\overset{\circ}{=}\{\gamma^{m},\hat{\overline{\omega}}_{m}\},
\end{equation}
where
\begin{equation}
    \hat{\overline{\omega}}_m:=\frac{-i}{2}\hat{\overline{\omega}}_{\underline{bc}m}\sigma^{\underline{bc}}.
\end{equation}

The partial derivative term in Eq. \eqref{anti-commutator-TWD} reduces to
\begin{equation}
    \begin{split}
        \tilde{\overline{\Psi}}\tilde{\Gamma}{}^{M}(\partial_M\tilde{\Psi})
        &=\tilde{\overline{\psi}}\bar{\mathfrak{p}}{}^{-\mathfrak{w}}\tilde{\Gamma}{}^{M}\partial_M\left(\bar{\mathfrak{p}}{}^{\mathfrak{w}}\tilde{\psi}\right)\\
        &=\tilde{\overline{\psi}}\tilde{\Gamma}{}^{M}\partial_M\tilde{\psi}+\mathfrak{w}\tilde{\overline{\psi}}\tilde{\Gamma}{}^{M}(\partial_M\log\bar{\mathfrak{p}})\tilde{\psi}\\
        &=\tilde{\overline{\psi}}\tilde{\Gamma}{}^{m}\partial_m\tilde{\psi}+\mathfrak{w}\tilde{\overline{\psi}}\tilde{\Gamma}{}^{M}\tilde{g}_M\tilde{\psi},
    \end{split}
\end{equation}
since $\tilde{\psi}=\tilde{\psi}(x)$ is independent of $x^*$. The adjoint partial derivative term has a similar expansion,
\begin{equation}
    \begin{split}
        -(\partial_M\tilde{\overline{\Psi}})\tilde{\Gamma}{}^{M}\tilde{\Psi}&=-\partial_M\left(\tilde{\overline{\psi}}\bar{\mathfrak{p}}{}^{-\mathfrak{w}}\right)\tilde{\Gamma}{}^{M}\bar{\mathfrak{p}}{}^{\mathfrak{w}}\tilde{\psi}\\
        &=-(\partial_M\tilde{\overline{\psi}})\tilde{\Gamma}{}^{M}\tilde{\psi}+\mathfrak{w}\tilde{\overline{\psi}}(\partial_M\log\bar{\mathfrak{p}})\tilde{\Gamma}{}^{M}\tilde{\psi}\\
        &=-(\partial_m\tilde{\overline{\psi}})\tilde{\Gamma}{}^{m}\tilde{\psi}+\mathfrak{w}\tilde{\overline{\psi}}\tilde{g}_M\tilde{\Gamma}{}^{M}\tilde{\psi}.
    \end{split}
\end{equation}
We combine the above results and substitute them into an action formed from $\tilde{\mathcal{L}}_{TWD}$. Dividing the resulting expression by $x^*_0$ to acquire proper physical dimensions, we find
\begin{equation}
\begin{split}
    \tilde{S}_{TWD}&=(-1)^q\frac{i\eta_0}{2x^*_0}\int_{V\mathcal{M}}d^{m+1}x|\tilde{e}|\left(\tilde{\overline{\psi}}\tilde{\Gamma}{}^{m}\tilde{D}^-_m\tilde{\psi}-(\tilde{D}^-_m\tilde{\overline{\psi}})\tilde{\Gamma}{}^{m}\tilde{\psi}\right)\\
    &\quad\quad+(-1)^q\frac{i\eta_0}{2x^*_0}\int_{V\mathcal{M}}d^{m+1}x|\tilde{e}|\left(-\frac{1}{2}\tilde{\overline{\psi}}\tilde{I}_{M}\tilde{\Gamma}{}^{M}\tilde{\psi}+2\mathfrak{w}\tilde{\overline{\psi}}\tilde{g}_{M}\tilde{\Gamma}{}^{M}\tilde{\psi}\right),
    \end{split}
\end{equation}
where
\begin{equation}
    \tilde{D}^{-}_m:=\partial_m\pm\tilde{\Omega}^{-}_m\overset{\circ}{=}\partial_m\pm\hat{\overline{\omega}}_m
\end{equation}
is the gauge-covariant derivative with respect to the anti-symmetric connection, and
\begin{equation}
    \tilde{I}_{M}:=\tilde{Q}_{\underline{B}}{}^{\underline{B}}{}_{M}-2(m+1)\tilde{g}_{M}+2\tilde{\mathcal{T}}{}^{\underline{A}}\equiv \tilde{N}_M
\end{equation}
is the same interaction with the anti-symmetrized trace of the projective distortion, Eq. \eqref{distortion-interaction}. However, the interaction coefficient necessarily differs from the Hermitian scenario in Eq. \eqref{nlp-spinor-action}, and therefore provides an interaction with the trace of the projective Schouten form $\overline{\mathcal{P}}$. The choice of \cite{gen-struc} for the projective weight $\mathfrak{w}$ was the dimension-dependent, $\mathbb{R}$-valued $\mathfrak{w}=\frac{-(m+1)}{4}$. This choice of $\mathfrak{w}$ was imposed to eliminate the induced chiral mass, and moreover, appears to be equivalent to absorbing the $\tilde{g}$ factor from each kinetic term into the trace of the connection $\tilde{\Omega}$, Eq. \eqref{tw-omega}, and removing all $\bar{\mathfrak{p}}$-factors from all fields. This was the choice in \cite{gen-struc,proj-diff-spinors} for projective spinors, and coincides with the choice of weight for Lorentz \textit{vectors} in \textit{conformal} theories \cite{isham-salam-strathdee}. However, one must keep in mind both the non-Hermiticity of this model and the inherently \textit{non-projective} nature of inexistent $\bar{\mathfrak{p}}$-factors. This latter point follows from the fact that the projective factor accounts for the equivalence class inherent to fields residing in a projective space.

Due to the non-Hermiticity of the TWD model, the field equations are \textit{not} adjoints of one another. This property was likely missed in \cite{gen-struc} due to only providing the variation of the adjoint projective spinor field. Explicitly, in exterior form, the field equation resulting from variation of $\tilde{\overline{\Psi}}$ is 
\begin{equation}
i*\tilde{\Gamma}\wedge\tilde{D}\tilde{\Psi}=0,
\end{equation}
as was provided in \cite{gen-struc}. The field equation resulting from variation of $\tilde{\Psi}$, which was not provided, is found to be
\begin{equation}
i(\tilde{D}\tilde{\overline{\Psi}})\wedge*\tilde{\Gamma}+i\tilde{\overline{\Psi}}(\tilde{D}*\tilde{\Gamma})=0.
\end{equation}
These equations seem to imply the spinor and its adjoint are unrelated fields \cite{matter-torsion2}, or that the usual spinor adjoint definition fails to hold. This asymmetry leads to the adjoint field interacting with the projective Schouten trace and the projectively invariant distortion in a manner different from the spinor field. 

The interaction and induced mass are the non-Hermitian terms of this model. Such models, albeit inherently non-projective, have been studied, for example, in \cite{non-hermitian1,anti-herm-mass-4,anti-herm-mass,anti-herm-mass-3,non-hermitian2,nonhermitian-seesaw,nonhermitian3,non-herm-interaction}. Gravitationally induced (non-chiral) spinor masses, such as those observed in Poincar\'{e} gauge gravity, have also been studied \cite{induced-fermion-mass-poincare}. Due to the explicit coupling in the TWD theory between the projective spinors and $\overline{\mathcal{P}}$, a non-zero relaxed state for $\overline{\mathcal{P}}$ will contribute further to the induced chiral mass. Such processes resemble the dynamical axion-generated neutrino mass, as in \cite{dynamical-neutrino-mass}. According to \cite{chiral-neutrino-curvature-mass}, the chiral coupling of spinors to the curvature scalar is used to discuss lepton asymmetries with heavy neutrinos. This is interesting in the present context, since the projective Schouten tensor of a Levi-Civita connection is nothing but a linear combination of the Ricci curvature and its trace, and thus, in some sense, resembles a Pauli-type coupling \cite{pauli-coupling}.

\pagebreak
\thispagestyle{empty}
\phantomsection
\addcontentsline{toc}{section}{\hspace{-1.5em}\textbf{Discussion}}
\begingroup
\renewcommand{\addcontentsline}[3]{}
\vspace*{5cm}
\noindent
\vspace{+3em}
\makebox[\textwidth]{\Huge \textbf{Discussion}}
\vfill
\label{closing remarks}
\endgroup

\pagebreak

\section{Summary}

Part I of the thesis establishes the mathematical and conceptual groundwork for developing a projective theory of gravity and matter, with a strong emphasis on metric-affine gauge theory as a foundational framework. It begins with an in-depth review of affine geometry, detailing how general affine connections decompose into their Levi-Civita, disformation, and contorsion components. The role of curvature tensors and their contractions is examined, particularly in the context of projective and the more fundamentally motivated symmetric projective transformations. The Einstein-Hilbert action and its field equations are introduced and extended to include matter coupling and a cosmological constant. These ideas were then reformulated within the exterior Palatini formalism through the use of differential forms. A central focus is placed on the Metric-Affine Gauge theory, which seeks to generalize General Relativity by gauging the affine group. This setting treats the metric and connection as independent variables, allowing for a more flexible geometric description that naturally accommodates torsion and non-metricity---in an unpleasantly asymmetric fashion. This affine gauge-theoretic approach sets the stage for the projective gauge formalism, which emerges as a refinement of the affine theory, unifying projective transformations with spacetime reparameterizations. Motivated by insights from string theory, particularly the interpretation of the Diffeomorphism gauge potential as analogous to the electromagnetic gauge potential, the study extends to a higher-dimensional volume bundle that encapsulates projective symmetry. The Thomas-Whitehead (TW) model is examined in this broader setting, revealing how its projective Schouten tensor can induce a Higgs-like potential, generating both a cosmological constant and its mass through a dimensional reduction scheme.

Part II builds upon this foundation by systematically developing a General Projective Gauge Gravitational Theory. The construction begins with the formulation of projective space from the affine (co)-tangent space, with an emphasis on identifying its group of transformations, which is shown to be the Projective General Linear Group. The algebraic structure of this group is examined under Lorentz decomposition, and it is demonstrated that a group contraction leads to a pseudo-affine group. The projective volume bundle is extended to incorporate translational degrees of freedom, describing shifts in the contact point between spacetime and its associated projective tangent space, i.e., a point serving as an origin. A crucial mathematical tool, the factoring map, is introduced between the volume bundle and the projective spaces, and its minimal condition for generating a projective structure results in the emergence of a novel scalar field. It is then shown that this scalar field manifests as the projective Schouten tensor.

A key contribution of this part is the definition of novel projective symmetric teleparallel connections, which are characterized by being torsion-free, flat, and invariant under projective transformations. These connections provide a generalization of teleparallel spacetime geometries, providing new areas of investigation. By removing the inertial condition, a gauge connection is introduced and the nonlinear realization of gauge symmetries technique applied, providing local Lorentz covariance to the theory. The study further extends to the role of general projective Higgs fields, which serve to construct fundamental geometric objects, including \textit{projective $2$-frames} and the $V\mathcal{M}$ connection. These fields allow for a dynamical mechanism in which projective symmetry breaking leads to effective gravitational interactions, modifying the cosmological constant in a controlled way. 

The general projective field variations are developed systematically, providing a means for investigating canonical structures. Relationships were made with the Metric-Affine variational structures and shown to provide a means for dynamically incorporating a metric-like field in a symmetrical manner. A particularly novel result is the derivation of a projective Pontrjagin density, which contains a new, metric-independent and projectively invariant topological term not previously documented in the literature. The action principle governing the General Projective Gauge Gravitational Theory is analyzed in detail, with a Lovelock-inspired model proposed as the natural extension of standard curvature-based theories. Notably, this action is found to support only curvature (metric) dynamics once a particular (TW) gauge is chosen. The study of the resulting solution space reveals the presence of nontrivial projective torsion modes and degenerate co-frames, as well as a modified (A)-dS description of spacetime, where the bare cosmological constant is dynamically provided by the generalized projective Higgs vector, and modified by the rigidity of the projective Schouten field. The field equations are also shown to permit flat space solutions that reduce the number of free parameters in the theory by $1$.

Part III extends the general projective framework to include spinorial matter fields, constructing a formalism for projective spinors within the developed geometric setting. The study begins by defining gamma matrices using the Goldstone metric associated with the cosets of the projective linear group, ensuring consistency with the nonlinear setting where the local Lorentz symmetry provides access to the ordinary flat gamma matrices. In constructing the nonlinear projective spinor fields, some assumptions were made with respect to the spinor representation of the group, permitting connections to familiar constructs. A key mathematical feature of this approach is the introduction of a general spinor metric, which initially allows for a redundant complex projective phase. In developing the dynamics, it was found that an extended gauge-covariant derivative operator was required for Hermiticity, as well as gauge and coordinate invariance. However, by imposing the Hermiticity condition on the spinor action, the theory naturally eliminates any coupling between spinors and non-metricity. As a result, torsion emerges as the sole mediator of gravitational interactions with spinorial matter, leading to an effective reduction to a projectively invariant Einstein-Cartan-type theory.

One of the most significant physical implications of this construction is the emergence of an induced chiral mass term. Unlike previous attempts in TW theory, where CP violation posed a potentially major issue, the present projective formulation provides a chiral mass mechanism that avoids such inconsistencies. This may have profound implications for neutrino physics, suggesting a natural pathway for chiral mass generation within the projective framework. Moreover, the induction of a chiral mass and a cosmological constant, provided by a reduction of the projective geometry, points towards a suitable formalism for discussing black hole mass superpositions \cite{black-hole-mass-superposition}, and further, an origin story for the dynamical shearing effects encountered in Metric-Affine Gauge theories \cite{dynamical-shear-MAG}, which result in black hole shear charges. Part II concludes with a brief examination of projective spinor currents, laying the groundwork for future investigations into a projective description of the chiral anomaly.

Together, Parts II and III extend the foundational ideas laid out in Part I, demonstrating that projective geometry offers a rich and unifying structure for gravitational and matter interactions. By embedding traditional theories of gravity within a broader projective framework, the work provides access to new insights into fundamental physics, particularly in the context of mass generation, spacetime symmetries, and the role of topological invariants in material processes. These findings open the door to further research on the implications of projective gauge theories in cosmology, high-energy physics, and the structure of fundamental interactions.

\section{Prospects}

Throughout this document, numerous potential research directions have been proposed, each building upon the findings and concepts explored in this work. In this section, we provide a more detailed discussion of some of these prospective projects, outlining the key research objectives, anticipated challenges, and their broader significance. By expanding on these proposals, we aim to highlight their relevance to the current study and their potential impact on future advancements in the field.

\begin{enumerate}
    \item \textbf{Projective Symmetric Teleparallel Connections (PST)}: The geodesic trajectories with respect to the projective symmetric teleparallel connection were investigated in the $y=0$ sector, which was shown to recover the teleparallel equivalent of Thomas-Whitehead geodesics. In $1$ dimension, a relationship was found between the affine parameter and the redundancy scalar field, where the latter manifests as the projective Schouten tensor. The $y\neq0$ sector will likely generalize this relationship and furthermore, modify geodesic trajectories. Since $y=y_0$ provides a fixed point which may be identified with a reference point (origin), the $y\neq0$ sector permits an investigation into the space of geodesic paths through an arbitrary point. Such an investigation may prove fruitful in understanding trajectories that General Relativity (GR) cannot explain. Moreover, the teleparallel equivalents to General Relativity (TEGR), wherein the curvature is expressed in terms of the geometric deformations, will have its symmetric projective generalization in terms of $y$ and $\mathcal{P}$. Therefore, investigating the PST equivalent to GR is an investigation into a $\mathcal{P}$-modulated curved geometry without a fixed point of reference. This would be a very interesting area of research since it permits one to study the ``flat" geometries, in the context of a dynamical field known to provide $\Lambda\neq0$. These projects are simple to accomplish---they only require one to recalculate the geodesic relations with $y\neq0$, and construct the Einstein-Hilbert action and subsequent variation with the curvature solved for in terms of $y$ and $\mathcal{P}$. Lastly, it is of deep interest to find a relationship between PST and Berwald-Thomas-Whitehead (BTW) connections, since the vanishing of the Berwald curvature is necessary for the existence of an affine connection. Given the form of the BTW connection, it appears that $y$ is intimately related to the trace of the Berwald curvature. This investigation will likely provide further information regarding the relationship between the redundancy scalar and Finsler functions, which give rise to general path geometries. 

    \item  \textbf{Independent $\mathfrak{p}$ and $\vartheta$}: In order to simplify calculations and interpretations, many of the geometric objects constructed were assumed to depend on one another. For example, the gauge-connection $\Omega$, the generalized projective Higgs vector $Y$ and co-vector $A$, as well as the projective factor $\mathfrak{p}$, were all defined with respect to the translational connection-form $\vartheta$ and its logarithmic derivative $\alpha$. Additionally, homogeneity assumptions were made resulting in the orthogonality of $\alpha$ and the vector parameters, such as $\xi$. Relieving the theory of these assumptions may provide further information on the nature of the projective Schouten tensor and its role in modulating the shear defects in the geometry. Additionally, the generalized projective Higgs fields will be modified, and essentially taken independent of one another, possibly leading to a more direct relation between the co-vector, the projective Schouten, and non-metricity. This project simply requires all the work contained herein to be rewritten with these assumptions removed. It is quite likely there will be obstacles encountered, since the exact reason for introducing these assumptions was to overcome the obstacles encountered. In particular, writing the nonlinear projective Schouten form in a gauge-covariant manner appeared to require the orthogonality assumption. 

    \item \textbf{Inverse Higgs Theorem}: The underlying idea motivating this document is the abandonment of a metric in favor of the projective Schouten tensor (Diffeomorphism field). There are multiple pieces of information which seem to suggest such a replacement. In particular, the Inverse Higgs Theorem informs us that there must be a relationship between the nonlinear projective Schouten tensor and the gauge-covariant derivative of the symmetric coset parameters. The latter are commonly identified with the non-metricity, which results from their equivalence with the symmetric part of the connection. This process reduces the number of gauge degrees of freedom by reorganizing the relationships between the coset parameters and the connection-forms. Therefore, it is quite likely that in this process, the nonlinear projective Schouten form absorbs the $10$ degrees of freedom commonly ascribed to the metric. However, this absorption is not as straightforward as one would like, since no combination of (pseudo)-translational connection-forms and coset parameters form $10$ degrees of freedom. The investigation of this reorganization of degrees of freedom is essential to identifying the projective Schouten field's relationship to the metric. This may be accomplished by first removing all assumptions, following $(2)$ above, since each assumption will alter the available degrees of freedom. Then, carefully investigating the relationships between the connection-forms, coset parameters, and their gauge-covariant derivatives will likely provide the sought after distribution of degrees of freedom. 

    \item \textbf{Boundary Variations}: The total functional variation of an action provides, in addition to the field equations, a total derivative. This total derivative is typically omitted, since it may be integrated to the boundary where all field variations are considered to vanish. Thus, it does not modify the field equations. However, in the general projective setting, this boundary data provides access to the symplectic structure associated with the spacetime, and since it has been shown the antisymmetric parts of the projective Schouten tensor provide a symplectic $2$-form, an investigation into the boundary data is in some sense equivalent to an investigation into the antisymmetric parts of $\mathcal{P}$, and further, the projective homothetic curvature. Since the particular nonlinear realization procedure employed in this document forced this $2$-form to vanish, such an investigation would require one to either abandon the procedure all together, or choose a larger stability subgroup, such as $GL(m,\mathbb{R})$. In the metric-independent sector of the actions discussed, the projective Pontrjagin density, the boundary form is related to the Chern-Simons (CS) form and the closely related projective Nieh-Yan form, which both appear to contain information that is highly sensitive to the antisymmetric parts of $\mathcal{P}$. In the metric-dependent sector, the projective Lovelock theory, such identifications are less obvious. Understanding this symplectic structure is essential for understanding the phase space of spacetime geometries, and therefore the Hamiltonian structure. This project, as mentioned previously, must be accomplished either without the use of the nonlinear realization procedure or with a larger stability subgroup. Essentially, once the action functional is chosen, one simply computes the boundary forms. With the boundary forms available, investigating the symplectic structure can take many directions. For example, one may compute the corner charges and the associated corner algebras, which may be valuable in understanding the effects of a vanishing antisymmetric part of $\mathcal{P}$. In the case of the projective Lovelock theory, one obstacle is most obvious---the absence of a well-defined metric. One may certainly introduce a metric via various means, but it may prove more direct to rather consider a constitutive tensor \cite{constitutive-1,constitutive-2,constitutive-3}. 

    \item \textbf{Canonical Quantization}: In addition to studying the symplectic structure, it may be of significant interest to develop the canonical Hamiltonian formalism associated with the General Projective Gauge Gravitational Theory. This will provide a means of quantizing the theory, and thus the projective Schouten tensor, as well as determining if there are novel charges associated with the geometric defects contained in the projective theory. This project may allow for an avoidance of the many pitfalls associated with quantizing gravity, simply by viewing through the lens of projective geometry. This hypothesis is guided by the known resolutions which follow from using the Ashtekar variables, in tandem with their strikingly projective-esque form. In this project, once the field variations are inverted, and the Hamiltonian constructed, the quantization procedure may be followed as a recipe. Applying the ADM formalism to the spacetime aspects, within the projective aspects, would lead to a type of double foliation. It would certainly be interesting, and potentially quite fruitful, to also consider the space-time splitting in a projective manner. This may lead to a formal derivation of the Ashtekar variables from within a larger theory. 

    \item \textbf{Projective Spacetime Geometries}: As outlined in \ref{proj-spacetime-table}, we propose the possible existence of an extended class of standard spacetime geometries which are projective. These projective spacetime geometries may not all be permissible, and their existence likely to be model-dependent. Investigating their existence in the context of the projective Lovelock model may be accomplished simply by imposing the vanishing conditions for the projective spacetime of interest via Lagrange multipliers and reapplying the variational techniques. In this project, one may easily investigate each projective spacetime in this manner and tabulate the results. Since the various spacetime geometries provide the basis on which every gravitational theory is constructed, having access to a new class of spacetime geometries paves the way for decades of opportunity. An obvious obstacle one may encounter is in the invertibility of the constraint equations, since the projective Lovelock model is quadratic in the dimension of interest. The quadratic nature may also lead to a splitting into two distinct sectors, resulting from the two distinct roots in the inversion process. 

    \item \textbf{Projective Nieh-Yan and the Chiral Anomaly}: It is well known, and quite debated, that the Nieh-Yan form contributes to the chiral anomaly in addition to the Pontrjagin density. Since the projective spinor theory proposed in this document is initially massless, it could be viewed as having a chiral symmetry. However, the extended gamma matrices do not provide any simple means for constructing an associated chiral current. If one permits a complex projective weight, a chiral mass is induced via dimensional reduction. Since this mass is itself chiral, a standard chiral symmetry may still be seen to exist. Understanding this process deeper, and developing a formal program for describing chirality projectively is tantamount to understanding the potential chiral anomaly. In computing the projective chiral anomaly, one may utilize the square of the extended gauge-covariant derivative operator for a general projective spinor metric in a Fujikawa-type method. To do so, one would have to first ensure the validity of the method in this curved projective context. Once confirmed, the process is highly regimented, and may easily be followed step-by-step.

    \item \textbf{Coupled Projective Matter/Gravitation Theory}: This document provided a brief analysis of the projective gravitational and material sectors independently. A complete projective description of the universe would likely require one to consider the fully coupled projective model. The field variations for this were provided, but the subsequent analysis left for future investigation. Therefore, insofar as the projective Lovelock theory is concerned, this project consists only in analyzing the fully coupled theory. Such an analysis will first consider whether or not the same gravitational solutions exist, now in the presence of matter. In particular, determining what effect, if any, the inclusion of matter has on one's ability to dynamically arrive at the APV-gauge choice. Furthermore, since it known that matter may be viewed as sourcing torsion, the field strength of the co-frame, the non-vanishing torsion solutions found in this document would be an interesting area to investigate. This is because if one then imposes a vanishing torsion, one finds a degenerate co-frame. This statement is also likely to be modified in the presence of matter, especially in light of \cite{torsion-cancel-divergences} where, in the context of Einstein-Cartan gravity, torsion can be shown to contribute to the spinor's mass. Lastly, the fully coupled projective theory may provide insight into the ``coupling" of matter to the projective Schouten field, since there may be an indirect balancing that occurs between the two at the level of the equations of motion.
\end{enumerate}

\phantomsection
\addcontentsline{toc}{section}{\hspace{-1.5em}Appendix A: Useful Formulae}
\begingroup
\renewcommand{\addcontentsline}[3]{}
\section*{Appendix A: Useful Formulae}
\label{app-A}
\endgroup

\phantomsection
\addcontentsline{toc}{subsection}{\hspace{-1.5em}A.1: Matrix Identities}
\begingroup
\renewcommand{\addcontentsline}[3]{}
\subsection*{A.1: Matrix Identities}
\label{app-A:matrix-id}
\endgroup

For an $(m+1)\times (m+1)$ block matrix $M$, consider the $m+1$ block matrix form
\begin{equation}
\tag{A1.1}
    M=\begin{pmatrix}
        A&B\\C&D
    \end{pmatrix},
\end{equation}
for an $m\times m$ square matrix $A$, an $m\times 1$ matrix (vector) $B$, a $1\times m$ matrix (co-vector) $C$, and a $1\times 1$ scalar (density) $D$. The determinant of $M$, denoted $|M|$, has two equivalent expressions. Each equivalent expression depends on the existence of particular component inverses,
\begin{equation}\label{det-formula}
\tag{A1.2}
    A^{-1}\Rightarrow |M|=|A|\;|D-CA^{-1}B|,\quad\quad D^{-1}\Rightarrow |M|=|A-BD^{-1}C|\;|D|.
\end{equation}
Assuming the existence of $A^{-1}$, we form the \textit{Schur Complement} of $A$ \cite{schur},
\begin{equation}
\tag{A1.3}
    E=D-CA^{-1}B.
\end{equation}
This permits a decomposition of $M$ into
\begin{equation}
\tag{A1.4}
    M=\begin{pmatrix}
        \bm{1}_m&0\\CA^{-1}&1
    \end{pmatrix}\begin{pmatrix}
        A&0\\0&E
    \end{pmatrix}\begin{pmatrix}
        \bm{1}_m&A^{-1}B\\0&1
    \end{pmatrix}.
\end{equation}
The inverse of $M$ may easily be found from the above expression, resulting in
\begin{equation}
\tag{A1.5}
    M^{-1}=\begin{pmatrix}
        \bm{1}_m&-A^{-1}B\\0&1
    \end{pmatrix}\begin{pmatrix}
        A&0\\0&\frac{1}{E}
    \end{pmatrix}\begin{pmatrix}
        \bm{1}_m&0\\-CA^{-1}&1
    \end{pmatrix}.
\end{equation}
This is the standard \textit{projective decomposition} of a matrix. Executing the matrix multiplication produces
\begin{equation}\label{inverse-proj-matrix-gen}
\tag{A1.6}
    M^{-1}=\begin{pmatrix}
        A^{-1}+\frac{1}{E}(A^{-1}B)(CA^{-1})&\frac{-1}{E}A^{-1}B\\\frac{-1}{E}CA^{-1}&\frac{1}{E}
    \end{pmatrix}.
\end{equation}

\phantomsection
\addcontentsline{toc}{subsection}{\hspace{-1.5em}A.2: Orientation Symbol}
\begingroup
\renewcommand{\addcontentsline}[3]{}
\label{app-A:orientation}
\subsection*{A.2: Orientation Symbol}
\endgroup
\phantomsection

This section serves as a review of the basic identities associated with the orientation symbol and the closely related orientation tensor density. Much of this information can be found in any standard text on General Relativity \cite{carroll} or mathematical physics \cite{nakahara}. 

The convention utilized in this document for the completely anti-symmetric \textit{orientation symbol} of an $m$-dimensional space is
\begin{equation}\label{epsilon-definition}
\tag{A2.1}
    \hat{\epsilon}_{a_1a_2\dots a_m}=\begin{cases}
        +1\quad \text{for} \;\{a_1a_2\dots a_m\}\;\text{an even permutation of}\;\{1,2,\dots,m\},\\-1\quad \text{for} \;\{a_1a_2\dots a_m\}\;\text{an odd permutation of}\;\{1,2,\dots,m\},\\\;\;\;0\quad \text{otherwise}.
    \end{cases}
\end{equation}
Any even permutation of $12\dots m$ returns $+1$, while any odd permutation returns $-1$, and any repeated indices returns $0$. For example,
\begin{equation}
\tag{A2.2}
\begin{split}
\hat{\epsilon}_{1234\dots m}&=+1,\\
\hat{\epsilon}_{2134\dots m}&=-\hat{\epsilon}_{1234\dots m}=-1,\\
\hat{\epsilon}_{1\bm{1}34\dots m}&=-\hat{\epsilon}_{\bm{1}134\dots m}=0,
\end{split}
\end{equation}
where a bold-face $\bm{1}$ is used in the last expression to make explicit the action of permutation. 

For an arbitrary $m\times m$ square matrix $M^a{}_b$, we may use $\hat{\epsilon}_{a_1a_2\dots a_m}$ to access the determinant of $M^a{}_{b}$, since
\begin{equation}\label{epsilon-M}
\tag{A2.3}
\hat{\epsilon}_{a_1a_2\dots a_m}M^{a_1}{}_{b_1}M^{a_2}{}_{b_2}\dots M^{a_m}{}_{b_m}=|M|\hat{\epsilon}_{b_1b_2\dots b_m},\quad\quad\quad |M|:=\det(M^a{}_b).
\end{equation}
When $M^a{}_b$ is taken to be the inverse Jacobian of a coordinate transformation $\{x^a\}\rightarrow\{x^{a'}\}$, denoted $J^{a}{}_{b'}=\partial x^{a}/\partial x^{b'}$, we find the transformation behavior of $\hat{\epsilon}_{a_1a_2\dots a_m}$,
\begin{equation}\label{epsilon-transformation}
\tag{A2.4}
    \hat{\epsilon}_{b'_1b'_2\dots b'_m}=|J|\hat{\epsilon}_{a_1a_2\dots a_m}J^{a_1}{}_{b'_1}J^{a_2}{}_{b'_2}\dots J^{a_m}{}_{b'_m},\quad\quad\quad |J|:=\det(J^{a'}{}_b).
\end{equation}
Thus, $\hat{\epsilon}_{a_1a_2\dots a_m}$ is a rank-$(0,m)^1$ tensor, i.e., a tensor density of weight $+1$. If $M^a{}_b$ is instead taken as the frame field $e^a{}_{\underline{b}}$, then
\begin{equation}
\tag{A2.5}
\hat{\epsilon}_{\underline{b}_1\underline{b}_1\dots \underline{b}_m}=|e|\hat{\epsilon}_{a_1a_2\dots a_m}e^{a_1}{}_{\underline{b}_1}e^{a_2}{}_{\underline{b}_2}\dots e^{a_m}{}_{\underline{b}_m},\quad\quad\quad |e|:=\det(e^{\underline{b}}{}_a).
\end{equation}
In the above, we stick to the convention of using $|e|=\sqrt{|g_{mn}|}$ to denote the determinant of the \textit{co-frame} in order to avoid confusion. This follows from the identity
\begin{equation}
\tag{A2.6}
    |e|^{-1}=[\det(e^{\underline{b}}{}_a)]^{-1}=\det[(e^{\underline{b}}{}_a)^{-1}]=\det(e^a{}_{\underline{b}}).
\end{equation}

The relationship to the orientation symbol with all indices raised is found by simply raising all indices and using the identity in Eq. \eqref{epsilon-M},
\begin{equation}\label{epsilon-det-metric}
\tag{A2.7}
\begin{split}
    \hat{\epsilon}_{a_1a_2\dots a_m}g^{a_1b_1}g^{a_2b_2}\dots g^{a_mb_m}=|g|^{-1}\hat{\epsilon}^{b_1b_2\dots b_m}.
    \end{split}
\end{equation}
For an $m$-dimensional manifold with split-signature $(p,q)$, where $p$ denotes the number of positive entries and $q$ the number of negative entries, one has $|g|=|\eta|=(-1)^q$ in flat Cartesian coordinates. Therefore,
\begin{equation}
\tag{A2.8}
    \hat{\epsilon}_{a_1a_2\dots a_m}\eta^{a_1b_1}\eta^{a_2b_2}\dots \eta^{a_mb_m}=(-1)^q\hat{\epsilon}{}^{b_1b_2\dots b_m}.
\end{equation}
When the coordinate transformation defined in Eq. \eqref{epsilon-transformation} is applied to $\hat{\epsilon}^{b_1b_2\dots b_m}$, one finds a behavior inverse to that of $\hat{\epsilon}_{b_1b_2\dots b_m}$. Therefore, $\hat{\epsilon}^{b_1b_2\dots b_m}$ is a rank-$(m,0)^{-1}$ tensor, or a tensor density of weight $-1$.

Standard rank-$(m,0)^0$ and rank-$(0,m)^0$ tensors may be formed from each $\hat{\epsilon}$ by scaling with some factor of the opposite weight. The standard choices are
\begin{equation}\label{density-orientation}
\tag{A2.9}
    \epsilon_{b_1b_2\dots b_m}:=\sqrt{(-1)^q|g|}\hat{\epsilon}_{b_1b_2\dots b_m},\quad\quad\quad\epsilon^{b_1b_2\dots b_m}:=\frac{1}{\sqrt{(-1)^q|g|}}\hat{\epsilon}^{b_1b_2\dots b_m}.
\end{equation}
To check the validity of tensorality, we note that for a coordinate transformation $\{x^a\}\rightarrow \{x^{a'}\}$, the metric tensor transforms as
\begin{equation}
\tag{A2.10}
    g_{a'b'}=g_{ab}J^a{}_{a'}J^b{}_{b'}.
\end{equation}
Therefore, its determinant transforms as
\begin{equation}
\tag{A2.11}
    |g'|=|g||J|^{-2},
\end{equation}
and may be recognized as a rank-$(0,0)^{-2}$ tensor, or scalar density of weight $-2$. Thus, when working with the tensor densities of Eq. \eqref{density-orientation},
\begin{equation}
\tag{A2.12}
    \epsilon_{b'_1b'_2\dots b'_m}=\epsilon_{a_1a_2\dots a_m}J^{a_1}{}_{b'_1}J^{a_2}{}_{b'_2}\dots J^{a_m}{}_{b'_m},
\end{equation}
and 
\begin{equation}
\tag{A2.13}
    \epsilon^{a'_1a'_2\dots a'_m}=\epsilon^{b_1b_2\dots b_m}J^{a'_1}{}_{b_1}J^{a'_2}{}_{b_2}\dots J^{a'_m}{}_{b_m}
\end{equation}
are merely a re-labeling of indices.

For a spacetime manifold of $m=4$ dimensions and split-signature $(p,q)=(1,3)$, the pairing of two orientation symbols is easily deduced by noting the definition in Eq. \eqref{epsilon-definition}. Explicitly, this is the awful expression
\begin{equation}\label{4d-expanded-epsilon-epsilon}
\tag{A2.14}
    \begin{split}
\hat{\epsilon}^{abcd}\hat{\epsilon}_{ijkl}&=\delta^a{}_k\delta^b{}_l\delta^c{}_j\delta^d{}_i-\delta^a{}_l\delta^b{}_k\delta^c{}_j\delta^d{}_i+\delta^a{}_l\delta^b{}_j\delta^c{}_k\delta^d{}_i-\delta^a{}_j\delta^b{}_l\delta^c{}_k\delta^d{}_i\\
    &\quad +\delta^a{}_l\delta^b{}_i\delta^c{}_j\delta^d{}_k-\delta^a{}_j\delta^b{}_k\delta^c{}_i\delta^d{}_l+\delta^a{}_i\delta^b{}_l\delta^c{}_k\delta^d{}_j-\delta^a{}_k\delta^b{}_j\delta^c{}_l\delta^d{}_i\\
    &\quad-\delta^a{}_k\delta^b{}_l\delta^c{}_i\delta^d{}_j+\delta^a{}_j\delta^b{}_k\delta^c{}_l\delta^d{}_i-\delta^a{}_l\delta^b{}_i\delta^c{}_k\delta^d{}_j+\delta^a{}_l\delta^b{}_k\delta^c{}_i\delta^d{}_j\\
    &\quad-\delta^a{}_i\delta^b{}_k\delta^c{}_l\delta^d{}_j+\delta^a{}_k\delta^b{}_i\delta^c{}_l\delta^d{}_j-\delta^a{}_l\delta^b{}_j\delta^c{}_i\delta^d{}_k+\delta^a{}_j\delta^b{}_l\delta^c{}_i\delta^d{}_k\\
    &\quad+\delta^a{}_k\delta^b{}_j\delta^c{}_i\delta^d{}_l-\delta^a{}_i\delta^b{}_l\delta^c{}_j\delta^d{}_k+\delta^a{}_i\delta^b{}_j\delta^c{}_l\delta^d{}_k-\delta^a{}_j\delta^b{}_i\delta^c{}_l\delta^d{}_k\\
    &\quad+\delta^a{}_i\delta^b{}_k\delta^c{}_j\delta^d{}_l-\delta^a{}_k\delta^b{}_i\delta^c{}_j\delta^d{}_l+\delta^a{}_j\delta^b{}_i\delta^c{}_k\delta^d{}_l-\delta^a{}_i\delta^b{}_j\delta^c{}_k\delta^d{}_l.
    \end{split}
\end{equation}
Contracting one index in Eq. \eqref{4d-expanded-epsilon-epsilon} produces the much smaller expression
\begin{equation}
\tag{A2.15}
    \begin{split}\hat{\epsilon}^{abcd}\hat{\epsilon}_{ajkl}&=
    \delta^b{}_l\delta^c{}_k\delta^d{}_j-\delta^b{}_k\delta^c{}_l\delta^d{}_j-\delta^b{}_l\delta^c{}_j\delta^d{}_k\\
    &\quad+\delta^b{}_j\delta^c{}_l\delta^d{}_k+\delta^c{}_j\delta^b{}_k\delta^d{}_l-\delta^b{}_j\delta^c{}_k\delta^d{}_l,\\
    \end{split}
\end{equation}
and contracting two, three, and four indices produces
\begin{equation}\label{epsilon-epsilon-234}
\tag{A2.16}
\begin{split}
\hat{\epsilon}^{abcd}\hat{\epsilon}_{abkl}&=-2(\delta^c{}_k\delta^d{}_l-\delta^c{}_l\delta^d{}_k),\\
\hat{\epsilon}^{abcd}\hat{\epsilon}_{abcl}&=-2\cdot 3\delta^d{}_l,\\
\hat{\epsilon}^{abcd}\hat{\epsilon}_{abcd}&=-4!.
\end{split}
\end{equation}
The above may be regarded as (minus) the determinant of Kronecker deltas,
\begin{equation}
\tag{A2.17}
    \hat{\epsilon}^{abcd}\hat{\epsilon}_{ijkl}=-\begin{vmatrix}\delta^a{}_i&\delta^b{}_i&\delta^c{}_i&\delta^d{}_i\\\delta^a{}_j&\delta^b{}_j&\delta^c{}_j&\delta^d{}_j\\\delta^a{}_k&\delta^b{}_k&\delta^c{}_k&\delta^d{}_k\\\delta^a{}_l&\delta^b{}_l&\delta^c{}_l&\delta^d{}_l\end{vmatrix}.
\end{equation}
A significantly more concise form then follows,
\begin{equation}
    \tag{A2.18}
\hat{\epsilon}^{abcd}\hat{\epsilon}_{ijkl}=-4!\delta^{abcd}_{[ijkl]}=-4!\delta^{[abcd]}_{ijkl},
\end{equation}
where $\delta^{abcd}_{[ijkl]}$ is the generalized Kronecker delta, completely anti-symmetrized in its lower (upper) indices. Therefore, in an arbitrary $m$-dimensional spacetime manifold of split-signature $(p,q)$,
\begin{equation}
\tag{A2.19}
    \hat{\epsilon}^{a_1a_2\dots a_m}\hat{\epsilon}_{b_1b_2\dots b_m}=(-1)^qm!\delta^{a_1a_2\dots a_m}_{[b_1b_2\dots b_m]}.
\end{equation}
In general, for $p$ contracted indices,
\begin{equation}
\tag{A2.20}
    \hat{\epsilon}^{a_1\dots a_pa_{p+1}\dots a_m}\hat{\epsilon}_{a_1\dots a_pb_{p+1}\dots b_m}=(-1)^s(m-p)!\delta^{a_{p+1}\dots a_m}_{[b_{p+1}\dots b_m]}.
\end{equation}

Lastly, we may use the above identities to find the standard integration measure. We start with the definition of the non-invariant volume element
\begin{equation}
\tag{A2.21}
    \hat{\epsilon}:=\frac{1}{m!}\hat{\epsilon}_{a_1\dots a_m}dx^{a_1}\wedge\dots\wedge dx^{a_m}=\hat{\epsilon}_{1\dots m}dx^1\wedge\dots\wedge dx^{m}=:d^mx.
\end{equation}
Then, multiplying both sides (second and fourth equalities above) by $\hat{\epsilon}^{b_1\dots b_m}$, one finds
\begin{equation}
\tag{A2.22}
\begin{split}
    \hat{\epsilon}^{b_1\dots b_m} d^mx&=\frac{1}{m!}\hat{\epsilon}^{b_1\dots b_m}\hat{\epsilon}_{a_1\dots a_m}dx^{a_1}\wedge\dots\wedge dx^{a_m}\\
    &=\frac{-m!(-1)^q}{m!}\delta^{b_1b_2\dots b_m}_{[a_1a_2\dots a_m]}dx^{a_1}\wedge\dots\wedge dx^{a_m}\\
   &= dx^{b_1}\wedge\dots\wedge dx^{b_m}.
    \end{split}
\end{equation}
From these expressions, one may easily see that in the non-coordinate basis provided by $e^{\underline{a}}=e^{\underline{a}}{}_bdx^b$, one has
\begin{equation}
    \tag{A2.23}
     \hat{\epsilon}^{\underline{a}_1\dots \underline{a}_m} d^mx=e^{\underline{a}_1}\wedge\dots\wedge e^{\underline{a}_m}.
\end{equation}\\

\phantomsection
\addcontentsline{toc}{section}{\hspace{-1.5em}Appendix B: Calculations}
\begingroup
\renewcommand{\addcontentsline}[3]{}
\section*{Appendix B: Calculations}
\label{app-B}
\endgroup

\phantomsection
\addcontentsline{toc}{subsection}{\hspace{-1.5em}B.1: Abstract-Index to Exterior: Gravity}
\begingroup
\renewcommand{\addcontentsline}[3]{}
\subsection*{B.1: Abstract-Index to Exterior: Gravity}
\label{app-B1:abs-ind-exterior-form}
\endgroup

In this section, we present the explicit transitions between abstract-index and exterior form for the standard terms encountered in a typical gravitational action. For concreteness, we display these calculations for a $4$-dimensional spacetime manifold $\mathcal{M}$ of split-signature $(p,q)=(1,3)$. We utilize extensively the identities of Appendix \hyperref[app-A:orientation]{A.2}.

The Palatini action may be obtained from the Einstein-Hilbert (EH) action by transitioning to exterior form. The relationship between the two actions is most elegant when starting with the Palatini form, however, we follow the pattern of discussion and begin with the EH action,
\begin{equation}
\tag{B1.1}
\begin{split}
    S_{EH}&=\int_{\mathcal{M}}  d^4x\sqrt{-|g|}R\\
    &=\int_{\mathcal{M}} d^4x\sqrt{-|g|}R^{mn}{}_{[mn]}\\
    &=\frac{1}{2}\int_{\mathcal{M}} d^4x\sqrt{-|\eta||e|^2}(R^{mn}{}_{[mn]}-R^{mn}{}_{[nm]})\\
    &=\frac{1}{2}\int_{\mathcal{M}} d^4x\sqrt{-(-1)^3|e|^2}(\delta^p{}_m\delta^q{}_n-\delta^p{}_n\delta^q{}_m)R^{mn}{}_{[pq]}\\
    &=\frac{1}{2}\int_{\mathcal{M}} d^4x|e|(\frac{-1}{2}\hat{\epsilon}_{klmn}\hat{\epsilon}^{klpq})R^{mn}{}_{[pq]}\\
    &=\frac{-1}{4}\int_{\mathcal{M}} d^4x\hat{\epsilon}_{\underline{abcd}}e^{\underline{a}}{}_ke^{\underline{b}}{}_le^{\underline{c}}{}_me^{\underline{d}}{}_n\hat{\epsilon}^{klpq}R^{mn}{}_{[pq]}\\
    &=\frac{-1}{4}\int_{\mathcal{M}} d^4x\hat{\epsilon}_{\underline{abcd}}e^{\underline{a}}{}_ke^{\underline{b}}{}_l\epsilon^{klpq}R^{\underline{cd}}{}_{[pq]}\\
    &=\frac{-1}{4}\int_{\mathcal{M}} \hat{\epsilon}_{\underline{abcd}}e^{\underline{a}}{}_ke^{\underline{b}}{}_lR^{\underline{cd}}{}_{[pq]}dx^k\wedge dx^l\wedge dx^p\wedge dx^q\\
    &=\frac{-1}{2}\int_{\mathcal{M}} \hat{\epsilon}_{\underline{abcd}}(\frac{1}{2!}R^{\underline{cd}}{}_{[pq]}dx^p\wedge dx^q)\wedge(e^{\underline{a}}{}_{k}dx^k\wedge e^{\underline{b}}{}_{l}dx^l)\\
    &=-\int_{\mathcal{M}} R^{\underline{cd}}\wedge (\frac{1}{2!}\hat{\epsilon}_{\underline{abcd}}e^{\underline{a}}\wedge e^{\underline{b}})\\
    &=\int_{\mathcal{M}} R^{\underline{cd}}\wedge *(e\wedge e)_{\underline{dc}}.
    \end{split}
\end{equation}
This result is coordinate-independent and dubbed the \textit{Palatini Action}:
\begin{equation}
\tag{B1.2}
    S_{P}=\int_{\mathcal{M}} R^{\underline{ab}}\wedge *(e\wedge e)_{\underline{ba}}.
\end{equation}
The minus sign is absorbed by exchanging indices on the anti-symmetric dual of the $2$-form $e\wedge e$. This is the more natural presentation, since matrix products follow the index contraction convention, $A^a{}_bB^b{}_c$. 

The action $S_{\Lambda}$, the cosmological constant, may also be written in exterior form. Beginning with the abstract-index form, we find
\begin{equation}
\tag{B1.3}
    \begin{split}
        S_{\Lambda}&=\int_{\mathcal{M}}d^mx\sqrt{-|g|}(2\Lambda)\\
        &=\frac{-2}{4!}\int_{\mathcal{M}}d^mx|e|\Lambda\hat{\epsilon}_{mnpq}\hat{\epsilon}{}^{mnpq}\\
        &=\frac{-2}{4!}\int_{\mathcal{M}}d^mx\Lambda e^{\underline{a}}{}_me^{\underline{b}}{}_ne^{\underline{c}}{}_pe^{\underline{d}}{}_q\hat{\epsilon}_{\underline{abcd}}\hat{\epsilon}{}^{mnpq}\\
        &=\frac{-2}{4!}\int_{\mathcal{M}}\Lambda e^{\underline{a}}{}_me^{\underline{b}}{}_ne^{\underline{c}}{}_pe^{\underline{d}}{}_q\hat{\epsilon}_{\underline{abcd}}dx^m\wedge dx^n\wedge dx^p\wedge dx^q\\
        &=\frac{-2}{4!}\int_{\mathcal{M}}\Lambda \hat{\epsilon}_{\underline{abcd}}e^{\underline{a}}\wedge e^{\underline{b}}\wedge e^{\underline{c}}\wedge e^{\underline{d}}\\
        &=\frac{-1}{6}\int_{\mathcal{M}}\Lambda e^{\underline{a}}\wedge e^{\underline{b}}\wedge (\frac{1}{2!}\hat{\epsilon}_{\underline{abcd}}e^{\underline{c}}\wedge e^{\underline{d}})\\
        &=\int_{\mathcal{M}}\frac{\Lambda}{6} (e^{\underline{a}}\wedge e^{\underline{b}})\wedge *(e\wedge e)_{\underline{ba}}.
    \end{split}
\end{equation}

The Euler action $S_{\mathscr{E}}$, written in abstract-index form, contains the \textit{Gauss-Bonnet} action. Due to the $24$-term epsilon-epsilon contraction identity, Eq. \eqref{4d-expanded-epsilon-epsilon}, $S_{\mathscr{E}}$ is much simpler to handle in the opposite direction. Beginning with the exterior form, we find
\begin{equation}
\tag{B1.4}
    \begin{split}
S_{\mathscr{E}}&=\int_{\mathcal{M}}R^{\underline{ab}}\wedge*R_{\underline{ba}}\\
&=-\int_{\mathcal{M}}R^{\underline{ab}}\wedge(\hat{\epsilon}_{\underline{abcd}} R^{\underline{cd}})\\
        &=-\int_{\mathcal{M}}\hat{\epsilon}_{\underline{abcd}}(\frac{1}{2!}R^{\underline{ab}}{}_{[mn]} dx^m\wedge dx^n)\wedge (\frac{1}{2!}R^{\underline{cd}}{}_{[pq]}dx^p\wedge dx^q)\\
        &=\frac{-1}{4}\int_{\mathcal{M}}d^4x\hat{\epsilon}_{\underline{abcd}}R^{\underline{ab}}{}_{[mn]} R^{\underline{cd}}{}_{[pq]}\hat{\epsilon}^{mnpq}\\
        &=\frac{-1}{4}\int_{\mathcal{M}}d^4x\hat{\epsilon}_{\underline{abcd}}e^{\underline{a}}{}_ae^{\underline{b}}{}_be^{\underline{c}}{}_ce^{\underline{d}}{}_dR^{ab}{}_{[mn]} R^{cd}{}_{[pq]}\hat{\epsilon}^{mnpq}\\
    &=\frac{-1}{4}\int_{\mathcal{M}}d^4x|e|\hat{\epsilon}_{abcd}R^{ab}{}_{[mn]} R^{cd}{}_{[pq]}\hat{\epsilon}^{mnpq}\\
    &=\frac{-4!(-1)^3}{4}\int_{\mathcal{M}}d^4x|e|\delta^{mnpq}_{[abcd]}R^{ab}{}_{[mn]} R^{cd}{}_{[pq]}\\
    &=\int_{\mathcal{M}}d^4x\sqrt{-|g|}\left(R^{ab}{}_{[mn]} R^{mn}{}_{[ab]}-\Delta R^{m}{}_n\Delta R^n{}_m+R^2\right),
    \end{split}
\end{equation}
where
\begin{equation}
\tag{B1.5}
    \Delta R^m{}_n:=R^{[am]}{}_{[an]}=R^{am}{}_{[an]}-R^{ma}{}_{[an]}=R^{m}{}_{n}-\check{R}{}^{m}{}_{n}
\end{equation}
is the difference between the Ricci and co-Ricci curvature tensors. 

Lastly, the Pontrjagin term $S_{\mathscr{P}}$ transitions as
\begin{equation}
\tag{B1.6}
\begin{split}
    S_{\mathscr{P}}&=\int_{\mathcal{M}}R^{\underline{a}}{}_{\underline{b}}\wedge R^{\underline{b}}{}_{\underline{a}}\\
    &=\int_{\mathcal{M}}(\frac{1}{2!}R^{\underline{a}}{}_{{\underline{b}}[mn]}dx^m\wedge dx^n)\wedge(\frac{1}{2!}R^{\underline{b}}{}_{{\underline{a}}[pq]}dx^p\wedge dx^q)\\
    &=\frac{1}{4}\int_{\mathcal{M}}R^{\underline{a}}{}_{{\underline{b}}[mn]}R^{\underline{b}}{}_{{\underline{a}}[pq]}dx^m\wedge dx^n\wedge dx^p\wedge dx^q\\
    &=\frac{1}{4}\int_{\mathcal{M}}d^4xR^{\underline{a}}{}_{{\underline{b}}[mn]}R^{\underline{b}}{}_{{\underline{a}}[pq]}\hat{\epsilon}{}^{mnpq}\\
    &=\frac{1}{4}\int_{\mathcal{M}}d^4x|e|R^{\underline{a}}{}_{{\underline{b}}[mn]}R^{\underline{b}}{}_{{\underline{a}}[pq]}\epsilon^{mnpq}\\
    &=\int_{\mathcal{M}}d^4x\sqrt{-|g|}R^{\underline{a}}{}_{{\underline{b}}mn}R^{\underline{b}}{}_{{\underline{a}}pq}\epsilon^{mnpq}\\
    &=\int_{\mathcal{M}}d^4x\sqrt{-|g|}R^{a}{}_{bmn}R^{b}{}_{apq}\epsilon^{mnpq},
    \end{split}
\end{equation}
where $\epsilon^{mnpq}$ is the completely anti-symmetric \textit{tensor}. In the last step, we have exchanged indices $\underline{a}\rightarrow a$, so that the final form does not reference the local frame structure. 

The relationship between the the Pontrjagin $4$-form $\mathscr{P}=R^{\underline{a}}{}_{\underline{b}}\wedge R^{\underline{b}}{}_{\underline{a}}$ and the Chern-Simons $3$-form $\mathscr{C}$ may be developed as follows:
\begin{equation}
\tag{B1.7}
\begin{split}
    d\mathscr{C}&=d(\omega^{\underline{a}}{}_{\underline{b}}\wedge R^{\underline{b}}{}_{\underline{a}}-\frac{1}{3}\omega^{\underline{a}}{}_{\underline{b}}\wedge\omega^{\underline{b}}{}_{\underline{c}}\wedge\omega^{\underline{c}}{}_{\underline{a}})\\
    &=d\omega^{\underline{a}}{}_{\underline{b}}\wedge R^{\underline{b}}{}_{\underline{a}}-\omega^{\underline{a}}{}_{\underline{b}}\wedge d(d\omega^{\underline{b}}{}_{\underline{a}}+\omega^{\underline{b}}{}_{\underline{c}}\wedge\omega^{\underline{c}}{}_{\underline{a}})\\
    &\quad-\frac{1}{3}d\omega^{\underline{a}}{}_{\underline{b}}\wedge\omega^{\underline{b}}{}_{\underline{c}}\wedge\omega^{\underline{c}}{}_{\underline{a}}+\frac{1}{3}\omega^{\underline{a}}{}_{\underline{b}}\wedge d\omega^{\underline{b}}{}_{\underline{c}}\wedge\omega^{\underline{c}}{}_{\underline{a}}-\frac{1}{3}\omega^{\underline{a}}{}_{\underline{b}}\wedge\omega^{\underline{b}}{}_{\underline{c}}\wedge d\omega^{\underline{c}}{}_{\underline{a}}\\
    &=d\omega^{\underline{a}}{}_{\underline{b}}\wedge R^{\underline{b}}{}_{\underline{a}}-\omega^{\underline{a}}{}_{\underline{b}}\wedge d\omega^{\underline{b}}{}_{\underline{c}}\wedge\omega^{\underline{c}}{}_{\underline{a}}+\omega^{\underline{a}}{}_{\underline{b}}\wedge \omega^{\underline{b}}{}_{\underline{c}}\wedge d\omega^{\underline{c}}{}_{\underline{a}}\\
    &\quad-\frac{1}{3}d\omega^{\underline{a}}{}_{\underline{b}}\wedge\omega^{\underline{b}}{}_{\underline{c}}\wedge\omega^{\underline{c}}{}_{\underline{a}}-\frac{1}{3}d\omega^{\underline{b}}{}_{\underline{c}}\wedge\omega^{\underline{c}}{}_{\underline{a}}\wedge\omega^{\underline{a}}{}_{\underline{b}}-\frac{1}{3}d\omega^{\underline{c}}{}_{\underline{a}}\wedge\omega^{\underline{a}}{}_{\underline{b}}\wedge\omega^{\underline{b}}{}_{\underline{c}}\\
    &=d\omega^{\underline{a}}{}_{\underline{b}}\wedge R^{\underline{b}}{}_{\underline{a}}+d\omega^{\underline{a}}{}_{\underline{b}}\wedge\omega^{\underline{b}}{}_{\underline{c}}\wedge\omega^{\underline{c}}{}_{\underline{a}}\\
    &=R^{\underline{a}}{}_{\underline{b}}\wedge R^{\underline{b}}{}_{\underline{a}}-\omega^{\underline{a}}{}_{\underline{c}}\wedge\omega^{\underline{c}}{}_{\underline{b}}\wedge R^{\underline{b}}{}_{\underline{a}}+d\omega^{\underline{a}}{}_{\underline{b}}\wedge\omega^{\underline{b}}{}_{\underline{c}}\wedge\omega^{\underline{c}}{}_{\underline{a}}\\
    &=R^{\underline{a}}{}_{\underline{b}}\wedge R^{\underline{b}}{}_{\underline{a}}-\omega^{\underline{a}}{}_{\underline{c}}\wedge\omega^{\underline{c}}{}_{\underline{b}}\wedge d\omega^{\underline{b}}{}_{\underline{a}}+d\omega^{\underline{a}}{}_{\underline{b}}\wedge\omega^{\underline{b}}{}_{\underline{c}}\wedge\omega^{\underline{c}}{}_{\underline{a}}\\
     &=R^{\underline{a}}{}_{\underline{b}}\wedge R^{\underline{b}}{}_{\underline{a}},
\end{split}
\end{equation}
since the particular pairing of the wedge product and trace provides
\begin{equation}
\tag{B1.8}
    \omega^{\underline{a}}{}_{\underline{b}}\wedge\omega^{\underline{b}}{}_{\underline{c}}\wedge\omega^{\underline{c}}{}_{\underline{d}}\wedge\omega^{\underline{d}}{}_{\underline{a}}=-\omega^{\underline{d}}{}_{\underline{a}}\wedge\omega^{\underline{a}}{}_{\underline{b}}\wedge\omega^{\underline{b}}{}_{\underline{c}}\wedge\omega^{\underline{c}}{}_{\underline{d}}=0.
\end{equation}

\phantomsection
\addcontentsline{toc}{subsection}{\hspace{-1.5em}B.2: Exterior to Abstract-Index: Projective Gravity}
\begingroup
\renewcommand{\addcontentsline}[3]{}
\subsection*{B.2: Exterior to Abstract-Index: Projective Gravity}
\label{app-B2:exterior-abs-ind-form-proj-grav}
\endgroup

In this section, we make explicit the transition of the projective Lovelock action $S_{\bar{\mathfrak{p}}L}$ from exterior to abstract-index form. Much of this discussion parallels Appendix \hyperref[app-B1:abs-ind-exterior-form]{B.1}, with a few minor differences. For ease, we choose to work with the $(m+1)$-form Lagrangian density $\mathcal{L}_{\bar{\mathfrak{p}}L}$ and compute each term separately, identified by the coefficient label $\tilde{\alpha}_i$. Furthermore, we restrict the discussion to $m+1=5$ dimensions, since there exist more or less terms in $m+1\neq 5$ dimensions.

Beginning with the $\tilde{\alpha}_0$ term, the transition to abstract-index form is
\begin{equation}
\tag{B2.1}
\begin{split}
    \mathcal{L}^0_{\bar{\mathfrak{p}}L}&=\tilde{\mathcal{B}}{}^{\underline{AB}}\wedge*\tilde{\mathcal{B}}_{\underline{BA}}\\
    &=(\frac{1}{2}\tilde{E}{}^{\underline{A}}\wedge\tilde{E}{}^{\underline{B}})\wedge(\frac{-1}{3!}\hat{\tilde{\epsilon}}_{\underline{ABCDE}}\tilde{E}{}^{\underline{C}}\wedge\tilde{E}{}^{\underline{D}}\wedge\tilde{E}{}^{\underline{E}})\\
    &=\frac{-1}{12}\hat{\tilde{\epsilon}}_{\underline{ABCDE}}\tilde{E}{}^{\underline{A}}\wedge\tilde{E}{}^{\underline{B}}\wedge\tilde{E}{}^{\underline{C}}\wedge\tilde{E}{}^{\underline{D}}\wedge\tilde{E}{}^{\underline{E}}\\
    &=\frac{-1}{12}\hat{\tilde{\epsilon}}_{\underline{ABCDE}}\tilde{E}{}^{\underline{A}}{}_{M}\tilde{E}{}^{\underline{B}}{}_{N}\tilde{E}{}^{\underline{C}}{}_{P}\tilde{E}{}^{\underline{D}}{}_{Q}\tilde{E}{}^{\underline{E}}{}_{R}dx^{M}\wedge dx^{N}\wedge dx^{P}\wedge dx^{Q}\wedge dx^{R}\\
    &=\frac{-1}{12}|\tilde{e}|\hat{\epsilon}_{MNPQR}\hat{\epsilon}{}^{MNPQR}d^5x\\
    &=\frac{-1}{12}d^5x|\tilde{e}|((-1)^q\eta_0 5!)\\
    &=(-1)^{q+1}\eta_010d^5x|\tilde{e}|.
    \end{split}
\end{equation}
For a split-signature $(p,q)=(1,3)$, we have
\begin{equation}
\tag{B2.2}
     \frac{\tilde{\alpha}_0}{5}\mathcal{L}^0_{\bar{\mathfrak{p}}L}=2\eta_0\tilde{\alpha}_0d^5x|\tilde{e}|.
\end{equation}
For the $\tilde{\alpha}_1$ term, we find
\begin{equation}
\tag{B2.3}
\begin{split}
    \mathcal{L}^1_{\bar{\mathfrak{p}}L}&=\tilde{\mathcal{B}}{}^{\underline{AB}}\wedge*\tilde{\mathcal{K}}_{\underline{BA}}\\
    &=-(\frac{1}{2}\tilde{E}{}^{\underline{A}}\wedge \tilde{E}{}^{\underline{B}})\wedge(\hat{\tilde{\epsilon}}_{\underline{ABCDE}}\tilde{\mathcal{K}}{}^{\underline{CD}}\wedge\tilde{E}{}^{\underline{E}})\\
    &=\frac{-1}{4}\tilde{E}{}^{\underline{A}}{}_M \tilde{E}{}^{\underline{B}}{}_N\hat{\tilde{\epsilon}}_{\underline{ABCDE}}\tilde{\mathcal{K}}{}^{\underline{CD}}{}_{[PQ]}\tilde{E}{}^{\underline{E}}{}_Rdx^M\wedge dx^N \wedge dx^P\wedge dx^Q\wedge dx^R\\
    &=\frac{-1}{4}\tilde{E}{}^{\underline{A}}{}_M \tilde{E}{}^{\underline{B}}{}_N\hat{\tilde{\epsilon}}_{\underline{ABCDE}}\tilde{\mathcal{K}}{}^{\underline{CD}}{}_{[PQ]}\tilde{E}{}^{\underline{E}}{}_R\hat{\epsilon}{}^{MNPQR}d^5x\\
    &=\frac{-1}{4}\tilde{E}{}^{\underline{A}}{}_M \tilde{E}{}^{\underline{B}}{}_N\hat{\tilde{\epsilon}}_{\underline{ABCDE}}\tilde{E}{}^{\underline{C}}{}_S \tilde{E}{}^{\underline{D}}{}_T \tilde{\mathcal{K}}{}^{ST}{}_{[PQ]}\tilde{E}{}^{\underline{E}}{}_R\hat{\epsilon}{}^{MNPQR}d^5x\\
    &=\frac{-1}{4}d^5x|\tilde{e}|\tilde{\mathcal{K}}{}^{ST}{}_{[PQ]}\hat{\epsilon}_{MNSTR}\hat{\epsilon}{}^{MNPQR}\\
    &=\frac{-3!}{4}(-1)^q\eta_0d^5x|\tilde{e}|\tilde{\mathcal{K}}{}^{ST}{}_{[PQ]}(\delta^P{}_S\delta^Q{}_T-\delta^P{}_T\delta^Q{}_S)\\
    &=\frac{3}{2}(-1)^{q+1}\eta_0d^5x|\tilde{e}|(\tilde{\mathcal{K}}{}^{PQ}{}_{[PQ]}-\tilde{\mathcal{K}}{}^{PQ}{}_{[QP]})\\
    &=(-1)^{q+1}3\eta_0d^5x|\tilde{e}|\tilde{\mathcal{K}}{}^{PQ}{}_{[PQ]}\\
    &=(-1)^{q+1}3\eta_0d^5x|\tilde{e}|\tilde{\mathcal{K}}.
    \end{split}
\end{equation}
For a split-signature $(p,q)=(1,3)$, we have
\begin{equation}
\tag{B2.4}
    \frac{\tilde{\alpha}_1}{3}\mathcal{L}^1_{\bar{\mathfrak{p}}L}=\eta_0\tilde{\alpha}_1d^5x|\tilde{e}|\tilde{\mathcal{K}}.
\end{equation}
Lastly, we find for the $\tilde{\alpha}_2$ term,
\begin{equation}
\tag{B2.5}
\begin{split}
    \mathcal{L}^2_{\bar{\mathfrak{p}}L}&=\tilde{\mathcal{K}}{}^{\underline{AB}}\wedge *\tilde{\mathcal{K}}_{\underline{BA}}\\
    &=-\tilde{\mathcal{K}}{}^{\underline{AB}}\wedge (\hat{\tilde{\epsilon}}_{\underline{ABCDE}}\tilde{\mathcal{K}}{}^{\underline{CD}}\wedge\tilde{E}{}^{\underline{E}})\\
    &=-(\frac{1}{2}\tilde{\mathcal{K}}{}^{\underline{AB}}{}_{[PQ]}dx^P\wedge dx^Q)\wedge\hat{\tilde{\epsilon}}_{\underline{ABCDE}}(\frac{1}{2}\tilde{\mathcal{K}}{}^{\underline{CD}}{}_{[UV]}\tilde{E}{}^{\underline{E}}{}_T dx^U\wedge dx^V\wedge dx^T)\\
    &=-\frac{1}{4}d^5x\tilde{E}{}^{\underline{A}}{}_{M}\tilde{E}{}^{\underline{B}}{}_{N}\tilde{\mathcal{K}}{}^{MN}{}_{[PQ]}\hat{\tilde{\epsilon}}_{\underline{ABCDE}}\tilde{E}{}^{\underline{C}}{}_{R}\tilde{E}{}^{\underline{D}}{}_{S}\tilde{\mathcal{K}}{}^{RS}{}_{[UV]}\tilde{E}{}^{\underline{E}}{}_T \hat{\epsilon}{}^{PQUVT}\\
    &=-\frac{1}{4}d^5x|\tilde{e}|\tilde{\mathcal{K}}{}^{MN}{}_{[PQ]}\tilde{\mathcal{K}}{}^{RS}{}_{[UV]} \hat{\epsilon}_{MNRST}\hat{\epsilon}{}^{PQUVT}.
    \end{split}
\end{equation}
Using the lengthy epsilon-epsilon contraction identity, Eq. \eqref{4d-expanded-epsilon-epsilon} in $5$ dimensions, and choosing the split-signature $(p,q)=(1,3)$, we find
\begin{equation}
\tag{B2.6}
\tilde{\alpha}_2\mathcal{L}^2_{\bar{\mathfrak{p}}L}=\eta_0\tilde{\alpha}_2d^5x|\tilde{e}|\left(\tilde{\mathcal{K}}^{\underline{AB}}{}_{[\underline{CD}]}\tilde{\mathcal{K}}^{\underline{CD}}{}_{[\underline{AB}]}-\Delta\tilde{\mathcal{K}}^{\underline{A}}{}_{\underline{B}}\Delta\tilde{\mathcal{K}}^{\underline{B}}{}_{\underline{A}}+\tilde{\mathcal{K}}^2\right),
\end{equation}
where
\begin{equation}
\tag{B2.7}
    \Delta\tilde{\mathcal{K}}^{\underline{A}}{}_{\underline{B}}:=\tilde{\mathcal{K}}{}^{[\underline{CA}]}{}_{[\underline{CB}]}=\tilde{\mathcal{K}}{}^{\underline{CA}}{}_{[\underline{CB}]}-\tilde{\mathcal{K}}{}^{\underline{AC}}{}_{[\underline{CB}]}=\tilde{\mathcal{K}}{}^{\underline{A}}{}_{\underline{B}}-\check{\tilde{\mathcal{K}}}{}^{\underline{A}}{}_{\underline{B}}
\end{equation}
is the difference of projective Ricci and co-Ricci curvature tensors.\\

\phantomsection
\addcontentsline{toc}{subsection}{\hspace{-1.5em}B.3: Exterior to Abstract-Index: Spinor}
\begingroup
\renewcommand{\addcontentsline}[3]{}
\subsection*{B.3: Exterior to Abstract-Index: Spinor}
\label{app-B3:exterior-abs-ind-form}
\endgroup

It will be convenient to have the abstract-index form of $\mathcal{L}_{\bar{\mathfrak{D}}}$ available for discussion. To accomplish this, we utilize extensively the properties of the orientation symbol outlined in Appendix \hyperref[app-A:orientation]{A.2}. We consider a $(p,q)$ split-signature spacetime manifold $\mathcal{M}$ of dimension $m=2k$. 

We express $\mathcal{L}_{\bar{\mathfrak{D}}}$ in abstract-index form, beginning with the $\bar{\mathfrak{p}}{}^{\mathfrak{w}}$-spinor kinetic term, $\mathcal{L}_{\tilde{\Psi}}$. With each step explicit, we find
\begin{equation}\label{spinor-kinetic-indexform}
\tag{B3.1}
    \begin{split}
\mathcal{L}_{\tilde{\Psi}}&=\tilde{\overline{\Psi}}*\tilde{\Gamma}\wedge\tilde{D}\tilde{\Psi}\\
&=\tilde{\overline{\Psi}}\tilde{\Gamma}{}^{\underline{A}}\hat{\tilde{\epsilon}}_{\underline{A}}\wedge\tilde{D}\tilde{\Psi}\\
&=\frac{1}{m!}\tilde{\overline{\Psi}}\tilde{\Gamma}{}^{\underline{A}_1}\hat{\tilde{\epsilon}}_{\underline{A}_1\underline{A}_2\dots \underline{A}_{m+1}}\tilde{E}{}^{\underline{A}_2}\wedge\dots\wedge\tilde{E}{}^{\underline{A}_{m+1}}\wedge\tilde{D}\tilde{\Psi}\\
&=\frac{1}{m!}\tilde{\overline{\Psi}}\tilde{\Gamma}{}^{\underline{A}_1}\hat{\tilde{\epsilon}}_{\underline{A}_1\underline{A}_2\dots \underline{A}_{m+1}}\tilde{E}{}^{\underline{A}_2}{}_{M_2}\dots\tilde{E}{}^{\underline{A}_{m+1}}{}_{M_{m+1}}dx^{M_2}\wedge\dots\wedge dx^{M_{m+1}}\wedge dx^{M_1}\tilde{D}_{M_1}\tilde{\Psi}\\
&=\frac{1}{m!}(-1)^m\tilde{\overline{\Psi}}\tilde{\Gamma}{}^{\underline{A}_1}\hat{\tilde{\epsilon}}_{\underline{A}_1\underline{A}_2\dots \underline{A}_{m+1}}\tilde{E}{}^{\underline{A}_2}{}_{M_2}\dots\tilde{E}{}^{\underline{A}_{m+1}}{}_{M_{m+1}}dx^{M_1}\wedge dx^{M_2}\wedge\dots\wedge dx^{M_{m+1}}\tilde{D}_{M_1}\tilde{\Psi}\\
&=\frac{1}{m!}(-1)^md^{m+1}x\tilde{\overline{\Psi}}\tilde{\Gamma}{}^{\underline{A}_1}\hat{\tilde{\epsilon}}_{\underline{A}_1\underline{A}_2\dots \underline{A}_{m+1}}\tilde{E}{}^{\underline{A}_2}{}_{M_2}\dots\tilde{E}{}^{\underline{A}_{m+1}}{}_{M_{m+1}}\hat{\epsilon}{}^{M_1\dots M_{m+1}}\tilde{D}_{M_1}\tilde{\Psi}\\
&=\frac{1}{m!}(-1)^md^{m+1}x\tilde{\overline{\Psi}}\tilde{\Gamma}{}^{N}\tilde{E}{}^{\underline{A}_1}{}_{N}\hat{\tilde{\epsilon}}_{\underline{A}_1\underline{A}_2\dots \underline{A}_{m+1}}\tilde{E}{}^{\underline{A}_2}{}_{M_2}\dots\tilde{E}{}^{\underline{A}_{m+1}}{}_{M_{m+1}}\hat{\epsilon}{}^{M_1\dots M_{m+1}}\tilde{D}_{M_1}\tilde{\Psi}\\
&=\frac{1}{m!}(-1)^md^{m+1}x|\tilde{e}|\tilde{\overline{\Psi}}\tilde{\Gamma}{}^{N}\hat{\epsilon}_{NM_2\dots M_{m+1}}\hat{\epsilon}{}^{M_1\dots M_{m+1}}\tilde{D}_{M_1}\tilde{\Psi}\\
&=(-1)^{m+q}\eta_0d^{m+1}x|\tilde{e}|\tilde{\overline{\Psi}}\tilde{\Gamma}{}^{N}\delta^{M_1}{}_{N}D_{M_1}\tilde{\Psi}\\
        &=(-1)^{m+q}\eta_0d^{m+1}x|\tilde{e}|\tilde{\overline{\Psi}}\tilde{\Gamma}{}^{M}\tilde{D}_M\tilde{\Psi},
    \end{split}
\end{equation}
where the $(-1)^m$ results from permuting the $M_1$ index through $m$ spaces. For the physical case of interest, we take $(p,q)=(1,3)$ and $\eta_0$ arbitrary to find
\begin{equation}
\tag{B3.2}
    \mathcal{L}_{\tilde{\Psi}}=-\eta_0d^{5}x|\tilde{e}|\tilde{\overline{\Psi}}\tilde{\Gamma}{}^{M}\tilde{D}_M\tilde{\Psi}.
\end{equation}

The adjoint kinetic term $\overline{\mathcal{L}}_{\tilde{\Psi}}$ is expressed similarly. Omitting the first few equivalent steps, we find
\begin{equation}
\tag{B3.3}
        \begin{split}
\overline{\mathcal{L}}_{\tilde{\Psi}}&=\tilde{D}\tilde{\overline{\Psi}}\wedge*\tilde{\Gamma}\tilde{\Psi}\\
&=\frac{1}{m!}(\tilde{D}_{M_1}\tilde{\overline{\Psi}})\tilde{\Gamma}{}^{\underline{A}_1}\hat{\tilde{\epsilon}}_{\underline{A}_1\underline{A}_2\dots \underline{A}_{m+1}}\tilde{E}{}^{\underline{A}_2}{}_{M_2}\dots\tilde{E}{}^{\underline{A}_{m+1}}{}_{M_{m+1}}dx^{M_1}\wedge dx^{M_2}\wedge\dots\wedge dx^{M_{m+1}}\tilde{\Psi}\\
&=\frac{1}{m!}d^{m+1}x|\tilde{e}|(\tilde{D}_{M_1}\tilde{\overline{\Psi}})\tilde{\Gamma}{}^{N}\hat{\epsilon}_{NM_2\dots M_{m+1}}\hat{\epsilon}{}^{M_1\dots M_{m+1}}\tilde{\Psi}\\
&=(-1)^{q}\eta_0d^{m+1}x|\tilde{e}|(\tilde{D}_{M}\tilde{\overline{\Psi}})\tilde{\Gamma}{}^{M}\tilde{\Psi}.
        \end{split}
\end{equation}
Notice that $\overline{\mathcal{L}}_{\tilde{\Psi}}$ does not receive the additional factor of $(-1)^m$ present in Eq. \eqref{spinor-kinetic-indexform} due to the $1$-form derivative appearing on the left instead of the right. 

The interaction between the gauge-covariant derivative and dual gamma matrix, $\mathcal{L}_{\tilde{\Gamma}}$, is first split into its action on the gamma matrix and its action on the dual $m$-form,
\begin{equation}
\tag{B3.4}
    \begin{split}
        \tilde{\overline{\Psi}}(\tilde{D}*\tilde{\Gamma})\tilde{\Psi}&=\tilde{\overline{\Psi}}\tilde{D}(\tilde{\Gamma}{}^{\underline{A}}\hat{\tilde{\epsilon}}_{\underline{A}})\tilde{\Psi}\\
        &=\tilde{\overline{\Psi}}(\tilde{D}\tilde{\Gamma}{}^{\underline{A}})\wedge\hat{\tilde{\epsilon}}_{\underline{A}}\tilde{\Psi}+\tilde{\overline{\Psi}}\tilde{\Gamma}{}^{\underline{A}}(\tilde{D}\hat{\tilde{\epsilon}}_{\underline{A}})\tilde{\Psi}.
    \end{split}
\end{equation}
The interaction $\tilde{D}\tilde{\Gamma}{}^{\underline{A}}$ was related to the $\bar{\mathfrak{p}}{}^{-2}$-non-metricity in Eq. \eqref{gamma-non-metricity}. Expressing this in abstract-index form follows the same structure as $\overline{\mathcal{L}}_{\tilde{\Psi}}$, since the $1$-form derivative appears on the left,
\begin{equation}\label{D-Gamma-appendix}
\tag{B3.5}
    \begin{split}
        \tilde{\overline{\Psi}}(\tilde{D}\tilde{\Gamma}{}^{\underline{A}})\wedge\hat{\tilde{\epsilon}}_{\underline{A}}\tilde{\Psi}
        &=\frac{1}{2}\tilde{\overline{\Psi}}\tilde{\eta}{}^{\underline{AB}}\tilde{Q}_{\underline{BC}}\tilde{\Gamma}{}^{\underline{C}}\wedge\hat{\tilde{\epsilon}}_{\underline{A}}\tilde{\Psi}\\
        &=(-1)^q\frac{\eta_0}{2}d^{m+1}x|\tilde{e}|\tilde{\overline{\Psi}}\tilde{\eta}{}^{\underline{AB}}\tilde{Q}_{M\underline{BC}}\tilde{\Gamma}{}^{\underline{C}}(\tilde{E}{}^{-1})^M{}_{\underline{A}}\tilde{\Psi}\\
        &=(-1)^q\frac{\eta_0}{2}d^{m+1}x|\tilde{e}|\tilde{\overline{\Psi}}\tilde{Q}_{\underline{B}}\tilde{\Gamma}{}^{\underline{B}}\tilde{\Psi}.
    \end{split}
\end{equation}
In the above, we use the obvious notation $\tilde{Q}_{\underline{A}}{}^{\underline{A}}{}_{\underline{B}}=\tilde{Q}_{\underline{B}}$ to denote the unnatural, external trace of the $\bar{\mathfrak{p}}{}^{-2}$-non-metricity. This is unambiguous since the unnatural, internal trace, $\tilde{Q}_{M}{}^{\underline{A}}{}_{\underline{A}}=2(m+1)\tilde{g}_M$, provides the projective Weyl co-vector.

The second term in $\mathcal{L}_{\tilde{\Gamma}}$ to be expressed in abstract-index form contains $\tilde{D}\hat{\tilde{\epsilon}}_{\underline{A}}$. This further splits into two parts. The first yields the gauge-covariant derivative of the $\bar{\mathfrak{p}}{}^{-(m+1)}$-orientation symbol, as in Eq. \eqref{cov-der-nlp-orientation-symbol},
\begin{equation}
\tag{B3.6}
    \tilde{D}\hat{\tilde{\epsilon}}_{\underline{A}_1\dots\underline{A}_{m+1}}=-(m+1)\tilde{g}\hat{\tilde{\epsilon}}_{\underline{A}_1\dots\underline{A}_{m+1}},
\end{equation}
giving rise to an interaction with the natural trace of the $\bar{\mathfrak{p}}{}^{-2}$-non-metricity, i.e., the projective Weyl-form $\tilde{g}$. Since $\tilde{g}$ appears on the left, the transition to abstract-index form again only produces $(-1)^q$,
\begin{equation}\label{g-tilde-spinor-indexform}
\tag{B3.7}
    \begin{split}
\mathcal{L}_{\tilde{\Gamma}}&\supset\frac{1}{m!}\tilde{\overline{\Psi}}\tilde{\Gamma}{}^{\underline{A}_1}(\tilde{D}\hat{\tilde{\epsilon}}_{\underline{A}_1\dots\underline{A}_{m+1}})\wedge\tilde{E}{}^{\underline{A}_2}\wedge\dots\wedge\tilde{E}{}^{\underline{A}_{m+1}}\tilde{\Psi}\\
&=\frac{-(m+1)}{m!}\tilde{\overline{\Psi}}\tilde{\Gamma}{}^{\underline{A}_1}\tilde{g}\hat{\tilde{\epsilon}}_{\underline{A}_1\dots\underline{A}_{m+1}}\wedge\tilde{E}{}^{\underline{A}_2}\wedge\dots\wedge\tilde{E}{}^{\underline{A}_{m+1}}\tilde{\Psi}\\
&=\frac{-(m+1)}{m!}\tilde{\overline{\Psi}}\tilde{\Gamma}{}^{\underline{A}_1}\tilde{g}_{M_1}\hat{\tilde{\epsilon}}_{\underline{A}_1\dots\underline{A}_{m+1}}\tilde{E}{}^{\underline{A}_2}{}_{M_2}\dots\tilde{E}{}^{\underline{A}_{m+1}}{}_{M_{m+1}}dx^{M_1}\wedge\dots\wedge dx^{M_{m+1}}\tilde{\Psi}\\
&=\frac{-(m+1)}{m!}d^{m+1}x|\tilde{e}|\tilde{\overline{\Psi}}\tilde{\Gamma}{}^{N}\tilde{g}_{M_1}\hat{\epsilon}_{NM_2\dots M_{m+1}}\hat{\epsilon}{}^{M_1\dots M_{m+1}}\tilde{\Psi}\\
&=(-1)^{q+1}(m+1)\eta_0d^{m+1}x|\tilde{e}|\tilde{\overline{\Psi}}\tilde{\Gamma}{}^{M}\tilde{g}_{M}\tilde{\Psi}.
    \end{split}
\end{equation}
The second part of $\tilde{D}\hat{\tilde{\epsilon}}_{\underline{A}}$ yields the gauge-covariant derivative of the exterior product of $m$ $\bar{\mathfrak{p}}$-co-frames. Recalling that
\begin{equation}
\tag{B3.8}
    \tilde{\mathcal{K}}{}^{\underline{A}}{}_{\underline{B}}\tilde{\Upsilon}{}^{\underline{B}}=\tilde{D}\tilde{D}\tilde{\Upsilon}{}^{\underline{A}}=\tilde{D}\tilde{E}{}^{\underline{A}}=\tilde{\mathcal{T}}{}^{\underline{A}},
\end{equation}
we relate the expression of interest to the $\bar{\mathfrak{p}}$-torsion,
\begin{equation}
\tag{B3.9}
    \begin{split}
        \tilde{D}(\tilde{E}{}^{\underline{A}_2}\wedge\dots\wedge\tilde{E}{}^{\underline{A}_{m+1}})&=m\tilde{D}(\tilde{E}{}^{\underline{A}_2})\wedge\dots\wedge\tilde{E}{}^{\underline{A}_{m+1}}\\
        &=m\tilde{\mathcal{K}}{}^{\underline{A}_2}{}_{\underline{B}}\tilde{\Upsilon}{}^{\underline{B}}\wedge\tilde{E}{}^{\underline{A}_3}\wedge\dots\wedge\tilde{E}{}^{\underline{A}_{m+1}}\\
        &=m\tilde{\mathcal{T}}{}^{\underline{A}_2}\wedge\tilde{E}{}^{\underline{A}_3}\wedge\dots\wedge\tilde{E}{}^{\underline{A}_{m+1}}.
    \end{split}
\end{equation}
Passing to abstract-index notation,
\begin{equation}
\tag{B3.10}
    \begin{split}
        \mathcal{L}_{\tilde{\Gamma}}&\supset\frac{1}{m!}\tilde{\overline{\Psi}}\tilde{\Gamma}{}^{\underline{A}_1}\hat{\tilde{\epsilon}}_{\underline{A}_1\dots\underline{A}_{m+1}}\tilde{D}(\tilde{E}{}^{\underline{A}_2}\wedge\dots\wedge\tilde{E}{}^{\underline{A}_{m+1}})\tilde{\Psi}\\
        &=\frac{1}{(m-1)!}\tilde{\overline{\Psi}}\tilde{\Gamma}{}^{\underline{A}_1}\hat{\tilde{\epsilon}}_{\underline{A}_1\dots\underline{A}_{m+1}}\tilde{\mathcal{T}}{}^{\underline{A}_2}\wedge\tilde{E}{}^{\underline{A}_3}\wedge\dots\wedge\tilde{E}{}^{\underline{A}_{m+1}}\tilde{\Psi}\\
        &=\frac{1}{2(m-1)!}\tilde{\overline{\Psi}}\tilde{\Gamma}{}^{\underline{A}_1}\hat{\tilde{\epsilon}}_{\underline{A}_1\dots\underline{A}_{m+1}}\tilde{\mathcal{T}}{}^{\underline{A}_2}{}_{[M_1M_2]}\tilde{E}{}^{\underline{A}_3}{}_{M_3}\dots\tilde{E}{}^{\underline{A}_{m+1}}{}_{M_{m+1}}dx^{M_1}\wedge\dots \wedge dx^{M_{m+1}}\tilde{\Psi}\\
        &=\frac{1}{2(m-1)!}d^{m+1}x|\tilde{e}|\tilde{\overline{\Psi}}\tilde{\Gamma}{}^{N_1}\hat{\epsilon}_{N_1N_2 M_3\dots M_{m+1}}\tilde{\mathcal{T}}{}^{N_2}{}_{[M_1M_2]}\hat{\epsilon}{}^{M_1M_2\dots M_{m+1}}\tilde{\Psi}\\
        &=(-1)^{q}\frac{\eta_0}{2}d^{m+1}x|\tilde{e}|\tilde{\overline{\Psi}}\tilde{\Gamma}{}^{N_1}\tilde{\mathcal{T}}{}^{N_2}{}_{[M_1M_2]}(\delta^{M_1}{}_{N_1}\delta^{M_2}{}_{N_2}-\delta^{M_1}{}_{N_2}\delta^{M_2}{}_{N_1})\tilde{\Psi}\\
        &=(-1)^{q}\eta_0d^{m+1}x|\tilde{e}|\tilde{\overline{\Psi}}\tilde{\Gamma}{}^{N_1}\tilde{\mathcal{T}}{}^{N_2}{}_{[N_1N_2]}\tilde{\Psi}\\
        &=(-1)^{q}\eta_0d^{m+1}x|\tilde{e}|\tilde{\overline{\Psi}}\tilde{\Gamma}{}^{M}\tilde{\mathcal{T}}_{M}\tilde{\Psi},
    \end{split}
\end{equation}
where $\tilde{\mathcal{T}}{}^{N}{}_{[MN]}:=\tilde{\mathcal{T}}_{M}$ is the torsion vector, defined with contraction on the outside index. Note that as a result of the definition of $\tilde{\mathcal{T}}$, no factor of $2$ is introduced in this contraction. Given our conventions, 
\begin{equation}
\tag{B3.11}
    \tilde{D}\hat{\tilde{\epsilon}}_{\underline{A}}=-(m+1)\tilde{g}\wedge\hat{\tilde{\epsilon}}_{\underline{A}}+\tilde{\mathcal{T}}{}^{\underline{B}}\wedge\hat{\tilde{\epsilon}}_{\underline{AB}},
\end{equation}
in complete agreement with \cite{Hehl}. Returning each piece of $\mathcal{L}_{\tilde{\Gamma}}$ provides
\begin{equation}\label{abs-ind-interaction-gen}
\tag{B3.12}
      \mathcal{L}_{\tilde{\Gamma}}=(-1)^q\frac{\eta_0}{2}d^{m+1}x|\tilde{e}|\tilde{\overline{\Psi}}\tilde{\Gamma}{}^{M}\left(\tilde{Q}_M-2(m+1)\tilde{g}_{M}+2\tilde{\mathcal{T}}_M\right)\tilde{\Psi}.
\end{equation}

The final two terms to be expressed in abstract-index notation are $\mathcal{L}_{\tilde{m}\tilde{\Gamma}}$ and $\mathcal{L}_{\tilde{\Gamma}\tilde{m}}$. The first of these parallels Eq. \eqref{g-tilde-spinor-indexform} with $\tilde{m}$ in place of $-(m+1)\tilde{g}$. Therefore, 
\begin{equation}
\tag{B3.13}
\mathcal{L}_{\tilde{m}\tilde{\Gamma}}=\tilde{\overline{\Psi}}\tilde{m}\wedge*\tilde{\Gamma}\tilde{\Psi}=(-1)^{q}\eta_0d^{m+1}x|\tilde{e}|\tilde{\overline{\Psi}}\tilde{m}_{M}\tilde{\Gamma}{}^{M}\tilde{\Psi}.
\end{equation}
Since for $\mathcal{L}_{\tilde{\Gamma}\tilde{m}}$, the $1$-form $\tilde{m}$ appears on the right of the $m$-form $*\tilde{\Gamma}$, we pick up an additional factor of $(-1)^m$ as in Eq. \eqref{spinor-kinetic-indexform},
\begin{equation}
\tag{B3.14}
    \mathcal{L}_{\tilde{\Gamma}\tilde{m}}=(-1)^{q+m}\eta_0d^{m+1}x|\tilde{e}|\tilde{\overline{\Psi}}\tilde{\Gamma}{}^M\tilde{m}_M\tilde{\Psi}.
\end{equation}
Notice that there is a relative sign, $(-1)^m$, which is dimension dependent between these two terms. Therefore, if their coupling coefficients are equal, these terms cancel in all odd $m=2k+1$-dimensions, with $k\in\mathbb{N}_0$.\\

\phantomsection
\addcontentsline{toc}{subsection}{\hspace{-1.5em}B.4: Calculations: Spinor}
\begingroup
\renewcommand{\addcontentsline}[3]{}
\subsection*{B.4: Calculations: Spinor}
\label{app-B4:calc-spinor}
\endgroup

\subsubsection*{Interaction Term}

The interaction term in Eq. \eqref{abs-ind-interaction-gen}, 
\begin{equation}
\tag{B4.1}
      \mathcal{L}_{\tilde{\Gamma}}=(-1)^q\frac{\eta_0}{2}d^{m+1}x|\tilde{e}|\tilde{\overline{\Psi}}\tilde{\Gamma}{}^{M}\left(\tilde{Q}_M-2(m+1)\tilde{g}_{M}+2\tilde{\mathcal{T}}_M\right)\tilde{\Psi},
\end{equation}
may be written in terms of the projective distortion tensor $\tilde{N}$ of $V\mathcal{M}$. To show this, we first decompose the connection on the volume bundle,
\begin{equation}
\tag{B4.2}
    \tilde{\Gamma}{}^L{}_{MN}=(\tilde{E}{}^{-1})^L{}_{\underline{A}}\tilde{D}_N\tilde{E}{}^{\underline{A}}{}_M,
\end{equation}
into its Levi-Civita and distortion parts,
\begin{equation}
\tag{B4.3}
    \tilde{\Gamma}{}^L{}_{MN}=\hat{\tilde{\Gamma}}{}^L{}_{MN}+\tilde{N}{}^L{}_{MN}.
\end{equation}
In this expression,
\begin{equation}
\tag{B4.4}
    \hat{\tilde{\Gamma}}{}^{L}{}_{MN}:=\frac{1}{2}\tilde{G}{}^{LK}(\partial_N\tilde{G}_{KM}+\partial_M\tilde{G}_{KN}-\partial_K\tilde{G}_{MN})
\end{equation}
is the Levi-Civita connection of $V\mathcal{M}$, and 
\begin{equation}\label{VM-distortion}
\tag{B4.5}
    \tilde{N}{}^L{}_{MN}=\tilde{L}{}^L{}_{MN}+\tilde{K}{}^L{}_{MN}
\end{equation}
is the associated distortion tensor. The latter further decomposes into two independent $V\mathcal{M}$-tensors. The first is the disformation tensor $\tilde{L}$,
\begin{equation}
\tag{B4.6}
    \tilde{L}^K{}_{MN}:=\frac{1}{2}\tilde{G}{}^{KL}(\tilde{Q}_{MNL}+\tilde{Q}_{NLM}-\tilde{Q}_{LMN}),
\end{equation}
constructed from the non-metricity
\begin{equation}
\tag{B4.7}
    \tilde{Q}_{LMN}:=-\tilde{\nabla}_L\tilde{G}_{MN}=-\partial_L\tilde{G}_{MN}+\tilde{\Gamma}{}^P{}_{ML}\tilde{G}_{PN}+\tilde{\Gamma}{}^P{}_{NL}\tilde{G}_{PM}.
\end{equation}
This non-metricity is defined with respect to the connection on $V\mathcal{M}$, i.e., $\tilde{\nabla}=d\pm\tilde{\Gamma}$, and expresses the incompatibility between the connection $\tilde{\Gamma}$ and the metric
\begin{equation}
\tag{B4.8}
    \tilde{G}_{MN}:=\tilde{\eta}_{\underline{AB}}\tilde{E}{}^{\underline{A}}{}_{M}\tilde{E}{}^{\underline{B}}{}_{N}.
\end{equation}
From the definition of $\tilde{G}$, it is simple to show
\begin{equation}
\tag{B4.9}
    \tilde{Q}_{LMN}=\tilde{Q}_{L\underline{AB}}\tilde{E}{}^{\underline{A}}{}_{M}\tilde{E}{}^{\underline{B}}{}_{N}.
\end{equation}
The second object in the decomposition of Eq. \eqref{VM-distortion} is the contortion tensor $\tilde{K}$,
\begin{equation}
\tag{B4.10}
    \tilde{K}{}^L{}_{MN}:=\frac{1}{2}\tilde{G}{}^{LK}(\tilde{\mathcal{T}}_{MKN}+\tilde{\mathcal{T}}_{NKM}-\tilde{\mathcal{T}}_{KNM}),
\end{equation}
constructed from the torsion
\begin{equation}
\tag{B4.11}
    \tilde{\mathcal{T}}_{KNM}:=\tilde{G}_{KL}\tilde{\mathcal{T}}{}^L{}_{NM}=\tilde{G}_{KL}(\tilde{\Gamma}{}^{L}{}_{NM}-\tilde{\Gamma}{}^L{}_{MN}).
\end{equation}

The relationship between the interaction $\mathcal{L}_{\tilde{\Gamma}}$ and the distortion is easily found by taking the anti-symmetrized trace of $\tilde{N}$,
\begin{equation}
\tag{B4.12}
    \begin{split}
        \tilde{I}_{M}&:=\tilde{N}_{[ML]}{}^L\\
        &=\tilde{N}_{ML}{}^L-\tilde{N}_{LM}{}^L\\
        &=(\tilde{G}{}^{RQ}\tilde{G}_{MP}-\delta^Q{}_M\delta^R{}_P)\tilde{N}{}^P{}_{QR}\\
        &=(\tilde{Q}_M-(m+1)\tilde{g}_M+\tilde{\mathcal{T}}_M)-((m+1)\tilde{g}_M-\tilde{\mathcal{T}}_M)\\
        &=\tilde{Q}_M-2(m+1)\tilde{g}_{M}+2\tilde{\mathcal{T}}_M.
    \end{split}
\end{equation}
Conveniently, this interaction may also be written as a commutator when expressed in the basis $B_{\tilde{\Gamma}}$. This can be seen by explicitly calculating
\begin{equation}
\tag{B4.13}
    \begin{split}
        [\tilde{N}_{M},\tilde{\Gamma}{}^{M}]&:=[\frac{1}{4}\tilde{N}_{\underline{AB}M}\tilde{\Gamma}{}^{\underline{A}}\tilde{\Gamma}{}^{\underline{B}},\tilde{\Gamma}{}^{M}]\\
        &=\frac{1}{4}\tilde{N}_{\underline{AB}M}\tilde{\Gamma}{}^{\underline{A}}\tilde{\Gamma}{}^{\underline{B}}\tilde{\Gamma}{}^{M}-\frac{1}{4}\tilde{N}_{\underline{AB}M}\tilde{\Gamma}{}^{M}\tilde{\Gamma}{}^{\underline{A}}\tilde{\Gamma}{}^{\underline{B}}\\
        &=\frac{1}{4}\tilde{N}_{\underline{AB}M}\tilde{\Gamma}{}^{\underline{A}}\tilde{\Gamma}{}^{\underline{B}}\tilde{\Gamma}{}^{M}+\frac{1}{4}\tilde{N}_{\underline{AB}M}\tilde{\Gamma}{}^{\underline{A}}\tilde{\Gamma}{}^{M}\tilde{\Gamma}{}^{\underline{B}}
        -\frac{1}{2}\tilde{N}_{\underline{AB}}{}^{\underline{A}}\tilde{\Gamma}{}^{\underline{B}}\\
        &=\frac{1}{2}\tilde{N}_{\underline{AB}}{}^{\underline{B}}\tilde{\Gamma}{}^{\underline{A}}
        -\frac{1}{2}\tilde{N}_{\underline{AB}}{}^{\underline{A}}\tilde{\Gamma}{}^{\underline{B}}\\
        &=\frac{1}{2}\tilde{I}_{\underline{A}}\tilde{\Gamma}{}^{\underline{A}}.
    \end{split}
\end{equation}
Using the information developed in the previous section, Appendix \hyperref[app-B3:exterior-abs-ind-form]{B.3}, along with the two relations above, we may thus express the interaction between the gauge-covariant derivative and the dual gamma matrix as
\begin{equation}
\tag{B4.14}
    \tilde{D}*\tilde{\Gamma}=\tilde{N}\wedge *\tilde{\Gamma}-*\tilde{\Gamma}\wedge\tilde{N},
\end{equation}
where
\begin{equation}
    \tag{B4.15}
    \tilde{N}:=\tilde{N}_{M}dx^M=\frac{1}{4}\tilde{N}{}^{\underline{A}}{}_{\underline{B}M}\tilde{\Gamma}_{\underline{A}}\tilde{\Gamma}{}^{\underline{B}}dx^M
\end{equation}
is the distortion $1$-form.

\subsubsection*{Axial Currents}

In this section, we display the explicit calculations associated with the $\bar{\mathfrak{p}}{}^{\mathfrak{w}}$-spinor current densities. We restrict the following calculations to $m+1=5$ dimensions. The exterior form of the anti-symmetric generator is
\begin{equation}
\tag{B4.16}
    \tilde{\Sigma}=\frac{i}{2}\tilde{\Gamma}\wedge\tilde{\Gamma}=\frac{i}{2}\tilde{\Gamma}_{\underline{A}}\tilde{\Gamma}_{\underline{B}}\tilde{E}{}^{\underline{A}}\wedge\tilde{E}{}^{\underline{B}}=\frac{1}{2}\tilde{\Sigma}_{\underline{AB}}\tilde{E}{}^{\underline{A}}\wedge\tilde{E}{}^{\underline{B}}.
\end{equation}
Having the gauge-covariant derivative of $\tilde{\Gamma}$ available, Eq. \eqref{D-Gamma-appendix}, we may easily calculate the covariant derivative of $\tilde{\Sigma}$,
\begin{equation}
\tag{B4.17}
\begin{split}
    \tilde{D}\tilde{\Sigma}&=\frac{i}{2}\tilde{D}(\tilde{\Gamma}_{\underline{A}}\tilde{\Gamma}_{\underline{B}})\wedge\tilde{E}{}^{\underline{A}}\wedge\tilde{E}{}^{\underline{B}}+\frac{i}{2}\tilde{\Gamma}_{\underline{A}}\tilde{\Gamma}_{\underline{B}}\tilde{D}(\tilde{E}{}^{\underline{A}}\wedge\tilde{E}{}^{\underline{B}})\\
    &=\frac{-i}{4}(\tilde{Q}_{\underline{AC}}\tilde{\Gamma}{}^{\underline{C}}\tilde{\Gamma}_{\underline{B}}+\tilde{Q}_{\underline{BC}}\tilde{\Gamma}_{\underline{A}}\tilde{\Gamma}{}^{\underline{C}})\wedge\tilde{E}{}^{\underline{A}}\wedge\tilde{E}{}^{\underline{B}}+\tilde{\Sigma}_{\underline{AB}}\tilde{\mathcal{T}}{}^{\underline{A}}\wedge\tilde{E}{}^{\underline{B}}\\
    &=\frac{1}{2}\tilde{Q}_{\underline{AC}}\tilde{\Sigma}{}^{\underline{C}}{}_{\underline{B}}\wedge\tilde{E}{}^{\underline{A}}\wedge\tilde{E}{}^{\underline{B}}+\tilde{\Sigma}_{\underline{AB}}\tilde{\mathcal{T}}{}^{\underline{A}}\wedge\tilde{E}{}^{\underline{B}}.
    \end{split}
\end{equation}
Taking
\begin{equation}
\tag{B4.18}
    \tilde{\mathcal{A}}:=\frac{-1}{3!}\tilde{\overline{\Psi}}\tilde{\Sigma}\wedge\tilde{\Sigma}\tilde{\Psi},
\end{equation}
we find the components of the covariant derivative of $\tilde{\mathcal{A}}$ via
\begin{equation}\label{star-D-A}
\tag{B4.19}
    \begin{split}
        *\tilde{D}\tilde{\mathcal{A}}&=\frac{-1}{3!}*\left((\tilde{D}\tilde{\overline{\Psi}})\wedge\tilde{\Sigma}\wedge\tilde{\Sigma}\tilde{\Psi}+\tilde{\overline{\Psi}}\tilde{\Sigma}\wedge\tilde{\Sigma}\wedge(\tilde{D}\tilde{\Psi})\right)\\
        &\quad+\frac{-1}{3!}*\left(\tilde{\overline{\Psi}}(\tilde{D}\tilde{\Sigma})\wedge\tilde{\Sigma}\tilde{\Psi}+\tilde{\overline{\Psi}}\tilde{\Sigma}\wedge(\tilde{D}\tilde{\Sigma})\tilde{\Psi}\right)\\
        &=\mp\frac{i}{\sqrt{\eta}_0}(-1)^q\left((\tilde{D}_{\underline{A}}\tilde{\overline{\Psi}})\tilde{\Gamma}{}^{\underline{A}}\tilde{\Psi}+\tilde{\overline{\Psi}}\tilde{\Gamma}{}^{\underline{A}}(\tilde{D}_{\underline{A}}\tilde{\Psi})\right)\\
        &\quad-\frac{1}{4!}\tilde{\overline{\Psi}}\left(\tilde{Q}_{\underline{C}}{}_{\underline{EF}}+2\tilde{\mathcal{T}}_{\underline{F}}{}_{\underline{EC}}\right)\{\tilde{\Sigma}_{\underline{AB}},\tilde{\Sigma}{}^{\underline{F}}{}_{\underline{D}}\}\hat{\tilde{\epsilon}}{}^{\underline{ABCDE}}\tilde{\Psi}.
    \end{split}
\end{equation}
To arrive at this result, we utilized 
\begin{equation}
\tag{B4.20}
    \tilde{\Gamma}_{\underline{A}}=\pm\frac{i}{\sqrt{\eta}_0}\frac{1}{4!}\hat{\tilde{\epsilon}}_{\underline{ABCDE}}\tilde{\Gamma}{}^{\underline{B}}\tilde{\Gamma}{}^{\underline{C}}\tilde{\Gamma}{}^{\underline{D}}\tilde{\Gamma}{}^{\underline{E}}.
\end{equation}
Evaluated on the space of solutions, $\tilde{\mathbb{F}}=\tilde{\overline{\mathbb{F}}}=0$, the first two terms in the result of Eq. \eqref{star-D-A} combine to form
\begin{equation}
\tag{B4.21}
    (\tilde{D}_{\underline{A}}\tilde{\overline{\Psi}})\tilde{\Gamma}{}^{\underline{A}}\tilde{\Psi}+\tilde{\overline{\Psi}}\tilde{\Gamma}{}^{\underline{A}}(\tilde{D}_{\underline{A}}\tilde{\Psi})=-\frac{1}{2}\tilde{\overline{\Psi}}\tilde{I}_{\underline{A}}\tilde{\Gamma}{}^{\underline{A}}\tilde{\Psi}.
\end{equation}
Using
\begin{equation}
\tag{B4.22}
    \left\{\tilde{\Sigma}_{\underline{AB}},\tilde{\Sigma}{}^{\underline{F}}{}_{\underline{D}}\right\}\hat{\tilde{\epsilon}}{}^{\underline{ABCDE}}=-3!\tilde{\Gamma}_{\underline{A}}\tilde{\Gamma}_{\underline{B}}\hat{\tilde{\epsilon}}{}^{\underline{ABFCE}}-2\tilde{\Gamma}{}^{\underline{F}}\tilde{\Gamma}_{\underline{D}}\tilde{\Gamma}_{\underline{A}}\tilde{\Gamma}_{\underline{B}}\hat{\tilde{\epsilon}}{}^{\underline{ABCDE}},
\end{equation}
and the fact that $\tilde{Q}_{\underline{A}}{}_{[\underline{BC}]}=0$, we find
\begin{equation}\label{star-D-A-2}
\tag{B4.23}
\begin{split}
     *\tilde{D}\tilde{\mathcal{A}}&\overset{\circ}{=}\pm\frac{i}{2\sqrt{\eta}_0}(-1)^q\tilde{\overline{\Psi}}\tilde{I}_{\underline{A}}\tilde{\Gamma}{}^{\underline{A}}\tilde{\Psi}+\frac{1}{12}\tilde{\overline{\Psi}}\tilde{Q}_{\underline{C}}{}{}_{\underline{EF}}\tilde{\Gamma}{}^{\underline{F}}\tilde{\Gamma}_{\underline{D}}\tilde{\Gamma}_{\underline{A}}\tilde{\Gamma}_{\underline{B}}\hat{\tilde{\epsilon}}{}^{\underline{ABCDE}}\tilde{\Psi}\\
    &\quad+\frac{1}{2}\tilde{\overline{\Psi}}\tilde{\mathcal{T}}_{\underline{f}}{}_{\underline{ce}}\tilde{\Gamma}_{\underline{d}}\tilde{\Gamma}{}^{\underline{f}}\tilde{\Gamma}_{\underline{a}}\tilde{\Gamma}_{\underline{*}}\hat{\tilde{\epsilon}}{}^{\underline{adce*}}\tilde{\Psi}\\
     &\overset{\circ}{=}\pm\frac{i}{2\sqrt{\eta}_0}(-1)^q\tilde{\overline{\Psi}}\tilde{I}_{\underline{A}}\tilde{\Gamma}{}^{\underline{A}}\tilde{\Psi}\pm\frac{4i}{\sqrt{\eta}_0}(-1)^q\tilde{\overline{\Psi}}\tilde{g}_{\underline{A}}\tilde{\Gamma}{}^{\underline{A}}\tilde{\Psi}-\frac{ix^*_0\eta_0}{2}(-1)^q\tilde{\overline{\Psi}}\gamma^{\underline{5}}\overline{\mathcal{P}}{}^{+}\tilde{\Psi}\\
     &\quad+\frac{1}{2}\tilde{\overline{\Psi}}\tilde{\mathcal{T}}_{\underline{f}}{}_{\underline{ce}}\tilde{\Gamma}_{\underline{d}}\tilde{\Gamma}{}^{\underline{f}}\tilde{\Gamma}_{\underline{a}}\tilde{\Gamma}_{\underline{*}}\hat{\tilde{\epsilon}}{}^{\underline{adce*}}\tilde{\Psi}\\
     &\overset{\circ}{=}\pm\frac{i}{\sqrt{\eta}_0}(-1)^q\tilde{\overline{\Psi}}\tilde{\mathcal{T}}_{\underline{a}}\tilde{\Gamma}{}^{\underline{a}}\tilde{\Psi}
     +\frac{1}{2}\tilde{\overline{\Psi}}\tilde{\mathcal{T}}_{\underline{f}}{}_{\underline{ce}}\tilde{\Gamma}_{\underline{d}}\tilde{\Gamma}{}^{\underline{f}}\tilde{\Gamma}_{\underline{a}}\tilde{\Gamma}_{\underline{*}}\hat{\tilde{\epsilon}}{}^{\underline{adce*}}\tilde{\Psi}.
     \end{split}
\end{equation}
The remaining simplification comes from substituting
\begin{equation}
\tag{B4.24}
    \tilde{\Gamma}_{\underline{*}}=\pm\frac{i}{\sqrt{\eta}_0}\frac{1}{4!}\hat{\tilde{\epsilon}}_{\underline{*bcde}}\tilde{\Gamma}{}^{\underline{b}}\tilde{\Gamma}{}^{\underline{c}}\tilde{\Gamma}{}^{\underline{d}}\tilde{\Gamma}{}^{\underline{e}}
\end{equation}
into the final term of Eq. \eqref{star-D-A-2}, and using the identity from Eq. \eqref{epsilon-epsilon-234} for
\begin{equation}
\tag{B4.25}
    \hat{\tilde{\epsilon}}{}^{\underline{aefg*}}\hat{\tilde{\epsilon}}_{\underline{abcd*}}=\hat{\epsilon}{}^{\underline{aefg}}\hat{\epsilon}_{\underline{abcd}}.
\end{equation}
The result is
\begin{equation}
\tag{B4.26}
    *\tilde{D}\tilde{\mathcal{A}}\overset{\circ}{=}\pm i(-1)^q\left(\frac{1}{\sqrt{\eta_0}}+\sqrt{\eta_0}\right)\tilde{\overline{\Psi}}\tilde{\gamma}{}^{\underline{a}}\overline{\mathcal{T}}_{\underline{a}}\tilde{\Psi},
\end{equation}
which vanishes for $\eta_0=-1$.

\phantomsection
\addcontentsline{toc}{section}{\hspace{-1.5em}Appendix C: Supplementary Material}
\begingroup
\renewcommand{\addcontentsline}[3]{}
\section*{Appendix C: Supplementary Material}
\label{app-C}
\endgroup

\phantomsection
\addcontentsline{toc}{subsection}{\hspace{-1.5em}C.1: \texorpdfstring{$\bar{\mathfrak{p}}$}{\bar{\mathfrak{p}}}-Development}
\begingroup
\renewcommand{\addcontentsline}[3]{}
\subsection*{C.1: \texorpdfstring{$\bar{\mathfrak{p}}$}{\bar{\mathfrak{p}}}-Development}
\label{app-C1:p-bar-development}
\endgroup

Following \cite{proj-conn}, we would like to consider the development of a curve in $V\mathcal{M}$ into a curve in $G/H$, where
\begin{equation}
\tag{C1.1}
    G/H:=GL^{+}(m+1,\mathbb{R})/\mathbb{R}^+\cong SL(m+1,\mathbb{R})\cong PGL(m,\mathbb{R}).
\end{equation}
However, in the general projective setting, the translational connection-form $\vartheta$ may not be identified with the co-frame. This follows from its behavior under local gauge transformations, Eq. \eqref{gauge-trans-of-PGL-conn}. We therefore consider, rather,
\begin{equation}
\tag{C1.2}
    G/H:=PGL(m,\mathbb{R})/SO(m,\mathbb{R}).
\end{equation}

Let $\dot{X}(\tau)\equiv dx/d\tau$ be a curve in $V\mathcal{M}$. This defines a curve $C(\tau)$ in $\mathfrak{g}$, the Lie algebra of $G$, by
\begin{equation}
\tag{C1.3}
    C(\tau)=\tilde{\Omega}{}^{\underline{A}}{}_{\underline{B}M}\dot{X}{}^M\bm{L}{}^{\underline{B}}{}_{\underline{A}}.
\end{equation}
Suppose $g(\tau)$ is a curve in $G$ satisfying
\begin{equation}
\tag{C1.4}
    \dot{g}(\tau)=gC(\tau).
\end{equation}
In abstract-index notation,
\begin{equation}
\tag{C1.5}
    \frac{d}{d\tau}g{}^{\underline{A}}{}_{\underline{C}}=g^{\underline{A}}{}_{\underline{B}}C^{\underline{B}}{}_{\underline{C}}.
\end{equation}
As a matrix, the components of $C(\tau)$ are simply
\begin{equation}\label{curve-appendix}
\tag{C1.6}
    [C(\tau)]^{\underline{A}}{}_{\underline{B}}=\tilde{\Omega}{}^{\underline{A}}{}_{\underline{B}M}\frac{dx^M}{d\tau}=\begin{pmatrix}
        \overline{\omega}{}^{\underline{a}}{}_{\underline{b}m}\dot{X}{}^m&\frac{1}{\lambda_0}\overline{\vartheta}{}^{\underline{a}}{}_{m}\dot{X}^m\\\lambda_0\overline{\mathcal{P}}_{\underline{b}m}\dot{X}^m&0
    \end{pmatrix}.
\end{equation}
If $\tilde{\Upsilon}(\tau)=g(\tau)\tilde{\Upsilon}_{0}$, where $\tilde{\Upsilon}_0$ denotes the stability point (origin) for which $H=SO(m,\mathbb{R})$ is the stabilizer, then $\tilde{\Upsilon}(\tau)$ is gauge invariant. We have $\tilde{\Upsilon}(\tau)$ as the development of $\dot{X}(\tau)$. 

Recall that $\tilde{\Upsilon}$ is the general projective Higgs vector. By construction, this contains the equivalence class of points, usually denoted $[\tilde{\Upsilon}]$. Therefore, no additional function is needed to account for this. In the (A)PV-gauge, there is still a residual scalar density function, the projective factor $\bar{\mathfrak{p}}$, which will contribute a shift to the surface at infinity,
\begin{equation}
\tag{C1.7}
    \tilde{\Upsilon}{}^{\underline{A}}_0(\tau)=\bar{\mathfrak{p}}\tilde{\upsilon}{}^{\underline{A}}_0=\bar{\mathfrak{p}}\begin{pmatrix}
        0^{\underline{a}}\\x^*_0
    \end{pmatrix}
\end{equation}

For $\tilde{\Upsilon}(\tau)$ to be geodesic in $G/H$, it must develop into straight lines in the Klein geometry. Suppose $\tau$ is an affine parameter with respect to $\tilde{\Omega}$. Then, the straight lines are those curves which satisfy
\begin{equation}
\tag{C1.8}
    \ddot{\tilde{\Upsilon}}(\tau)=0.
\end{equation}
For sake of symmetry in calculation, let $C(\tau)\equiv\dot{\tilde{\Omega}}(\tau)$. The acceleration is found as
\begin{equation}
\tag{C1.9}
    \begin{split}
        \ddot{\tilde{\Upsilon}}(\tau)&=\frac{d^2}{d\tau^2}(g\tilde{\Upsilon}_0)\\
        &=\frac{d^2}{d\tau^2}(g\bar{\mathfrak{p}})\tilde{\upsilon}_0\\
        &=(\ddot{g}\bar{\mathfrak{p}}+2\dot{g}\dot{\bar{\mathfrak{p}}}+g\ddot{\bar{\mathfrak{p}}})\tilde{\upsilon}_0\\
        &=\left(\frac{d}{d\tau}(gC)\bar{\mathfrak{p}}+2gC\dot{\bar{\mathfrak{p}}}+g\ddot{\bar{\mathfrak{p}}}\right)\tilde{\upsilon}_0\\
&=\left(\dot{g}C\bar{\mathfrak{p}}+g\dot{C}\bar{\mathfrak{p}}+2gC\dot{\bar{\mathfrak{p}}}+g\ddot{\bar{\mathfrak{p}}}\right)\tilde{\upsilon}_0\\
&=\left(gC^2\bar{\mathfrak{p}}+g\dot{C}\bar{\mathfrak{p}}+2gC\dot{\bar{\mathfrak{p}}}+g\ddot{\bar{\mathfrak{p}}}\right)\tilde{\upsilon}_0\\
        &=g\bar{\mathfrak{p}}\left(\dot{\tilde{\Omega}}{}^2+\ddot{\tilde{\Omega}}+2\dot{\tilde{\Omega}}\dot{\tilde{g}}+\ddot{\tilde{g}}+\dot{\tilde{g}}{}^2\right)\tilde{\upsilon}_0\\
&=g\bar{\mathfrak{p}}\left((\dot{\tilde{\Omega}}+\dot{\tilde{g}})^2+\ddot{\tilde{\Omega}}+\ddot{\tilde{g}}\right)\tilde{\upsilon}_0,
    \end{split}
\end{equation}
where
\begin{equation}\label{lil-g-appendix}
\tag{C1.10}
    \dot{\tilde{g}}:=\frac{d}{d\tau}\left(\log\bar{\mathfrak{p}}\right)=\tilde{g}_M\dot{X}{}^M=\begin{pmatrix}g_m\dot{X}^m,&\frac{1}{x^*}\dot{X}^{*}\end{pmatrix}.
\end{equation}

From the definitions in Eqs. \eqref{curve-appendix} and \eqref{lil-g-appendix}, we find the explicit component matrix for $(\dot{\tilde{\Omega}}+\dot{\tilde{g}})^2+\ddot{\tilde{\Omega}}+\ddot{\tilde{g}}$, given by
\begin{equation}
\tag{C1.11}
    \begin{pmatrix}
            \dot{\overline{\omega}}{}^{\underline{a}}{}_{\underline{b}}\dot{\overline{\omega}}{}^{\underline{b}}{}_{\underline{c}}+\dot{\overline{\vartheta}}{}^{\underline{a}}\dot{\overline{\mathcal{P}}}_{\underline{c}}+\dot{\tilde{g}}\dot{\overline{\omega}}{}^{\underline{a}}{}_{\underline{c}}+\dot{\overline{\omega}}{}^{\underline{a}}{}_{\underline{c}}\dot{\tilde{g}}+\delta^{\underline{a}}{}_{\underline{c}}\dot{\tilde{g}}{}^{2}+  \ddot{\overline{\omega}}{}^{\underline{a}}{}_{\underline{c}}+\ddot{\tilde{g}}&\frac{1}{x^*_0}\left(\dot{\overline{\omega}}{}^{\underline{a}}{}_{\underline{b}}\dot{\overline{\vartheta}}{}^{\underline{b}}+\dot{\tilde{g}}\dot{\overline{\vartheta}}{}^{\underline{a}}+\dot{\overline{\vartheta}}{}^{\underline{a}}\dot{\tilde{g}}+\ddot{\overline{\vartheta}}{}^{\underline{a}}\right)\\x^*_0\left(\dot{\overline{\mathcal{P}}}_{\underline{b}}\dot{\overline{\omega}}{}^{\underline{b}}{}_{\underline{c}}+\dot{\tilde{g}}\dot{\overline{\mathcal{P}}}_{\underline{c}}+\dot{\overline{\mathcal{P}}}_{\underline{c}}\dot{\tilde{g}}+\ddot{\overline{\mathcal{P}}}_{\underline{c}}\right)&\dot{\overline{\mathcal{P}}}_{\underline{b}}\dot{\overline{\vartheta}}{}^{\underline{b}}+\dot{\tilde{g}}{}^{2}+\ddot{\tilde{g}}
        \end{pmatrix}.
\end{equation}
Imposing the affine condition $\ddot{\tilde{\Upsilon}}(\tau)=0$, left-multiplying by $g^{-1}\in G$, and dividing by $\bar{\mathfrak{p}}\neq0$, we find
\begin{equation}\label{affine-geo-upsilon}
\tag{C1.12}
0=\left((\dot{\tilde{\Omega}}+\dot{\tilde{g}})^2+\ddot{\tilde{\Omega}}+\ddot{\tilde{g}}\right)\tilde{\upsilon}_0
        =\begin{pmatrix}
            \dot{\overline{\omega}}{}^{\underline{a}}{}_{\underline{b}}\dot{\overline{\vartheta}}{}^{\underline{b}}+\dot{\tilde{g}}\dot{\overline{\vartheta}}{}^{\underline{a}}+\dot{\overline{\vartheta}}{}^{\underline{a}}\dot{\tilde{g}}+\ddot{\overline{\vartheta}}{}^{\underline{a}}\\x^*_0\left(\dot{\overline{\mathcal{P}}}_{\underline{b}}\dot{\overline{\vartheta}}{}^{\underline{b}}+\dot{\tilde{g}}{}^{2}+\ddot{\tilde{g}}\right)
        \end{pmatrix}.
\end{equation}
In the above, we utilize the obvious notation:
\begin{equation}
\tag{C1.13}
\begin{split}
    \dot{\overline{\vartheta}}{}^{\underline{a}}&:=\overline{\vartheta}{}^{\underline{a}}{}_m\dot{X}^m=\overline{\vartheta}{}^{\underline{a}}{}_m\frac{dx^m}{d\tau},\\
\dot{\overline{\mathcal{P}}}_{\underline{b}}&:=\overline{\mathcal{P}}_{\underline{b}m}\dot{X^m}=\overline{\mathcal{P}}_{\underline{b}m}\frac{dx^m}{d\tau},\\
\dot{\overline{\omega}}{}^{\underline{a}}{}_{\underline{b}}&:=\overline{\omega}{}^{\underline{a}}{}_{\underline{b}m}\dot{X}^m=\overline{\omega}{}^{\underline{a}}{}_{\underline{b}m}\frac{dx^m}{d\tau}.
    \end{split}
\end{equation}
As for the second derivative, we consider
\begin{equation}
\tag{C1.14}
    \ddot{\overline{\vartheta}}{}^{\underline{a}}=\dot{\overline{\vartheta}}{}^{\underline{a}}{}_m\dot{X}^m+\overline{\vartheta}{}^{\underline{a}}{}_m\ddot{X}^m=\overline{\vartheta}{}^{\underline{a}}{}_m\frac{d^2x^m}{d\tau^2}.
\end{equation}
The first set of $m$ equations in the result of Eq. \eqref{affine-geo-upsilon} are
\begin{equation}\label{geo-dev}
\tag{C1.15}
    0=\overline{\vartheta}{}^{\underline{a}}{}_m\ddot{X}^m+\overline{\omega}{}^{\underline{a}}{}_{\underline{b}n}\overline{\vartheta}{}^{\underline{b}}{}_m\dot{X}^n\dot{X}^m+2(\tilde{g}_M\dot{X}^M)\overline{\vartheta}{}^{\underline{a}}{}_n\dot{X}^n,
\end{equation}
while the remaining equation is 
\begin{equation}\label{pbar-f-eq}
\tag{C1.16}
    0=\overline{\mathcal{P}}_{mn}\dot{X}^m\dot{X}^n+\frac{d}{d\tau}(\tilde{g}_M\dot{X}^M)+(\tilde{g}_M\dot{X}^M)^2.
\end{equation}

The above geodesic equations may be related to those containing $\Pi$ and $\mathcal{D}$, by simply removing the factors of $\overline{\vartheta}$, and noting that we have effectively been working in the PV-gauge. For the first $m$ equations,
\begin{equation}
\tag{C1.17}
    \frac{d^2x^r}{d\tau^2}+\left((\overline{\vartheta}{}^{-1})^r{}_{\underline{a}}\overline{\omega}{}^{\underline{a}}{}_{\underline{b}n}\overline{\vartheta}{}^{\underline{b}}{}_m+g_n\delta^r{}_m+g_m\delta^r{}_n\right)\frac{dx^n}{d\tau}\frac{dx^m}{d\tau}=\frac{-2}{x^*}\frac{dx^*}{d\tau}\frac{dx^r}{d\tau}.
\end{equation}
Since we are assuming $\dot{\overline{\vartheta}}{}^{\underline{a}}{}_{m}=0$, the terms in parenthesis combine to form
\begin{equation}\label{appendix-pi}
\tag{C1.18}
    \Pi^r{}_{mn}\frac{dx^n}{d\tau}\frac{dx^m}{d\tau}=\left((\overline{\vartheta}{}^{-1})^r{}_{\underline{a}}\overline{\omega}{}^{\underline{a}}{}_{\underline{b}n}\overline{\vartheta}{}^{\underline{b}}{}_m+g_n\delta^r{}_m+g_m\delta^r{}_n\right)\frac{dx^n}{d\tau}\frac{dx^m}{d\tau}.
\end{equation}
We thus arrive at the final result,
\begin{equation}
\tag{C1.19}
    \frac{d^2x^r}{d\tau^2}+\Pi^r{}_{mn}\frac{dx^n}{d\tau}\frac{dx^m}{d\tau}=\frac{-2}{x^*}\frac{dx^*}{d\tau}\frac{dx^r}{d\tau}.
\end{equation}

For the remaining equation, Eq. \eqref{pbar-f-eq}, we first point out an interesting interpretation. Define the dimensionless $V\mathcal{M}$-tensor $\tilde{H}_{MN}$ as the metric-like object
\begin{equation}
\tag{C1.20}
    \tilde{H}_{MN}:=\begin{pmatrix}
        -(x^*_0)^2\overline{\mathcal{P}}_{mn}-(x^*_0)^2g_mg_n&\frac{-(x^*_0)^2}{x^*}g_m\\\frac{-(x^*_0)^2}{x^*}g_n&\left(\frac{x^*_0}{x^*}\right)^2
    \end{pmatrix}.
\end{equation}
We then form the infinitesimal length element with respect to $\tilde{H}$ as
\begin{equation}
\tag{C1.21}
    d\tilde{s}{}^2=\tilde{H}_{MN}dx^Mdx^N.
\end{equation}
Expanding the right-hand side, we find
\begin{equation}
\tag{C1.22}
    d\tilde{s}{}^2=-(x^*_0)^2\left(\overline{\mathcal{P}}_{mn}dx^mdx^n+(\tilde{g}_Mdx^M)^2\right).
\end{equation}
Parameterizing $d\tilde{s}{}^2$ with $\tau$, this is simply
\begin{equation}
\tag{C1.23}
    \left(\frac{d\tilde{s}}{d\tau}\right)^2=-(x^*_0)^2\left(\overline{\mathcal{P}}_{mn}\dot{X}^m\dot{X}^n+(\tilde{g}_M\dot{X}^M)^2\right).
\end{equation}
We may therefore view the scalar geodesic equation, Eq. \eqref{pbar-f-eq}, as expressing the parameterized displacement of the surface in $V\mathcal{M}$, defined by $\tilde{g}_M\dot{X}^M$, since
\begin{equation}
\tag{C1.24}
    \frac{d}{d\tau}(\tilde{g}_M\dot{X}^M)=\frac{1}{(x^*_0)^2}\left(\frac{d\tilde{s}}{d\tau}\right)^2=\frac{1}{(x^*_0)^2}\tilde{H}_{MN}\dot{X}^M\dot{X}^N.
\end{equation}
The additional factors of $x^*_0$ may be viewed as necessary to enact the unit-sensitive transition between coordinate types, Eq. \eqref{x-X-map}.

Lastly, to relate Eq. \eqref{pbar-f-eq} to the standard $\mathcal{D}$-containing form, we simply note that
\begin{equation}
\tag{C1.25}
\begin{split}
    0&=\overline{\mathcal{P}}_{mn}\dot{X}^m\dot{X}^n+\frac{d}{d\tau}(\tilde{g}_M\dot{X}^M)+(\tilde{g}_M\dot{X}^M)^2\\
&=\overline{\mathcal{P}}_{mn}\dot{X}^m\dot{X}^n+\left(\dot{X}^m\dot{X}^n\partial_mg_n+g_m\ddot{X}^m+\frac{1}{x^*}\ddot{X}^*-\left(\frac{1}{x^*}\dot{X}^*\right)^2\right)\\
&\quad+\left(g_mg_n\dot{X}^m\dot{X}^n+\frac{2}{x^*}g_m\dot{X}^*\dot{X}^m+\left(\frac{1}{x^*}\dot{X}^*\right)^2\right)\\
&=\frac{1}{x^*}\ddot{X}^*+\left(\overline{\mathcal{P}}_{mn}+\partial_mg_n+g_mg_n\right)\dot{X}^m\dot{X}^n+\left(g_m\ddot{X}^m+\frac{2}{x^*}g_m\dot{X}^*\dot{X}^m\right).
    \end{split}
\end{equation}
Then, using Eq. \eqref{geo-dev} to substitute for $\ddot{X}^m$ in the last set of parenthesis, we find
\begin{equation}
\tag{C1.26}
        0=\frac{1}{x^*}\ddot{X}^*+\left(\overline{\mathcal{P}}_{mn}+\partial_mg_n-g_l\Pi^l{}_{mn}+g_mg_n\right)\dot{X}^m\dot{X}^n,
\end{equation}
where Eq. \eqref{appendix-pi} was also used. Noting that
\begin{equation}
\tag{C1.27}
    \mathcal{D}_{mn}=\overline{\mathcal{P}}_{mn}+\partial_mg_n-g_l\Pi^l{}_{mn}+g_mg_n,
\end{equation}
we arrive at the final result,
\begin{equation}
\tag{C1.28}
    \frac{d^2x^*}{d\tau^2}+x^*\mathcal{D}_{mn}\frac{dx^m}{d\tau}\frac{dx^n}{d\tau}=0.
\end{equation}


\phantomsection
\addcontentsline{toc}{section}{\hspace{-1.5em}References}

\begin{spacing}{1}
\begingroup
\renewcommand{\addcontentsline}[3]{}
\bibliographystyle{plain}
\bibliography{refs.bib}
\endgroup
\end{spacing}

\end{onehalfspacing}

\end{document}